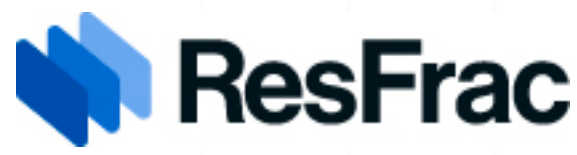

# The A to Z Guide to Accelerating Continuous Improvement with ResFrac

9th Version

Mark McClure, Garrett Fowler, Chris Hewson, Charles Kang,
Chris Ponners, and Aileen Zebrowski

ResFrac Corporation

info@resfrac.com

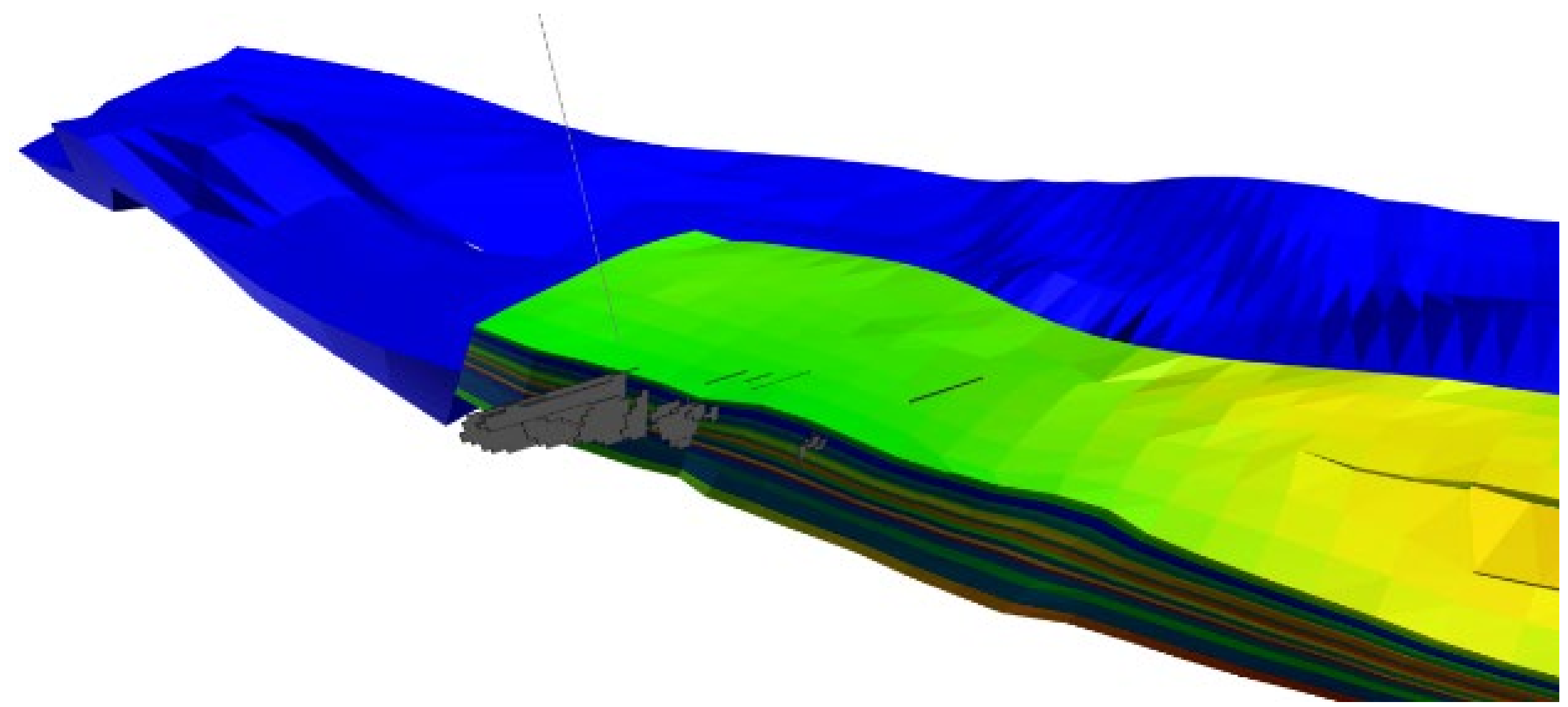

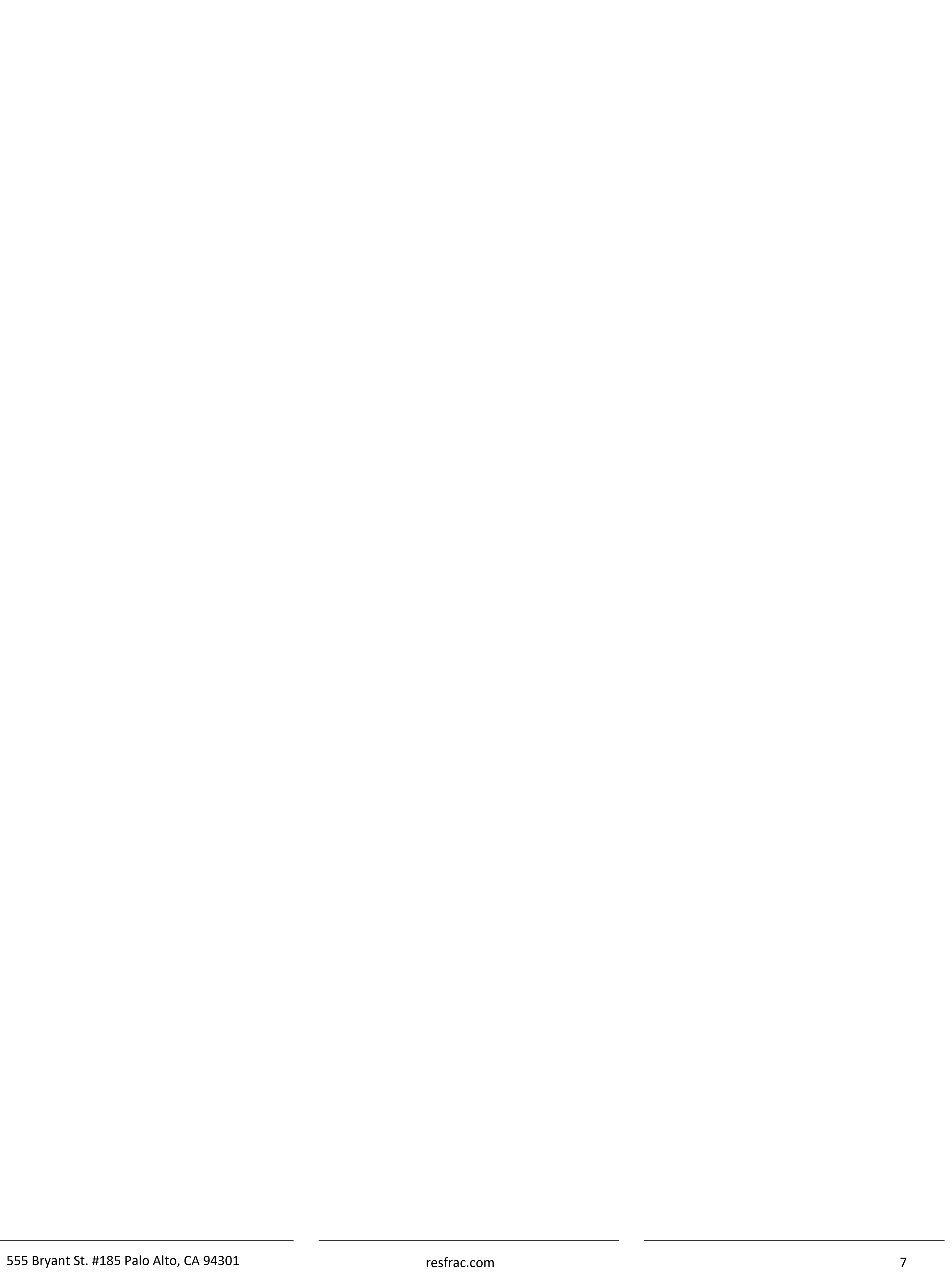

# 1. Introduction

## 1.1 The purpose of this document

We built ResFrac to help operators make optimal decisions regarding hydraulic fracture design, well placement, and reservoir engineering. Information is valuable when it: (a) impacts a decision, and (b) that decision impacts your objectives. For example, an operator's objective may be to maximize net present value (NPV), internal rate of return (IRR), or investment efficiency (NPV divided by spending). Every ResFrac project should begin by planning a workflow that will help your company make better-decisions and improve its performance measured against its objectives. Focus on "your company's" objectives, not "your" objectives; align your personal objectives with your company's objectives!

Economic optimization is part of operators' process of continuous improvement (McClure, 2021). Operators evolve their completion and well designs over time. Over years, this has led to dramatic improvement in production and lower cost. This evolution has largely been accomplished with trialing changes, measuring performance, and iterating. ResFrac allows you to trial a design digitally before testing it in the field. It can help optimize the design, inspire good ideas, and rule out bad ones.

This document is designed to give a new ResFrac user the tools that they need to succeed. It describes what ResFrac does and how to use it. We cover much more than just 'button pressing.' Using ResFrac successfully requires high-level engineering skills: how to design and execute a modeling workflow, how to think critically about problems, and how to communicate the results.

Our company philosophy is to constantly stive for improvement. As part of that, we record and update 'lessons learned' along the way. This document relates nuggets of hard-earned wisdom.

The ResFrac simulator embodies a set of physical relations and a conceptual model for what is happening in the subsurface and why. This approach is a synthesis of the engineering and scientific knowledge available in the literature, as well as our experience applying the tool to practical problems and field datasets. Because of the limited information inherent to subsurface engineering, there is not always universal consensus on every issue. In the ResFrac Technical Writeup (McClure et al., 2021) and our recent "Frequently Asked Questions" paper (McClure et al., 2020), we explain the key physics embodied in ResFrac and explain why certain choices have been made. ResFrac offers a wide variety of options and physics, depending on user preference and geologic setting. But ultimately, we do our best to lay out for users how we believe the tool should be best applied.

ResFrac's unique capability is that it fully integrates a 'true' hydraulic fracturing simulator within a multiphase reservoir simulator (as well as including a wellbore simulator). In many applications, such as in shale and Enhanced Geothermal Systems (EGS), there is a tight relationship between fracture processes (propagation, stress shadowing, proppant transport, and flow) and reservoir processes (depletion, multiphase flow, thermal drawdown). It is extremely advantageous to describe everything in a single integrated package, rather than breaking the problem off into pieces solved by separate simulators. In legacy workflows, transfers of information between different simulators involve inconvenience and loss of information (for example, they typically use different meshes). But more important, many real-life processes involve all the key physics happening at the same time. They cannot be modeled by separate software packages that break off different pieces of the physics. For example, frac hits involve fracture mechanics (stress shadowing from fracture reopening) simultaneous with multiphase flow (water displacing oil/gas in the parent frac), proppant remobilization, and cross-flow through the wellbore. This process cannot be modeled with either a pure fracturing simulator, nor a pure reservoir simulator. Also, ResFrac is a ‘true’ hydraulic fracturing simulator. We mesh cracks as cracks and solve transport and deformation equations designed for cracks. Sometimes, codes perform continuum based geomechanics simulation and call it a hydraulic fracturing simulator; that is *not* a hydraulic fracturing simulator! Hydraulic fractures are not bands of high permeability rock.

Since it was first commercialized in 2018, ResFrac has been applied in projects with over 30 companies – including most of the top 10 oil producers in the US. ResFrac has been applied in the shale plays: Midland Basin, Delaware Basin, Duvernay, Eagle Ford, Bakken, SCOOP/STACK, Montney, Marcellus, Utica, Powder River, Haynesville, and Vaca Muerta; for optimization of EGS resources; and in projects for hydraulic fracturing in conventional reservoirs.

Even if you are not a ResFrac user, you may still find this document to be a valuable resource. It covers material that is broadly applicable for anyone working in subsurface engineering: (a) structuring and executing a modeling study and (b) the key physics and issues around hydraulic fracture design and optimization in shale (and EGS). Everything is focused on our overriding goal: to make optimal, practical design decisions that impact economic performance.

## 1.2 List of resources

Here are the resources available to you as a ResFrac user:

1. This document – what you need to know to be a successful ResFrac user
2. The ResFrac fundamentals online training videos (https://www.resfrac.com/onlinetraining)
3. Help content built into the user-interface – detailed documentation of every single input parameter (hover over or click on the “?” next to every parameter in the simulation builder)
4. The ResFrac technical writeup – a detailed description of the equations solved by ResFrac, along with citations to the literature (https://arxiv.org/abs/1804.02092)
5. The ResFrac FAQ – Q&A structured discussion with answers to common questions (https://drive.google.com/file/d/18NQvHD-dZESBiVWUf_0fxShalaTBunYl/view?usp=sharing)
6. ResChat – an interactive ResFrac assistant trained on all of our documentation to help you answer questions and find information (https://www.resfrac.com/reschat)
7. Built-in templates – when you create a new simulation or workflow in ResFrac, there are options to start from prebuilt templates. These give starting models for different basins, and also for different application areas. Each template has a PPT deck with supporting explanation. These PPT decks can be found with your ResFrac installation in the folder “<ResFrac folder>/resources/documentation/Workflow templates”, or by selecting the gear icon in the job manager, and clicking the ‘Documentation’ button.
8. Other ResFrac videos (https://www.resfrac.com/videos)
    a. ‘ResFrac Users Guide’ – a one-hour intro video
    b. ‘The Case for Planar Fracture Modeling’ – the technical basis for why we use planar fracture modeling, rather than complex fracture network modeling
    c. ‘Best Practices in Interpretation of DFIT Tests for Shmin, Permeability, and Pore Pressure’ – our recommended procedure for interpreting DFIT tests

d. ‘Office Hour Highlights’ and ‘Full Office Hour Recordings’ – recordings of our regularly scheduled ResFrac ‘office hours’ sessions with users
9. The ResFrac blog – short articles on a variety of topics relevant to users (https://www.resfrac.com/blog)
10. The ResFrac paper library – ResFrac-related technical papers coauthored by ResFrac personnel (https://www.resfrac.com/papers).

Aside from our ResFrac content, there are several reference books that we recommend for learning the fundamentals. They are: *Hydraulic Fracturing* by Smith and Montgomery, *Reservoir Stimulation* edited by Economides and Nolte, *Reservoir Geomechanics* by Zoback, and *Fundamentals of Rock Mechanics* by Jaeger, Cooke, and Zimmerman.

# 2. Onboarding Process for New ResFrac Users

If you are looking for a quick overview, refer to Section 2.1. If you are planning to be a serious ResFrac user, refer to Section 2.2. Please invest the time to review these materials. It will pay off!

## 2.1 Time-efficient overview

For a time-efficient, general overview of ResFrac, refer to Sections 1.1 (general overview), 3 (applications), 4 (the workflow), 5 (the user interface), and 7.1 (overview of technical capabilities). This procedure is intended for someone who wants to be more familiar with ResFrac, but plans to be only a casual user.

## 2.2 Onboarding process for a ResFrac user

If you are planning to use ResFrac for a modeling study, it is strongly recommended that you follow the onboarding procedure outlined in this section. We want to equip you with all the tools needed to be successful. If you are already an experienced ResFrac user, you will still find it useful to go through this material.

We recommend a four-week process. You can go faster or slower, depending on your schedule. As you go through the material, keep a list of questions or topics for discussion. Email us at support@resfrac.com, and we will schedule weekly calls. On the calls, ask the questions that you wrote down. Optionally, you could prepare short PowerPoint presentations to review your results. Check out Sections 6.5 and 10 for guidelines on making presentations.

**Week 1**

Read Sections 1-7 of this document. While reviewing Section 5, open the ResFrac user interface and follow along. Experiment and run a few different simulations and visualize the results. Watch Modules 2-4 of the ResFrac Simulation Training at <https://www.resfrac.com/resfrac-fundamentals-simulation-training>. Optionally, also watch Module 5.

**Week 2**

Go through the detailed worked tutorial on setting up a simulation in Section 11 of this document. Read the material in Section 8 and begin the history matching process. If you prefer to watch a video, optionally watch Modules 6-10 from the recorded ResFrac Simulation Training at < https://www.resfrac.com/resfrac-fundamentals-simulation-training>.

**Week 3**

Complete the history match. Read Section 9 and plan the design optimization sensitivities.

**Week 4**

Perform the design optimization sensitivities and analyze the results.

**Week 5+**

If you didn't finish the full tutorial in 5 weeks, don't give up! Keep going, and ask for us help as needed.

# 3. Applications of ResFrac

ResFrac is designed to help companies make critical decisions regarding hydraulic fracture design and field development. These decisions drive the overall economic performance of the play. Literally, billions of dollars are at stake.

Always start a modeling study by asking what question you want to answer and develop a game plan. The modeling study will be valuable if it impacts a decision, and that decision is important to your objectives (typically, economic performance).

## 3.1 Design optimization

### *3.1.1 Physics-based and data-driven approaches*

Why should you use ResFrac? Why should you use a simulator at all?

The following passage (quoted from McClure et al., 2020) discusses the different ways that operators make decisions:

> "In shale, the main parameters for hydraulic fracture design and optimization are perforation cluster spacing, well spacing (vertical and horizontal), perforation cluster design, proppant mass, fluid volume, injection rate, fluid type, and well/cluster sequencing. When considering parent/child wells, additional considerations include preloading, protection fracs, and modification of frac design and well spacing to account for prior depletion. These design decisions interact with complex, tightly coupled physical processes, such as fracture propagation, proppant placement, wellbore dynamics, geomechanical stress changes, and multiphase fluid flow. These processes are strongly impacted by the petrophysical properties of the formation. Design optimization integrates all of these factors with economic variables related to cost and revenue (Kaufman et al., 2019; Fowler et al., 2019).
>
> In practice, a wide range of strategies exist to optimize design. 'Physics-based' and 'data-driven' concepts are used to predict outcomes and make decisions. These approaches are not mutually exclusive, and nearly all companies incorporate elements of both.
>
> Data-driven approaches do not attempt to understand the underlying causal processes that drive behavior. Instead, they review prior data and experience, and identify correlations. A 'data-driven' approach could be as simple as relying on the experience of an individual engineer, or as complex as a large-scale machine learning project. A 'physics-based' approach tries to understand why things happen, and use that insight to predict response to change. Physics-based approaches typically use fundamental physical laws (such as conservation of mass), and constitutive equations drawn from the scientific/engineering literature. Physics-based approaches can involve varying levels of complexity: a mass balance calculation in a spreadsheet, the physical intuition of an individual engineer, or a multiphysics numerical simulator. Physics-based approaches are calibrated to ensure consistency with actual data (for example, history matching), but relative to data-driven models, typically involve less emphasis on the precision of the 'match' to data. There is a tension between 'matching data' and 'model predictivity.' A simple, physically unrealistic model is likely to be flexible and therefore easy to match to data, but this does not necessarily mean that the model will be predictive when applied to problems where the answer is not already known.

The relative value of data-driven and physics-based models depends on the availability of data and the degree of understanding of the physics (Starfield and Cundall, 1988). For optimizing fracture design in shale, data-driven models are useful because the physics are complicated and not fully understood. Data-driven models go directly to the inputs and outputs that matter, integrate information, and provide practical recommendations. However, data-driven models cannot extrapolate outside the training dataset and predict response to actions that have not previously been tested. Operators rarely, if ever, have the time and money to perform a systematic program of varying the many relevant fracture design parameters, and also, they need to perform enough trials to overcome random variance. Further, trials may be confounded by geologic variability, uncontrolled changes in operating practice, and parent/child interactions. Thus, training datasets are incomplete, do not sample the full range of possible designs, and usually have correlated inputs. Because each well is expensive, improvement over time requires careful, targeted testing of new concepts. Prior data and physical insight both play a role. Physics-based models can predict response to new approaches, for which prior data is not available. These insights guide future development strategy, inspire new ideas, and help discard ideas that are less promising. This allows operators to shorten the learning curve as they test new approaches and iterate (using field trials) to continuously improve economic performance. Physics-based approaches are also used to help interpret diagnostic data and guide future data collection."

Physics-based and data-driven approaches are complementary. Data-driven approaches look at past experience and observe what worked best. Physics-based approaches help identify opportunities for future improvement – evaluating ideas that have not yet been tried previously. In shale development, physics based approaches can usually address more detailed engineering questions than data-driven approaches, which are better at evaluating overall trends. There is usually insufficient data to use data-driven approaches to answer granular shale fracture design questions.

Physics-based approaches improve results from data-driven approaches by identifying the types of relationships to seek in datasets, identifying potential confounding covariates in data, and 'gut-checking' conclusions drawn from data-driven models. Similarly, data-driven approaches improve results from physics-based approaches by providing 'prior knowledge' to constrain calibration and history matching.

We like to say that our goal is to 'shorten the learning curve.' Engineering analysis should be applied in an iterative fashion. Perform an analysis to help make decisions -> test new designs in the field and gather new data -> feed that new information back into the engineering analysis and iterate. Frac designs keep getting better (Baihly et al., 2015). Frac designs from 2010 were much less effective than frac designs in 2015. Frac designs from 2015 were much less effective than they are today. Why has it taken the industry so long to iterate towards the current designs? Data-driven look-backs on past production will result in repeating designs that have already been tried, and they may miss granular engineering details that are critical to success. A physics based model asks 'why?' It can arrive at unexpected results and lead you to genuine innovation.

Depending on your company's risk preference, you have more or less freedom to try new things. If your company plans to drill 100 wells next year, the *value of information* from discovering an improved design is high. Testing an alternative design on the next pad could lead to an improvement that can subsequently be rolled out on all future wells. It is a calculated 'risk' with a big potential payout. Our goal with ResFrac is to maximize the success rate of these calculated risks. Rule out bad ideas, inspire new ones, and select which ideas to pursue further.

Also, ResFrac can help you quantitatively optimize as you progress through field development. Are you gradually moving out to lower quality rock with lower permeability or thinner pay? Has the price of oil changed? Should you be tweaking your design to account for these changes?

### 3.1.2 The bread and butter – frac design and well placement in shale

The #1 most common application of ResFrac is to optimize frac design and well placement in shale. We optimize: cluster spacing, well spacing (vertical and horizontal), well landing depth, proppant mass, fluid volume, and perforation design (limited entry). McClure et al. (2020) discusses some of the basic considerations in these optimizations.

Tighter cluster spacing reduces fracture spacing, which helps maximize recovery. However, tighter cluster spacing (holding proppant fluid per lateral ft constant) reduces fracture length. It also builds up more stress shadow, resulting in more height growth and more irregularly shaped fractures. Increasing perforation pressure drop (limited-entry) helps maintain good cluster efficiency as you tighten cluster spacing, but increases injection pressure and horsepower requirements. If clusters get too close together, then flow behind casing (between the clusters or across the plug back to the previous stage) can start to become significant (Cramer et al., 2020), and no amount of limited entry can overcome this effect.

Tightening well spacing increases the total production per acre, but reduces the production per well. Because most plays have multiple benches, well spacing decisions relate to vertical – as well as lateral – placement. A key challenge for well spacing is to characterize propped fracture length. Proppant does not reach the actual crack tip, and typically in shale is far behind the actual crack tip. Interference tests can help characterize the true propped length (Cipolla et al., 2020; Fowler et al., 2020a; Shahri et al., 2021).

An accurate permeability estimate is critical for optimizing cluster and well spacing. Fowler et al. (2019) give an optimization example from the Utica shale. Optimizing cluster and well spacing leads to significant increase in NPV, but only if an accurate estimate is used for permeability. Permeability and fracture length can be constrained from a combination of RTA, DFIT, and interference tests, among others.

With proppant and fluid volume, pumping more results in more production, but more cost. Typically, there is a point of diminishing return, and so there is an optimum amount to pump to maximize NPV. Depending on the proppant transport mechanisms and localized screenout, fluid volume or proppant volume might be the most important driver.

These parameters are all connected. The 'global' optimum requires optimization of everything simultaneously. However, because of practical constraints, we often will optimize one or a few things, while holding the rest constant. For example, hold the basic frac design constant, but optimize limited-entry and cluster spacing.

### 3.1.3 Parent/child issues

Child wells tend to underperform relative to parent wells, and in many formations, frac hits cause substantial loss of production for the parent wells (Miller et al., 2016). A variety of mitigation strategies are used in the industry: modifying well spacing or job design, preloading or refracturing the parent well, 'cube development' schemes designed to avoid fracturing near parent wells, chemical treatment, and far-field diverter. These approaches must all be evaluated on the basis of their cost and benefit. The optimal approach depends on context, such as the geologic setting and the age and frac design of the parent well. Once an approach is selected, the approach must be quantitatively optimized. For example, how large should the preload be? What chemical formulation? When should the preload be pumped?

ResFrac captures the key physics of these processes: multiphase flow as injection fluid reinflates hydrocarbon-filled producing fractures, fracture stress shadowing, and fracture asymmetry due to poroelastic stress changes (Fowler et al., 2020b). Also, ResFrac includes a variety of 'fracture damage mechanisms' to describe processes such as the formation of 'gummy bear' gunk (Rassenfoss, 2020; the section 'Fracture Damage Mechanisms from McClure et al., 2021).

These fracture damage mechanisms affect productivity of wells in various ways. Simply as a function of repressurization and increased fracture connections, parent well productivity can increase from a frac-hit. Depending on whether the child well is within the depletion halo and/or loses frac energy to the parent well, it may be more or less effected by the offset depletion. In basins and formations where parent and child well underperformance cannot be explained solely by pressure interference, so an additional process or processes are occurring. There are several damage mechanisms built into ResFrac: conductivity damage, skin damage, relative permeability reduction, and water block damage.

Conductivity damage: this is modeled as a conductivity restriction in the proppant pack as a function of a reaction rate constant, mass of reactant (specified by user, and usually something like HVFR), and optionally, the hydrocarbon saturation.

Relative permeability damage: similar to the conductivity damage, this process occurs within the fractures but selectively restricts the relative permeability of the hydrocarbon phases (versus conductivity damage which would impact all phases equally).

Skin damage: this is modeled as a skin at the interface between the rock and the fracture and is formed as a function of the volume of fluid leaked off and skin permeability multiplier.

Water block: is similar to skin damage and restricts flow between the matrix and the fracture; however, water block only restricts the hydrocarbon phase.

The image below compares the reciprocal productivity index for the four damage mechanisms and the base case.

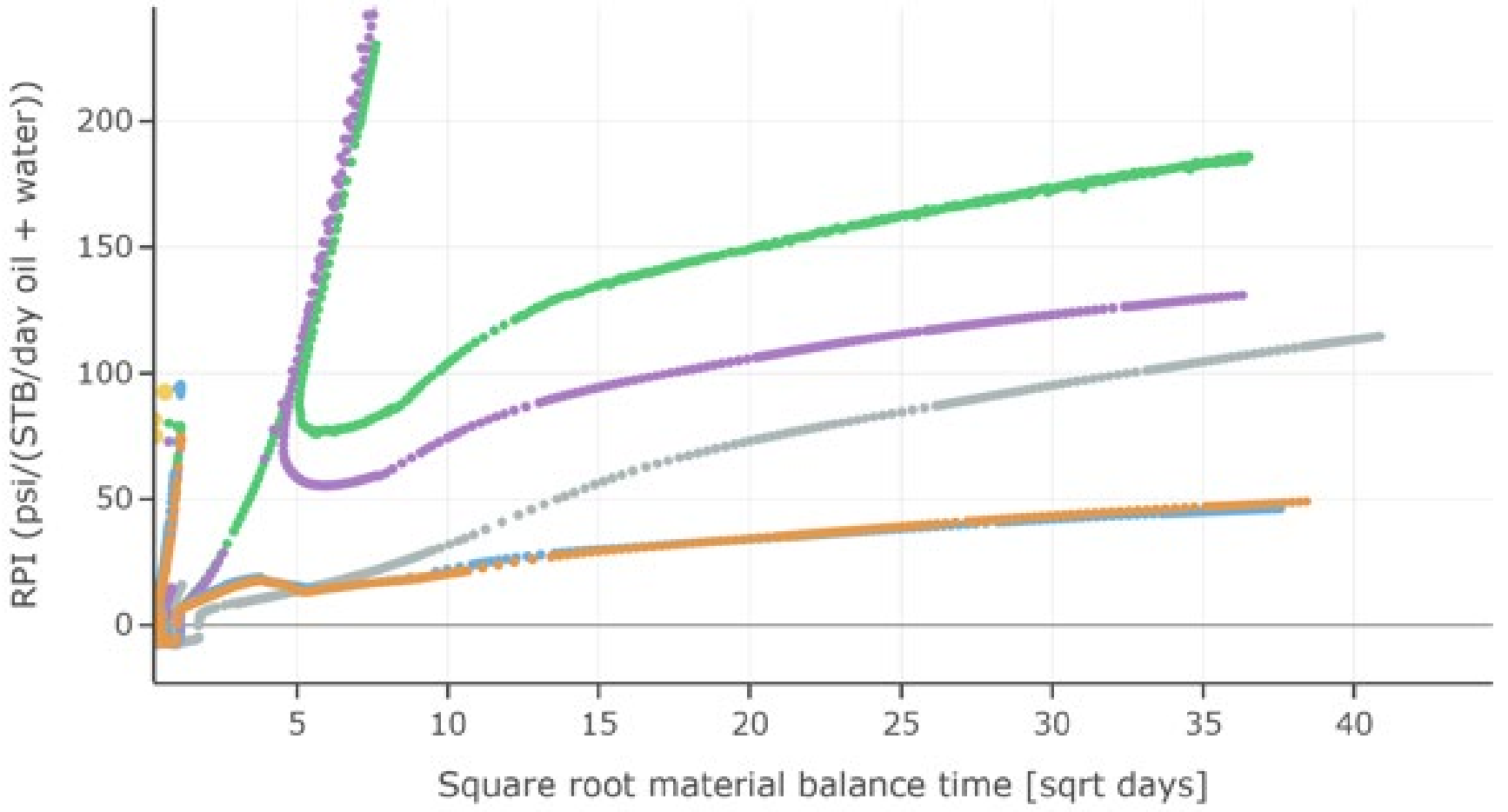

This chart can be helpful in examining your field data and determining what processes you think might be in effect. Processes within the fracture network (relative permeability or conductivity damage) manifest as a translation (shift upward) of the RPI trend; whereas processes affecting the flow from the matrix into the fracture manifest as a steepening of the RPI trend. Note that in this base case, the mobility of the water phases was such that when water block was used, the total volumetric deliverability of the well was unchanged, but water cut increased significantly (hence no slope change). For a more detailed discussion, see SPE-22194-MS by Fowler et al., 2022.

In field data we have seen conductivity damage (Ratcliff et al., 2022; McClure et al., 2023) and skin damage (McClure et al., 2023).

### 3.1.4 Other shale optimization topics

ResFrac is used to optimize a variety of other decisions regarding shale development. Some common applications are: design and evaluation of refracturing, diverter placement and timing, and enhanced oil recovery in shale.

ResFrac can model EOR because it includes an option for compositional fluid model. The diverter model is described by McClure et al. (2020).

### 3.1.5 Enhanced Geothermal Systems

The term 'Enhanced Geothermal Systems' (EGS) refers to the use of hydraulic stimulation to improve production for geothermal energy production. Conventionally, EGS was performed from vertical wells, in a single openhole stage, without proppant. But in recent years, there has been attention to designs using multiple stages, horizontal wells, and potentially proppant (Shiozawa and McClure, 2014). Companies like Fervo Energy and Deep Earth Energy Production, and the FORGE project sponsored by the US Department of Energy, are testing these designs in the field.

ResFrac has several features that make it uniquely well-suited for optimization and design of an EGS. An EGS involves hydraulic fracturing and then long-term circulation of fluid between wells. During long-term circulation, there will be thermal cooling, and thermoelastically driven reduction in stress. Fracture conductivity will evolve over time, and it is possible that fractures could even reopen and propagate due to the stress reduction. The problem is tightly coupled with flow in the wellbore, since the overall circulation rate and the distribution of flow between each fracture along the wells will be affected by friction in the wellbore. Heat conduction between the wellbore and surrounding rock occurs all the way up to the surface. ResFrac seamlessly includes all of these processes – from the wellbore to the fractures and back up – and can simulate fracturing and circulation in one continuous simulation.

EGS projects are often strongly affected by flow in preexisting faults. McClure and Horne (2014) review observations from historical projects. Often, flow diverts strongly into conductive, large preexisting features. Unlike in shale, where we generally conceptualize flow pathways in natural fractures as occurring at a smaller scale than the hydraulic fractures (see Section 7), the natural fracture flow pathways in *many or most* EGS project appear to be of similar size or often larger than the hydraulic fractures that form (McClure et al., 2014; Schoenball et al., 2020).

ResFrac is not fully-featured for true 'discrete-fracture network' (DFN) modeling for flow through a network of natural fractures. For example, users can specify preexisting fractures, but cannot give them dip. Therefore, if the goal is to perform a detailed simulation of the specifics of this so-called 'mixed-mechanism' stimulation, ResFrac is not the best choice. However, if the goal is to perform an overall, high-level optimization of an EGS design, rather than a detailed investigation of specific processes, ResFrac is very well-suited. Section 7 discusses why we do not use the DFN approach for hydraulic fracturing in shale.

### 3.1.6 Conventional reservoirs

ResFrac can also be used to design frac jobs in conventional (high permeability) formations, such as from a vertical well. Key optimization considerations include job size, injection rate, and fluid type. When optimizing a single fracture from a vertical well in a conventional reservoir, the job size is optimized to ensure that you balance production uplift from creation of a 'negative skin' against the cost of pumping the job.

ResFrac is also very well-suited for modeling long-term fluid injection. Long-term fluid injection causes porothermoelastic stress changes from pressurization and cooling. The thermal reduction in stress may cause crack initiation, and stable growth tracking the cooling front. This crack may cause negative skin in the injection well, and the entire process is coupled over time. ResFrac has all the capabilities to model this process: full 3D reservoir simulation capabilities, the ability to model propagation and opening of a crack, thermoporoelastic stresses, and flow and heat transfer in the wellbore.

## 3.2 Study a specific topic

ResFrac can also be used to study topics that are not directly design optimization. This may include studies to improve interpretation of diagnostic data, understand underlying physics, or to address a specific question.

For these applications it is especially important to apply the 'scientific method.' You should lay out hypotheses to explain physical observations. Set up a ResFrac model of the system, and then carefully investigate the simulation results to test whether and why the hypothesis was validated or not. We recommend pulling up the 3D visualization of the simulation results, zooming in, looking carefully at different properties, and carefully identifying exactly what is happening in the simulation. When you run a ResFrac simulation, you are performing a 'computational experiment.' You are asking (and answering) the question: if we assume XYZ physics and problem setup, then what will be the result? The discussion in next section gives an example of how this works in practice.

### 3.2.1 Improve interpretation and understanding of diagnostic data

ResFrac is excellent for designing interpretations to diagnostic data. Our study on diagnostic fracture injection tests (DFIT) (McClure et al., 2019) is a great example.

In a DFIT test, fluid is injected for 5-10 minutes and then the well is shut-in and pressure is monitored. Trends in pressure over time are interpreted to estimate stress, pore pressure, and permeability. The DFIT test is not *itself* an optimization of frac design. But the interpretation of the DFIT has direct, significant consequences for the optimization (Fowler et al., 2019). McClure et al. (2019) set up various DFIT scenarios and simulated them. The simulations generated synthetic pressure transients, which were compared with: (a) actual data, and (b) existing interpretation procedures, in order to evaluate whether the model could represent reality, and whether existing interpretation procedures are accurate. The results led to surprising results. For example, ResFrac predicted that if you perform a DFIT in a gas shale (but not oil shale), there is usually an apparent 'false radial' signature caused by the interaction of the viscosity contrast, the changing fracture stiffness over time, and the transition to impulse flow. This insight is supported by field data. But previously, the cause of surprising 'radial flow' signatures in gas shale DFITs was not known, and there was not agreement in the industry about whether to use them to estimate permeability. The new insight came from ResFrac because it *combined* all the key physics in a way that had not been done previously.

McClure et al. (2019) used ResFrac to develop, test, and refine interpretation methods. The recommended interpretation procedure does not use ResFrac, but it could not have been developed without it.

ResFrac has been used to study all kinds of diagnostic data: fiber optic responses in offset wells (Shahri et al., 2021), step-rate test results in unconsolidated sandstone (Kellogg and Mercier, 2019), rate-transient analysis techniques (Fowler et al., 2020b). The process is:

(1) Lay out a series of hypotheses.
(2) Simulate these hypotheses.
(3) Compare the simulation results with real data.
(4) Draw conclusions regarding the hypotheses – some are confirmed and others are falsified (Oreskes et al., 1994).
(5) Systematize the results into an interpretation procedure.

### 3.2.2 Improve understanding of the physics

You can use ResFrac to improve understanding of the physics. The 'false radial' DFIT signature in gas shales (discussed above) is an example of a physical phenomenon that arises from the combination of multiple physics simultaneously. ResFrac provides a unique combination of physics, and this provides many opportunities to investigate novel questions and learn new things.

McClure et al. (2020) discusses several other examples of physical insights drawn from ResFrac simulations. These include: fracture symmetry, proppant trapping, and ISIP trends along the well.

Researchers in academia often do not have access to high-quality datasets. They can nevertheless generate useful results from computational modeling tools by investigating the physics and deepening our understanding of important physical processes.

### 3.2.3 Answer a specific question

Sometimes, ResFrac is used to address a specific question, usually with practical importance. For example, Kellogg and Mercier (2019) needed to investigate how to interpret step-rate tests in unconsolidated sandstone to satisfy regulatory requirements. Or, perhaps your company is interested in investigating wellbore integrity during fracturing, and so you want to understand how stress evolves around the wellbore over time. Point is – don't feel constrained by the items listed in this section!

## 3.3 How often should we optimize, and at what scale?

Most commonly, we are performing ResFrac simulations to optimize the next 3-6 months of operations. We typically calibrate to data from a pad or DSU – supplementing with general knowledge from the overall development. The first calibration is the most complex and open-ended. Once this calibration/optimization cycle has been completed, subsequent recalibration and updates in neighboring wells are much easier.

As you step out spatially (move further from the original calibration pad), it helps to periodically update the calibration and optimization. If layer thicknesses are changing, permeability is changing, or fluid saturations are changing, the optimum frac design may vary.

Similarly, the optimum frac design may change over time because of changes in price. For example, at higher oil price, you can justify drilling wells at tighter spacing. Recent oil price volatility means you may find that the optimization you did 6 months ago now needs to be updated.

Finally, your calibration might change over time as you learn more. As you get production data from your most recent designs, update the calibration, and then update the optimization. Has your optimum design changed based on these new results?

Instead of optimizing at the pad scale, should you consider optimizing each well individually, or even optimizing each stage? If you have the data to support that kind of optimization, that is great. For example, you may have done surface reflection seismic and carefully mapped changes in rock properties across the pad. If you have that resolution of data, then it is absolutely a good idea to consider modifying the design at the well or stage scale.

## 3.4 Why use a model?

Physics based models are rational "logic machines" that elucidate the mechanics of a problem, assist in identifying dominant parameters/processes, and provide forecasted results.

In 2008, TNO launched a modeling competition where they used a model to create a data set including the production subjection of 30 wells for the first ten years of production, and required geologic data (with uncertainty). TNO then invited nine modeling groups to construct, tune, and optimize simulation models to the data set provided by TNO. During phase one, competitors were tasked to optimize NPV over the next 20 years of the field's life. Each group's optimized development plan was then tested in the original, ground-truth model. In phase two of the study, competitors were able to re-optimize their models after each year of production over that 20 year forecast (i.e. competitors would create a optimized plan looking 20 years into the future, then update again with 19 years to go, again with 18 years to go, etc.).

There are two findings from the Brugge study that are impressive. From phase one, they found that the top competitor (with imperfect information) created a field development plan over the next 20 years that achieved only 3% less NPV than the organizers achieved with perfect information, demonstrating in light of uncertainty, model optimizations still provide meaningful performance uplift. And from phase two, they show that continual optimization (year after year) continues to yield incremental gains each year, demonstrating that performance improves with frequency of model use.

So why do some facets of the industry still struggle to accept and value modeling? It may be the result of poor application of models, resulting in the ubiquitous phrase "garbage in, garbage out." The assumption of "garbage in, garbage out" is that if the modeling inputs are uncertain, the modeling results are uncertain; and therefore, models can't add value. As it applies to oil and gas, our reservoir data is inherently uncertain as our rocks are two miles underground, and so modeling skeptics devalue or negate the value of modeling as results are necessarily uncertain.

However, here we counter with another popular aphorism, "All models are wrong, but some are useful." Everyone has heard George Box's famous words and it is quoted widely. The expanded context of the quote is that Box goes on to use the ideal gas law ($PV = RT$) as an example of a model that is imprecise for a real gas; however, he extolls the values and insights provided from the ideal gas law. Because the relation is physics-based, it remains directional, even if slightly imprecise. Of course, for applications that need precision or where the gas behavior deviates significantly from ideal behavior, we can use a more detailed calculation to include the Z-factor. Starfield and Cundall (1988) expand upon this modeling philosophy in their work, "Towards a Methodology for Rock Mechanics Modelling."

Starfield and Cundall introduce Holling's classification of modeling problems, as in Figure 1. Holling's diagram separates modeling problems into four quadrants:

1. Good data and lacking understanding
2. Lacking data with good understanding
3. Good data and good understanding
4. Lacking data and lacking understanding

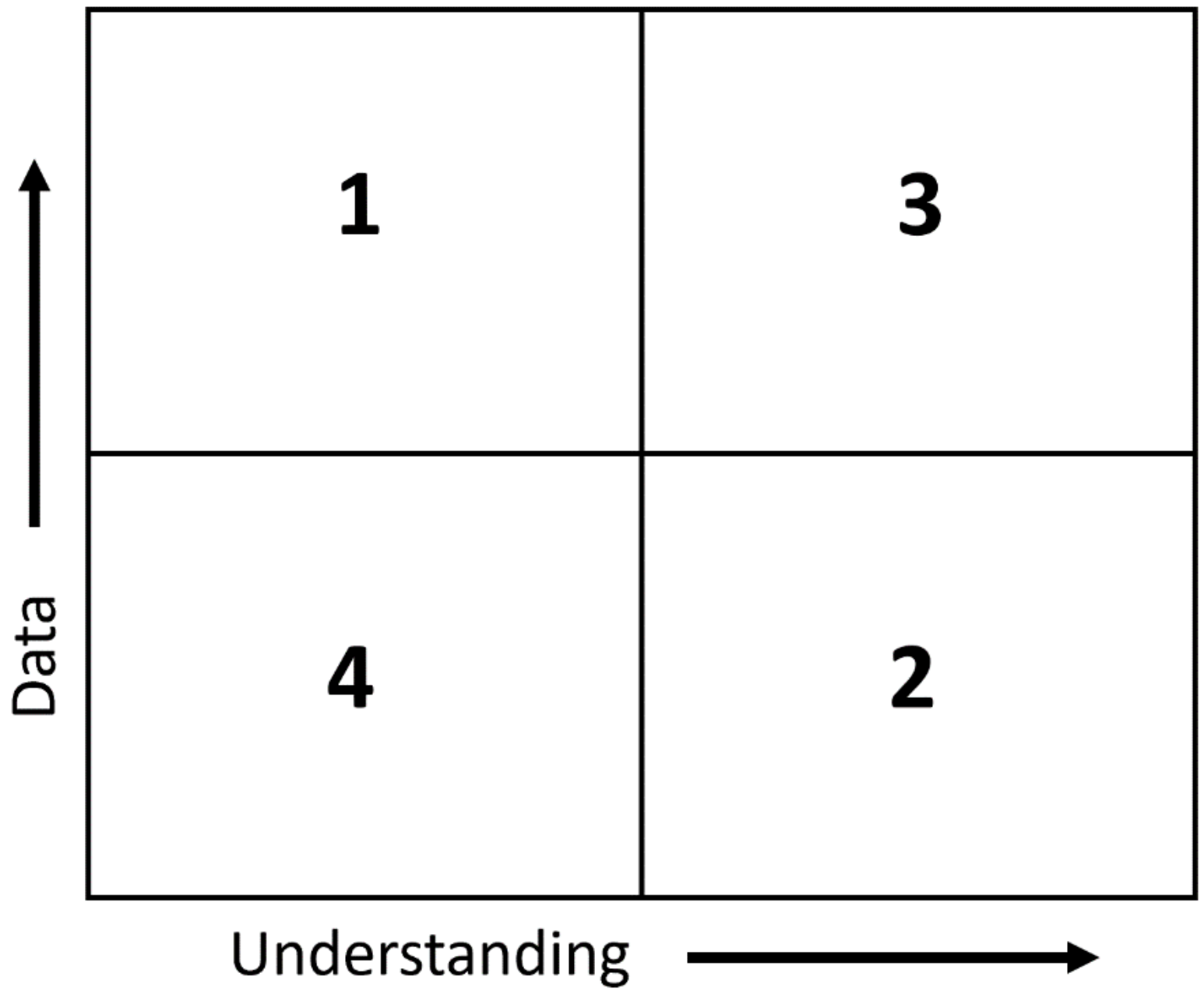

The appropriate modeling approach is a function of the quadrant wherein the model problem resides. Quadrant 1 is the quintessential application of data-driven modeling, such as machine learning and data analytics. Many surface modeling problems fall into this category - things like optimizing weight on bit or predicting component failure due to vibrations. In the subsurface, we occasionally receive data sets in Category 3 (a very high-quality science pad), but most subsurface petroleum engineering problems fall into Category 2. Recognizing which quadrant your subsurface problem falls into and the questions your model is trying to solve are critical for the correct application of modeling workflows. Applying the correct modeling framework to Category 2 and 4 problems counters the “garbage in, garbage out” objection posed by skeptics.

In Category 2 and 4 problems, data is uncertain (“garbage in”) and many times the problems themselves are ill-posed (number of uncertainties exceed the number of constraints). In these categories, data-driven models are inappropriate and can be prone to produce “garbage out.” On the other hand, physics-based models are slaves to the confines of physics, and all outputs must obey these confines. Applying physics based models to data-limited problems allows for identification of data gaps, falsification of hypotheses, and forecasts with reasonable error bars. Quoting Starfield and Cundall, “a system of interacting parts often behaves in ways that are surprising to those who specified the rules of interaction” (Starfield et al., 1988). This is exactly what we’ve done with ResFrac. Integrating the physics of fracturing, production, and wellbore often interact in complex and nuanced ways - consistently providing use with new insights. A transparent model (where the causational relationships are clear and exposed to the user, as they are in ResFrac), allows the user to explain model predictions, thereby identifying combinations of complex processes that yield unexpected results.

So how do we apply this to ResFrac modeling? “Modelling in a cautious and considered way leads to new knowledge or, at the least, fresh understanding” (Starfield et al., 1988).

- ResFrac is an interdisciplinary model, requiring input and reconciliation of geologic, petrophysical, geomechanical, and production data.
- Embrace differences/uncertainties in input data. Form hypotheses for the inferences supported by each data. Test these in the model. You will quickly find some hypotheses are not supported.
- Identify gaps in the data (sources of major uncertainty). If you were to collect more data, where would be most valuable to do so?
- Remain focused on the questions being asked of the model.
- Don't overfit beyond the constraints available. Category 3 models allow for higher resolution of parameter fitting then Category 2 models.
- The greater the uncertainty, the greater the focus should be on global parameters to match the calibration data.

# 4. The workflow

## 4.1 Structuring a project

Modeling studies should typically follow this workflow:

1. 'Kickoff' - Establish scope and objectives; plan the workflow
2. 'Initial simulation' - Gather relevant data and set up an initial simulation; communicate back to stakeholder to check communication; present a list of 'key observations to match'
3. 'History matching - Calibrate to field data
4. 'Sensitivities' - Run numerical experiments on the calibrated model

Each of these steps represents one or more 'checkpoint' meetings between you and the other stakeholders. These stakeholders may be your manager, others on your team, or the third party that has hired you to perform the modeling study. If you do not have other 'stakeholders' to keep updated, you should still go through the process laid out by these checkpoints. DO NOT proceed to a step until you have completed the checkpoint that proceeds it.

Between checkpoint meetings, you should keep in regular contact with the stakeholders. Naturally, questions will come up as you go through the data. For quick, easy questions, email works great. For more complicated questions, best to wait until you are having a conversation.

The first checkpoint meeting is a kickoff meeting involving all key parties. Usually, in advance of this meeting, a statement of work has already been drawn up. In the meeting, you should review the statement of work – the scope, objectives, and workplan of the project – and resolve any questions or ambiguities. The modeling study will need data, and you should lay out exactly what data you need and determine who is going to gather it. Often, the data may come in different forms. For example, the geologic properties may be a table of properties versus depth, a geocellular model, or just a well log. This kickoff meeting is an opportunity to talk through these kinds of issues and make a concrete plan for data collection.

In a ResFrac modeling study, we usually model sections of the lateral, rather than the full well(s) (discussed more in Section 11). The kickoff meeting is a good time to look at maps and decide which wells and stages to include.

At the kickoff meeting, you should review process. Outline checkpoint meetings and the overall workflow that will be followed.

You start setting up an 'initial model.' Do not attempt to history match yet –focus on ingesting all the information provided to you by the other stakeholders. As you get data, if you see any discrepancies or have any questions or doubts, pull on that thread! The base case model is critically important. If you make mistakes in the setup of the base case model, this compromises all subsequent results. Follow up on loose ends. You do not need to be a perfectionist – keep in mind the project objectives, and if something is not important or can be neglected, that's ok. But if something does make a significant difference to the project objectives, follow-up and make sure everything is pinned down. You should not yet start calibrating to data, but you should get the calibration process set up. For example, if history matching to production data, load the production data into the visualization tool and prepare a template that allows easy comparison between actual and simulated data (Section 5.12).

As you are building the initial model, you should also be examining the data to compile a list of 'key observations to match.' This a bulleted list of 5-15 items that summarize the key characteristics of the dataset that you will use as calibration (Section 6.3.1).

In the second checkpoint meeting, you present the list of 'key observations to match' and the 'initial model' back to all stakeholders. Emphasize that you have not calibrated to data yet. Until calibration is performed, the simulated results may be quite different from the actual data. Regurgitate back all the information that as provided to you – the geologic properties, the well landing depth, the frac schedule, etc. The main purpose of this meeting is to check and confirm that everything has been communicated correctly. Encourage stakeholders to ask questions and ask for clarification: 'pull on the thread.' These sorts of questions can uncover miscommunication; you want to get that figured out now, and not have it come out later! Most of the time, something arises during this meeting that leads to a change in the initial model.

Also, preview your plan for history matching. Explain what you are planning to change in order to get a match (Section 8). This may nudge stakeholders to remember additional helpful data that they could provide. Or, they may reveal constraints. For example, perhaps a company is very confident in their stress profile or permeability estimate, and they do not want you to change that as part of calibration. If so, you need to know that now, prior to starting detailed calibration.

Calibration to field data is the most open-ended and challenging part of doing numerical modeling. In Section 8, we lay out our procedures for history matching. For most projects, following these procedures will get you to a good match.

You may want to do an intermediate 'preliminary history match' checkpoint meeting during the history match: after the 'initial simulation' meeting, but before the final 'history matching' meeting. At the intermediate checkpoint meeting, you present the path that you are following as part of the history match. You want to get stakeholders' approval before you spend time polishing the history match down to a final match.

For example, we sometimes find that a company's initial permeability estimates are too high. If we used these permeability estimates, this would imply that the effective producing fracture length is too low, and they may be inconsistent with the DFIT pressure transients (Fowler et al., 2019). So – we may come to the preliminary history matching meeting with the objective of getting stakeholders' approval to use a permeability significantly lower than their initial estimate. We lay out the basis for our recommendation – DFIT interpretations, our own experience, publications (Fowler et al., 2019; McClure et al., 2019), etc. This preliminary history matching meeting is also one last chance for stakeholders to raise any issues, or identify miscommunication, before we move forward.

Once we have confidence in the planned approach, we move on to performing calibration to data (Section 8). Once the calibration is complete, we have a 'history matching' checkpoint meeting in which we present the final history match to stakeholders. If you have done a good job of communicating, and done the checkpoint meetings, this meeting should not have any surprises! If it does, to fix the issue, you will have go back and make changes.

At this 'History matching' meeting, you finalize the plans for the numerical experiments. For example, you may intend to optimize cluster spacing and stage length. You and the stakeholders concretely decide the range of cluster spacing values to be considered, whether to hold constant lbs/ft or lbs/cluster, etc.

Now, perform the numerical experiments. Before scheduling the 'Sensitivities' checkpoint meeting, it may be a good idea to review the results in an 'informal update' meeting with a smaller group (such as with your direct contact with the client company). Often, the results confirm what they were expecting. But maybe not! If results are surprising – it is good to know that prior to presenting to the full group. First, with surprising results, double check to make sure you did not make a mistake. Your results found that the optimum cluster spacing was XYZ. Go back – are you sure that you tabulated the cluster spacing correctly? Once you are sure the results are solid, then the next step is to ask 'why?'

Surprising results can be a great thing. They often provide an opportunity to help the client improve performance. But – if the client is going to actually act on surprising results, you need to be prepared to justify them. As discussed more in Section 6.3.3, zoom-in on the 3D visualization and look carefully at what is happening. Try writing down simple 'back of the envelope' calculations. If you really can not decipher the 'why,' email us at support@resfrac.com and ask our opinion! We have seen hundreds of ResFrac simulations and have a strong intuition into why things happen. Ask yourself what model inputs affect these results. Maybe try some sensitivities changing model inputs and see if the results hold up. Very often, once you dig in, you will realize that the surprising model results make perfect sense.

The model surprises our intuition, but once we understand it, our intuition changes. The 'compliance method' of picking fracture closure is an example of a model result that was surprising and initially confusing, but subsequently turned out to be right on target (McClure et al., 2019).

You should also consider how the company arrived at their current views. Perhaps the company previously did this optimization with a different method (RTA? Comparison with offset wells?). For example, maybe they did an optimization using RTA calculations and a higher permeability than you used. If so, this could have led them to find a different optimum (Fowler et al., 2019).

Once you have done this legwork, you will be much more prepared to present the results to the full group. If the modeling shows that their current design is already optimal, great! That will make for an easy meeting. If you advise changes, be prepared for (understandable) skepticism, and be ready to explain the reason for the model results and for your recommendations. Be ready to assess your overall confidence in the finding. Do not get your feelings hurt if they don't completely adopt your recommendations. Your job is to perform the ResFrac analysis, provide the results, and then explain *why* those were the results. Critically analyze the problem to pinpoint (as much as possible) the root cause of differences with their current practice. You give that information to stakeholders, and then it is their responsibility to make decisions based on your analysis and all of the other information available to them.

ResFrac should be used in an iterative cycle with field data. Perform an analysis to suggest an improved design or to identify key uncertainties that could be reduced with data collection. Test the new design and/or gather new data, compare with predictions, and iterate. The goal with ResFrac is to *shorten the learning curve* – iterate towards the best designs sooner – and to facilitate fine-tuning of the design over time in response to new geologic conditions, market conditions, or technology.

Field trials are best performed as A/B testing. As much as possible, operators should change one thing at a time, and then systematically compare the impact on production. Practically, this is not always possible. But it is good to try!

## 4.2 'Checkpoint' meetings and 'informal update' meetings

It is important to differentiate between 'checkpoint' meetings and 'informal update' meetings. In a checkpoint meeting, you are presenting results that have been vetted and are ready to be communicated across the full group. Between these checkpoint meetings, you communicate regularly with stakeholders, and may be sharing and discussing results. Go out of your way to make sure that everyone understands that results in-between 'checkpoint' meetings are only preliminary. We follow the process in Section 4.1 in order to make sure that model setup and results are carefully vetted. In meetings with stakeholders, it can be very tempting to short-circuit the process and start prematurely drawing conclusions from the model results before you have finished. Do not do this. As you go through the process – checking, calibrating, analyzing, and communicating with the group – you will almost certainly be making changes to the model and your interpretation. Everyone needs to know – if you are still in the middle of the process, then everything is still subject to change. This can be communicated by establishing up-front that you have 'checkpoint' and 'informal update' meetings, and by labeling the calendar invites and slides decks as an 'informal update' or a 'checkpoint meeting.'

Delineating 'checkpoint' meetings also helps make sure that the right people attend the meetings. Managers are busy and may not want to be involved in every single meeting or informal communication. By delineating 'checkpoint 'meetings, we are communicating to them which meets are highest value for them to attend, and which are lower priority. Try to encourage managers to attend all of the checkpoint meetings (but not necessarily the informal meetings). Checkpoint meetings are an opportunity for stakeholder to express their priorities, give direction, and raise issues. If a senior manager does not attend any of the checkpoint meetings, and then joins for the final project wrap-up, they might bring new perspective and priorities that haven't been raised previously. This can be an unwelcome surprise and reduce their satisfaction with the project. Checkpoint meetings structure the interactions so that stakeholder preferences are identified *as early as possible*. By keeping interactions 'informal' until you are ready to formally present at a checkpoint meeting, you ensure that when that meeting occurs, and the full group attends, you are prepared to lead a high-quality meeting. With meetings, emphasize *quality* over *quantity*!

Supplementing this section, Sections 6 and 10 have more recommendations on the nuts and bolts of how to execute a successful study. Please check them out!

# 5. Getting acquainted with ResFrac

The purpose of this section to get acquainted with the software. For a tutorial going through the full workflow, refer to <https://resfracsoma.blob.core.windows.net/collateral/ResFrac_ModelSetupTutorial_Bakken.zip>.

## 5.1 Overview

ResFrac simulations are run on the Microsoft Azure cloud computing server. This allows you to run a large number of simulations simultaneously, without needing to bog down your own personal computer.

Simulations are set up, organized, and visualized using a locally installed user-interface. It runs on Windows and should work fine on any modern PC or laptop. One caveat – your computer needs to have an adequate graphics card (GPU) to render the 3D visualizations of the results. In order to submit and download simulations, you need access to the internet.

## 5.2 Versions and updating

We push out updates regularly. The user-interface (locally installed) and the simulator (on the cloud) are two different pieces of software and are updated separately and have different version numbers. When you submit a simulation, it automatically submits to run on the most recent available version. It happens automatically, and you do not have to do anything. However, you have the option to submit simulations using older versions (Section 5.4). This can be useful because updates can cause minor changes to simulation results, and so if you are in the middle of running a set of simulations, you may want to keep the version number consistent. When you run a simulation, the version number is recorded in the 'comments' file generated as output (Section 5.6).

When we push out an update to the user-interface, this requires an update to the software installed on your computer. If an update is available, you will be automatically prompted to update from within the UI. If you click ok, the update goes through automatically. On some corporate IT systems, autoupdate capabilities are disabled. In those environments, we email your IT person when the update is ready, and they put out the update. To check your user-interface version number, refer to the 'settings' screen in the job manager (described below).

## 5.3 Installation and setup

We will provide you with a link to download an installer for the ResFrac user interface. When you run the installer, it will ask you for the location to make the installation. Make sure to install it in a place where you will not need special administrator privileges to access it. The default location, C:\ResFracPro, is usually fine. We recommend installing it on a drive with at least 50 GB of free space, as simulations are stored locally, by default in the installation directory, and can each be multiple GB. It is fine to use an external hard drive. If you are interested in doing a network installation, please have your IT team contact us at support@resfrac.com and we can work with them on how to set this up.

You can always use the ResFracPro UI to set up simulations and view results. But to submit simulations and download results, you need to be able to log in. To do that, we need to add you to the user list for your company. Please contact support@resfrac.com and we will work with you to set this up. If you are the first user at your company, we may need to work with your IT team to get things configured on your end. If your company uses Microsoft Active Directory or Google G-Suite for its accounts, we can set it up so that you use your company account to log in to ResFrac. Alternatively, we can provide you with a separate account managed within our system.

On initial setup, we may encounter a variety of issues with corporate IT systems. We have seen practically everything at this point and have done everything possible to make ResFrac robust to work on many different systems. Still, companies throw us curveballs. Also, some companies explicitly require ResFrac to be ‘whitelisted’ in their IT systems before allowing it to access the internet. Therefore, if you have any trouble logging in or downloading results, please contact us at support@resfrac.com, and include an IT security or general IT contact if possible. We will ask you for a description of the behavior you’re observing, and for you to send us the ResFracPro detailed log file. Section 14.3 describes how to turn on detailed logging to produce this file.

## 5.4 The job manager

When you first open ResFrac, the program opens to the ‘job manager.’ The job manager is used to keep track of your simulations. Your simulations are organized into ‘Projects,’ which are basically just folders that you can use to keep your simulations organized.

You will be prompted to log in (see Section 14.2 for details). Note that even if your email address is not a Microsoft or Google email address, you can still ‘register’ it with them and use it for logging into ResFrac.

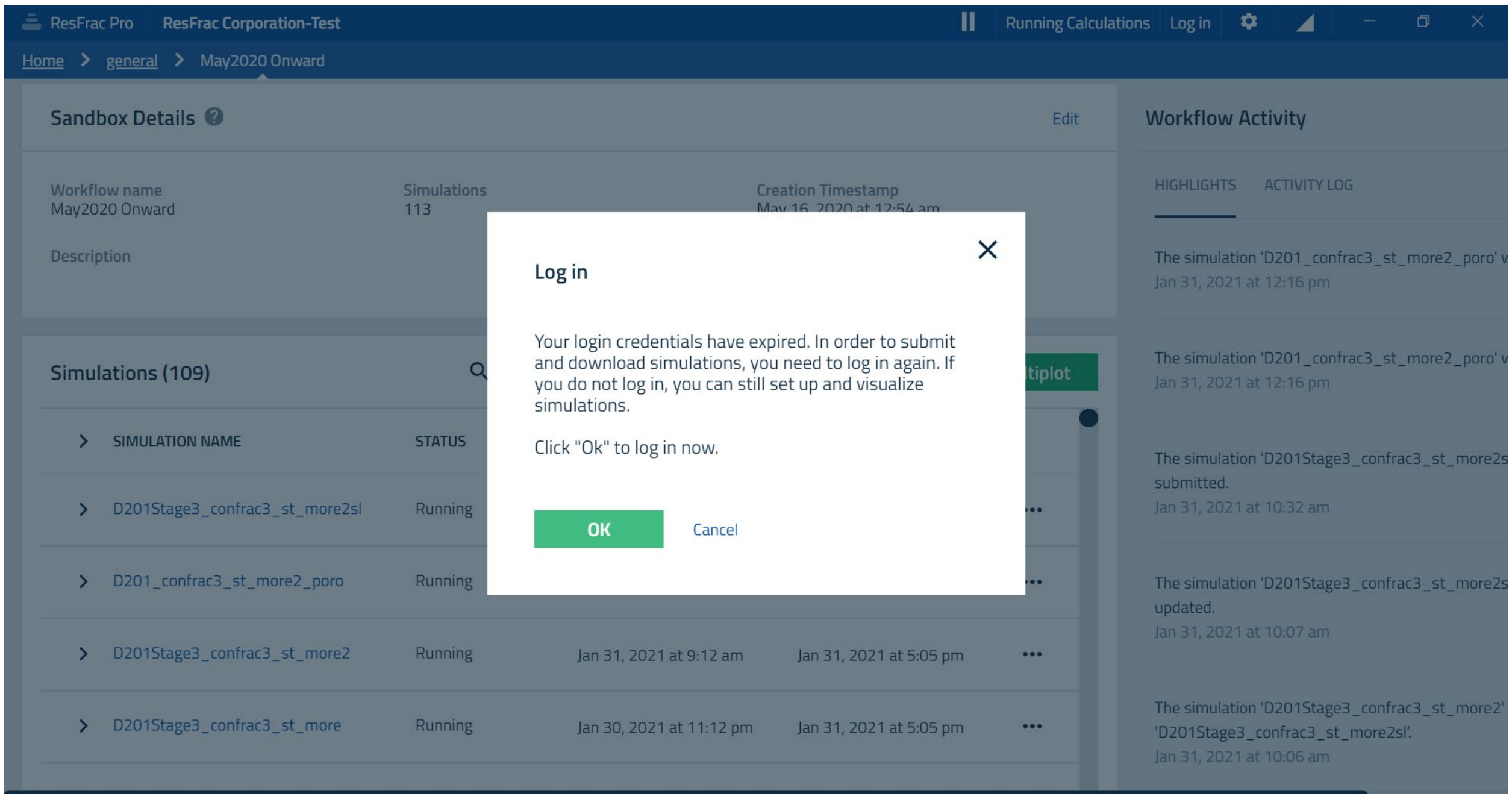

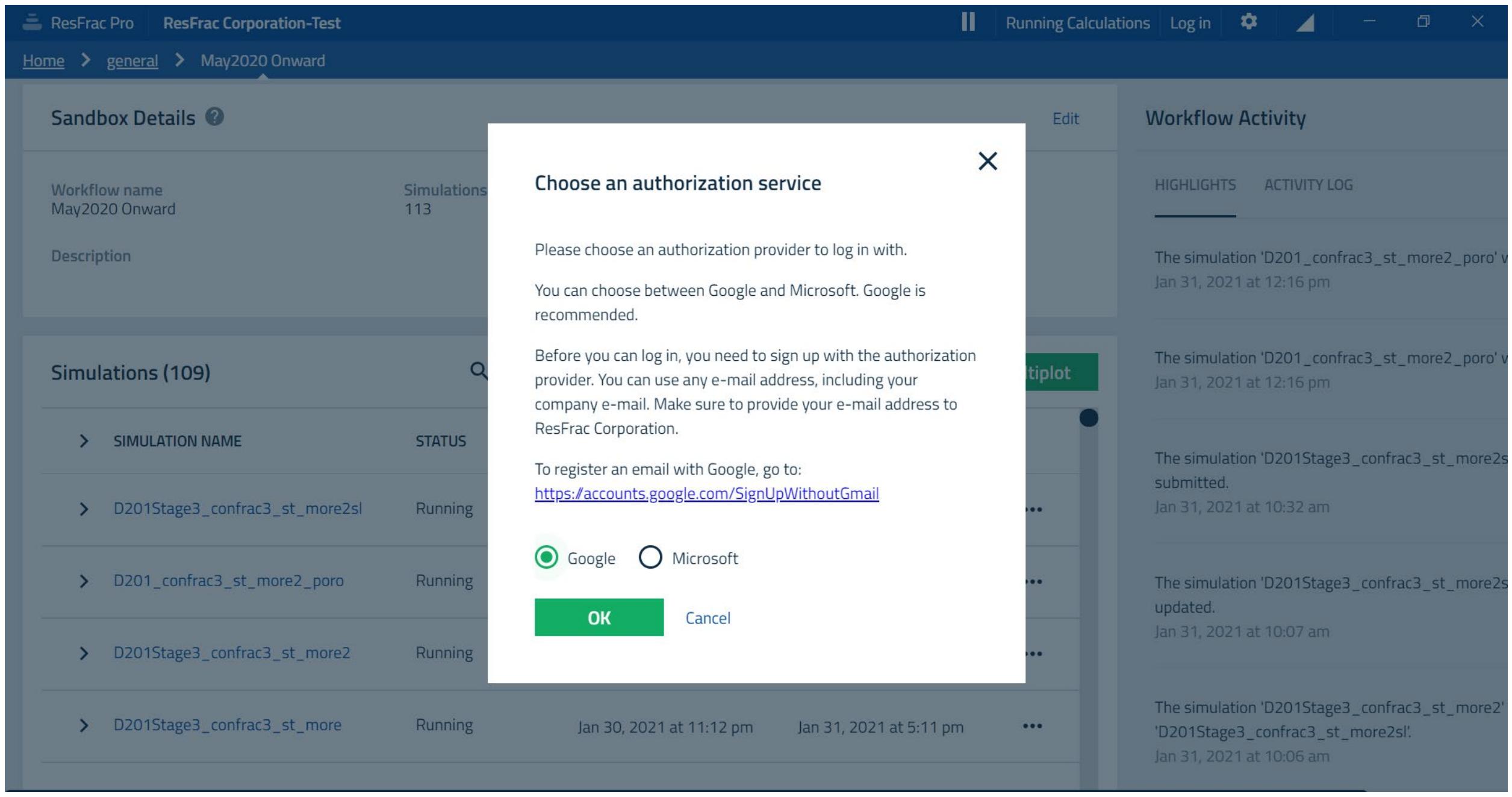

Click on the ? button in the upper left corner of the window. It pulls up help content – both written and video. If you click 'Show more,' it goes full-screen.

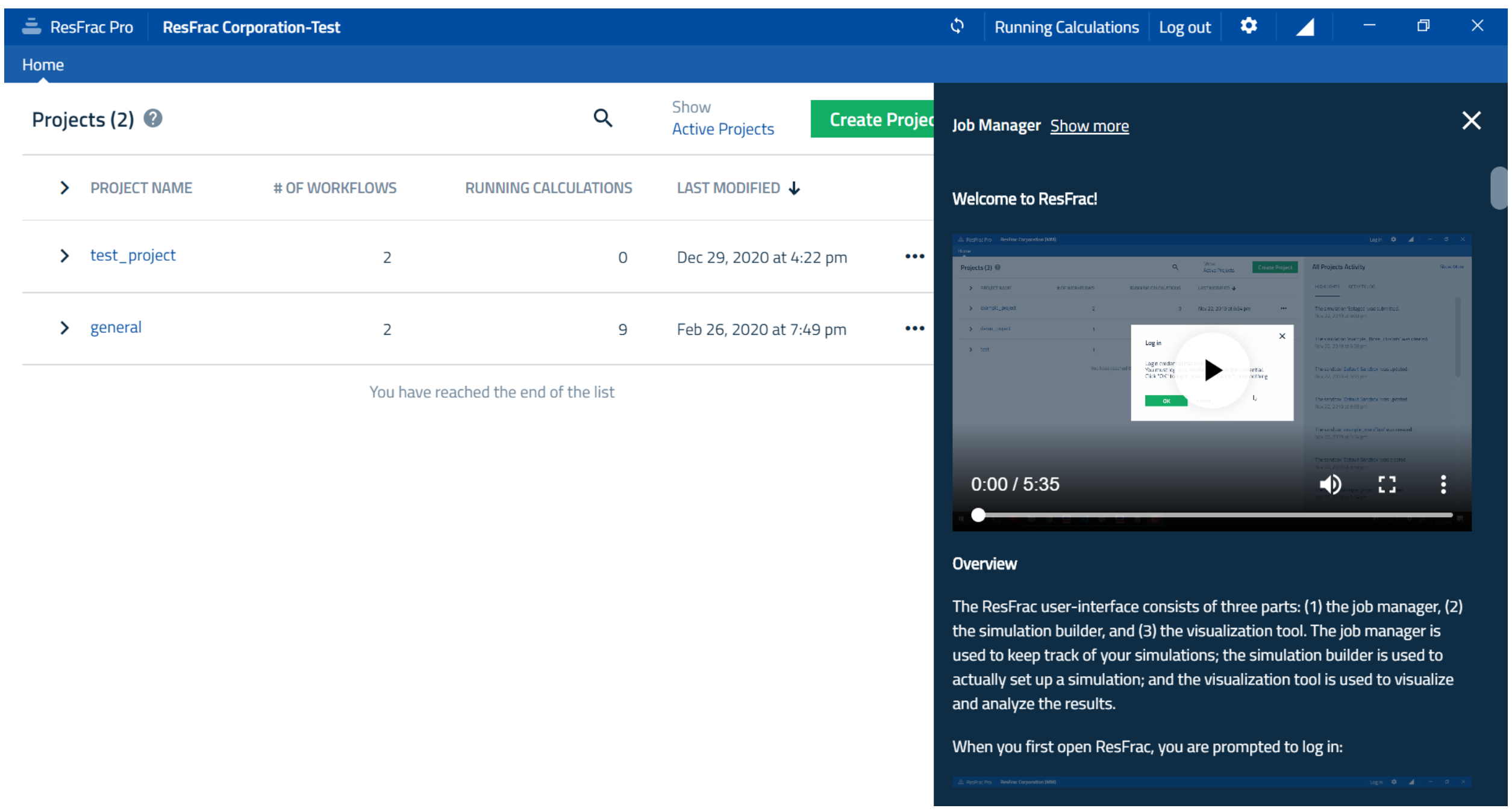

Click 'Create Project,' to make your first project.

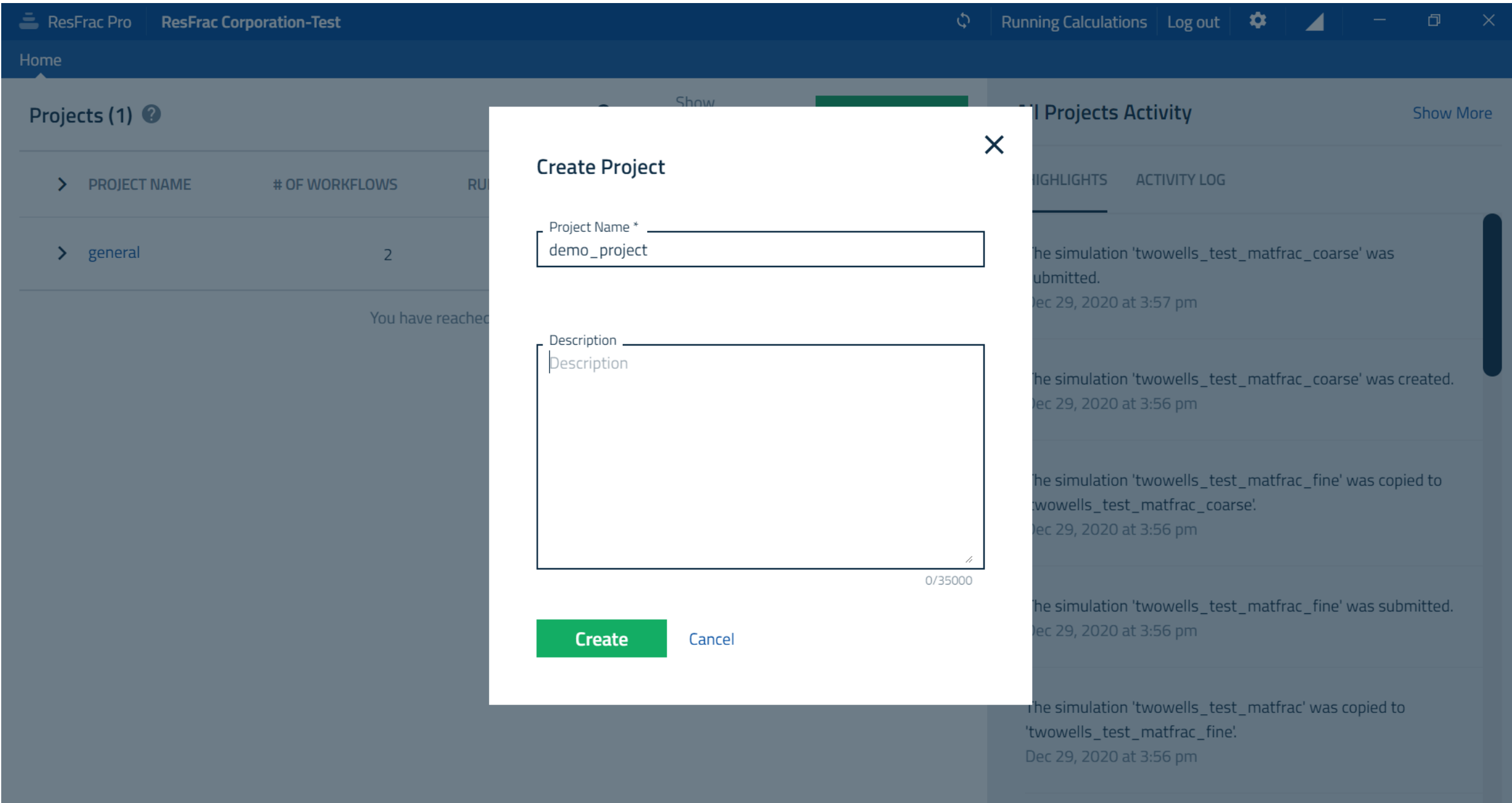

Within each project, there are a set of 'workflows.' The available workflow types are 'Sandbox', 'Sensitivity Analysis', 'History Matching', or 'Optimization'.

A sandbox is just a folder full of simulations. Click the button for 'Create Sandbox,' and create a new one. Within the sandbox, press the button for 'Create Sim.' ResFrac comes preloaded with a default simulation of a shale well with three frac stages. Alternatively, a base simulation can be selected from a growing repertoire of templates (see Section 12 for more details on simulation templates). When you click 'Create Sim,' the program opens up the simulation builder.

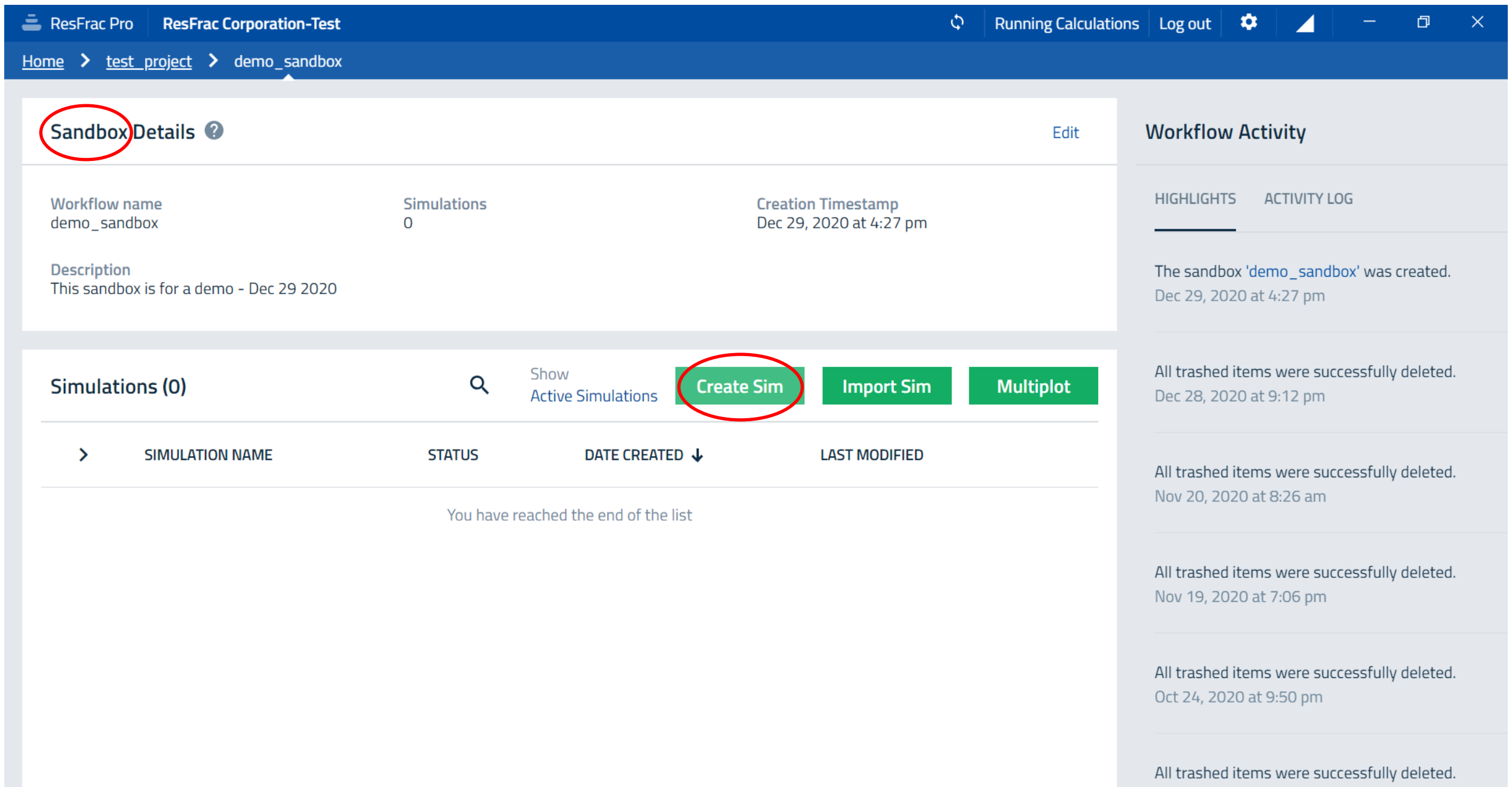

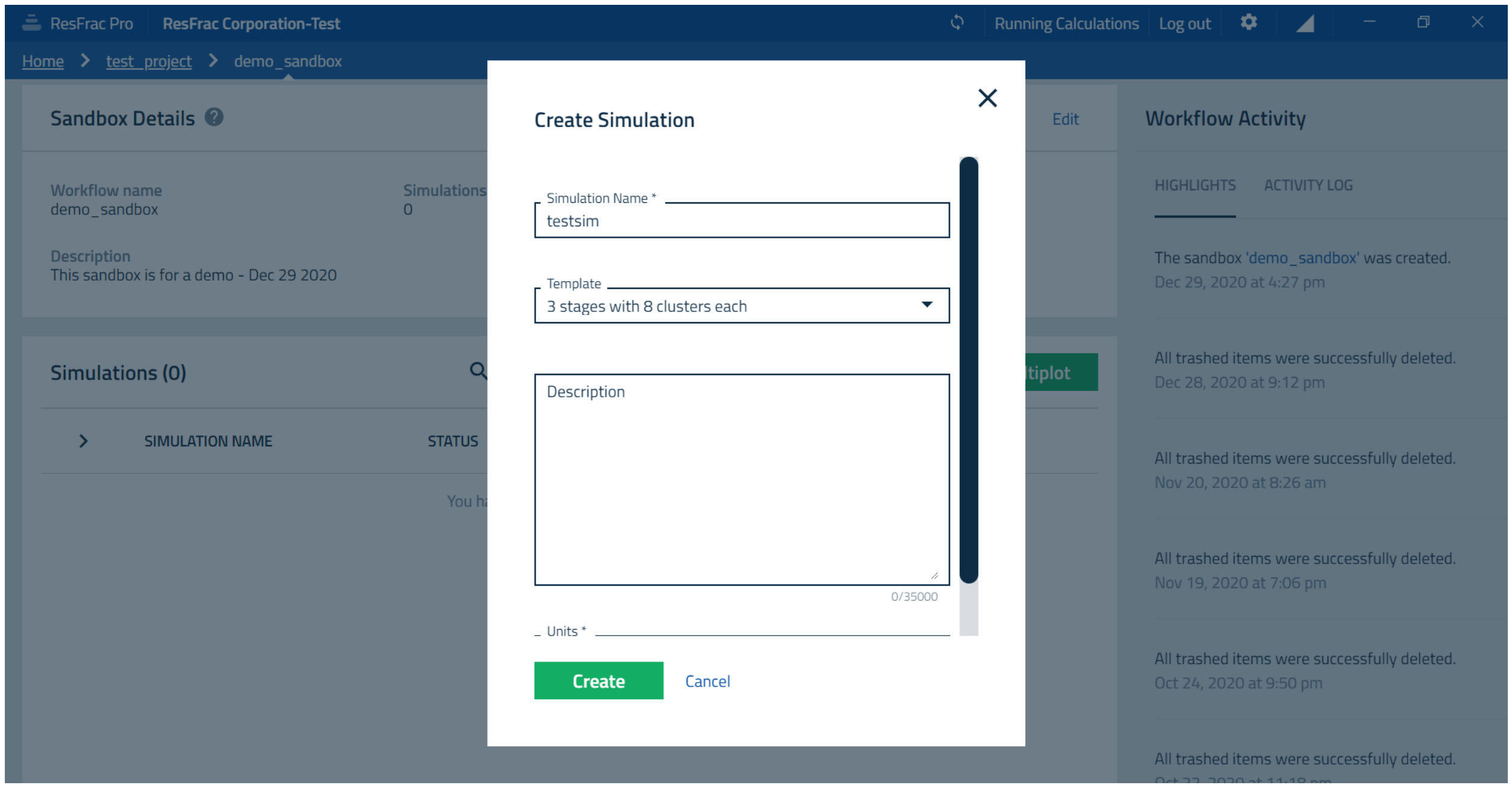

Before we discuss the simulation builder, let's go back to the job manager and discuss a few other details. So, go ahead and click "Exit" to go back to the job manager.

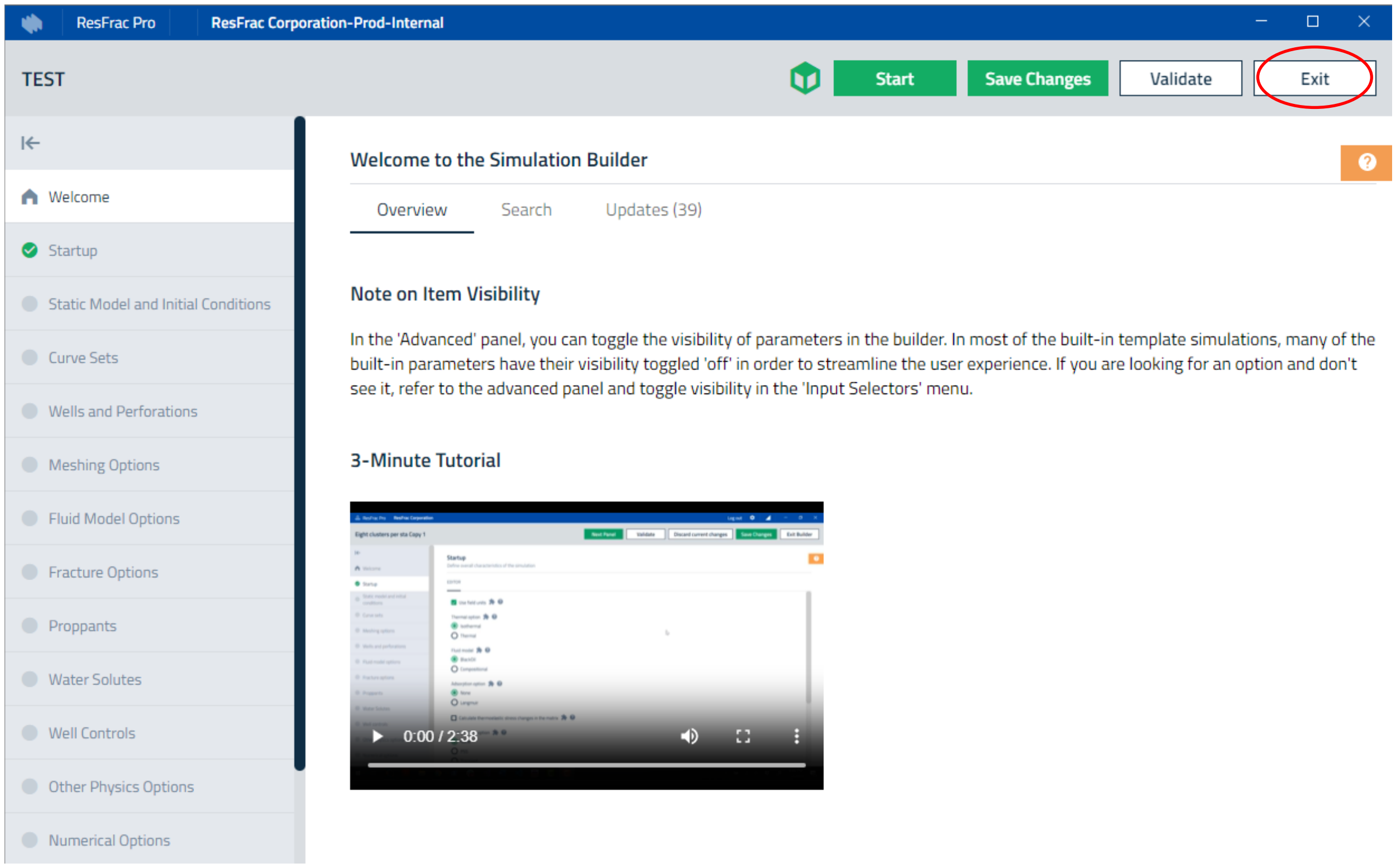

If you click on the simulation name, it pulls up a menu with a list of actions. You can edit the simulation details (rename the simulation or modify the 'description'), edit the simulation setup in the builder, etc.

For the purposes of this demo, go ahead and select 'run simulation.' When you clicked 'Create Sim' it loaded a reasonable default simulation, so if you submit this simulation, it will run.

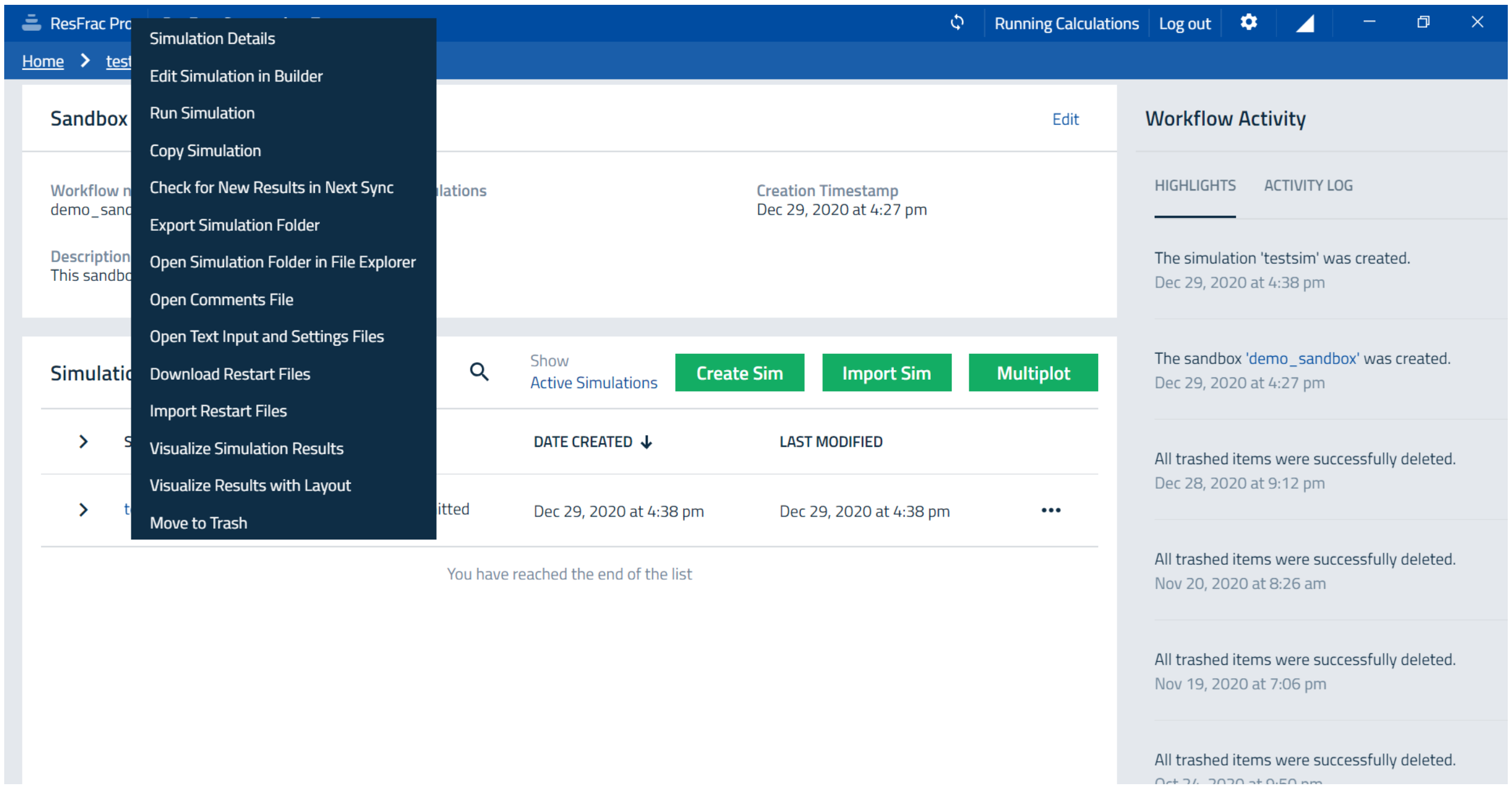

When you select 'Run Simulation,' you will have the option to select simulator version, job priority, and Vm size. It defaults to use the newest available simulator version, which is recommended unless comparing to jobs run on a previous simulator. Job priority dictates which simulations start running first, and Vm size indicates the speed of the run. Note, the faster runs also cost more in simulation hours on your license, and so are recommended only for time sensitive projects or for large simulations. The default Vm size can be set in the User Settings. Go ahead and click 'Run Simulation.' As the simulation runs on the cloud, the results automatically download. We will be able to track the simulation progress and view the results as the simulation progresses. However, it will take about 15 minutes for the simulation to start up on the cloud, and so in the meantime, let's look at the simulation builder. Click on your simulation and click 'Copy Simulation.' Then, click on that simulation, and select 'Edit Simulation in the Builder.'

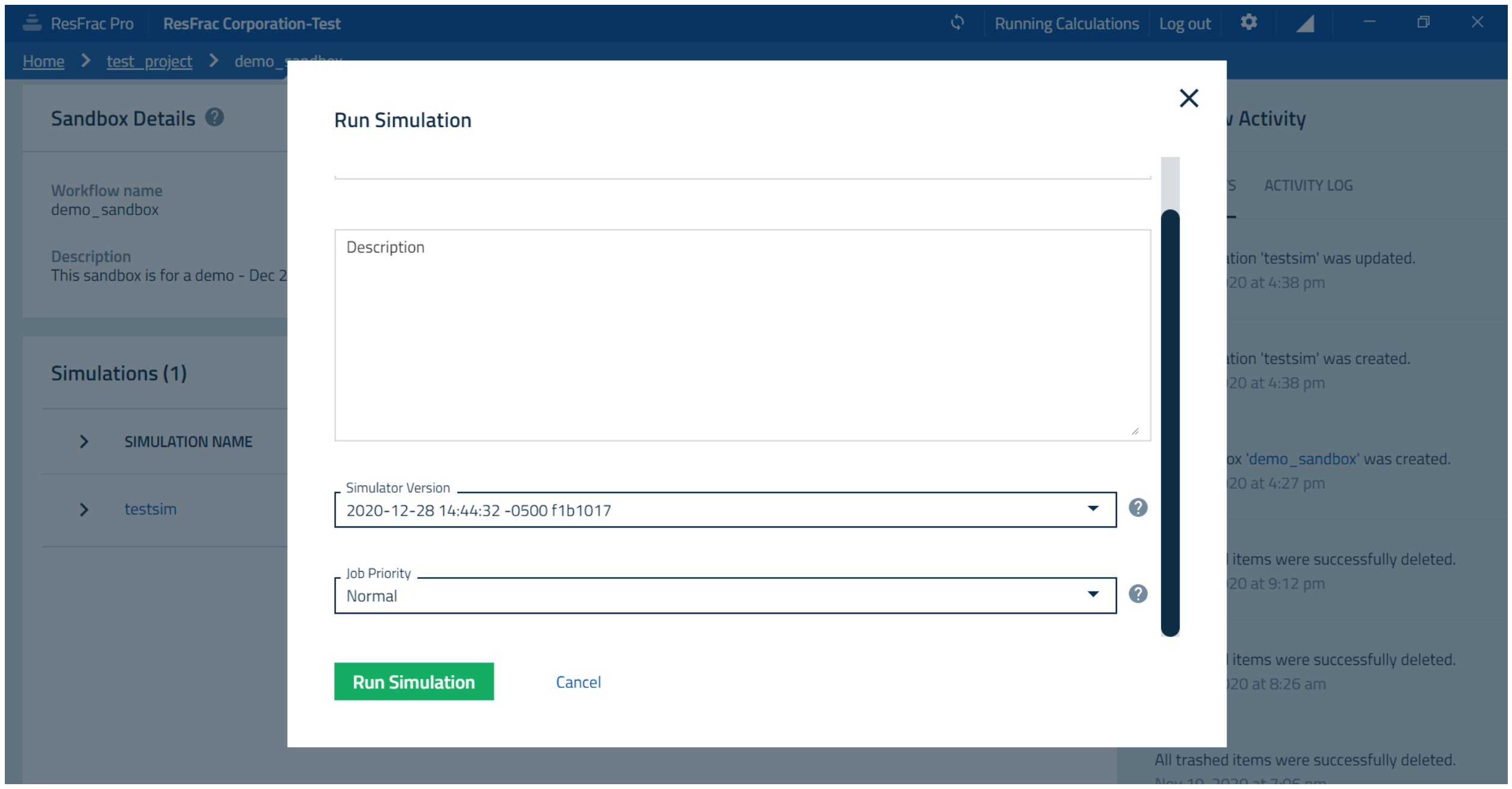

## 5.5 The simulation builder

The simulation builder is organized into a series of panels, as shown along the left side of the screen. To set up a simulation, you generally will progress from the top panel downwards.

The welcome screen provides a 3-minute tutorial video to go through the builder. Also, every screen in the builder has detailed help content built in. To access the help content, click on the orange ? button in the top right corner. If you click on the 'Search' button in the central welcome panel, it pulls up a search capability that allows you to search all available parameters.

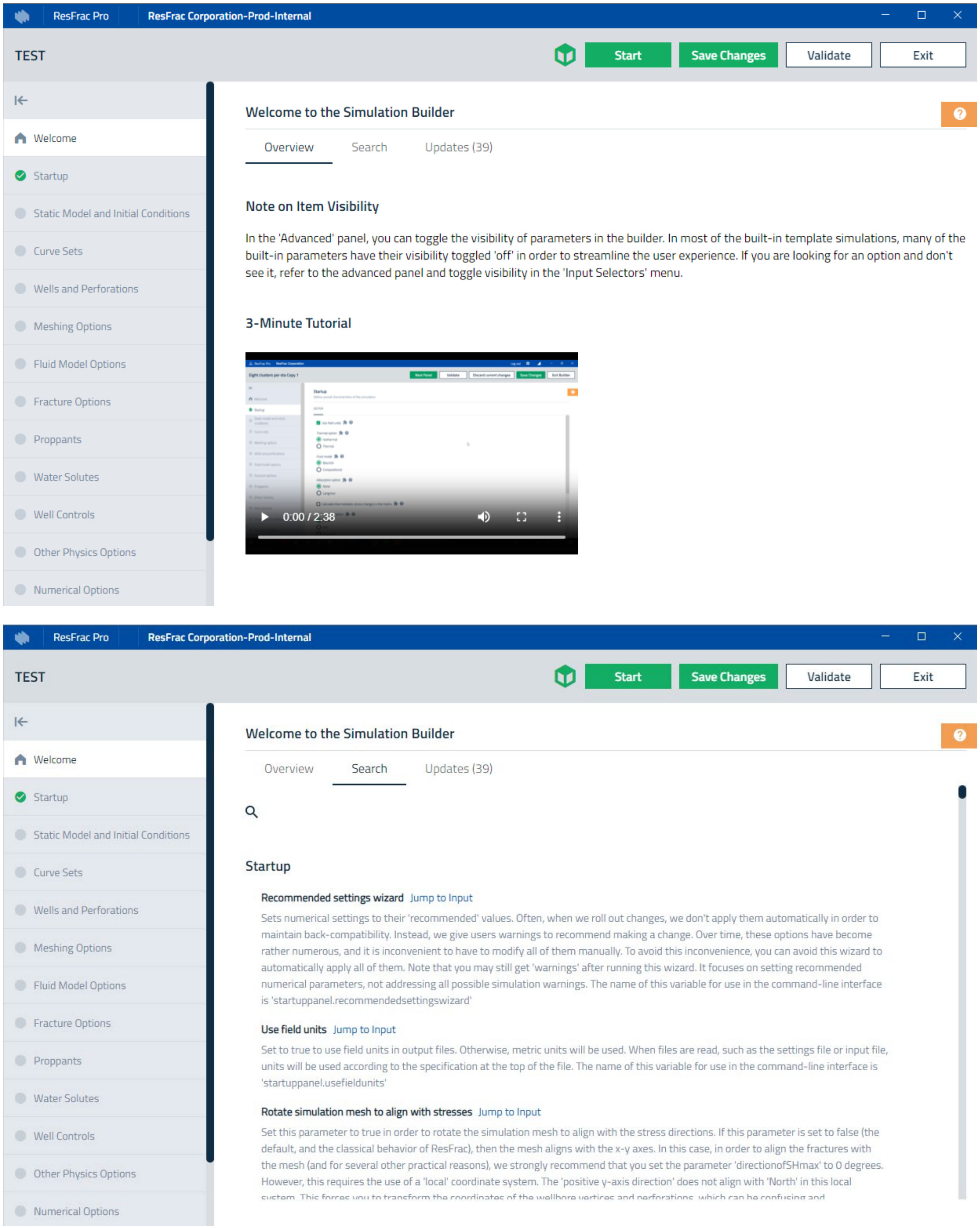

Move on by clicking on the “Startup” panel. In the startup panel, you select the governing physics for your simulation. Do you want to run a thermal or isothermal simulation? Compositional or black oil? For more detail about these selections, do not forget that you can click on the question button! For example, if you click on the ? button for "Fluid model", it pulls up an explanation of the black oil and compositional fluid models. You can also see the help content by hovering the mouse over the ? button. Clicking the button pulls up the content in pane view, which may contain a more detailed explanation.

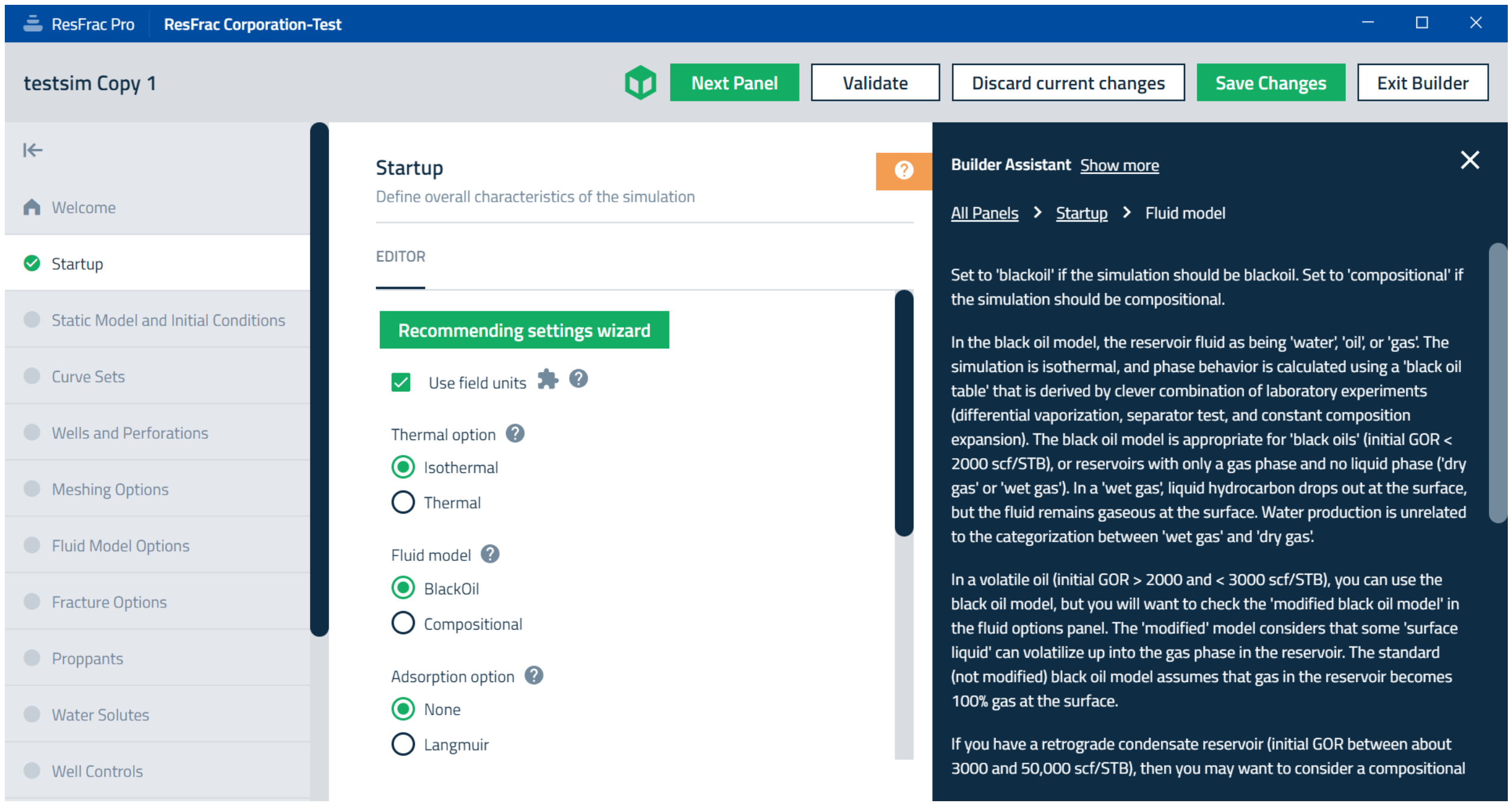

Depending on your selections in this panel (and other panels), the visibility of different inputs changes throughout the simulation builder. For example, if you select 'Thermal,' then in subsequent panels, you will be prompted to specify information such as the thermal conductivity of the rock, the initial temperature distribution, and the temperature of injection fluid. If you select 'Isothermal,' then you will not see entry boxes for this information. Similarly, if you select 'Black oil', then the 'Fluid Model Options' panel will request black oil model information. If you select 'Compositional,' then the panel will request information to specify a compositional fluid model – a completely different set of inputs.

'Validation functions' provide extremely useful functionality to assist in setting up simulations. Every single entry in every single box in the builder is checked by a validation function. The validation function returns an error if the input is impossible, and returns a warning if the input is acceptable, but possibly problematic. You will see visual clues for these validation functions. If a panel contains an error, a red X will appear next to the panel name on the left. Additionally, a red marker will appear next to the parameter giving the error, while an orange marker will appear where there is a warning.

At any time the 'Validate' button can be selected to check for any errors or warnings within the simulation. A window pops up with a summary of validation results for the entire simulation. A validation check is also automatically run whenever the simulation is saved or closed. For example, scroll down to 'maximum wallclock time' in the 'Startup' panel and set it to "-1". This will trigger a validation error. The screenshot below shows the error printed after clicking 'Save Changes.'

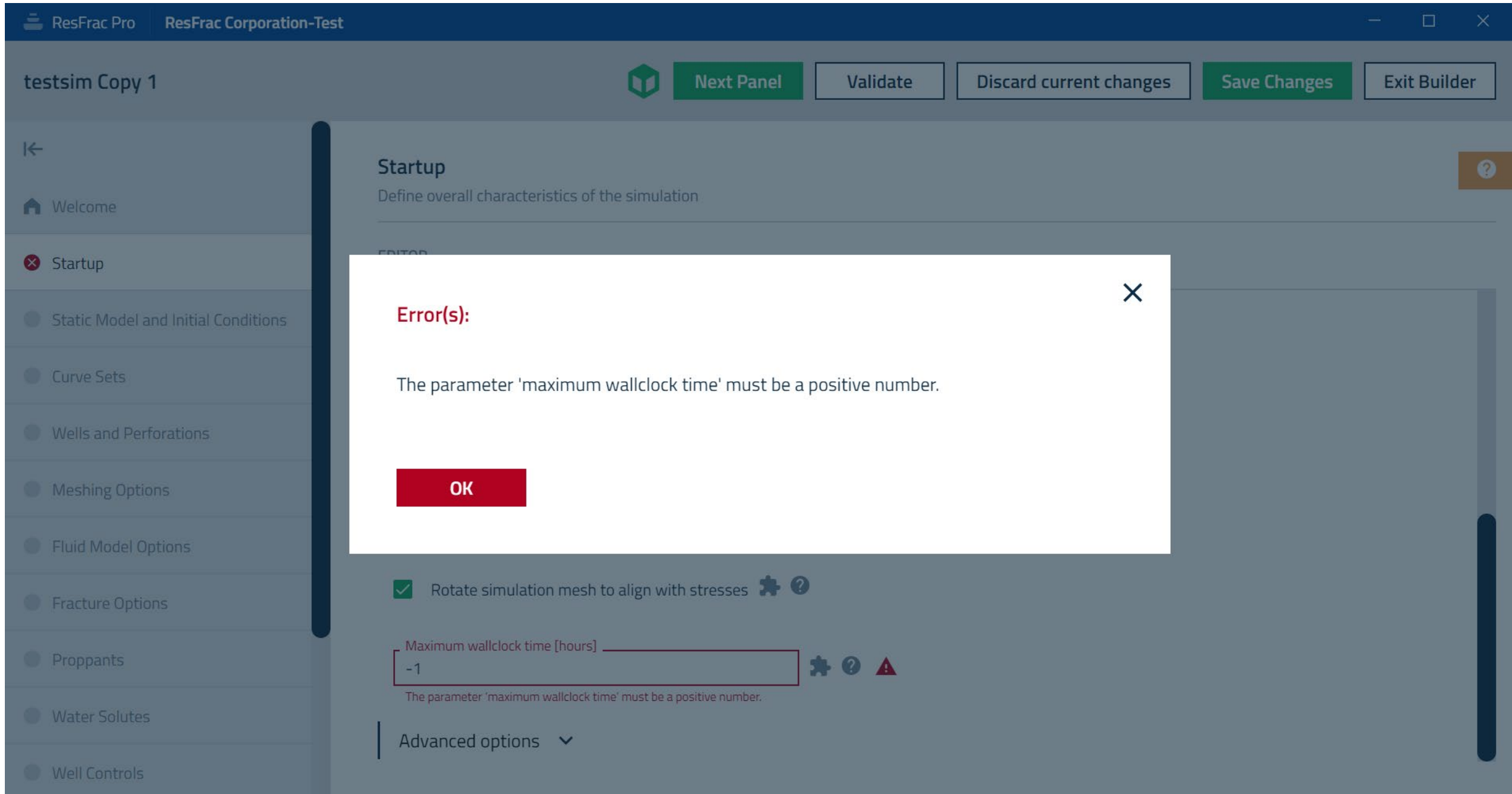

Every time you leave a panel and go to a different panel, ResFrac autosaves. If you exit the builder, you will be prompted to save. If you do not save, the simulation will autosave and you will be prompted to reload the autosaved version next time you open the simulation. To load the last manual save, do not recover the autosave – instead select 'Cancel' or X-out of the prompt.

ResFrac does not have a generalized 'undo' capability, but you do have some ability to bulk 'undo' by clicking 'Discard current changes.' This button goes back to the last autosave. If you go ahead and press that button, in this example the 'Maximum wallclock time' will revert to its previous value of 120 hours.

At the top of the 'Startup' panel, there is a large green button called 'Recommended settings wizard.' Wizards like this are scattered throughout the simulation builder. Wizards are designed to automate or simplify a certain aspect of setting up simulations.

In this case, the 'Recommended settings wizard' has a rather unique, but quite useful, role. We frequently roll out new updates. Often, these updates impact the behavior of the simulator in some way. In order to maintain back-compatibility, we strive to avoid rolling out changes that modify simulator behavior *by default*. Instead, we implement new keywords, and give the user a *validation warning* if the new keywords are not activated, and suggest that they update *when convenient*. If an update may impact simulation behavior somewhat, it may not be convenient to make the change right away because history matched simulation models may need to be tweaked to regain the history match. If quite a few of these keywords accumulate over time, and a user decides to update, the search capability in the welcome screen is very useful in finding the keywords in the builder. To save that inconvenience, the 'Recommended settings wizard' automatically applies all 'recommended' simulation settings.

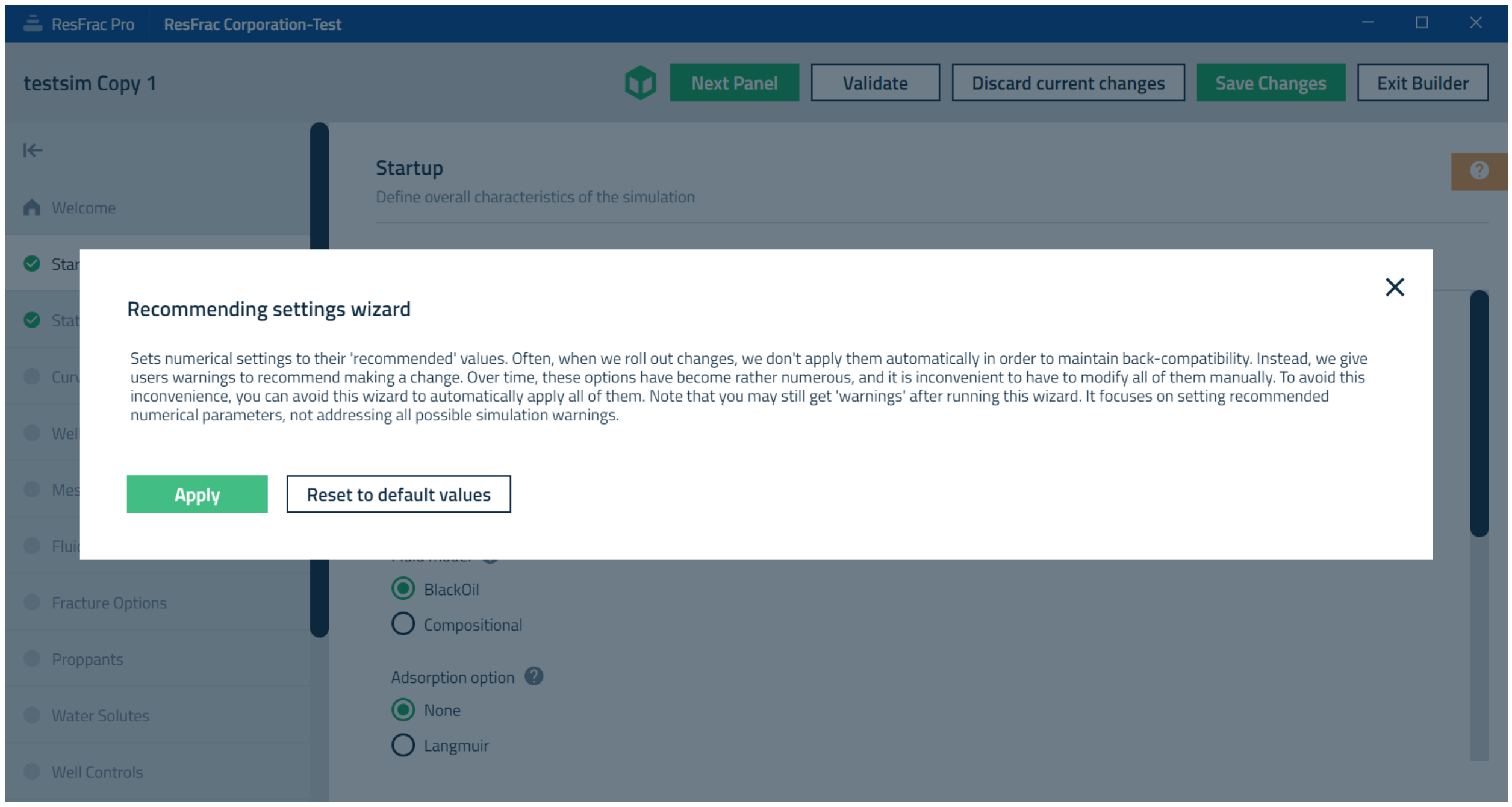

The purpose of this section is to give an overview of the code, not to provide a detailed description of every panel. For that, you can refer to the help content built into the builder, and refer to Section 11 for step-by-step tutorials on history matching and optimizing frac design.

Before moving on, let's skip ahead to the 'Fluid Model Options.' Click on the 'Black oil wizard' for an example of a more detailed wizard. This wizard asks you to input information such as initial producing GOR, and will generate an entire black oil fluid model. Note that the wizards also have built-in help content and validations. No need to actually run this wizard – the simulation comes preloaded with a black oil model – so click the X in the upper right corner to exit the wizard.

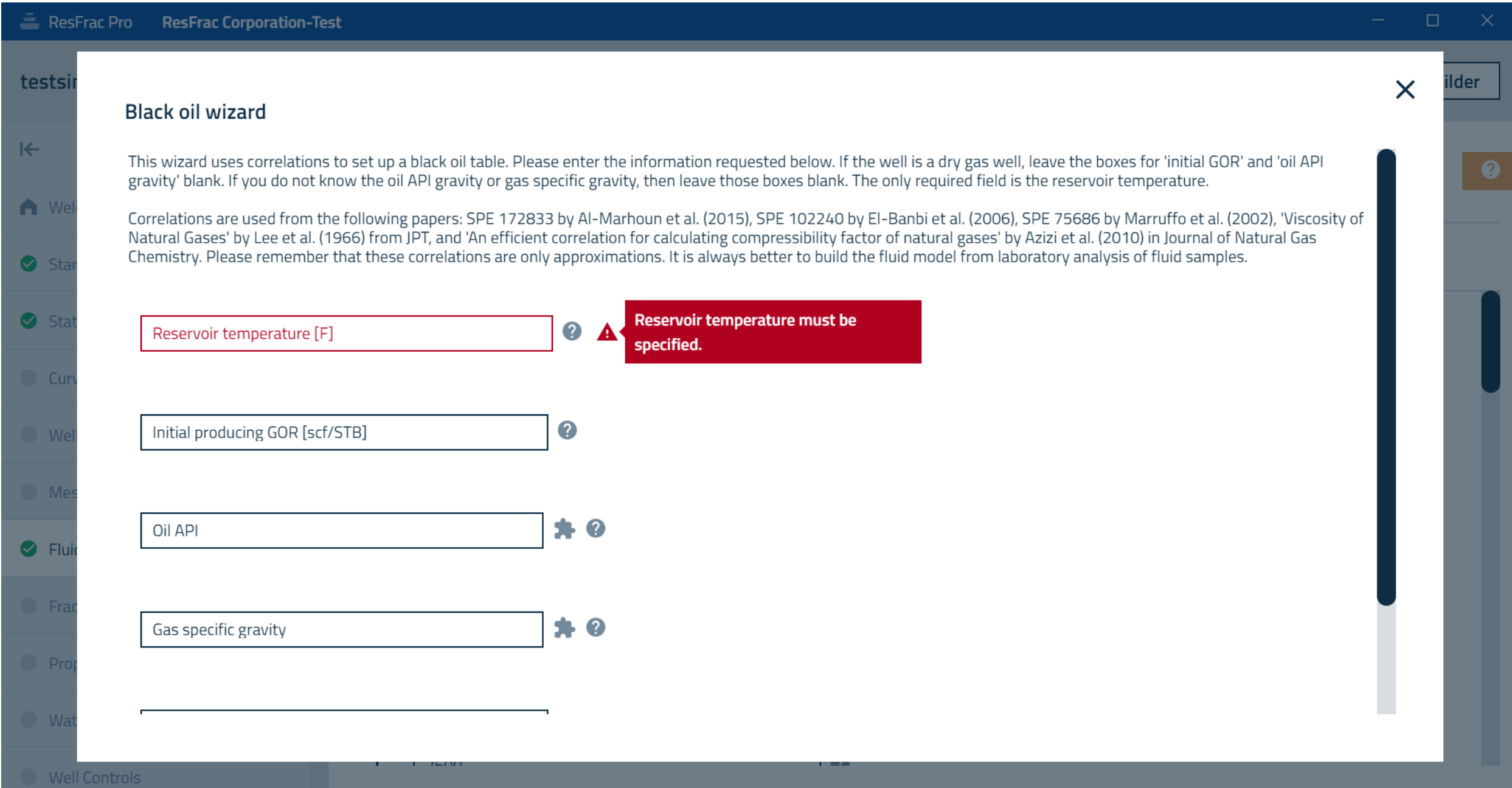

Scroll down to the 'Black oil model property table' and click on the three-bar icon next to the help icon. This brings up a preview line plot of the properties inputted into the table. These three-bar icons are scattered throughout the builder to provide preview line plots.

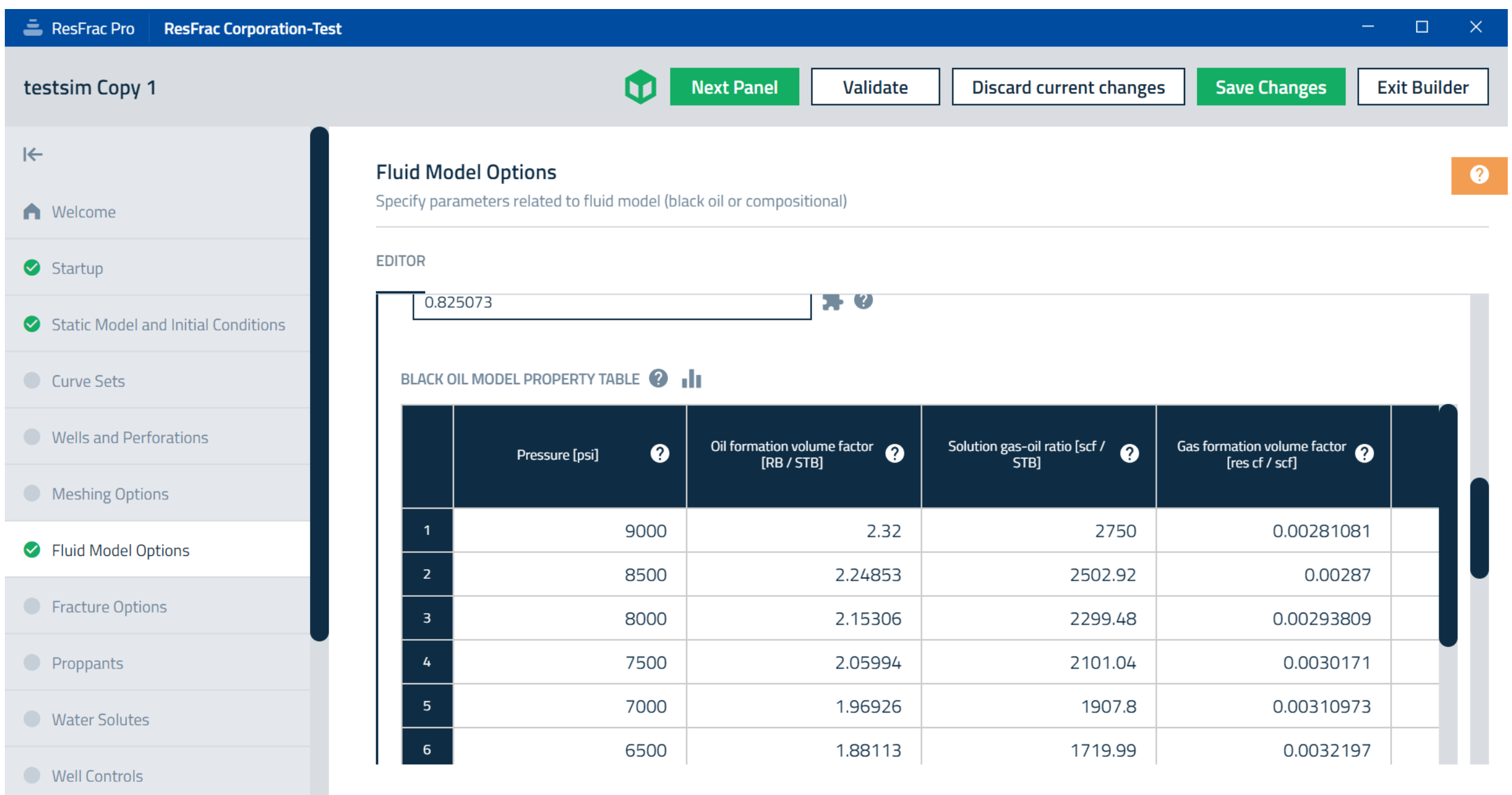

You can modify what is visible on the plot, view other properties, and otherwise customize the plot by clicking the gear icon in the upper right corner. Note that if you change the values in the black oil table, this will not automatically update the preview plot if it is already open. To update the plot, click the red 'Refresh' button in the upper right corner.

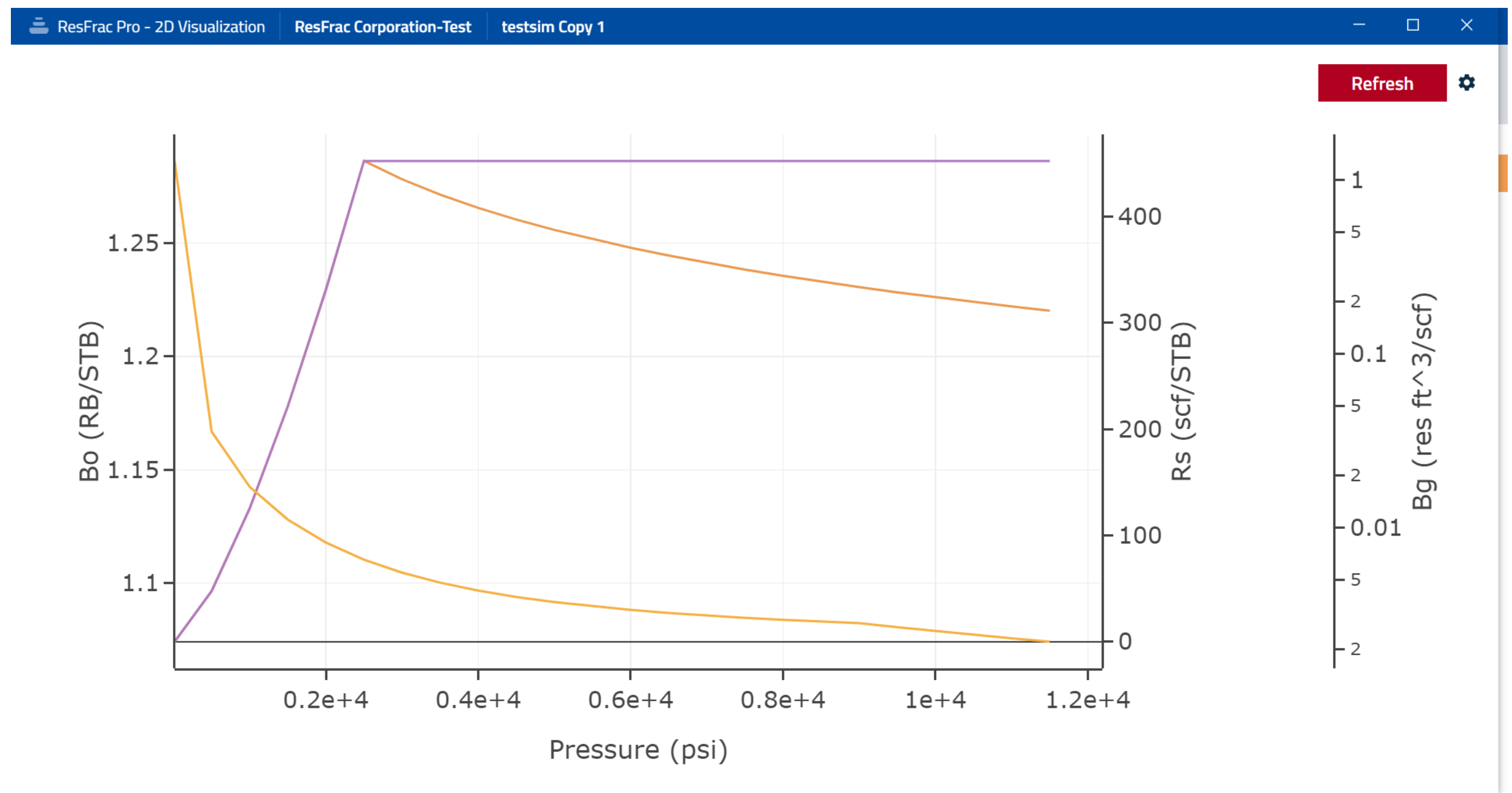

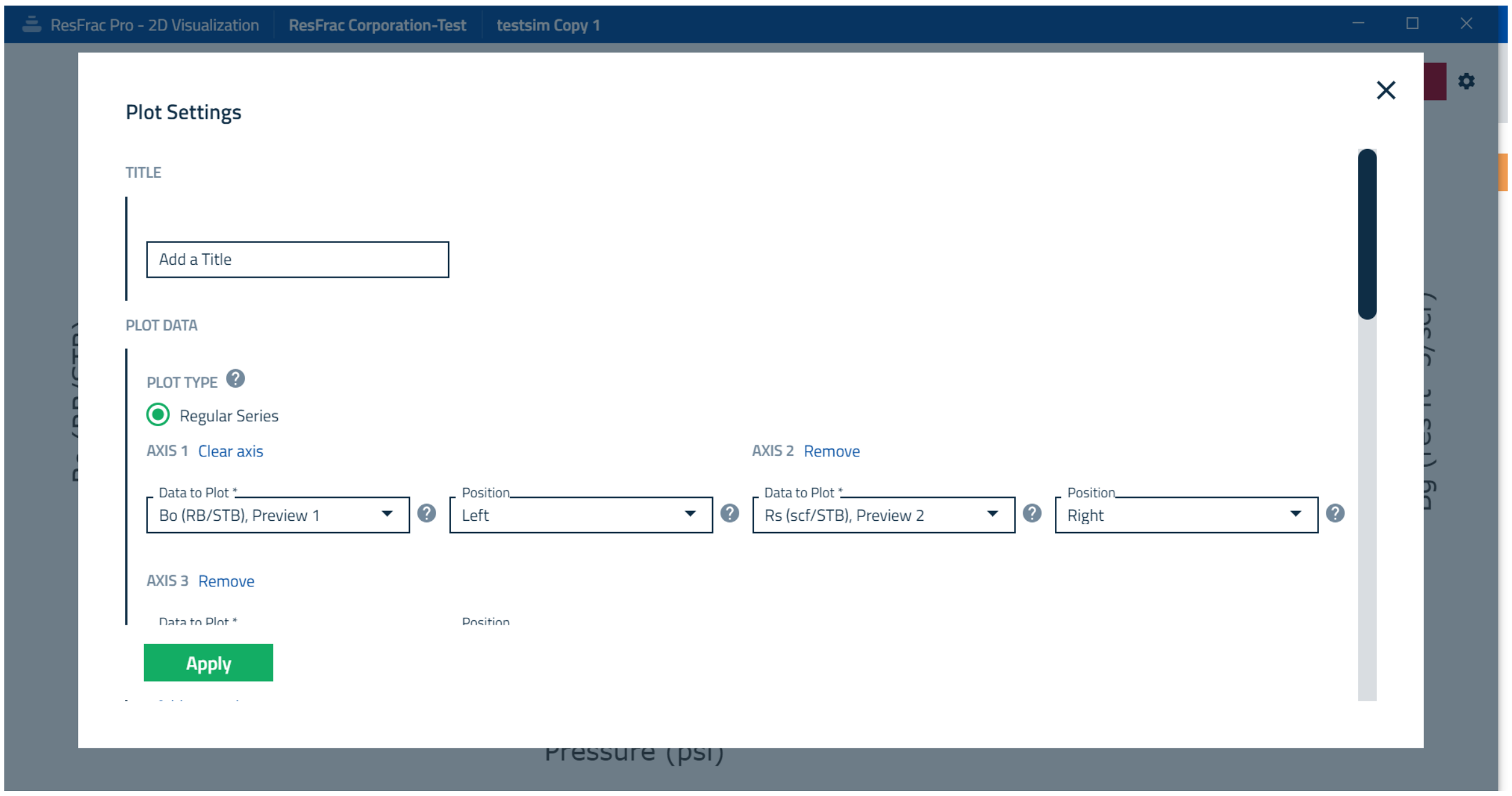

Close the preview plot and return to the simulation builder. From any panel, you can click on the green cube along the top of the screen to pull up a 3D preview plot. This 3D preview plot is a simplified version of the 3D visualization tool that you use to view simulation results. In the builder context, you can view the simulation region, the mesh, the well, the formation properties, and other simulation inputs. Left and right click, along with pressing shift and control with the clicks to manipulate the camera controls. You can toggle visibility of things like the matrix region, stretch the axis, change the color scale, etc.

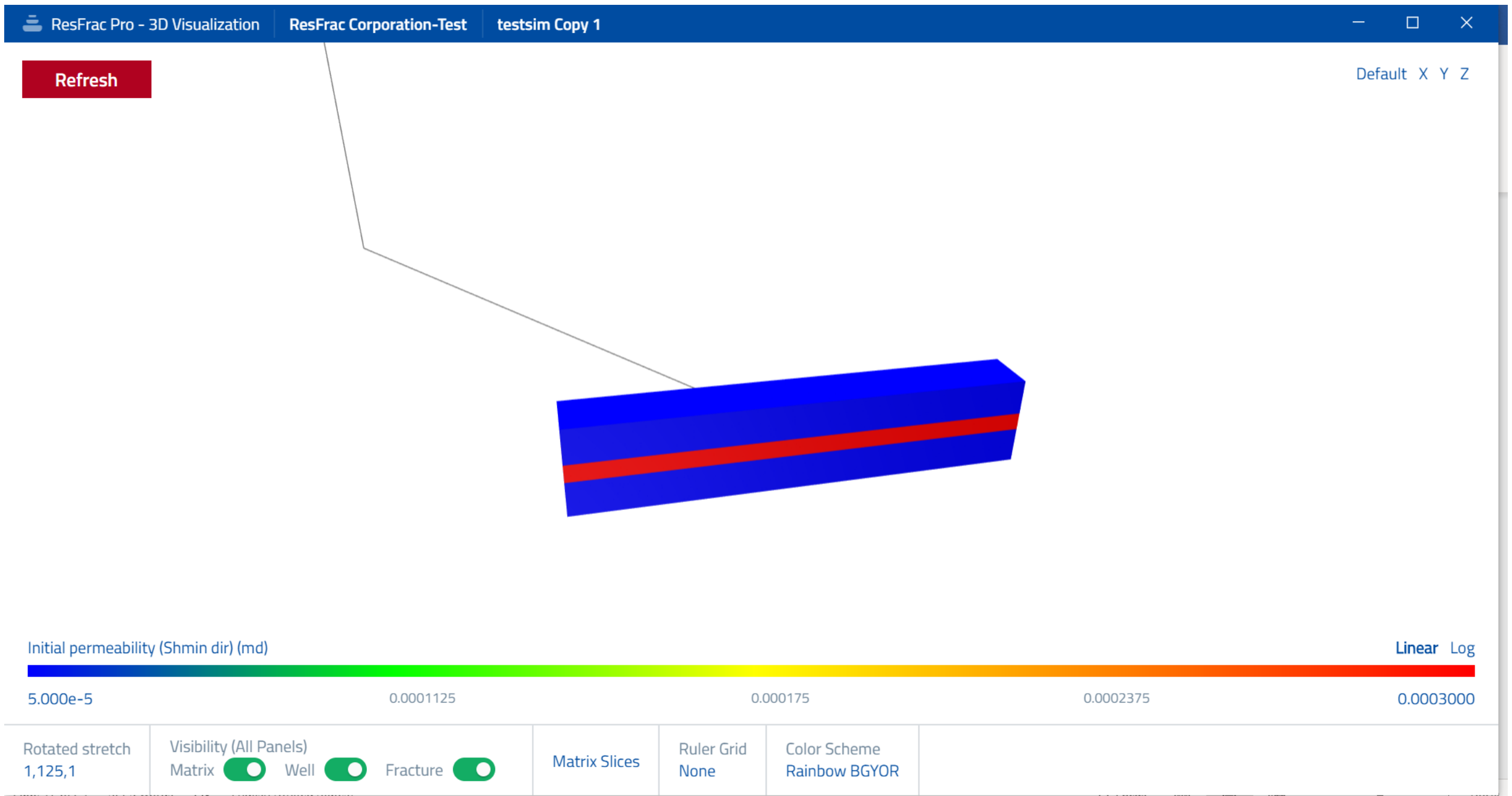

The preview plots will not show things that fail validation. For example, go to the Wells and Perforations panel and scroll down to the ribbon for 'Well_Demo', and click on it to show the vertices defining the well location. Delete a few of them. Now, the 'Well vertices' are failing validation, since one of the vertices is undefined. If you reopen the 3D preview (may need to click 'Refresh'), the well will no longer be visible, and a ribbon notification will pop up in the upper right corner to explain that the wells cannot be shown due to a validation problem.

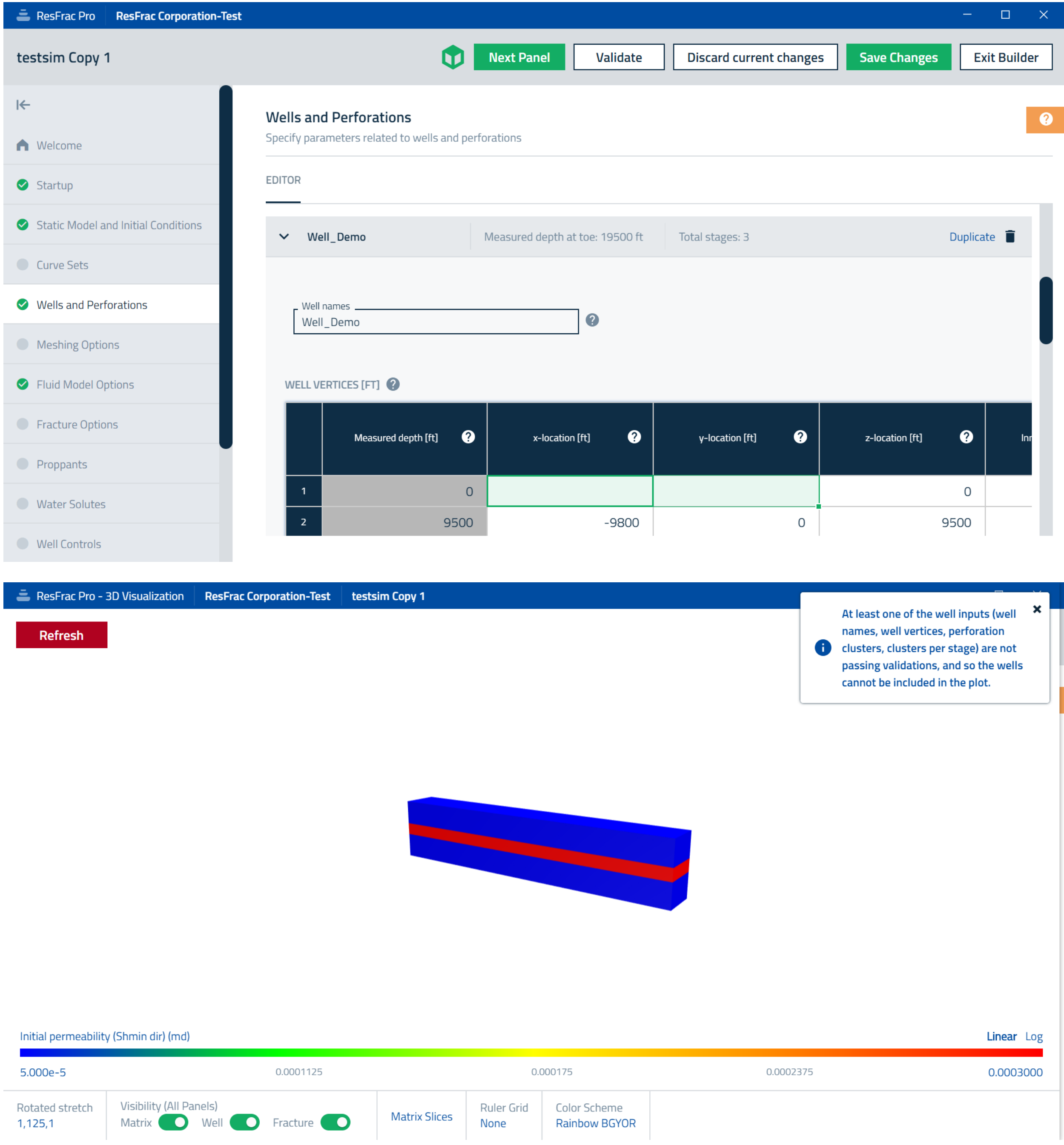

If you scroll down to the bottom of the panel, you will see a ribbon to unfold 'Advanced options.' Most of the time, users do not need to modify advanced options. Therefore, we have hidden them from your view to avoid them being distracting!

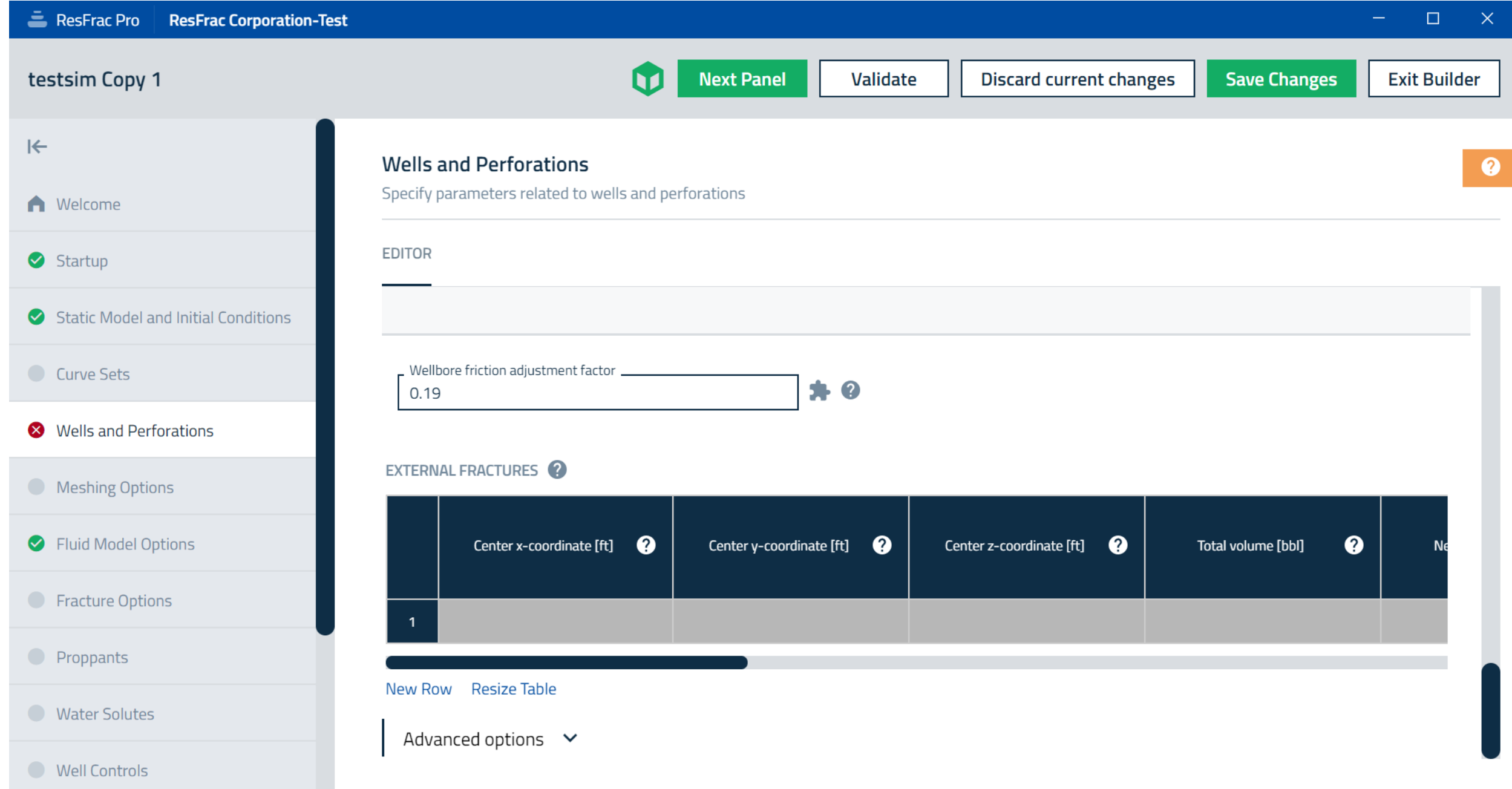

Hopefully, this section has given you a good overview of the simulation builder. For more detailed step-by-step tutorials on setting up simulations, history matching, and optimizing design, refer to Sections 6-10. Exit the builder and return to the job manager.

## 5.6 Tracking simulations in progress, file formats, and other job manager details

By now, the simulation that you submitted has probably started running on the cloud. Before we look at the results in the visualization tool, let's explore more capabilities in the job manager.

In the Sandbox view, you can see a list of your simulations. The simulation that you submitted earlier should now show a status of 'Running.' If you click on the carrot icon on the left of the row, it pulls up a 'short status' description providing the timestep number, simulation duration, and other information.

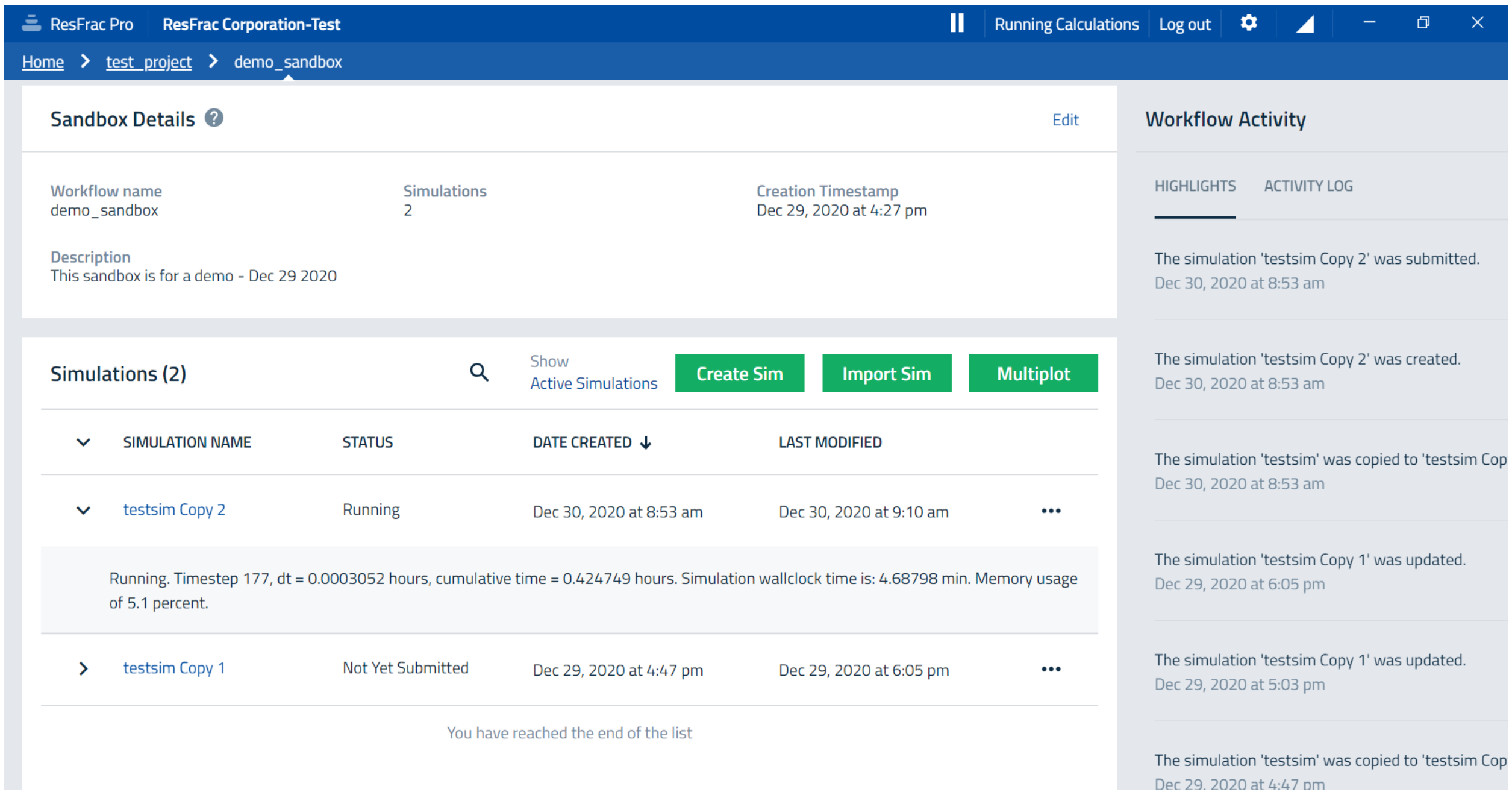

Now, let's talk through the remaining options in the menu that pulls up when you click on the name of the simulation.

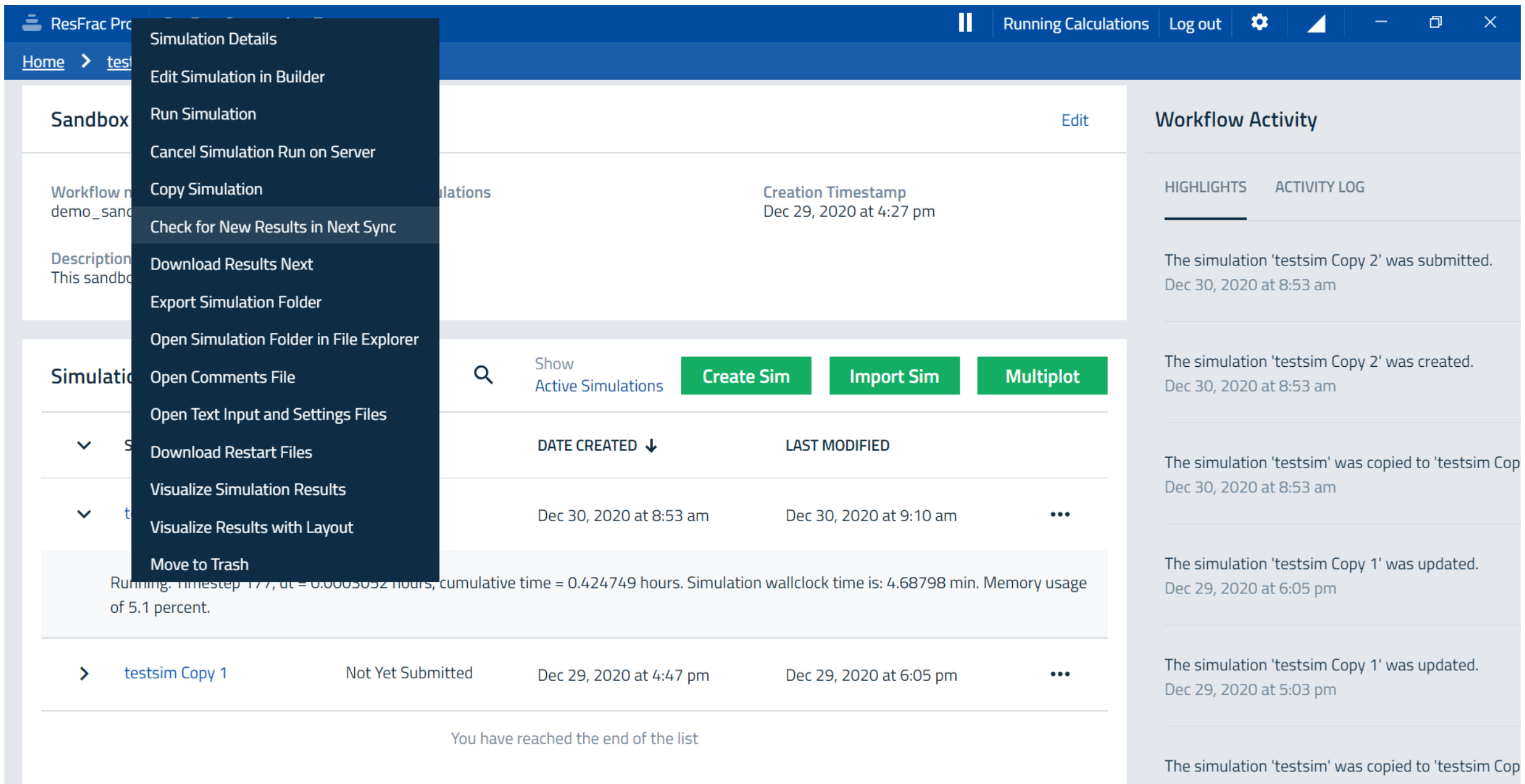

If you click 'Cancel Simulation Run on Server,' then obviously, the simulation is stopped. However, the results that have already been generated will not be deleted from your computer.

Simulations run on the server are stored there for about 30 days, and then they are deleted. You are expected to keep track of your own simulations using your own local hard drive. We do not store simulation results on the cloud long-term.

If there are multiple simulations running at the same time, then there may be some delay as the computer downloads all the results at the same time. If you want to prioritize the downloading of a particular simulation (or multiple simulations) click on each and select 'Download Results Next.' The next the download cycle, the simulation(s) that you select will be prioritized and downloaded first.

The button 'Check for New Results in Next Sync' should typically not have any effect. It is a backstop. If there is an error in the refresh cycle and results from a simulation are not fully downloading, then the 'Check for New Results in Next Sync' button forces the UI to try again and check for new results that need to be downloaded.

Next click 'Download Restart Files.' A restart file allows you to create a new simulation that picks up and runs from a certain point in time from another simulation. In the restart, you can change things like the permeability or relative permeability. This makes restart files convenient for history matching. We also use restart files for reproducing and fixing simulator problems. Restart files are intended to be 'perfect,' such that they reproduce the previous simulation identically, to machine precision. However, code changes over time occasionally introduce small flaws in the restarts. If you notice that a restart is not perfectly reproducing results, please report this to us (see Section 5.10 on reporting simulation problems) so that we can track it down!

The restart files are generated automatically by the simulator at timestep intervals (controlled by 'Restart time step interval' in the simulation builder). The files are stored on the server for about 30 days and then deleted. If you want to access a restart file, you need to download it to your hard drive within 30 days of running the simulation. If more than 30 days have passed, you can rerun the simulation to make new ones (and to ensure consistency with an old simulation, make sure that you run with the same version number as the original simulation).

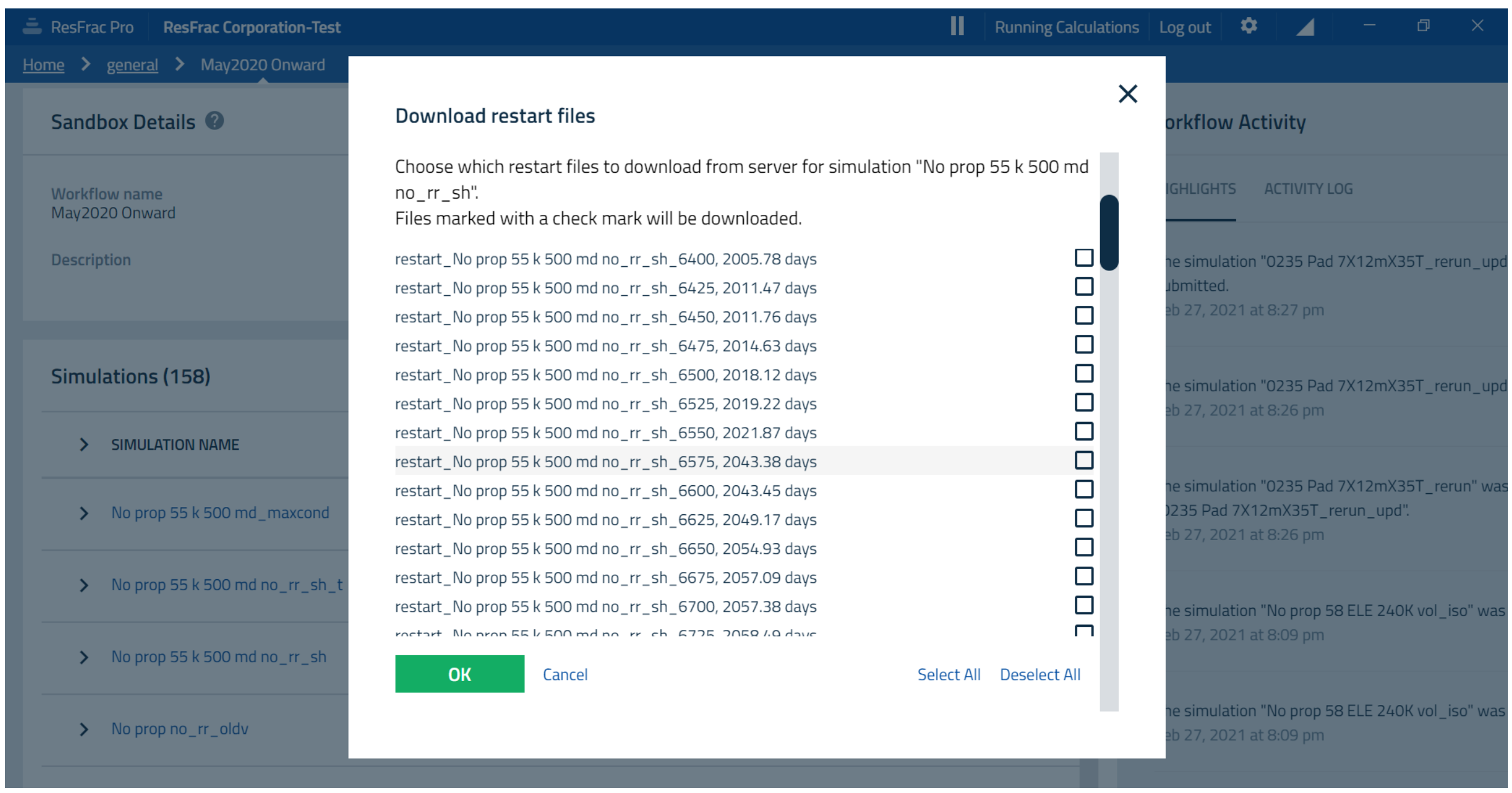

Go ahead and download a restart file by clicking one of the rows, and then clicking OK. The restart files are listed by 'timestep number.' More recent versions of the code may also list the simulation time corresponding to each restart file. If you do not see simulation time, you can refer to the comments file to determine which timestep number corresponds to the point in time that you are interested in.

After the restart file has downloaded, click on the simulation and copy it. You will be prompted to ask if you want to also copy the restart file. If you click 'yes,' then the restart file will be copied along with the simulation files themselves, and when you run the new simulation, you will have the option to run a new simulation from scratch or a restart using the restart file. If a restart file is available, the program will default to use it when you submit the simulation. In the example below, I downloaded the restart file corresponding to timestep 100. So if I run a restart, the first timestep in the new simulation will be timestep 101 of the simulation.

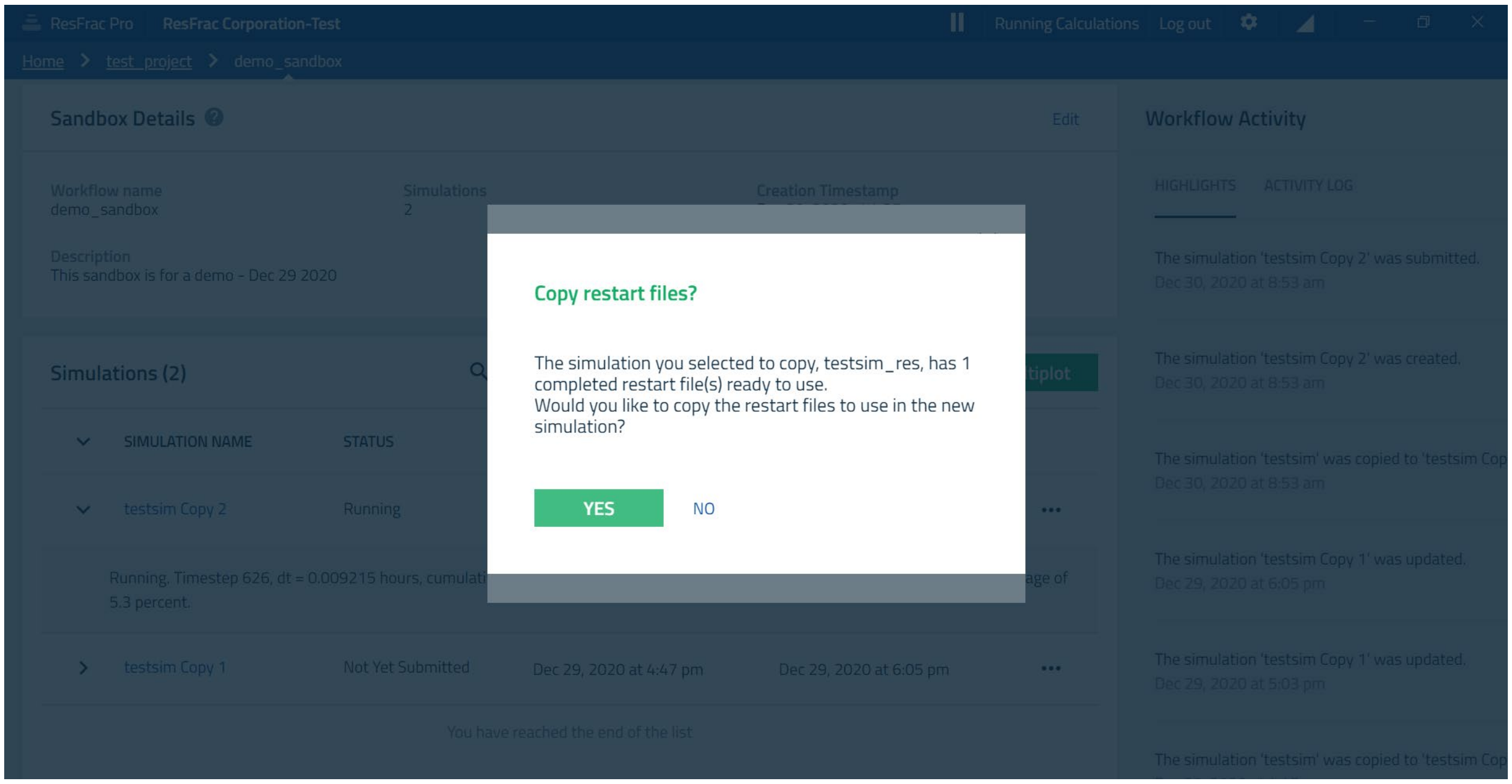

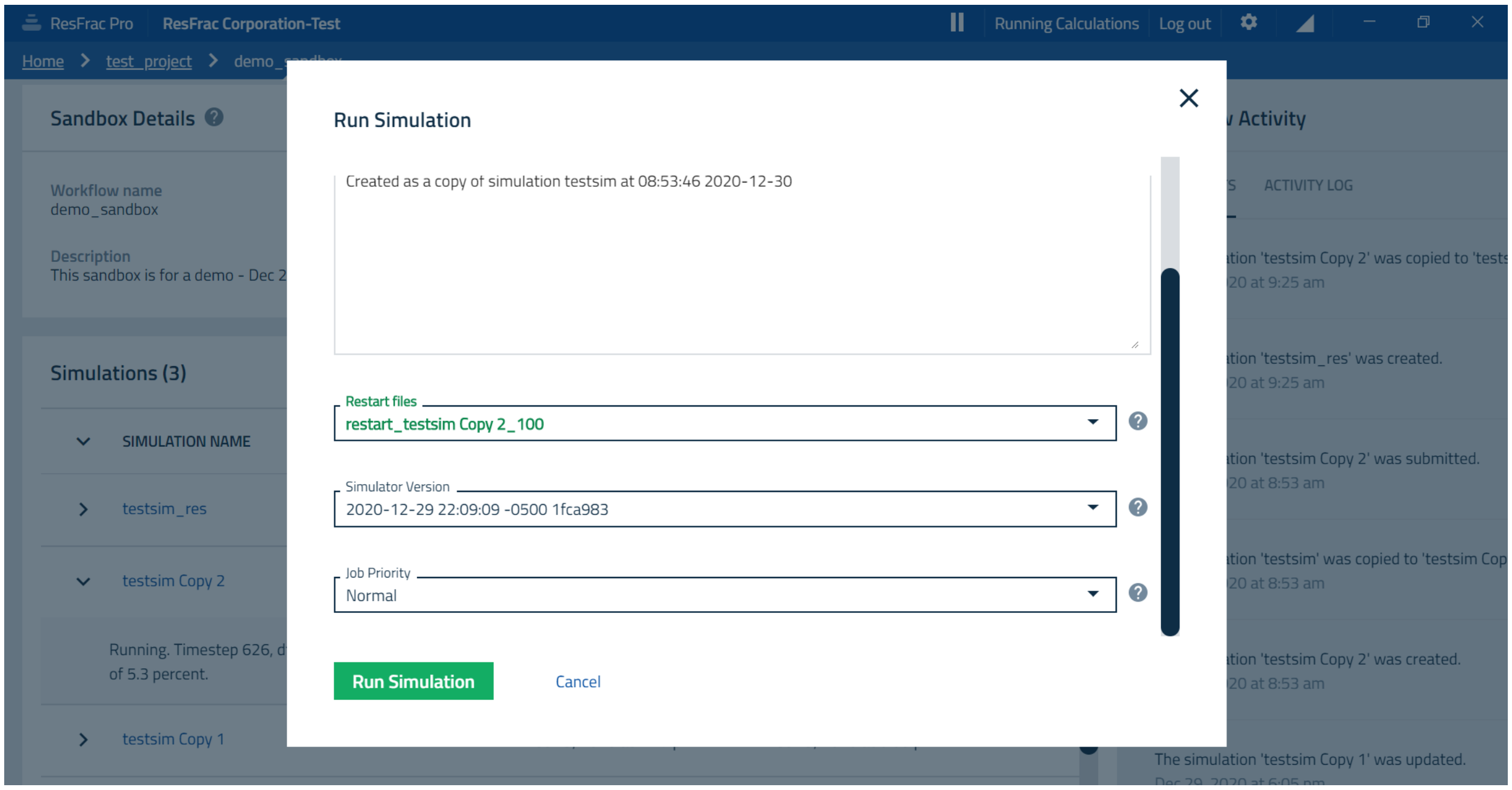

## 5.7 Simulation folders

Next, click ‘Open Simulation Folder in File Explorer.’ This pulls up an actual folder on your computer in the Windows File Explorer. Every simulation in the sandbox has its own self-contained folder.

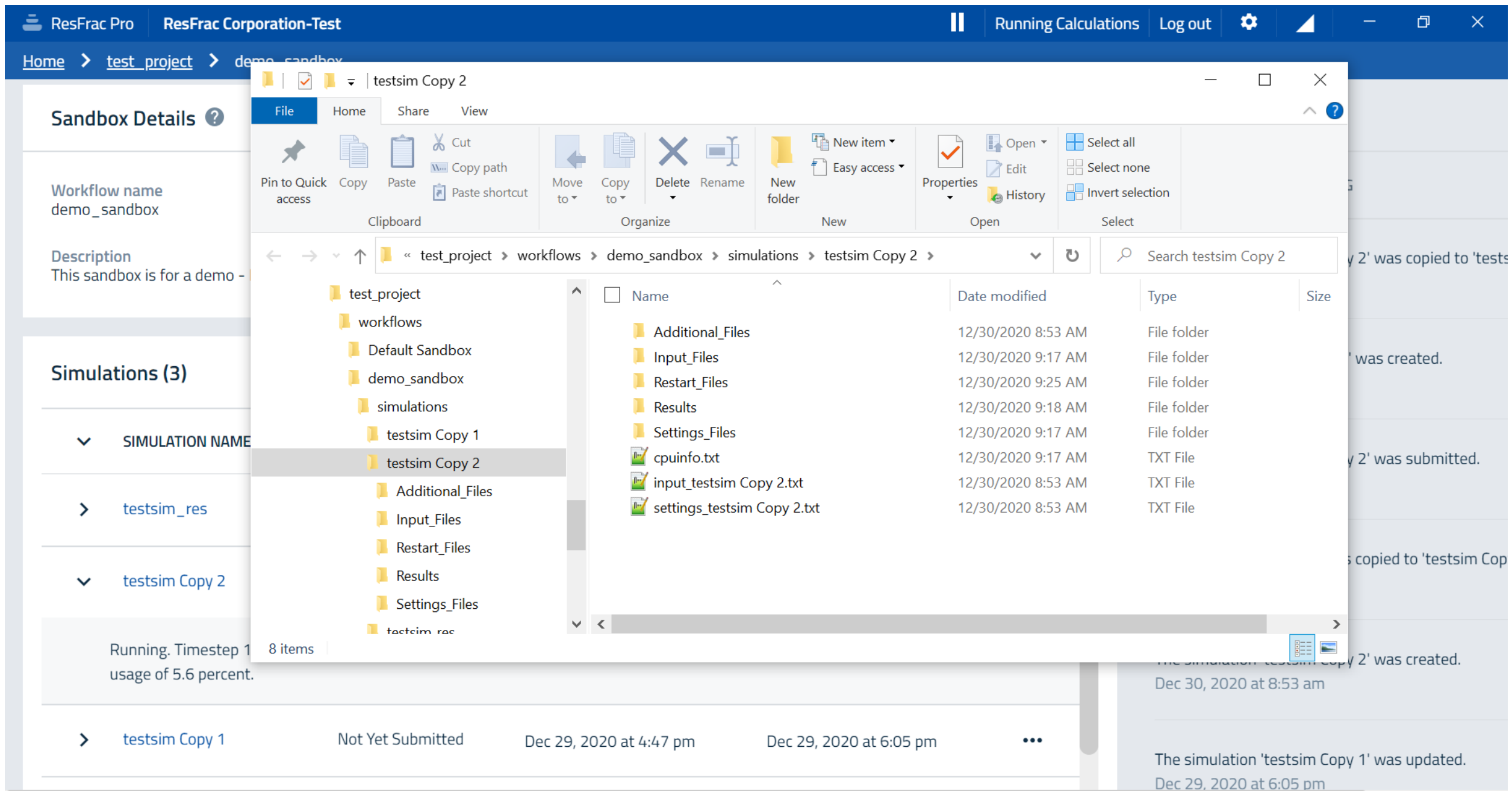

The ‘input’ and ‘settings’ files contain the simulation-specific input parameters. These are human-readable text files in a ResFrac-specific, simple format. You never need to modify these files because they are generated by the simulation builder. However, you may find it convenient to directly modify the text files once you become a more advanced user. Go ahead and open the settings file, for example. The file consists of a series of keywords. Each has the format of ‘Variable name:”, ‘Length:’ and ‘Value(s):”. Optionally, you can abbreviate these as ‘name’, ‘length’, and ‘val’. Lines starting with two slashes “//” are ‘comment’ lines that are ignored by ResFrac as it parses the file. The ResFrac UI automatically inserts comments above each variable that explain what it is. Some keywords require several ‘tab-delineated’ values to be entered in each line. ResFrac always uses the ‘tab’ to separate entries in a settings file. This has to be a ‘tab’, not four spaces. We agree with Richard Hendricks – tabs are better than four spaces!

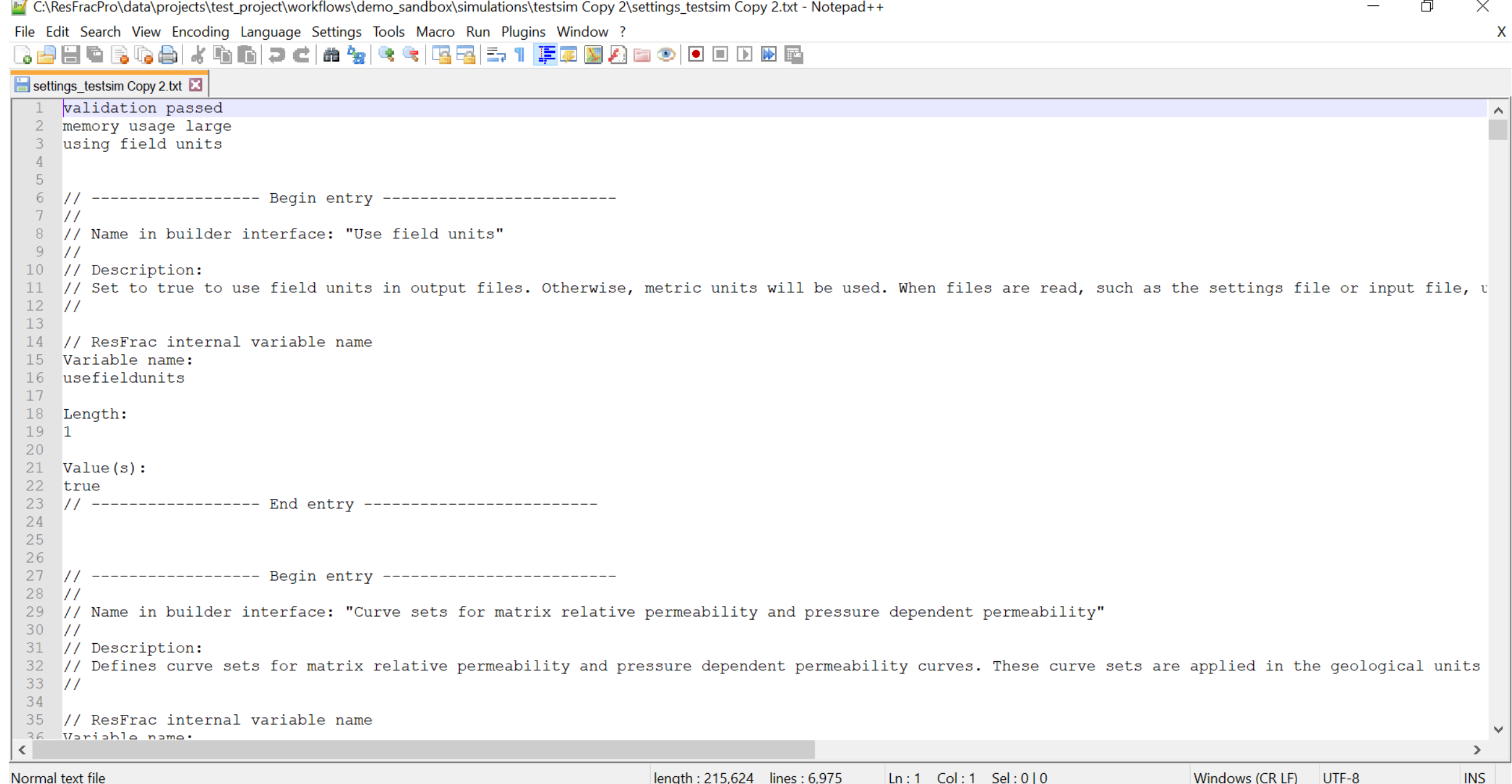

Returning to the simulation folder, you only really need to worry about the ‘Additional_Files’ folder and the ‘Results’ folder. Additional_Files is used for ‘extra’ simulation files, such as used with corner point simulations. The Results folder contains all the simulation results; let’s open that folder.

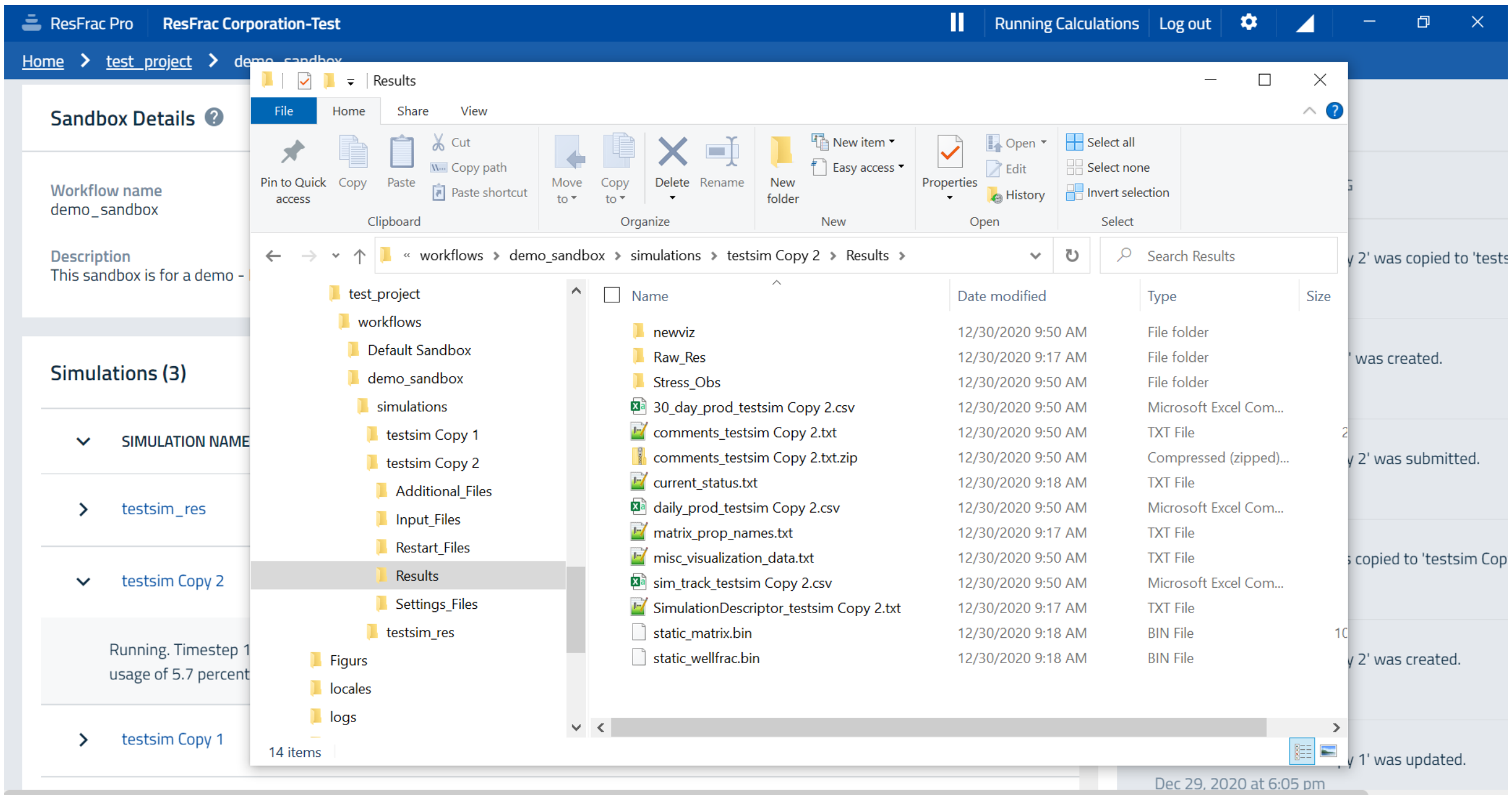

Some of the files and folders in the Results folder are for internal use by the UI, but some of them are useful for you to be aware of.

The comments file is very important and useful and discussed in Section 5.9 below. The ‘sim_track’ file outputs an Excel-readable format (csv) with a variety of ‘line-series’ data: WHP, BHP, injection rates, production rates, etc., for every timestep. For convenience, the ‘30_day_prod’ and ‘daily_prod’ files output production rates per well on regularly spaced increments. This is useful because the sim_track files can have thousands of rows and many columns with a lot of different output properties. Users find it useful to have the basic daily and monthly production data outputted in small, convenient files.

The 'Raw_Res' folder contains various human-readable summary files. They provide information about production and frac length by layer, a gun barrel view of proppant placement, and element-by-element properties. The 'A-README' file in that folder gives more detail.

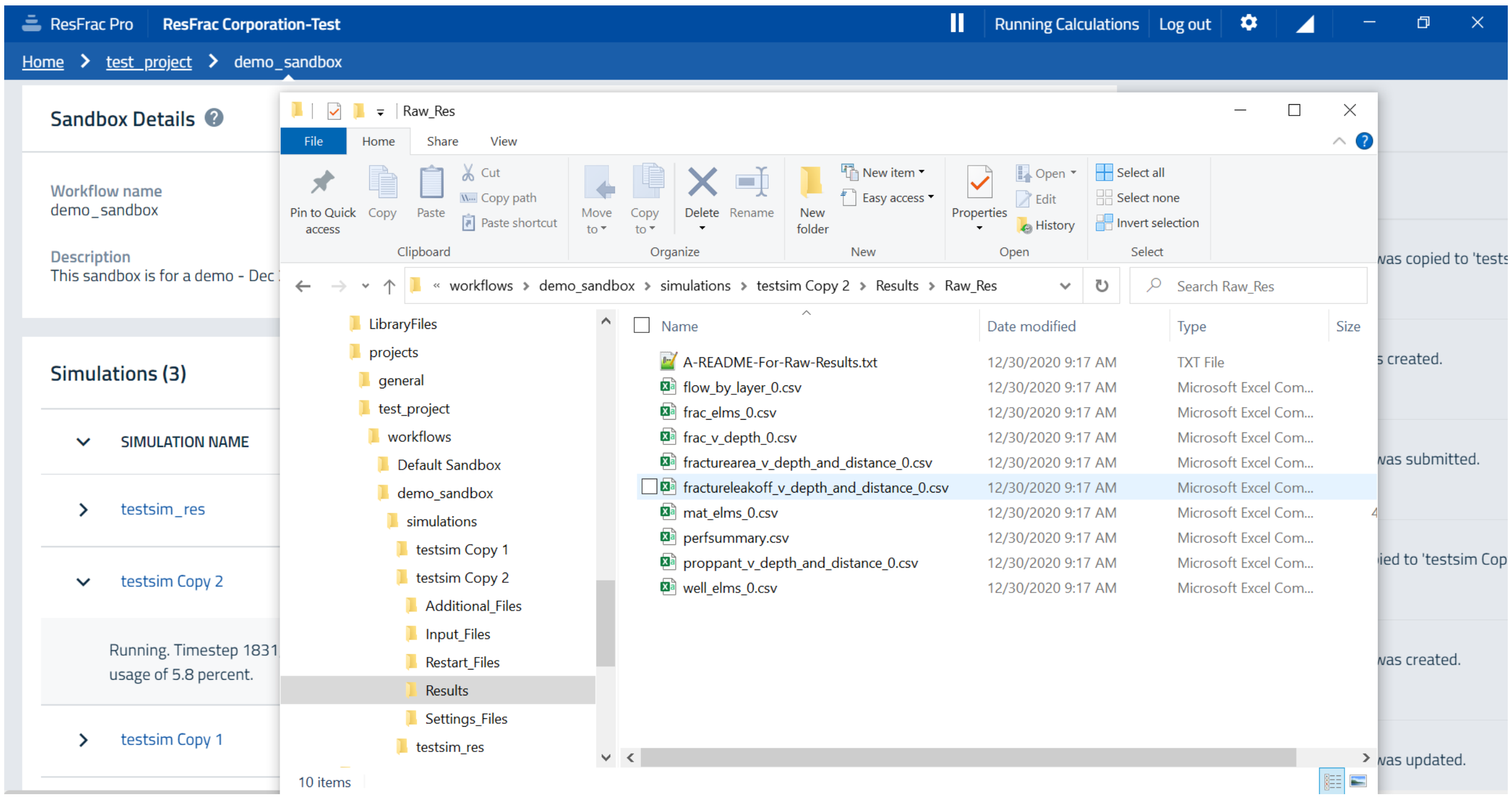

## 5.7.1 Manual manipulation of input and settings text files

As mentioned at the beginning of this section, some users will choose to open the input and settings files in a text editor and manually edit the values within. This can be convenient for users who are experienced with ResFrac and other simulation software. However, please exercise caution when doing this, as doing so can cause a variety of problems. Changes within the input and settings files sometimes need to be coordinated across several locations for the files to be self-consistent.

Additionally, there is one pitfall related to additfor settings that users should be aware of. If you edit the settings file manually, do not delete or modify the "`// #----------------# End Additional Manual Settings #----------------------#`" line at the end of the additionalmanualsettings field. Even though this is a commented-out line, it is essential to the functioning of additional manual settings. Removing or changing the line (specifically, the words "End Additional Manual Settings") will cause the settings file to be read incorrectly by the builder and simulator.

```
// #----------------# Begin Additional Manual Settings  #----------------------#
//
// Name in builder interface: "Additional manual settings"
//
// Description:
// Any text you write here will appear directly in the settings file. This is intended to be used as a way to manually write
settings into the settings file. No validation checks are performed on text entered here.
//

Variable name:
additionalmanualsettings
<Additional manual settings text here>

// #----------------# End Additional Manual Settings #----------------------#
```

## 5.8 Sharing and importing simulations

You may want to share a simulation with someone else. Click on the simulation and select 'Share simulation.' Enter a recipient email address (which by default is support@resfrac.com, though this can be changed), select a restart file to include, and add an optional message. The recipient will receive a link to download the simulation folder.

Alternatively, you may select 'Export simulation folder.' You will be prompted to select a location. The entire folder, along with an additional metadata file with information such as the 'simulation description,' will be copied to that location.

To import a simulation folder, click on 'Import Sim' from the job manager screen in your chosen Sandbox. You have three options for importing a simulation. First, you can navigate directly to an input and settings file. Those two files, alone, contain all the information required to set up and run a simulation. If you are running a corner point simulation or a simulation that otherwise has additional data files as part of the setup, they must be placed in a separate folder for 'Additional_Files', and so you are asked to optionally specify the location of that folder for the import. The second option is to select an entire self-contained simulation folder, such as the folder that you created with the 'Export simulation' folder. Note that you could also simply copy the simulation folder directly from the Windows File Explorer (without using 'Export simulation') and that would work fine, except that it would not copy some of the supplementary data, such as the 'Simulation Description' stored in the job manager. Finally, you can import a 'folder of folders.' For this, you could have copied several 'simulation folders' over from another computer, and placed them all in the same folder. You could import all of the folders at once using the 'folder of folders' option.

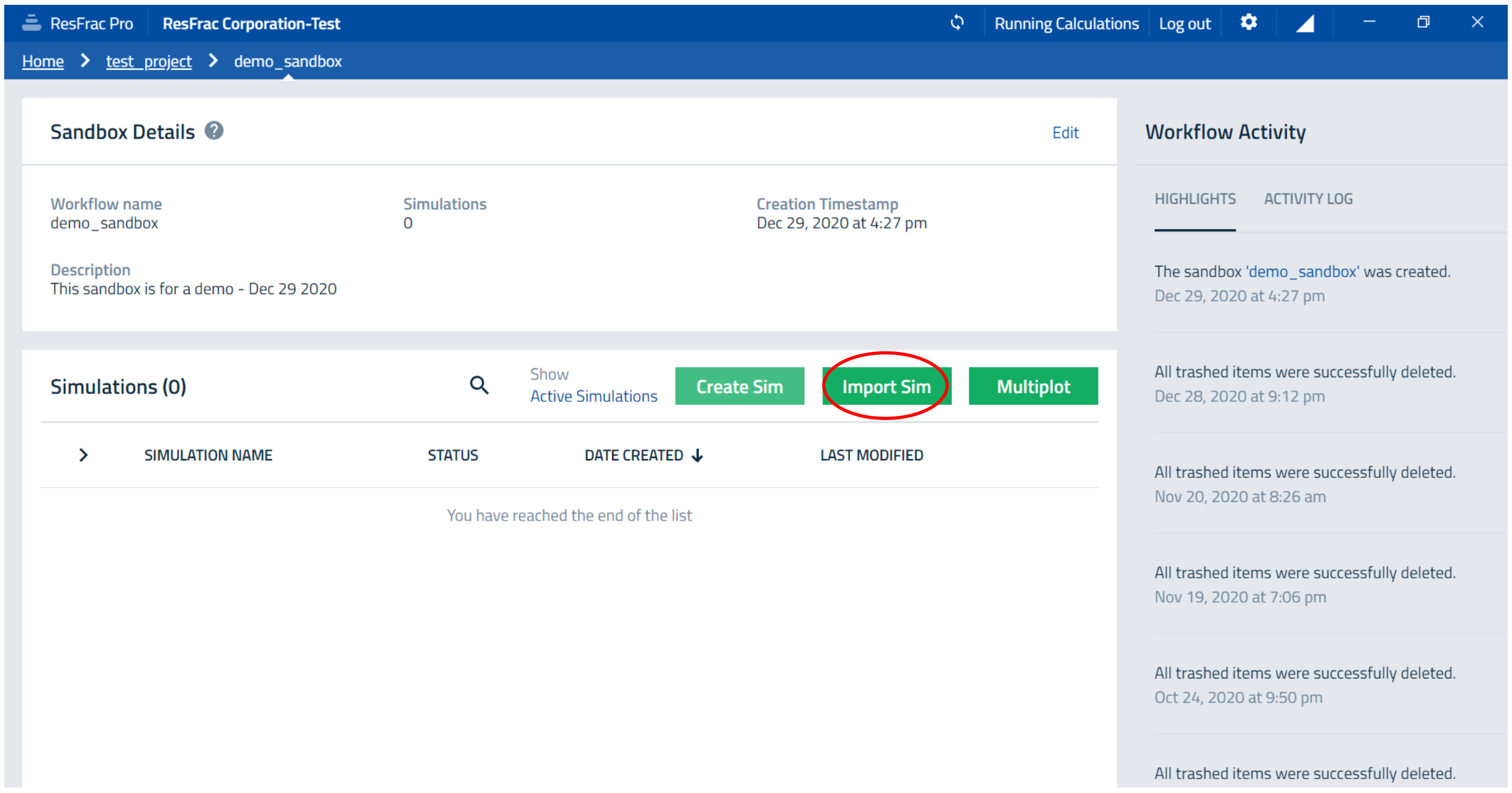

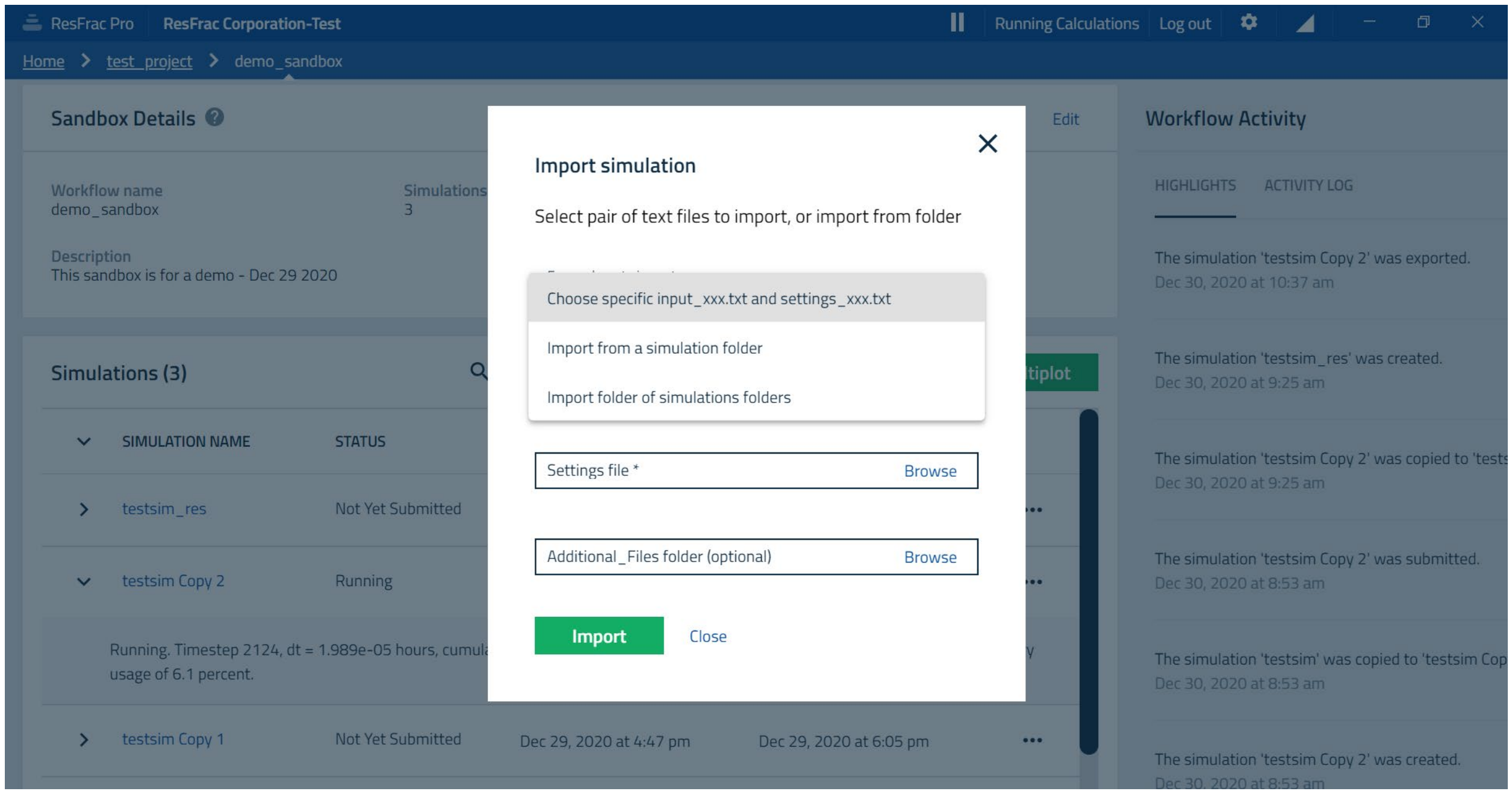

Select 'Import from a simulation folder' and select the folder created when you previously did 'Export simulation.' Note that you can do either 'Cut/paste' or 'Copy.' The former option will remove the folder from its current location when the import is performed. The copy option copies the folder so that it remains in its current location when the import is performed.

When I attempted the import, it gave me an error that the simulation name was already being used (since it has the same name as the simulation we exported from), and so I had to change the name as part of the import.

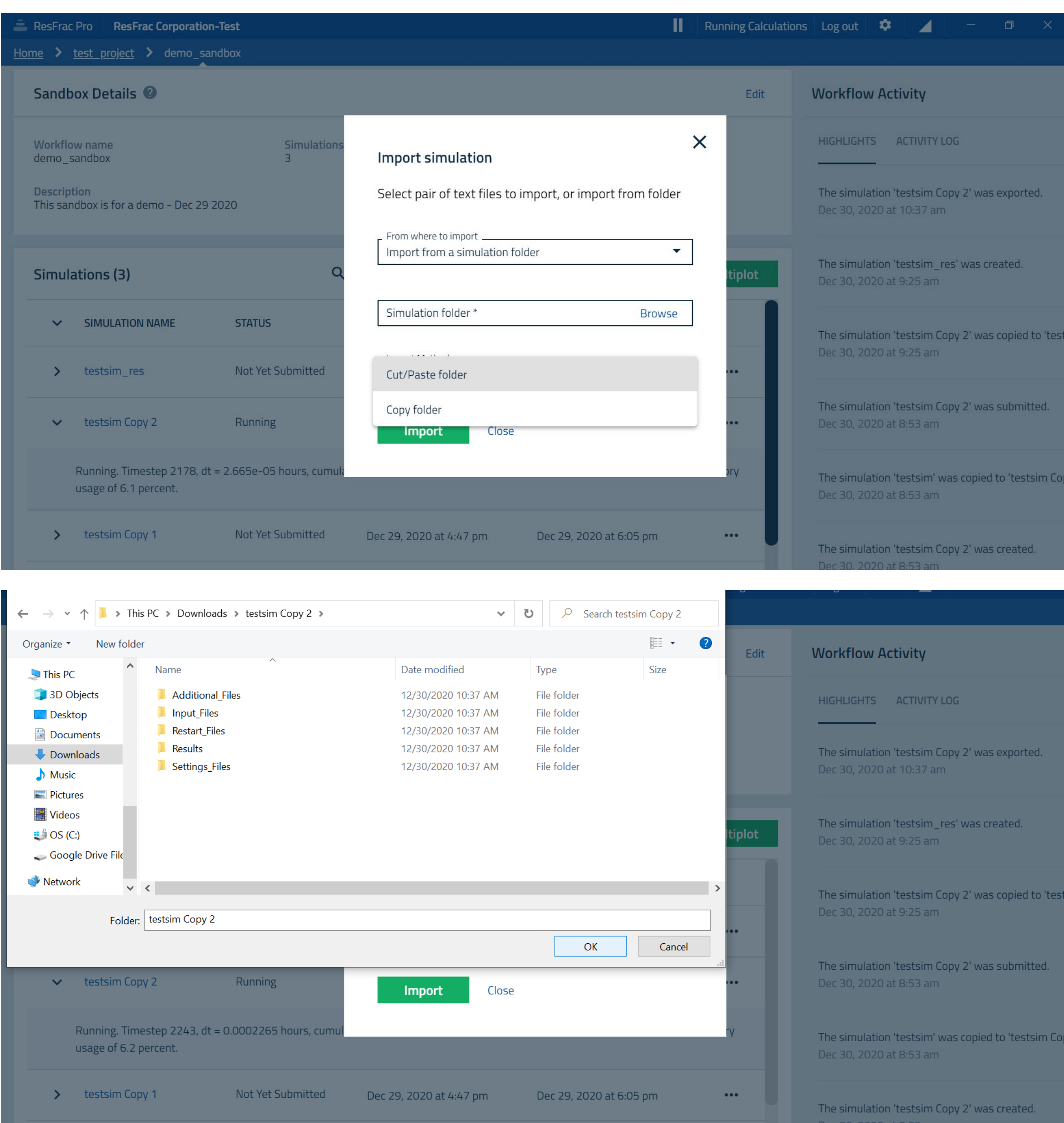

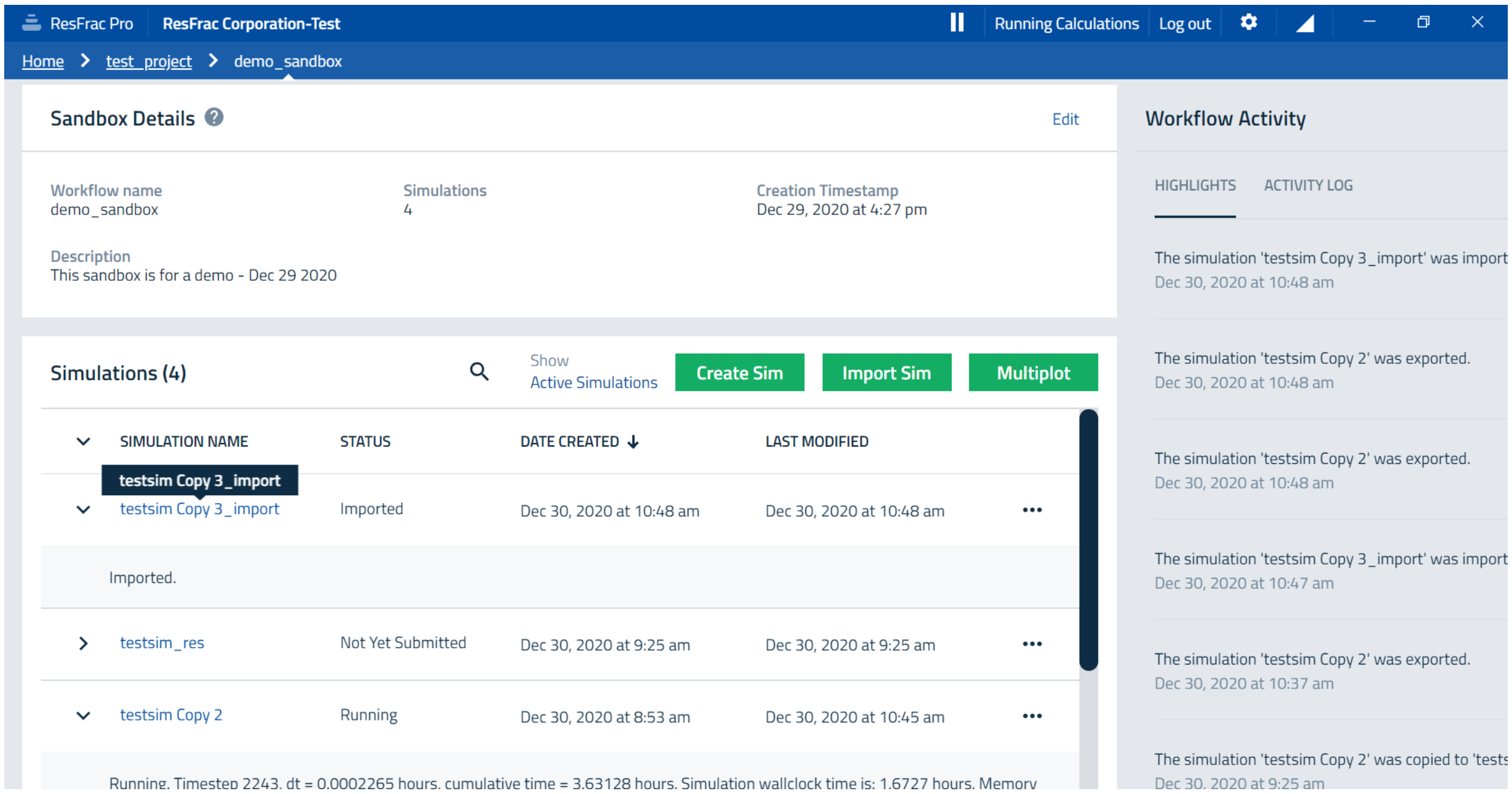

## 5.9 The comments file

The comments file is a text file with a detailed log of everything the simulator is doing. It can be useful if you want to get a more detailed look at your simulation's progress. It is placed in the 'Results' folder. Or you can access it directly from the options accessed by clicking on the name of a simulation.

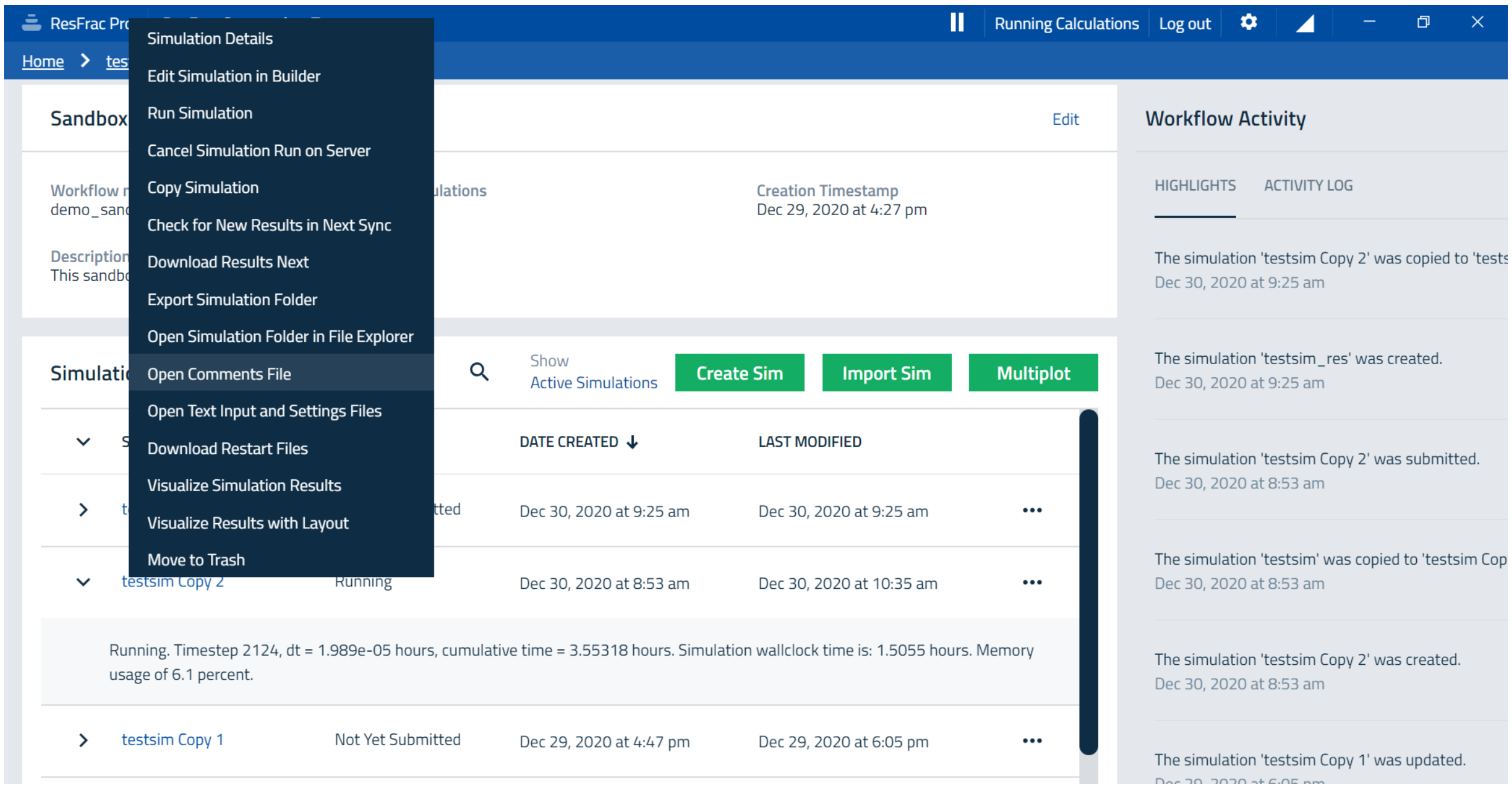

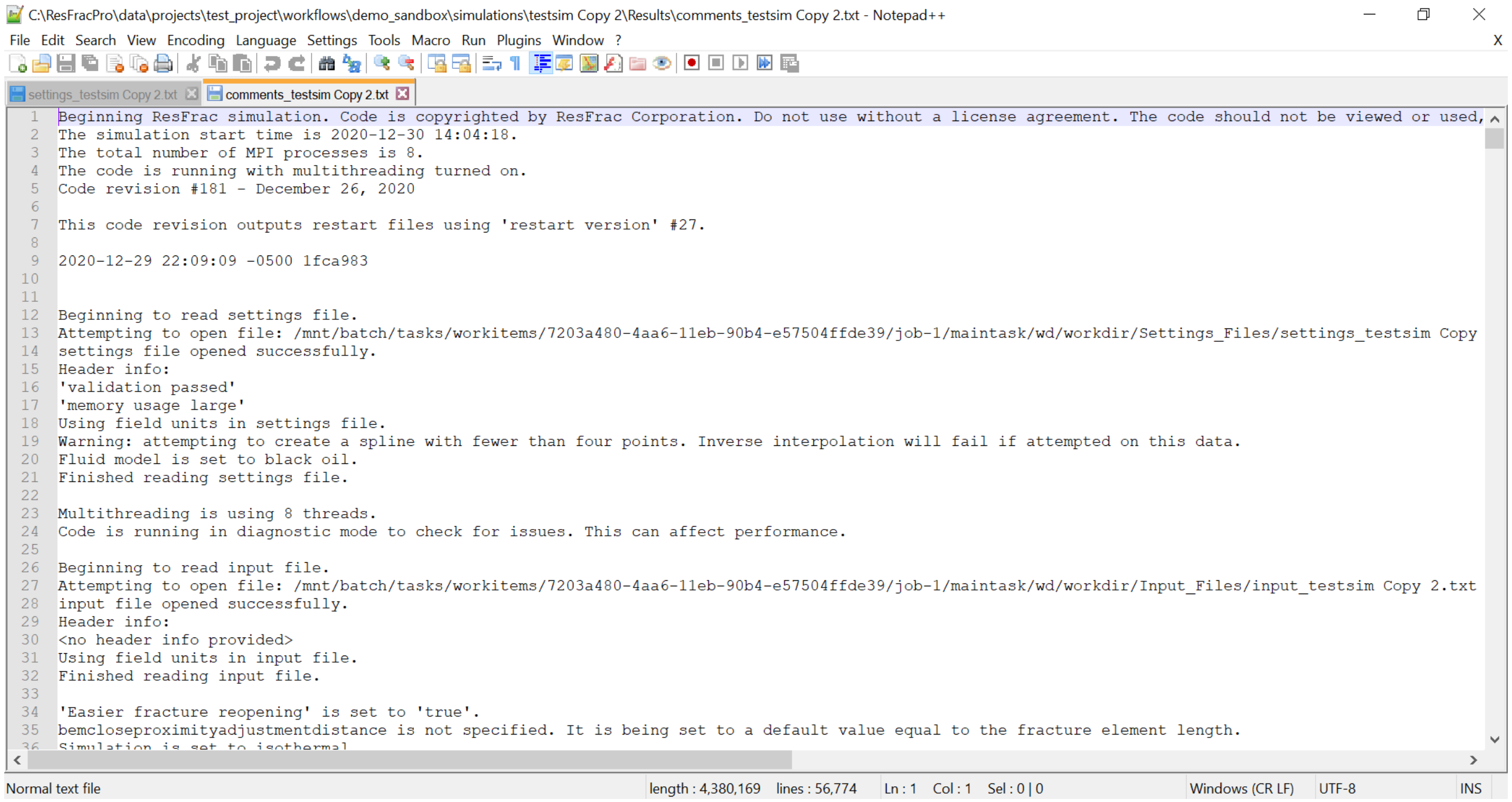

The comments file starts by providing the date/time of the simulation. The version number is given as 'Code revision #181'. However, the *true* version number is given with the hexadecimal pattern below that – 1fca983 in the example above.

For each layer defined in the simulation, the comments file prints a table of fracture conductivity versus effective normal stress for different values of proppant mass/area. The far-left column is for an unpropped fracture, and the subsequent columns are for increasing amounts of proppant.

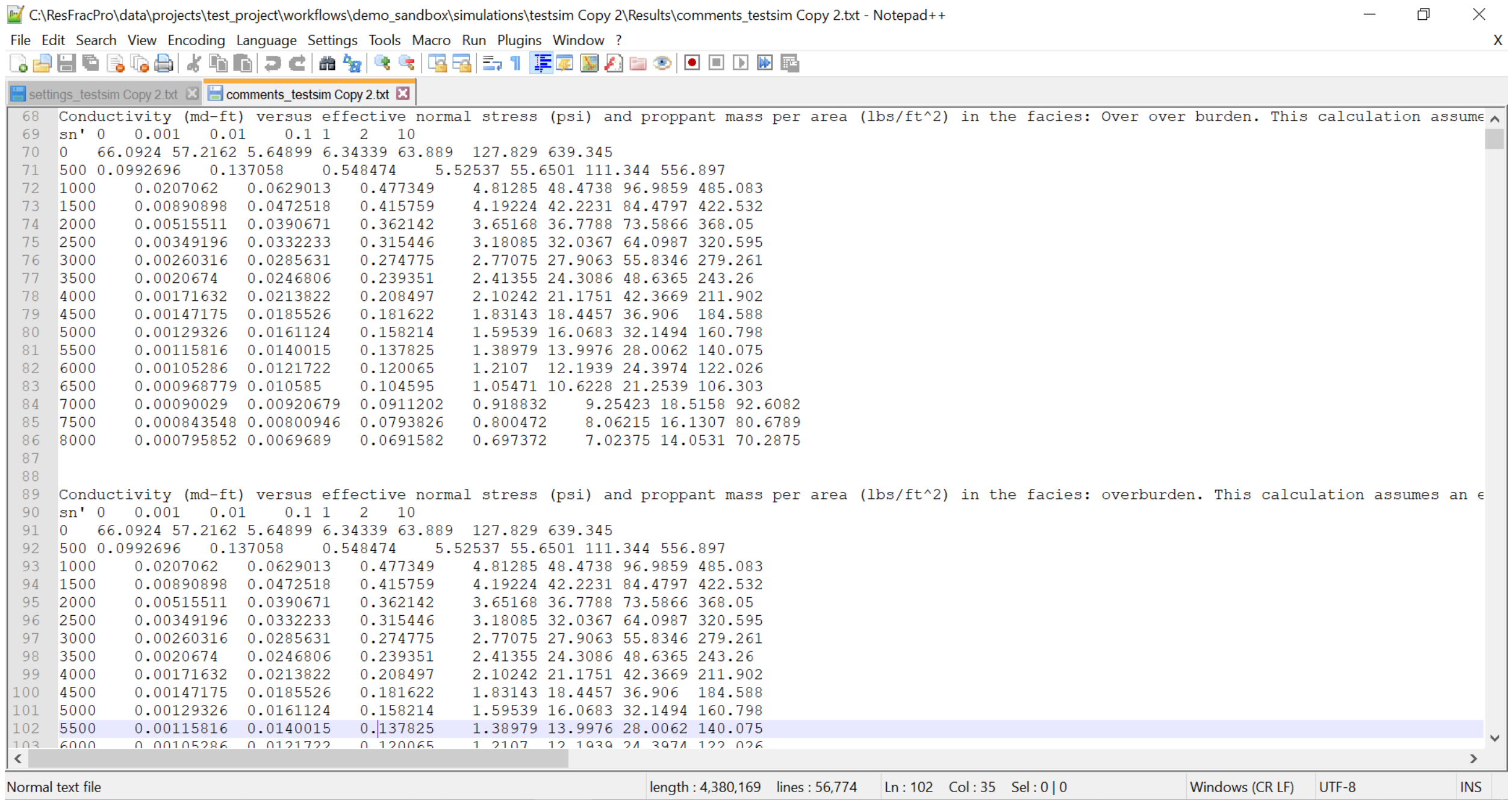

Scroll down to the bottom of the file to get the latest update on what the simulation is doing. In the screenshot below, we can see that it is beginning timestep #1992. The previous timestep had a duration of 0.0027 hours (about 10 seconds), and the cumulative time elapsed is 3.3 hours. There is a line printing the wellbore behavior – WHP, BHP, and injection rate. There is also a line printing the number of unsuccessful, discarded timesteps. This summary information allows you to get a quick look at where the simulation is, and what is happening.

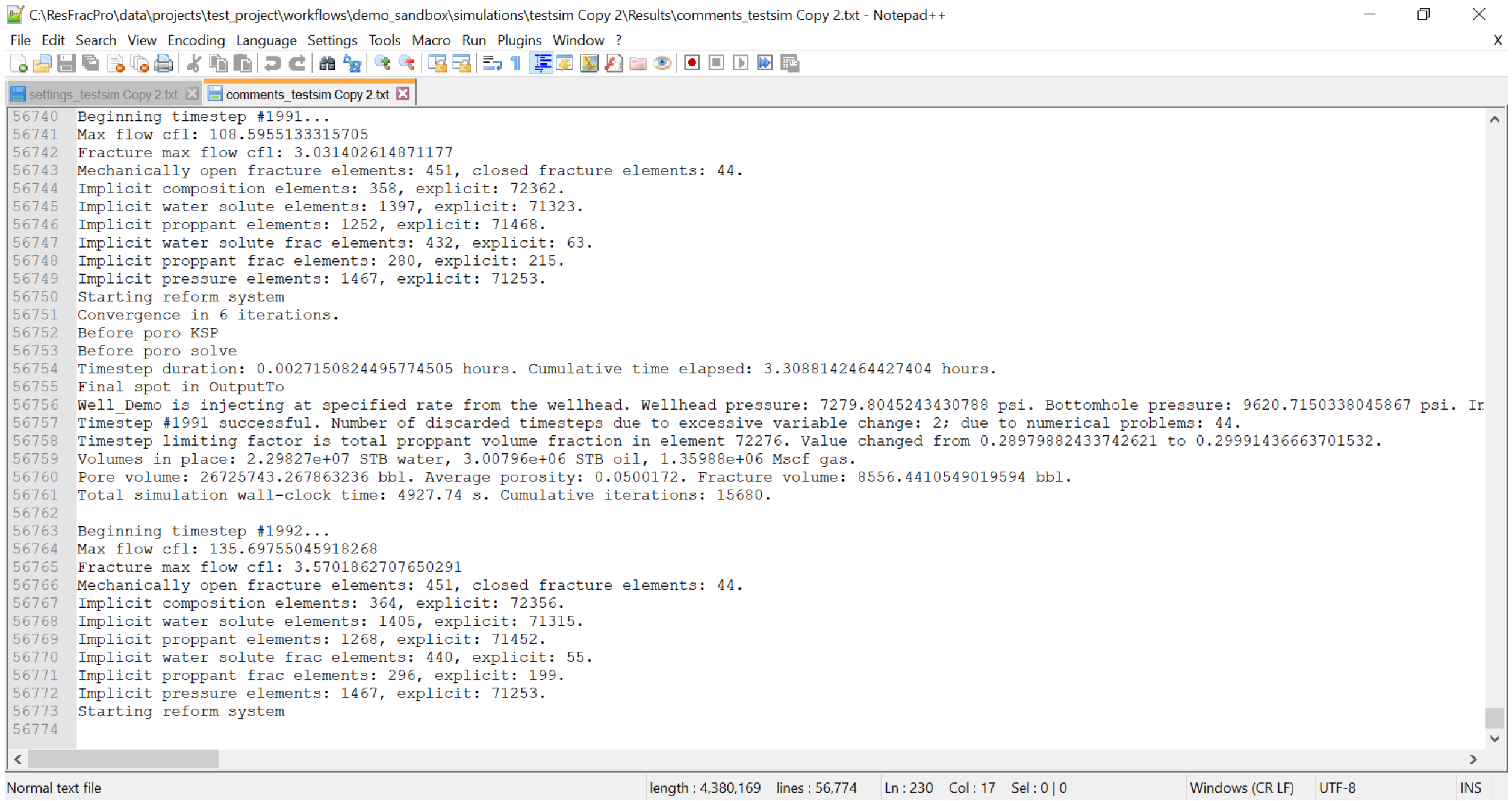

```
56740 Beginning timestep #1991...
56741 Max flow cfl: 108.5955133315705
56742 Fracture max flow cfl: 3.031402614871177
56743 Mechanically open fracture elements: 451, closed fracture elements: 44.
56744 Implicit composition elements: 358, explicit: 72362.
56745 Implicit water solute elements: 1397, explicit: 71323.
56746 Implicit proppant elements: 1252, explicit: 71468.
56747 Implicit water solute frac elements: 432, explicit: 63.
56748 Implicit proppant frac elements: 280, explicit: 215.
56749 Implicit pressure elements: 1467, explicit: 71253.
56750 Starting reform system
56751 Convergence in 6 iterations.
56752 Before poro KSP
56753 Before poro solve
56754 Timestep duration: 0.0027150824495774505 hours. Cumulative time elapsed: 3.3088142464427404 hours.
56755 Final spot in OutputTo
56756 Well_Demo is injecting at specified rate from the wellhead. Wellhead pressure: 7279.8045243430788 psi. Bottomhole pressure: 9620.7150338045867 psi. Ir
56757 Timestep #1991 successful. Number of discarded timesteps due to excessive variable change: 2; due to numerical problems: 44.
56758 Timestep limiting factor is total proppant volume fraction in element 72276. Value changed from 0.28979882433742621 to 0.29991436663701532.
56759 Volumes in place: 2.29827e+07 STB water, 3.00796e+06 STB oil, 1.35988e+06 Mscf gas.
56760 Pore volume: 26725743.267863236 bbl. Average porosity: 0.0500172. Fracture volume: 8556.4410549019594 bbl.
56761 Total simulation wall-clock time: 4927.74 s. Cumulative iterations: 15680.
56762
56763 Beginning timestep #1992...
56764 Max flow cfl: 135.69755045918268
56765 Fracture max flow cfl: 3.5701862707650291
56766 Mechanically open fracture elements: 451, closed fracture elements: 44.
56767 Implicit composition elements: 364, explicit: 72356.
56768 Implicit water solute elements: 1405, explicit: 71315.
56769 Implicit proppant elements: 1268, explicit: 71452.
56770 Implicit water solute frac elements: 440, explicit: 55.
56771 Implicit proppant frac elements: 296, explicit: 199.
56772 Implicit pressure elements: 1467, explicit: 71253.
56773 Starting reform system
56774
```

Despite our best efforts, sometimes simulations crash or run slowly due to numerical problems. The comments file can help you identify whether this is happening. If the simulation is running slowly because of a numerical problem, then the number of discarded timesteps due to numerical problems will be high – typically greater than 10%. In the example shown above, about 2% of the simulations have been discarded due to numerical problems. A percentage this low is normal and part of correct operation of the code.

## 5.10 Reporting simulation problems

A simulation crash should be easy to identify. In the job manager, the ‘Status’ will be listed as ‘Terminated.’ The simulation detailed status will be listed as something like ‘The simulation terminated prematurely.’ If you open the comments file and scroll to the bottom, you should see that it ends before reaching the prescribed maximum simulation time. Simulation crashes are not common, but they still occur occasionally. Most of the time, they are caused by problems with recently added features. Older functionality has been thoroughly tested by experience, but we update ResFrac roughly monthly, and we not always catch every possible problem prior to release.

Sometimes, a simulation does not crash, but a numerical problem causes it to run more slowly than it should. This is less obvious to diagnose, but as discussed above, this typically is indicated by more than 10% of the timesteps being discarded due to numerical problems. Also, if you scroll up and notice that the timestep duration is persistently small – less than 0.0001 hours – this probably means that timestep failures are forcing timestep durations to be limited.

A third type of problem is simply the simulation does something that it should not. For example, you thought that you set the well to inject at 90 bpm for 3 hours, but it actually injected for 5 hours. Usually, these kinds of apparent simulator problems are just a mix-up. So, start by carefully reviewing your simulation setup in the builder, and double check that the simulator really is not doing what you told it to do.

If you encounter any of these kinds of issues, please report to us by emailing support@resfrac.com. You are not annoying us – we really want to find and resolve all possible issues! We really appreciate help finding them.

To report a problem, it is important to always do the following:

1. Download the last available restart file. Or, if the problem is not occurring at the end of the simulation, download a restart file from a point in time during (or right before) the problem occurs.
2. Open the simulation folder and find the input file, the settings file, the comments file, and the restart file. The restart file is not needed if the problem occurs on startup.
3. Send us an email at support@resfrac.com. Attach the four files. We *always* need the four files, for a variety of reasons. For example, the comments file prints the exact version number, and we need to know version number to reproduce the problem.
4. In the email, include a description of the problem. Phrase the description as: "I expected the simulation to do XYZ. But actually, it did ABC." A 'bad' description would be something like 'The injection pressure looks weird.' Alternatively, a 'good' description would be: "I expected the injection pressure of Well HJK to do XYZ at 123 hours because of ABC. Instead, it did EFG."

We do our best to resolve simulator issues ASAP. We take this very seriously! Because the simulations are performed on the cloud, bug fixes can be rolled out very quickly.

## 5.11 The job manager settings screen

Click on the gear icon in the upper right corner to pull up a settings screen. The 'Refresh Database' button is useful when you want to clean out old simulations. Rather than deleting them individually from the job manager, you can delete the folders corresponding to each simulation directly from the Windows File Explorer. This is a bit faster and more convenient if you're deleting a lot of simulations. Once you have done this, press the 'Refresh Database' button and the simulations that you deleted will be automatically removed from the list of simulations in the sandbox(es).

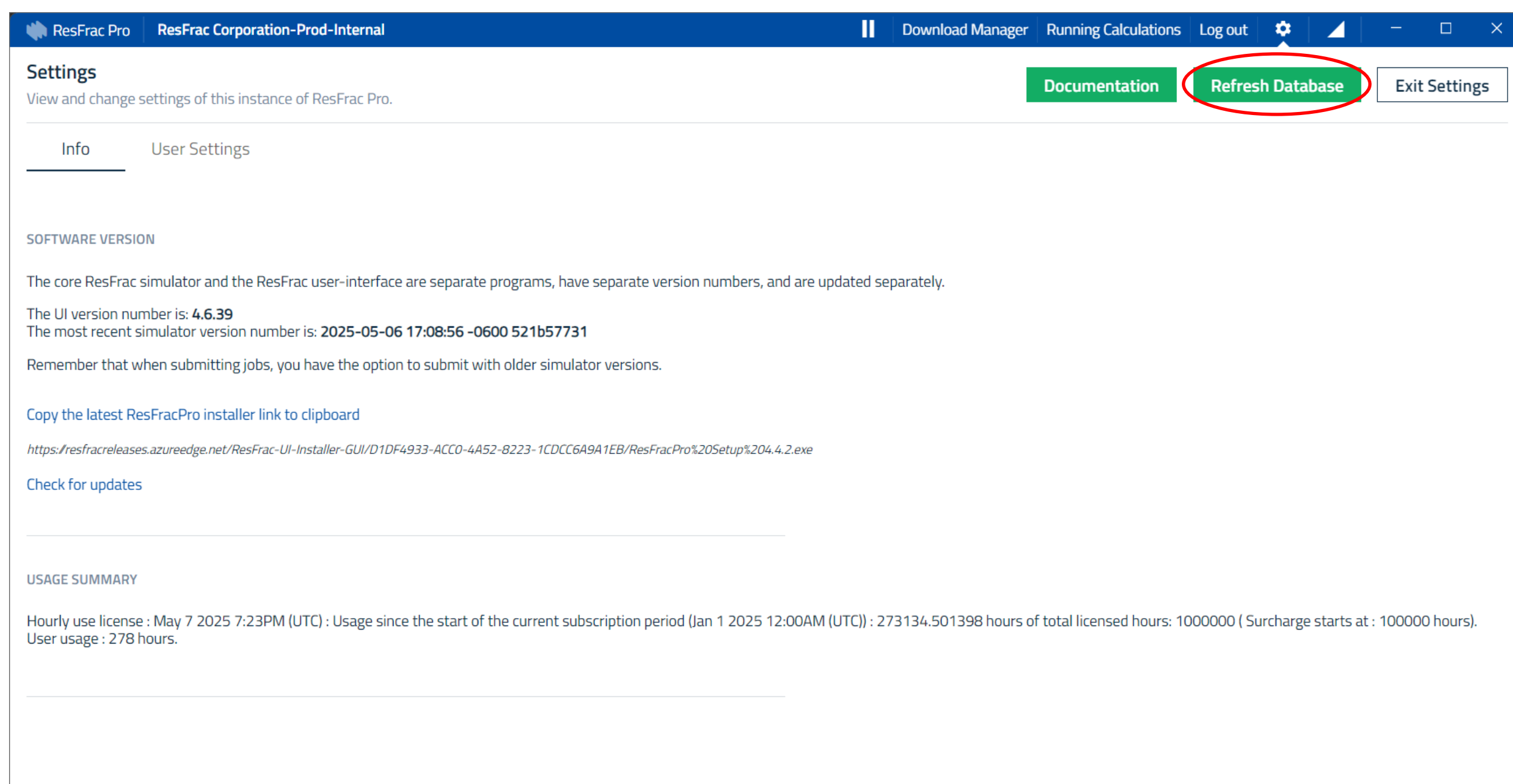

ResFrac Pro
ResFrac Corporation-Prod-Internal
Download Manager
Running Calculations
Log out
Settings
View and change settings of this instance of ResFrac Pro.
Documentation
Refresh Database
Exit Settings
Info
User Settings
SOFTWARE VERSION
The core ResFrac simulator and the ResFrac user-interface are separate programs, have separate version numbers, and are updated separately.
The UI version number is: 4.6.39
The most recent simulator version number is: 2025-05-06 17:08:56 -0600 521b57731
Remember that when submitting jobs, you have the option to submit with older simulator versions.
Copy the latest ResFracPro installer link to clipboard
https://resfracreleases.azureedge.net/ResFrac-UI-Installer-GUI/D1DF4933-ACC0-4A52-8223-1CDCC6A9A1EB/ResFracPro%20Setup%204.4.2.exe
Check for updates
USAGE SUMMARY
Hourly use license : May 7 2025 7:23PM (UTC) : Usage since the start of the current subscription period (Jan 1 2025 12:00AM (UTC)) : 273134.501398 hours of total licensed hours: 1000000 ( Surcharge starts at : 100000 hours).
User usage : 278 hours.

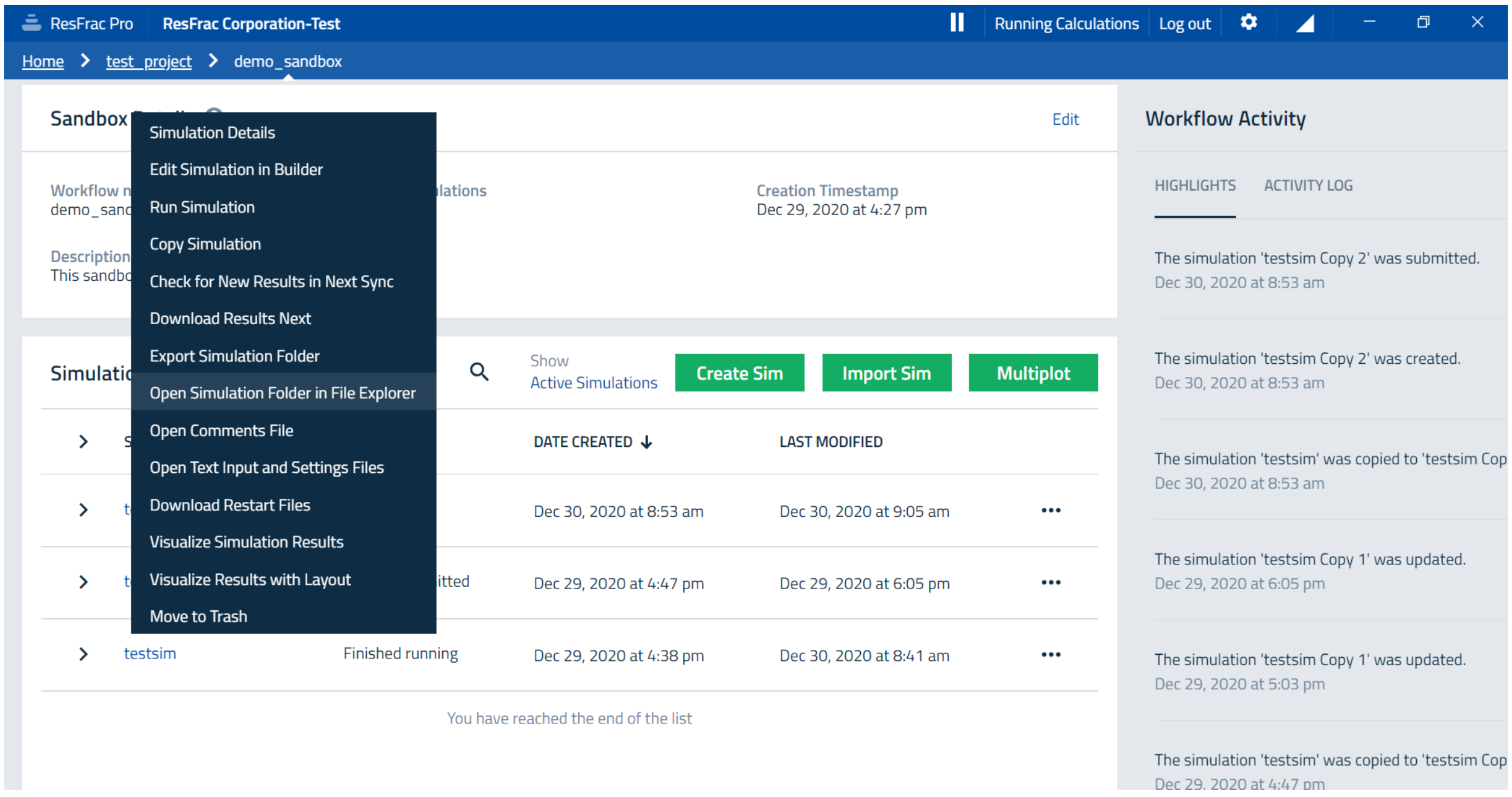

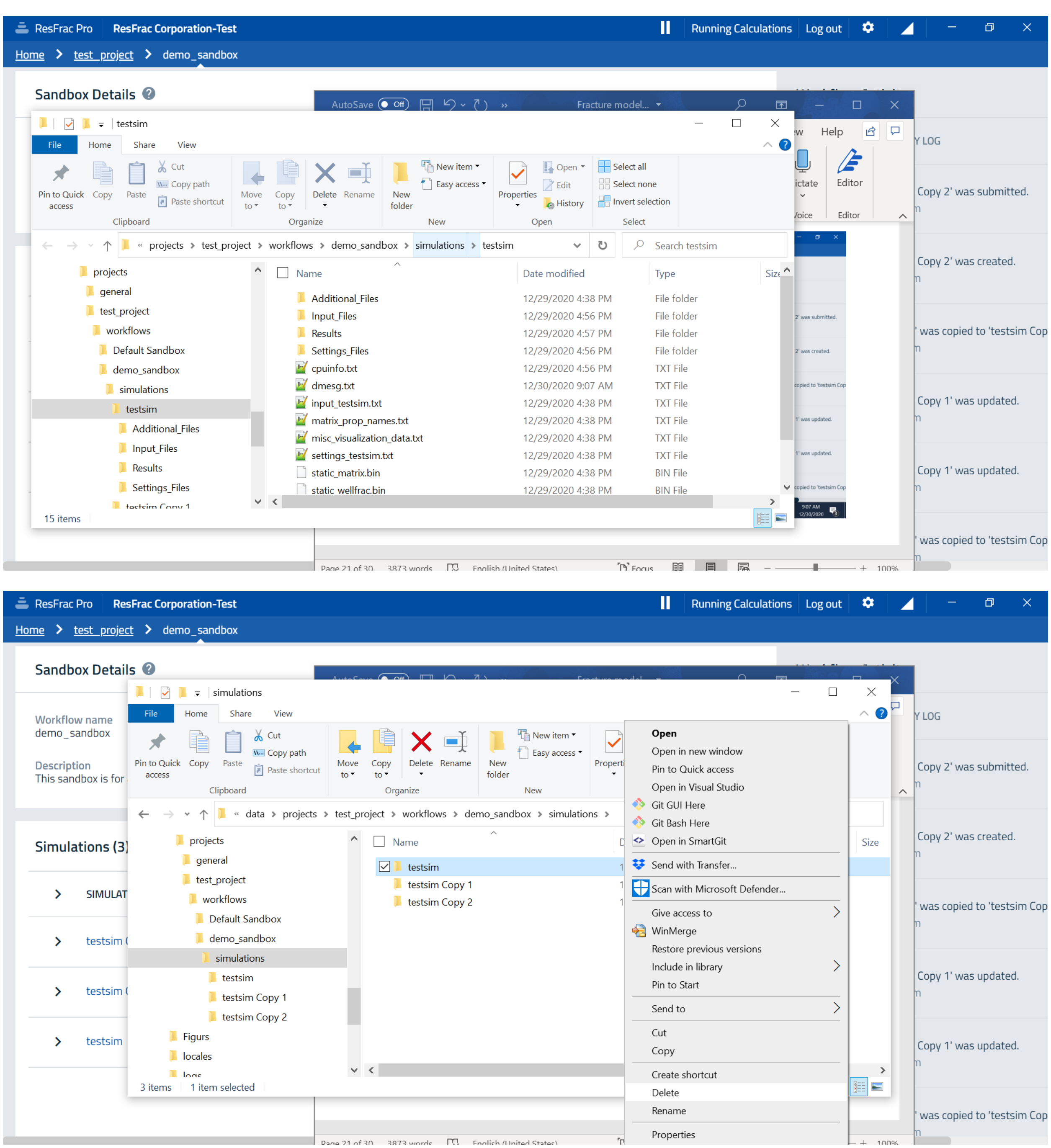

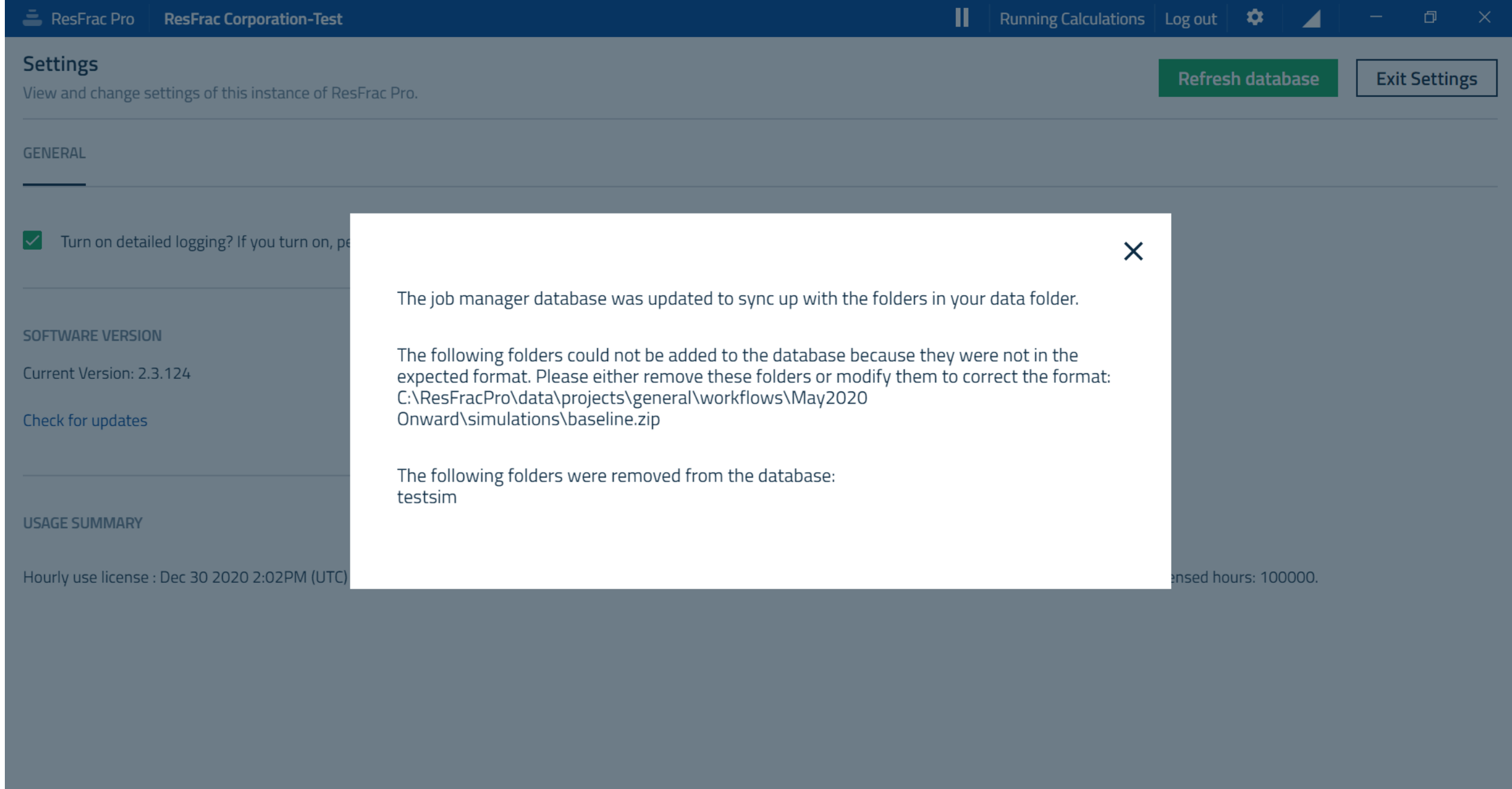

You should typically leave the 'detailed logging' button unchecked. Detailed logging is helpful for us to track down the source of problems that users report. However, it causes some antivirus software to severely slow down the responsiveness of the user-interface. It is possible to tell the antivirus software to ignore ResFrac as it writes to the log, but this is a hassle, so we do not ask users to do it.

The software version lists the user-interface software version. In newer user-interface versions, this screen may print the version number for both the UI and the simulator on the cloud (which as discussed above, are separate pieces of software with separate version tracking). If you are not sure if you have the latest version of the UI, click 'Check for updates.'

## 5.12 The visualization tool

To visualize simulation results, click on the name of the simulation and select 'Visualize Simulation Results.'

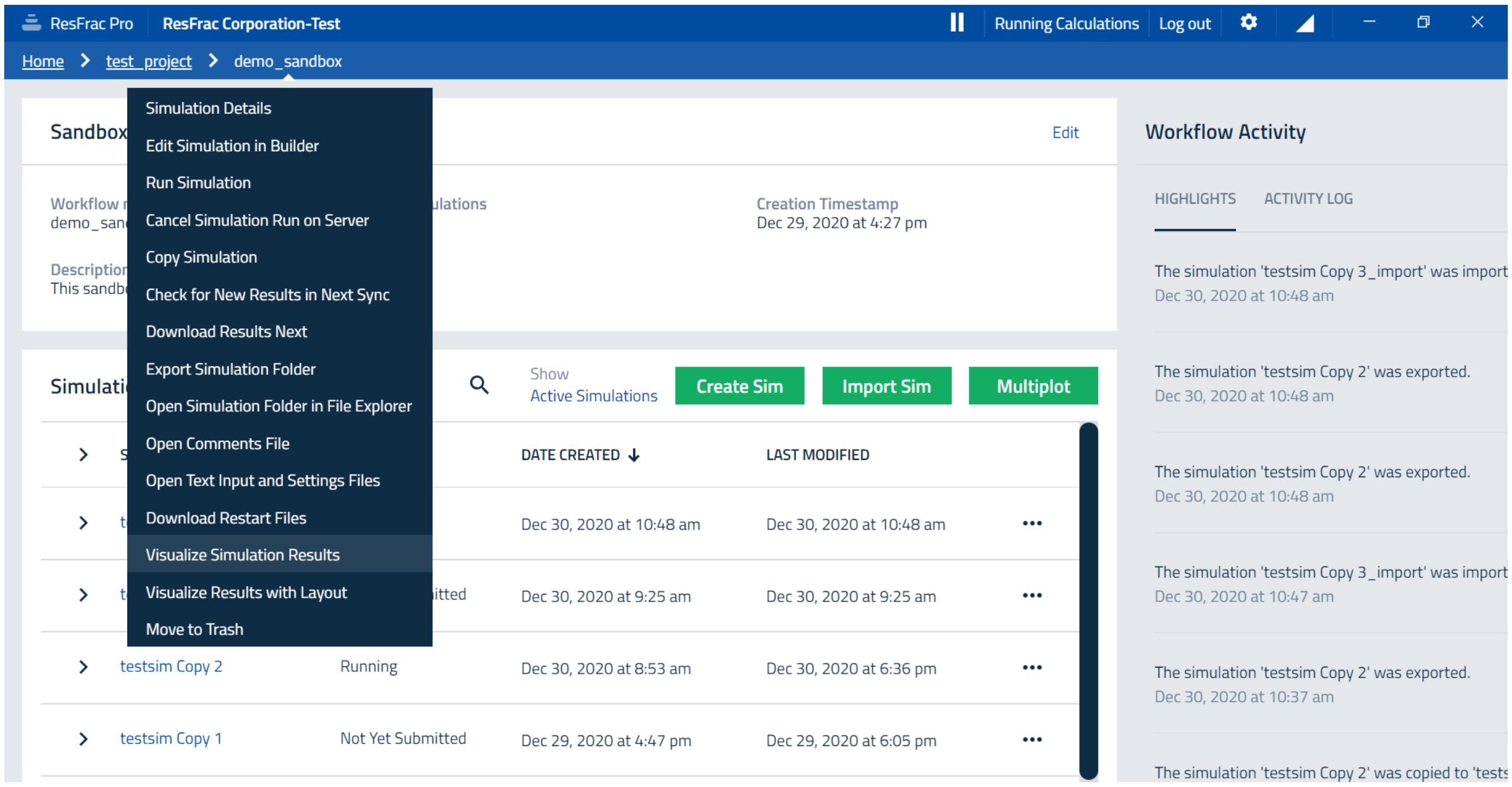

The visualization tool defaults to open with four visualization panels – a line plot of pressure versus time in the upper left, and three 3D viewer panels showing pressure, aperture, and proppant mass per area. The 3D camera controls can be manipulated using left and right click, the shift and control keys, and optionally with the scroll wheel. The specific controls can be viewed by clicking the grey ? icon in the upper left corner of each panel window, and an info box like the one below will appear.

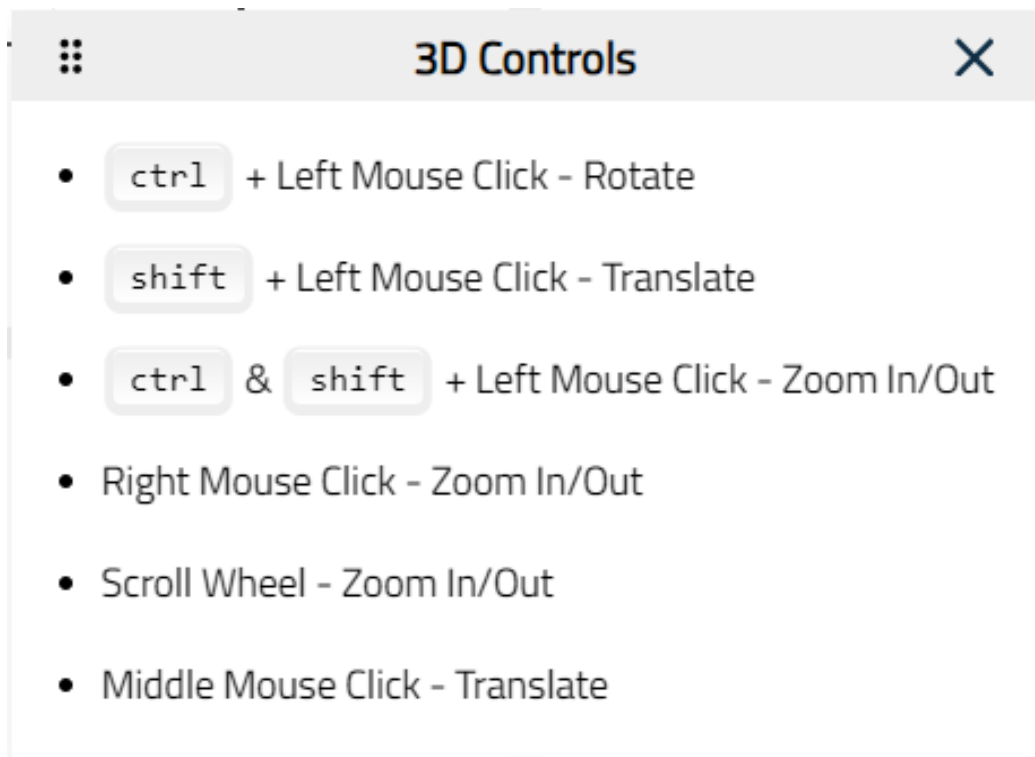

You can also toggle visibility of the model components on or off (like the matrix region), stretch the axis, change the color scales, etc. Click on the Matrix toggle to make the matrix invisible, and test out the 3D controls to view the fractures.

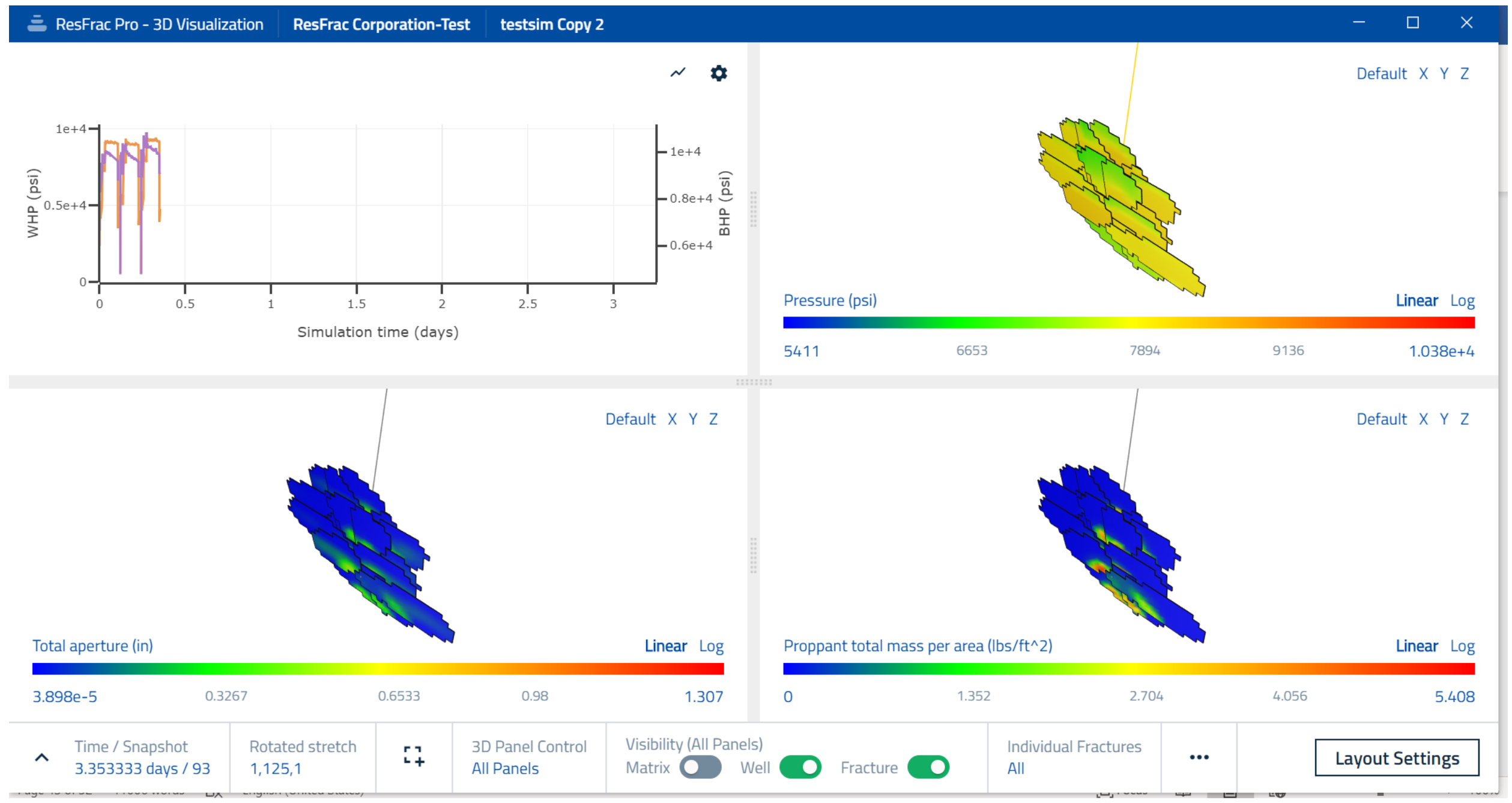

This is a simulation of three stages along the well. The well is not quite drilled perpendicular to Shmin, and so the fractures are propagating at an angle. The fractures are a bit hard to see since they are all so close-together. To resolve this issue, you can use the 'rotated stretch' feature to stretch the visualization in a direction (or optionally, vertical). The default stretch direction is the direction of Shmin, which is nearly always the direction that you actually do want to stretch. Click on 'Rotated Stretch' and input '5' for the horizontal stretch, for a 5x stretch. Now the fractures should all be visible. Note that the matrix region was automatically aligned with the direction of Shmin/SHmax. This is controlled by a setting in the 'Startup' panel in the simulation builder.

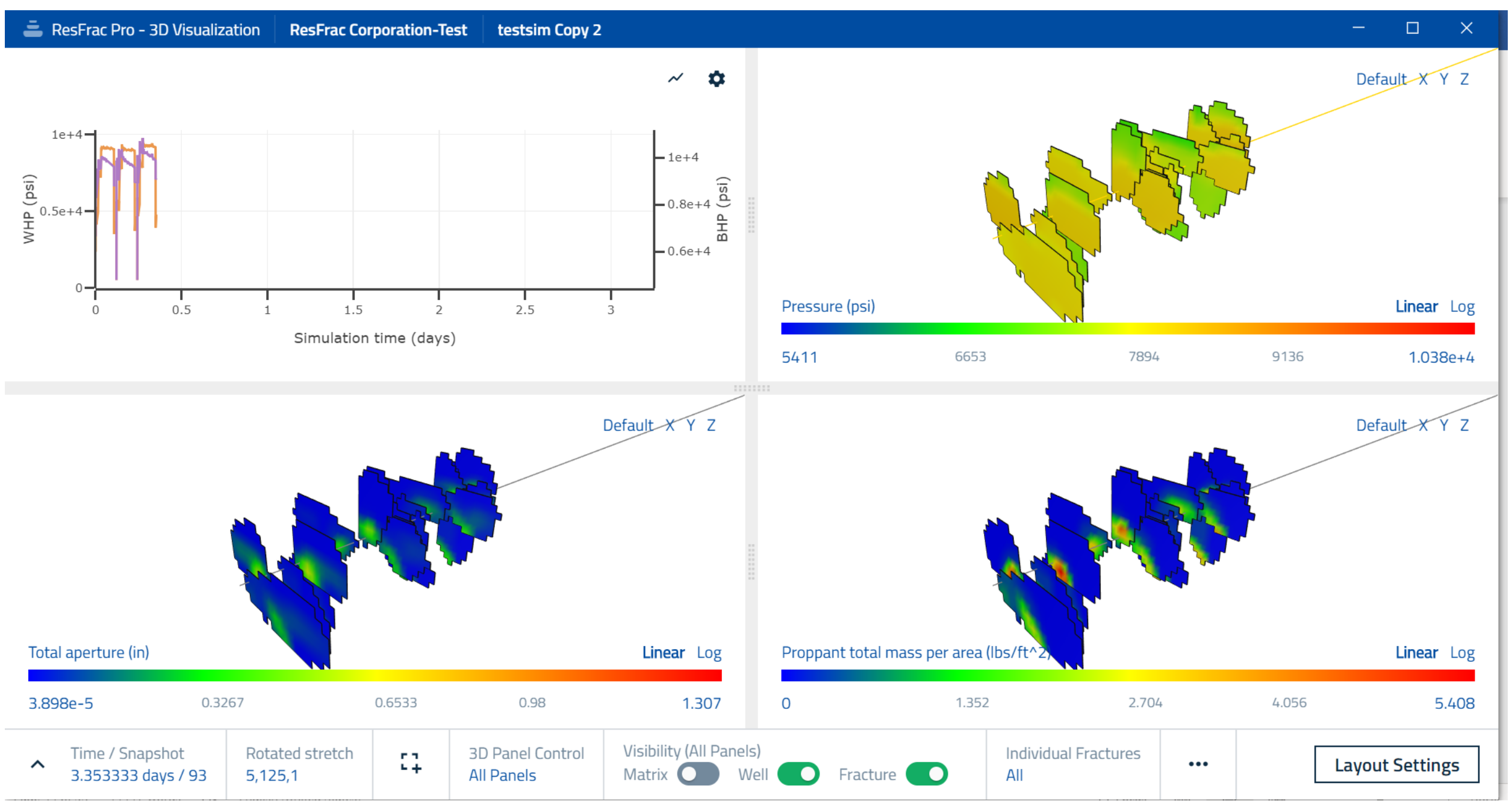

Select ‘Panel 1’ in ‘3D Panel Controls’ and then toggle back on the visibility of the matrix region. This will change the visibility of the matrix region only in the upper left panel. Click the three-dot button and select ‘Matrix Slices’ and do a matrix slice in the ‘y’ direction (which is the SHmax direction) from 1 to 13. Now, you can see a cross-section of the pressure distribution in the matrix. During fracturing, pressure will be elevated in the region surrounding the fractures. During production, pressure will be depleted.

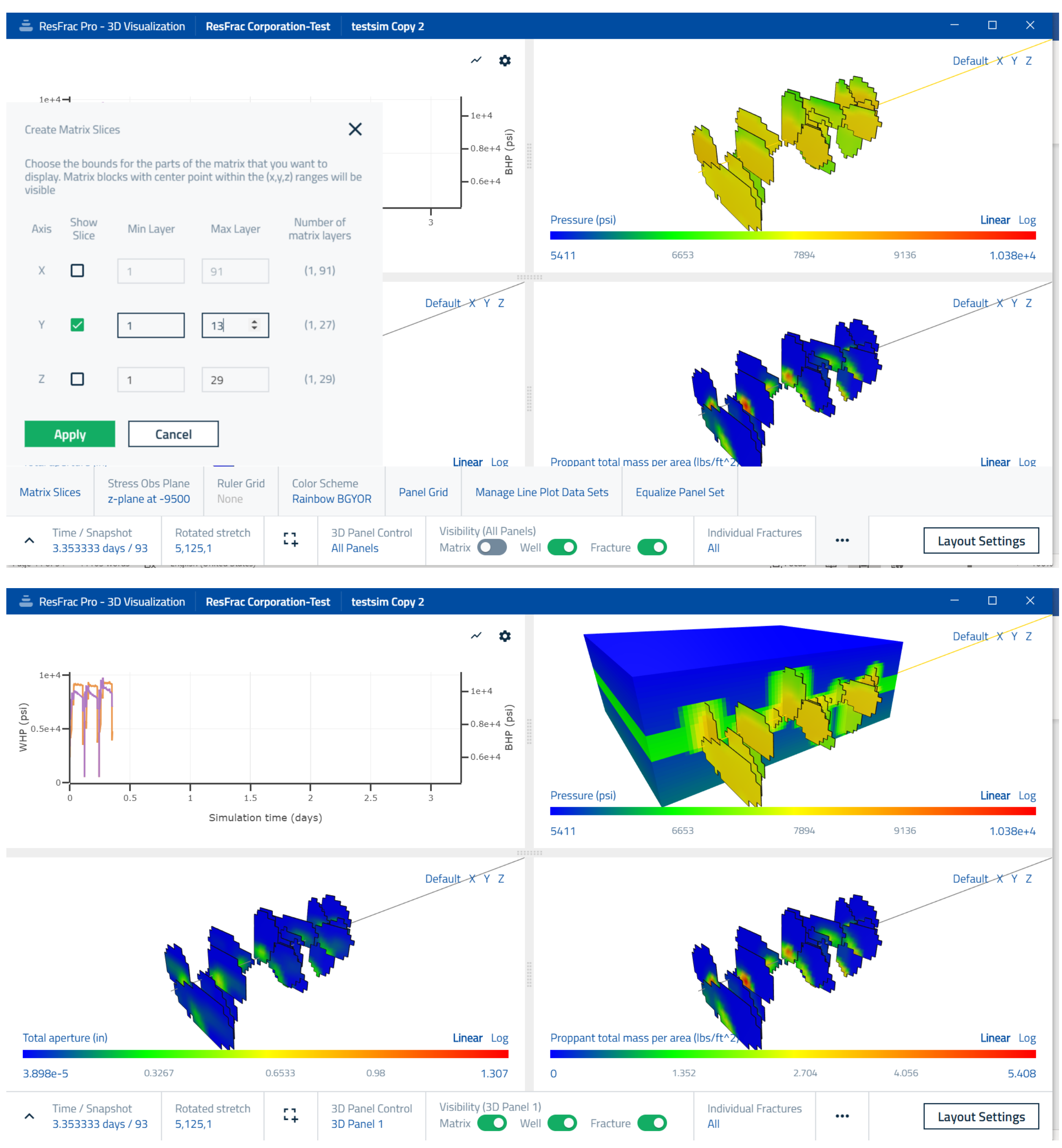

Click the three dot menu and select ‘Panel Grid.’ Here, you can decide to have any combination of panels, from one to six. They can be line plots or 3D plots. The 3D plots can use different camera angles. Enable the two panels on the right and set the upper middle plot to be in ‘Camera group 2.’

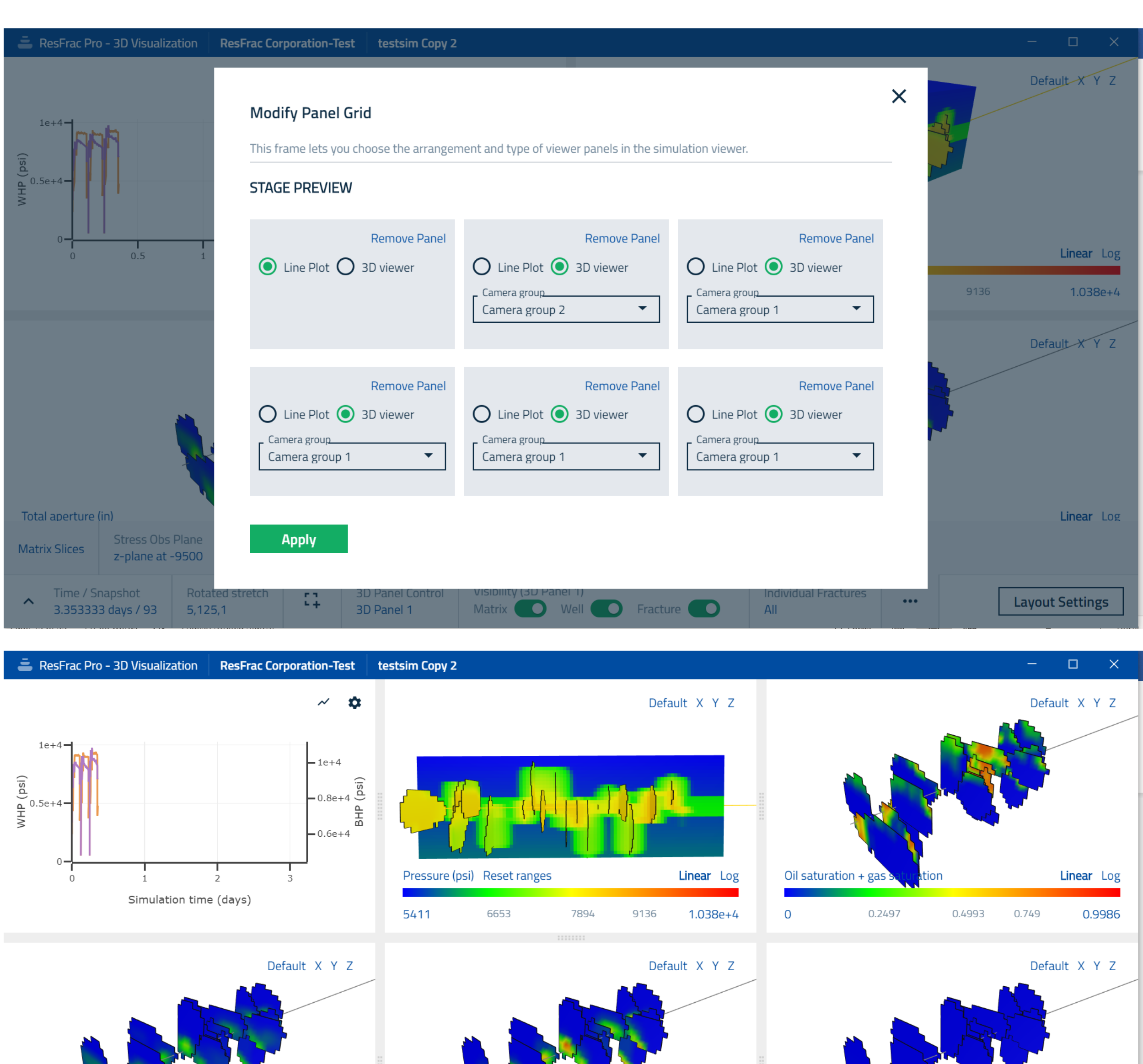

In the upper left panel, click the gear icon to bring up plot settings. Add a new axis and add water injection rate. Scroll through the menu and check out the other plotting options. Once the new line plot comes up, you can click and drag inside the line plot box to change the axes scales. You can also click and drag on the numbers on the axes to change the scale. Double click to zoom back out.

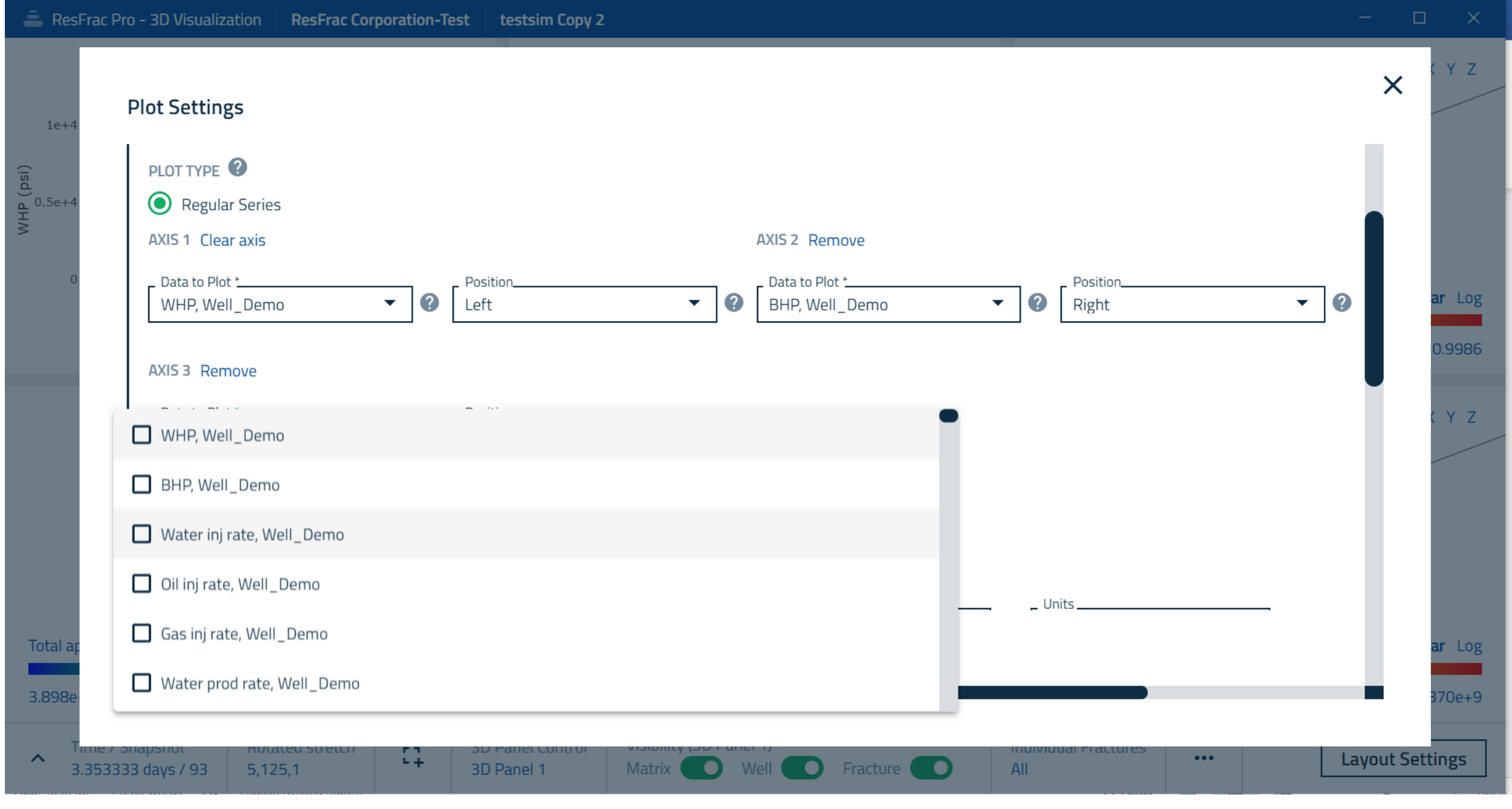

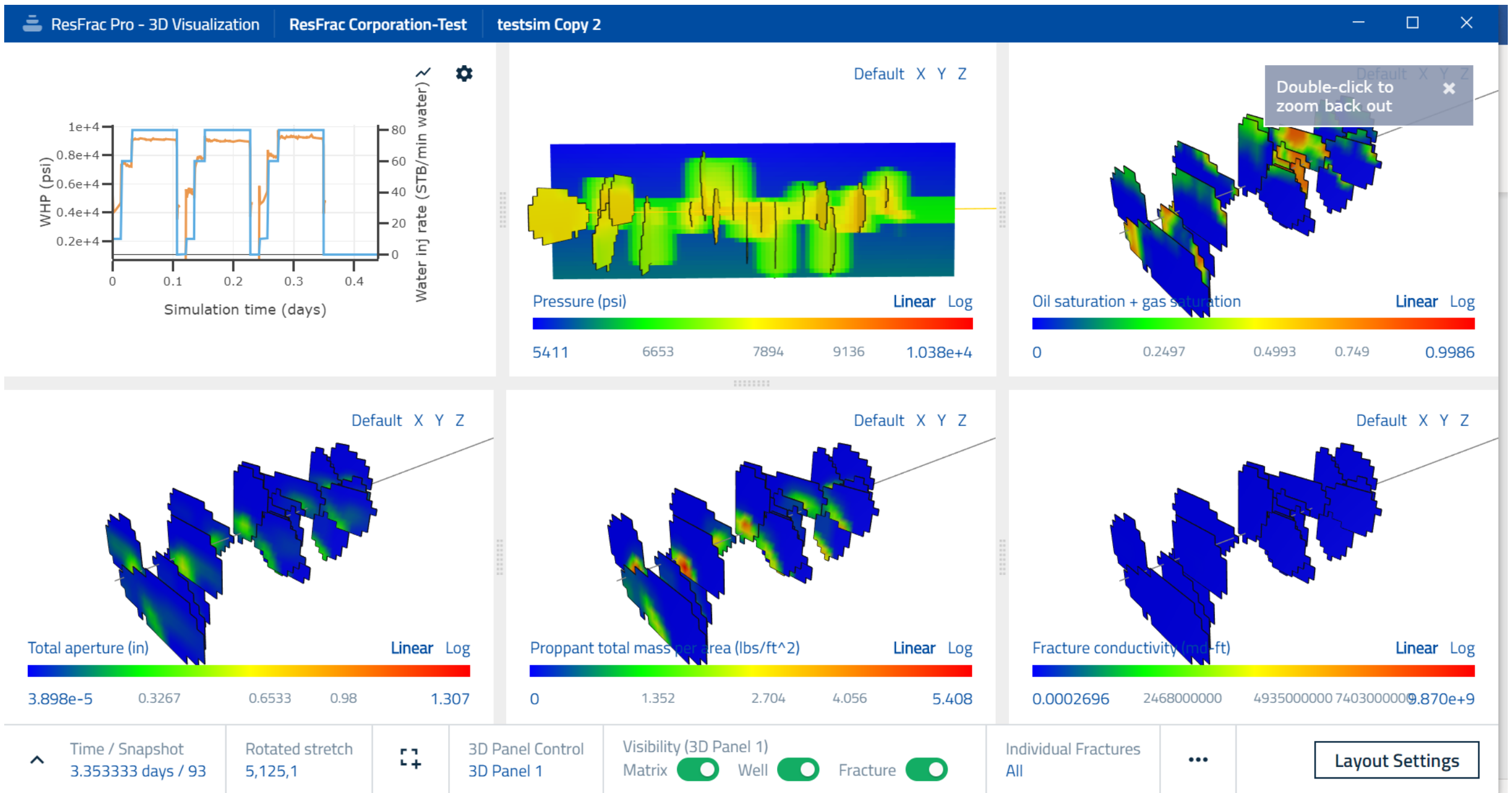

You change the color scales by clicking on the numbers on the far left or right of the color bar. You can also move through time.

Click on 'Time/Snapshot' and a variety of options come up. In the screenshot below, there have been 93 snapshots – points in time when the simulator outputted data to make a 3D visualization. There are boxes to type in the snapshot number, or to type in a 'time' to skip to a different snapshot. You can move through time using the slider bar, or by clicking the buttons for 'back,' 'next,' etc.

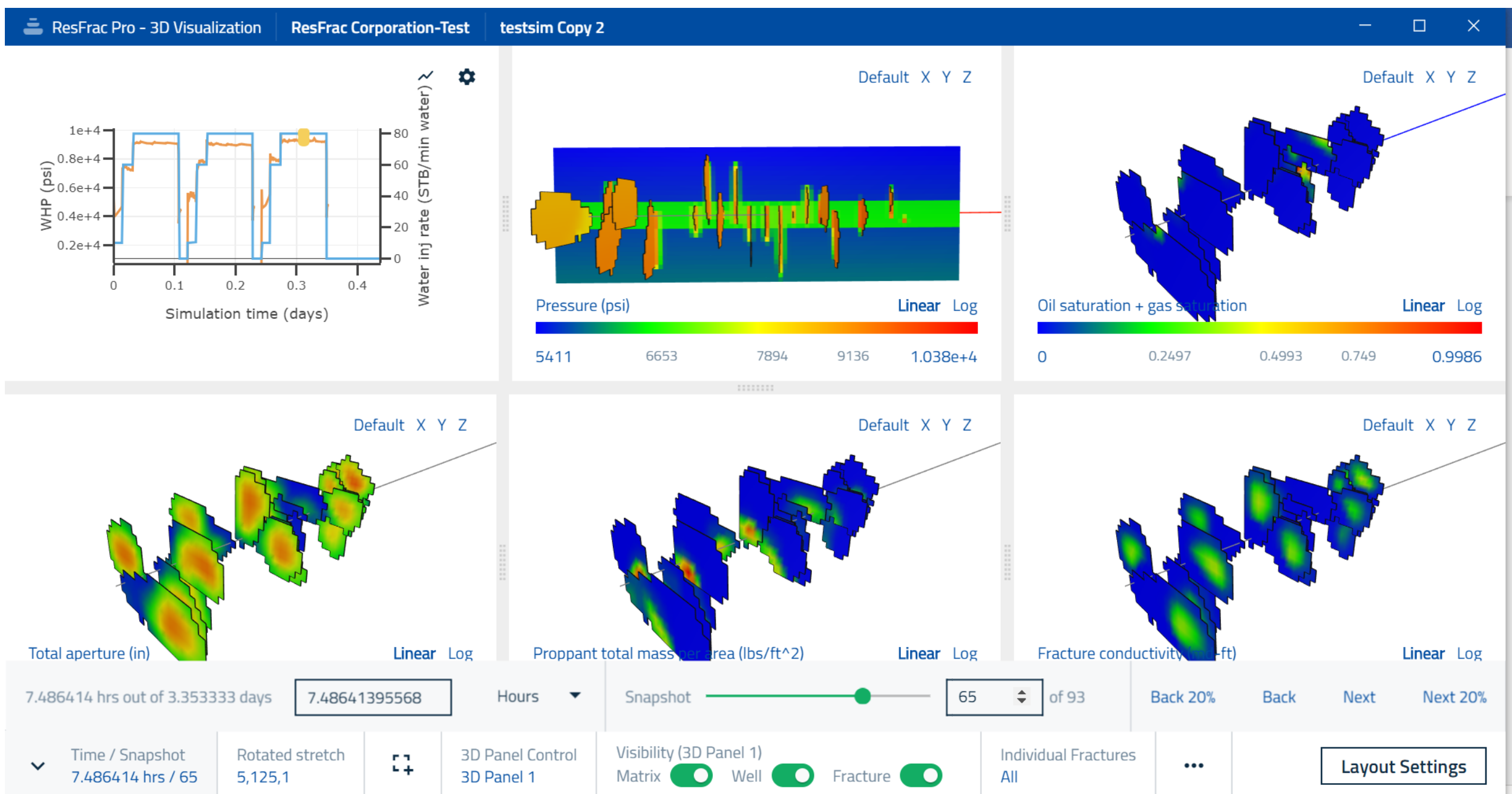

The line plot(s) show a yellow dot at the point in time corresponding to the snapshot being shown in the 3D image.

Click on the name of the property in the lower left panel, and a list of properties comes up. You can visualize any of these properties in the 3D panel. Select 'Delta stress Shmin direction.' This shows contours of stress change around the fractures, created by stress shadow from the cracks opening (including locked-in stress shadow from the proppant). With the rotated stretch, the image looks a bit odd, so go back and set the rotated stretch back to '1' so that the image is shown without stretch. The stresses are shown along a 'stress observation plane.' You can define any number of the observation planes, but they have to be specified in the builder prior to running the simulation (their size and location).

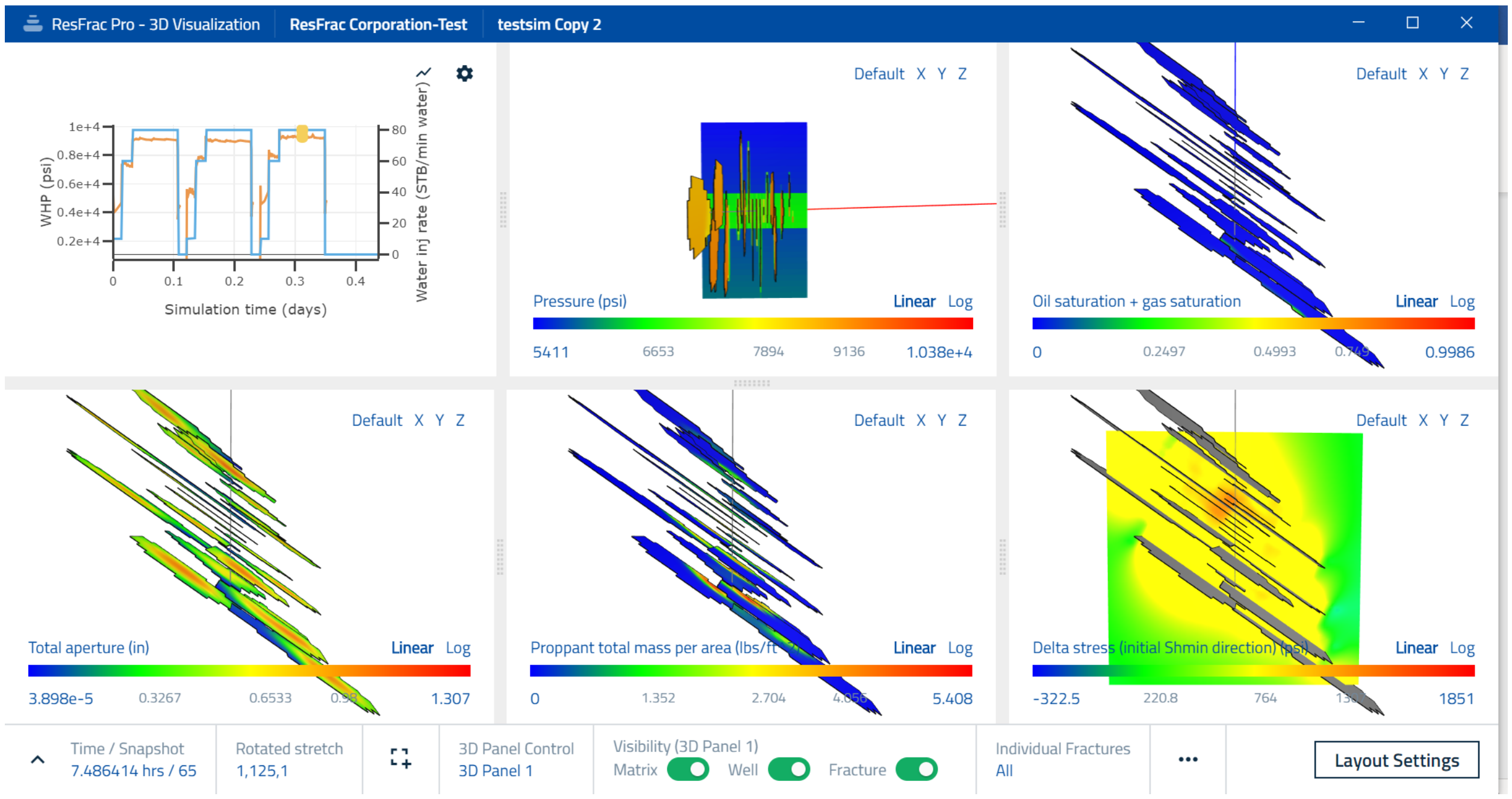

There are also options to make a screenshot or movie (the dashed box icon), toggle visibility of individual fractures, change the color scale, and import external data for the line plots.

You can save this setup as a ‘layout’ and reopen the same visualization later (with the same simulation or with a different simulation). Click on ‘Layout Settings.’ Click ‘Save Current Layout,’ and give your layout a name. Now, exit out of the visualization tool and return to the job manager.

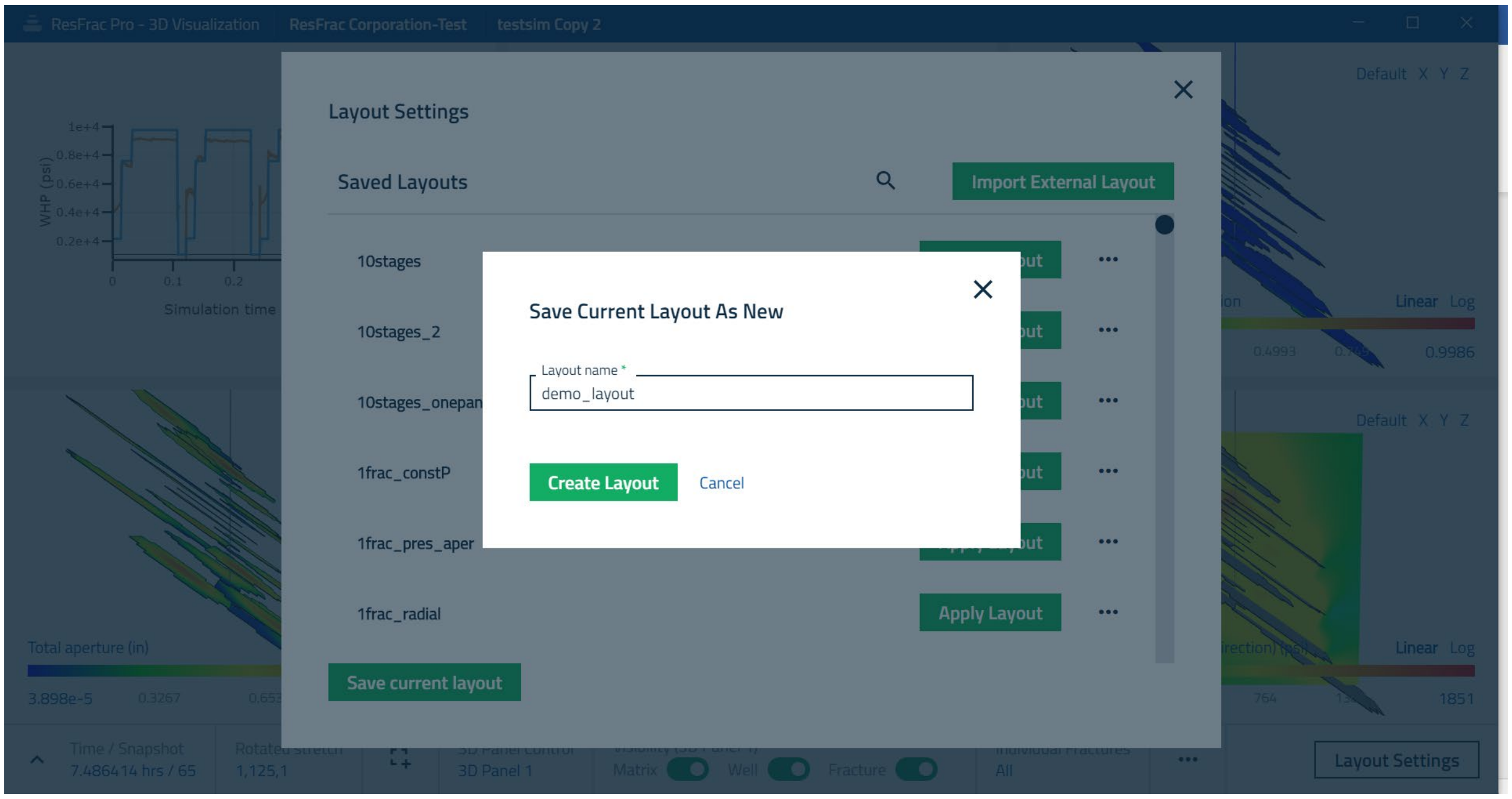

Click on your simulation and select ‘Visualize Results with Layout.’ A window pops up to select files. It should open to the folder that contains your saved layout. Select it, and this will reopen the visualization tool. The visualization should come up exactly like it was when you saved the layout.

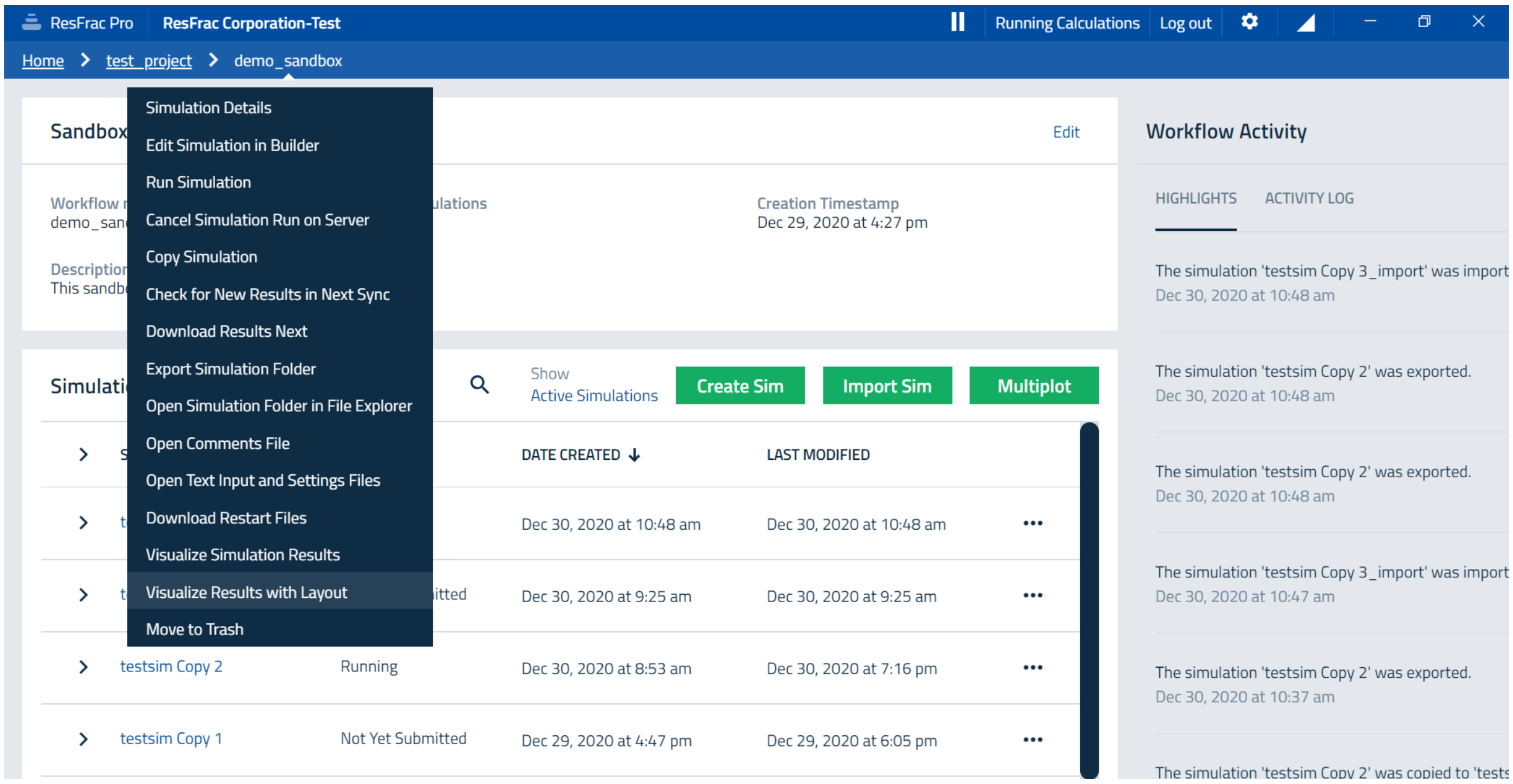

Close the visualization tool again and return to the job manager. Click on the 'Multiplot' button. The option for 'Batch Create Plots' will use a single layout to create multiple versions of the same figure for different simulations. If you have been following along with this demo, you should have at least two simulations in your list – the one that you submitted to run, and the one that you exported and then imported (which is a version of that simulation). Click 'Batch Create Plots,' and you will be prompted to select a layout and which simulations you would like to apply with the layout.

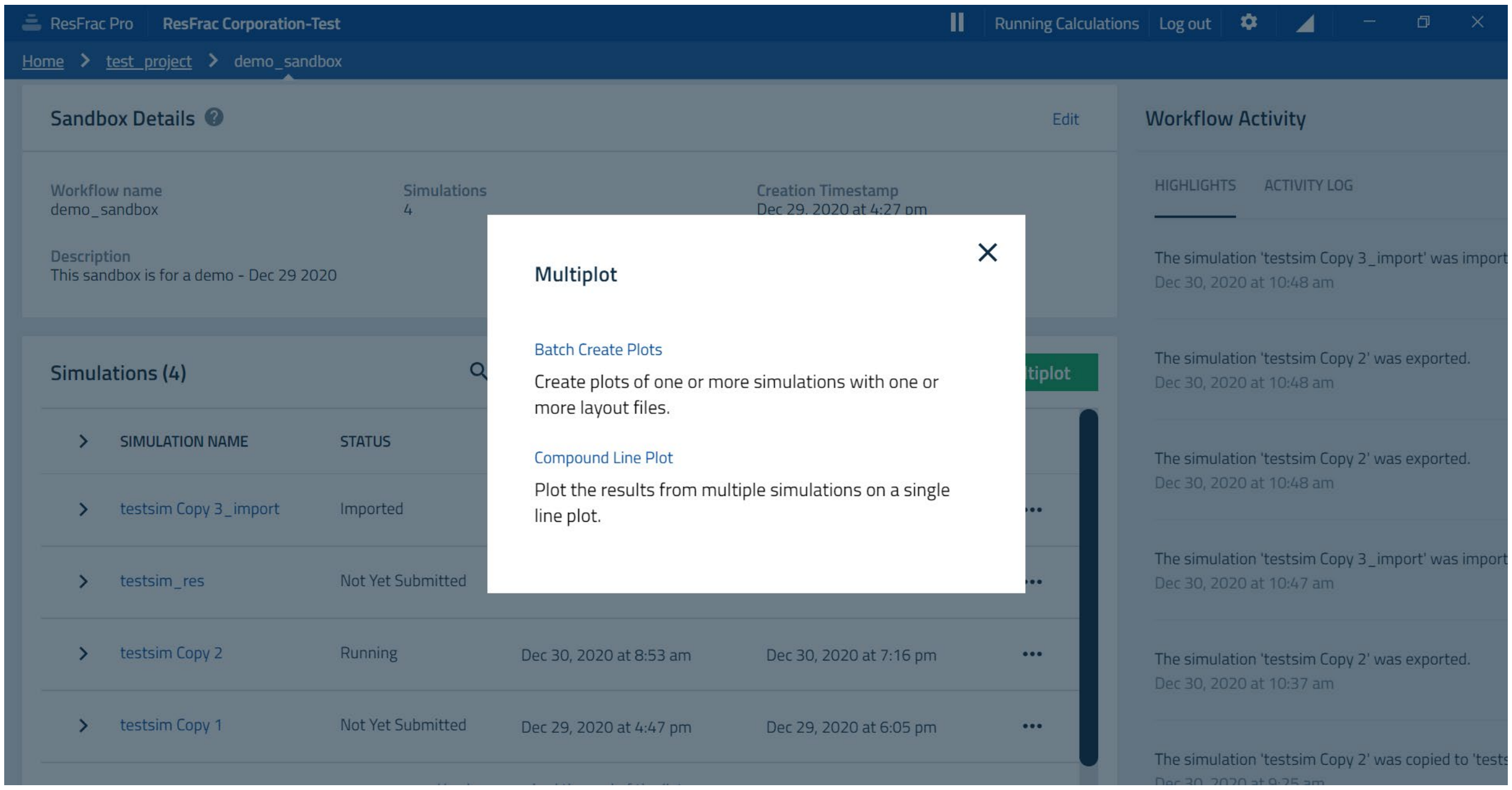

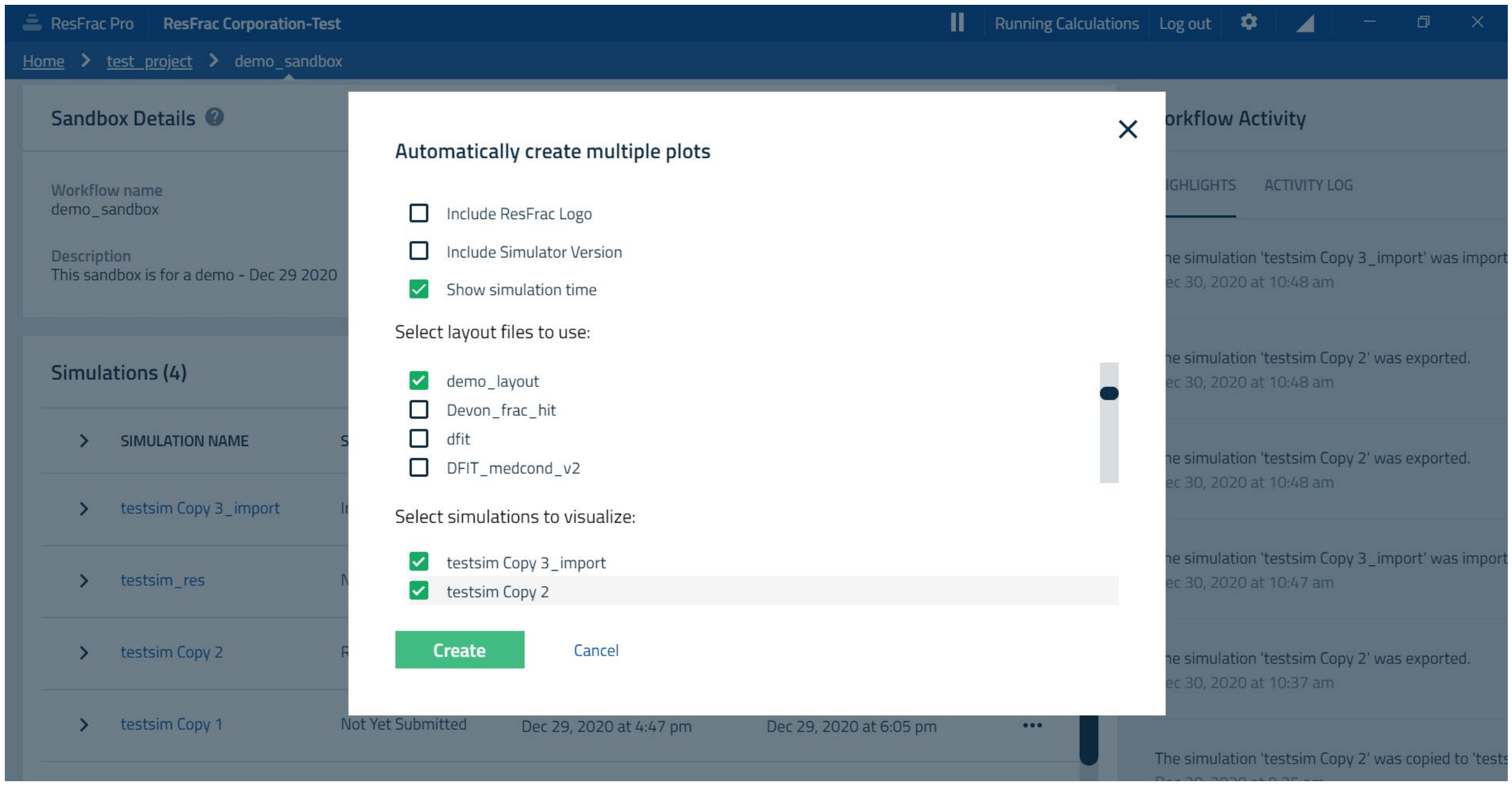

After a minute or so, it will finish making the plots and open a folder on your computer containing the screenshots.

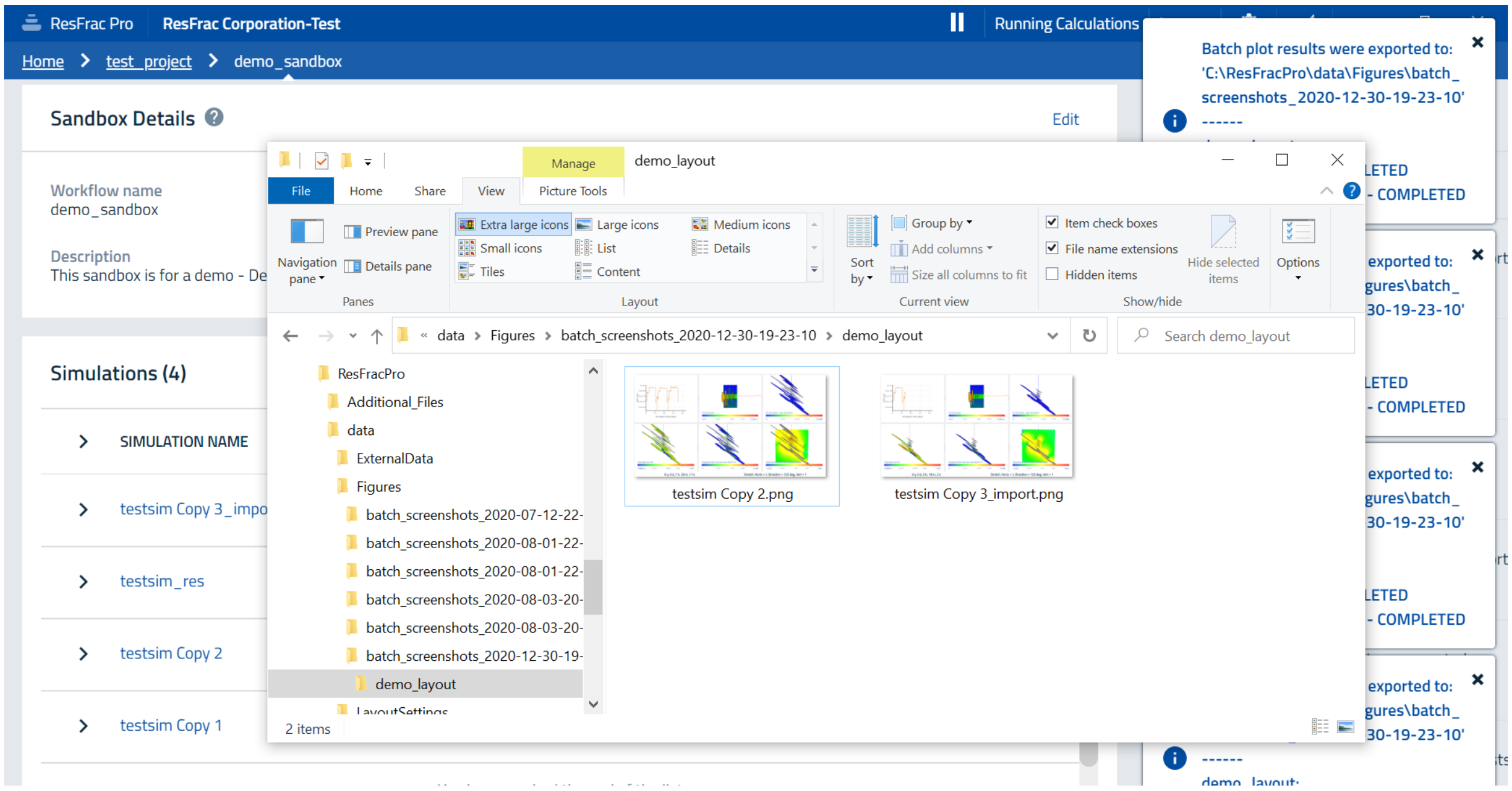

To use the 'Compound Line Plot' option, create a visualization template that contains only line plots. Then, use the compound line plot to automatically make a line plot containing the results for multiple simulations. This same thing could also be accomplished by opening a single simulation in the visualization tool, and then using the 'Import External Data' feature to import the results from other simulations into the visualization tool. Note that these can then be saved in a template and reused.

For any questions on what various visualization parameters are showing (either line plot or 3D) please reference the <ResFrac simulation outputs.xlsx> in the resources > documentation folder of your ResFrac installation.

## 5.13 Troubleshooting problems with the user-interface

The most common issue that occurs when viewing results is that you are trying to open a simulation with incompletely downloaded results. Incomplete results downloads can result in error messages when attempting to open a simulation in the visualization tool, limited functionality of the visualization tool after it is open, 'odd' results displayed in the visualization tool, and even inability to open results in the visualization tool. This section is a troubleshooting guide for resolving common download problems.

The ResFracPro UI constantly downloads results for all your running and recently completed simulations. This means that you should see new results for your simulations every few minutes or so. If you find that results for your simulations are not accumulating even after an hour, then something is going wrong with results download behavior. In that case, we suggest taking the following steps to troubleshoot.

1. Check to see if your simulation has a `Results` folder, a Results/newviz folder, and that these folders have multiple files inside. If this is the case, then the most likely problems are that (a) the UI is taking a while to download because you have a lot of simulations, (b) the UI is stuck on something and so is failing to get the newest results for this simulation, or (c) the UI is confused about the timeliness of the local copy of certain files vs the ones available on the server.
    - If problem (a) is the case, then if you wait for at least 10 minutes or so you should see more items accumulate in the results folders for your currently running or recently completed simulations.
    - If problem (b) is the case, then exiting/reopening the app and restarting the computer will help to overcome the 'stuck' state.
    - If problem (c) is the case, then you will want to try triggering a re-download, which is one of the later steps in this guide. This guide will walk you through the steps to troubleshoot all these potential problems.

2. The first thing to check is to make sure you do not have downloads paused – use the "Pause" / "Sync" toggle icon in the top bar. After you unpause downloads, the automatic download cycle will kick in within a few minutes and new results should start to accumulate for your currently running and recently completed simulations.

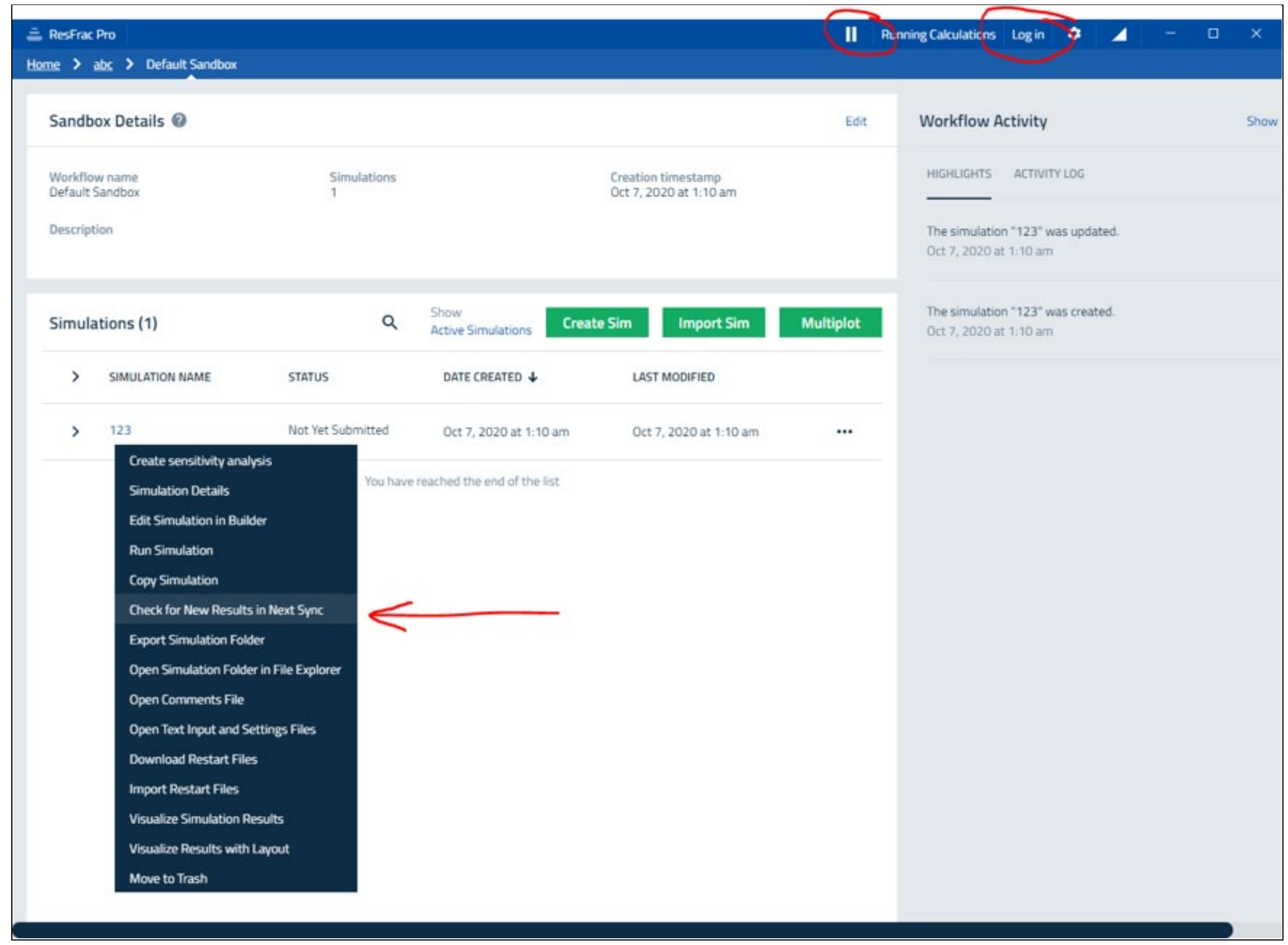

3. If you have lots of results to download, it can take a while for the UI to get everything. For example, if you submitted 20 jobs and then turned off your computer for a couple days, there could be 50 GB of files waiting to download. That could take a long time even with a fast internet connection. You can check to see how much bandwidth ResFracPro and other apps are using in the Windows Task Manager. Look for the “Network” column on the “Processes” tab to see by application, and then look at “Ethernet” or “Wi-Fi” under the “Performance” tab to see total system usage.

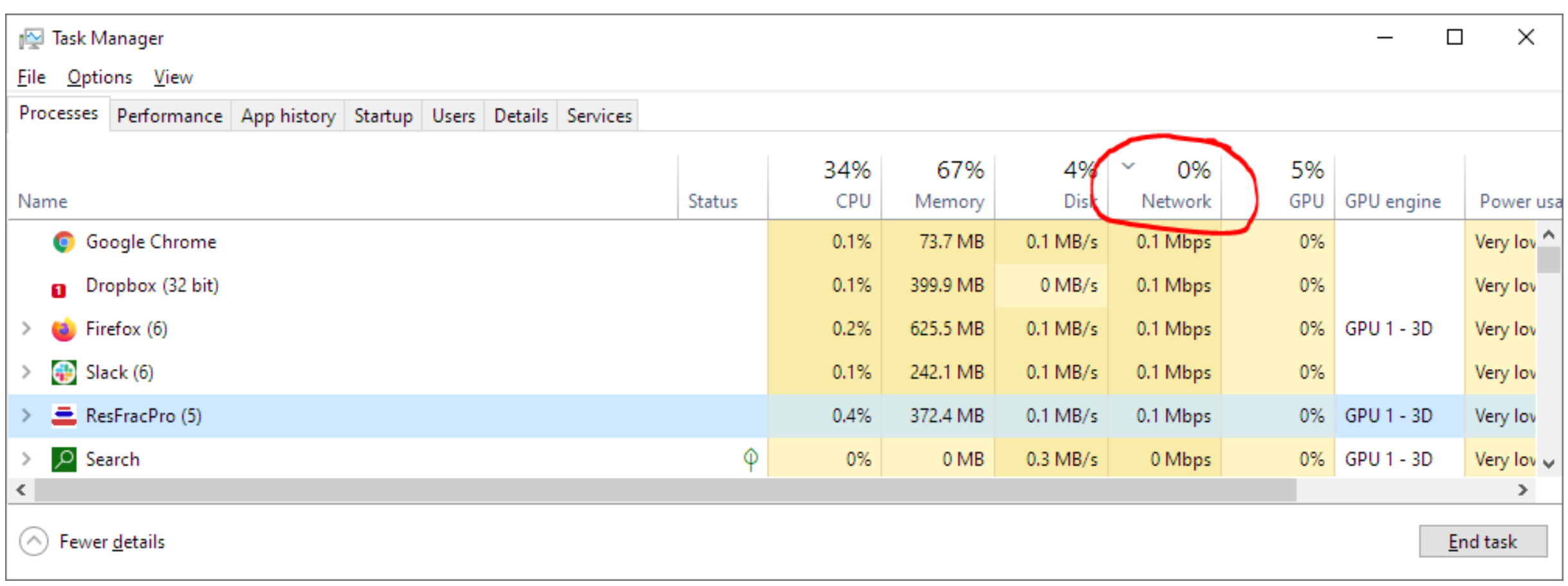

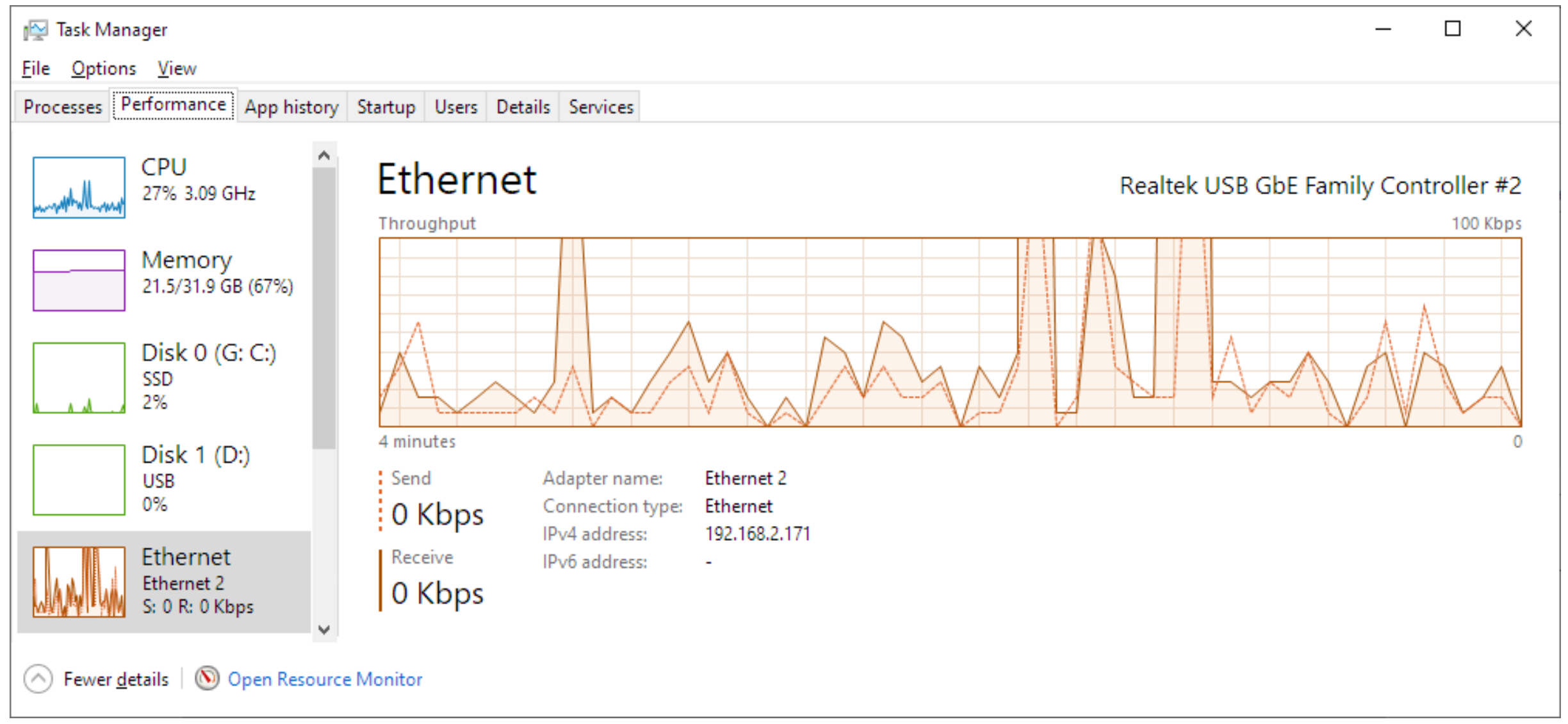

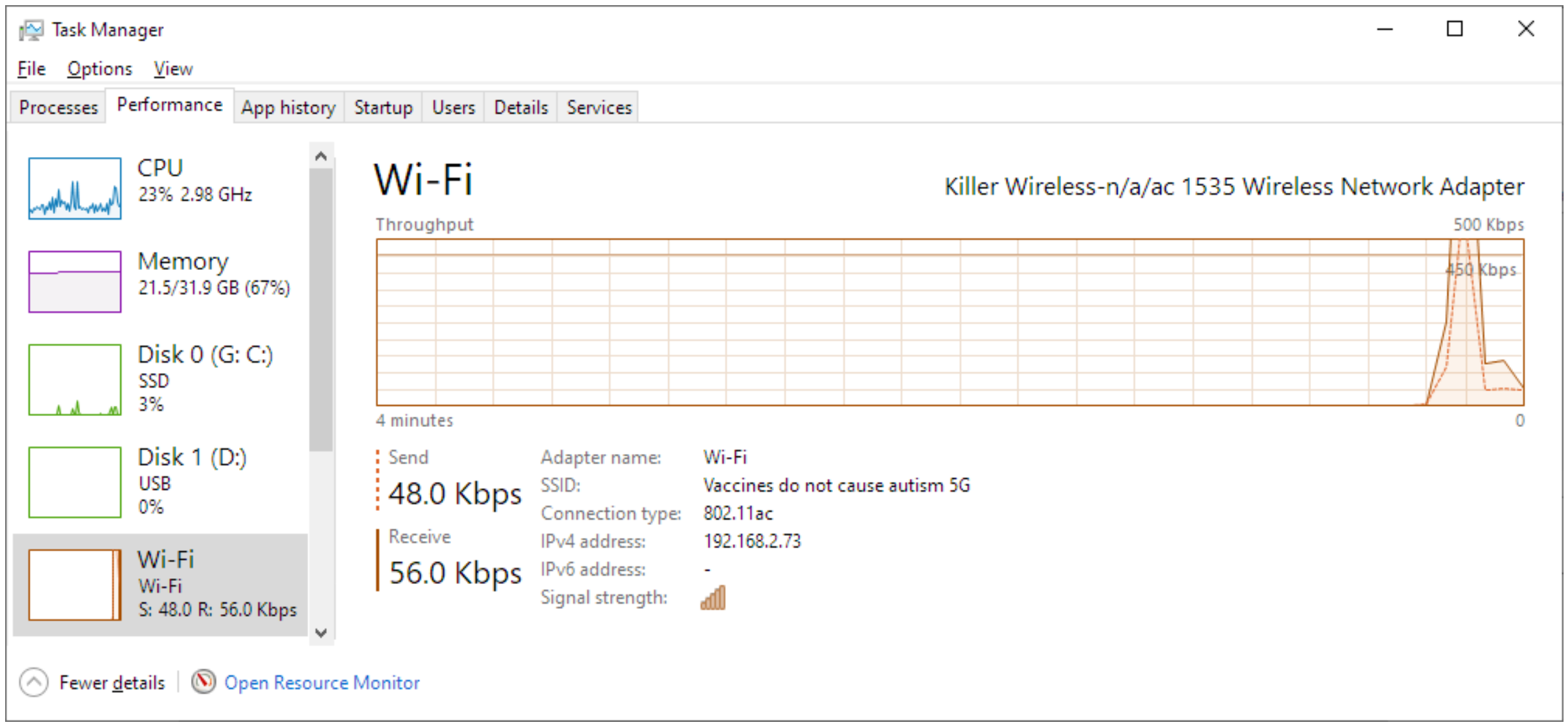

If the network usage values displayed are high for ResFracPro – i.e. hundreds of KB/s or more, depending on your internet connection – then most likely you just have a lot of results waiting to download and it will take time to get your results.

You can prioritize downloading results for a particular simulation by using the “Download Results Next” menu item. Using this on a simulation will cause that simulation to be the next simulation to download results for. Results for that simulation should start downloading within a few minutes.

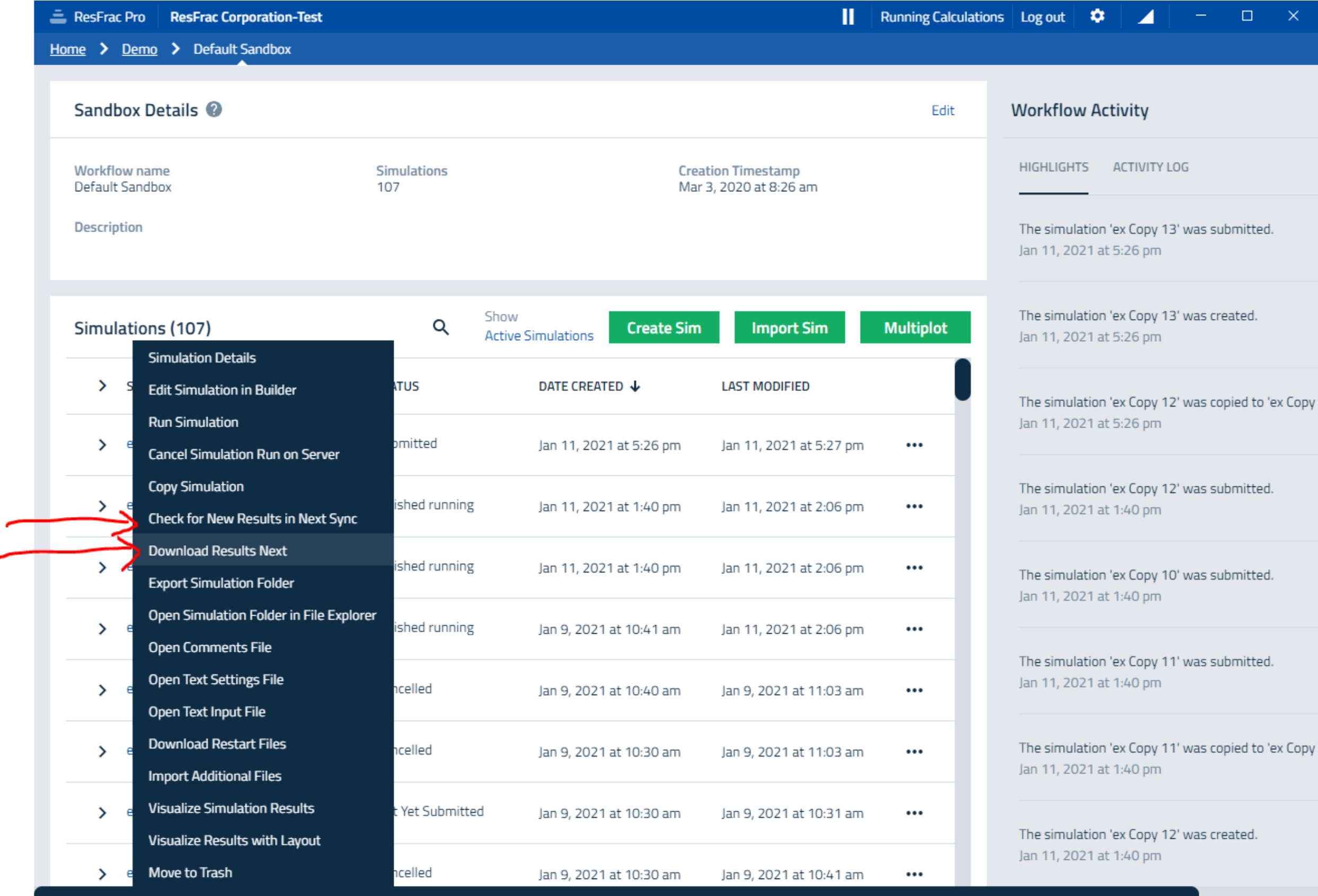

After you've given your simulation some more time to download, you can check the `Results` and `Results/newviz` folders for that simulation and see the date modified values for the files contained within. If new files are recently modified (i.e., within the last 15 minutes), that indicates that downloading is working properly and the results should be viewable after they accumulate.

4. Log out and log back in using the "Log out" / "Log in" link in the top bar.

5. Exit and reopen the app. If there is no improvement, restart your computer.

6. Trigger a re-download.

   One way to trigger a re-download is by clicking on the simulation name and choosing "Re-Download All Results." This is recommended for simulations that have been completed already.

   An alternative way to trigger a re-download is by renaming/moving/deleting the repeatedly modified files (such as comments_xxx.txt or sim_track_xxx.txt) and then right-clicking on the simulation name and choosing, "Check for new results in the next sync." This is recommended for simulations that are still running.

7. After triggering a re-download, upon the next download cycle, which occurs every few minutes, the UI should download the latest version of the repeatedly updated files locally. This works the same way for any file that gets downloaded repeatedly, including the following:

- sim_track_xxx.csv (line plot data in the UI)
- newviz/snapshots.txt (used for 3D plots in the UI)
- newviz/snapshottimes.txt (used for 3D plots in the UI)
- newviz/sobs_summary.txt (used for stress observation plane 3D plots in the UI)
- comments_xxxx.csv (not used in UI; for the user to read)
- 30_day_prod_xxx.csv (not used in UI; for user to analyze manually)
- daily_prod_xxx.csv (not used in UI; for user to analyze manually)

8. If steps 1-7 are not sufficient to resolve your issue, then we will need to work with you to resolve the issue. To facilitate that, you will need to gather information for troubleshooting. Section 5.14 "What to provide when requesting support for a problem with results" provides instructions on data to include that will help with this process.

As a backup, you can use the 'share simulation' option to share the simulation with yourself, download, and then import.

## 5.14 What to provide when requesting support for a problem with downloading

If you make a support request related to problems with the job manager and results downloading, please include the following items if you can. Doing so will help us to work through your problem more efficiently.

- `logs.log` – Turn on detailed logging by following the instructions in Section 14.3 "Running ResFracPro with detailed logging." After turning on detailed logging, wait ten minutes so that data can accumulate in the file. By default this is located at `C:\ResFracPro\logs\logs.log`
- `main.db` – By default located at `C:\ResFracPro\data\main.db`
- `input_xxx.txt` – Located in the simulation folder
- `settings_xxx.txt` – Located in the simulation folder
- `comments_xxx.txt` – Located in the "Results" folder for the simulation
- `sim_track_xxx.txt` – Located in the "Results" folder for the simulation
- Screen capture of the full ResFracPro UI window opened to the page that lists the simulation that is having trouble
- Screen capture of Windows Explorer opened to the simulation folder
- Screen capture of Windows Explorer opened to the "`Results`" folder for the simulation
- Screen capture of Windows Explorer opened to the "`Results/newviz`" folder for the simulation

Frequently some of the above files are quite large, 100s of MB. If that's the case, you may not be able to attach them in the e-mail, so we will work out a way for you to send the files. If the `logs.log` file is too big, one trick is to rename/remove the existing file and then turn on detailed logging, which will create a new `logs.log` file in the same place that is much smaller. After detailed logging is on for 10-15 minutes, the new `logs.log` file should be relatively small (a few MB at most), small enough to attach in an e-mail.

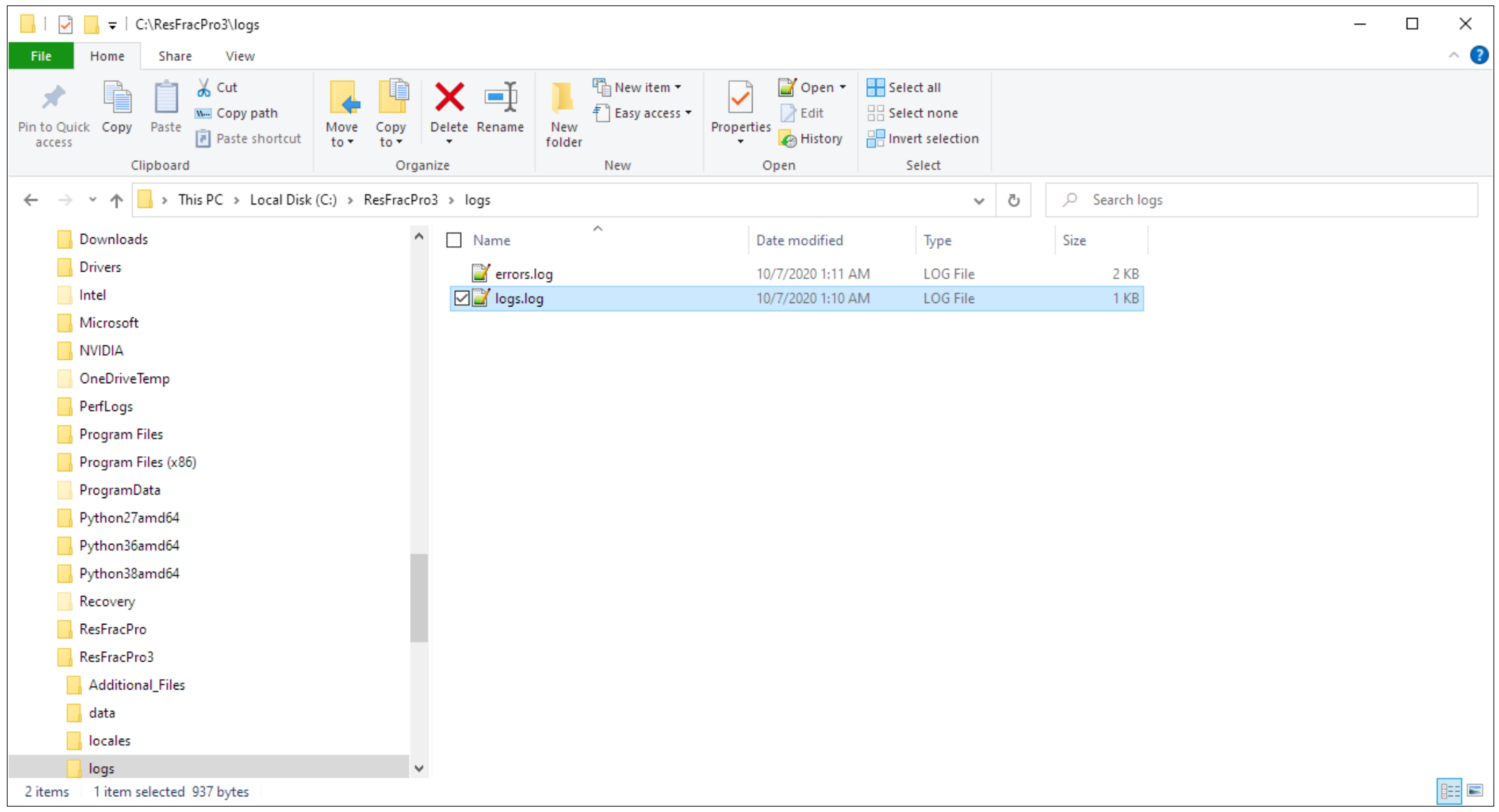

`logs.log` file

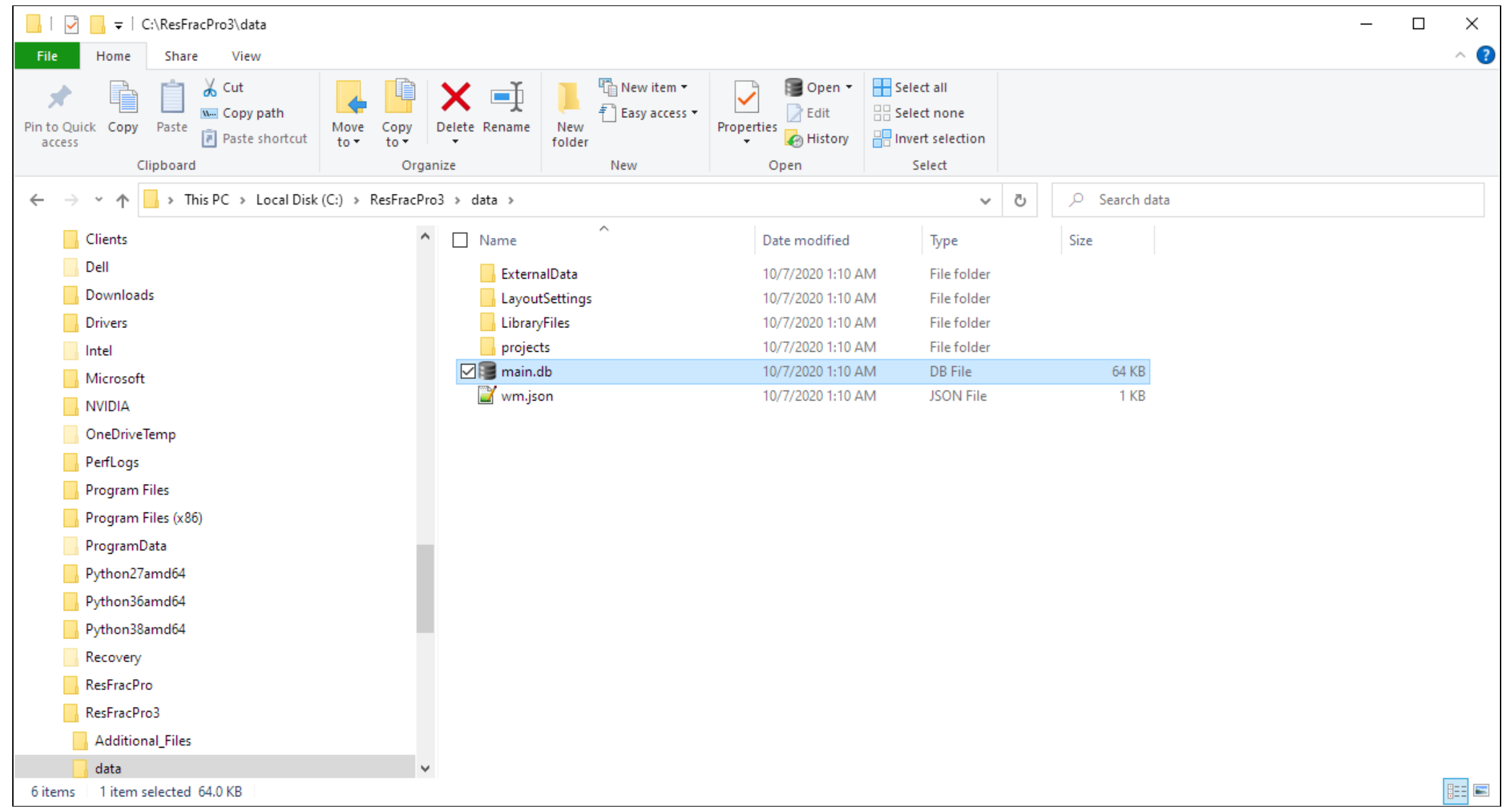

`main.db` file

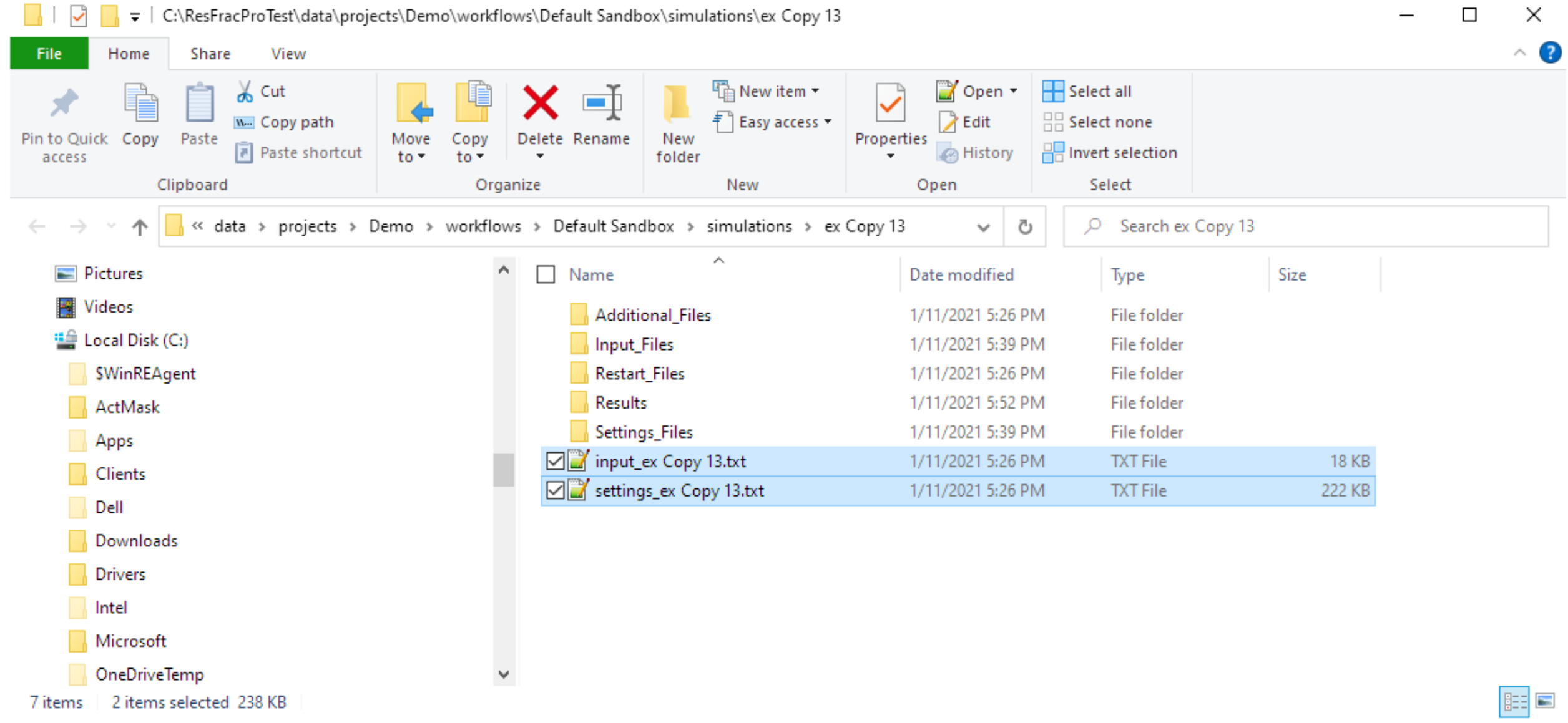

`input_xxx.txt` and `settings_xxx.txt` files and simulation folder screen capture

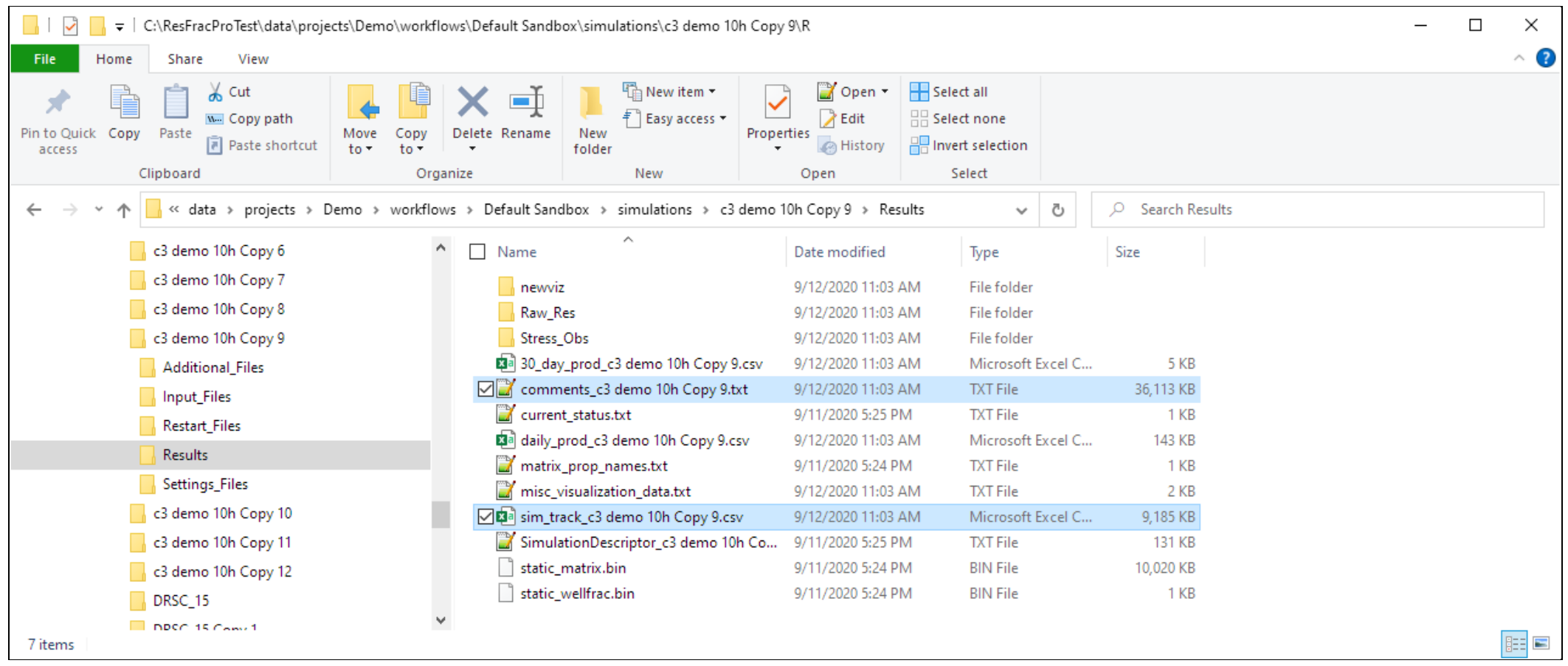

`comments_xxx.txt` and `sim_track_xxx.txt` files, and “`Results`” folder screen capture

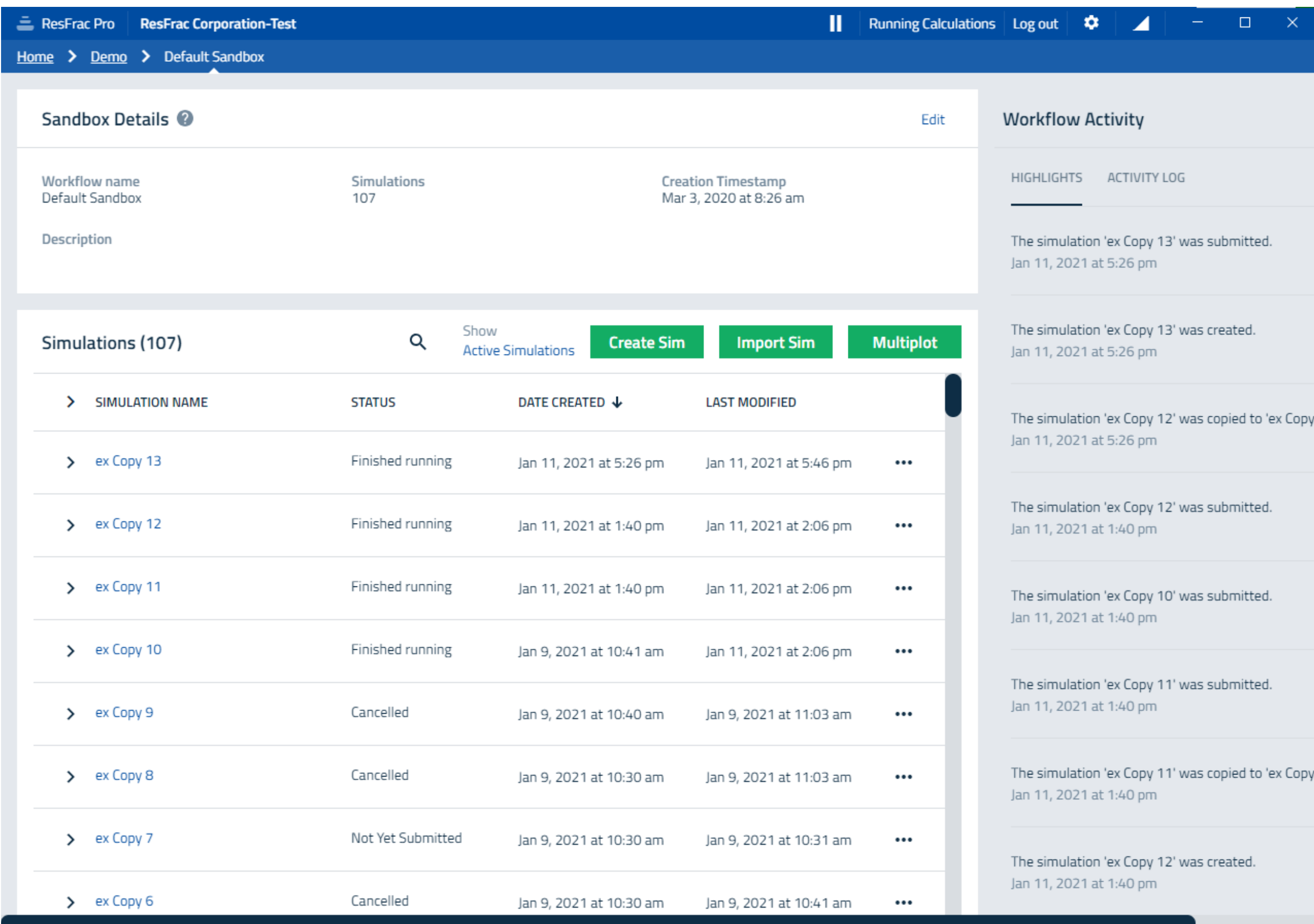

Screen capture of the full ResFracPro UI window displaying simulation of interest

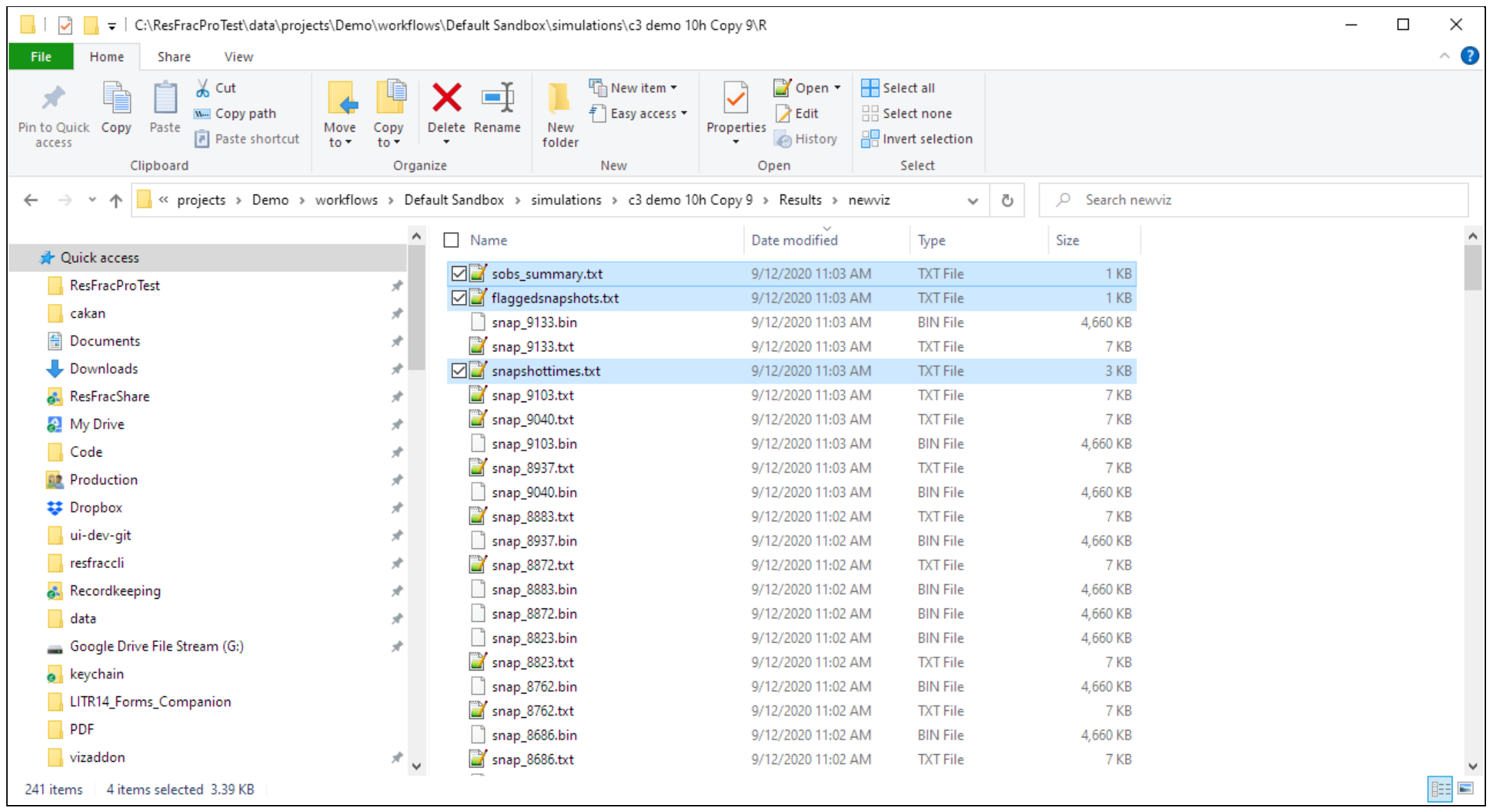

“Results/newviz” folder screen capture

## 5.15 Wrap-up

This is the end of the ResFrac user-interface demo! You should now have a familiarity with the capabilities of the user-interface. This demo did not cover the nuts and bolts of actually setting up a simulation or performing a modeling study. For those details, refer to Sections 6-10.

# 6. Practical tips for performing a modeling study

Within ResFrac, we do not only develop and support our software: we are its #1 user! We have a team of consultants running a variety of simultaneous projects. We've built a lot of experience running ResFrac modeling studies, and over time, have internally put together a 'best practices' document to memorialize our hard-earned learnings. This section distills key nuggets from our internal 'best practices' guide.

Before reading this section, read Section 4. It describes the workflow and structure of a modeling project, and explains the value of dividing the project into a series of 'checkpoint' meetings. The key steps of a project are:

1. 'Kickoff' - Establish scope and objectives; plan the workflow
2. 'Initial simulation' - Gather relevant data and set up an initial simulation; communicate back to stakeholder to check communication
3. 'History matching - Calibrate to field data
4. 'Sensitivities' - Run numerical experiments on the calibrated model

Section 11 provides a detailed step-by-step tutorial.

## 6.1 Establish a strong 'initial simulation'

Before you start calibrating to data, it is critical to make sure that you start from a firm 'initial simulation.' If there is an uncaught problem in your 'initial simulation' prior to history matching, then you could spend a lot of time history matching, and then later learn that the work needs to be redone because of a problem.

As the client gives you data, if you identify any discrepancies or have any doubt, always 'pull on the thread' and ask for clarification. Once you run the model, pull it up in the 3D viewer. Zoom in, step through time, and look at the simulation carefully. Does anything look amiss or surprising?

At the 'initial simulation' checkpoint meeting, you regurgitate the model setup back to stakeholders to give them the opportunity to ask questions and catch any miscommunications. Also, at this meeting, confirm with stakeholders which observations you will be matching as part of calibration. Show them the actual plots and talk through these observations. You do not want to get deep into the history match and discover that you aren't calibrating to quite the right data!

## 6.2 Address priorities and constraints

Keep in mind your project objectives. What parts of the model impact your objectives? Focus on the history match and data collection on the things that matter.

Keep track of project timeline. If you are at risk of slipping behind the initial time-estimate, communicate with stakeholders. Sometimes, stakeholders would rather you take a bit longer if that means getting the best possible answer. Other times, there is a deadline created by the timing of field operations or other considerations, and it is important to make sure that you finish on time, even if you can not make everything perfect. Either is fine, as long as you speak with stakeholders and assess priorities.

## 6.3 History matching

### 6.3.1 List 'observations to match' and develop hypotheses

To perform model calibration, start by making a list of the key observations to match. This should be a bulleted list of roughly 6-12 key points. In ResFrac, we prefer to calibrate against 'typical' results. For example, if a company has run downhole imaging and has estimates for perforation erosion, we would ask for a histogram showing the distribution of erosion across the well(s). We would calibrate to that distribution. We would *not* typically look really specifically at an individual stage, and try to match the exact details of that stage. When you look at smaller-scale, data become increasingly random – due to natural variability in the Earth and randomness – and while you can calibrating to noise, that does not mean you should. Our goal is usually to build a 'representative' model that can be used to optimize the next well, pad, or pads. A super detailed match to individual stage(s) in individual wells(s) does not serve that goal.

For example:

a. Match the fracture half-lengths and height inferred from microseismic and other diagnostics
b. Match the shape and slope of the rate-transient analysis plots of the wells
c. Match the water cut and GOR trends in the wells over time
d. Match that the outer wells on the pad outperform the inner wells by 20%
e. Match that geochemistry says that 90% of production comes from a certain interval
f. Match interference tests showing that wells interfere at a distance of 700 ft
g. Match downhole fiber estimates that perforation efficiency is 80-90%
h. Match that the parent well's production dropped sharply after the frac hit, and then recovered, but remained 50% lower long-term after the frac hit

When developing the list of key observations, here is a list of questions to ask. *This list is not exhaustive.* It is just a starting point. Every dataset is different, and you need to think critically to identify the key observations in your particular dataset. Depending on the data available in your dataset, you may or may not know the answers to all these questions.

With each observation, it is helpful to seek context from the operator's broader experience. In other words, is each observation specific to just this dataset, or is it consistent with observations from other pads nearby? This important to consider because real data has random variability, and you do not want to spend effort by overfitting to something that happened by chance.

1. What is the relative ranking of production between the wells? Is there any apparent pattern? In other words, are outer wells outproducing inner wells?
2. Make RTA plots of the production data. Is there a significant y-intercept? Do the lines bend upward, and if so, when and by how much? Does the upward bend correspond to the timing of pressure going below the bubble point pressure and an increase in GOR? *If so, the upward bend could be related to rel perm loss due to the gas coming out of solution*. Is the formation highly overpressured? *If so, the upward bend could be related to pressure/stress dependent permeability loss*.

3. What is the GOR behavior over time?

4. What is the water cut trend during flowback and long-term?

5. What is the typical fracture length? *Look at microseismic, offset fiber, tracer, frac hits, and sealed-wellbore pressure response. Keep in mind that in some datasets, microseismic significantly underestimates frac length.*

6. What is the typical fracture height? *Look at microseismic, tracer, and frac hit data. Stress logs can help infer height but are not always reliable.*

7. What is the typical propped length? *This is hard to estimate with high confidence. Your best method is to look at interference tests and/or relative performance of inner/outer wells. Proppant tracer recovery in offset wells is sometimes available.*

8. What is the typical propped height? *Oil fingerprinting can help estimate proportion of production by layer. Interference tests between wells in different intervals can also help. Proppant tracer recovery in offset wells is sometimes available.*

9. What is the typical perf efficiency and distribution of flow between the clusters? Is there a heel or toe side bias? *Look at fiber in the injection well or borehole imaging.*

10. What is the typical ISIP and the range of values observed? How much does ISIP change from the first stage to the subsequent stages? *The ISIP change from stage 1 to later stages helps diagnose the magnitude of stress shadowing, which is related to the effective fracture length. ISIP overall is related to Shmin and stress shadowing. However, keep in mind that it can also be elevated by near-wellbore tortuosity. Problems with ISIP matching are discussed by McClure (2020).*

11. What is the approximate WHP during pumping? *We recommend against trying to calibrate to every bit of the WHP versus time data. There are too many things that can cause subtle changes, and they may be related to wellbore friction and near-wellbore effects that don't have a major effect on the well's long-term production. Nevertheless, you would like to be in the ballpark, which you mainly accomplish by calibrating wellbore friction. Refer to McClure (2020) for more discussion.*

12. Are there any known geohazards, like major faults, near the wells?

13. Were there any operational problems to be aware of?

If you have a parent/child dataset, you should ask these additional questions:

1. How does the 'child' production compare to the parent production? How are their frac designs different? Spacing with other wells? What other differences are there?

2. How did the parent well production change after the child well(s) were completed and put on production?

3. How was the parent well 'prepared' for the child frac? Shut-in? Preloaded? Etc.

4. How strong was the pressure response at the parent well? Look at this statistically - frequency and magnitude.

5. Was the child well hydraulic fracture propagation asymmetrical? At what spacing? Did it change as you went down the lateral?

6. What do we know about the spatial extent of the drainage region? Spatial extent of the propped region?

7. Was there an apparent ISIP difference between the parent and child wells?

8. How does pressure versus time during pumping compare between the parent and child wells?

Once you have made the list of observations, make a list of hypotheses about how you could match these observations. For example, you might look at an RTA plot and observe that the real data has a steeper line than the simulated data. You might hypothesize this could be matched by decreasing permeability.

It is also helpful to fill out a quick summary table for each well. This table can look something like this:

| Property | Well 1 | Well 2 |
|---|---|---|
| Landing zone | MB | TF |
| POP Date | 3/16/2012 | 6/12/2019 |
| Proppant loading (lbs/ft) | 750 | 1500 |
| Stage length (ft) | 300 | 220 |
| Openhole or cemented? | Openhole | Cased |
| Cluster spacing (ft) | NA | 30 |
| Perf diam/shots per cluster | NA | 6 shots; 0.34'' |
| Proppant type | 40/70 | 40/70 and 100 |
| Fluid type | Crosslinked | HVFR |
| Total prod at one year (MSTBO) | 150 | 200 |
| Lateral length (ft) | 10000 | 10000 |

### 6.3.2 Change one thing at a time

Do not try to save time and change multiple variables at the same time. If you change two or more variables simultaneously and get a different result, you can not tell which change caused the different behavior. You can *guess* which caused the change, but if you guess wrong, you can be led astray. This strategy will cost you more time than it can save.

Once you have made your list of hypotheses, run a series of simulations – changing one thing at a time – to validate whether those changes will have the effect that you want, and if they will be able to change the results by *enough* to achieve a match.

### 6.3.3 Bracket the solution and look at the 3D visualizations

Next, test your hypotheses by bracketing the solution. Change the parameter by *more than is reasonable*. Did the model change in the way that you expected? For example, you observe that the real data shows increasing GOR over time, and the simulated GOR is mostly flat. You might hypothesize that you need to modify the gas relative permeability curve in the formation. To bracket the solution, you could change the gas relative permeability try an extreme – try a Brooks-Corey exponent of 1.0 and a residual gas saturation of 0.001. You might run that simulation and find out that the model *still* does not show the observed GOR increase. The model has falsified your hypothesis – no matter how much you change gas rel perm, *that change alone* will not be sufficient to match the observed GOR. If that happens, take a step back and ask why your hypothesis was wrong. What assumption are you making that may not be valid?

To understand what is happening in the simulation and why, look carefully at the 3D visualizations. For example, if the simulation is not predicting an increase in GOR, look at the distribution of fluid pressure in the hydraulic fracture. Is fluid pressure below the bubble point? If the fracture conductivity is not high enough, then even if the BHP goes below the bubble point, the pressure in the fracture may not. Regardless of the gas relative permeability curves, if pressure in the formation never goes below the bubble point, then you will not see an increase in GOR. Thus, in this example, you might be able to get an increasing GOR trend by increasing the conductivity of the proppant pack. Test this new hypothesis by again bracketing the solution – make the proppant pack conductivity 100x bigger (by increasing k0). Do you now see an increasing GOR trend? Do you now see the pressure in the fracture going below the bubble point?

As another example, we were recently history matching a well-to-well interference test, and the simulation showed no pressure interference. We pulled up the simulation the 3D viewer, and we could see that there was adequate proppant coverage and conductivity between the wells. Further, pressure appeared to be drawing down in the fracture near the observation well. So why wasn't the wellhead pressure changing? We panned the camera uphole along the well, looking at the distribution of pressure, and found a point where there was a sharp discontinuity in pressure. This turned out to be a spot where proppant had flowed into the well during a frac hit, sealing off flow and blocking the well. This was confirmed by plotting 'proppant volume fraction' in observing that a section of well had reached a volume fraction of 0.66 (the default proppant packing volume fraction in ResFrac). Proppant blocking the wellbore can happen in real life, and so was not a simulator 'bug.' Instead, we realized that we needed to specify a 'wellbore cleanout' to occur after the offset frac job, and before the interference test. Without looking carefully in the 3D image, it would have been impossible to diagnose what was happening in the simulation. This was also a meaningful result for the simulation project – the model was predicting the likelihood of serious proppant flowback during the frac hit, under these conditions.

Once you've confirmed that an approach is viable and you've bracketed the solution, it is relatively straightforward to run several simulations, iterating until you have a reasonable match.

### 6.3.4 Take the model out to 30 years

You may only have a few years – or even months – of production data to calibrate against. Nevertheless, run the model out to 30 years and compare with the EUR estimated from DCA. By no means should we consider a DCA estimate of EUR to be highly reliable – DCA is nothing more than a heuristic extrapolation. But this comparison provides one more error-check on the model setup and may help catch issues.

Pull up the result in the 3D visualization and look at the distribution of pressure depletion. You can also plot the distribution of 'net cumulative oil production' along the fractures. Is production coming from the layers that you expect? Is there a large mismatch between the DCA and simulation result?

Also, take a look at GOR and water cut trends. Do they look reasonable in late time?

### 6.3.5 Degrees of freedom and non-uniqueness

In conventional reservoir simulation, history matching is notorious for being vulnerable to non-uniqueness. History matching algorithms are sometimes run with millions of degrees of freedom – varying permeability, saturation, and other properties on a block-by-block basis. With so many degrees of freedom, the problem is severely underdetermined.

During development of a conventional reservoir, you drill wells and start producing, and then you will continue to drill more wells and optimize secondary/tertiary recovery in that *same reservoir*. Therefore, it is useful to use rate/pressure data to infer the distribution of geologic properties. Improved knowledge of the distribution of geologic properties within the reservoir helps make future development decisions. But in shale, each well drains only its immediate surroundings, and most major decisions are made *prior to putting a well on production*. When we calibrate to production data, our goal is usually not to further optimize production on the calibration wells, but rather, to optimize fracture design and well placement for future pad(s), which of course, involve a *different physical location*. Thus, there is less value in attempting to match every single little detail. Even if we did observe that the stress was slightly higher in Stages 15-18 along a particular well, how would that help us optimize the next pad over? Instead, the goal of the calibration is to establish a model that is reasonably representative, so that we can model the next pad. We should history match to 'typical' or 'overall' observations, not to stage-by-stage data.

If it is possible to use geologic mapping or geophysical imaging to predict the formation properties in the next pad over, and this is different from the calibration pad, then this should be used in the workflow. First, calibrate the model to the 'overall' behavior of the calibration dataset. Then, adjust the model as needed to account for the site-specific details of the upcoming pad, and use that modified model for the design optimization.

Because we are calibrating to 'overall' observations, keep the 'degrees of freedom' in the calibration low. We may only vary 5-12 parameters in a typical history match. For example, we usually vary permeability with a single global permeability multiplier, rather than varying permeability element-by-element or even layer by layer. By keeping the degrees of freedom low, you greatly reduce the risk of arriving at a model overfit and being affected by non-uniqueness.

On the other hand, limiting degrees of freedom makes it *more difficult* to match the data. If you have two wells, and one produced more than the other, perhaps this is caused by completion design/ordering, or perhaps one well is just in slightly higher permeability rock? If the latter, then you could not possibly match all the data perfectly, just by varying a single global permeability multiplier. At this point, you have two options. The first option is to increase the degrees of freedom in the model. You could specify permeability on an element-by-element basis, and actually put higher permeability around the better producing well. The second option is to accept that with only 5-10 global parameters, you are not going to match every detail of the calibration data. That is ok! Your goal is to build a 'representative' model for optimizing future pads. If you pursue the first option too often – increase degrees of freedom – this is slippery slope. It is easy to start 'cheating' and simply crank up/down permeability around wells to force a match. This does not help design the frac job in future pads (since they are located elsewhere an unaffected by small-scale heterogeneity in the calibration pad) and may lead you to miss something meaningful in the data. Maybe the overproducing well was a better performer because it was an outer well in the pad, or completed differently, or in a different sequence. You should always pursue these causal explanations first.

When seeking causal explanations for the observations, do not forget that real data does exhibit random variability. This is when it useful to draw on information from outside just your calibration dataset. Talk to stakeholders – "I noticed that in this pad, the outer wells outproduced the inner wells by XYZ%. Is this typical of your wells using this spacing?" If they confirm that this is a systematic trend observed in the company's overall experience, then you should weight this as a calibration observation. On the other hand, if they report that the behavior is inconsistent or often different from what is seen in the calibration dataset, then you should be tolerant of not matching that bit of the data. Maybe it really is just a bit of random variance or formation heterogeneity.

### 6.3.6 Other strategies for QC'ing a simulation

Here are some things to check with your simulation:

a. Look at the distribution of pressure in the fractures during production. If propped fracture conductivity is too low, you'll see a large difference in pressure between the well and the adjacent fracture elements. If propped fracture conductivity is too high, you'll see only a few 100 psi of pressure drop from the fracture elements at the well to the edge of the propped region. Ideally, you'd like to see a substantial pressure drop – perhaps 1000-2000 psi – from the fracture element at the well to the edge of the propped region. This will typically correspond to fracture conductivity in the propped region of a few md-ft. You may find that decreasing conductivity to achieve that pressure drop *within* the propped region leads to excessive pressure drop between the well and adjacent fracture element. In that case, you may increase the value of the 'wellboretofractureconnectionconductivitymultiplier' (and also, make sure that 'updatetowellboretofractureconnectionconductivitymultiplier' is set to 'true'). Don't be shy – it's ok to make it 10 or even 100x. The idea is that the near-wellbore region has significantly more complexity and proppant conductivity than the far-field fractures.
b. Make sure that your mesh in the direction along fracture strike is sufficiently refined between the wells. You may want to use the 'logarithmic spacing wizard' to make sure that the mesh size between the wells is on the order of 50-100 ft, but then allow that mesh to widen out to 100s of ft as you get 1000s of ft away from the wells on either side.

c. Think about what is limiting your fracture size. If fracture lengths are < 1000 ft, then most likely, some process is limiting fracture length (of course, depending on factors like height growth and injection volume). Most commonly in shale, length is limited by toughness. If so, then with lengths <1000 ft, you'll probably see asymmetrical fracture propagation (ie, fractures going one direction or another), and you'll need significant limited-entry to get good perf efficiency. If you do not, check on your leakoff – how much fluid is leaking off? If you run the simulation with 'extrafractureinfointrackingfile' set to true, it will output the 'efficiency' of each fracture in the tracking file (so you can plot in the visualization tool). That is, the percentage of fluid that has entered the fracture that is still in the fracture. If efficiency is dipping below 80% by the end of a stage, this suggests that leakoff is having a significant impact. If it is below 50%, then it has had a major impact. If you are in shale, you probably don't want such low efficiency. If you see that in your simulation – check your PDP tables, and see if maybe there's more PDP than needed. If the fracture lengths are longer, like 1000s of ft, then you probably have low toughness and low leakoff. This is fine – most (but not all) shale plays are like this.

d. Pay attention to the toaster warnings that pop up when you open in the visualization tool. For example, if you hit maximum injection WHP and don't pump the designed job, a toaster notification will say so.

e. Plot 'total cumulative leakoff' and 'total cumulative production' of different phases in the 3D visualization. This can be helpful because sometimes, simulations are set up so that they inadvertently produce or leakoff too much fluid from a given zone. For example, perhaps you have inadvertently made the permeability of a layer too high.

## 6.4 Software tools

We recommend using several free software tools.

For viewing large text files (such as the comments file), use glogg. It does not allow you to edit files, which means that it can be highly optimized for viewing files. It can rapidly open and search through very large text files.

Once you are an experienced user, you might find it convenient to directly edit the input or settings files, instead of always using the simulation builder. For editing text files, we recommend Notepad++. It allows you to keep multiple tabs open simultaneously, and it is a bit easier to use than the built-in Windows Notepad. You can set Windows to automatically open text files with Notepad++. Right-click on a text file and select 'Open with' and then 'Choose another app.' Select Notepad++ and check the box for 'Always use this app to open .txt files.'

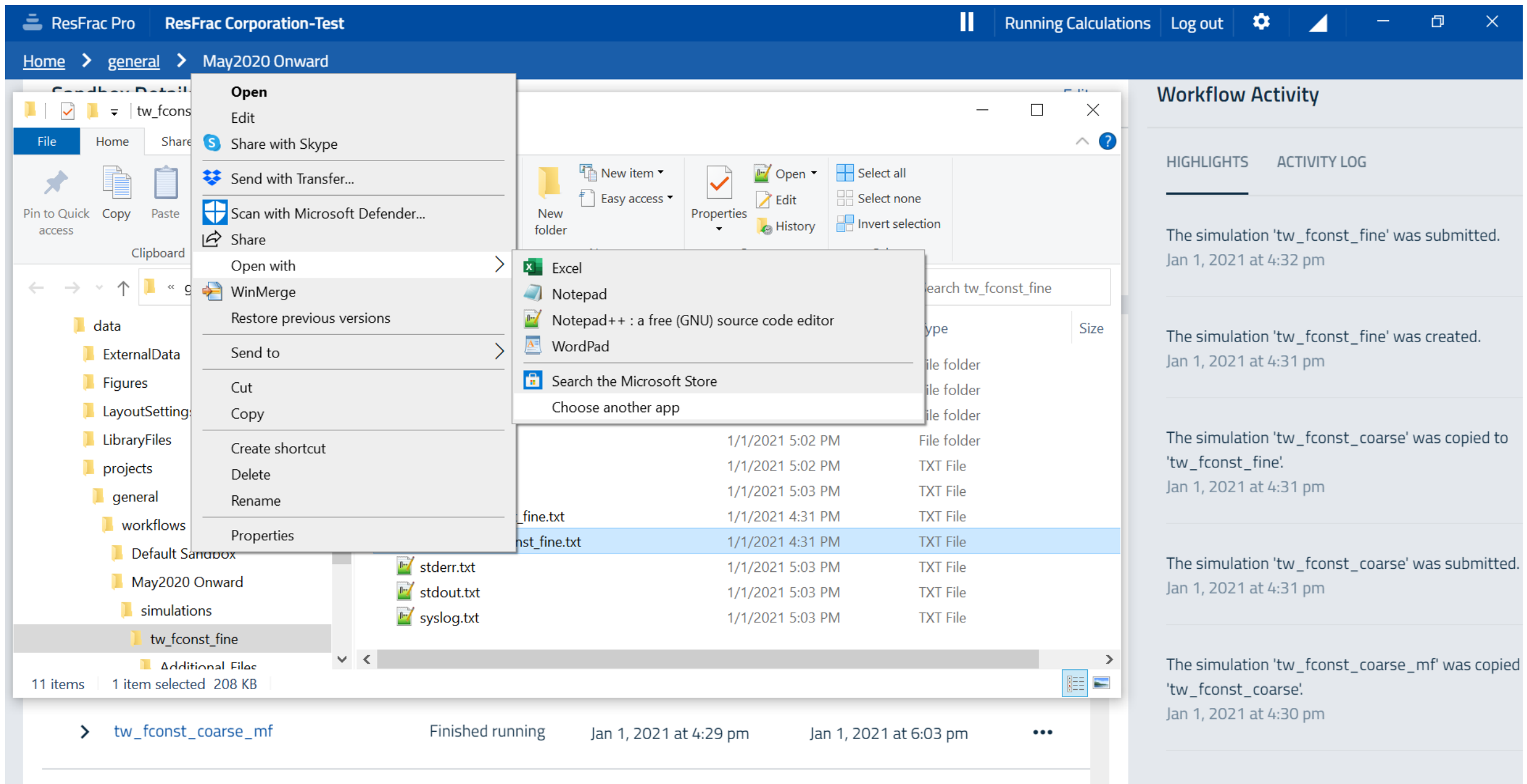

Winmerge can be used to compare text files. This is extremely useful for comparing simulations. As you run different simulations, you should be naming them and/or typing notes in the 'description' box to keep track. But sooner or later, you are going to lose track and forget the details of how you set up a simulation. You can use Winmerge to compare the settings files of two different simulations and identify what is different.

Sometimes, as a user, you'll run two different simulations and get surprisingly different results. If this happens, run Winmerge on their settings files. Very often, you'll discover that one of the simulations had something set differently, which you had forgotten about.

Winmerge asks you to navigate in Windows to the location of the files that you want to compare. To find them, click on them in the job manager and select 'Open Simulation Folder in File Explorer.'

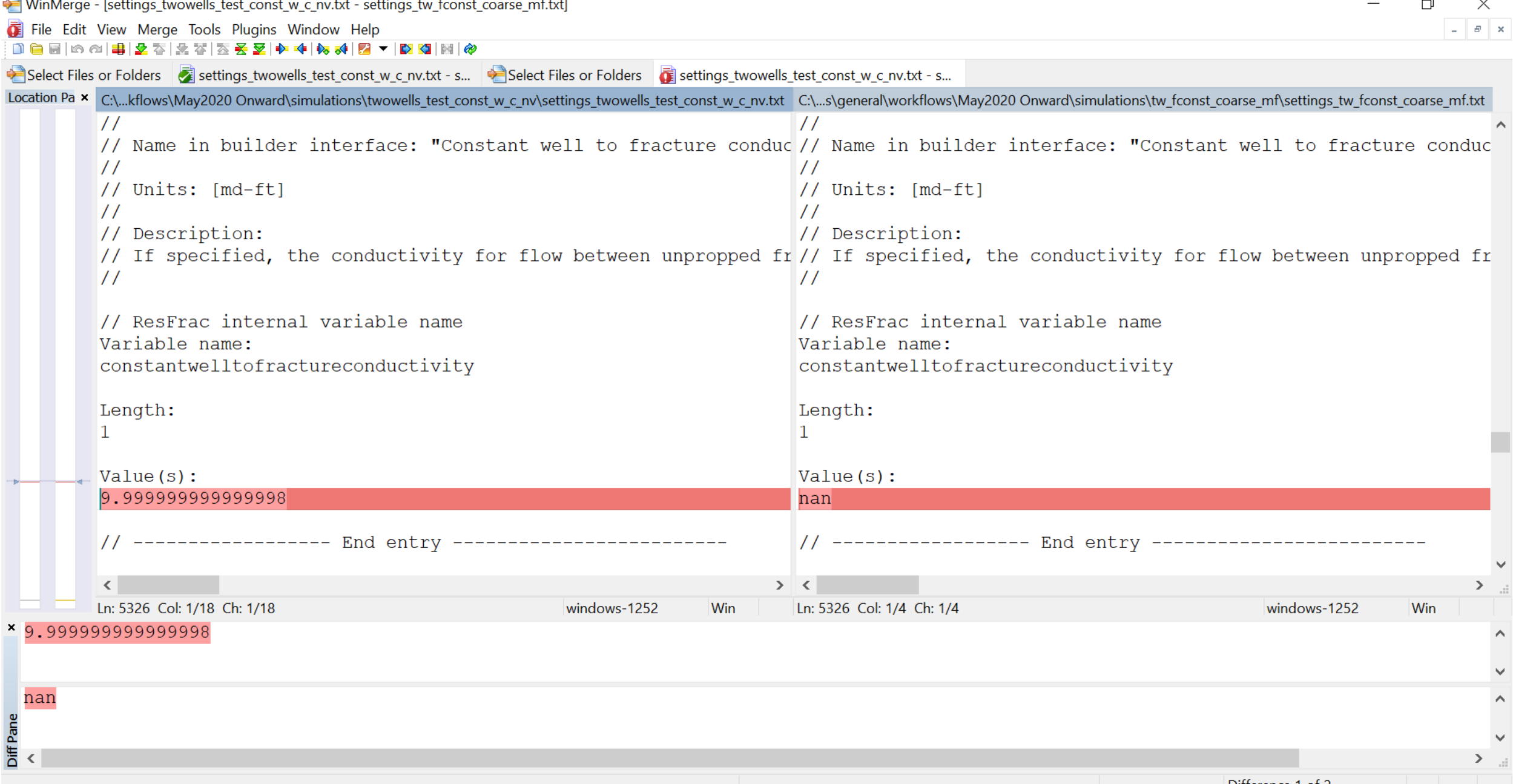

## 6.5 Presentations

This section has guidelines on making good PowerPoint presentations. A coherent, well planned presentation is crucial for effective projects.

Do not present results until you understand them. Your audience is likely to be full of smart, inquisitive people. If there is anything about the data that is anomalous or surprising, they will probably notice and ask about it. You should have analyzed the results in detail, anticipated questions, and already done the leg-work to figure out what happened. When you get those questions, you should be prepared. Also, if you get surprising results, there might just be an error. Maybe you ran a series of simulations, and one of the results was surprising. Open up Winmerge and compare that simulation's settings file with your baseline simulation. Was the simulation with the surprising result set up correctly? Maybe you made a typo when setting up that simulation? Or, maybe your simulation results are correct, and this result could be really useful to share. As discussed in Section 6.3.3, look in the 3D viewer to understand what is happening and why. If you present to stakeholders and can not explain *why*, then you will not have a lot of credibility. Even if the result is dead right, if you can not explain, then you will not present the finding effectively, and the stakeholders likely will not act on it.

Start every presentation with an overview slide. With the overview slide, determine the few key messages that you want to convey with this presentation. You are creating a mental scaffolding for the audience. As much as possible, you never want the audience to feel uncertain or confused. Everything you say in the presentation should fit into that mental scaffolding that you created with the overview slide. They should be thinking "ok, this makes sense because he started off by saying XYZ, and this fits in with that." As you go along, they should be feeling satisfied and comforted that everything is hanging together.

Maybe you are a great storyteller, and you understand that good stories end with a punchline, maybe even a surprise. That is great for telling stories, but terrible for engineering presentations. Your goal is to communicate information in a way that is clear and concise. This is not a mystery novel. No surprise endings!

Starting with an overview slide forces you to synthesize your thoughts into a few key messages. What is the point? You should do that critical thinking and analysis prior to the meeting. If you can not express your main points in a few pithy bullets, maybe you have not yet clearly thought through the problem.

You may have spent 80% of your time on 20% of what matters. That is ok – you needed to get that 20% right. But do not spend 80% of the time talking about it! Allocate time – and craft the key messages from the overview slide – by considering the audience and what they care about. What are the project objectives? What is the overall workflow and what are the key checkpoints? If you are spending a lot of time talking about something that does not clearly tie into the project objectives or workflow (or has only minor impact), is this something that you should be prioritizing time to talk about?

Consider the audience's preexisting knowledge. It never hurts to give a bit of background and context. For example, at a checkpoint meeting presenting the history match, you could start by quickly reviewing the project objectives, workflow, and the names/locations of the wells. Then, move on to the overview slide. An agenda/outline is useful if you have a longer meeting with multiple, mostly unrelated topics with clear breakpoints between them. Otherwise, it may be unnecessary.

In an informal update meeting, you may still be developing results and do not have any firm conclusions. If you are working on the history match, pull up a table with your list of 'key observations' (Section 6.3.1), and the status for each. This reminds the audience of the overall structure and status of the calibration process. Then, move onto the summary slide, where you list off what you have done since the last meeting, the outcome, and what you plan to do next. This lets the audience know your most recent results, and your near-term plan for what is next. For example: "(a) Varied gas rel perm, and still could not match the GOR trend. (b) Am now in the process of testing simulations with higher fracture conductivity, to get low pressure further out into the formation."

Generally, try to think about the presentation from the perspective of the audience – what they know and understand going in – to evaluate what to talk about, how much background information to provide.

As you go through results, slides should have about one figure and one or a few bullets that convey the main point of the slide. Do not try to convey multiple different messages with the same slide. For example, you might start with a slide that shows an image from the 3D visualization of the base case simulation – a line plot showing production and one or a few 3D panels showing pressure, proppant, etc. That slide could have just one bullet point – this is the base design. The next slide has the identical figure, but with different cluster spacing. You have one bullet point: "with wider cluster spacing, we see XYZ; the impact on production is ABC." Maybe you use an arrow to point to a location on the 3D image that demonstrations your point. The next slide has the same figure for yet another simulation, etc. The points that you are making in each slide fits into the 'mental scaffolding' that you set up in the summary slide. An audience-member might be thinking: "This makes sense that the 10 ft cluster spacing simulation did worse, because he started the presentation by saying that the 20 ft spacing design was the best." After going through each simulation (3D images with one image per slide), perhaps you have a bar chart comparing results between the simulations. This bar chart is not providing new information, just reinforcing the same messages that you have already repeated several times.

When making the figures in the 3D viewer, do not forget that the figure is going to shrink down when it's put in a PowerPoint. In order to keep the text legible, you need to make the font size rather cartoonishly big when it is shown full-screen in the visualization tool. If you use multiple panels in the visualization tool, the size of each individual panel shrinks, and it gets harder to squeeze the figure and large font all into the panel. If this is an issue, then use fewer panels. Do not forget that you can save visualization templates and use the multiplot button in the job manager to easily make the same figure for multiple simulations (see Section 5.12).

## 6.6 Optimizing design

Because of the inherent randomness of the fracture stress shadowing and geometry, there is perhaps 5% of variability in production. When optimizing design, run enough simulations to distinguish noise from signal. For example, to optimize cluster spacing, you could run simulations at 10, 12.5, 15, 17.5, 20, 22.5, 25, 27.5, 30, 32.5, 35, 37.5, and 40 ft cluster spacing. The results will bounce around slightly, but with this many data points, you will clearly see the trend and be able to identify the optimum value. The figure below from Fowler et al. (2019) is an example of testing a sufficiently wide range of well spacings, such that the optimum for each permeability assumption is clear.

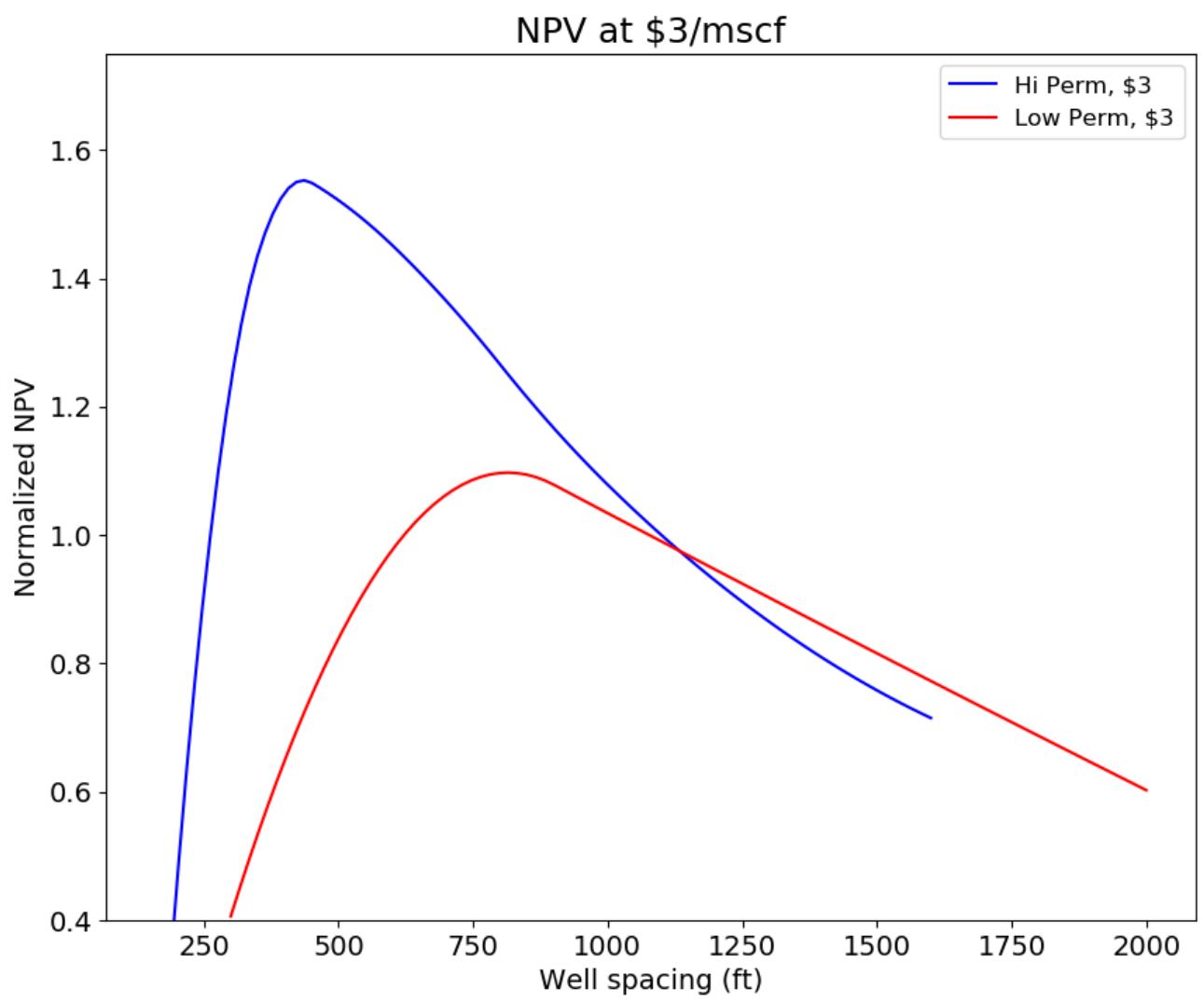

If an optimum is not reached within your range of values, keep extending the range until the curve breaks over. For example, if you ran 30, 32.5, 35, 37.5, and 40 ft spacing, and the 30 ft spacing had the best production, then run more simulations at tighter spacing – 20, 10, and 5 ft - until you see production go back down. If you do not, then you have not 'bracketed the solution' (Section 6.3.3). With some parameters, like 'proppant per ft of lateral,' you will see production go up forever. But even so, you could keep increasing proppant loading until you find a point of diminishing benefit. Or you could calculate a parameter that includes cost, like NPV, to find a metric that exhibits an optimum.

If you are changing a design variable that cannot be changed continuously (for example, switching from one type of proppant to another), it does not hurt to run a few versions of each simulation and take the average. This is done by varying the 'Random seed' parameter in the 'Other physics' panel.

## 6.7 Mindset

Strive to create a collaborative, intellectually humble atmosphere. Stakeholders have a diverse range of experiences and perspectives to share. We should strive to make everyone comfortable to put out ideas, get reactions, and have an open discussion. Be willing to be your own Devil's advocate.

Take ownership. You are likely to be collaborating with a variety of stakeholders. But if this is your project, you are ultimately responsible for the outcome. If you notice something that might be an issue, pull on the thread. Follow-up with people if you need something from them, follow the procedure (Section 4), and keep an eye on timeline. Come to meetings prepared, having already done the follow-up work to investigate the simulation results and see what is going on.

As the project progresses, constantly evaluate how it is going, and think about what you can improve the next time. Write these ideas down so you do not forget them, and you are reminded to follow up. Checklists can be useful.

## 6.8 Send us a note – support@resfrac.com

If you are unsure about what is happening in a simulation or why, send us an email. Explain what you expected to happen, what actually happened, and any relevant figure or plot that might help communicate. Send the input file, settings file, comments file, and (if applicable) a restart file from right before the problem. (See Section 5.14 on what to provide when sending us simulation files)

# 7. Technical Approach

## 7.1 Overview of technical capabilities

The ResFrac technical writeup (McClure et al., 2021) gives a detailed description of the equations solved by the ResFrac simulator. Modules 2-5 of the ResFrac Training Course (https://www.resfrac.com/resfrac-fundamentals-simulation-training) cover key physics. McClure et al. (2020) addresses nuances and frequently asked questions.

Rather than repeating that content here, we provide a quick overview of ResFrac's key capabilities. ResFrac fully integrates a wellbore-flow simulator, a 'true' hydraulic fracturing simulator, and a multiphase-flow reservoir simulator into a single, continuous simulation.

Quoting from Fowler et al. (2019):

> "Mass balance equations are solved for fluid components (water, oil, and gas in the black oil model; or pseudocomponents in the compositional model), water solute components (such as high viscosity friction reducer), and defined proppant types. Mechanical equilibrium equations are solved to calculate stress changes due to crack opening and due to fluid pressure changes in the matrix. The equations are solved in a fully coupled approach; all governing equations are satisfied in every element in every timestep. The hydraulic fractures are meshed as cracks with aperture on the order of microns to millimeters, and the cubic law is used to calculate conductivity. Proppant transport is calculated considering gravitational settling, gravitational convection, hindered settling, bed slumping, proppant trapping, and other effects. As the proppant volume fraction approaches 0.66, the proppant becomes immobilized into a packed bed of particles, and the fluid flow equations transition to equations appropriate for flow through porous media. The flow equations consider relative permeability, gravitational effects, non-Darcy pressure drop, and non-Newtonian fluid rheology."

Sometimes, we see simulators calling themselves 'coupled hydraulic fracturing and reservoir simulators,' but they are not 'true' hydraulic fracturing simulators. They are geomechanical reservoir simulators that try to mimic hydraulic fractures as zones of high permeability rock. In contrast, a 'true' hydraulic fracturing simulator meshes cracks as cracks. It should use transport equations appropriate for flow through a crack, solve the equations of crack mechanics, and handle proppant transport through a crack. Integrating a 'true' hydraulic fracturing simulator with a wellbore simulator and a multiphase, numerical reservoir simulator requires a variety of special considerations. The simulator must smoothly transition between equations appropriate for flow through an open crack and equations for flow through a closed crack (with and without proppant), and potentially with proppant screenout, non-Newtonian fluid rheology, and multiphase flow. Prior to ResFrac, constitutive equations enabling these transitions were not available in the literature. Drawing on the literature, we developed constitutive equations to handle these transitions (McClure et al., 2020).

We strongly recommend that you refer to Modules 2-5 of the ResFrac Training Course (https://www.resfrac.com/resfrac-fundamentals-simulation-training) for an overview of the key physics.

## 7.2 Planar fracture modeling and complex fracture network modeling

ResFrac is usually used to perform 'planar fracture' modeling of hydraulic fracturing. An alternative approach is the 'complex fracture network' modeling approach. In 'The Case for Planar Fracture Modeling' (https://www.resfrac.com/videos), Dr. McClure provides a detailed discussion of why we chose to focus on planar fracture modeling. The quote below from McClure et al. (2020) summarizes:

> "Planar fracture modeling is the conventional approach to hydraulic fracture modeling, and it is the primary approach used in ResFrac. Hydraulic fractures are assumed to be mostly linear, spatially continuous features. 'Complex fracture network' (CFN) modeling is an alternative approach that has sometimes been used over the past decade (Weng et al., 2011). This approach is started by seeding a network of preexisting natural fractures. Propagating hydraulic fracture are assumed to sometimes terminate against natural fractures, creating branching, zig-zagging flow pathways.
>
> McClure et al. (2020b) provides a detailed discussion of this topic. A few key points are repeated in this section.
>
> In-situ observations (from core across studies and offset well fiber) indicate that hydraulic fractures are propagating quite linearly, in a consistent orientation, and in a relatively narrow band (Raterman et al., 2017, 2019; Gale et al., 2018; Ugueto et al., 2019a; 2019b). This is fundamentally at odds with the zig-zagging fracture networks conceptualized by the CFN approach.
>
> In-situ observations show that fractures are complex at small-scale. However, when we zoom out to the reservoir scale, these fractures look like linear, planar features. Core suggests fractures have small-scale bifurcations and jobs, and multiple (subparallel) strands. The effect of these features can be captured using constitutive relations (such as ResFrac's proppant trapping/immobilization model and scale dependent fracture toughness). Complex fracture network models attempt to explicitly represent this small-scale complexity with a DFN. But there is too much complexity to truly reproduce the geometry of the fractures, and so CFN models are also grossly simplifying reality. At the same time, because adding a huge DFN has heavy computational cost, DFN models are forced to sacrifice on physical realism. For example, the model from Weng et al. (2011) is not fully 3D. Not only are CFN models also simplifying reality, they appear to be actually incorrect in most applications. The zig-zagging flow pathways of CFN models are directly contradicted by in-situ observations in major shale plays, which suggest subparallel hydraulic fractures dominate flow. The CFN approach is perhaps most useful in applications like Enhanced Geothermal Systems, or in shallow formations, where there is low stress anisotropy and fractures are less likely to be mineralized shut.
>
> An important result from the core across studies is there are very numerous (subparallel) hydraulic fracture strands. However, Raterman et al. (2019) found that only a small percentage of these fractures contain proppant, and only the propped fractures are associated with pressure depletion a significant distance from the well. Thus, during production, it is reasonable to model production as occurring from a relatively small number of major propped fractures – a planar fracture model. We routinely history match to production in shale with this modeling approach. During fracturing, core indicates that there are many water filled fractures strands. These fracture strands increase the surface area for leakoff, and so cause an accelerated leakoff. In ResFrac, we mimic this increase in leakoff area as an increase in leakoff permeability. This is done with a user-input table of pressure dependent permeability (PDP) multipliers. Thus, during fracturing, leakoff is accelerated by PDP. But production is dominated by the much more sparsely distributed set of propped hydraulic fractures, which is mimicked as the PDP multiplier goes back down to 1.0 as pressure depletes. Occasionally, the history matching process leads us to use a pressure dependent permeability loss to decrease effective permeability as depletion occurs."

ResFrac does allow you to specify the locations of preexisting fractures. This is recommended when there are a relatively small number of major fractures that create dominant flow pathways. For example, if a large fault connects between adjacent wells, you may observe that fluid localizes into the fault and creates anomalous frac hits.

On the other hand, if there are believed to be a large number of smaller-scale preexisting fractures, this is likely a situation where the fractures can be handled with an 'effective continuum' approximation and not represented explicitly.

Hydraulic fractures may experience temporary terminations and jogs when encountering an open natural fracture. For example, Schoenball et al. (2020) observed two closely-spaced hydraulic fractures terminating against a conductive natural fracture, jogging slightly, and then reinitiating and propagating on the other side of the fracture. If modeling just that process – with only a handful of fractures in the model – it might be appropriate to model the natural fracture and each individual hydraulic fracture strand. On the other hand, if performing a field-scale model – with many more hydraulic fractures and much less detailed data availability – it would be recommended to approximate the entire system as a single fracture, and account for the effect of multistranded bands of propagation and/or temporary terminations and jogs with constitutive relations. For this purpose, ResFrac includes constitutive equations for proppant holdup and trapping, elevated apparent fracture toughness, enhanced leakoff and viscous pressure drop due to multiple fracture strands, etc. (McClure et al., 2020, 2021).

## 7.3 Geologic model and gridding

There are two options available for gridding – rectilinear and corner point. In a rectilinear grid, elements are rectangular cuboids. The mesh is specified with a length, width, and height, and element lengths along each direction. In a corner point grid, elements are still defined with (i,j,k) indices. However, the elements can have much more complex shapes. They can be distorted, dipping, nonconforming, etc. Corner points are powerful because they can much more realistically capture the dip and structure of the subsurface. However, they are considerably more complicated to build and use.

Regardless of the options selected for gridding and heterogeneity, you can always press the cube icon at the top of the screen within the builder to get a 3D preview.

You have the option to specify each property either 'by depth/index' or on an element-by-element basis. If specifying by depth with a rectilinear grid, we strongly recommend that you align the element tops with the layer boundaries. There is a vertical mesh alignment wizard available in the Meshing Panel to facilitate this. If specifying by depth with a corner point grid, you specify properties by z-index, rather than by depth, since depth is not uniform within each layer.

If specifying properties on an element-by-element basis, you must provide a separate file that provides the properties. For details on the format, check out the built-in help content for 'Use general heterogeneity.'

To assist in setting up the rectilinear grid, we provide a 'Meshing wizard.' To set up a corner point grid, you need to use third party software, such as Petrel. We provide a wizard for importing corner point grid, and any available element-by-element properties. Refer to the built-in help content for the 'Static Properties' panel for details.

Typically, we are running pad-scale simulations of shale formations that have minimal structural complexity and modest dip. Also, at this scale, we typically do not have sufficient data to populate a detailed 3D distribution of formation properties. In these cases, it can considerably simplify the workflow to use a rectilinear grid with 'properties versus depth,' and without meaningful loss of accuracy. The 'Well landing depth' adjusts well depth up and down to ensure that the wells are landed in the correct zones.

If you have a geomodel in Petrel and you want to extract formation properties versus depth, the correct steps are:

1. Go to your model tab, and make sure the correct model that you want to extract the property from is **bold,** just click on it
2. Then go to your wells, right click on a well, go to settings, then make logs, from property
3. A list of possible logs will appear, so just check the one you want
4. There is an option at the bottom to 'override the existing synthetic well log'. I usually check that box, unless you don't want more than one
5. Then just hit ok
6. The well log will be created in your well, and will have 'synthetic' next to it.

# 8. History matching

## 8.1 Why history match models?

Physics-based models such as ResFrac use independent variables (permeability, water saturation, stress, etc.) as inputs into equations describing physical processes in order to produce forecasts of dependent variables (oil, water, gas rates, fracture lengths, etc.). History matching is the process of modifying and adjusting the independent variables to match real-life observations that correspond to the model's dependent variables. If the physics are known, why is it necessary to perform history matching?

There are two primary reasons for history matching:

1) Our independent variables often are uncertain. For instance, permeability may have significant uncertainty, depending on the data source. To home in on the correct, representative permeability, we vary permeability in the model to test which value/s are supported by the measured observations in the field.
2) Some independent variables are difficult or impossible to measure a priori either from field measurements (like logs) or in the laboratory. For example, effective fracture toughness is very difficult to measure robustly, and instead, is adjusted in the model to match observations of fracture propagation from fiber, microseismic, or offset frac hits.

During the history match process, we collect our observations from the field (data such as production rates and pressures, microseismic, perforation imaging, etc.) and calibrate the model input parameters to replicate these observations.

## 8.2 How to structure the history matching process

Prior to history matching, it is imperative to have a robust base case model. Be sure to use the checklist and validation process to ensure that your base case provides a solid foundation. After confirming, the history matching process can be separated into four stages:

1) Familiarize yourself with the dataset
2) List key observations and characteristics
3) Form and validate hypotheses
4) Iterate to match observations

### 8.2.1 Familiarize yourself with the dataset

Prior to modifying parameters, we recommend performing a detailed analysis of the historical data. The historical review of the data should examine the differences in completion designs, geology, and performance of the wells in your data set.

It is often easiest to compile a summary of these data in tabular form. Below are common data that we compile and compare across the wells in our data sets:

a) Landing zone

b) Well spacing
c) Cluster spacing
d) Stage length
e) Perforation friction
f) Proppant/fluid per lateral length
g) Differences in proppant and/or fluid types
h) Timing and sequence
i) Average ISIP and magnitude of ISIP escalation (if observed)
j) IP 30, 60, 180 or other production metric
k) Notes of production characteristics
    i) Is there evidence of interference from well to well?
    ii) Are there performance differences between landing zones?
    iii) Are there performance differences between generations of wells?
l) Other data-set specific observations (fracture velocity from fiber, more/less erosion as measured in a variety of ways, etc.)

Helpful tip: as you proceed through a modeling project, keep this table and a gun-barrel schematic handy. Those two slides should be in the "back-up" section of every presentation - if not in the main body of the presentation itself.

After tabulating the differences in the wells, we proceed to thinking critically about the differences and listing key observations from the data.

### 8.2.2 List key observations and characteristics

There are innumerable observations to make in a data set. Part of the art of history matching is distilling a huge quantity of data down to the key observations that characterize the performance of the system and serve as the objectives functions for the calibration. If the observations are conflicting, you must reconcile them and decide which to weight more than the others.

The raw data may be contradictory. With the list of key observations, you are not simply listing facts about the data. You are digesting the data, doing critical thinking, and deciding what you really think happened. You are making a list of attributes that you will build a model to match.

For example, sometimes, data from offset fiber shows that fractures are longer than implied by microseismic. The microseismic half-length may be 800 ft, but the fiber optic data shows that it's more like 1500-2000 ft. In that case, you might decide that the fiber is more trustworthy, because it is based on direct observation, and so your 'key observation' is that fracture half-length is 1500-2000 ft. You would not list a 'key observation' saying 'microseismic half-length is 800 ft,' because that's not what you think the half-length actually is, and that's not what you're going to build a model to match.

The list of key observations should be presented to *all relevant stakeholders* at a checkpoint meeting. It is critical for everyone to be aware of the judgment calls that you are making. You don't want to checkpoint meeting presenting the final history matching and discover that your stakeholders believe that the microseismic is more accurate, and you should have built a model with 800 ft half-length. This is why the list of 'key observations' is your digested synthesis of all the data, not just a list of facts. You need to make sure that everybody is on-board with your interpretation prior to embarking on the detailed calibration.

In nearly all data sets, ISIPs and cumulative production curves form the core of the key observations. Note that we typically do not recommend trying to match each wiggle in net pressure or production rate (more on net pressure below); however, the overall trends of these data provide valuable and important calibration data.

Additional key observations may be present depending on the diagnostics acquired and particularities of the data sets. In the sections below we describe examples of these and the trends to be extracted, though keep in mind that these are rarely all available.

**Well performance ranking** - In the wells modeled, is there a difference in performance between wells? Do inside wells underperform outside wells? Do child wells underperform parent wells? Guard against random variance by also asking questions more broadly about behavior in the field. For instance, if you observe that the well with the tightest cluster spacing in your dataset has the highest water cut, look elsewhere in the field and see whether that pattern is consistent, if your data is a one-off.

**Hydrocarbon fingerprinting / salinity testing** - Analysis of the constitutive chemical makeup of production fluids can provide information on where production is coming from. An example summary statistic might be that 80% of production is coming from Zone 1 and 20% from Zone 2. Following this observation, we might form a hypothesis that propped fracture area is predominantly contained in Zone 1 and use that to match the production allocation between zones in the model.

**Perforation imaging** - Many technologies exist today to image and measure perforation diameter. These data form the basis for observations of the degree of perforation erosion, toe/heel erosional bias, and perforation efficiency.

For example, if perforation imaging shows relatively even perforation hole sizes after fracturing, we would hypothesize that limited entry is high enough and stress shadowing low enough, that the fluid and proppant distribute evenly across the stage. Alternatively, if perforation imaging shows severe and variable perforation erosion, we need to include this in the model calibration.

**Proppant tracer** - Similar to perforation imaging, proppant tracer can be used to generate observations about the distribution of sand across the stage. There may also be data from traced proppant pumped at different periods during the stage. This allows you to draw inferences on whether proppant pumped early in the stage ends up in different locations than the proppant pumped later.

**Oil/Gas/Water tracers** - Oil, water, and gas tracers may provide information on where fluids go and where they come from, as well as information about the connectivity of fracture networks between wells.

For example, a water tracer may be injected into one well and then produced out of another well. This suggests that the fracture networks of those two wells must be connected.

**Microseismic** - Microseismic can be collected and interpreted in a variety of ways and used to provide approximate dimensions of fracture height and length. When evaluating microseismic, be sure to assess the precision (for example, a surface array will have less precision of fracture height than a downhole array will have). From microseismic, we typically try to extract average fracture geometries. Sometimes, we have encountered certain evidence of fracture propagation at much greater length than shown by the microseismic. This appears to be an artifact of the data collection, processing, or interpretation. It is important to be aware of this potential risk when using microseismic.

**Cross-well fiber optics** - Cross-well strain measured with either “dip-in” temporary fibers or permanent fibers is great data that can show fractures from adjacent wells crossing over the measurement well. We typically try to match models to aggregate statistics from these measurements: number of fractures crossing per stage, min/max/average crossing time, etc. Matching these aggregate statistics will produce a more robust model than if overfit the model by trying to match observations such as “the third cluster of the second stage crossed at 21 minutes.”

**Permanent fiber optics** - Along with the cross-well strain measurements discussed above, permanent fiber may provide data on flow distribution during injection/production and how the production distribution changes when offset wells go on or off production.

**Interference testing** - Interference testing provides valuable information on the degree of communication between wells which connotes a propped length. The best interference tests are repeated at several intervals as communication between wells can diminish as net pressure on fractures increases.

**Downhole pressure gauges** - Downhole pressure information removes many of the uncertainties associated with wellbore dynamics (hydrostatic head, friction, slugging, etc.). In some cases, pressure gauges along the lateral may even be able to characterize the distribution and variability of drainage.

**Frac hits during injection** - Frac hits, often identified by pressure spikes in the hit well, connote a minimum length of the offsetting fractures. They can also be used for sealed-well diagnostics (Haustveit et al., 2020).

**Other diagnostics** - Operators and service companies are continually innovating and creating new diagnostics to provide insights into the subsurface. These insights can almost always be imparted into fully coupled models, like ResFrac, because these models must include all physics to create a coherent representation of the subsurface.

Please refer to Section 6.3.1 for a list of questions to ask as you compile your list of key observations.

Once done brainstorming your key observations, take some time to reflect on the key observations.

- Review the guiding questions in Section 6.3.1 to get ideas of observations and differences between wells.
- Which are most important to match, and to what degree? The answer to this question will be informed by your modeling objectives and what questions you want to ask with the calibrated model.
- Are the observations consistent with broader experience? If your objective is to build a predictive model to optimize future performance, you do not want to calibrate to an outlier. Alternatively, your objective may be to *explain* this outlier. In that case, we recommend first history matching a model to the expected behavior. *Then* test what parameter/s you need to change in the base case to produce the outlier observation.

### 8.2.3 Form and validate hypotheses; plan the order that the match will be performed

Now that we have key observations listed and prioritized, we need to construct a single, coherent story to explain these observations. This "story" should be organized as a series of hypotheses to explain each observation. With each hypothesis, list the model parameters you will need to vary to implement the hypothesis. Before performing any simulations, finish listing *all* of your hypotheses, and make a preliminary plan for the history match. The preliminary plan may change, but at least you have formulated a plan that considers the totality of the data. This plan can be updated as you go along.

Which variables do you plan to vary, and in what order? The order is very important. For example, effective fracture toughness affects fracture length. Fracture length affects production. Therefore, you do not want to vary permeability to match production, and *then* vary toughness to match length. Because varying toughness will cause you to lose the production match. Conversely, in shale, permeability has a modest effect on fracture length. So vary toughness to match length first, and then vary permeability to match production.

Below is a simple example of a table of key observations and an initial plan for history matching. The observations are listed in the order that you will do the history match. Sometimes, observations need to be matched simultaneously, and so multiple rows may belong to the same group. The second column is a 'watch group.' While matching group 1, you might want to keep in mind the observations in group 2 and make sure that they aren't too far off. As discussed above, the key observations are a list of 'what you want the model to be like,' not a 'description of the raw data.' So you may add an additional column called 'Source' to record which data was used to make the interpretation that went into the key observation. You may want to refer to a specific document or PPT deck.

| Group | Key an eye on group(s) | Key observation(s)/History matching objectives | Source | What to vary - initial plan for history matching |
| --- | --- | --- | --- | --- |
| 1 | 2 | Fractures span from upper Three Forks Bench to lower part of Lodgepole | Microseismic from downhole array | Vertical toughness and/or stress layering |
| 1 | 2 | From 8 clusters, see 6 frac hits on average at 700 ft (30 min), 4 at 1400 ft (75 min), and 2 at 2100 ft (150 min) | Downhole fiber | First try strands, then modify toughness and/or viscous pressure drop |
| 1 | 2 | ISIP escalates about 300 psi from the first stages to subsequent stages | Pressure versus time data | First try strands, then modify toughness and/or viscous pressure drop |
| 2 | 1 | Perf efficiency is 90+% | Fiber in injection well | Effective tensile strength |
| 3 | | Propped half-length is at least 700 ft within zone. | Interference tests | Maximum trapped proppant per area |
| 4 | 5,6 | The RTA plots start flat, but eventually start to curve up gradually | Production/BHP versus time data | Perm multiplier, rel perm curves or PDP |
| 4 | 5,6 | The RTA curve has zero y-intercept | Production/BHP versus time data | Make sure propped conductivity is sufficiently high |
| 5 | 5,6 | Water cut is 60% long-term, with temporary spike during start of production | Production/BHP versus time data | Residual water sat, water banking parameter |
| 6 | 5,6 | GOR starts at 1500 sft/STB, and then increases gradually to 3000 | Production/BHP versus time data | Gas rel perm (residual and exponent) |

After forming your hypotheses, and before launching an extensive effort to "dial in" the correct parameters, it is best to validate the hypothesis by testing the extremes of the model parameter ranges and confirming that the observed data falls within the model prediction range.

Examples of production trends we sometimes see:

- Outside wells outperform inside wells
    - Example hypothesis: propped fracture length is at least half the well spacing, so inside wells feel interference on both sides, but edge wells can drain unimpeded on one side.
    - Governing model parameters: toughness, toughness scaling factor, proppant immobilization.
- Tighter cluster spacing and/or tighter well spacing yielding more height growth
    - Example hypothesis: more fractures in the same volume increases stress shadowing which incentivizes height growth.
    - Governing model parameters: toughness, toughness anisotropy, toughness scaling, Shmin profile
- Asymmetric fracture propagation toward depletion

- Example hypothesis: poroelastic stress changes make it easier for fractures to propagate in areas of depletion versus areas of virgin stress.
- Governing model parameters: Biot's coefficient, toughness, toughness scaling factor, proppant immobilization, permeability

Sometimes there are contradictory observations. Subsurface data is imperfect. What is the confidence ranking of each observation? An example might be that a microseismic data set shows fractures only extending 600 feet from the well, but downhole fiber shows frac hits in a well 2000 feet away. If this is the case, which observation is the more important match? Which is most representative of your data? This will vary by dataset, so be sure to communicate with all stakeholders.

Try to keep your list of key observations down to about 6-12. Section 6.3.1 gives an example of a list of 'key observations to match.'

**Recommended history matching progression:**

Just as there is no single, exact solution to a history match, there is no single perfect progression. However, following the steps below streamline the history matching process. Keep in mind that at times history matching is non-linear and you may need to regress to prior steps if you falsify a prior hypothesis. The sequence of steps below is designed to make the process mostly linear. The changes made in each step will probably not significantly impact the match that you already performed at a previous step.

1) Set up the model with geologic properties, wellbore and completion designs, and fracturing and production schedules. Add a stress observation plane at wellbore depth.
    a) Run the initial models using BHP control. You can change to specified rate once GOR and WC are close, but simulating with rate control when permeability is way off can cause non-uniqueness in the data that complicate history matching.
2) Use microseismic, frac hit data, DAS, and other data to constrain fracture geometries
3) If available, use step down test, camera, or fiber data to calibrate perforation efficiency
4) Add external fracture and tune to match stress shadowing observed in stress observation plane
5) Check that ISIPs and net pressures match field observations within reason*
6) Adjust wellbore friction factor to match surface treating pressures
7) Use interference tests to constrain the size of the propped area (controlled in ResFrac by varying 'maximum proppant trapping lbs/ft^2"
8) Tune relative permeability (endpoints and exponents) to match water cut
    a) Plot relative permeability curves (DRAW THEM OUT). Mark your initial water saturation. Plot water fractional flow.
9) Use RTA to home in on permeability and fracture conductivity in early time
    a) Adjust relative permeability to match GOR behavior
    b) Check 3D image. Is frac pressure = BHP?

10) Use RTA characteristics to identify later time phenomena
    a) Interference and no-flow boundaries
    b) Dropping below bubble point
    c) Pressure-dependent permeability
    d) Time-dependent conductivity loss

*Several processes (wellbore friction, near- mid-field tortuosity, perforation erosion) impact net pressures and even ISIPs (McClure, 2020). Therefore, an exact match to ISIPs or net pressures is not necessary.

Sometimes a simple observation has complex implications on prior steps in the history matching progression. Thus, even as you go through early steps, it is good to keep in mind the full set of observations. Examples:

- Interference testing shows communication between two wells: This will require the propped fracture lengths from each well to be *at least* half the well spacing (and likely more due to heterogeneity). Chronologically, you will likely match ISIPs before production interference - but do so knowing that there is not much utility to matching ISIPs if the fracture lengths are less than half the well spacing!
- GOR increases when BHP drops below bubble point: This will require high conductivity fractures (minimal pressure drop between the wellbore and fracture). Again, matching GOR will come later in your progression, but be sure to keep an eye on fracture conductivity while matching fracturing observations and early-time production.
- ISIP consistently 1000s of PSI above Shmin (as measured by DFIT): This will require high near-wellbore tortuosity. High near-wellbore tortuosity can elevate perforation efficiency due to the increased back-pressure. Keep an eye on perforation efficiency as you vary near-well complexity (simulation parameter to affect tortuosity). To match near-wellbore tortuosity that lingers for more than a few minutes (which is typical), you may need to use the 'pressure dependent NW tortuosity' options.

The final step is to validate your hypotheses. The most effective tactic to test hypotheses is to "bracket the solution", i.e. quickly test each hypothesis by testing the extremes of the parameter range. For example, if one hypothesis is that well interference causes inner wells to underperform outer wells, begin by testing the extreme case with effective fracture lengths *much longer* than what you think is reasonable (achieved by lowering effective fracture toughness and using a lower value of proppant immobilization - more on that below). If this test, with maximum interference between inner wells, does not produce the observation you are trying to replicate (that inner wells produce less than outer), then you have invalidated that hypothesis, and need to form a new hypothesis!

Helpful tip: whenever evaluating a simulation that does not yield the result you desired, start by interrogating the 3D images. If your hypothesis was that lowering proppant immobilization would create longer modeled propped fractures, but when you did so, production went down, start by looking at the 3D image. Was propped length longer? Check conductivity, does spreading the proppant over the larger area decrease propped conductivity, and so the pressure in the propped area is now too high? The answer is always in the 3D images!

It is usually recommended that you change only one thing at a time. If you change multiple design parameters simultaneously, you cannot be sure which change was responsible for the difference in model output.

If you are struggling to understand a result, do not hesitate to reach out to us within ResFrac, and we can provide our perspective.

With all hypotheses validated, it is time to proceed to iterating model parameters to minimize the mismatch between model predictions and observed data.

### 8.2.4 Guidance on how to iterate to match observations

The final step of the process is to fine-tune model parameters to match your observations. Below, we discuss various model outputs and what parameters may impact the output. Refer to the worked example to read about specific tactics to quantitatively use relative permeability curves and RTA plots to match data. We suggest frequently returning to your hypotheses and evaluating whether matching one observation contradicts another.

**Parameter recommendations**

**Fracture length and height**

Simulators should not be relied-upon to predict fracture lengths in absence of calibration. Fracture length depends on field-scale parameters that cannot be reliably predicted without calibration to field-scale data. These field-scale parameters are consistent between adjacent pads but vary considerably between shale basins.

In history matching, we tune fracture toughness to produce fractures of the expected length. “Effective” fracture length can be backed into using RTA. However, note that the effective fracture length is often much less than the total fracture length (which is what would be measured by microseismic or offset fiber optic strain). Below are parameters often used to modify fracture length and height.

- Fractures strands per swarm: The first step in tuning fracture length and height is to decide on the number of fracture strands you assume per propagating fracture. Recent core-through studies have shown that each dominant propped fracture is associated with a large number of unpropped fractures (Raterman et al., 2019; URTeC-263-2019). ResFrac mimics this effect by adjusting the toughness, viscosity, and leakoff as proposed by Fu et al. (2020) in SPE-199689. More strands = higher toughness, higher viscous pressure drop, and more leakoff (all of which cause shorter fractures).

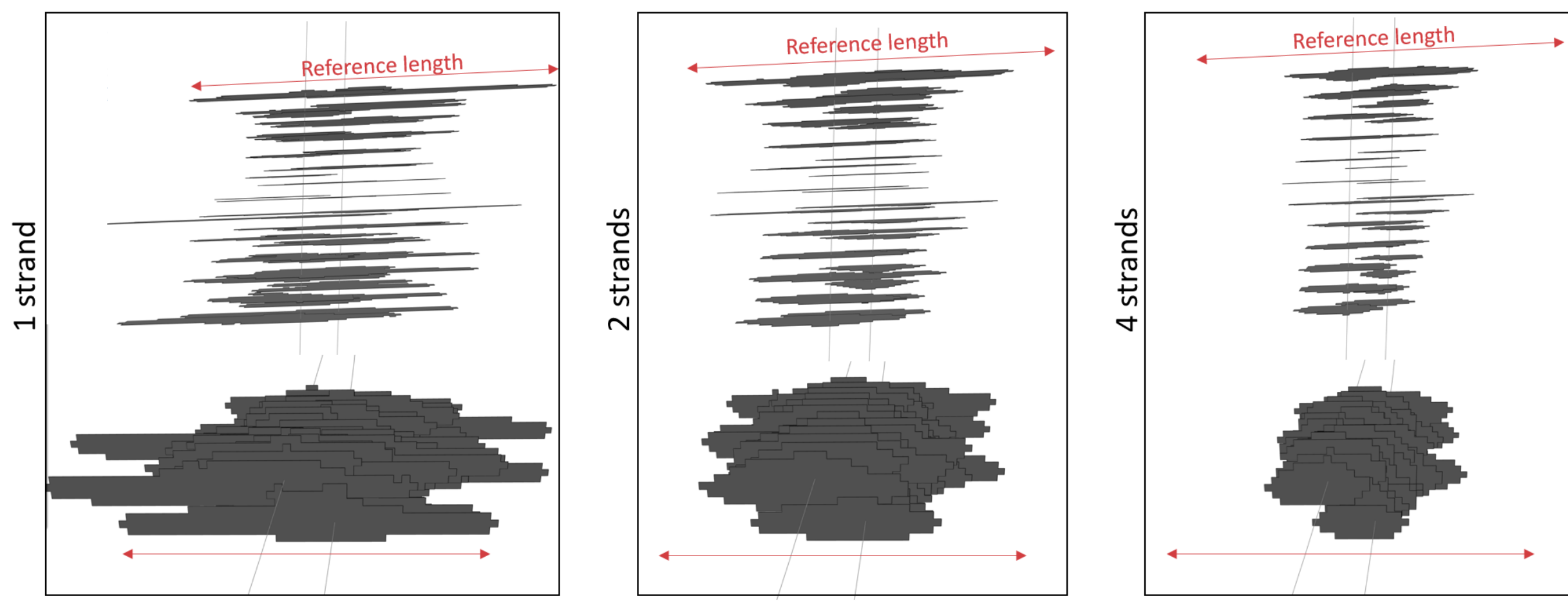

Figure 8.1 – Change in fracture dimensions with fracture strands per swarm.

- Fracture toughness: Higher toughness values will result in shorter fractures. ResFrac allows for anisotropic toughness. Sometimes a higher vertical fracture toughness than horizontal fracture toughness is used to control height growth.
- Toughness scaling factor: Elevated fracture toughness at field scale has been documented many times in the literature (Delaney et al., 1986; Shlyapobersky and Chudnovsky, 1994; Gale et al., 2018; Hurt, 2019). In ResFrac, the user can specify "Relative fracture toughness per square root fracture size," which will scale the effective fracture toughness as a function of either fracture length or fracture height. Increasing this scaling factor will result in shorter fractures and higher net pressures.

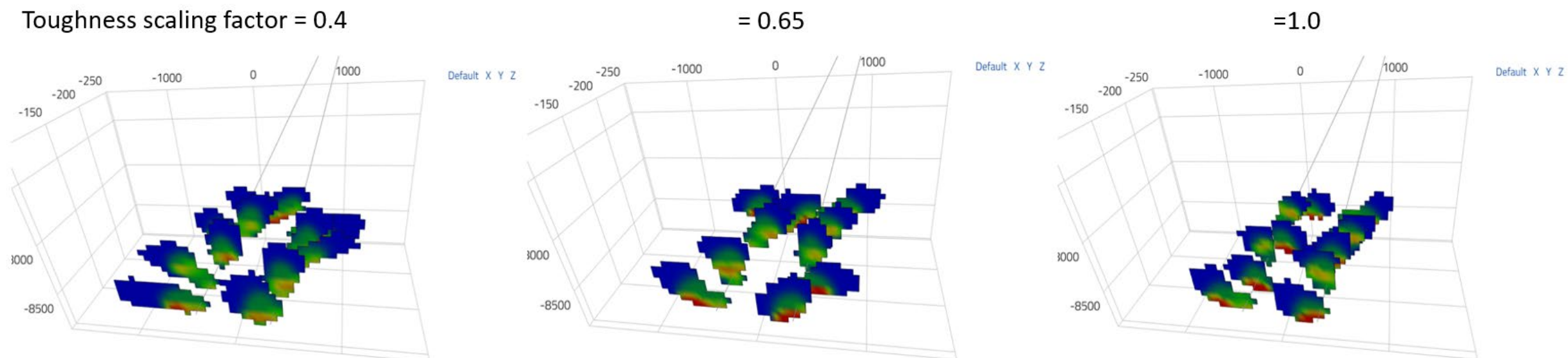

Figure 8.2 – Change in fracture dimensions with toughness scaling factor.

- Pressure dependent permeability (PDP): Increasing PDP will result in more leakoff. In most shale plays, fracture geometry is toughness-dominated, not leak-off dominated; however, in some formations (particularly higher permeability formations), leak-off will impact fracture geometries. Higher PDP will result in shorter fractures if leak-off is sufficiently high.
- Shmin profile: More variability and higher contrasts in Shmin between layers will result in greater confinement (fractures that do not grow as high). If diagnostic data shows fractures terminating at a specific depth, there is a strong likelihood that a stress barrier (high contrast) exists at that depth.

**Propped length**

Typically, the significant majority of production will come from the propped fracture region. If two or more wells are observed to interfere with each other, then the propped fracture regions of the fractures emanating from the respective wells are likely intersecting and/or overlapping.

- Maximum immobilized proppant per area: Fracture wall roughness and a variety of other effects may hold up proppant from settling and/or being transported as it would in an open crack. Higher values of max proppant immobilization will "hold-up" more proppant near the wellbore, resulting in a denser proppant pack near the well, but often shorter.
- Fracture toughness and toughness scaling factor: A higher effective fracture toughness will curtail total hydraulic fracture size, and may also curtail how far out proppant can travel (as proppant rarely can make it all the way to the fracture tip).
- Fluid viscosity: Higher viscosity fluids will transport proppant further from the wellbore. Sometimes there is uncertainty in the downhole viscosity of a fluid, and this might be modified to sensitize proppant transport impacts.

**Perforation efficiency**

Perforation efficiency is a loose term. The definition can vary depending on context. For instance, perforation efficiency may mean either the percentage of clusters that initiate and source a fracture (as measured during injection with downhole optical imaging, proppant tracer, or step-rate testing), or it could mean the percentage of clusters contributing to production (as measured during production with something like permanent fiber). Generally, in this discussion we treat perforation efficiency to mean the percentage of clusters that initiate.

- Net pressure: While not a discrete parameter in ResFrac, any parameter that increases net pressure may decrease perforation efficiency (McClure et al., 2020). The primary parameters impacting net pressure are fracture strands per swarm, toughness, toughness scaling factor, and pressure dependent permeability.
- Perforation erosion: Perforation friction, also known as limited entry, can overcome stress shadowing. Perforation friction is predominantly a function of the well design: number of perforations, perforation diameter, injectant viscosity, and injection rate. Well design parameters are generally treated as known variables and not modified during the history match; however, perforation friction will degrade as a function of perforation erosion. ResFrac uses the Long and Xu (2017) erosion model to calculate perforation erosion. If you have data to constrain erosion, you can adjust the model parameters *Perforation erosion alpha* and *Perforation erosion beta* to match erosion. Higher erosion can decrease perforation efficiency.
- Perforation diameter standard deviation: By default, ResFrac initializes all perforations with the diameter specified in the well trajectory table. However, there may be variability in true perforation diameter, so the user may specify a variance in diameter using this parameter.
- Randomized tensile strength: If the fracture breakdown pressure varies from cluster-to-cluster, then perforation efficiency may drop.

**Treating pressure**

Treating pressure, net pressure, and fracture geometries are tightly coupled. We recommend first matching expected fracture geometries, then proceeding to treating pressures (keeping in mind that treating pressure naturally has a lot of variability).

- Fracture toughness: Increasing fracture toughness will increase net pressure and treating pressure.
- Toughness scaling factor: Similar to absolute fracture toughness, higher values of the toughness scaling factor will result in higher net pressure and treating pressure.

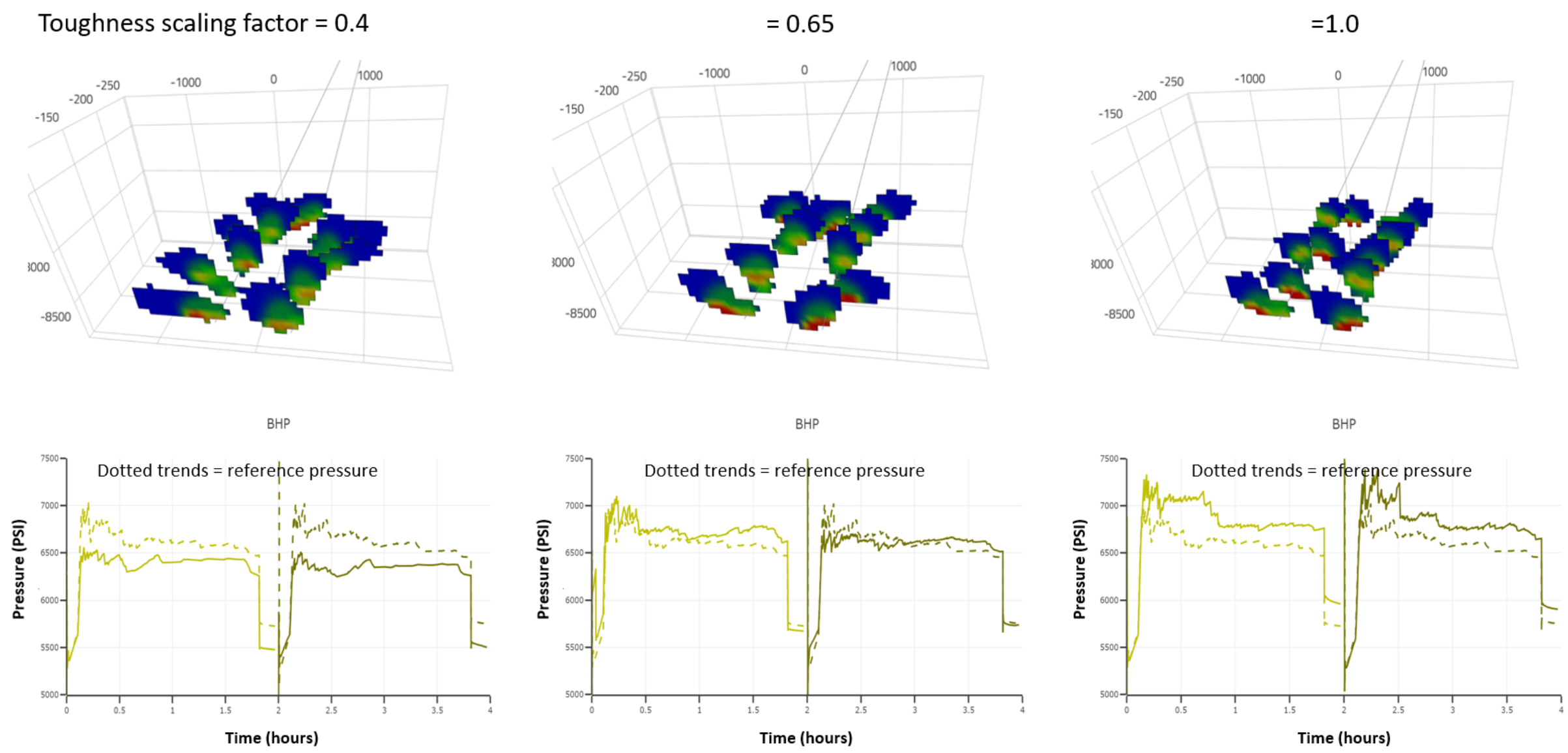

Figure 8.3 – Change in fracture dimensions and treating pressure with toughness scaling factor.

- Near-well (NW) complexity: There are two implementations of near-well complexity in ResFrac that the user can choose from. The options are mutually exclusive, and both are used to replicate the effect of near-well tortuosity. Higher values of near-well complexity will elevate treating pressure for a given injection rate, but not impact net pressure (i.e. pressure in the fracture will be unchanged).
- Pressure dependent permeability (PDP): PDP is used in ResFrac models to account for the propagation of multiple fracture strands and the associated fluid losses. Higher values of PDP will result in lower treating pressures and faster pressure fall-off after shut-in.
- Wellbore friction adjustment factor: Higher values of wellbore friction adjustment factor (generally < 1.0) will increase treating pressure while leaving BHP unchanged. You can vary a 'global' wellbore friction adjustment factor, or you have the option to specify a different friction adjustment factor for each type of water solute.
- Wellbore proppant friction adjustment factor: Higher values of wellbore proppant friction adjustment factor will cause the frictional pressure component of WHP to increase. Modifying wellbore proppant adjustment factor will not affect BHP.

**Production volumes**

The combination of cumulative production plots, GOR plots, and RTA plots can help identify the mechanisms responsible for production trends. The effective permeability and conductivity of the fracture system are the two most important aspects of a volumetric production match. Analyzing the RTA signature of the simulation and historical data is invaluable in determining which to modify.

- Permeability: Permeability may be modified either in the static model (which allows for the permeability of each layer to be independently modified) or using the global permeability multiplier (which multiplies all static model layers by the same factor). The slope of the RTA trend is inversely proportional to the product of effective fracture area (propped area) and the square root of matrix permeability. If propped fracture area is assumed the same between the model and the historical data, the user can calculate the required permeability change necessary to match the data by comparing the slopes of the two trends.
- Propped fracture conductivity: The higher the conductivity of a fracture, the smaller the pressure drop between the wellbore and the fracture. Increasing $k_o$ in the proppant properties table will increase the conductivity of the proppant pack, and is the primary parameter used to modify the conductivity of a fracture in a ResFrac simulation. The y-intercept of an RTA trend is inversely proportional to the fracture conductivity, i.e. a higher y-intercept corresponds to a lower fracture conductivity. By comparing the historical and simulated RTA trends, the user can deduce the required change in fracture conductivity. Note that $k_o$ is a linear multiplier of proppant conductivity. Very large values of $k_o$ may *increase* propped area by making very low concentrations of proppant highly conductive. Always gut check the 3D image to make sure the conductive area of a fracture makes sense.
- Pressure dependent permeability (PDP): In some reservoirs, permeability decreases as a function of pore pressure depletion (Heller et al., 2014; King et al., 2018). This is particularly common in over-pressured gas shales. ResFrac supports PDP and can be affected by inputting negative delta pressure values and multipliers less than one into the PDP table. An upward bending RTA trend may indicate PDP. As the slope of the RTA trend is inversely proportional to the square root of matrix permeability, one can compute the degree of PDP by analyzing the RTA plot.
- Relative permeability: The effective permeability of a given phase is equal to the product of the matrix permeability and the relative permeability of the phase. Check the relative permeability of the flowing phases in a simulation to make sure that the relative permeability is not excessively low or high. An upward curvature in the RTA plot may also be caused by a loss of relative permeability when gas comes out of solution. To match this trend, you need to shape the relative permeability curve so that when oil saturation goes down from the initial saturation, oil relative permeability promptly decreases.
- Unpropped fracture conductivity: When fractures close, they retain some conductivity even if there is no proppant in the fracture. The conductivity and stress dependence of this unpropped fracture conductivity is controlled by E0max, 90% closure stress, and Eresmax in the static model. This is usually a small effect and neglected in most simulations.

- Time-dependent conductivity loss may explain many upwardly bending RTA curves. This is based on: (1) observations from laboratory experiments suggesting time-dependent conductivity, and (2) experience with history matching suggesting that time-dependent conductivity loss may help with matching late-time EUR observations and RTA curvature. Time-dependent conductivity loss could happen due to deposition of organic or inorganic scaling in the proppant pack, fines migration, or time-dependent embedment or crushing. Refer to Section 19.7 of the ResFrac Technical Writeup (May 2022 version or newer) for additional details.

**Matching GOR**

GOR is a function of the fracture pressure and relative permeability curves. Fracture pressure is a function of the fracture conductivity. Fracture pressure must drop below bubble point in order for GOR to rise above the solution gas-oil-ratio ($R_s$)

- Proppant pack conductivity: When building a model, ResFrac allows for the user to create a proppant conductivity versus normal stress curve from a data sheet or use defaults. $k_o$ in the proppant properties table linearly multiplies this curve up or down, changing the conductivity of the propped fracture area. Higher conductivity results in less pressure drop between the wellbore and the fracture. If the wellbore pressure is below bubble point, but the pressure in the fractures is not, then the GOR will not increase.
  Hint: look at the 3D image. If the fracture pressure adjacent to wellbore is the same pressure as the wellbore, then increasing proppant conductivity is unlikely to help match GOR. If the fracture pressure is above the bubble point, but the wellbore pressure is below the bubble point, then higher conductivity will increase the GOR increase over time.

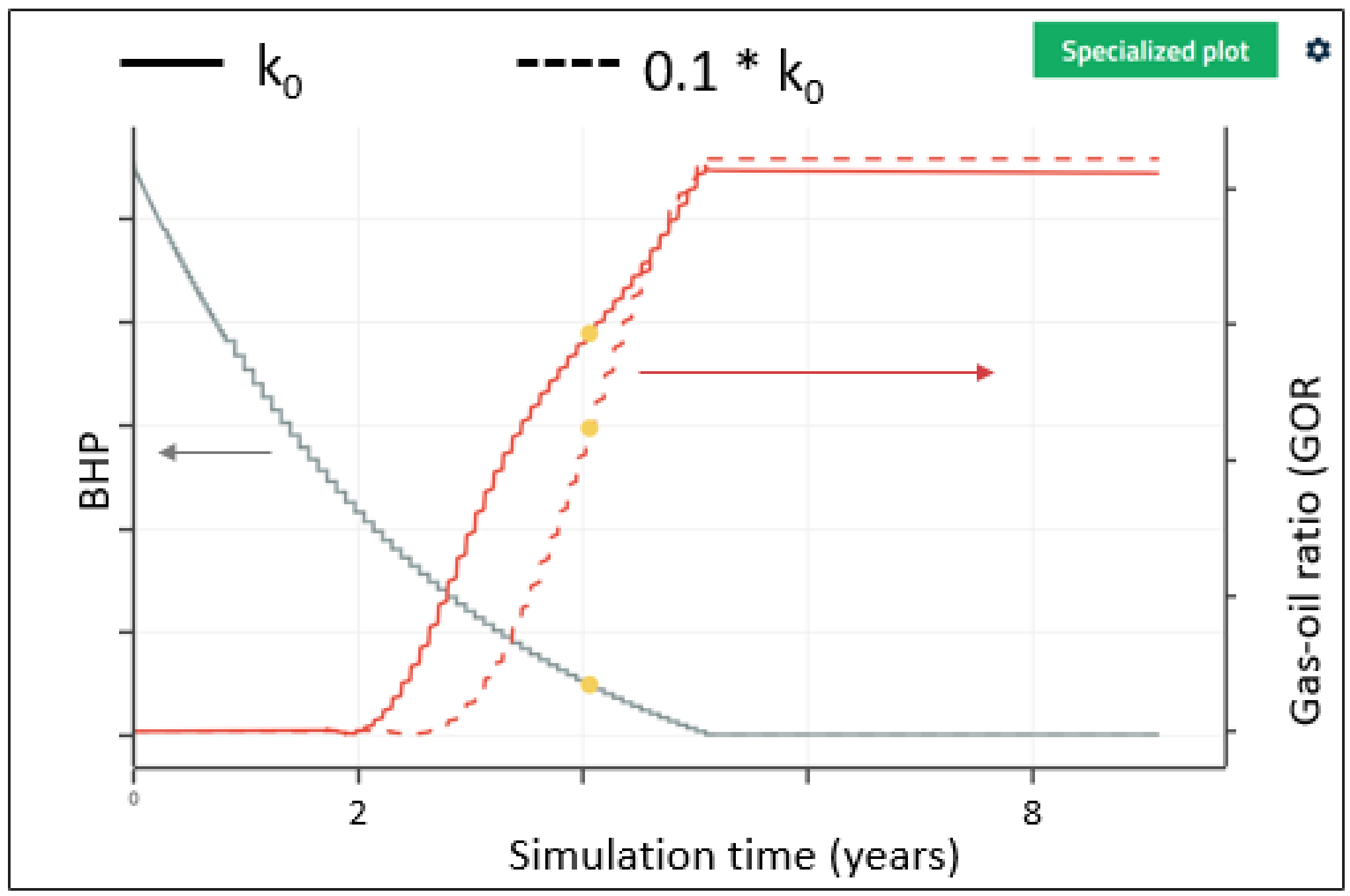

Figure 8.4 - Demonstrating how a lower propped fracture conductivity can delay the GOR rise, even when BHP is the same between the two models.

- Maximum immobilized proppant per area: In order for fracture pressure to efficiently be depleted, there must be a conductive pathway between the wellbore and the fracture. If proppant settles out below the wellbore, then a pinch point will develop between the wellbore and the fracture and there will be a pressure drop between the two. Increasing the maximum immobilized proppant per area will keep more proppant next to the wellbore, thus eliminating a pinch point. When should you use $k_o$ versus proppant immobilization? Check the 3D image and examine the density of the proppant pack near the wellbore. If the proppant pack is 0.25 lb/ft or greater, then there is likely sufficient proppant to eliminate a pinch point, so $k_o$ will be a more effective parameter to target.

- Relative permeability: The mobility of the gas phase, controlled by the gas relative permeability curve, will impact GOR. If the residual gas saturation is too high, matrix pressure may be dropping below $P_b$, but the evolved gas is "trapped" in the matrix. Increasing the relative permeability of the gas phase at a given saturation will also increase GOR as the free gas will be able to travel more easily through the matrix. Very often, we use aggressive relative permeability curves for gas, such as residual saturation of 0.03 and Brooks-Corey exponent of 1.5. For numerical reasons, we recommend keeping the residual saturations at least slightly above zero, no lower than 0.001, and ideally no lower than 0.03.

## 8.3 History matching tactics

In this section we cover ResFrac features helpful for history matching, helpful third party tools, and a worked example using relative permeability and Rate Transient Analysis (RTA) plots to quantitatively match data.

### 8.3.1 Useful ResFrac features

Several features in ResFrac facilitate expeditious history matching. These are briefly covered below, but be sure to navigate to the specific section in this manual for a more detailed review.

- Plotting external data
  ResFrac users can use the 'Import External Data' feature in the visualization tool to import the time series data from another simulation or the time series data from the field. You can plot the actual and simulated data side-by-side.

- Saving visualization layouts
  The ResFrac visualization tool allows you to save a particular layout (including any imported data) and use that same layout to open similar simulations. This allows the user to open several history match iterations with the exact same plots and 3D visuals, expediting the process of reviewing simulation runs. See Section 5.12.

- Multiplot
  The Multiplot button in the job manager allows you to select a layout and multiple simulations to automatically make the same figure for multiple simulations. There is also an option to automatically create a line plot that shows the results from multiple simulations. See Section 5.12.

- Using restart files
  A restart file is a binary file that saves the exact state of the simulation at a point in time. Using a restart file, the user can modify model parameters and restart the simulation from that point in time. This can greatly reduce runtimes. However, be cognizant that any changes made to the model could impact the portion of the model already run. See Section 13.1.

- Using simulation Raw Results
  Every ResFrac simulation creates results readily visible in the visualization as well as additional outputs available in the Raw Results. The Raw Results files include outputs such as production versus depth and summarized fracture statistics.

- Definitions of results trends
  For any questions on what various visualization parameters are showing (either line plot or 3D) please reference the <ResFrac simulation outputs.xlsx> in the resources > documentation folder of your ResFrac installation.

### 8.3.2 Third-party tools for history matching

Several open source, free software solutions can help you manipulate and compare simulation files and results. This section lists tools specifically for history matching. Refer to Section 6.4 for additional suggestions.

*Excel or other spreadsheet tools*
As you progress through a history match, we recommend you document the parameters that you have changed and their impact on simulations. In the image below, each row represents a separate simulation, and each column records either the parameter change or resulting impact. Using a spreadsheet to organize your iterations will help elucidate trends and prevent unneeded simulations.

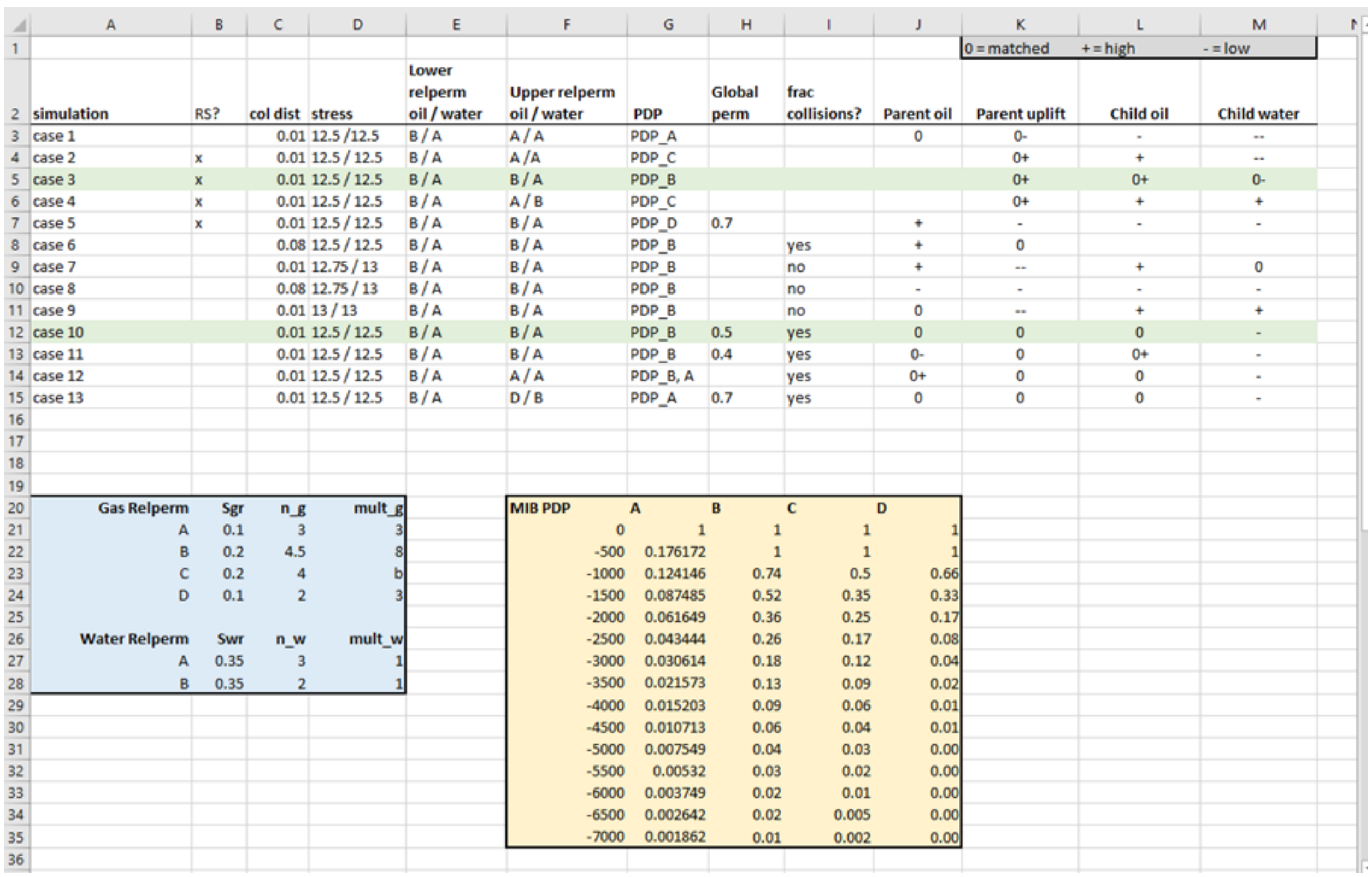

0 = matched + = high - = low

| simulation | RS? | col dist | stress | Lower relperm oil / water | Upper relperm oil / water | PDP | Global perm | frac collisions? | Parent oil | Parent uplift | Child oil | Child water |
|---|---|---|---|---|---|---|---|---|---|---|---|---|
| case 1 | | 0.01 | 12.5 /12.5 | B / A | A / A | PDP_A | | | 0 | 0- | - | -- |
| case 2 | x | 0.01 | 12.5 / 12.5 | B / A | A /A | PDP_C | | | | 0+ | + | -- |
| case 3 | x | 0.01 | 12.5 / 12.5 | B / A | B / A | PDP_B | | | | 0+ | 0+ | 0- |
| case 4 | x | 0.01 | 12.5 / 12.5 | B / A | A / B | PDP_C | | | | 0+ | + | + |
| case 5 | x | 0.01 | 12.5 / 12.5 | B / A | B / A | PDP_D | 0.7 | | + | - | - | - |
| case 6 | | 0.08 | 12.5 / 12.5 | B / A | B / A | PDP_B | | yes | + | 0 | | |
| case 7 | | 0.01 | 12.75 / 13 | B / A | B / A | PDP_B | | no | + | -- | + | 0 |
| case 8 | | 0.08 | 12.75 / 13 | B / A | B / A | PDP_B | | no | - | - | - | - |
| case 9 | | 0.01 | 13 / 13 | B / A | B / A | PDP_B | | no | 0 | -- | + | + |
| case 10 | | 0.01 | 12.5 / 12.5 | B / A | B / A | PDP_B | 0.5 | yes | 0 | 0 | 0 | - |
| case 11 | | 0.01 | 12.5 / 12.5 | B / A | B / A | PDP_B | 0.4 | yes | 0- | 0 | 0+ | - |
| case 12 | | 0.01 | 12.5 / 12.5 | B / A | A / A | PDP_B, A | | yes | 0+ | 0 | 0 | - |
| case 13 | | 0.01 | 12.5 / 12.5 | B / A | D / B | PDP_A | 0.7 | yes | 0 | 0 | 0 | - |

| Gas Relperm | Sgr | n_g | mult_g |
|---|---|---|---|
| A | 0.1 | 3 | 3 |
| B | 0.2 | 4.5 | 8 |
| C | 0.2 | 4 | b |
| D | 0.1 | 2 | 3 |
| **Water Relperm** | **Swr** | **n_w** | **mult_w** |
| A | 0.35 | 3 | 1 |
| B | 0.35 | 2 | 1 |

| MIB PDP | A | B | C | D |
|---|---|---|---|---|
| 0 | 1 | 1 | 1 | 1 |
| -500 | 0.176172 | 1 | 1 | 1 |
| -1000 | 0.124146 | 0.74 | 0.5 | 0.66 |
| -1500 | 0.087485 | 0.52 | 0.35 | 0.33 |
| -2000 | 0.061649 | 0.36 | 0.25 | 0.17 |
| -2500 | 0.043444 | 0.26 | 0.17 | 0.08 |
| -3000 | 0.030614 | 0.18 | 0.12 | 0.04 |
| -3500 | 0.021573 | 0.13 | 0.09 | 0.02 |
| -4000 | 0.015203 | 0.09 | 0.06 | 0.01 |
| -4500 | 0.010713 | 0.06 | 0.04 | 0.01 |
| -5000 | 0.007549 | 0.04 | 0.03 | 0.00 |
| -5500 | 0.00532 | 0.03 | 0.02 | 0.00 |
| -6000 | 0.003749 | 0.02 | 0.01 | 0.00 |
| -6500 | 0.002642 | 0.02 | 0.005 | 0.00 |
| -7000 | 0.001862 | 0.01 | 0.002 | 0.00 |

*WinMerge*

WinMerge is a free text file comparison tool. If you ever forget what you changed in a simulation, or want to figure out the difference in two simulations, WinMerge allows you to compare the two files and it will highlight all the differences.

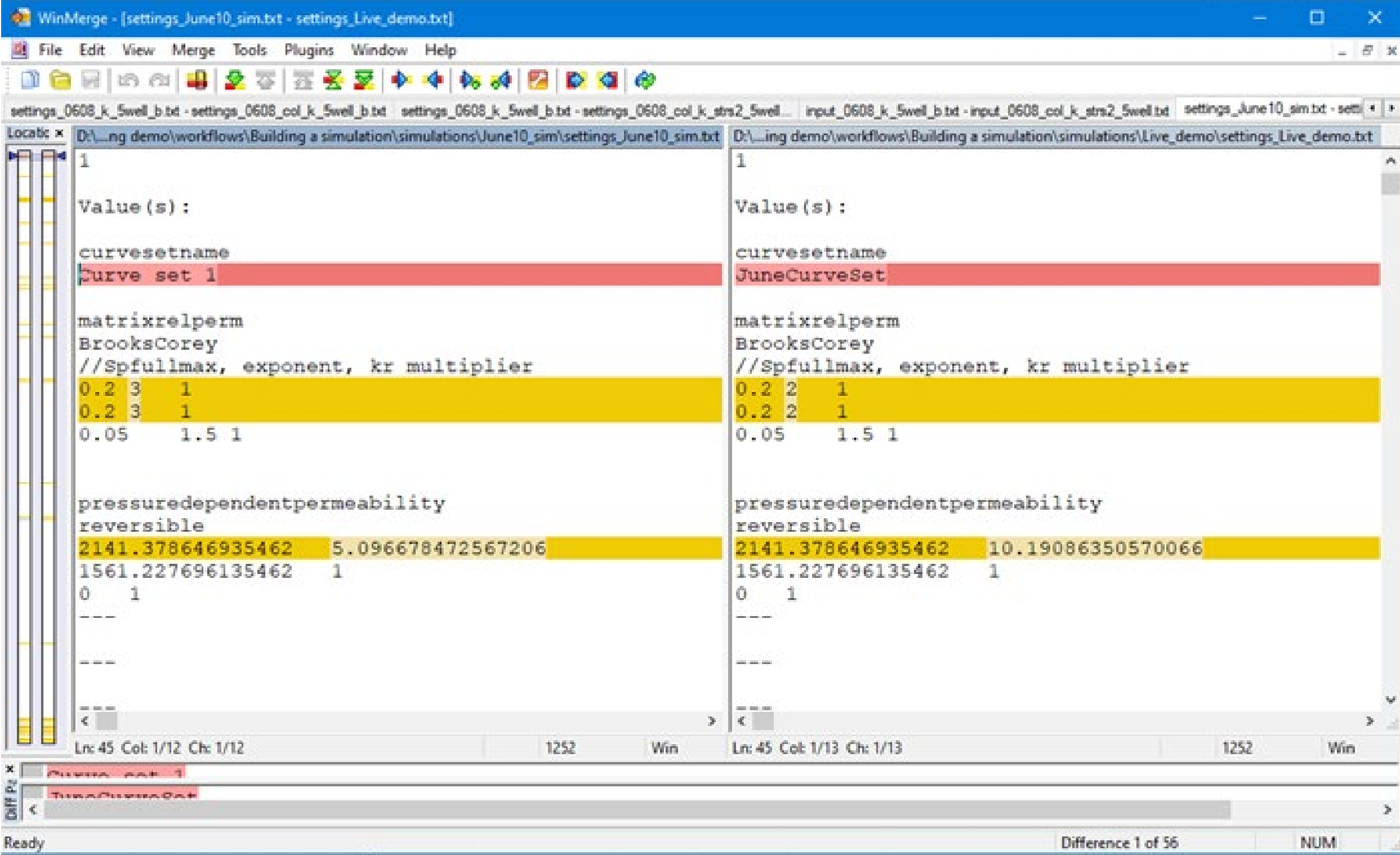

Figure 8.6 – Image of example WinMerge comparison.

*Python*

Advanced ResFrac users use Python or other programming languages to methodically modify and manipulate simulator inputs and outputs. All ResFrac inputs and outputs are available as human readable ASCII files to enable programmable pre or post processing. ResFrac even offers a command-line interface that can be used to enable Python scripts to automatically run batches of simulations.

## 8.4 Production history match: a worked example

In the example below, we walk through tactics that use relative permeability and RTA plots to streamline the production history match.

Leveraging the recommended history matching workflow in the section above, production matching starts at step seven:

7) Tune relative permeability (endpoints and exponents) to match water cut
    a) Plot relative permeability curves (DRAW THEM OUT). Mark your initial water saturation. Plot water fractional flow.
8) Use RTA to hone-in on permeability and fracture conductivity in early time
    a) Adjust rel perms to match GOR behavior
    b) Check 3D image. Is frac pressure = BHP?
9) Use RTA characteristics to identify later time phenomena
    a) Interference and no-flow boundaries

b) Dropping below bubble point
c) Pressure-dependent permeability
d) Time-dependent conductivity loss

The figure below shows the water cut, GOR, and oil rate for the historical data (or “actual” data) to match in dashed trends and simulated data in solid trends.

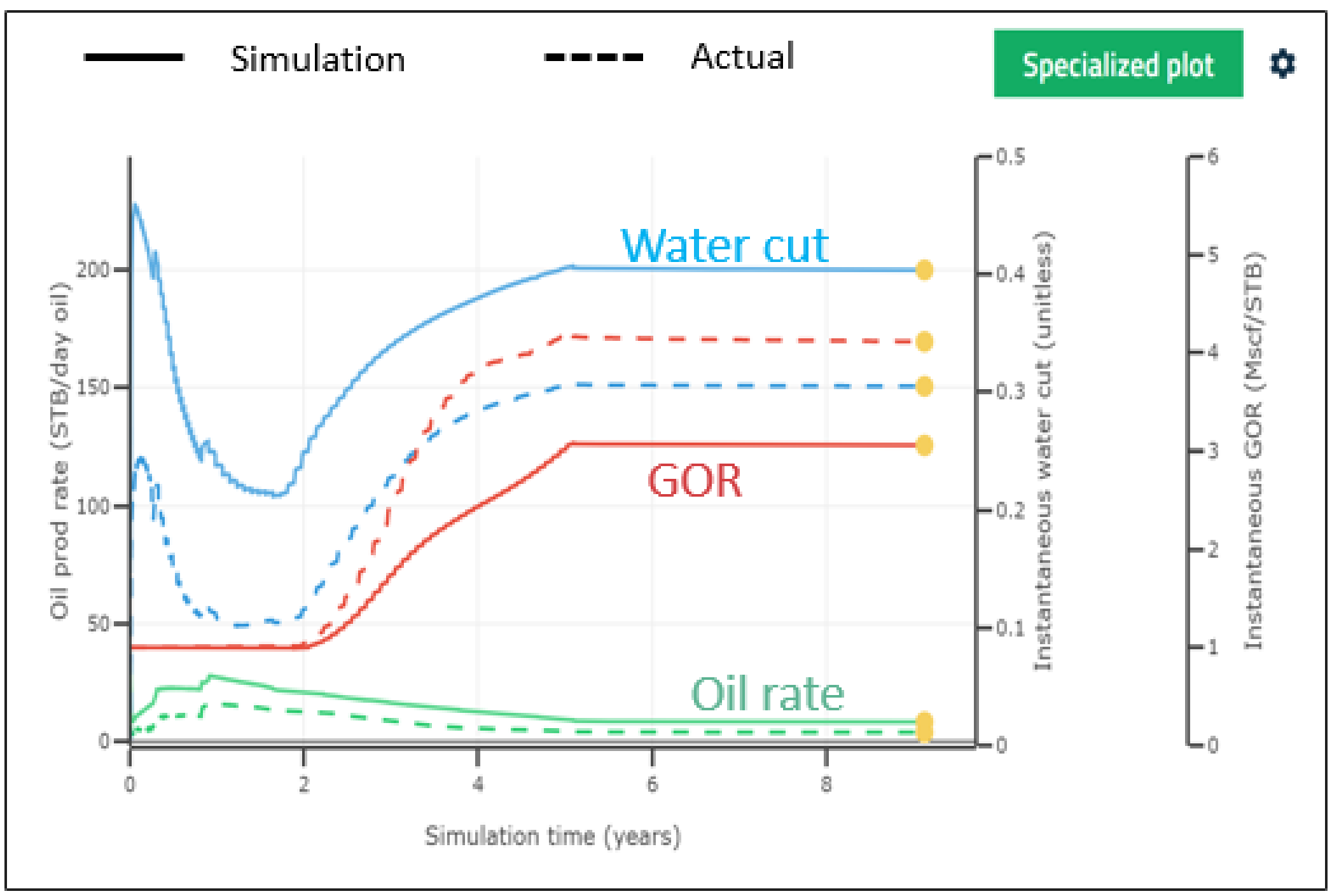

The first step to matching this production data is to roughly dial in the water cut. The simulated water cut levels off at about 0.2-0.22 before increasing rapidly. Note that the rapid increase in water cut corresponds to GOR rising and thus is associated with the reservoir dropping below bubble point.

In figure 8.9 we plot the relative permeability curves used for this simulation.

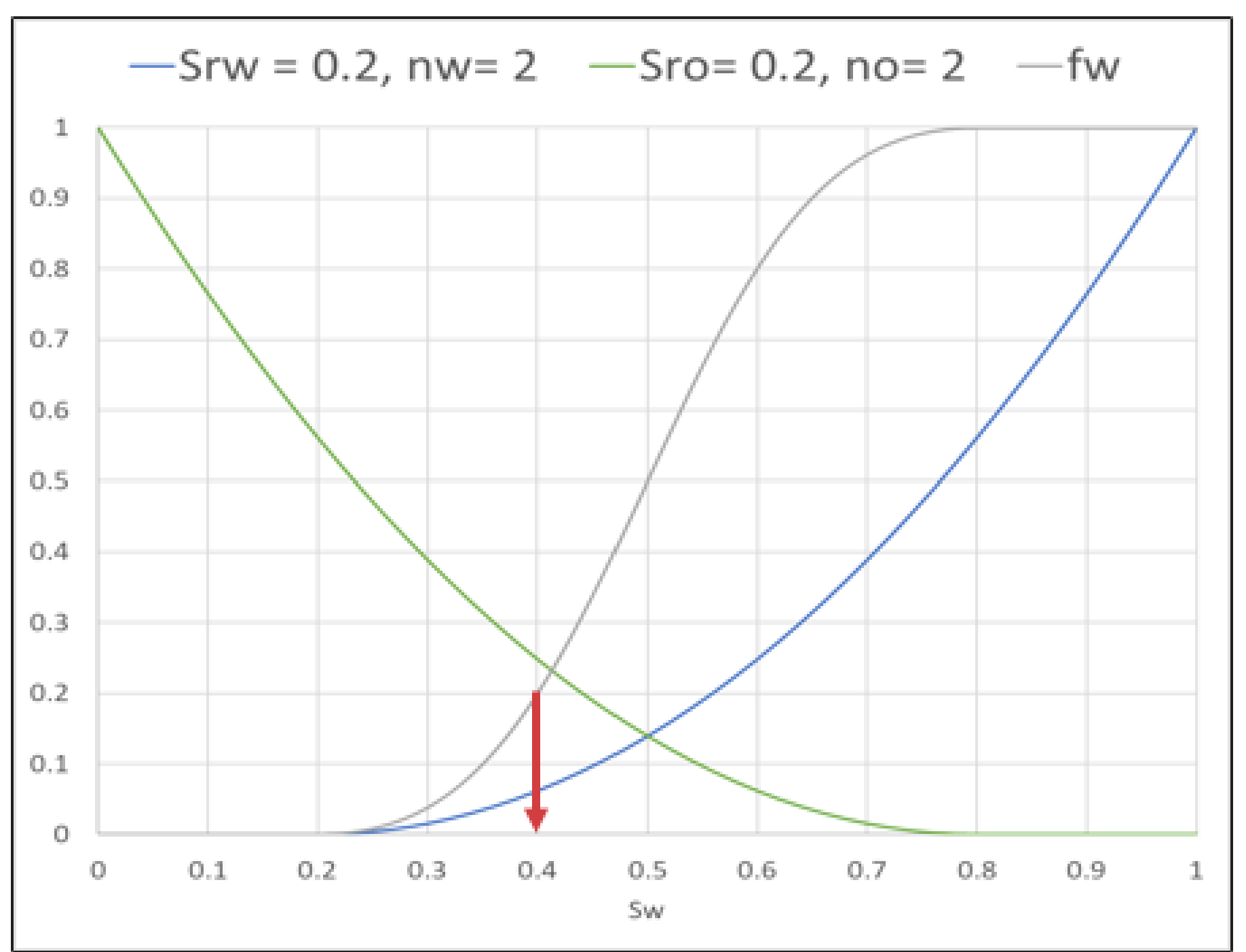

The fractional flow of water (fw) is equivalent to the water cut, so we can see that at observed stabilized water cut of ~0.2, the effective water saturation of the system is 0.4. Looking back to the historical data, we observe that our desired water cut at this point in time is 0.1. We adjust our relative permeability curves such that the fractional flow is equal to 0.1 at a water saturation of 0.4 (prior relative permeability curves are shown in dashed trends).

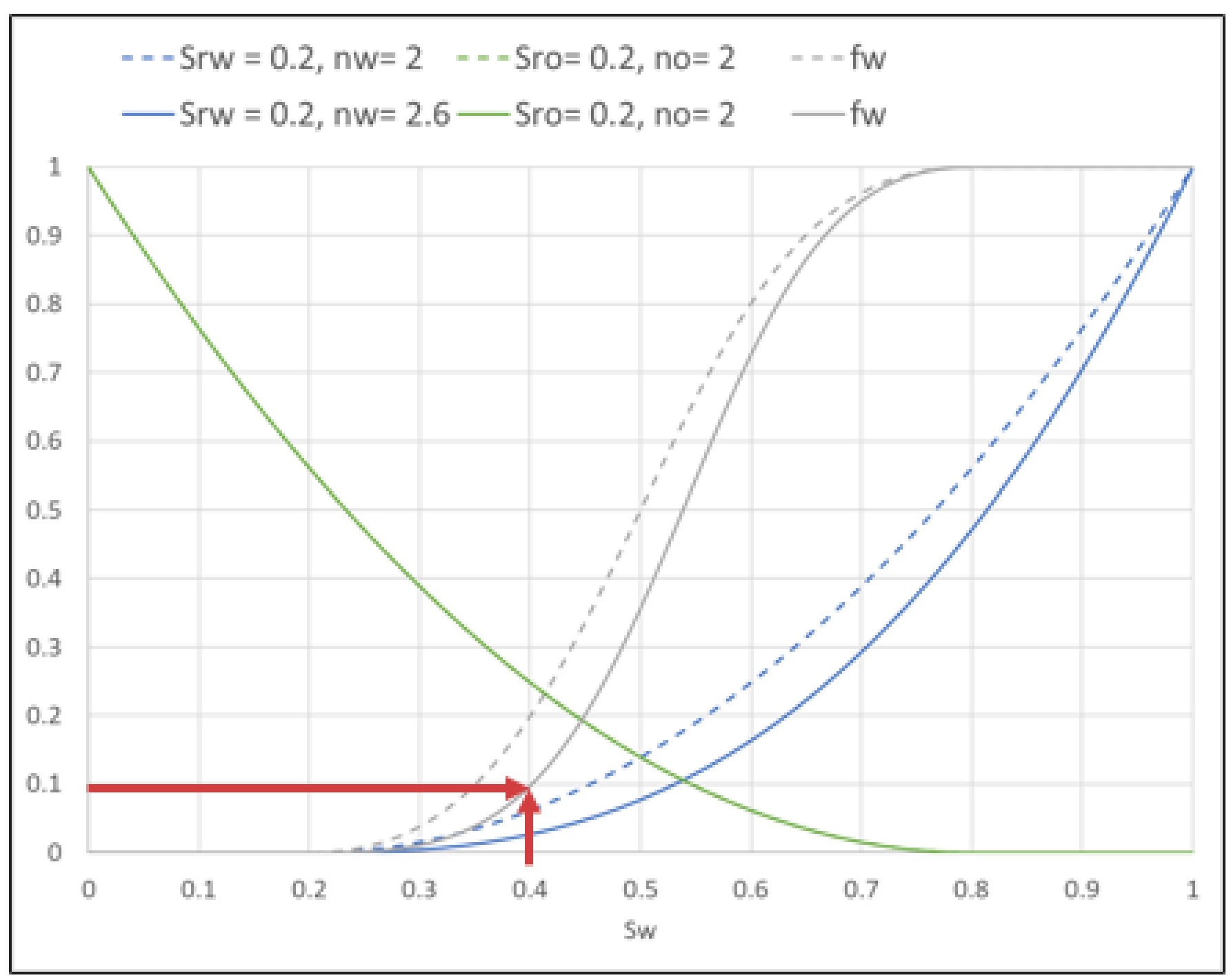

The resulting simulation result in shown in the figure below.

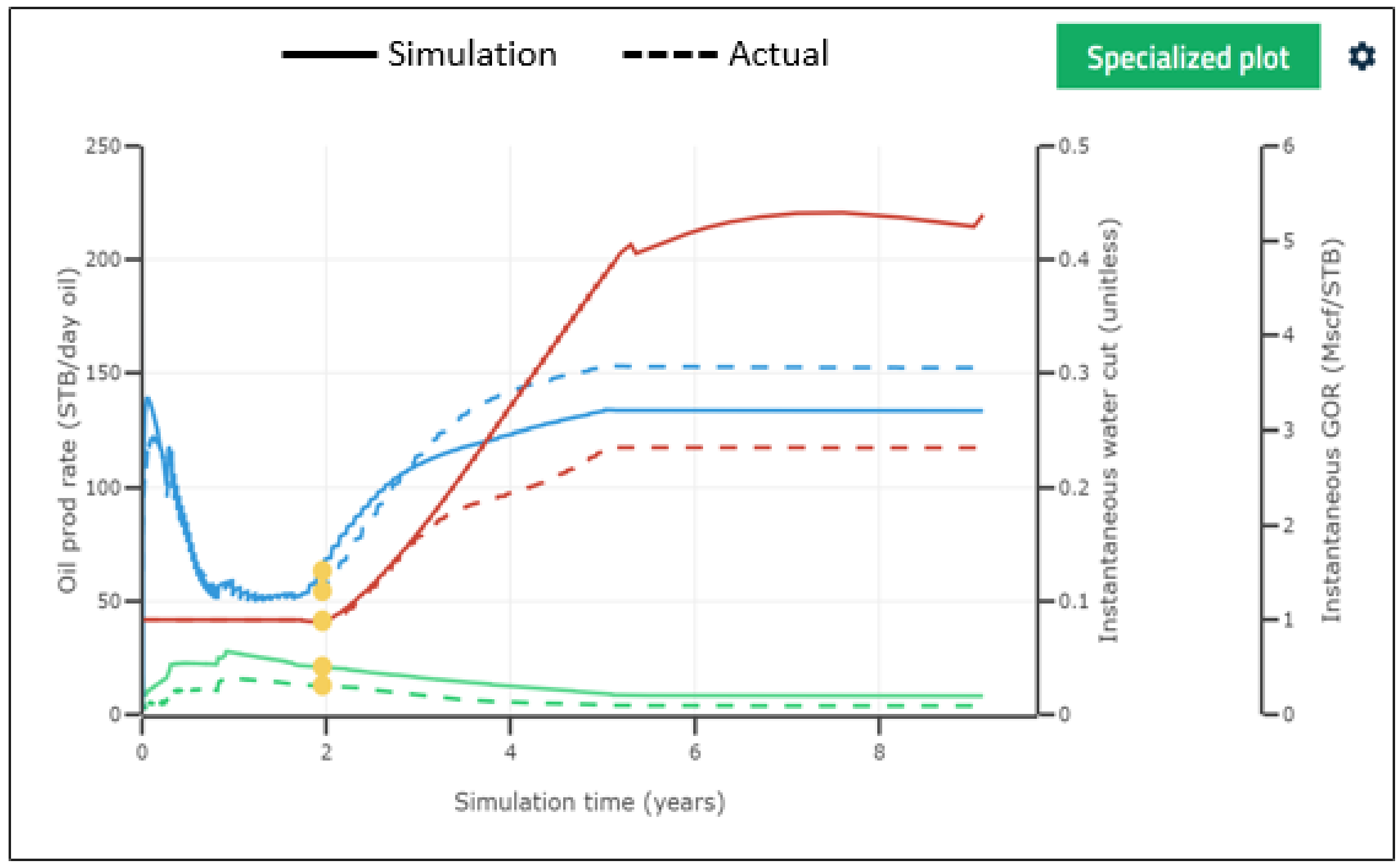

We see that in the resulting simulation, the simulated and actual water cuts match for early time (before the well is drawn down below bubble point).

Proceeding to step eight of our history matching workflow, we can use an RTA plot to calibrate permeability. Figure 8.12 shows the RTA plot for the most recent simulation case and historical data.

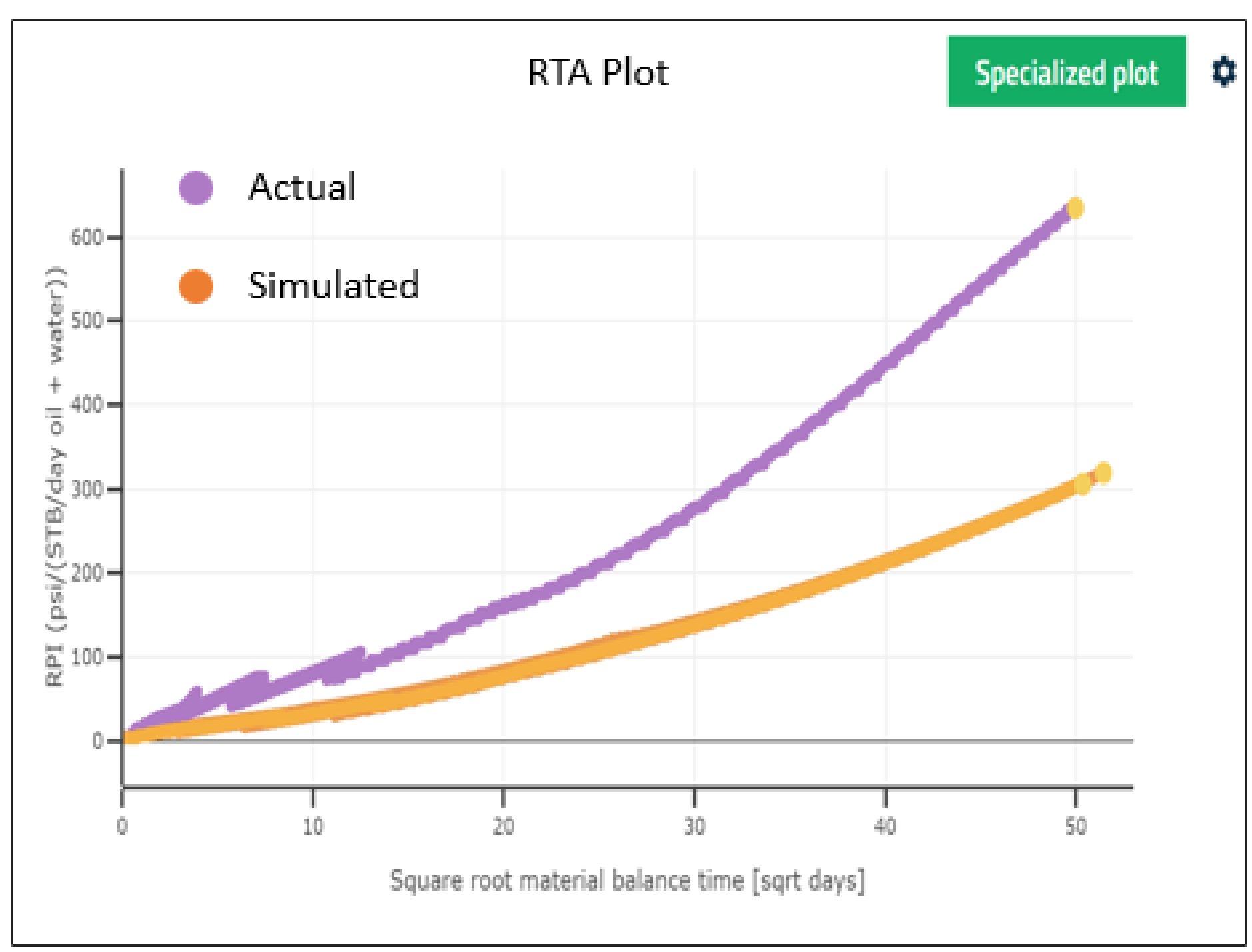

We can see that the actual data exhibits a steeper initial slope and bends upward more aggressively than the simulated data. The slope of an RTA trend is inversely proportional to the product of fracture area and the square root of permeability. Figure 8.13 shows how assuming the fracture area is matched between the actual and simulated data, we can calculate the permeability change required in our model.

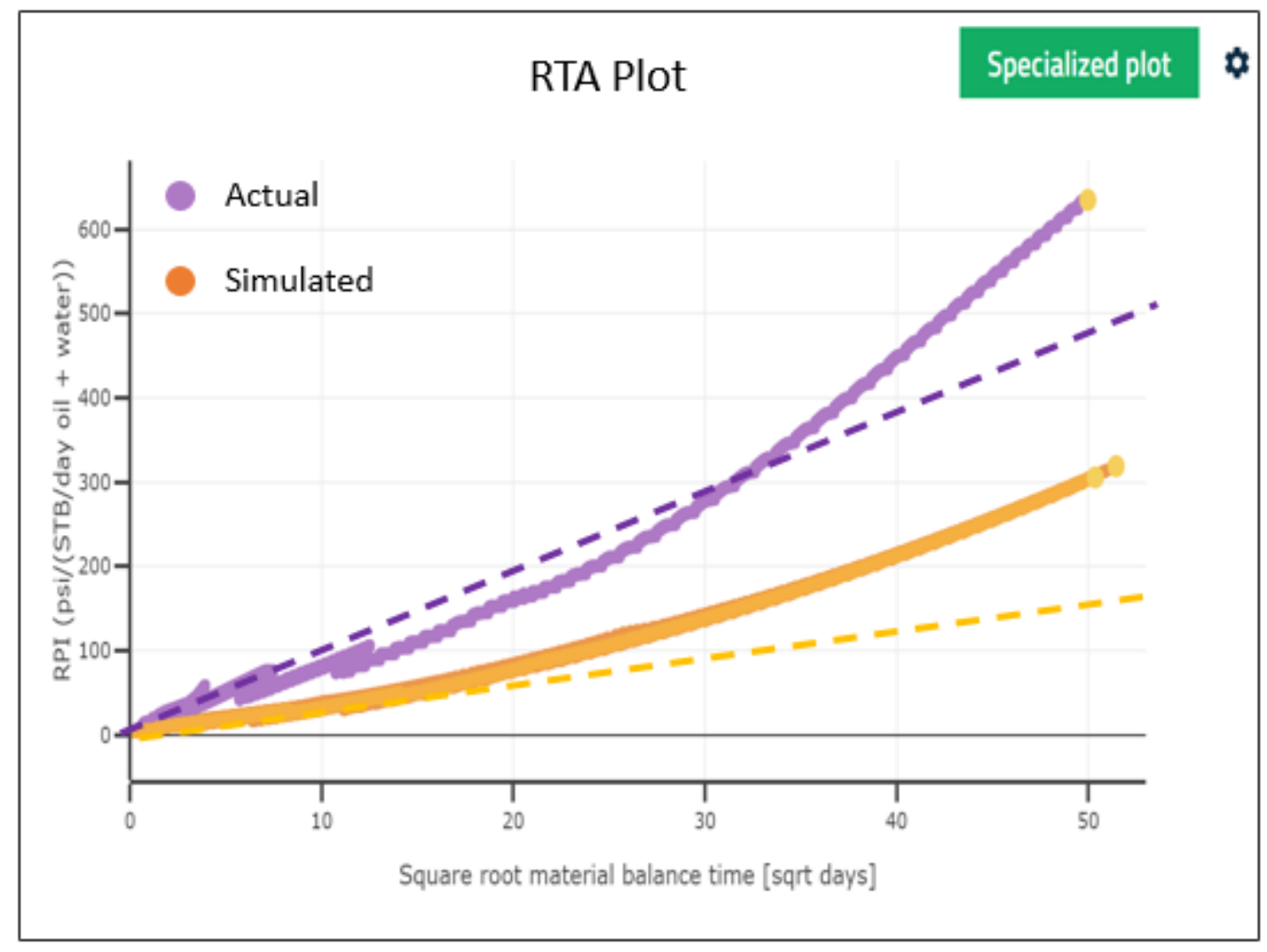

- **RTA slope** of simulation = 160/50 = 3.2
- **RTA slope** of actual data = 340/50 = 9.8
- Ratio of the slopes is **9.8** / **3.2** = 3.1
- Permeability is proportion to the square of the slope, 3.1^2 = ~10

After reducing our model permeability by 10x (as calculated above), we see that the updated simulation oil rate now matches in early time along with the water cut in Figure 8.14.

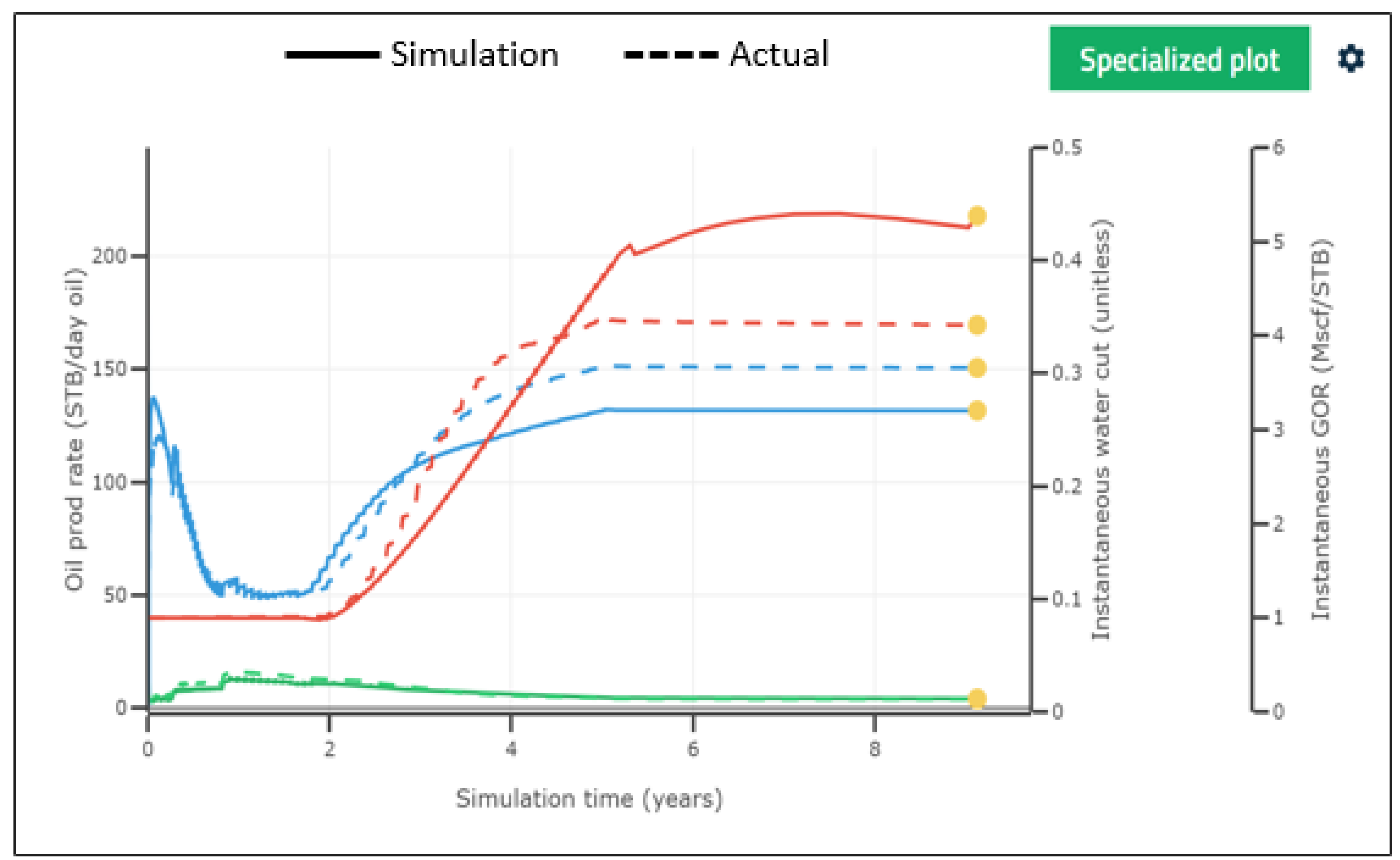

At this point we can proceed to step nine of our history match process and use the RTA plot to calibrate model parameters to match late time production behavior. Figure 8.15 shows the RTA of the updated simulation and actual data.

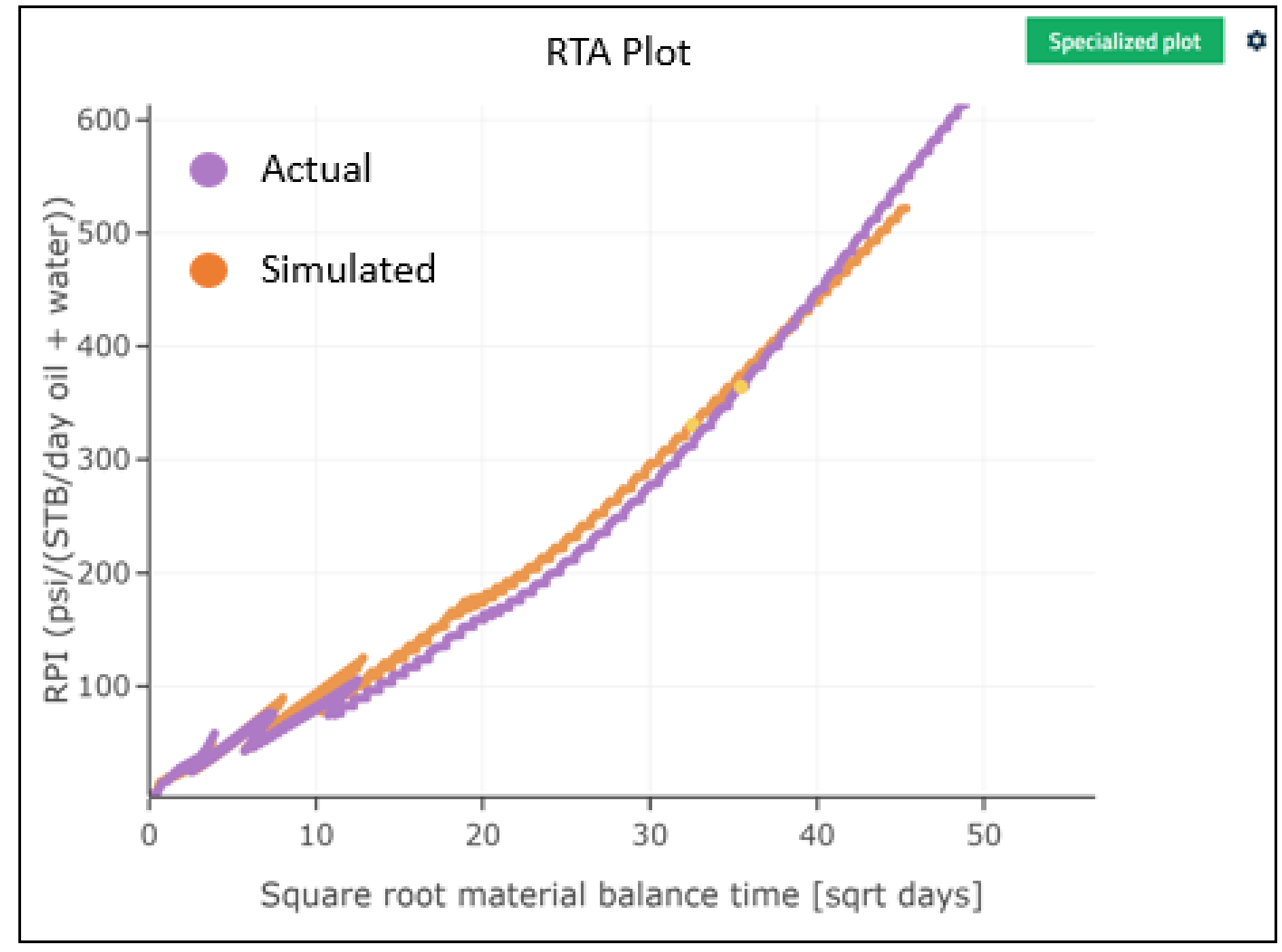

As calibrated in the previous step, we see that the initial slope of the actual and simulated trends are equal; however, in later time the two trends deviate. The actual RTA trend still bends upward more aggressively that the simulated data. Fowler et al. (2020a) discuss various phenomena that can explain the upward curvature of RTA trends. In this case, we observe that the mismatch is occurring after GOR starts to rise which suggests that the curvature is associated with relative permeability changes below bubble point.

Because the RTA trend slope is proportional to the inverse of the square root of permeability and because the actual RTA trend is *steeper* than the simulated in late time, we know we must make the relative permeability of oil steeper, such that relative permeability falls more rapidly when gas comes out of solution. Figure 8.16 shows how we update the water and oil relative permeability curves to maintain the same 0.1 fractional flow of water at 0.4 water saturation while also making the oil curve steeper.

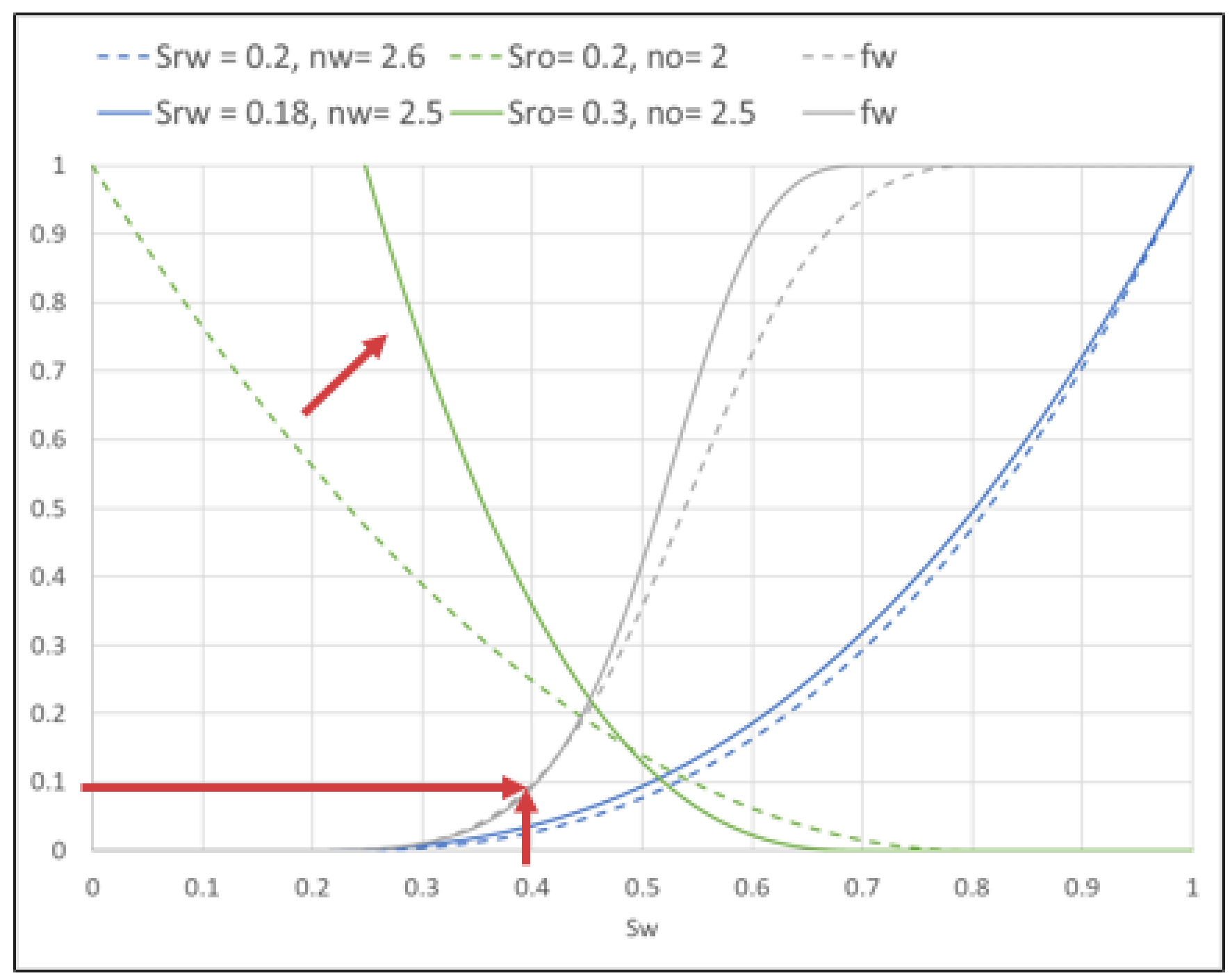

Figure 8.16 – Making relative permeability adjustments to make oil permeability more sensitive to saturation changes.

Figure 8.17 below shows the resulting RTA and production curves for the simulation.

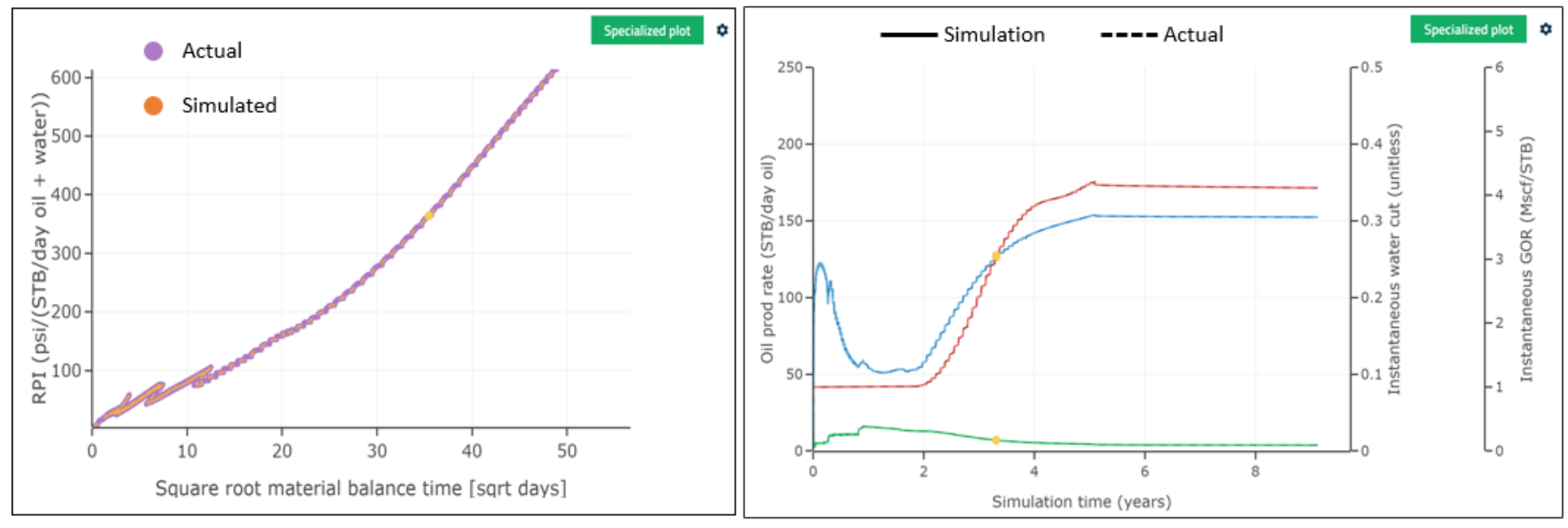

By drawing out the relative permeability curves and making RTA plots we are able to expeditiously history match the production data. We use the plots to make informed, quantitative adjustments to history matching parameters thereby reducing the number of iterations.

## 8.5 History matching cheat sheet

This cheat sheet lists field observations that you typically use for model calibration and lists the parameters that we recommend that you vary to achieve a match to these parameters. The list is not exhaustive. However, if you find yourself varying parameters that are *not* on this cheat sheet, you may consider checking with us for advice, to confirm that you are on the right track.

The order in which you perform the calibration is important. Typically in shale applications, you want to calibrate to fracture geometry parameters first, and then calibrate to production data. That way, you don't find that later changes mess up your matches from earlier stages in the calibration.

The parameters are listed in the rough order in which they should be calibrated. These parameters are discussed in more detail in the subsections below the table.

### 8.5.1 Fracture length

Relative fracture toughness – larger values make the fractures shorter. We typically use values of 0 – 0.15 ft^(-1/2).

Effective fracture aperture conductivity factor – smaller values make the fractures shorter and, especially, more symmetrical. The default value is 1.0. Values may be reduced to be as low as 0.3.

To keep an eye on: If the value of permeability times 'pressure dependent permeability' is greater than 0.1 md, it can start to significantly limit fracture size.

### 8.5.2 Fracture height

Stress layering – fracture height growth is inhibited by high stress layers. To reduce height growth, you can increase stress in zones above and below.

Toughness anisotropy – if stress logs do not give any indication of stress laying, but you need to reduce height growth, increase the vertical fracture toughness to be higher than the horizontal fracture toughness. The default is for vertical fracture toughness to be approximately equal to horizontal fracture toughness. To limit height growth, we often increase vertical toughness by 20-50%. In a handful of cases, we have made vertical toughness 2-3x higher than horizontal toughness.

Less often, you may choose to decrease the 'vertical open and roughness fracture conductivity multiplier.' This parameter reduces the fracture conductivity during propagation in the vertical direction. Values such as 0.1-0.3 can significantly reduce fracture height.

### 8.5.3 Perforation erosion

Perforation erosion alpha – higher values result in more perforation erosion.

If you select 'BuiltIn' or 'Custom' for the 'Perforation phasing option,' this turns on the special proppant transport from the well capabilities that are implemented in StageOpt. To utilize these parameters, we recommend starting with StageOpt and then bringing the parameters over to ResFrac. You can follow the 'StageOpt' history matching cheat sheet, which is available at www.resfrac.com/onlinetraining.

### 8.5.4 Perforation efficiency

In this context, for 'perforation efficiency,' we are referring to the number of perforation clusters that break down.

Perforation efficiency is affected by fracture net pressure/toughness. Therefore, it should be calibrated *after* calibrating fracture length.

Perforation efficiency is strongly affected by cluster spacing and perforation design, but those are usually taken as 'given,' and cannot change them during calibration.

NW complexity coefficient (in well vertices table) – higher values tend to increase perforation efficiency (as well as increasing injection pressure overall).

Tensile strength (in the geologic units table), in conjunction with setting 'Randomized tensile strength' to true – higher values tend to decrease perforation efficiency. This setting is used to account for the nonuniform breakdown pressure in each cluster along a well. There is empirical evidence to support this approach. Figure 2 from SPE-204185-MS suggests that breakdown pressure can vary from 300 psi to 2300 psi within the same stage.

### 8.5.5 ISIP

Shmin or fracture gradient (in the geological units table) – higher values tend to increase ISIP. However, these values should typically be constrained by DFITs, and should not be set to values significantly different from the DFIT results.

NW complexity coefficient (in well vertices table) – higher values tend to increase ISIP and WHP during fracturing.

Pressure dependent NW dP – these four parameters, at the bottom of the advanced section in well vertices, can be used as an *alternative* to the NW complexity coefficient. The NW complexity coefficient technique tends to dissipate NW tortuosity too quickly after shut-in. The pressure dependent NW dP parameters give more flexibility to extend the duration of NW pressure drop after shut-in. However, they are more complicated to use. Refer to the built-in help content for more information. This model does tend to provide somewhat more realistic models of post-closure transients. However, it is more complicated to use than the standard method using the NW complexity coefficient. Therefore, in most cases, we advise against using them, unless you have a special application where a close match to the post shut-in pressure transient is needed.

Relative fracture toughness affects ISIP, but should be used to calibrate fracture length, not for the ISIP calibration.

Generally, we recommend not being overly concerned with matching the details of ISIP and the pressure behavior in the few minutes after shut-in. They are affected by complex (and relatively unimportant) near-wellbore effects that are difficult to characterize. Therefore, you may not gain a significant better model by working hard to improve the match on ISIP.

### 8.5.6 Treating pressure

Treating pressure should be calibrated after you are satisfied with the ISIP. Refer to <https://www.resfrac.com/blog/calibration-wellbore-pressure-during-and-after-fracturing> for a discussion of this topic.

Wellbore friction adjustment factor – lower values reduce wellbore friction and decrease WHP. Friction adjustments are necessary because friction reducer chemicals in the wellbore cause wellbore friction to be much lower (ballpark 85%) than would be predicted from standard pipeline friction correlations for Newtonian fluids.

Friction adjustment factor (in the water solutes table) – has the same effect as ‘wellbore friction adjustment factor,’ but can be specified separately for each type of water solute.

We usually recommend leaving ‘wellbore friction adjustment factor’ at 1.0, and instead using the ‘per water solute’ friction adjustment factors.

Viscosity multiplier per 0.001 mass fraction (in the well) (in the water solutes table) – specifies an effective viscosity parameter for water solutes specifically for when they are in the wellbore. Useful for modeling cross-linked fluids that do not crosslink and increase viscosity until they have exited the wellbore.

Wellbore proppant friction adjustment factor – lower values reduce wellbore friction and decrease WHP, specifically related to when you increase proppant concentration. For example, if you increase proppant concentration, hydrostatic changes tend to make WHP go down, but friction changes may tend to make WHP go up. This parameter allows you to tune the degree to which proppant affects wellbore friction. This parameter is varied rarely.

Generally, we recommend achieving only a general match to WHP during injection. WHP can show a variety of upward or downward trends during injection, and they can have so many different causes, that it is risky to overinterpret them. Trying to match them can lead to a model overfit.

More recently, we have released an option to specify wellbore friction tables, where frictional gradient is specified directly as a function of velocity, pipe diameter, and fluid concentration. This may be used as an alternative strategy for matching pipe friction.

### 8.5.7 Propped height

Vertical proppant flow holdup factor – Increasing this value reduces propped height. The default is zero. To significantly decrease propped height, you may find it necessary to increase this parameter to 0.9 or even higher.

### 8.5.8 Propped length

Maximum immobilized proppant mass per area – larger values result in a smaller propped fracture surface area. A reasonable default value is 0.15 lbs/ft^2. To match field data, we have gone as low as 0.05 lbs/ft^2, and as high as 0.5 lbs/ft^2.

### 8.5.9 Production volumes, RTA trends, and GOR

#### *8.5.9.1 Overall RTA trends*

For history matching, we strongly recommend that you specify BHP controls and match rate, rather than specifying rate and matching pressure. Once you have finished the match, you will get the same answer either way. But as you go through the process, you will find that specifying BHP allows the process to go much more smoothly. This happens because many simulation behaviors are affected by BHP. If you specify rate, then errors in one parameter can cause error in BHP, which causes other problems to arise that are unrelated to the original parameter error. For example, excessively high permeability could cause BHP to be too high, which will cause GOR to not behave properly. Conversely, if you specify BHP, then error in permeability will not affect the GOR trend. Thus, specifying BHP, not rate, avoids nonlinear relationships that can confuse the process. We also recommend against changing well controls excessively frequently. This will slow down simulator runtime and will not really assist with performing the match. Instead, we recommend you specify a simple, smooth BHP control (assuming this is reasonably consistent with reality).

To assist with the production history match, it is often helpful to first construct an RTA plot of reciprocal productivity index (rate normalized pressure) versus sqrt material balance time. The slope of the line is inversely proportional to propped area times the square root of permeability. If the fracture is significantly finite conductivity, then the plot will have a y-intercept.

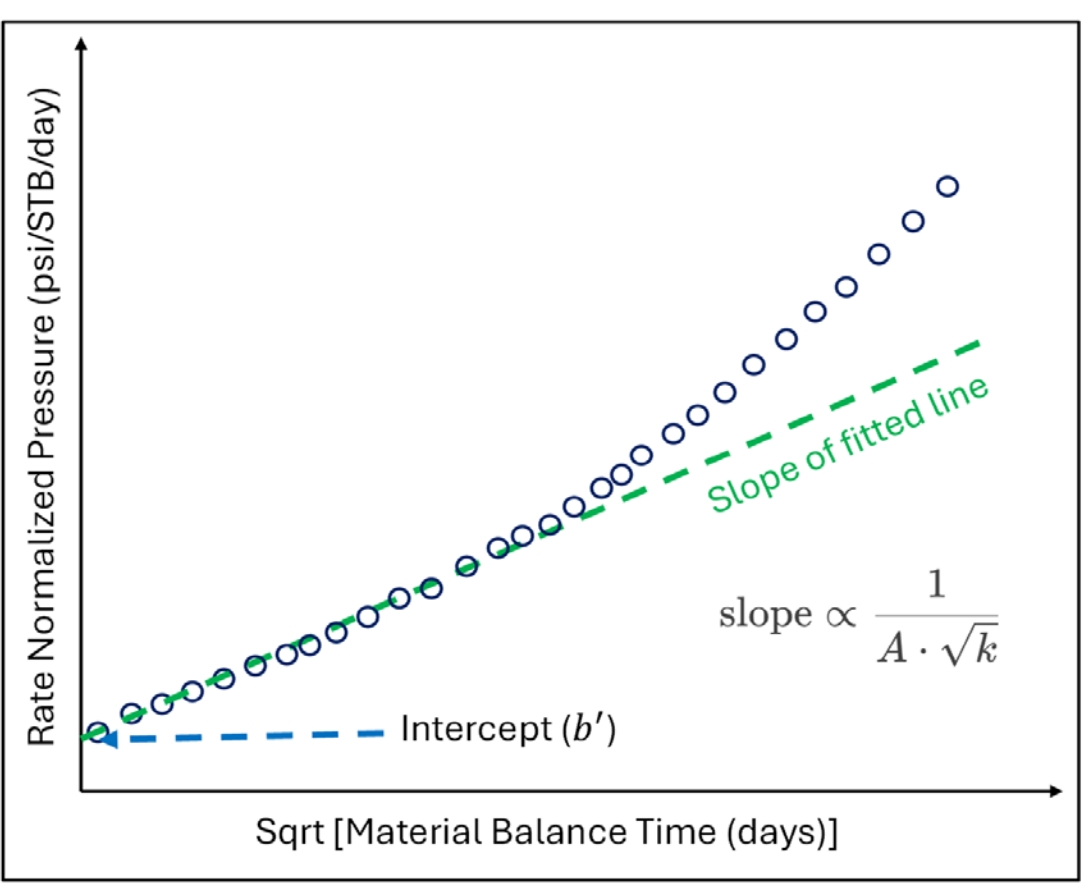

*RTA plot to interpret permeability, propped fracture area, fracture conductivity.*

You can use ResFrac's built-in specialized plot tool for assistance.

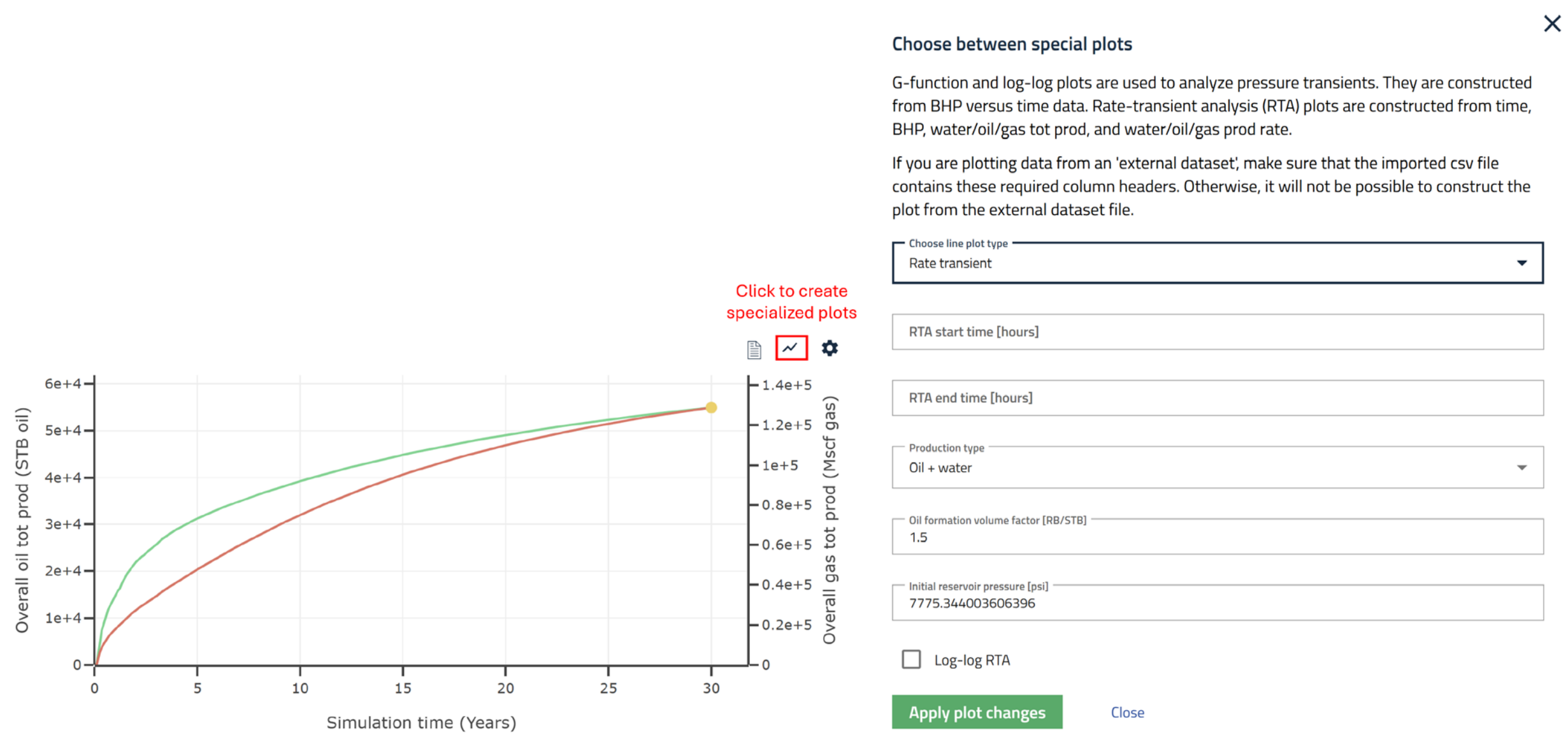

*Use ResFrac's specialized plots functionality to perform RTA analysis*

A summary of RTA history matching basics is given in the blog post: <https://www.resfrac.com/blog/using-rta-aid-resfrac-history-matching>.

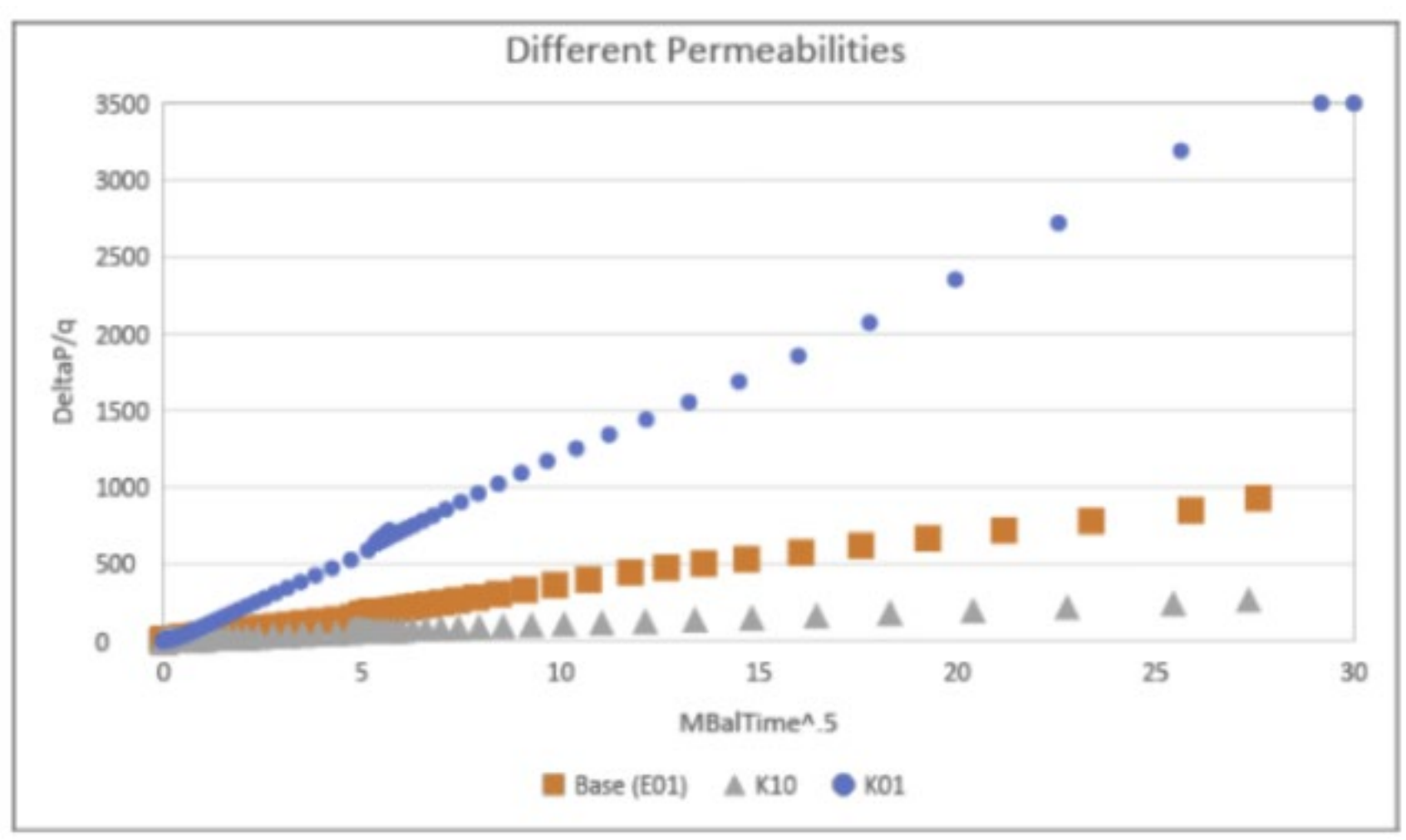

Global permeability multiplier – higher values increase production and reduce the slope of the RTA plot. Your initial guess value for permeability can be taken from DFIT analysis.

Larger propped area increases production and reduces the slope of the RTA plot (see section above on propped area). It is best if you have independent constraint on propped length from interference tests. If you do not, you may consider modifying propped area as part of a match.

k0 (in the proppants table) – Usually we do *not* recommend changing proppant pack conductivity a part of the history matching process. As discussed by Singh et al. (2025) in SPE 223567, published field observations give good guidance on proppant pack conductivity. We recommend using parameter values such that proppant pack conductivity starts in the 10s of md-ft at the start of production, down to single digit md-ft during long-term production, and perhaps even going down below 1 md-ft over time. From the Kozeny-Carmen equation, k0 should equal 1/180 = 0.0055. We often use a somewhat lower value as a starting point, around 0.002-0.006. We set proppant gamma to .00025 psi^-1.

### *8.5.9.2 GOR trends with time*

Typically, for oil shales, we use gas rel perm curves with residual saturation around 0.01, and Brooks-Corey exponent around 1.3. These aggressive curves are needed to capture the rapid increase in GOR that is seen when BHP goes below saturation pressure. For more information on setting up the oil rel perms, check out the worked example in Section 8.4 of the A to Z Guide.

Generally, to match GOR and GOR trend over time, we recommend modifying residual gas saturation and exponent. However, note that GOR can only increase if the fluid pressure is less than the saturation pressure. If the BHP is above the saturation pressure – or if the proppant pack conductivity is so low that the pressure in the fracture is much higher than the BHP – GOR rise will be suppressed.

If you are using the 1D submesh, which is typical, there are two numerical settings to be aware of. First – "multiphase 1D submesh version." For this parameter, we recommend using whatever is the most recent recommended version (which is version 8 as of April 2026).

There is also a parameter called "Submesh multiphase flow numerical option" (which has a similar name, and be careful to not be confused by this). This parameter defaults to '0'. If set to '1', a change is made to the calculation that causes it to be considerably more aggressive in predicting a GOR increase. To match datasets that have very large GOR increase, this parameter could be set to 1.

The GOR trend has an impact on how you choose to match the RTA curvature, as described in the next section.

#### *8.5.9.3 RTA curvature*

At some point, the RTA curve will begin to bend upwards. For single phase flow and constant permeability, this occurs because of interference between adjacent fractures. The radii of investigation of adjacent fractures begin to overlap and so they produce less than they would if they were not in proximity.

It is widely recognized that fracture-to-fracture interference *alone* is insufficient to explain upward bending RTA curves (which correspond to anomalous loss of productivity, relative to basic single phase linear flow).

To enhance the upward bend in the RTA curve, you may use: (a) a fracture swarm fractal scaling parameter, (b) relatively permeability loss, (c) pressure dependent permeability in the formation, (d) pressure/stress dependent proppant pack conductivity loss, or (e) time-dependent conductivity loss.

Usually, it’s best to avoid having one single process dominate – this can lead to surprising or unrealistic results. For example, if you make pressure dependent permeability loss extremely strong, you may find that lowering BHP *reduces* production. To avoid such artifacts, it’s generally best to have multiple processes operating at the same time.

The GOR trend can help you diagnose which of these properties. The figure below is reproduced from Lamidi et al. (2026). It shows how different physical processes are associated with different GOR responses. In data with muted GOR increase, you should focus on pressure dependent permeability (PDP) reduction with drawdown and time-dependent conductivity loss (TDC). If GOR goes up significantly, relative permeability loss may be an important effect in the RTA plot. Fracture swarms cause a very early and sharp GOR trend, and in our experience, this process has proven to be less-often used. In formations with intermediate GOR increase, relative permeability, PDP, and TDC might all be present. We use the ‘fractal fracture swarm’ capability less frequently.

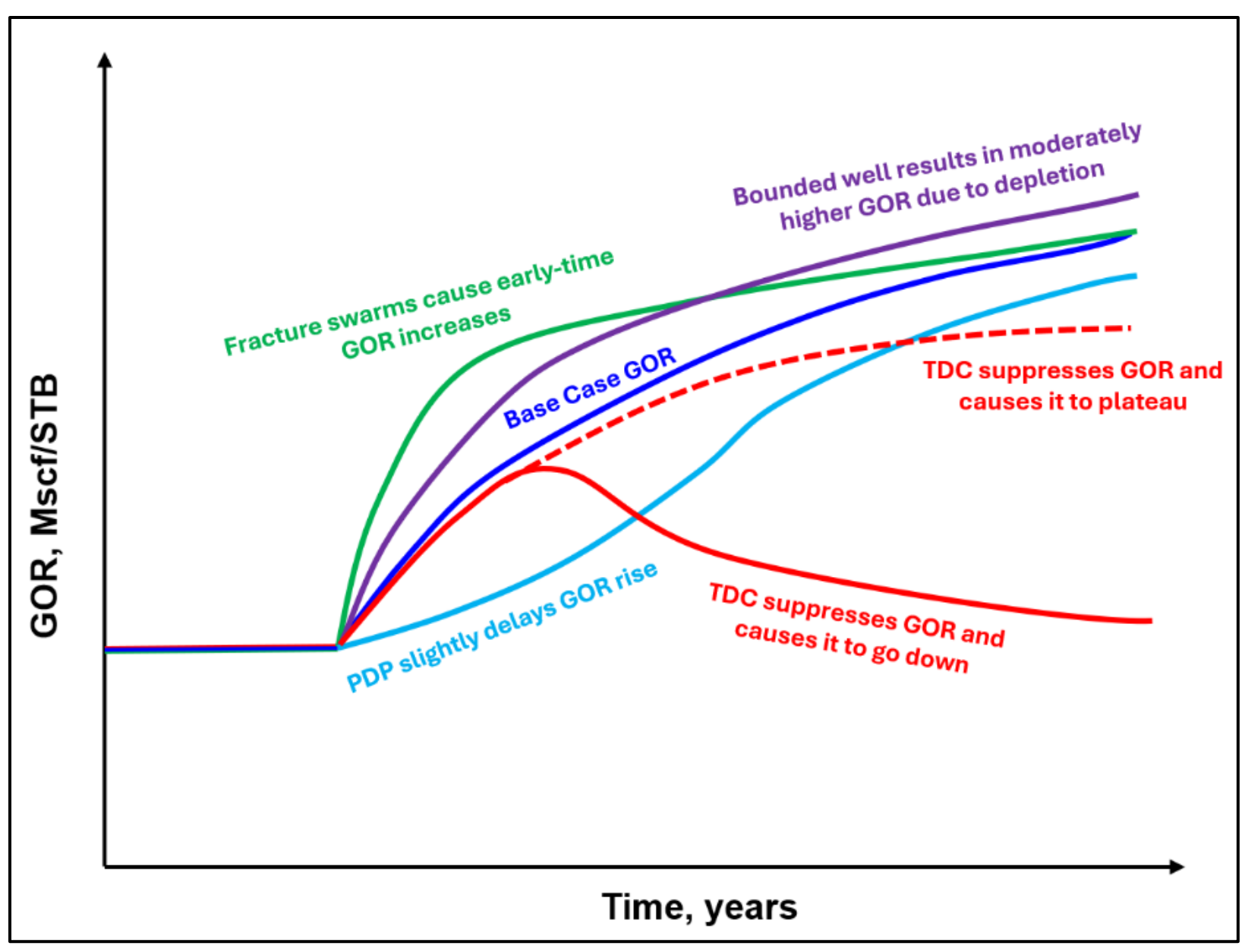

*Figure reproduced from Lamidi et al. (2026)*

### *8.5.9.4 Multiphase flow effects*

Very often, we observe that RTA curves begin to bend upwards when fluid pressure goes below the saturation pressure. The upward bend seems to coincide with an increase in GOR.

Note that if your simulation does not have a significant increase in GOR, then multiphase flow is unlikely to be one of the primary causes of the bend in the RTA plot.

To match this effect, we set up the oil (and possibly also the gas) rel perm curves so that they drop off relatively quickly as saturation drops. Also, we use aggressive gas rel perm curves. As the fluid in the formation goes below the saturation pressure, gas forms. That gas can flow more easily than the oil, and so is preferentially produced, causing an increase in GOR.

What should you do if you see BHP go below saturation pressure, but GOR does not increase? As noted above, in order for GOR to increase, fluid pressure needs to go below the saturation pressure *in the formation*. So first, open the 3D image and confirm whether the fluid pressure in the fracture is below saturation pressure. If the proppant pack conductivity is not high enough, then the fluid pressure in the fracture may not be as low as the pressure in the well.

It is useful to understand how ResFrac calculates rel perm for production. For each fracture-matrix element connection, the code performs a special flash calculation using the composition of the matrix element and the pressure of the fracture element. Or more precisely, the code does not actually use the pressure of the fracture element, but uses a pressure from the 1D submesh method that is close to the fracture element. Newer versions of ResFrac perform an even more complicated calculation that simulates a 'differential vaporization' in order to provide a better approximation to the 'fine mesh' multiphase flow solution. Nevertheless, it is not perfect. You should not expect an exact correspondence between the multiphase 1D submesh solution and what you would get from a highly refined mesh up against the fractures. We lack a practical method for estimating shale relative permeability, anyway, and so the relative permeability curves are history matching parameters either way.

#### *8.5.9.6 Pressure dependent permeability loss in the matrix*

In some formations, especially those that have strong overpressure (P is within 500-1000 psi of Shmin), it appears that pressure dependent permeability loss is a factor in RTA curvature.

For this parameter, go to the Curve Sets panel, and go to the pressure dependent permeability table. Set up a curve so that the permeability goes below 1.0 for values of dP less than 0. It is best to make the curve continuous and avoid abrupt changes. For example, you could use:

| dP (psi) | Permeability multiplier |
|---|---|
| 0 | 1 |
| -1000 | 0.562341 |
| -2000 | 0.316228 |
| -3000 | 0.177828 |
| -4000 | 0.1 |
| -5000 | 0.056234 |
| -6000 | 0.031623 |
| -7000 | 0.017783 |
| -8000 | 0.01 |
| -9000 | 0.005623 |

Alternatively, it can be easier to click the checkbox for 'Use parameterized PDP curves.' With that, you only need to specify a quantity for '10x irreversible permeability loss per pressure increment.' So for example, if you specify 5000 psi, then permeability will drop 10x for every 5000 psi that pressure drops from its initial value.

In these PDP tables and settings, you have options to select 'reversible' or 'irreversible.' Typically, we use 'irreversible' for permeability loss with drawdown.

#### *8.5.9.7 Time-dependent conductivity loss*

Time-dependent conductivity loss is another very useful way to simulate an upwardly bending RTA curve. This is motivated by: (1) observations from laboratory experiments suggesting time-dependent conductivity, and (2) experience with history matching suggesting that time-dependent conductivity loss may help with matching late-time EUR observations and RTA curvature. Time-dependent conductivity loss could happen due to deposition of organic or inorganic scaling in the proppant pack, fines migration, or time-dependent embedment or crushing. Refer to Section 19.7 of the ResFrac Technical Writeup (May 2022 version or newer) for additional details.

In the table for 'time-dependent proppant conductivity loss,' we recommend setting the 'damage type' to 2. For the 'rate constant,' a good initial value is 0.0005 days^-1. If you make it too high, then you may see GOR drop off rapidly over time, because of the loss of connection from the well to the proppant pack. So, while it is very useful to have *some* time-dependent conductivity loss, don't make it too rapid.

It is usually advised that you set the 'Minimum conductivity multiplier' to a value around 0.01-0.05. This prevents time-dependent conductivity from making the conductivity exceptionally low at very late time, like shown below.

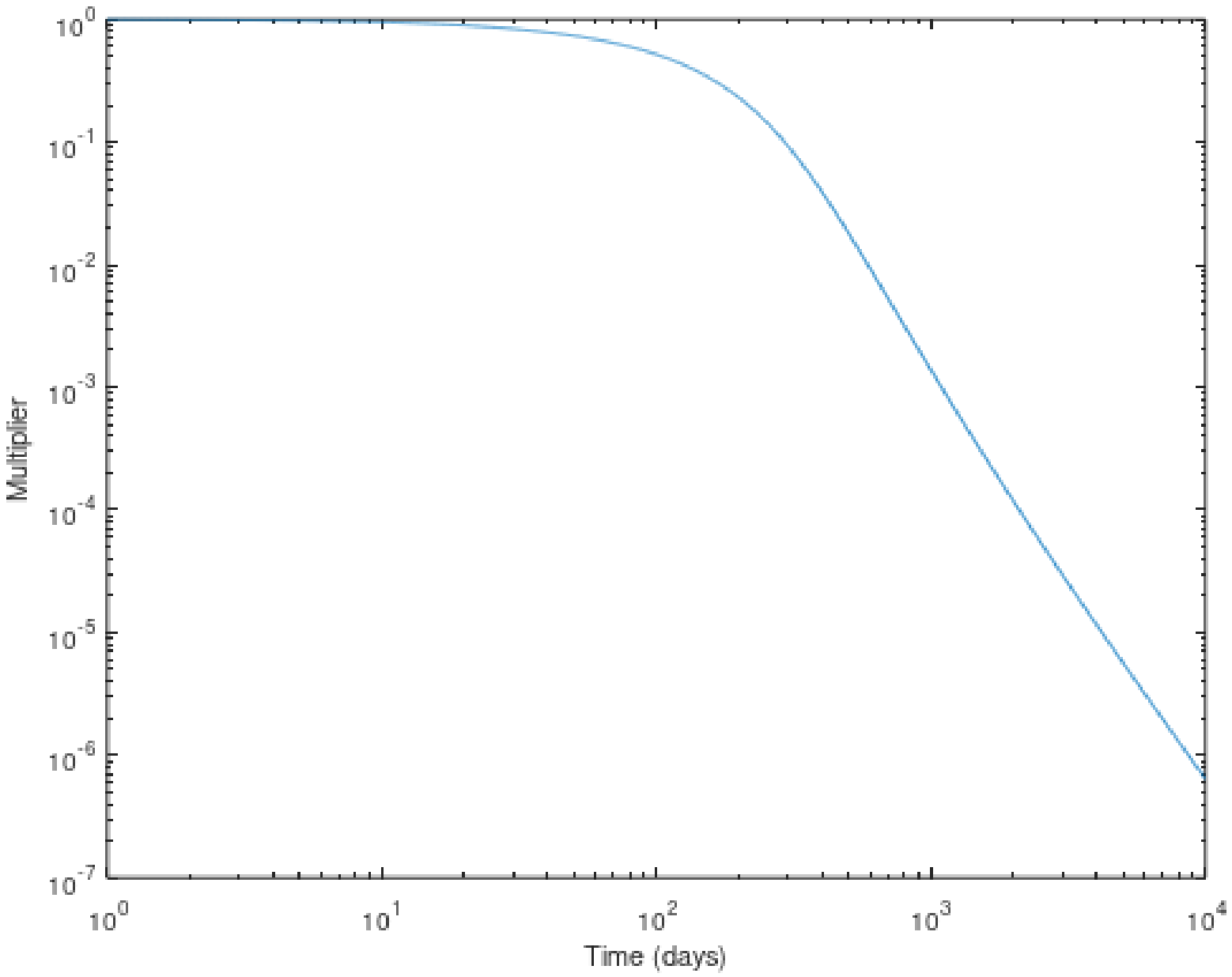

*Conductivity multiplier versus time using time-dependent conductivity loss with Option 2.*

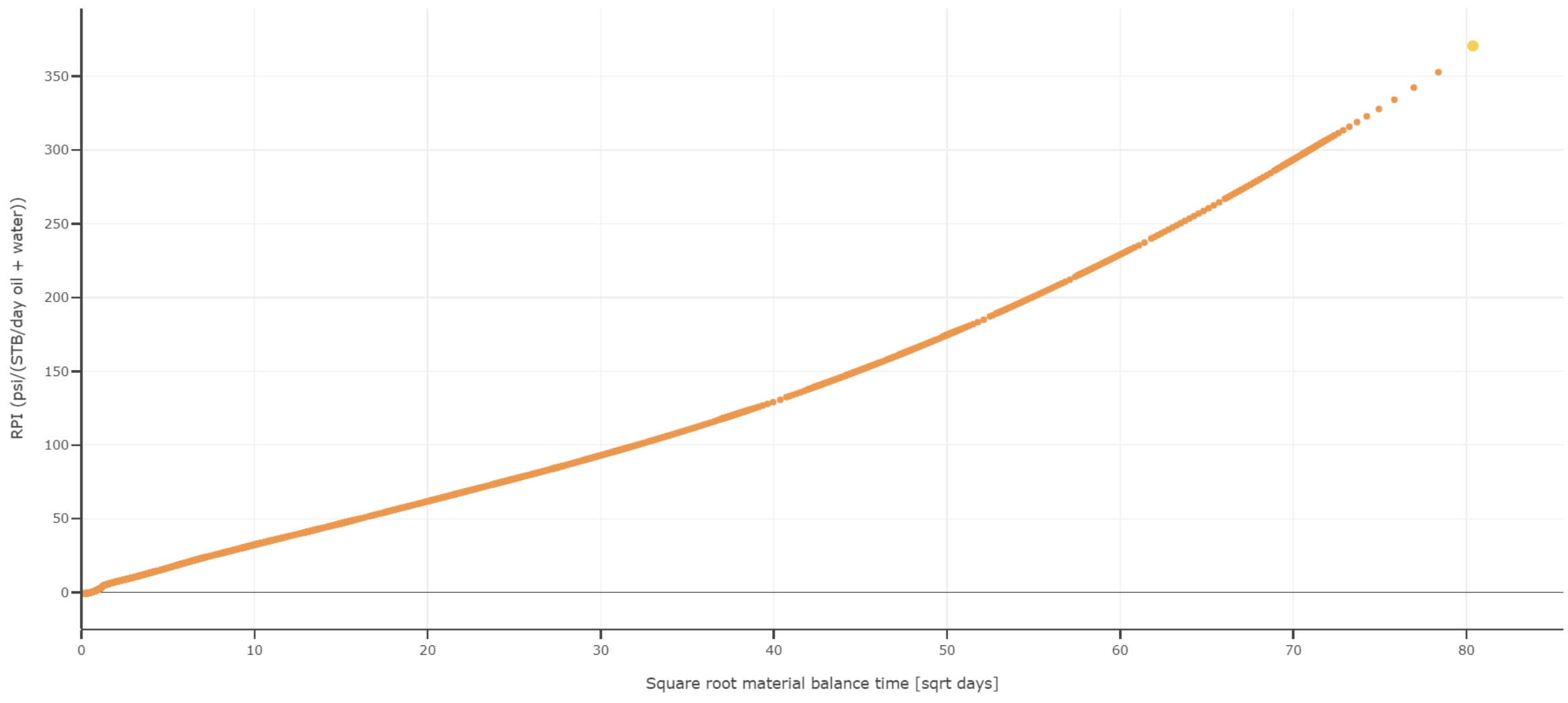

*Rate-transient analysis plot from a simulation using time-dependent conductivity loss Option 2. The upward curvature is caused primarily by the conductivity loss.*

To match long-term GOR trends and cumulative gas, we recommend modifying the residual gas and gas relative permeability exponent.

### 8.5.9.3 Submesh fractalD

This parameter is available, but we find that it is not used in most history matches.

You can use the 'submesh fractalD' parameter to mimic the effect of fracture swarming. This parameter is described in the blog post: <https://www.resfrac.com/blog/simulating-fractal-fracture-swarms-in-a-general-purpose-reservoir-simulator>. Usually, it's safe to set this parameter to around 0.3, in order to get a modest upward trend in the RTA curve. For reference, a default value of 0 has no effect, and a value of 0.6 or greater has a strong effect. If you use this parameter, make sure that the matrix elements are at least 20 ft or so wide in the direction perpendicular to the fractures. To reiterate – you need the mesh to be sufficiently *coarse*; you should not use a highly refined mesh. This is because the submesh fractalD parameter relies on using the 'submesh' created within the elements for fracture-matrix flow. If you increase the fractalD submesh parameter, you may need to apply a 'global permeability multiplier' to decrease permeability overall, since increasing the fractalD submesh parameter on its own increases production.

This parameter tends to cause GOR to spike sharply, early on.

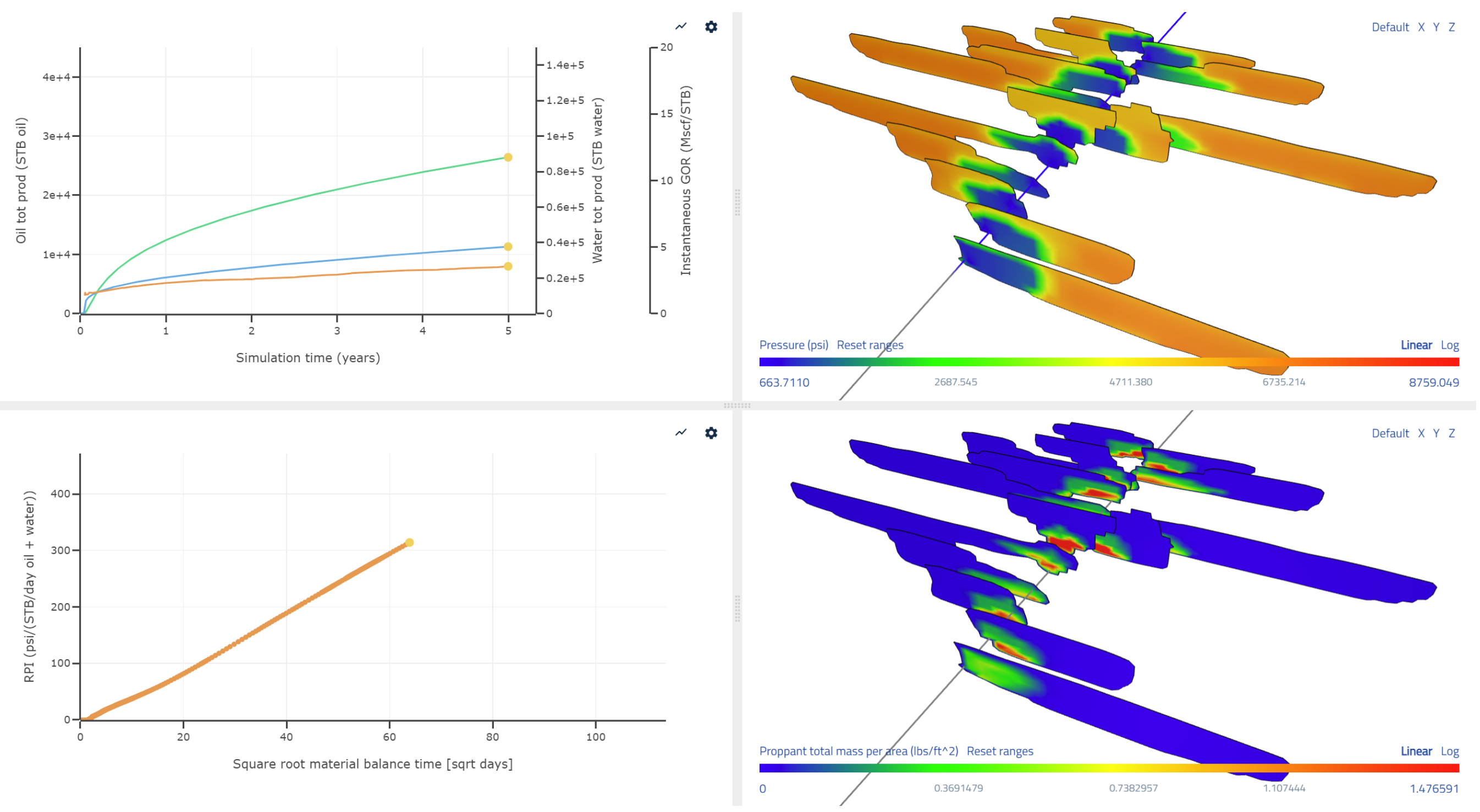

*Example simulation that does not use the 'submesh fractal D' setting.*

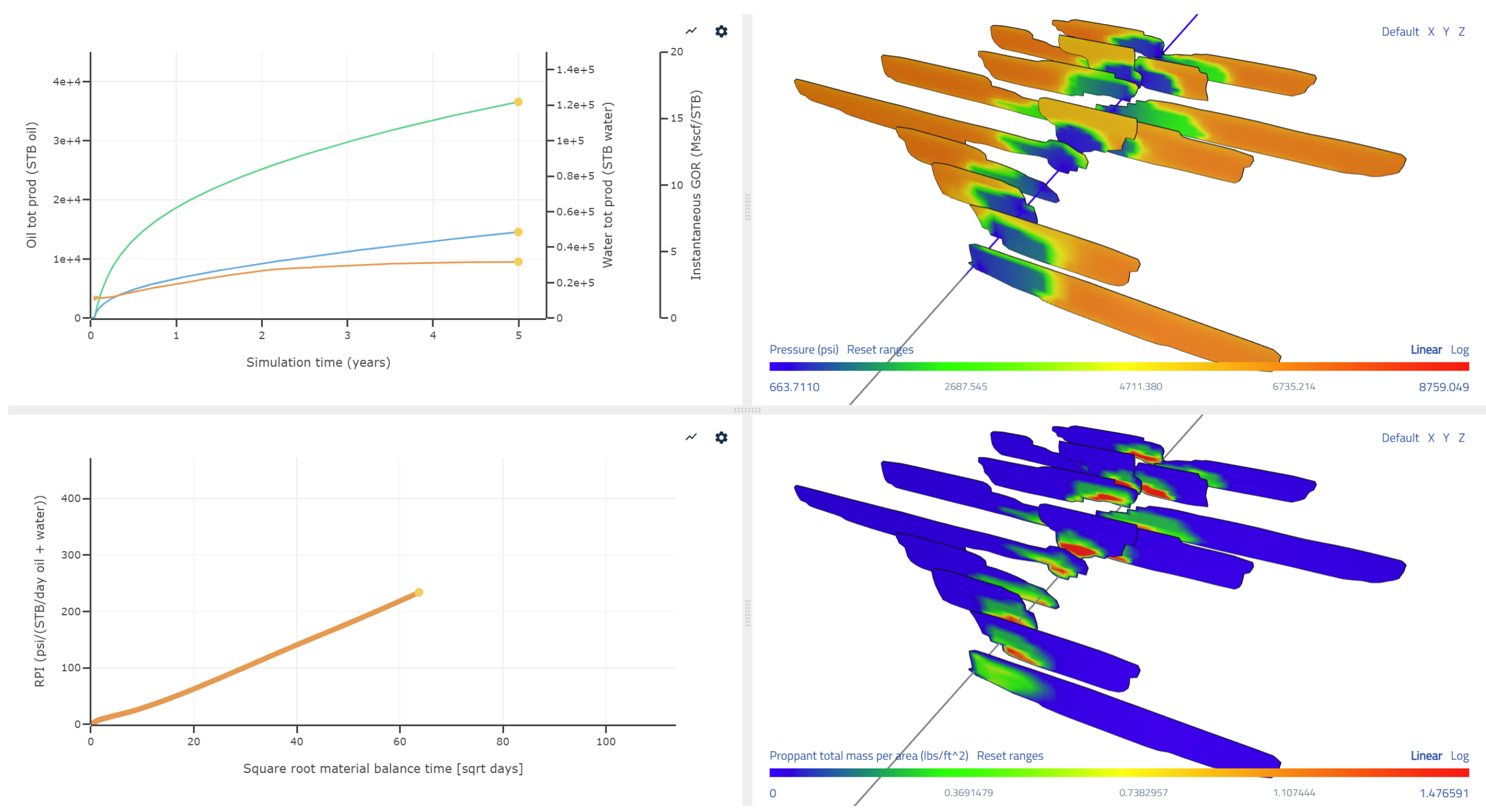

*Example simulation with 'submesh fractal D' set to 0.3.*

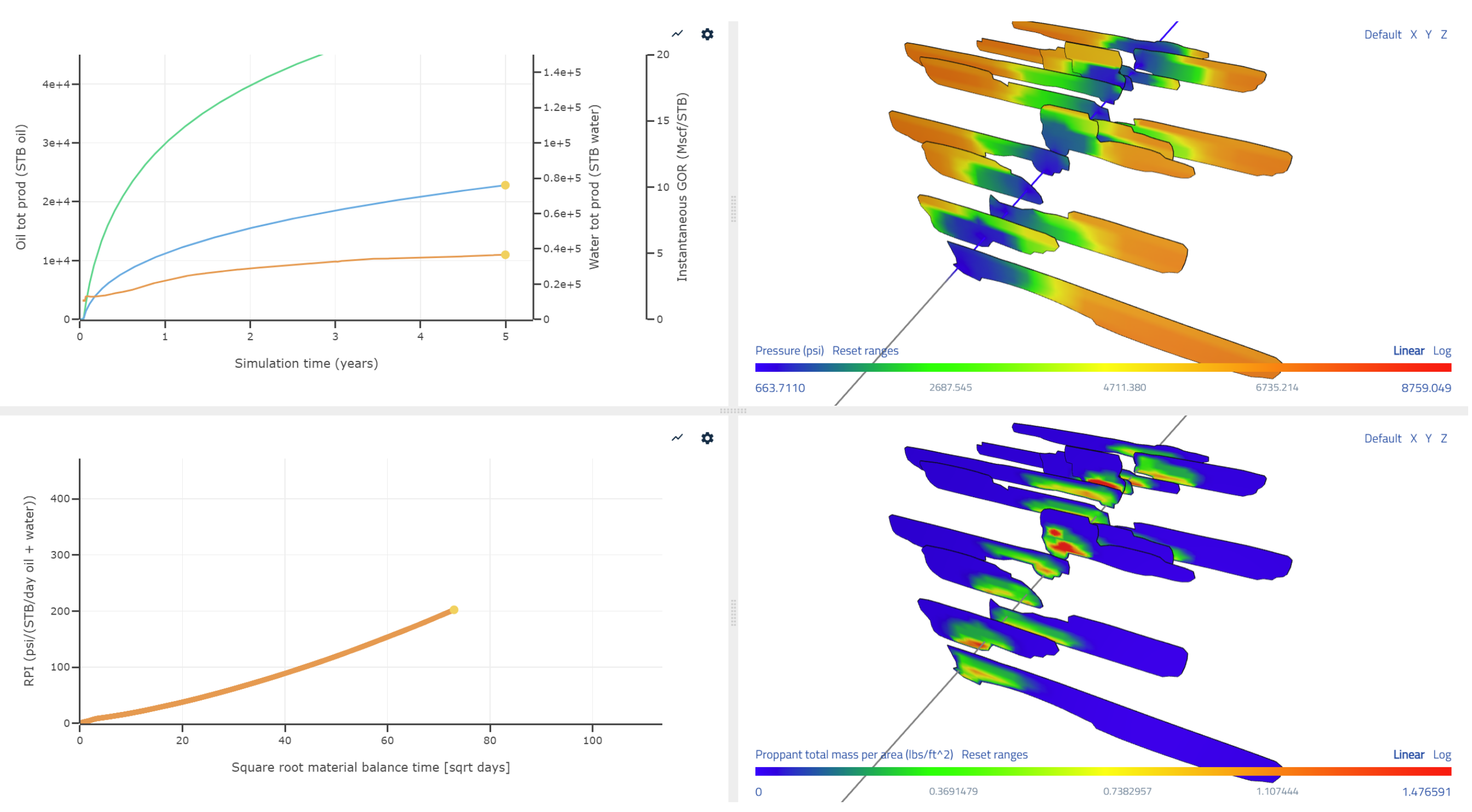

*Example simulation with 'submesh fractal D' set to 0.6.*

### 8.5.10 Water flowback

Water flowback at the start of production comes from two places – the fractures themselves, and the formation surrounding the fractures.

If the fractures have not yet mechanically closed at the start of flowback, then they will have very high conductivity and you will see a brief period of extremely high flowback rate at high water cut. To avoid this happening, make sure that you set up the 'pressure dependent permeability' table to accelerate leakoff during fracturing. This mimics the effect of unpropped fracture strands on water leakoff (but not on production). Use the pressure dependent permeability wizard to easily set this up. Use the 3D image to confirm that the fractures are mechanically closed when fluid production starts. Open cracks have Eopen value greater than zero.

As an easier alternative, rather than setting up a PDP table, you can check the box for 'Use parameterized PDP curves.' Then, you should enter the 'lower deltaP threshold for reversible permeability increase,' the 'upper deltaP threshold,' and the 'perm multiplier for reversible permeability increase.' For the lower threshold, pick a number around 300 psi less than (Shmin – initial pressure). For the upper threshold, pick a number around 300 psi greater than (Shmin – initial pressure). Then, for the multiplier, a good typical value is 50.

Second, water may flow back after leaking off into the formation. ResFrac uses a special 'water bank' treatment to keep track of how much water has leaked off and accumulated in the near-fracture region. The water bank grows as water leaks off, and depletes as water is produced back. Water flowback is accelerated as long as the water bank region thickness is not zero.

Water bank 'rel perm increase' scaling thickness – lower values cause water to flow back faster, and so cause a higher water cut early on, but cause a more steeply decreasing water cut over time. Higher values cause lower water cut early on, but more slowly decreasing water cut over time.

Water bank 'rel perm decrease' scaling thickness – this parameter defaults to be unspecified so that it has no effect. If specified, it causes early-time water cut to be even higher by reducing the rate that oil/gas can be produced, based on the thickness of the water bank. Smaller values cause higher water cut early on, with oil/gas production increasing over time. Larger values may cause less water cut increase at early time, but it will take longer for oil/gas production to recover. Overall, we typically recommend that you leave this parameter blanks.

### 8.5. 11 Long term water cut

Once the water bank thickness has reduced back to near zero, the water cut is affected by the mobility ratio of the oil/gas and water phases.

Lower value of Swr (curve sets panel) – Lower values cause more water production and higher water cut.

We primarily recommend modifying Swr. As a supporting option, you may choose to modify 'water phase exponent' (curve sets panel). Lower values cause more water production and higher water cut.

Also, sometimes, it's justifiable to change water saturation itself. If a formation has high water cut, but the petrophysical properties provided have low water saturation, it's fair to ask if the water saturations are reliable!

When modeling water production, it is useful to assess – in the first year of production, do you produce more or less water than was injected? If less, then Swr may be higher than initial Sw, suggesting that some of the injected water is lost to the formation indefinitely. If greater, then Swr may be less than initial Sw. This suggests that you will eventually flow back the injection water, and that you will continue to produce water from the formation indefinitely.

### 8.5.12 Crossflow between clusters – cluster spacing and fracture performance

It is widely recognized that flow outside casing (either through the cement sheath or an axial fracture) can cause crossflow between fractures. These crossflow processes are modeled explicitly in ResFrac. The user specifies a 'cased well and fracture collision distance.' If a fracture physically intersects a well within that distance of a perforation, it forms a hydraulic connection directly into that perforation. These connections occur both for fractures propagating from other wells and for fractures initiating from the same well.

The simulator permits for flow directly between fracture elements initiating from the same well that are within the "Cased well fracture collision distance". These crossflow terms are restricted by a 'flow barrier' treated like flow in a series (harmonic averaging) and with strength proportional to the distance between the fractures. The strength of the flow barrier is given by the parameter "Connect frac through 'cased well fracture collision distance' transmissibility barrier," which is given in units of md-ft.

We do NOT recommend that you typically use this parameter as a primary way of matching field data. In most simulations, you should use the default parameters, as described below. There are a few special cases when you may want to modify crossflow between clusters as part of a history match.

First, you have the following set of observations: (a) high perf efficiency from the injection well (as measured by downhole imaging or fiber), and (b) low far-field efficiency as measured from offset fiber. In that special case, then extreme cross-flow outside casing is possible.

Second, varying crossflow outside casing may be useful in matching refrac datasets. However, as described by Morsy et al. (2025; 2026) and Garbino et al. (2026) (URTeC 4245581, 4457839, and 4488451), the default parameters work well in many refrac datasets, without any modification needed. URTeC 4488451 shows that the best value for crossflow conductivity during refracs depends on the annular area between the original and the newly run liner.

If we run a refrac simulation with completely unrestricted crossflow between the new perfs and the original fractures (and each other), we find that simulations tend to *underpredict* production because there is *too much* localization of flow. However, if we run a simulation with perfect isolation between the fractures, we tend to overpredict production. Thus, we need to have some kind of intermediate case – where crossflow can occur, but it is not entirely unrestricted.

Also, what would happen if you tried to use 5 ft cluster spacing? There would be severe crossflow between the clusters, and you would not truly initiate a fracture very 5 ft.

The figure below shows results from a simulation with 'cased well and fracture collision distance' equal to 50 ft, and cluster spacing equal to 20 ft. The value of the 'transmissibility barrier' is set to 3e6 md-ft. As you can see, there are many clusters that never break down, which is because of the crossflow during stimulation. Also, in this simulation, the third stage (top of the screenshot) is never stimulated. The line plot shows the cumulative production versus time for this unstimulated third stage. This production occurs almost exclusively from cross-flow from the perforations of the third stage to the adjacent stimulated fractures from the second stage.

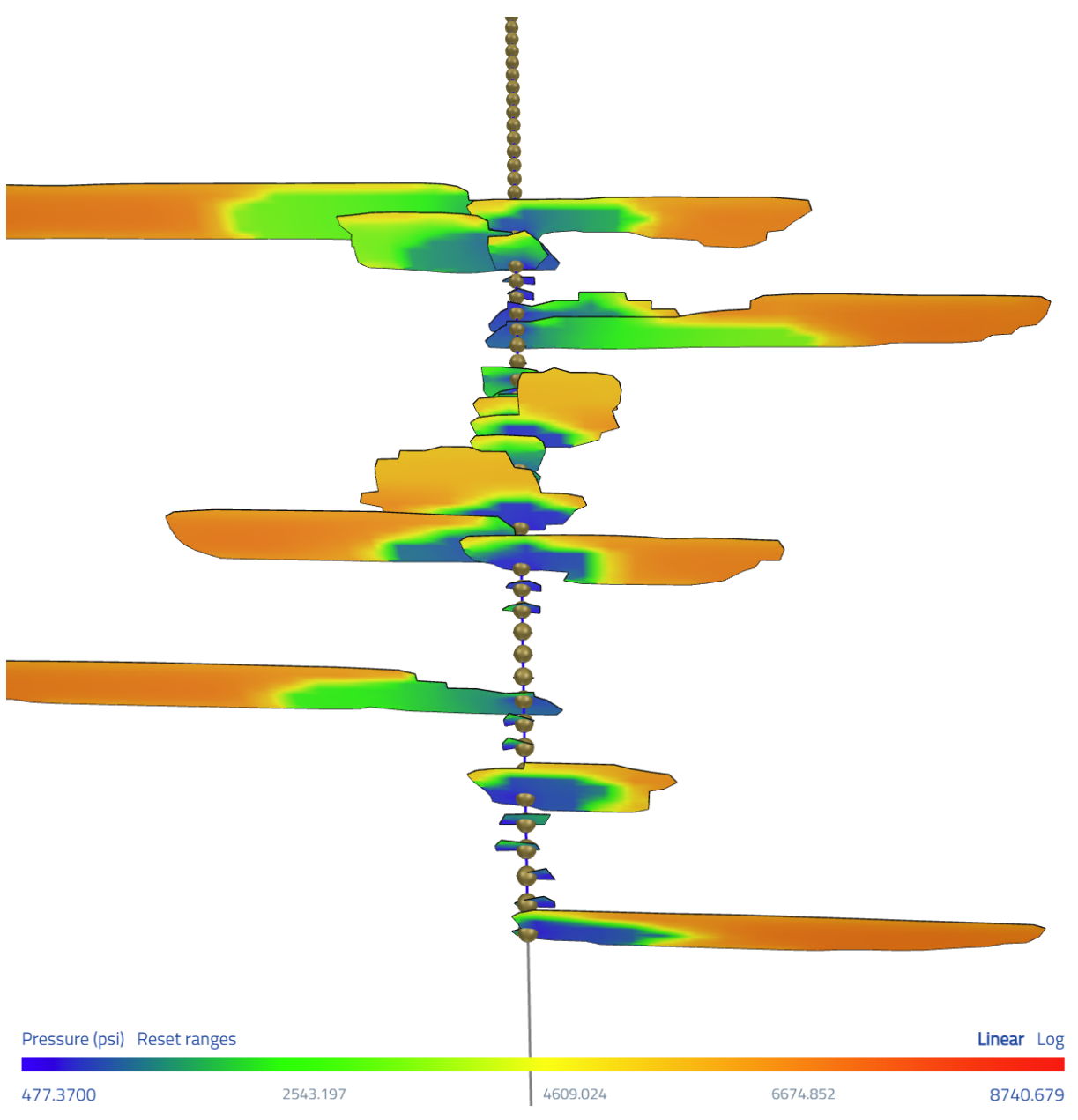

This figure shows the results with the 'transmissibility barrier' set to 3e5 md-ft.

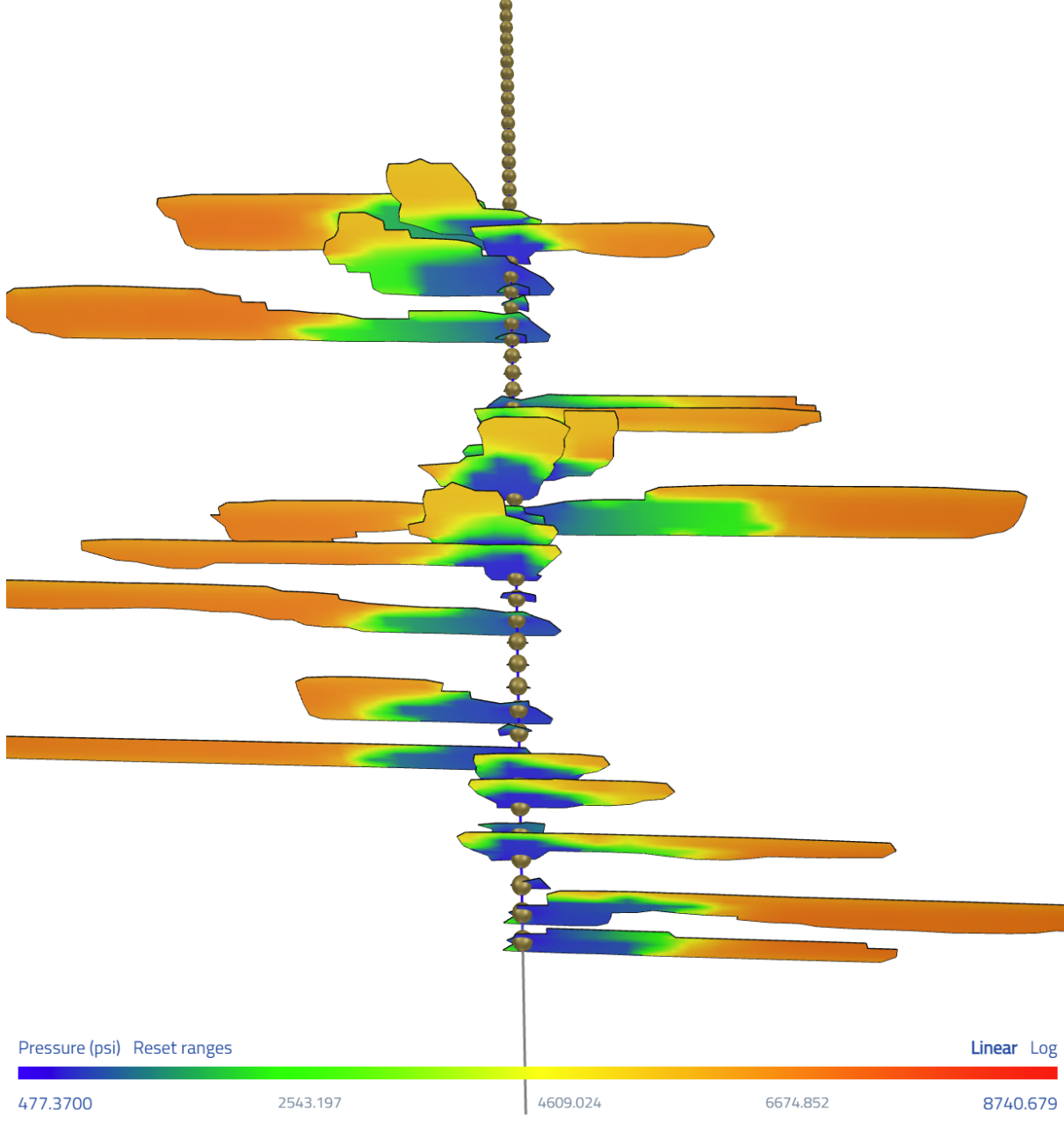

This figure shows the results with the 'transmissibility barrier' set to 3e4 md-ft.

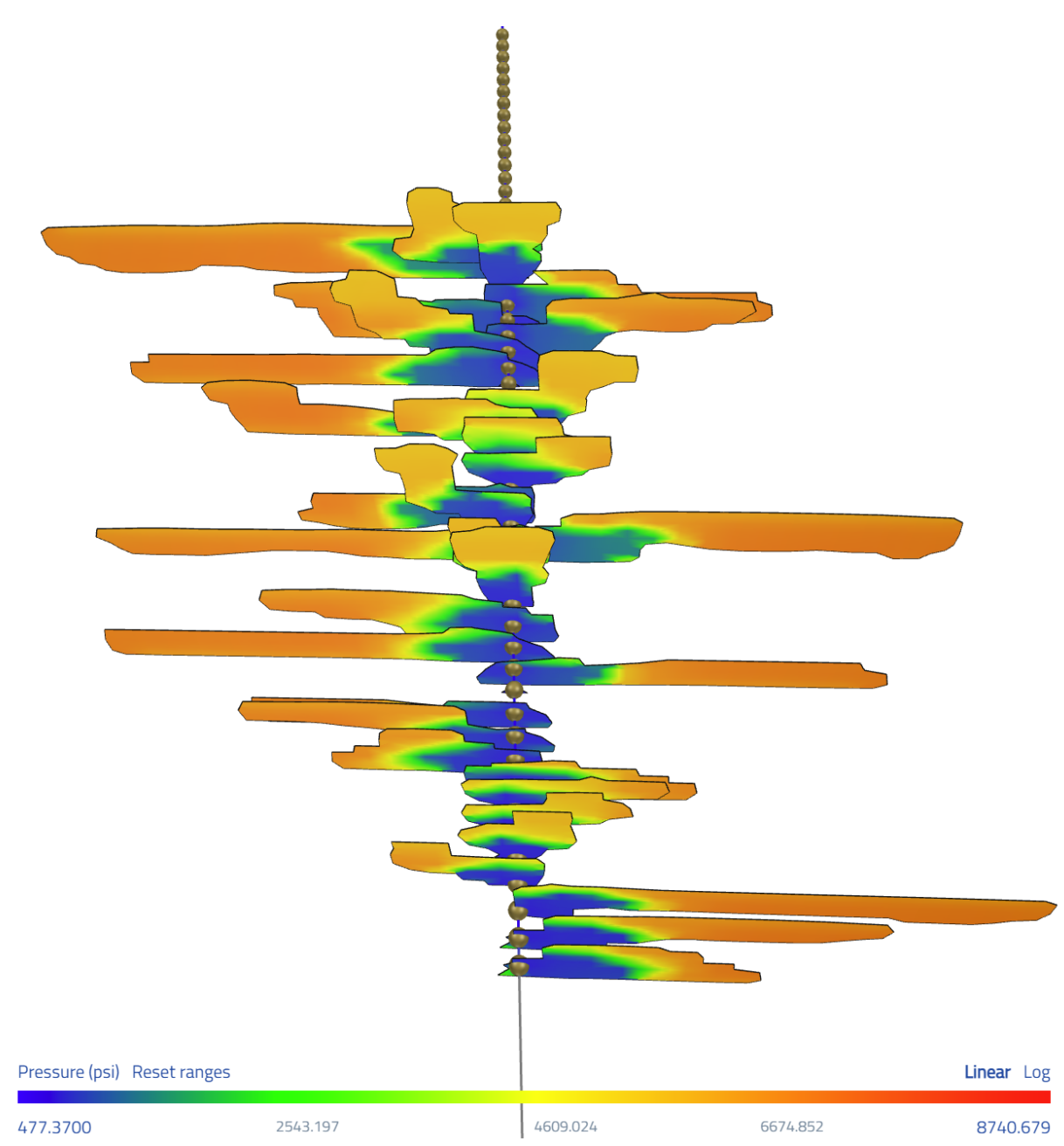

This figure shows the results with the ‘transmissibility barrier’ set to 3000 md-ft.

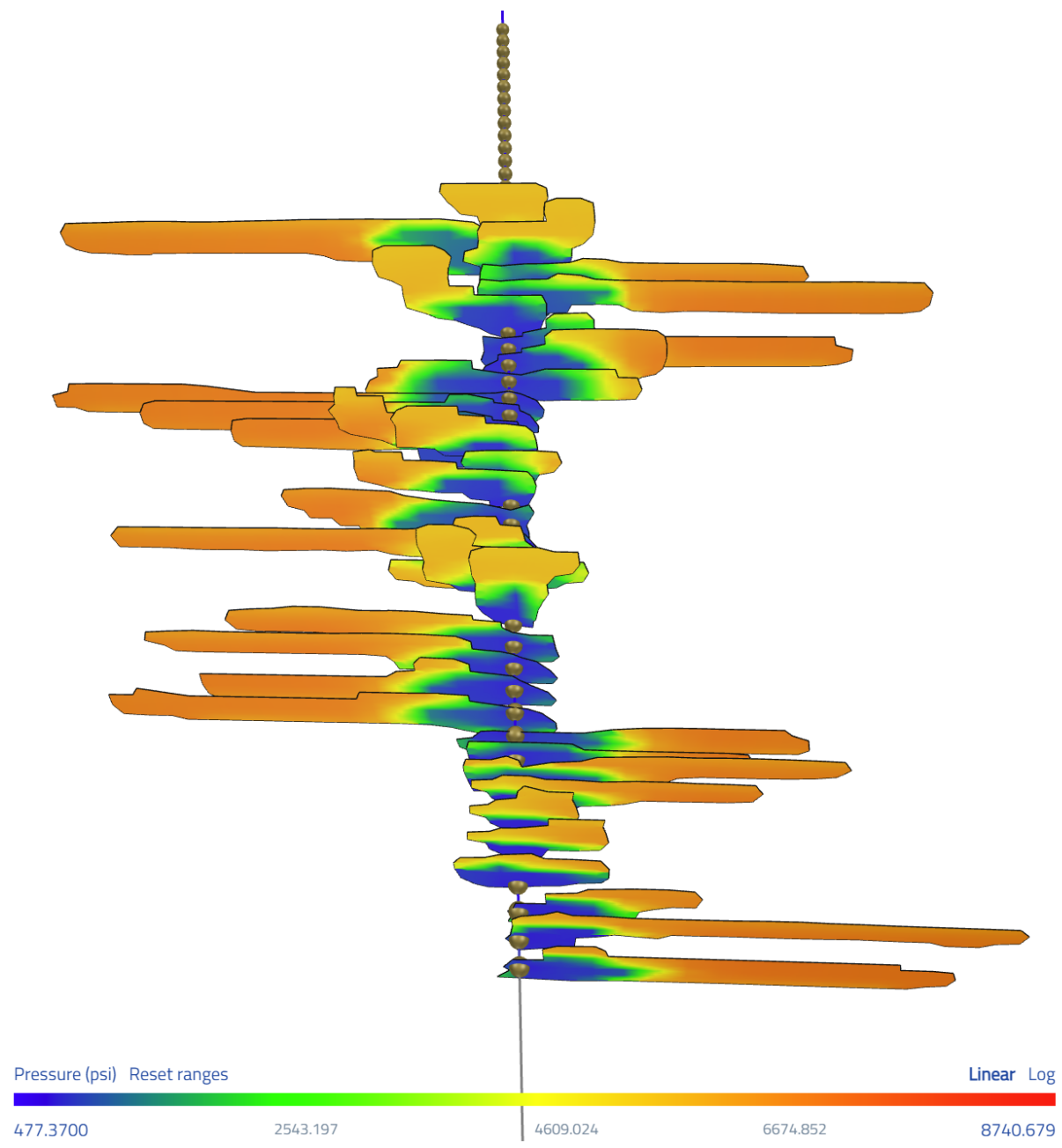

This figure shows the results with the ‘transmissibility barrier’ set to 30 md-ft.

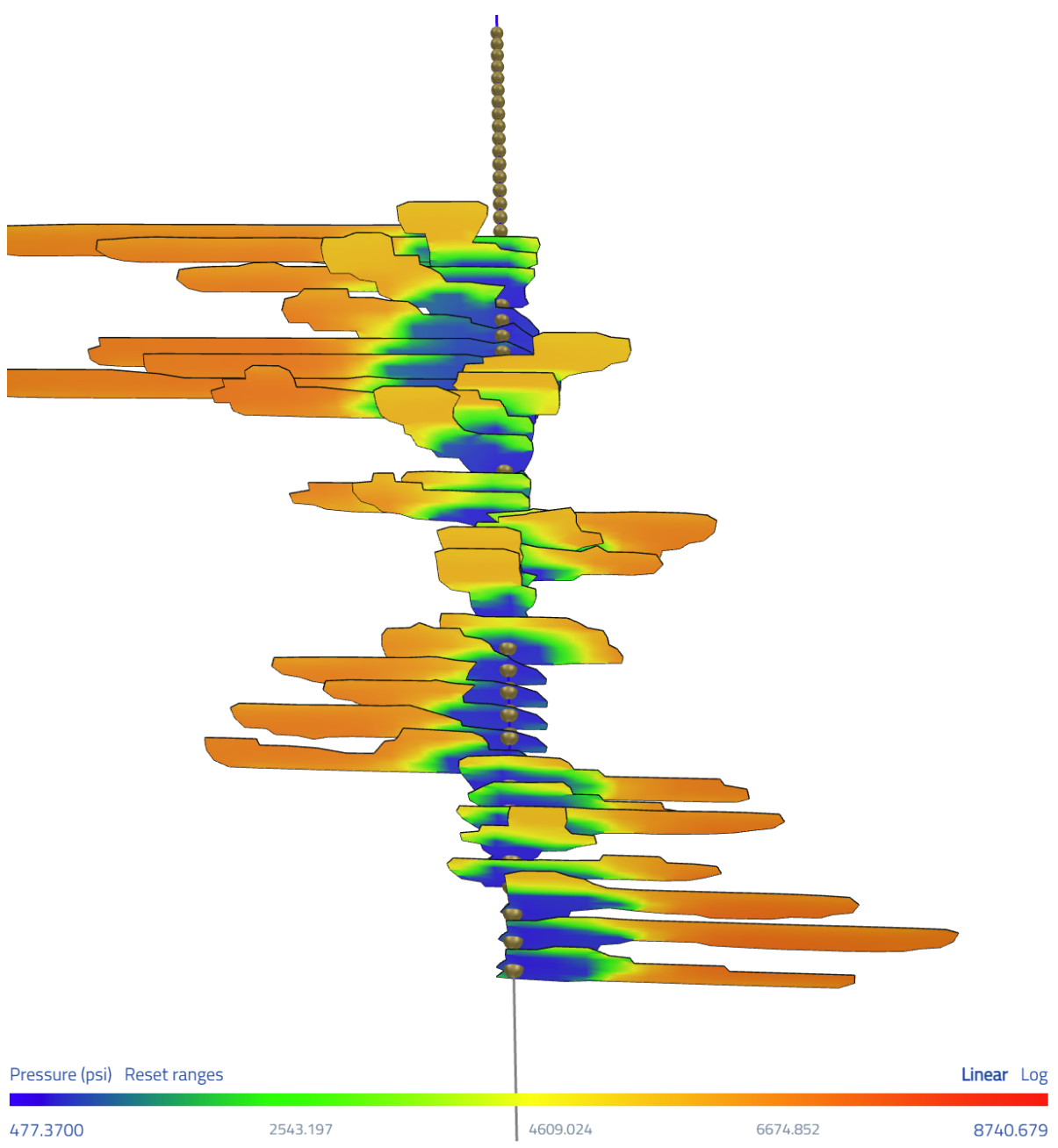

The results indicate that cross-flow during stimulation is significant at 300,000 md-ft, and relatively limited at 30,000 md-ft, and very limited at 3,000 md-ft. During production, flow rates are lower than during stimulation, and so while 3,000 md-ft is enough to prevent cross-flow during stimulation, it does not prevent significant cross-flow during production. On the other hand, going down to 30 md-ft restricts flow sufficiently that cross-flow during production is also significantly reduced.

The default value in ResFrac is 10,000 md-ft, which implies a modest amount of crossflow. As noted by Morsy et al. (2025; 2026) and Garbino et al. (2026), this value appears to work well in Bakken refrac simulations. However, in Permian refrac simulations, which typically involve wells with larger annular area, larger values may be needed.

# 9. Running sensitivities

History matching can be the most time-consuming part of a modeling study; however, the objective of most studies is to make improvements on future design. Running sensitivities and optimizations provides insight into how to improve the next generation of wells.

The most common completion design sensitivities to investigate are:

- Well spacing / landing zone
- Cluster spacing and perforation design
- Stage length
- Proppant loading
- Fracture sequencing

For each sensitivity above we list specific considerations when setting up the simulation as well as specific ResFrac features that are useful. At the end of the chapter there is a final section on evaluation metrics - common plots and images to compare sensitivity runs.

Remember: examination of the 3D visualization is crucial for understanding the implications and causes of production/performance results at different well spacings. Whenever interpreting a simulation result, start with the 3D images and seek to understand "why" things happen.

## 9.1 Structuring sensitivities

### 9.1.1 Well spacing / land zone sensitivities

Well spacing is one of the most important parameters for determining economic performance in shale. If the wells are spaced too far apart, there will be undrained reservoir between them. If wells are spaced too close together, they will experience excessive production interference. Optimal well spacing often involves *some* degree of interference between wells, with hydrocarbon price dictating how much interference is optimal.

**Setup considerations:**

- Number of wells and layout
  Often, you begin a well spacing sensitivity from a history matched model. That history matched model may have several wells all with different designs. For the first round of well spacing sensitivity analysis, we recommend standardizing on a single well design for all wells - or if multiple benches, perhaps a single well design in each zone.

  We recommend including at least three wells in the simulation models to capture stress effects (stress shadowing) from adjacent wells, as well as production interference. In multi-bench developments, more than three wells may be necessary in order to capture the three dimensionality of the problem. 'Zero permeability cubes' can then be used to account for additional offsetting wells (more below).

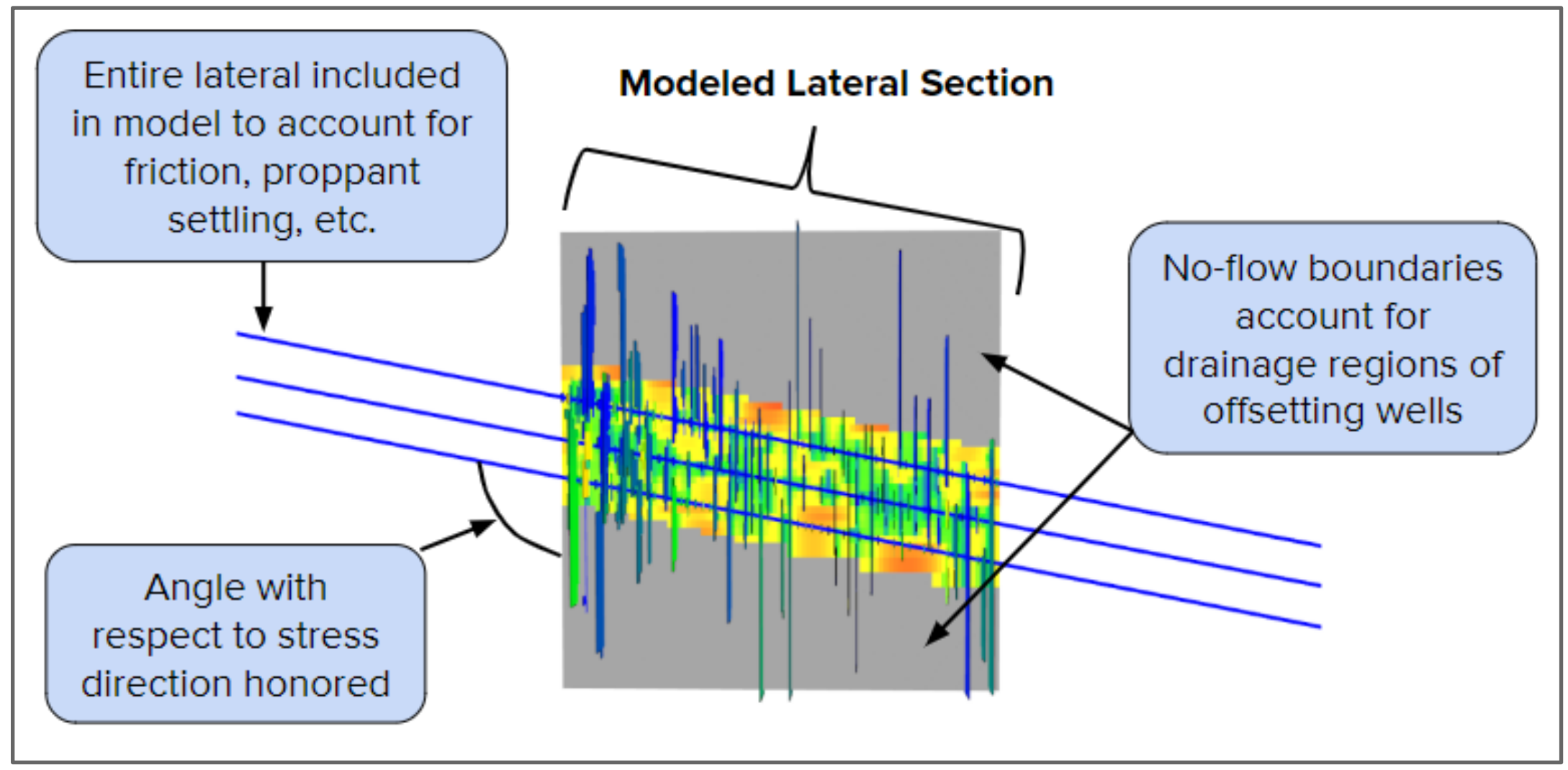

Using these tools, one can isolate wells in different configurations to test well spacing. For instance, with a single pay zone, one might consider a three-well layout like the one below:

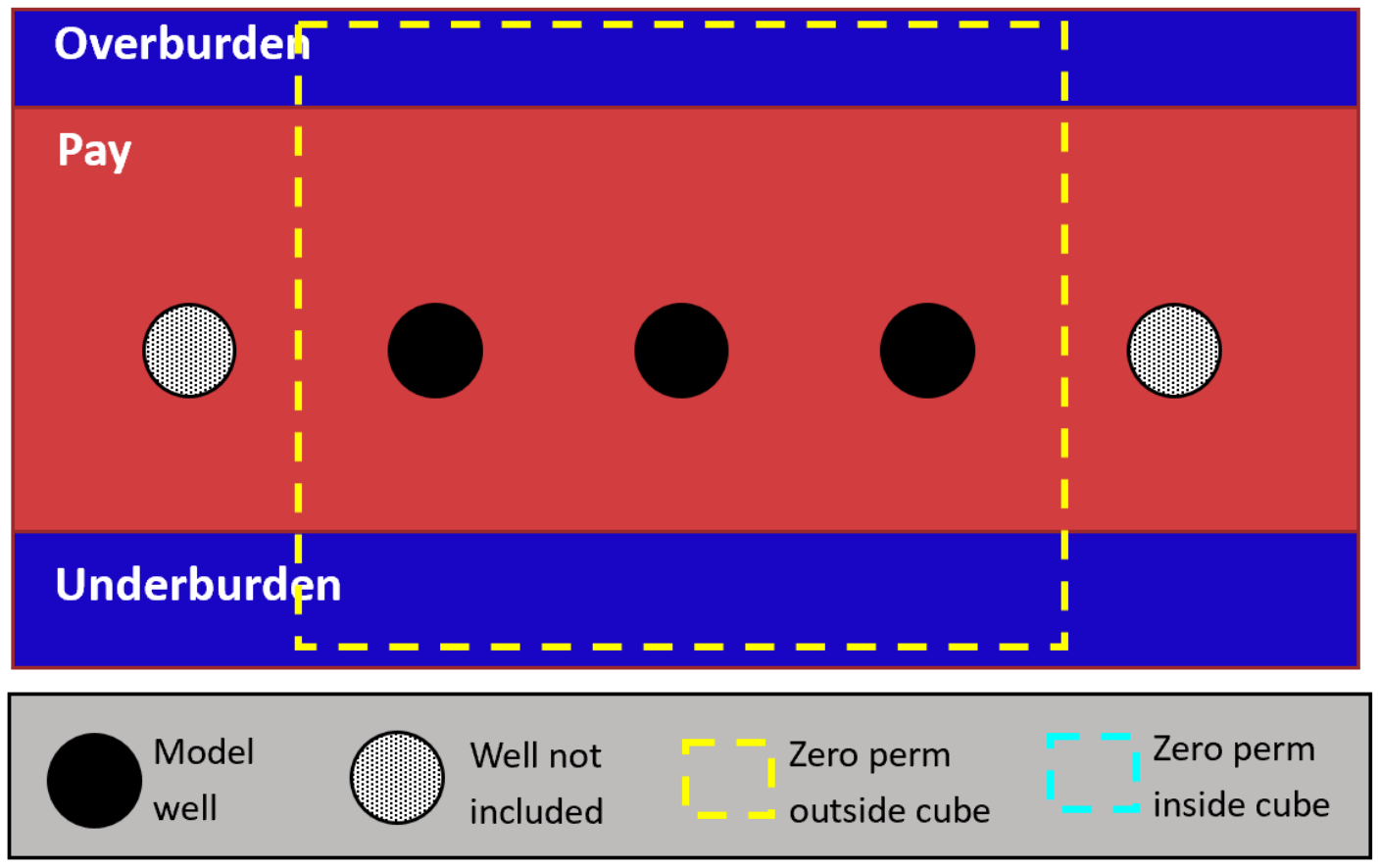

Alternatively, in a multiple bench development one might pick a quick and simple layout like this one:

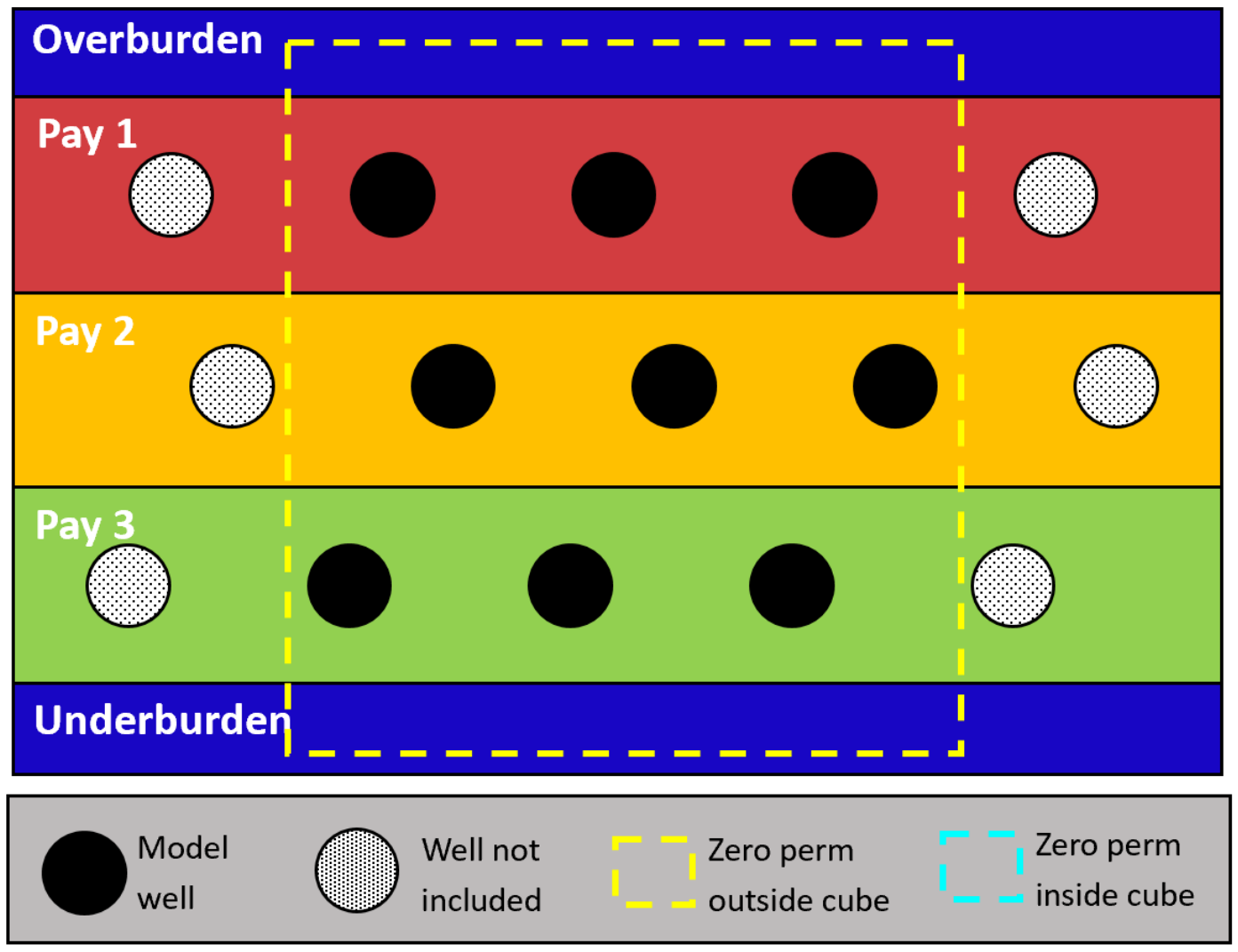

Or, one could make a more complicated layout such as the one below:

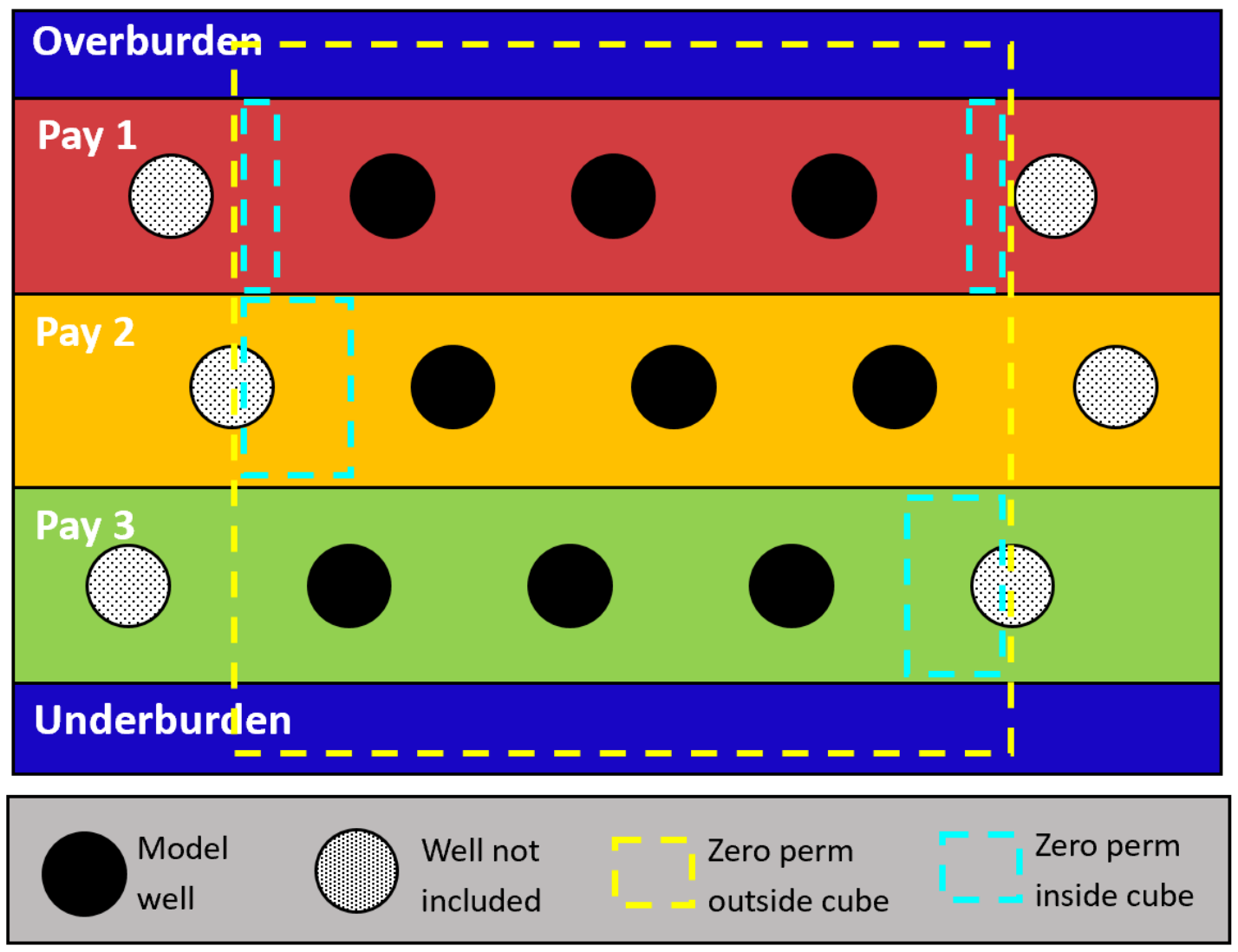

Picking the initial layout to test different well spacings can be challenging. Don't hesitate to reach out and run ideas past the ResFrac team!

- Number of stages
  We always need to balance computational time and scale in ResFrac simulations. Depending on the number of wells, you may choose to model a different number of stages. For instance, with only three wells, you might model three stages per well. With nine wells, it may only be tractable to model one or two stages per well. Keep in mind that external fractures (see the Useful ResFrac Features discussion below) can be used to simulate stress shadowing from stages *not included* in the model, and can be time dependent to mimic zippering sequence, etc.

- Placing wells
  Wells are not always aligned with a cardinal direction, so adding new wells at a given spacing in a model can seem daunting. Don't worry, the 'Well shift wizard' in the 'Wells and Perforations' panel can do this for you. If your new well has the same trajectory as an existing well, you can simply duplicate the well and then use the 'Well shift wizard' to apply the desired horizontal well spacing. The geometry of this procedure is shown in the image below. There are two shift options: shift in the direction of SHmax (represented by the blue line in the image below), or shift perpendicular to the well azimuth (represented by the green line below).

  From within the simulation builder, press the 3D preview plot to ensure that the wells are in the correct position.

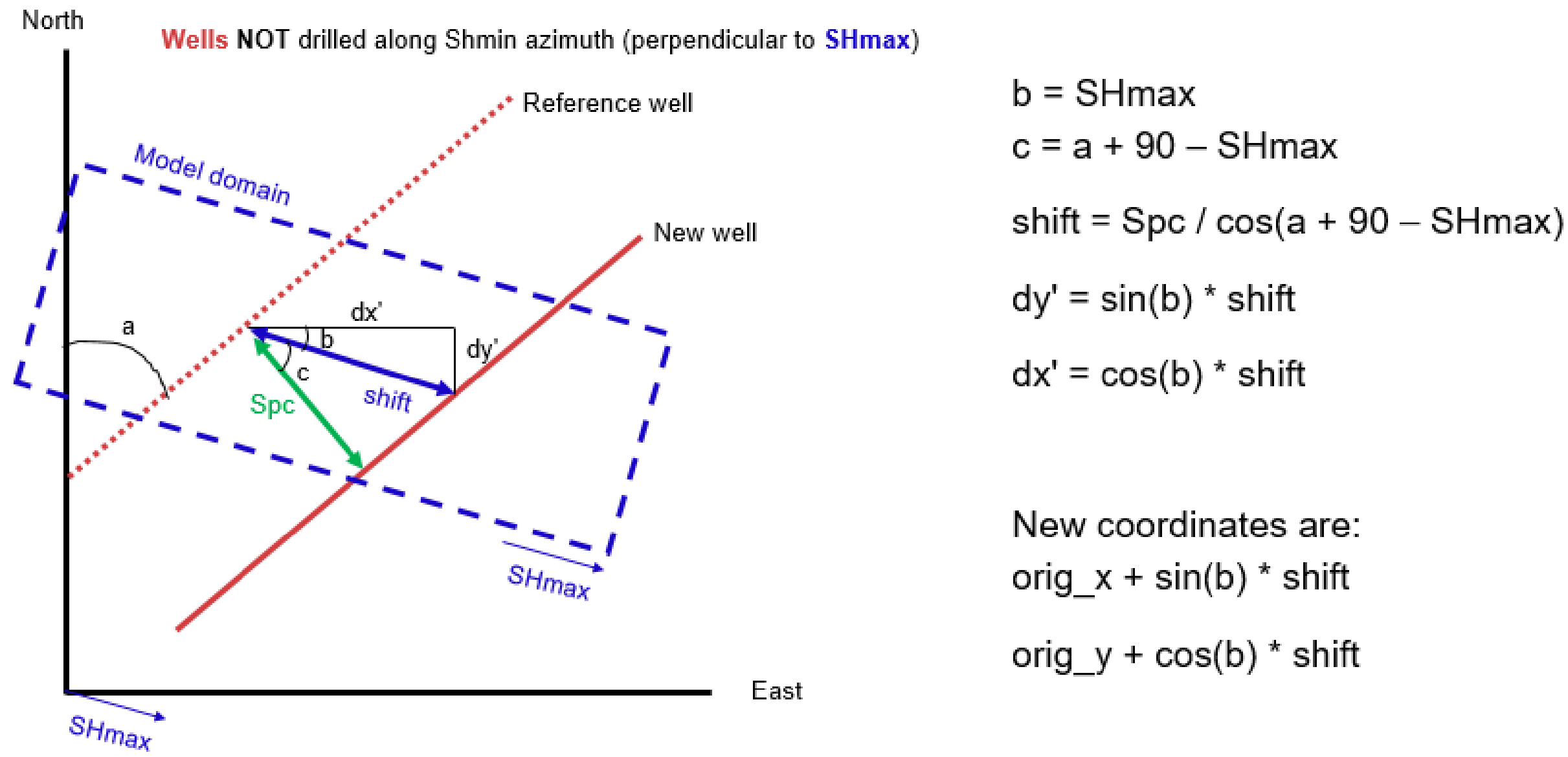

- Mesh size
  The same mesh resolution used during the history match should be used for all sensitivity analyses. If simulating larger well spacing, it may be necessary to expand the matrix - but you should do so using the same element size.

  For example, if you history matched with a matrix mesh 5000 ft wide (in the direction of SHmax), made up of 100 elements, then that means each element is 50 ft wide. If you then expand the matrix region, to accommodate larger well spacing, to 7000 ft, then you should increase the number of elements to 140 (to maintain the same 50 ft element width).

- Cluster spacing
  It is best to maintain the same cluster spacing and number of clusters as your base case.

- Pumping schedule
  Keep the same pump schedule as your base case.

**Useful ResFrac features:**

- Duplicate well

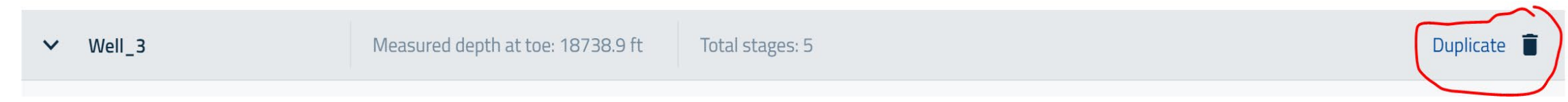

As noted above, placing additional wells can sometimes be daunting, but it is significantly simplified by duplicating an existing well and then using the 'Well shift wizard' to apply the desired horizontal spacing. This is done in a few easy steps.

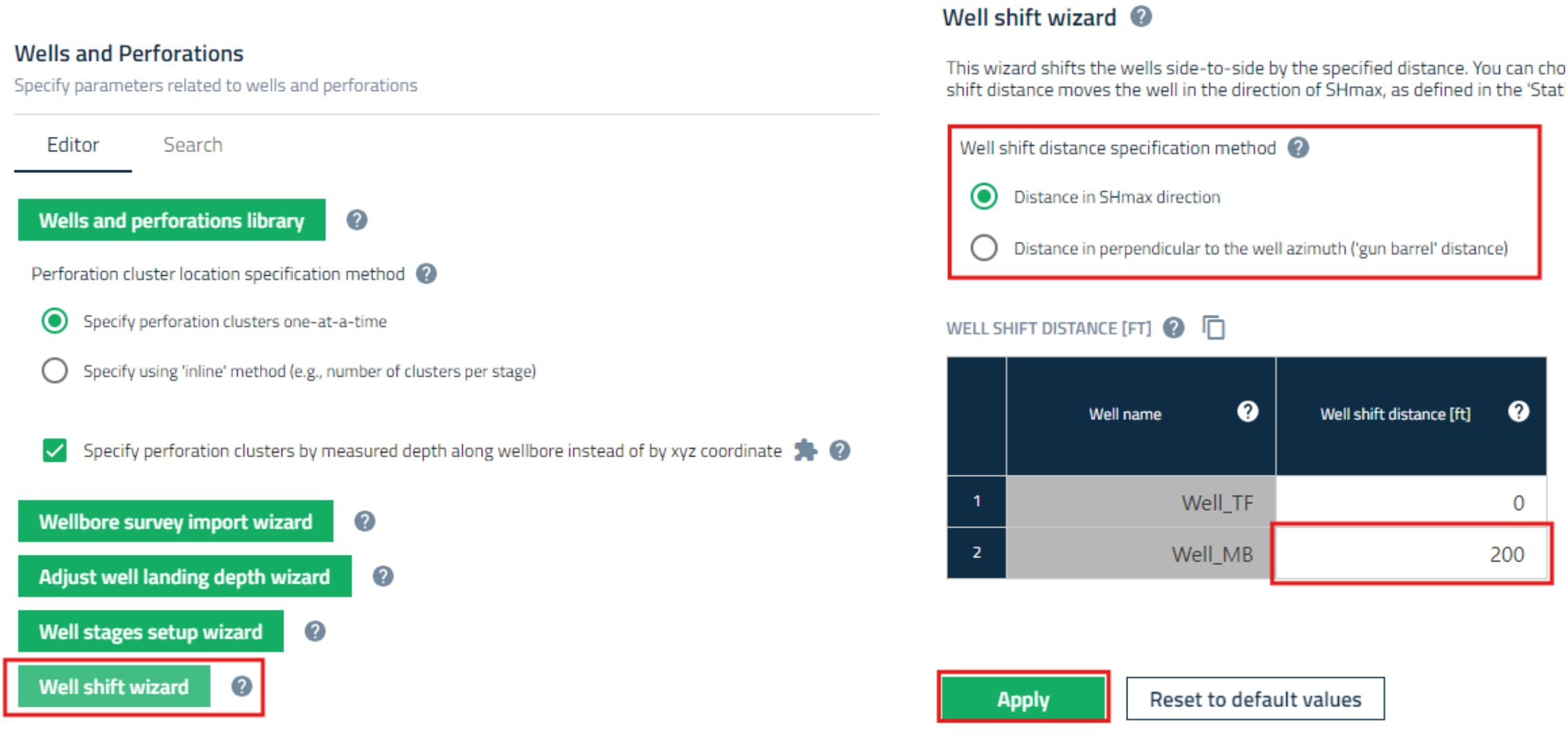

- Duplicate well controls

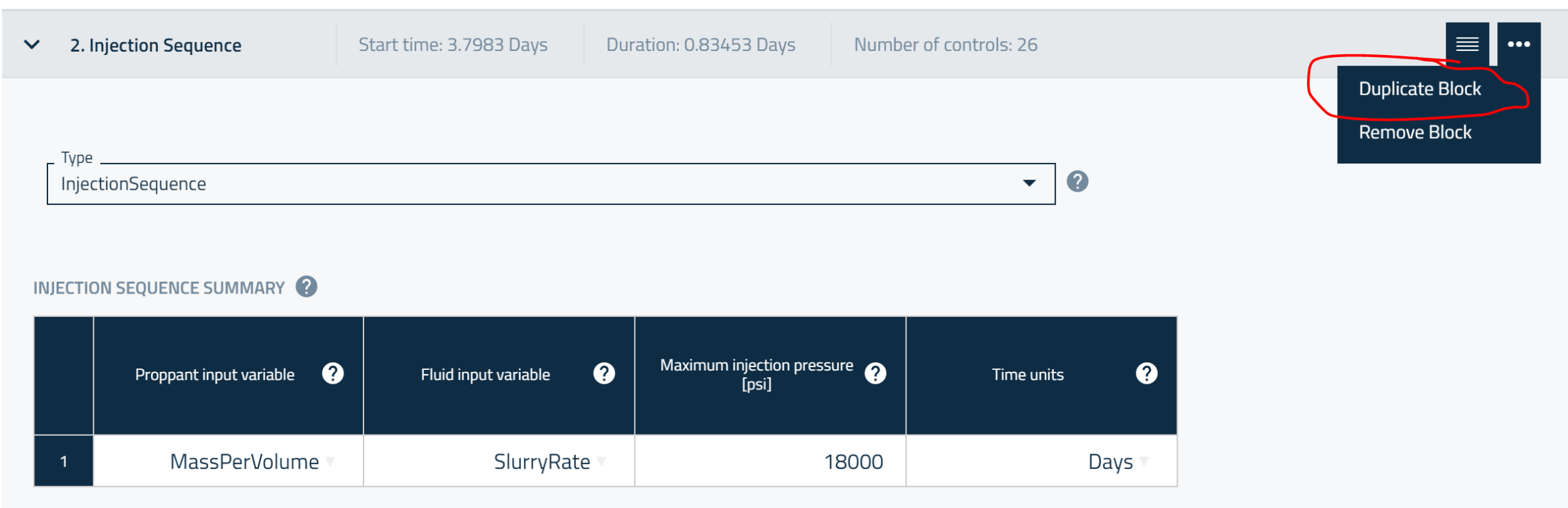

When populating controls for a new well, you can duplicate control sequence blocks from another well. This will save lots of typing and reduce the propensity for errors.

- Zero permeability ‘inside’ and ‘outside’ cubes
  Zero permeability cubes deactivate matrix and fracture elements, so that they have no flow in or out. In this way, zero permeability cubes can be used to enforce boundary conditions to replicate the no-flow boundary occurring between wells or between stages.

  There are two types of zero permeability cubes:
  A) Zero permeability outside this cube: zeros out permeability outside of the cube dimensions specified (gray in picture):

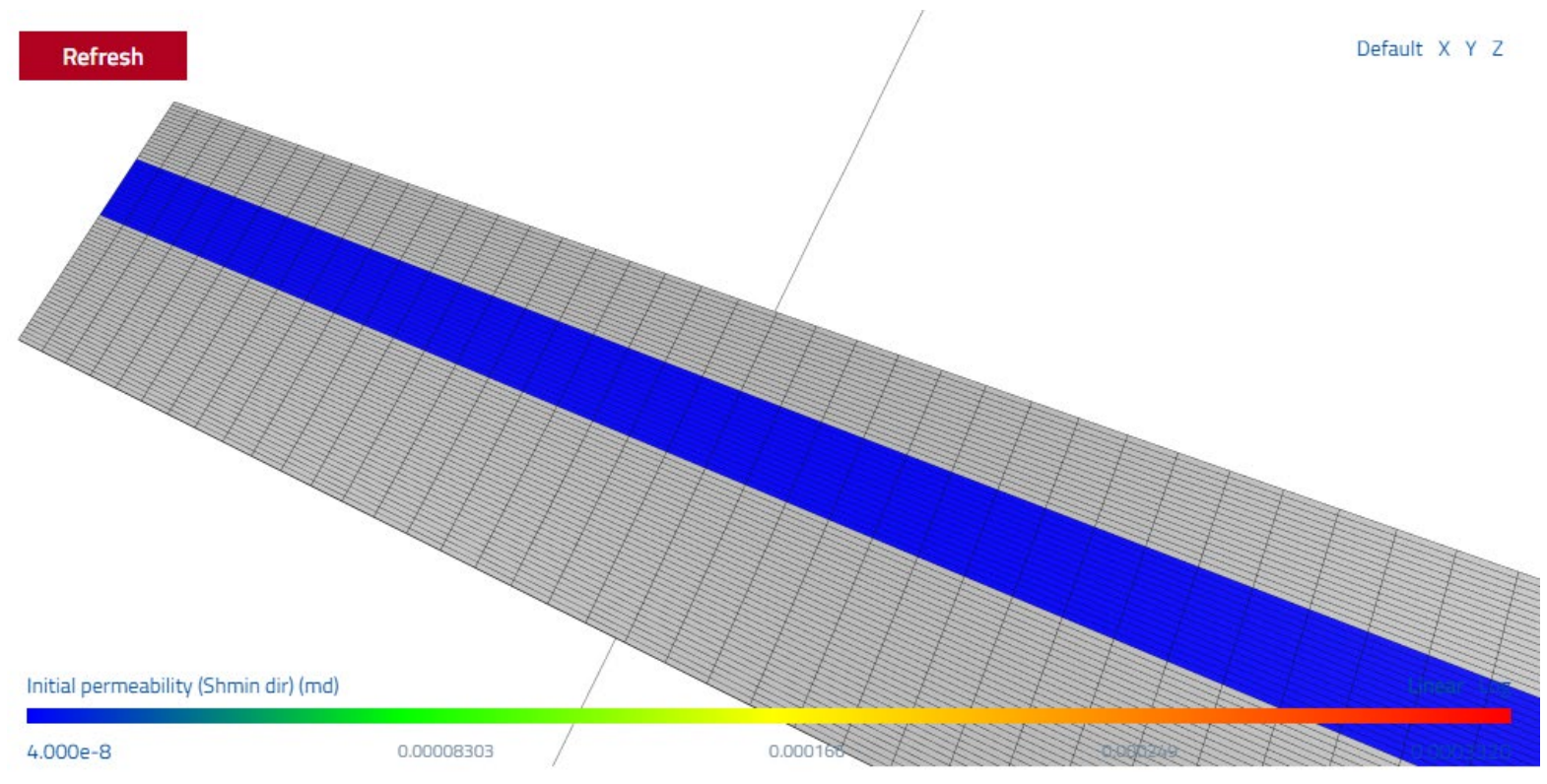

  B) Zero permeability inside cube: zeros out permeability within the cube dimensions specified:

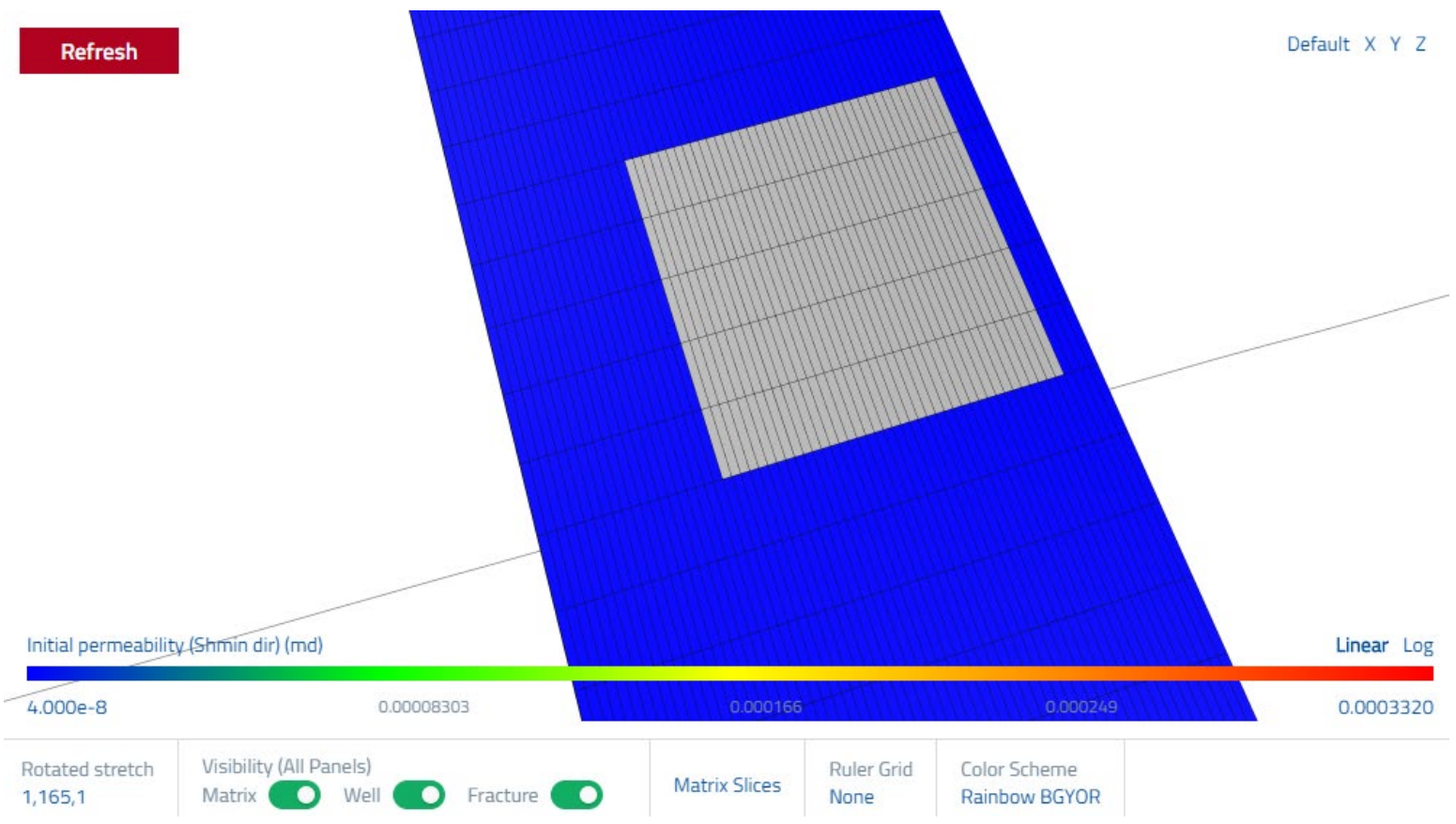

If the stage lengths on adjacent wells are mismatched, you can use the ‘zero perm *inside* cube’ setting to adjust. For example, the figure below shows two wells: one with three stages of length 200 ft each, and another with two stages of 250 ft each. The model width needs to be 600 ft for the first well and 500 ft for the second. So, two 'zero perm inside cubes' have been specified, shown as the two grey strips along the edge of the model. They effectively 'cut out' a section of the matrix region so that it is not included, so that the mismatched stage lengths can be simulated in a single model.

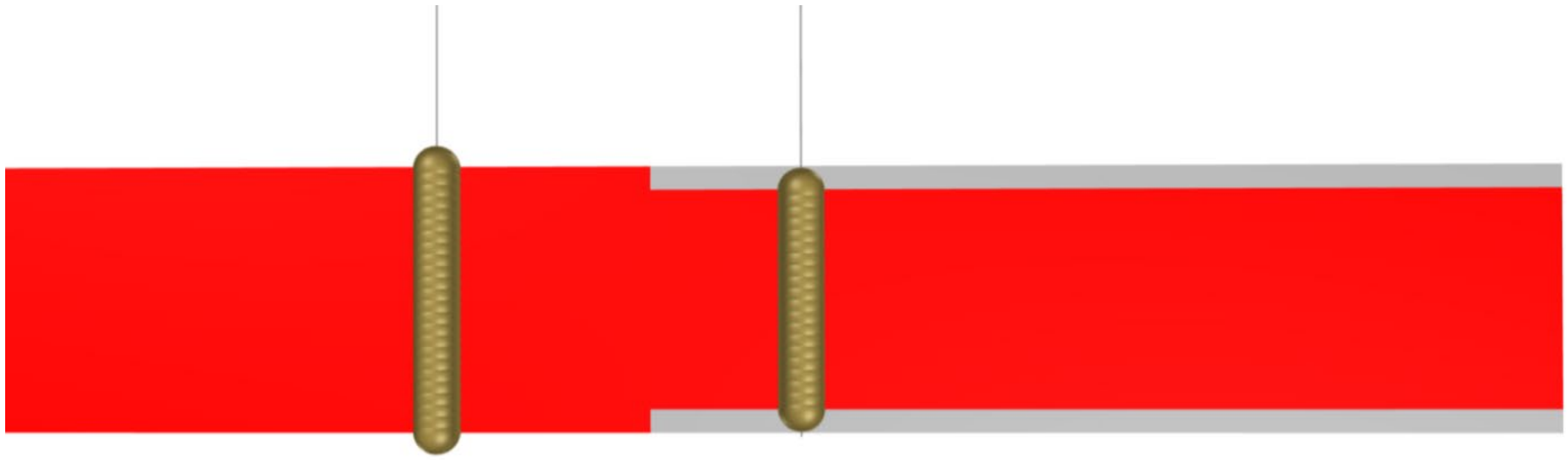

- External fractures
  Found on the 'Wells and Perforations' tab, External Fractures allow the user to specify a fracture *outside* of the matrix domain with a geometry and time-dependent net pressure (i.e. external fracture net pressure can be zero until a specified time, peak to some value, then decay to a residual net pressure - just as stress shadowing observed in an offsetting stage). Don't try to create an external fracture to represent all the fractures from all previous stages. Instead, set up a single fracture, just outside the model, to represent the stress shadow from all previous stages. Set up the external fracture length and height to be equal to approximately the length and height of the fractures created by the model. To select an appropriate value for 'net pressure', one strategy is to run a single stage model without any external fracture, and calibrate it to approximately match the actual fracture length from the data. Specify a stress observation plane, and observe the magnitude of the cumulative stress shadow built up along the stage. This can be used as the net pressure for the external fracture. Alternatively, you could set the net pressure equal to the amount of cumulative ISIP buildup seen in the actual data from the first stage to the second, third, and fourth.

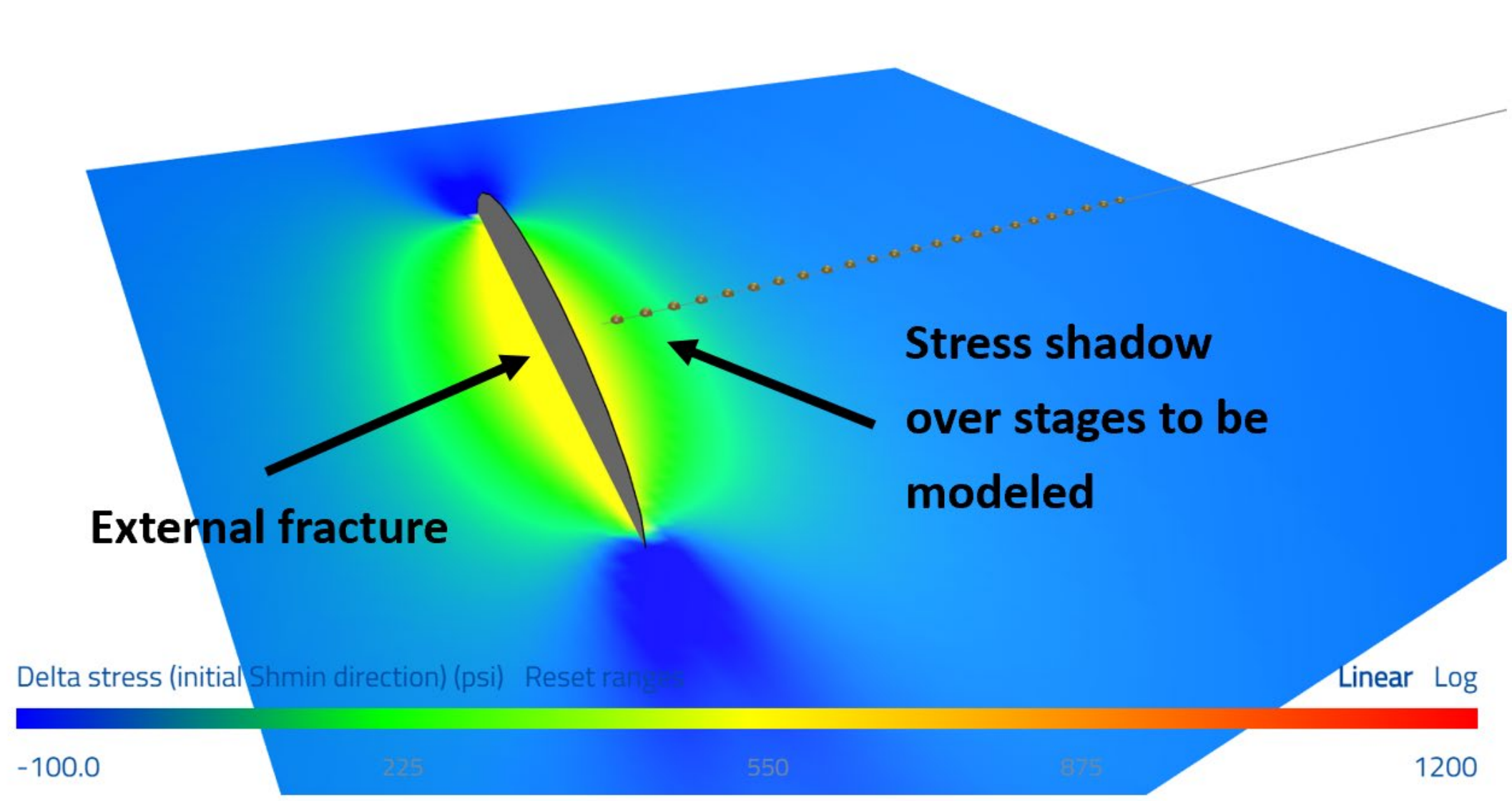

## 9.1.2 Cluster spacing sensitivities

Cluster spacing governs the density of the hydraulic fractures around a well. Benefits of decreasing cluster spacing can include more near-well fracture area, higher recovery factors, and more productive wells. However, when cluster spacing is *too* tight, well performance can suffer due to adverse stress shadowing and pressure interference between clusters.

Optimal cluster spacing is a function of geologic and petrophysical parameters (permeability, hydrocarbon compressibility, porosity, stress) as well as completion design limits (perforation friction, pressure limits).

**Setup considerations:**

- Matrix mesh size
  The same mesh resolution used during the history match should be used for all sensitivity analyses. Ideally simulations are set up to have at least one matrix mesh element in the direction of Shmin for every cluster. For example, if you are simulating three stages with eight clusters per stage, it is best to have at least 24 matrix elements (3 x 8) in the direction of Shmin. So, when first setting up your model it is advised to think about what the *maximum* number of clusters you will eventually have (including sensitivity analysis).

  Sometimes, you may forget to history match with a fine mesh, or refining the mesh may be computationally prohibitive. **Do not worry.** The code is setup to handle a coarse mesh. You should still maintain the <u>same</u> mesh for your sensitivity analysis as your history match, even if it means you will have more than one cluster in a matrix element.

- Wellbore mesh size
  Similar to the matrix mesh, it is best to have at least one wellbore element for each cluster. If you anticipate testing cluster spacing down to 10 ft, when setting up your initial model, make sure that the wellbore element length is 10 ft or less.

- Stage length
  It is easiest to keep stage length the same as the base case (often the history match) as this way you can also keep the injection schedule the same (which ensures the same proppant/fluid per lateral foot). This means to achieve tighter cluster spacing you add more clusters per stage, and sparser cluster spacing is synonymous with fewer clusters per stage.

- Limited entry or perforation friction
  Limited entry or perforation friction is the amount of "back pressure" generated by the perforations. Conceptually, the smaller the area the fluid has to exit the casing, the higher the pressure is needed to force it out at a given rate. While high perforation friction ("back pressure" and "$p_{pf}$" below) may require more horsepower at the surface, it has the advantage of more evenly distributing fluid across all clusters in a stage and increasing perforation efficiency.

$$P_{pf} = \Delta P_p = 0.2369\, Q^2 \frac{\rho}{{N_p}^2\, D_p^4\, {C_d}^2}$$

$P_{pf}$ = Perforation Friction (psi)
$\Delta P_p$ = Pressure drop across a perforation(s) (psi)
$Q$ = Total flow rate (bbl/min)
$\rho$ = Density of fluid (lb/gal)
$N_p$ = Number of open perforations
$D_p$ = Diameter of perforations (in)
$C_d$ = Coefficient of discharge

Figure 9.10 – Limited entry equation

When setting up your sensitivity cases, keep this concept of perforation friction in mind:

  - Often it is best to start with normalized perforation friction across all cases. So, if your base case has an injection rate of 80 bpm and 8 clusters per stage with four 0.4" holes per cluster, your base perforation friction is (assuming fluid density of 10 lb/gal and coefficient of discharge of 0.85):

$$p_{pf} = 0.2369 * 80^2 * \frac{10}{(8*4)^2 * 0.4^4 * 0.85^2} = 800\ psi$$

    Then, if you want to sensitize on 12 clusters per stage, you should **maintain the same perforation friction** by adjusting the number of perforations and/or the perforation diameter, such as using three perforations per cluster with a diameter of 0.377 in (which gives a perforation friction of 801 psi). Alternatively, if you created a sensitivity with only four perforations per stage, you would change the number of perforations per cluster to eight, each with a diameter of 0.4 inches, to maintain the same 800 psi perforation

friction.

  - Increasing perforation friction can help to more evenly distribute injection across the stage and overcome stress shadowing (which may reduce perforation efficiency). If you observe diminished perforation efficiency at tighter cluster spacings, then it may make sense to sensitize on *increasing* perforation friction.

    For example, in the 12 clusters per stage case above, if you observe that perforation efficiency drops to 67% at the base perforation friction value (800 psi), then the well may benefit from increasing perforation friction by reducing the number of shots and/or shot diameter per cluster or increasing injection rate - just make sure to check with the completions team to ensure that they can handle the additional pressure/rate requirements on surface! Also, if you increase rate, be sure to decrease injection duration to keep total fluid and total proppant the same as the base case so that you are only sensitizing on one variable (perforation friction) at once!

    How high can you go? There are published cases in literature of operators designing wells with 8000+ psi of perforation friction; however, 1500-2500 psi is a more common "extreme limited entry" design.

- Behind pipe communication
  In the advanced settings on the fracture options tab, we recommend checking the "Connect frac through 'cased well fracture collision distance'" option.

Connect frac through 'cased well fracture collision distance' 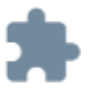

  This parameter accounts for behind-pipe flow that may occur due to poor cement or a longitudinal fracture in the annulus. Why might that matter? That flow from one cluster to another behind the casing effectively defeats limited entry and may result in the propagation of dominant fractures.

  Note this "Connect frac through 'cased well fracture collision distance'" needs to be used in conjunction with the "Cased well fracture collision distance." Based on the work of Ugueto and others, we recommend using 10 ft for this parameter.

- Perforation schemes
  Most well designs use the same number of shots in every cluster; however, some designs may "taper" perforations across the lateral (e.g. more shots in toe clusters than heel clusters, or vice versa).  To specify different perforation design for each cluster, make sure you are specifying clusters one at a time (see below) and pay attention to what the total perforation friction is for the stage (above).

- Proppant and fluid volumes
  You may also want to optimize proppant and fluid volumes, but in the spirit of testing one variable at a time, we suggest keeping the injection schedule the same, and vary proppant/fluid volume by modifying the duration of the final proppant ramp or by modifying the duration of each ramp.

**Useful ResFrac features:**

- In-line versus one-at-a-time cluster specification

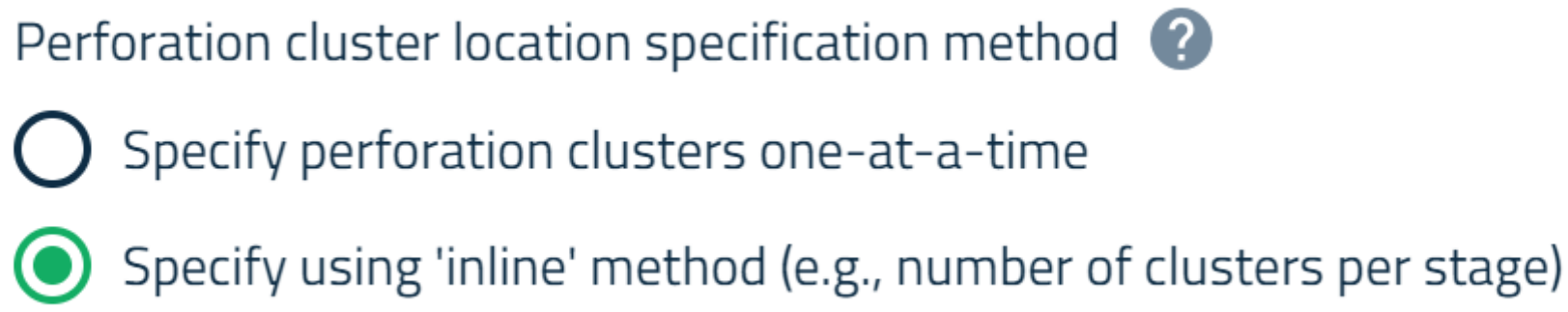

There are a couple ways to enter a completion (cluster/perforation) design in ResFrac. The "inline" method evenly distributes clusters along the stage and assumes every cluster has the same perforation design.

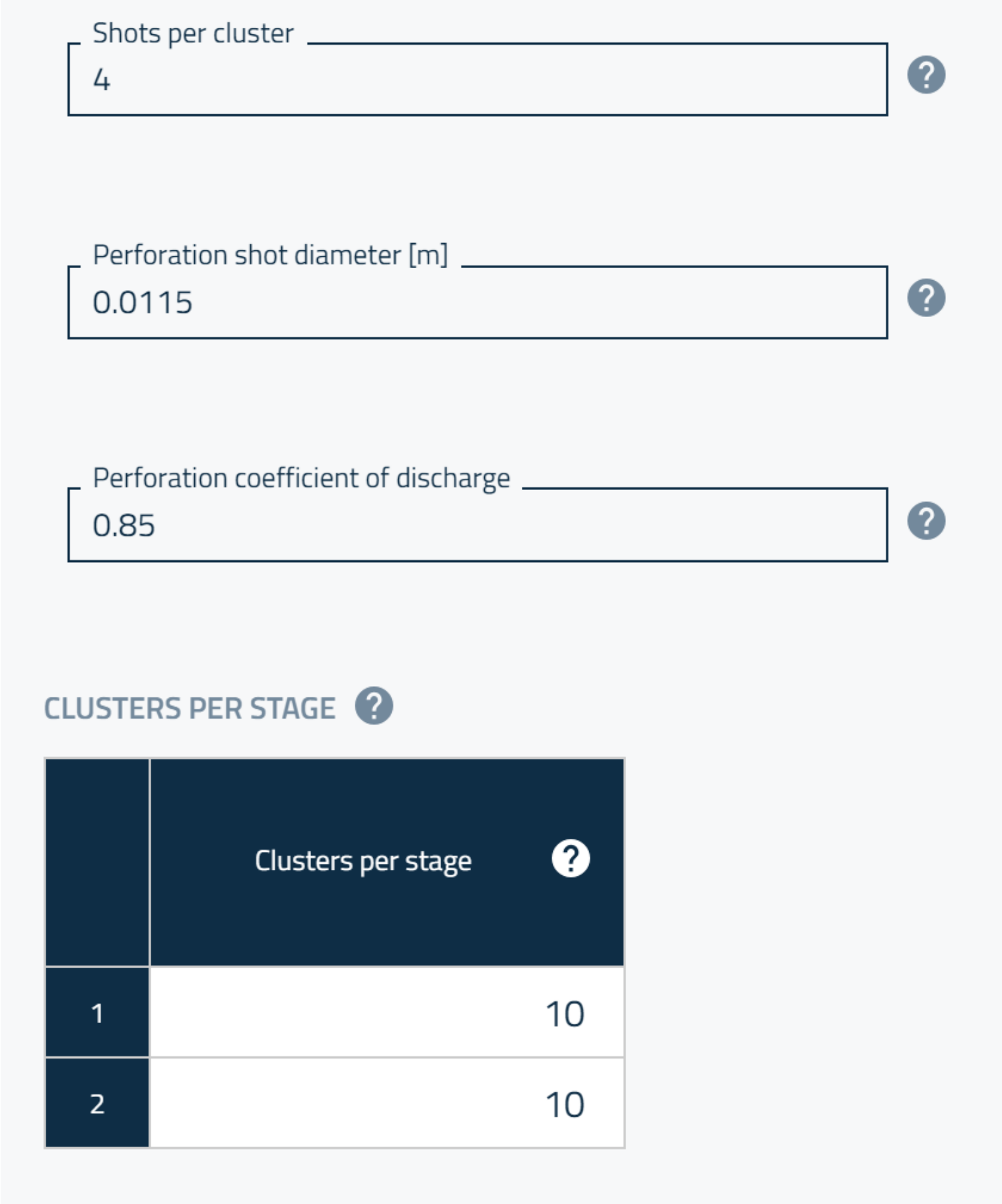

The “one-at-a-time” method allows for uneven spacing of clusters along a stage and/or perforation designs that vary by cluster:

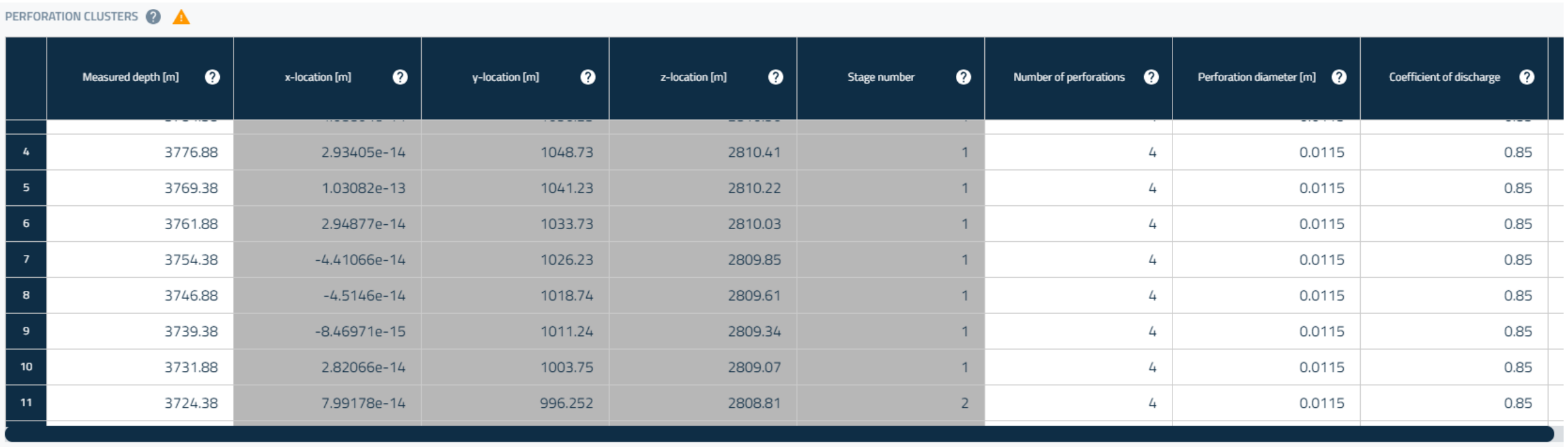
PERFORATION CLUSTERS

| | Measured depth [m] | x-location [m] | y-location [m] | z-location [m] | Stage number | Number of perforations | Perforation diameter [m] | Coefficient of discharge |
|---|---|---|---|---|---|---|---|---|
| 4 | 3776.88 | 2.93405e-14 | 1048.73 | 2810.41 | 1 | 4 | 0.0115 | 0.85 |
| 5 | 3769.38 | 1.03082e-13 | 1041.23 | 2810.22 | 1 | 4 | 0.0115 | 0.85 |
| 6 | 3761.88 | 2.94877e-14 | 1033.73 | 2810.03 | 1 | 4 | 0.0115 | 0.85 |
| 7 | 3754.38 | -4.41066e-14 | 1026.23 | 2809.85 | 1 | 4 | 0.0115 | 0.85 |
| 8 | 3746.88 | -4.5146e-14 | 1018.74 | 2809.61 | 1 | 4 | 0.0115 | 0.85 |
| 9 | 3739.38 | -8.46971e-15 | 1011.24 | 2809.34 | 1 | 4 | 0.0115 | 0.85 |
| 10 | 3731.88 | 2.82066e-14 | 1003.75 | 2809.07 | 1 | 4 | 0.0115 | 0.85 |
| 11 | 3724.38 | 7.99178e-14 | 996.252 | 2808.81 | 2 | 4 | 0.0115 | 0.85 |

You can easily toggle between the two input methods at the top of the ‘Wells and Perforations’ tab in the builder. If using the ‘one-at-a-time’ method, we recommend also checking the option ‘Specify perforation clusters by measured depth…”, also located at the top of the ‘Wells and Perforations’ tab.

### 9.1.3 Stage length sensitivity

For a set lateral length, longer stages result in fewer total stages and less time on location, which can reduce completion cost. In general, operators seek to maximize their stage length so long as production/stimulation performance is not impaired. Performance impairments could be caused by diminished perforation friction (limited entry), low flow-rates in the toe of the stage causing screen outs, or poor distribution of sand.

**Setup considerations:**

- Mesh size
  The same mesh resolution used during the history match should be used for all sensitivity analyses. If simulating longer stages than the history match, increase the number of matrix elements; and similarly, if simulating shorter stages, decrease the number of matrix elements.

  For example, if you history matched to two 250 ft stages (500 ft total matrix width) using 20 matrix elements in the Shmin direction (so 25 ft wide elements), and now you are sensitizing on the effect of 300 ft stages (600 ft total matrix width), you should have 24 matrix elements in the Shmin direction (you can leave the SHmax direction unmodified). Sometimes the math isn’t that clean, but do your best to keep element width the same between cases!

- Cluster spacing
  It is best to maintain the same cluster spacing as your “base case” such that you are only testing the impact of one variable (stage length) at a time. To this end, it is best to sensitize stage length in increments of cluster spacing. For example, if your base cluster spacing is 20 ft, sensitize for stage lengths of 240, 260, 280, 300, 320, etc. Note that it is not necessary to do *every* stage length increment, and for the sake of time you may choose to simulate 240, 300, 360 - or some other sampling scheme.

- Limited entry or perforation friction
  Limited entry or perforation friction is the amount of “back pressure” generated by the perforations. Conceptually, the smaller the area the fluid has to exit the casing, the higher the pressure is needed to force it out at a given rate. While high perforation friction (“back pressure” and “$p_{pf}$“ below) may require more horsepower at the surface, it has the advantage of more evenly distributing fluid across all clusters in a stage.

$$P_{pf} = \Delta P_p = 0.2369\ Q^2 \frac{\rho}{{N_p}^2\ D_p^4\ {C_d}^2}$$

$P_{pf}$ = Perforation Friction (psi)
$\Delta P_p$ = Pressure drop across a perforation(s) (psi)
$Q$ = Total flow rate (bbl/min)
$\rho$ = Density of fluid (lb/gal)
$N_p$ = Number of open perforations
$D_p$ = Diameter of perforations (in)
$C_d$ = Coefficient of discharge

  When setting up your sensitivity cases, keep this concept of perforation friction in mind:

  - Often it is best to start with normalized perforation friction across all cases. So, if your base case has an injection rate of 80 bpm and 8 clusters per stage with four 0.4” holes per cluster, your base perforation friction is (assuming fluid density of 10 lb/gal and coefficient of discharge of 0.85:

$$p_{pf} = 0.2369 * 80^2 * \frac{10}{(8*4)^2 * 0.4^4 * 0.85^2} = 800\ psi$$

    Then, if you want to sensitize on a stage that is 50% longer (ie 12 clusters per stage at the same spacing), you should **maintain the same perforation friction** by adjusting the number of perforations and/or the perforation diameter, such as using three perforations per cluster with a diameter of 0.377 in (which gives a perforation friction of 801 psi).

- Proppant settling in the wellbore
  As stages are extended longer and longer, the velocity of the slurry toward the toe of the stage decreases, increasing the propensity for proppant to settle in the wellbore. This can cause proppant to settle out and screen off the toe-section of the stage. Make sure you turn on this physics in the advanced section of the Proppant tab:

Include proppant settling in laterals 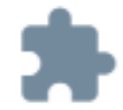

- Pumping schedule
  A pumping schedule is a combination of rates, proppant/fluid types, concentrations, and durations. Our objective is to normalize all parameters other than stage length. To do this, adjust the duration of each ramp in the pumping schedule by the proportional change to stage length. For example, if you make a stage twice as long, you should make the duration of each ramp in your injection schedule twice as long. This ensures that the same amount of proppant and fluid is pumped per unit lateral.

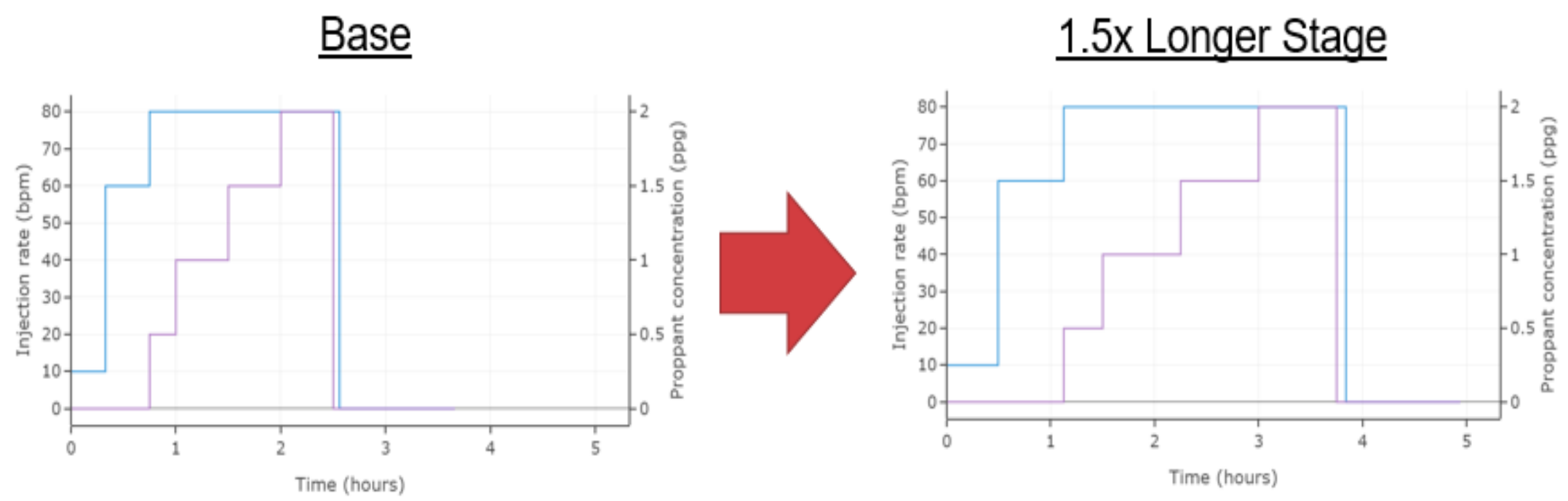

If the example above, where stage length is doubled, we don't recommend doubling the injection rate (which would be another way to maintain the same proppant/fluid per unit lateral). If rate is increased, then the perforation friction will also be impacted - and it would be easy to quickly exceed practical limitations.

**Useful ResFrac features:**

- Stage setup wizard
  The well stages setup wizard (found on the Wells and Perforations tab) is very helpful for setting up simulations with different stage lengths:

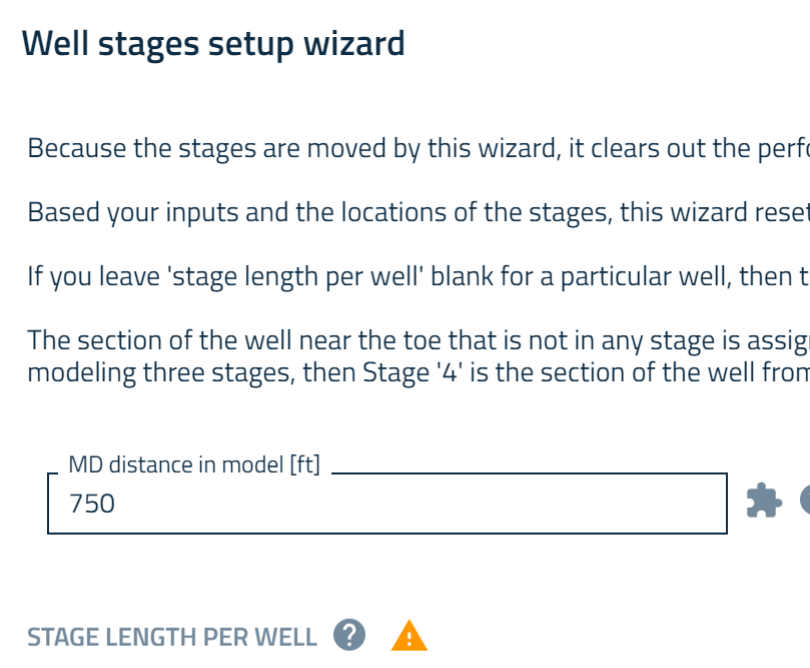

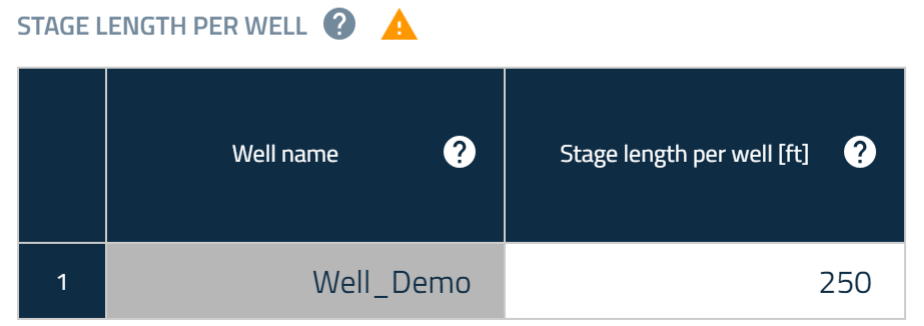

| | Well name | Stage length per well [ft] |
|---|---|---|
| 1 | Well_Demo | 250 |

After running the wizard to adjust stage length, make sure to recheck your clusters per stage (if using the inline cluster method), respecify cluster MDs (if specifying clusters one-at-a-time), and the mesh resolution on the meshing tab (ie, maintain same matrix block size).

- In-line versus one-at-a-time cluster specification

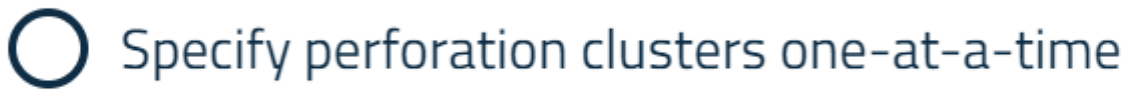

Specify perforation clusters one-at-a-time

Specify using 'inline' method (e.g., number of clusters per stage)

There are a couple ways to enter a completion (cluster/perforation) design in ResFrac. When changing stage lengths, the “inline” cluster specification method will be easiest. The inline method evenly distributes clusters along the stage and assumes every cluster has the same perforation design.

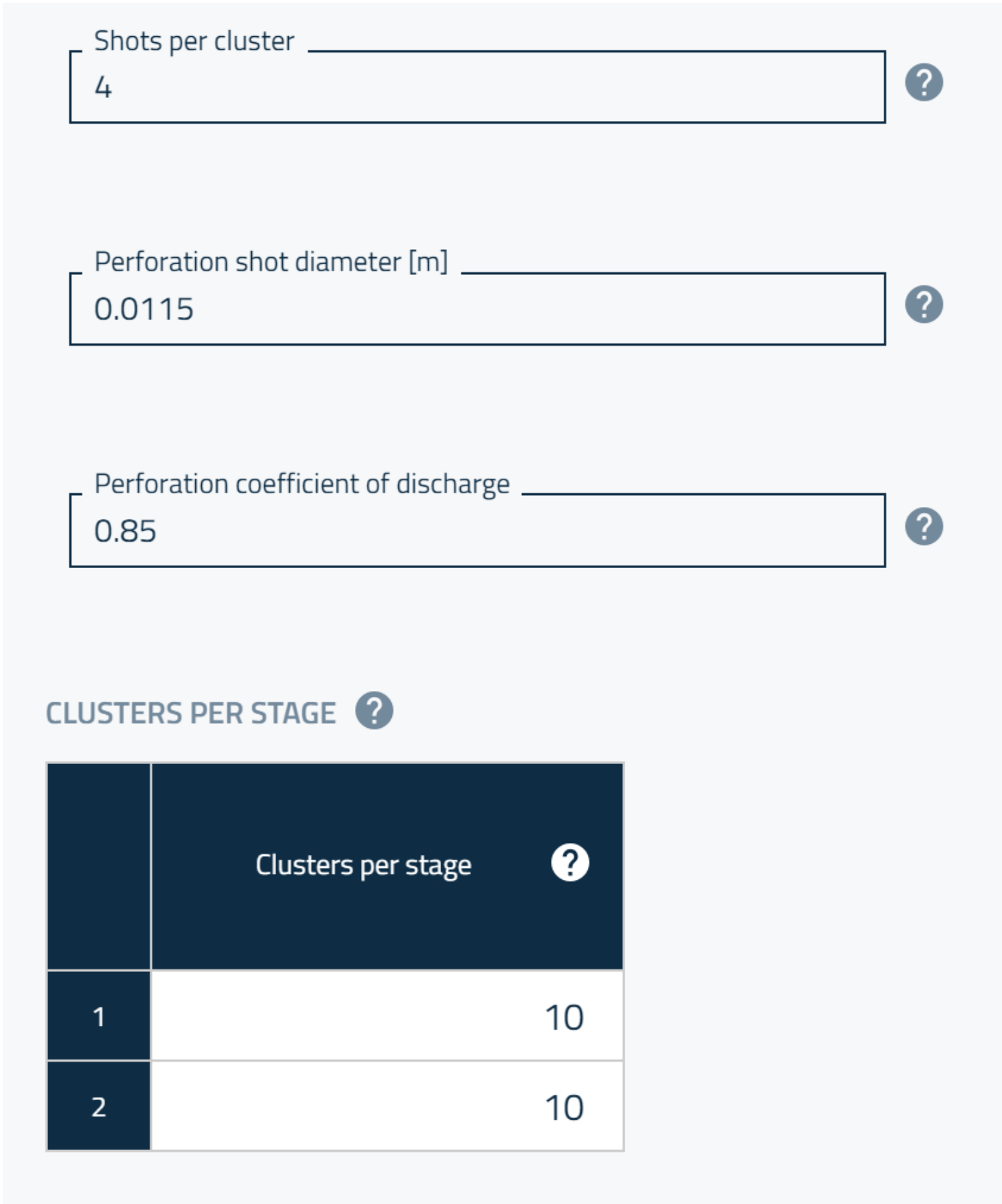

The “one-at-a-time” method allows for uneven spacing of clusters along a stage and/or perforation designs that vary by cluster. The disadvantage of this method is that it requires you to redefine your perf depths each time you run the stages setup wizard:

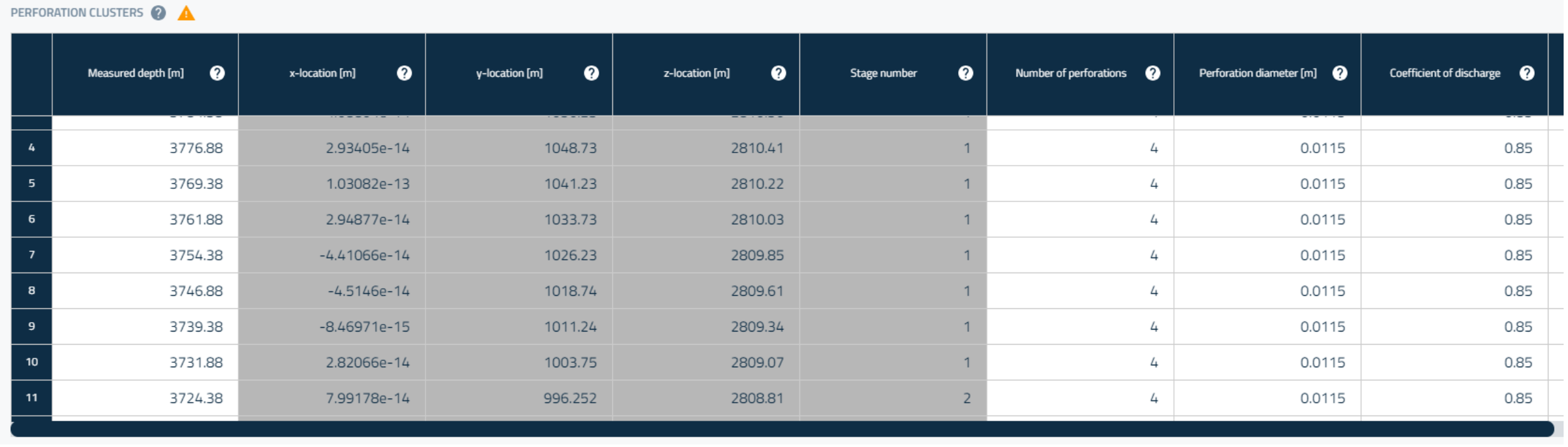
PERFORATION CLUSTERS

| | Measured depth [m] | x-location [m] | y-location [m] | z-location [m] | Stage number | Number of perforations | Perforation diameter [m] | Coefficient of discharge |
|---|---|---|---|---|---|---|---|---|
| 4 | 3776.88 | 2.93405e-14 | 1048.73 | 2810.41 | 1 | 4 | 0.0115 | 0.85 |
| 5 | 3769.38 | 1.03082e-13 | 1041.23 | 2810.22 | 1 | 4 | 0.0115 | 0.85 |
| 6 | 3761.88 | 2.94877e-14 | 1033.73 | 2810.03 | 1 | 4 | 0.0115 | 0.85 |
| 7 | 3754.38 | -4.41066e-14 | 1026.23 | 2809.85 | 1 | 4 | 0.0115 | 0.85 |
| 8 | 3746.88 | -4.5146e-14 | 1018.74 | 2809.61 | 1 | 4 | 0.0115 | 0.85 |
| 9 | 3739.38 | -8.46971e-15 | 1011.24 | 2809.34 | 1 | 4 | 0.0115 | 0.85 |
| 10 | 3731.88 | 2.82066e-14 | 1003.75 | 2809.07 | 1 | 4 | 0.0115 | 0.85 |
| 11 | 3724.38 | 7.99178e-14 | 996.252 | 2808.81 | 2 | 4 | 0.0115 | 0.85 |

You can easily toggle between the two input methods at the top of the ‘Wells and Perforations’ tab in the builder. If using the ‘one-at-a-time’ method, we recommend also checking the option ‘Specify perforation clusters by measured depth…”, also located at the top of the ‘Wells and Perforations’ tab.

### 9.1.4 Proppant loading sensitivities

It is commonly observed that larger frac jobs (more proppant, more fluid) produce more oil or gas. However, there are also diminishing returns. The benefit from pumping larger volumes is a function of geology, well spacing, completion design, and presence of “no-go” zones (zones that you do not want to frac into). With diminishing production returns, the optimal loading (proppant + fluid) is typically somewhere less than the maximum volume capable of being pumped.

**Setup considerations:**

The golden rule of sensitivity analyses is to only vary one parameter at a time. When investigating proppant and/or liquid loading sensitivities there are several ‘one parameter’ schemes to consider:

- Different proppant loads (total lbs) with the same liquid load (i.e., constant liquid load)
- Different liquid loads (total gals) with the same proppant load (i.e., constant proppant load)
- Vary proppant and liquid loads together (i.e., constant concentration)
- Keep proppant load the same, but change the pad or chase fluid volumes
- Keep proppant load the same, but ramp proppant concentration faster/slower

There are nearly limitless possibilities and all investigations can yield valuable insights into well/pad performance; however, before embarking on sensitivities it behooves you to think through each of these options before setting up your simulation cases and be consistent when setting up the sensitivity.

- Pumping schedules, well controls, and timing
  Pumping schedules in the Well Controls tab is where most of the changes will be made. Depending on whether you are following a constant proppant, constant liquid, or constant concentration sensitivity (or any other sensitivity), you will want to modify a different component of the pump schedule.

  If we were testing the effect of increasing the proppant load, there are several ways we could conceivably do so - some at a constant fluid load, some not. Below are examples of three different strategies that all increase proppant load by 50%.
  A - Base case
  B - Same proppant concentration as base case. This is implemented by multiplying the duration of each ramp stage by 1.5.
  C - Keep the same proppant ramping as the base case but extend the highest concentration ramps such that total proppant load is 150% of the base case.
  D - Same fluid load as the base case, but the concentration of each proppant ramp increased by 50%.

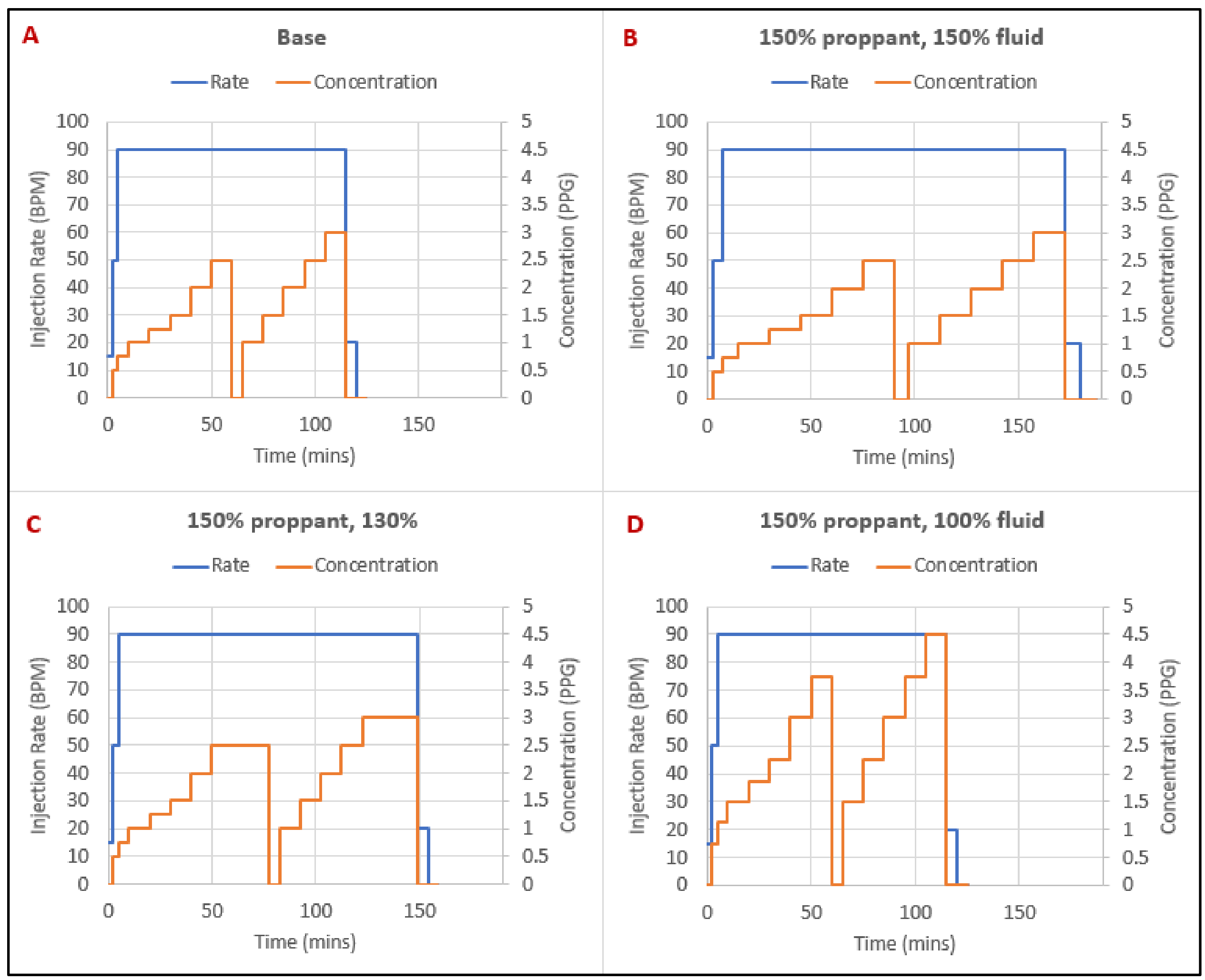

No single strategy is more correct than another; however, the crucial consideration is to be consistent in all sensitivities. If we want to test 50%, 75%, 100%, 125%, and 150% of base proppant load, we should be consistent in how we set up the simulations. For instance, if choosing to keep fluid volume constant and change proppant loading by modifying concentrations we would setup three cases as follows:

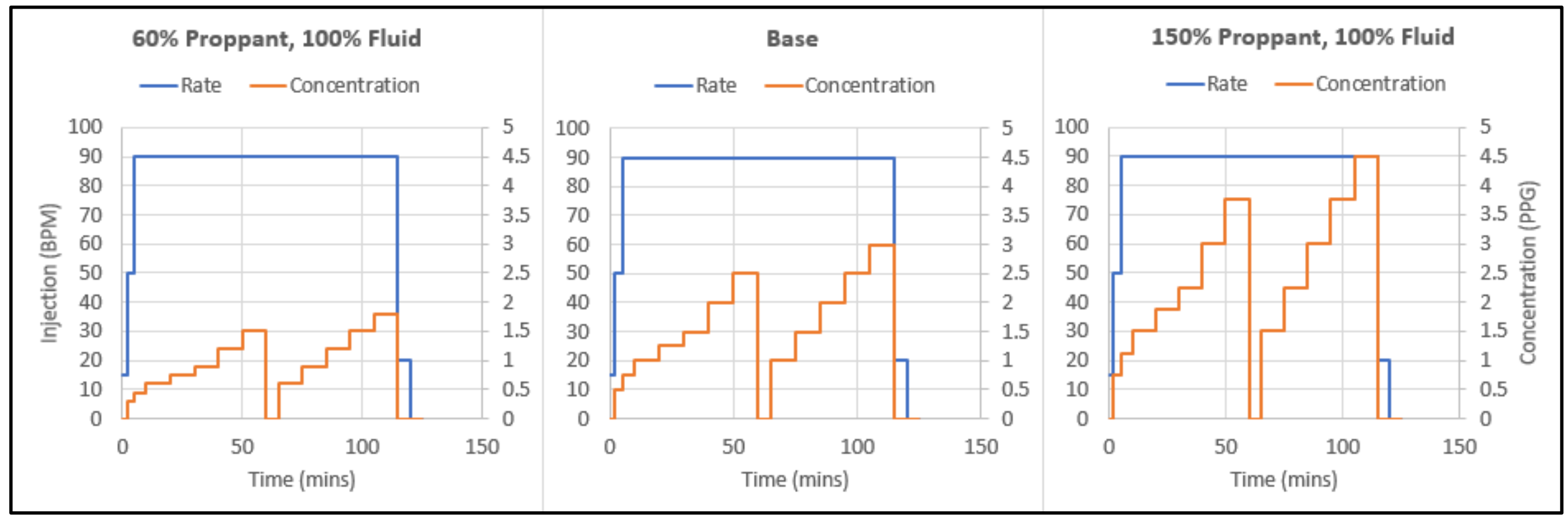

Do note that the high proppant cases reaches high concentrations, so if pursuing this strategy, make sure this is feasible in the field. Alternatively, you might run the simulation and find that the design with the highest concentration results in a screenout. Screenouts can definitely happen in ResFrac simulations!

Alternatively, we could test the same proppant loading while keeping concentration constant, as done in the following schedules:

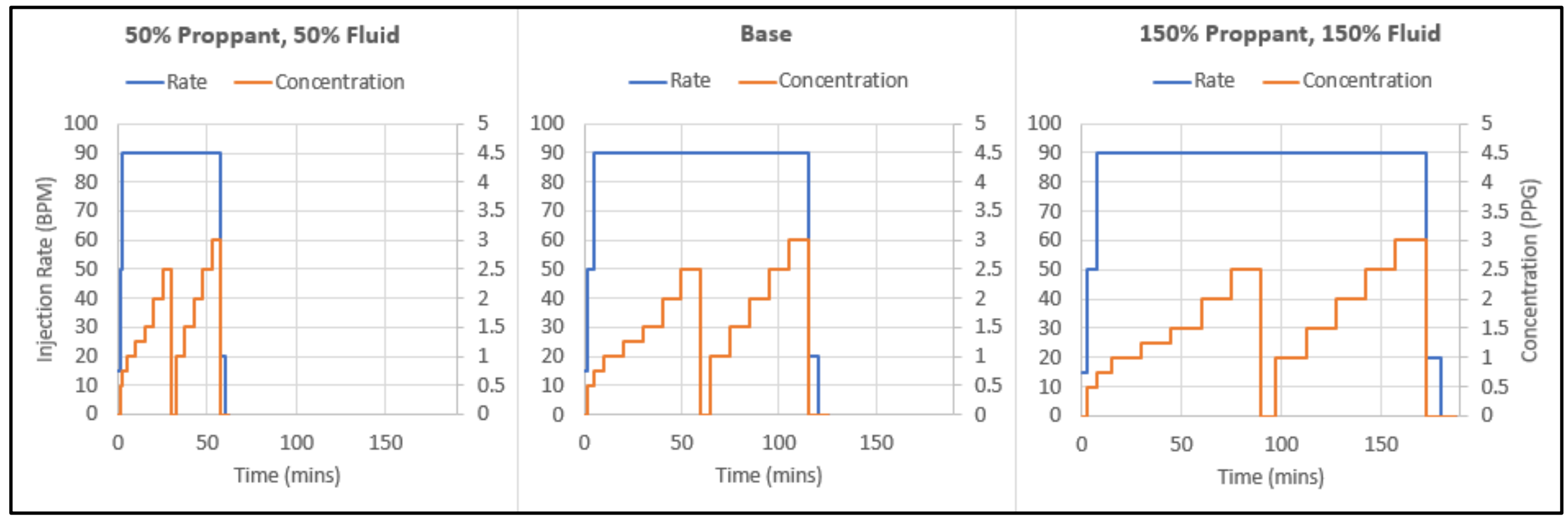

In this example, to achieve the desired variation of loading, the duration of each schedule segment is decreased or increased proportionally from the base case

Or, if we want to test the impact of fluid volumes while keeping the proppant load the same, then we could set up three sensitivities as follows:

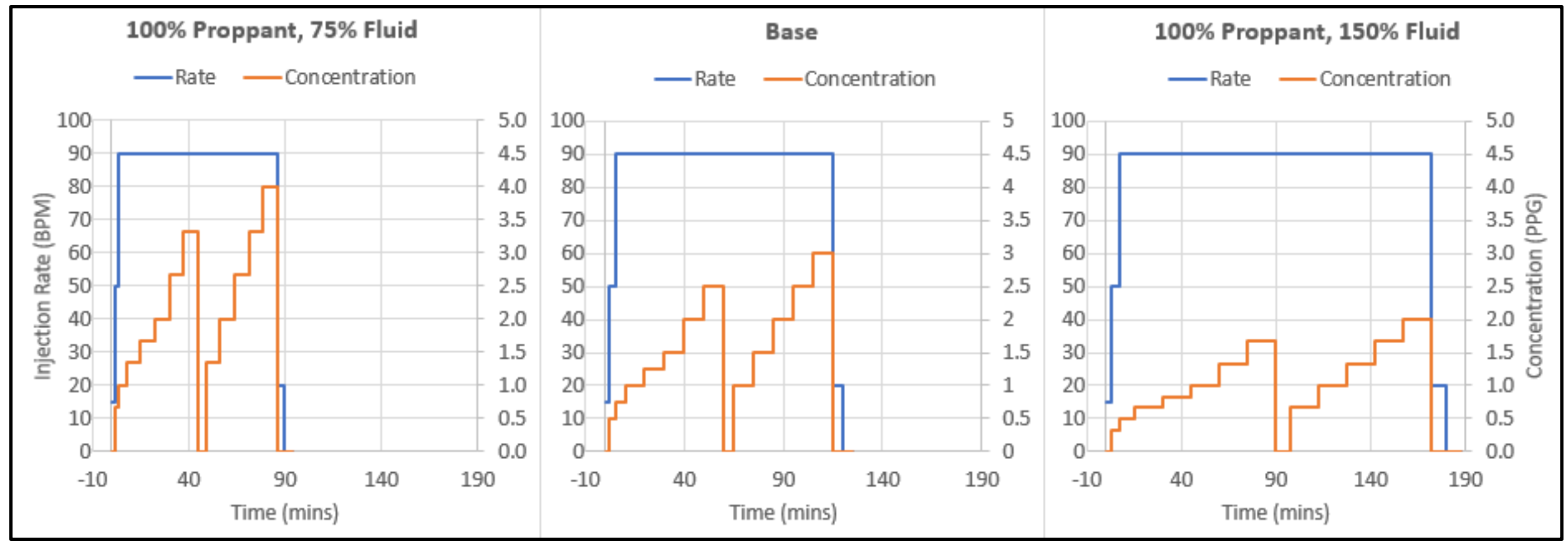

All three cases have the same volume of proppant pumped, but total fluid is changed by modifying proppant concentration.

There are also more nuanced investigations you may wish to pursue such as impact of pad/clean volumes or various schedules. As you do so, keep consistency in mind. For instance, to investigate ramping concentration faster. In the pump schedule below, we keep total proppant volume the same, but ramp concentration faster. This necessarily means that both the ramp schedule and the total fluid volume are different, so while this may be an insightful analysis, when analyzing remember that they may be two effects at play (volume and ramp speed):

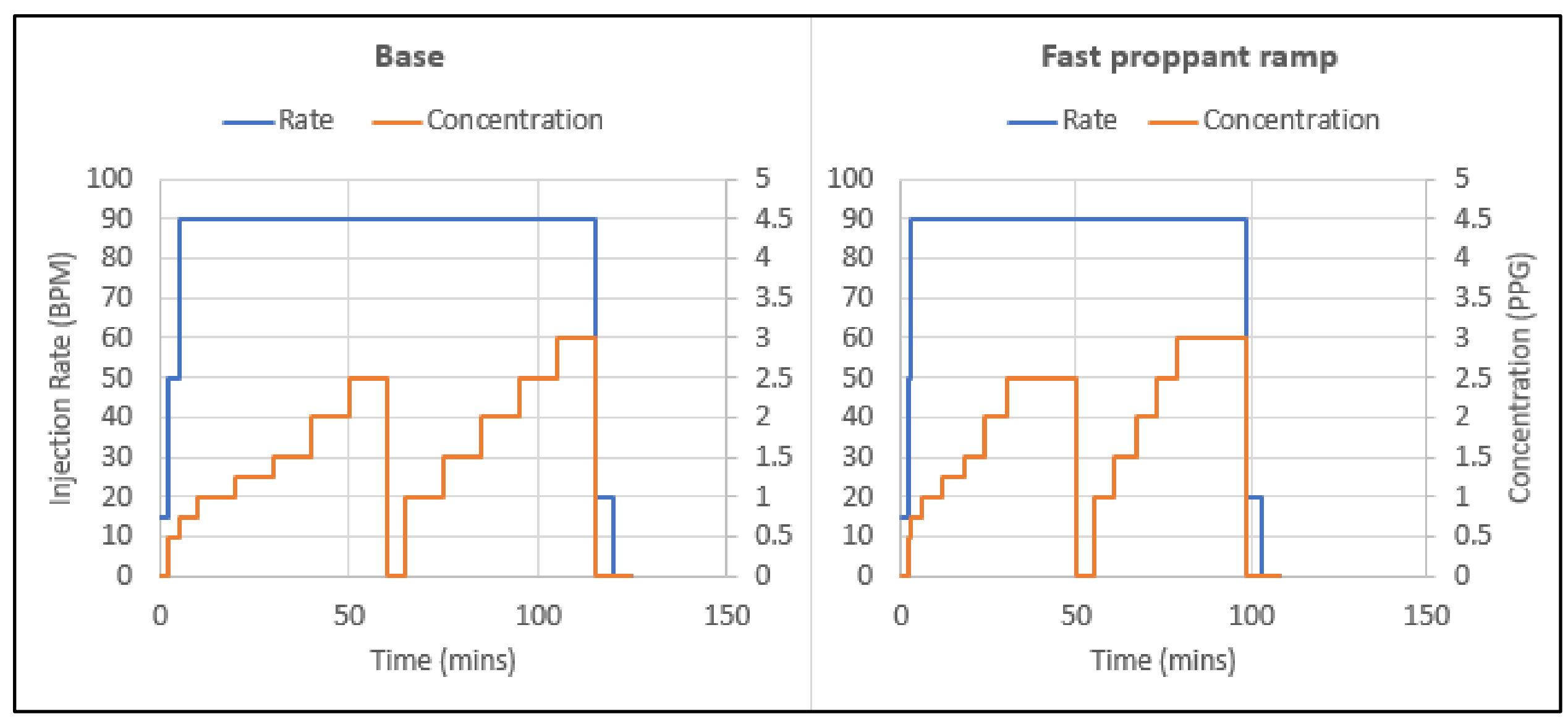

Moral of the story? Think through the effect you are trying to test ahead of time, then apply a consistent logic in generating your scenarios!

- Number of wells and layout
  Often, you begin a sensitivity analysis from a history matched model, that model contains several wells. These wells will form the basis of your analysis. Other times, your history match model may only have a single well, in which case, you will need to add additional wells to test the impact on multi-well setups. In either circumstance, once a layout is established, it is best to maintain the same layout for all sensitivities.

- Mesh size
  The same mesh resolution used during the history match should be used for all sensitivity analyses. If simulating larger well spacing, it may be necessary to expand the matrix - but you should do so using the same element size.

  For example, if you history matched with a matrix mesh 2000 ft wide (in the direction of SHmax), made up of 40 elements, then that means each element is 50 ft wide. If you then expand the matrix region, to accommodate more wells, to 5000 ft, then you should increase the number of elements to 100 (to maintain the same 50 ft element width).

- Cluster spacing
  It is best to maintain the same cluster spacing, number of clusters, perforation design, etc. as your base case.

**Useful ResFrac features:**

- Unit selection

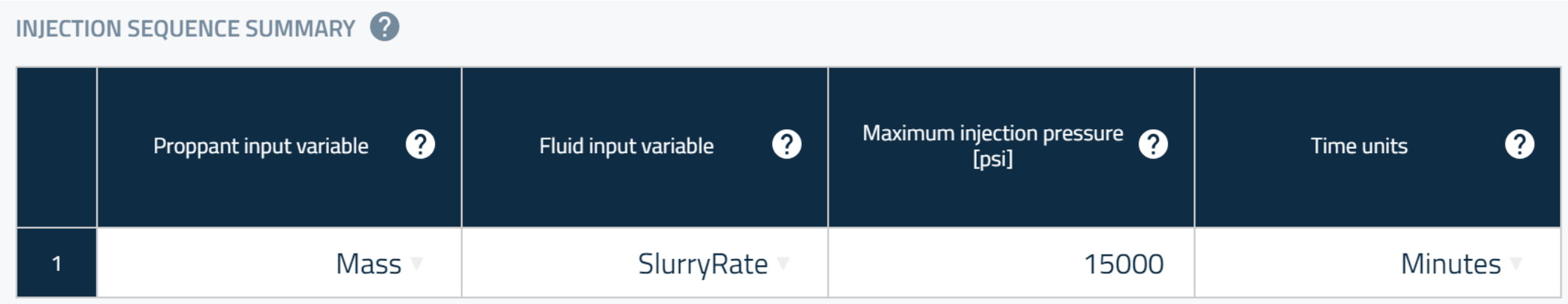

INJECTION SEQUENCE SUMMARY

| | Proppant input variable | Fluid input variable | Maximum injection pressure [psi] | Time units |
|---|---|---|---|---|
| 1 | Mass | SlurryRate | 15000 | Minutes |

After selecting 'InjectionSequence' as a Control Sequence Type in Well Controls, a table will appear allowing you to select units for the input table below. This is useful as you can match the units that are in the frac report or post-job report that you are using to populate the fracturing sequence.

After inputting the control sequence in, changing the units in the unit table will dynamically change the control table below. This can be helpful for modifying sensitivities. For example, the base case you are working from may have been populated using proppant pounds per gallon concentration (PPG). Subsequently, you can change the units to 'Mass', and the control table will show the proppant mass of each control step. This can help if you are trying to set up specific proppant mass targets (like 1000 lb/ft, 2000 lb/ft, and 3000 lb/ft).

- Excel
  We recommend modifying pump schedules in Excel. This has two advantages:
  1) the formulas in Excel let you manipulate schedules quickly and accurately
  2) the Excel sheet then provides a collective record of all pumping schedules you have created

### 9.1.5 Fracture sequencing sensitivities

Should you zipper wells or frac sequentially? Is there an optimal zippering sequence? Does simultaneous fracturing improve or hinder production? There are many different fracturing schemes out there. Sometimes the sequence matters, sometimes it does not. Below are some considerations for testing in a ResFrac model.

**Setup considerations:**

The golden rule of sensitivity analyses is to only vary one parameter at a time. In this vein, the only parameter to change should be the sequencing of fracs in Well Controls.

- Number of wells and layout
  Often, you begin a fracture sequencing sensitivity from a history matched model that contains several wells. These wells will form the basis of your analysis. Other times, your history matched model may only have a single well, in which case, you will need to add additional wells to test different sequencing options. In either circumstance, once a layout is established, it is best to maintain the same layout for all sensitivities.

- Mesh size
  The same mesh resolution used during the history match should be used for all sensitivity analyses. If simulating larger well spacing, it may be necessary to expand the matrix - but you should do so using the same element size.

  For example, if you history matched with a matrix mesh 2000 ft wide (in the direction of SHmax), made up of 40 elements, then that means each element is 50 ft wide. If you then expand the matrix region, to accommodate more wells, to 5000 ft, then you should increase the number of elements to 100 (to maintain the same 50 ft element width).

- Cluster spacing
  It is best to maintain the same cluster spacing and number of clusters as your base case.

- Pumping schedules, well controls, and timing
  We typically only model a portion of the lateral as in the image below. By using zero permeability cubes and external fractures (discussed in Section 9.1.1), modeling a portion of the lateral is representative of the larger system. When setting up fracture timing, keep in mind the physical setup of the model.
    - For example, if modeling sequentially frac'ing two wells, be sure to build in an adequate pause between the first well and the second well to account for the time it would take to frac the heel of the first well and toe of the second before getting to the modeled stages.

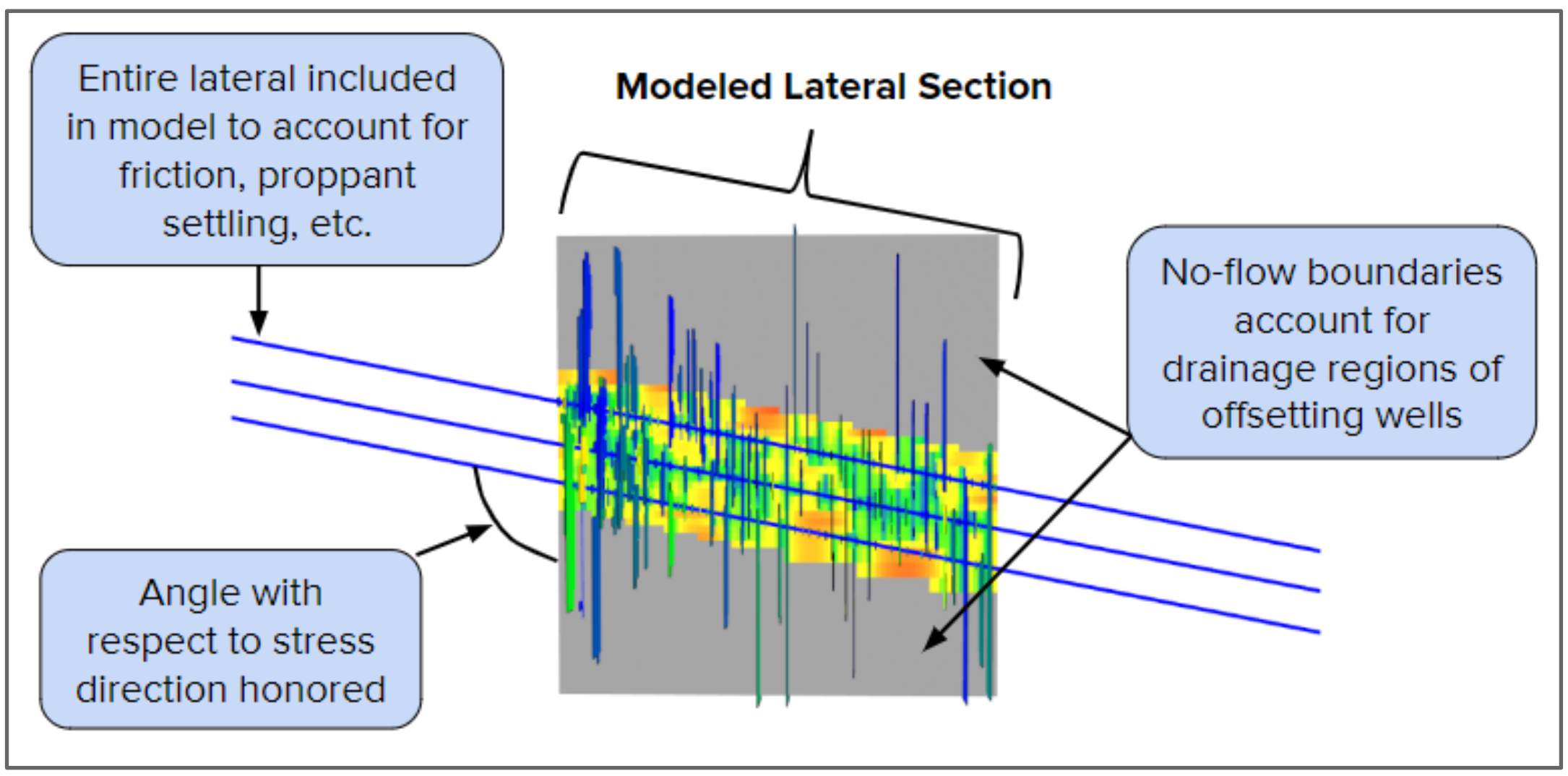

**Useful ResFrac features:**

- Duplicate well controls

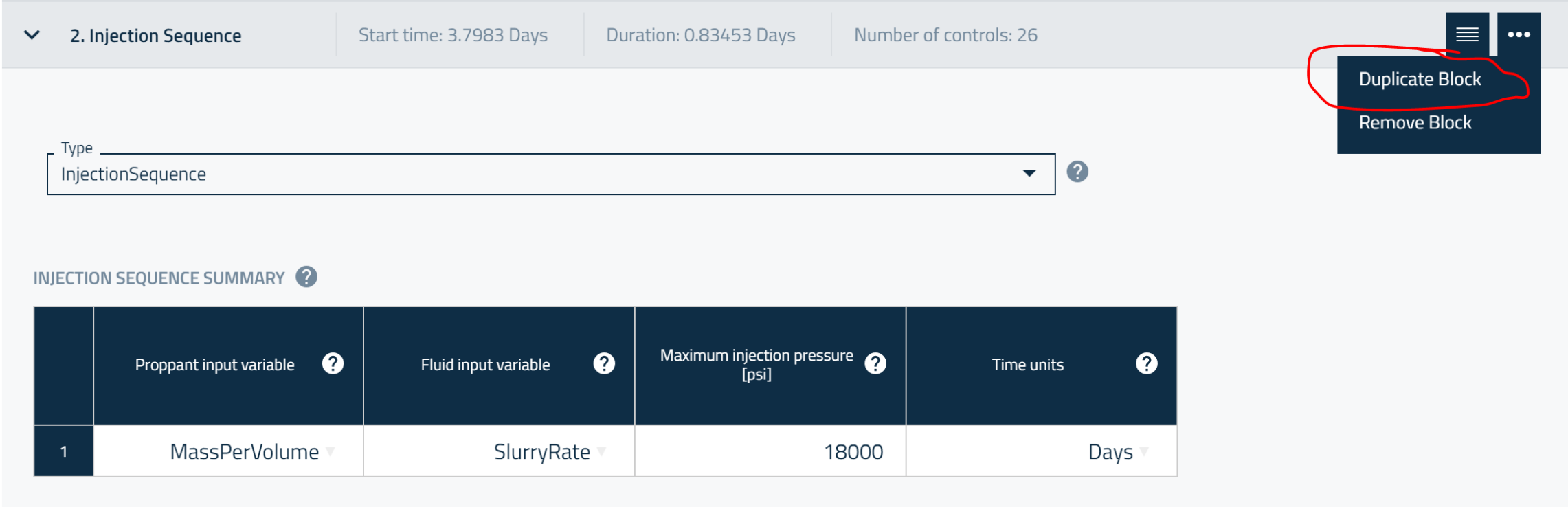

When populating controls for a new well, you can duplicate control sequence blocks from another well. This will save lots of typing and reduce the propensity for errors!

- External fractures
  Found on the ‘Wells and Perforations’ tab, External Fractures allow the user to specify a fracture *outside* of the matrix domain with a geometry and time-dependent net pressure (ie fracture net pressure can be zero until a specified time, peak to some value, then decay to a residual net pressure - just as stress shadowing observed in an offsetting stage).

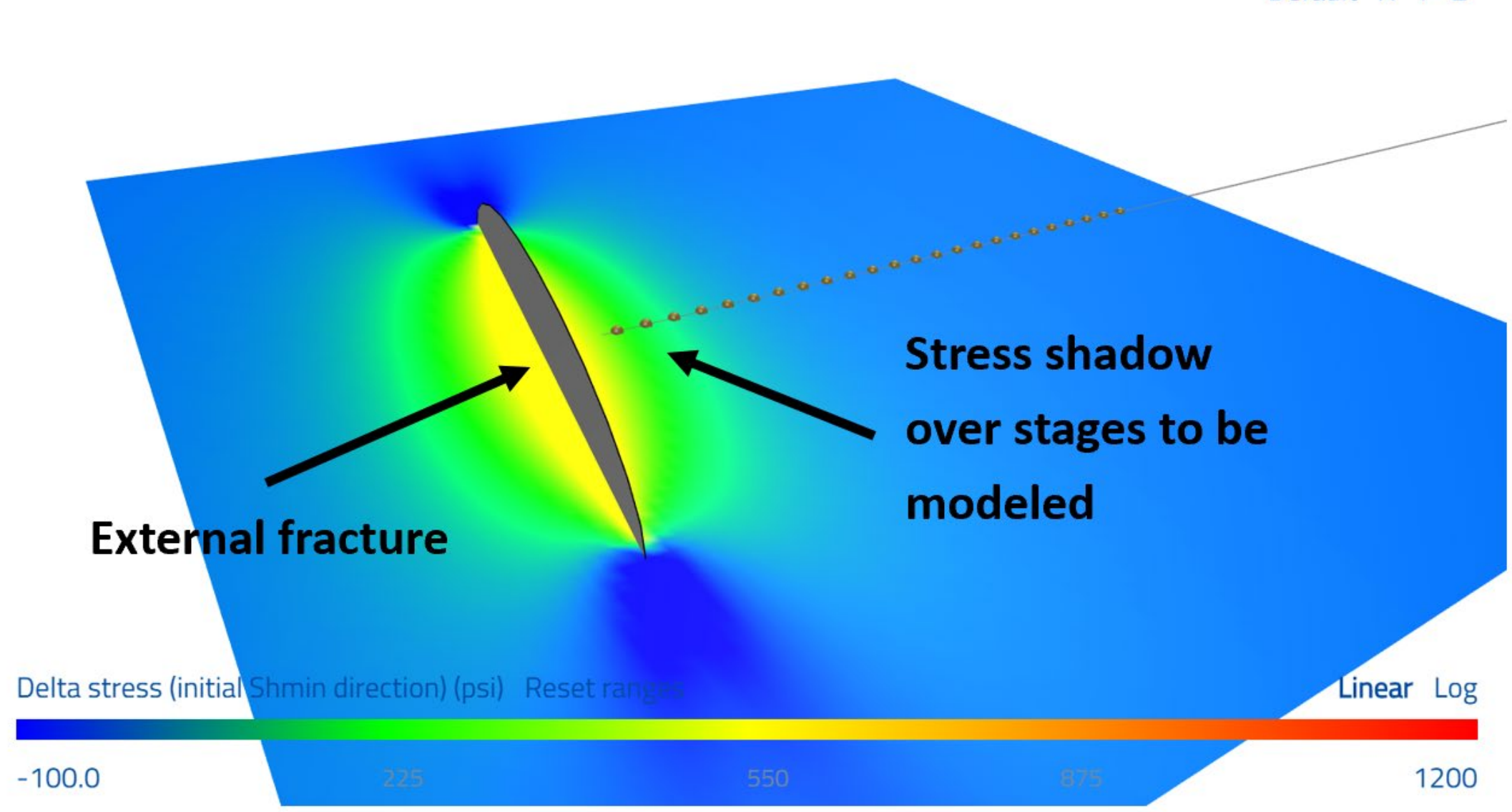

## 9.2 Evaluating sensitivities

The primary evaluation metric should be that which decisions are being made on. Often this is net present value (NPV), investment efficiency (net present value divided by spending), or internal rate of return (IRR). A chart of this evaluation metric versus the independent variable (like cluster spacing or well spacing) identifies where the optimum occurs. Then, a selection of additional plots and figures can be used to explain *why* the optimal scenario outperforms the others.

- NPV (or optimization objective) versus independent variable
  A chart of NPV versus your sensitivity variable is the best way to summarize a sensitivity in a single plot. Compile your simulation cases into excel or other plotting software, calculate your objective variable, and plot.

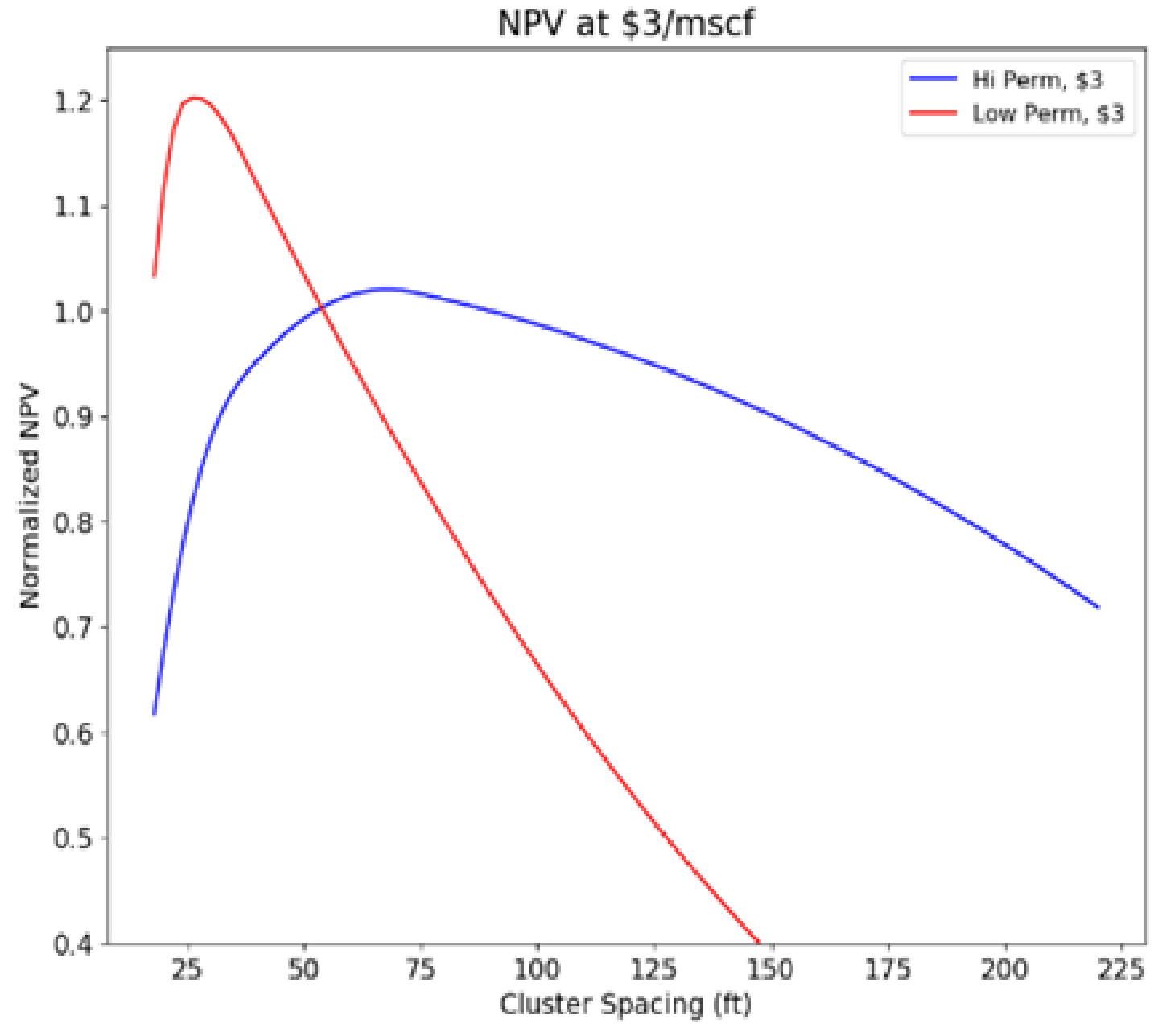

Fowler et al., SPE 195980

Ideally the range of sensitivities chosen results in a clear maximum/minimum. In the example above from Fowler et al., NPV maximums occur at 30 ft for the low permeability model and 80 ft for the high permeability model.

- Total production versus time
  This is the most common metric when comparing cases. Use the multiplot function in the ResFrac Job Manager to quickly plot multiple simulations on the same chart (see Section 5.12).

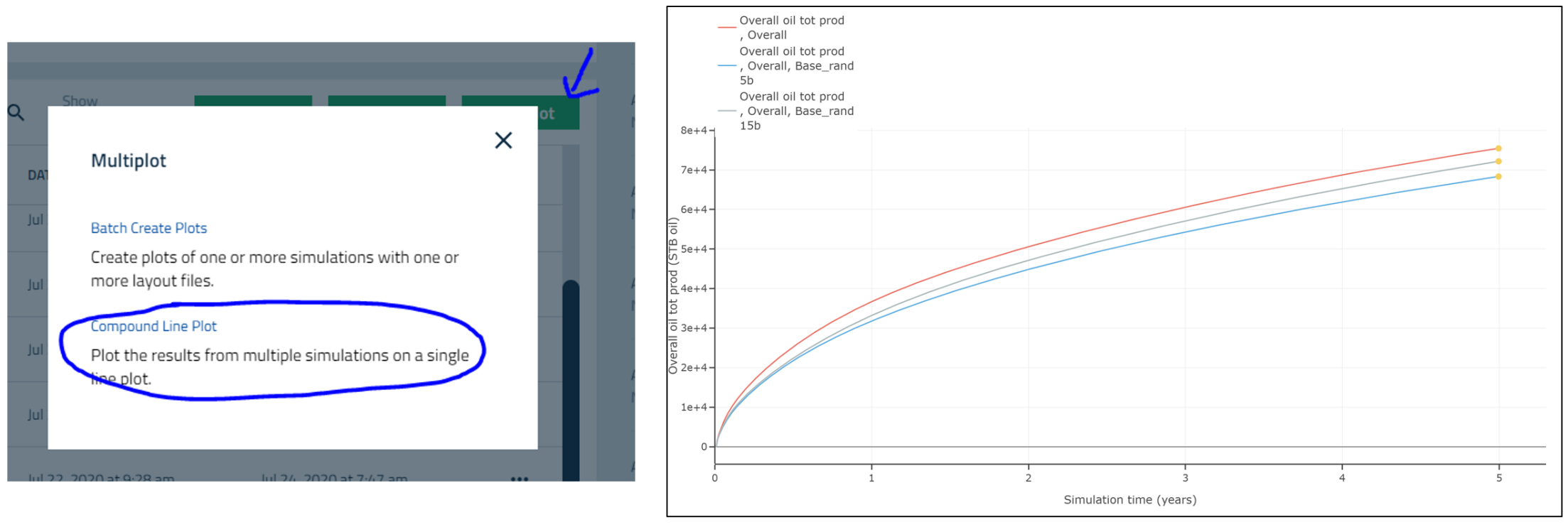

Once plotted, examine the trends. Are the various sensitivities evenly distributed or are there decreasing marginal returns beyond a point?

- RTA plots
  RTA plots are a great way to compare simulation cases. The slope of the RTA trend is inversely proportional to the product of effective fracture area and square root of permeability. Fowler et al. (2020a) discussed how RTA trends can deviate from linear for a variety of reasons. Examine your sensitivity cases. Are the initial slopes different? Is late time behavior different? Use the 'specialized plot' capability in the visualization tool to make RTA plots of production data.

In the image below, all three simulation cases have the same fracture area as observed by the same early time RTA trend. However, the green simulation case starts bending upward before the other two cases, indicating that production interference between clusters is occurring *earlier* and ultimately decreasing the late time production.

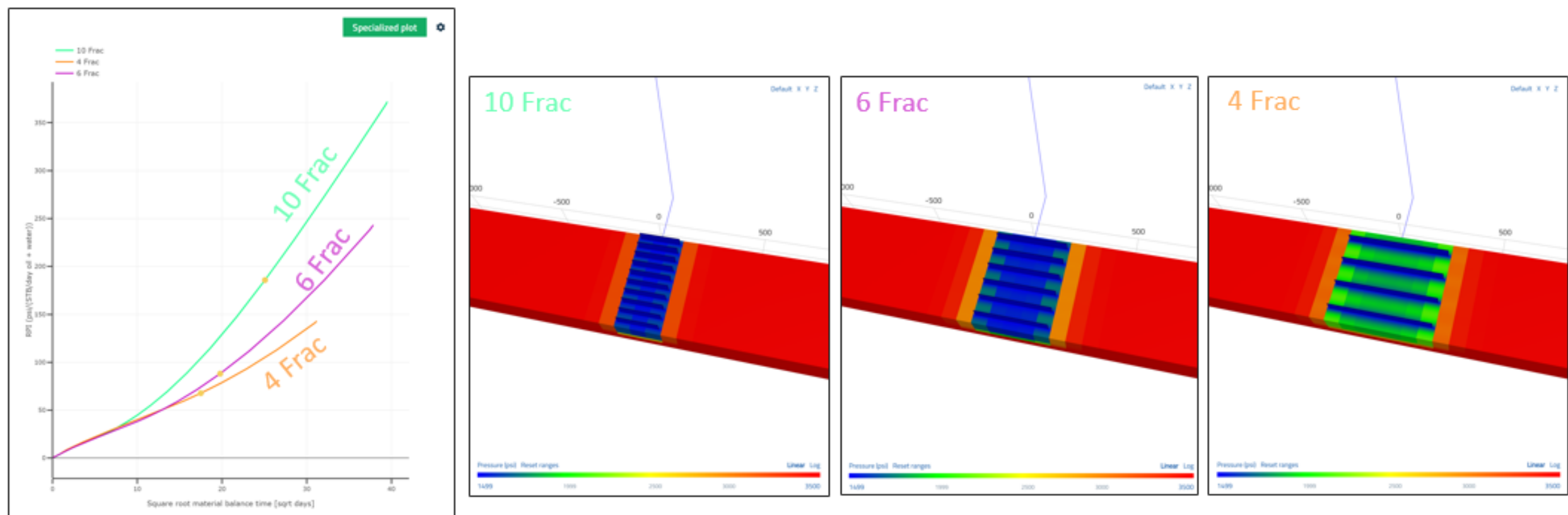

Keep in mind that less late time production is not always a bad thing (it is a question of NPV or ROR)!

If running a series of proppant sensitivities, look for differences in the initial RTA slope. Is more effective fracture area being created in one case versus another?

- Pressure depletion in the 3D viewer

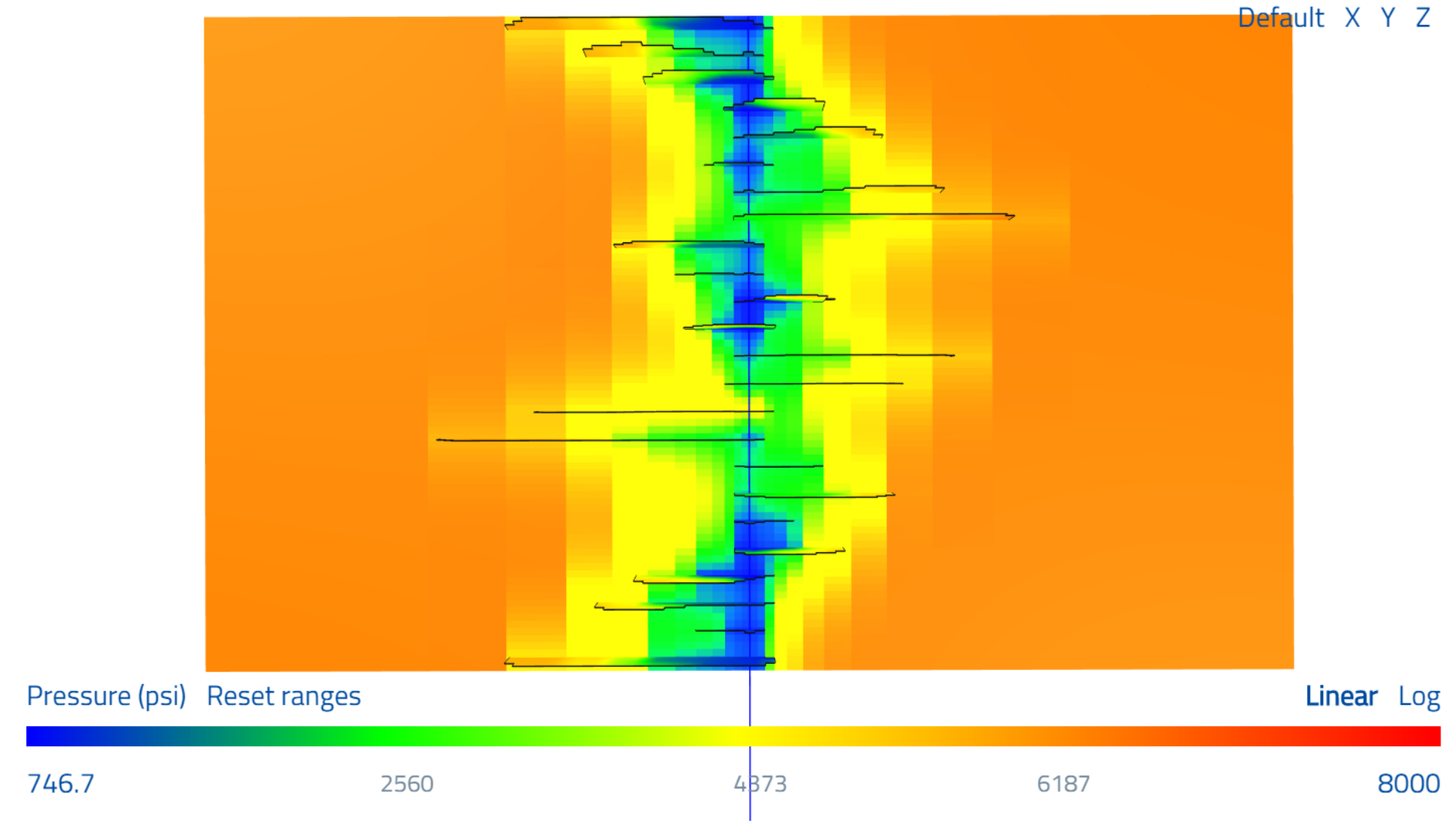

When looking at cluster spacing, ask are there gaps in pressure depletion between clusters? At what time do the pressure fronts from adjacent cluster intersect? If looking at well spacing or landing zone, look at pressure depletion over time. When or do depletion regions overlap (you can also take a vertical slice in the 3D viewer to see pressure in a cross-section)?

Increasing stage length can reduce the flow rate into any single cluster, which may impact how far proppant is carried. Does the width of depletion around the wellbore vary when stage length increased? Do any clusters screenout due to the low injection rate? Are there gaps in depletion along the wellbore?

Some level of interference is optimal. Where do the drainage regions of wells overlap (both horizontally and vertically)?

- Delta Shmin

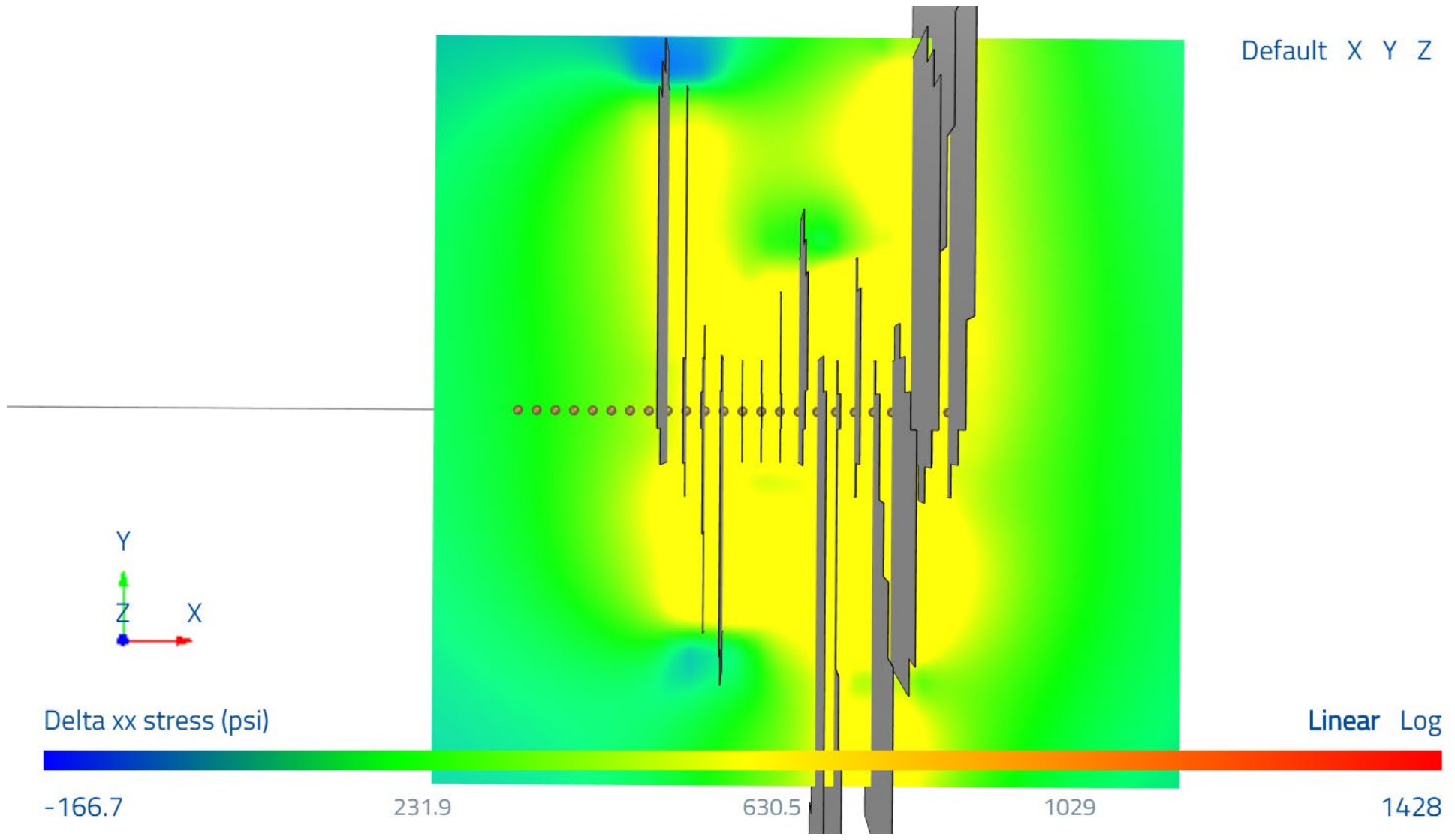

How does changing cluster spacing impact stress? Are there clusters that do not initiate because of stress shadowing? Do fracture networks from adjacent wells overlap and elevate stress in the region of intersection?

Make sure to look at timesteps during/right-after injection versus the end of time, as that is when stress will be highest.

- Superficial velocity
Superficial velocity is plotted in the 3D viz tool, and shows the velocity of the slurry in the wellbore:

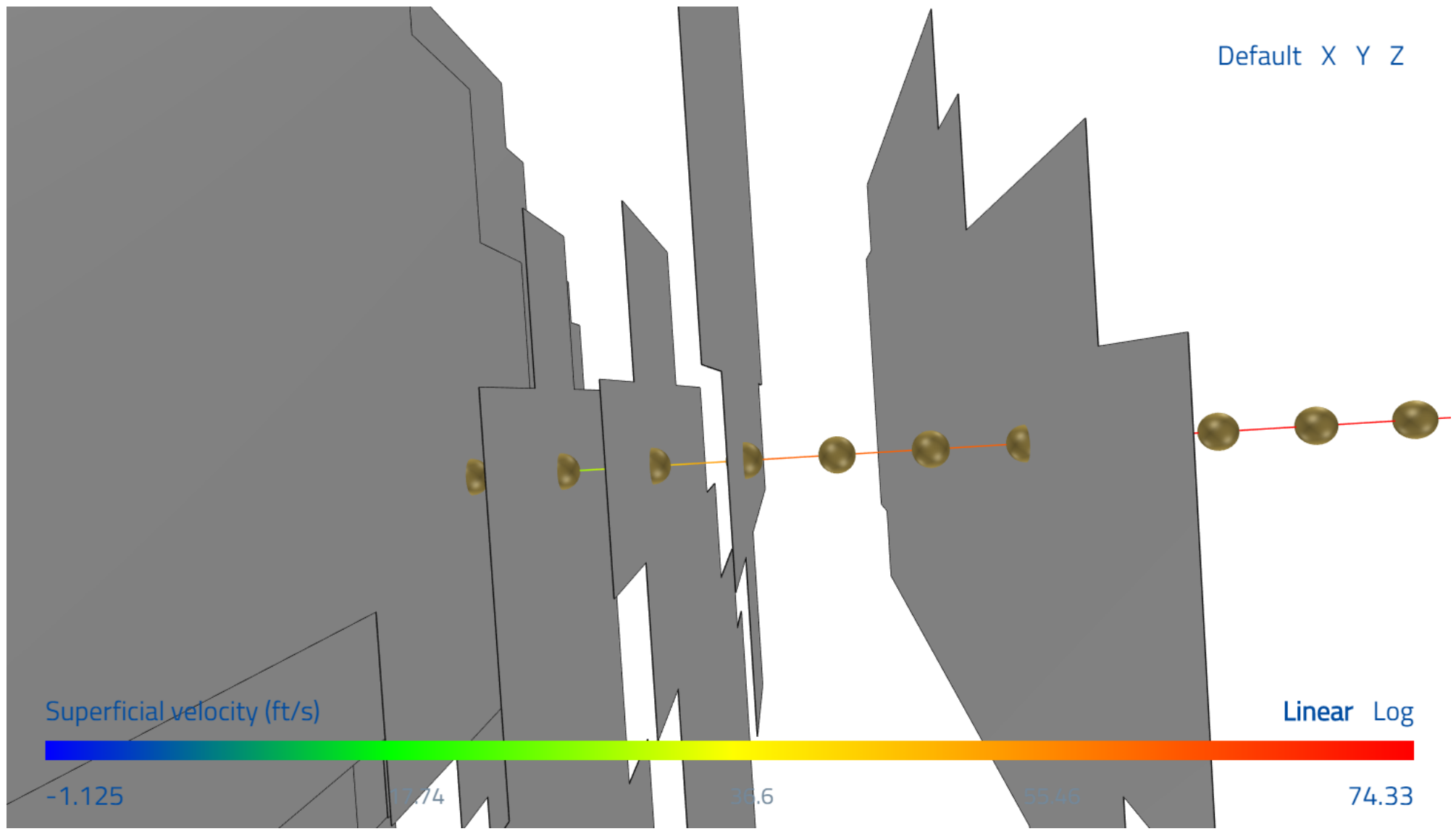

Is the velocity at the toe of the stage getting too low? Make sure to look at timesteps during injection versus the end of time.

- Proppant volume fraction
  If a cluster, or clusters, screenout, the proppant volume fraction will begin to build in the wellbore. In the image below, the proppant volume fracture in the wellbore is ~0.05 (dark blue/green). If a cluster/s screenout, you will see that volume fracture increase up to the maximum packing volume fraction, 0.66.

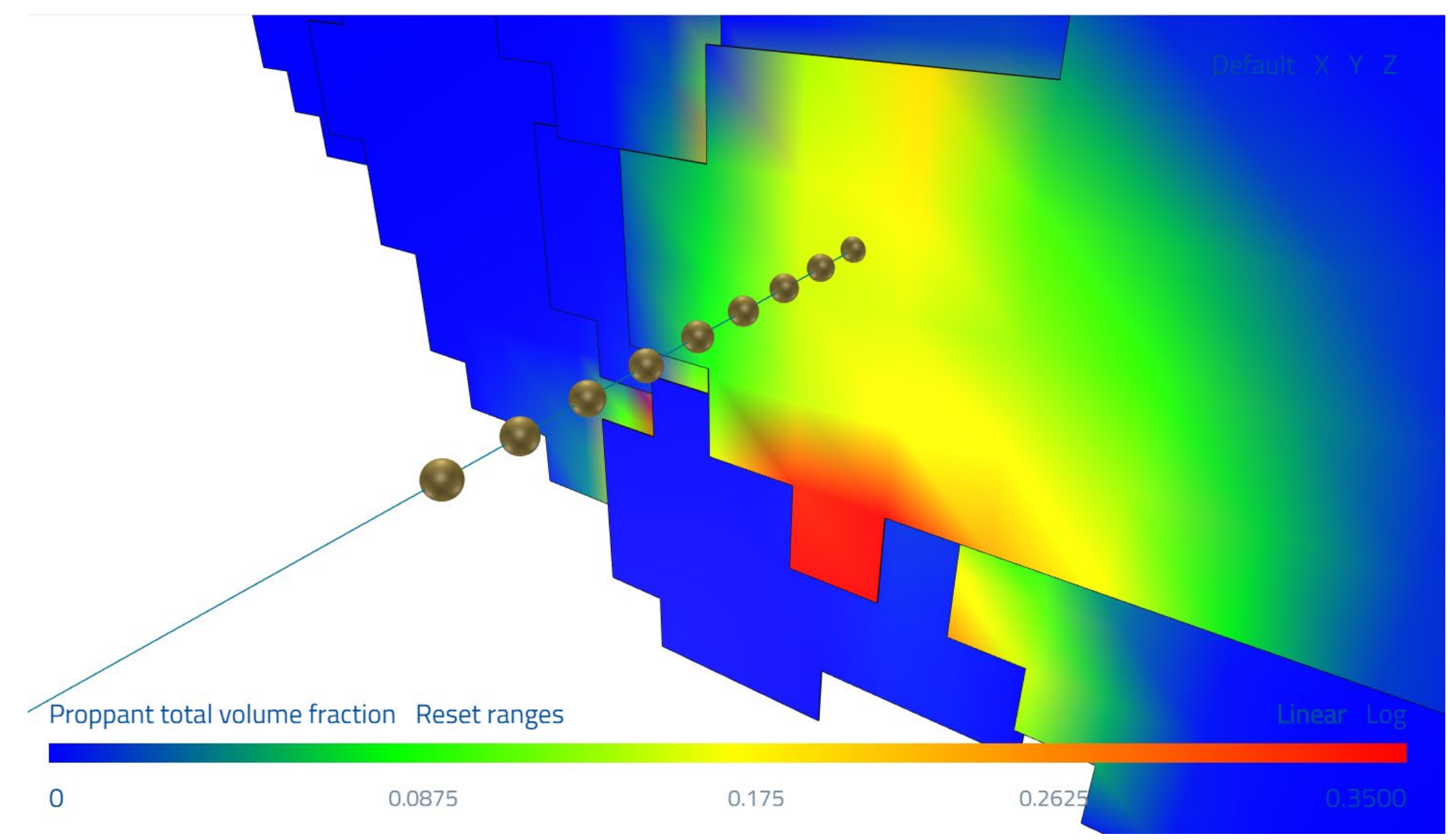

Make sure to look at timesteps during injection versus the end of time.

- Compare propped fracture area and Shmin profile. Are stress barriers confining fracs between stacked wells?

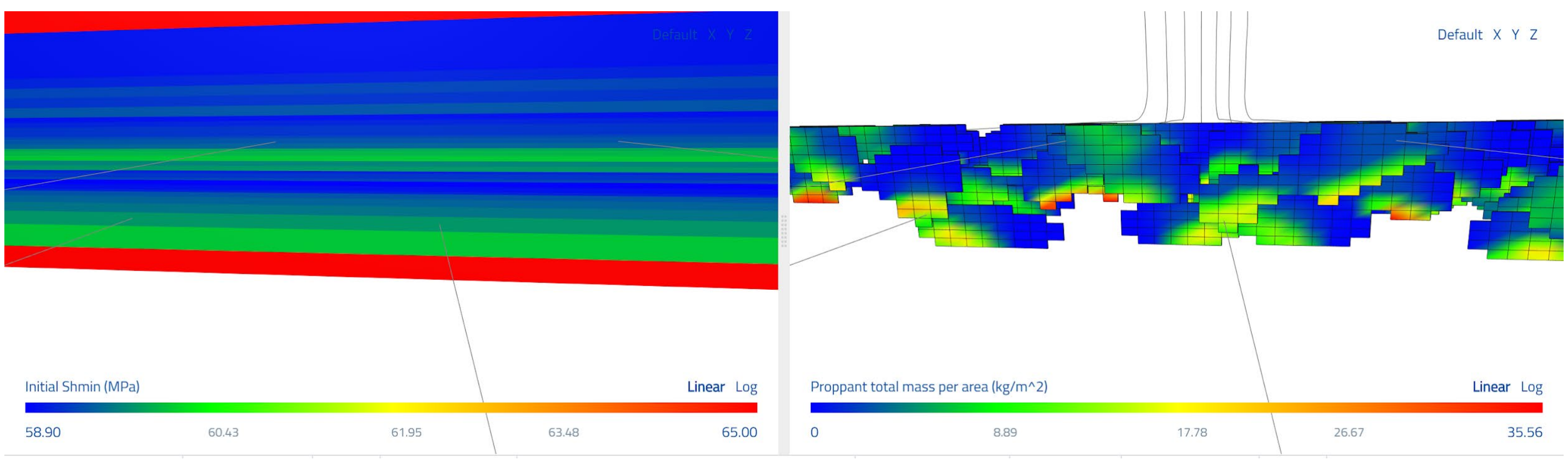

- 2D heat maps

  Sometimes it is hard to compare simulations in 3D and it's useful to get a 2D view of properties.

| | | | |
|---|---|---|---|
| frac_elms_16415.csv | 11/29/2020 8:32 AM | Microsoft Excel Com... | 750 KB |
| frac_elms_16436.csv | 11/29/2020 8:32 AM | Microsoft Excel Com... | 757 KB |
| frac_elms_16456.csv | 11/29/2020 8:32 AM | Microsoft Excel Com... | 757 KB |

  Cumulative production or propped area are two good properties to look at:

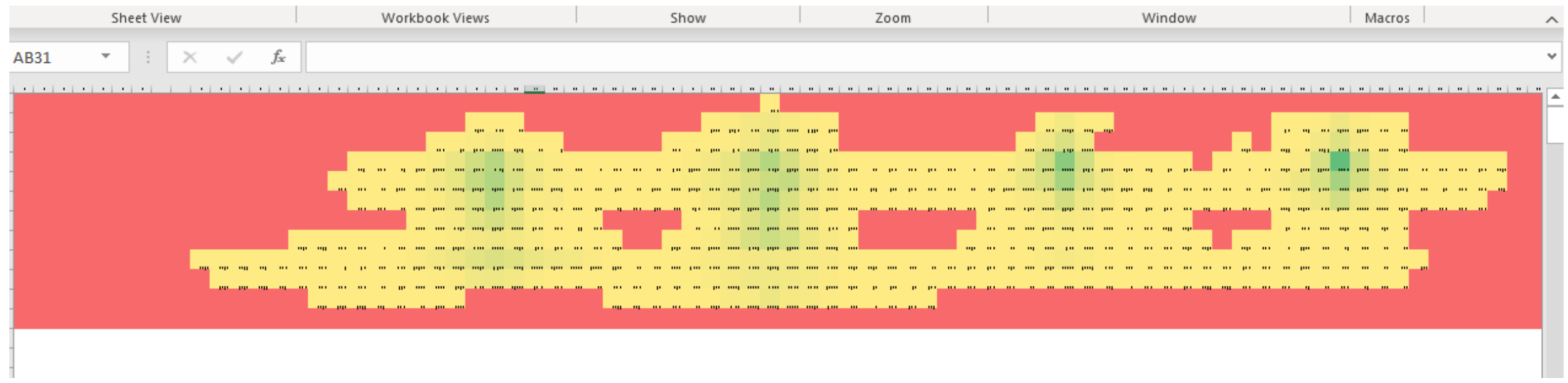

  Are fracture regions overlapping between wells? By how much? Are there any apparent frac barriers? Do wells in different landing zones produce more or less than others?

- Similar to the frac_elems_XXX.csv files, the frac_v_depth.csv files can be used to make 1D summaries of model outputs.

| | | | |
|---|---|---|---|
| frac_v_depth_12942.csv | 11/27/2020 12:59 PM | Microsoft Excel Com... | 5 KB |
| frac_v_depth_14097.csv | 11/29/2020 8:29 AM | Microsoft Excel Com... | 5 KB |
| frac_v_depth_14203.csv | 11/29/2020 8:30 AM | Microsoft Excel Com... | 5 KB |

  For example, you might compare total propped area versus depth for several simulation cases:

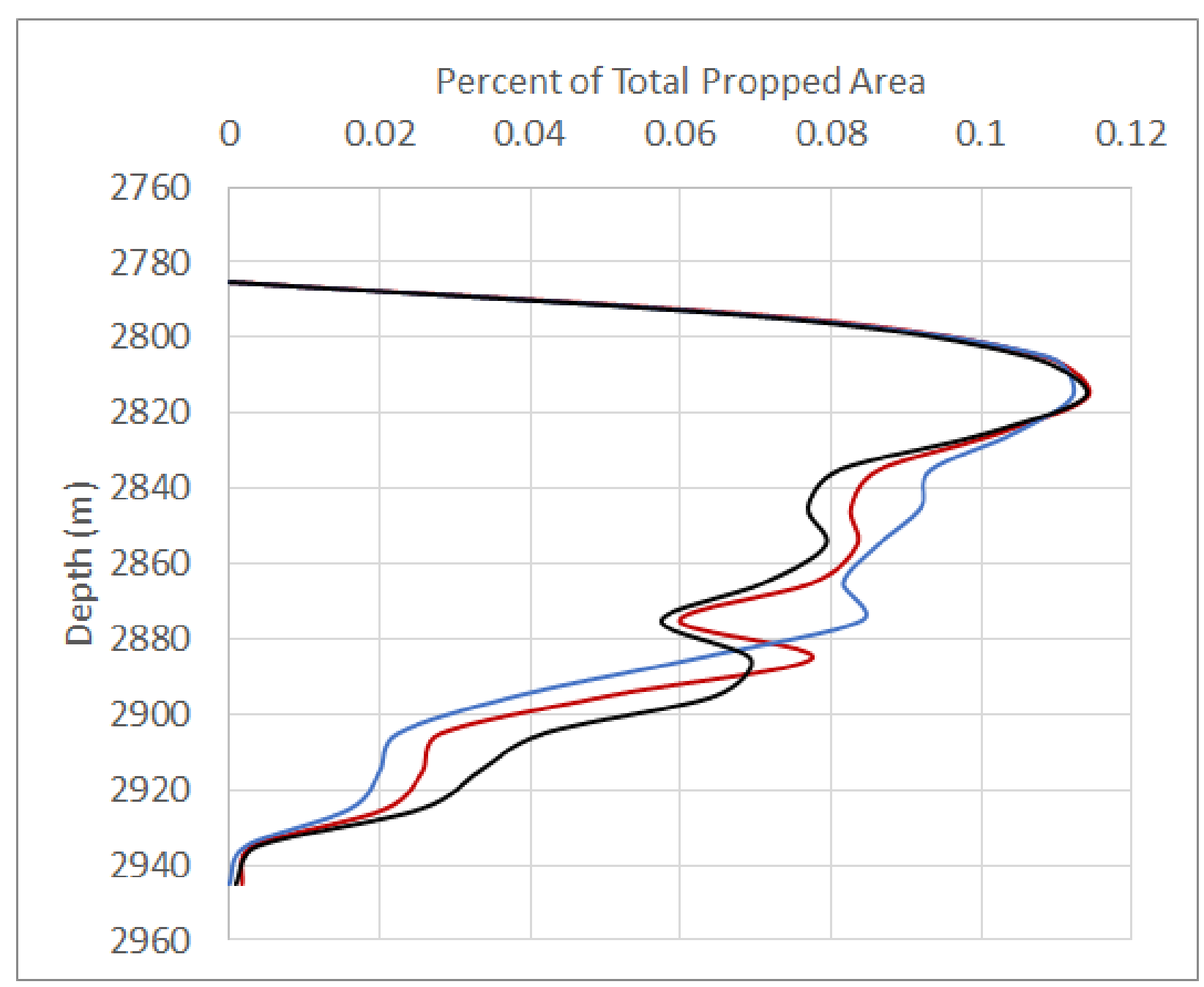

How does moving the wells left, right, up, down change the distribution of propped area? Are there well locations that could maximize otherwise by-passed pay?

# 10. ResFrac Decision Support tools: Sensitivity Analysis, Automated History Match, and Optimization

ResFracPro includes three Decision Support workflow tools: Sensitivity Analysis, Automated History Match, and Optimization. Users can easily create workflows using these tools. The three tools share a similar setup methodology, but are used for different objectives. Sensitivity Analysis allows the use to automatically generate a distribution of simulations within the parameter ranges specified. Automated History Matching allows the user to similarly define parameter ranges; however, the user also specifies objective function/s and the tool will run multiple generations of simulations to converge to the objective function/s. Finally, Optimization allows the user to specify parameter ranges like a Sensitivity Analysis or Automated History Match; however, the user can specify an aspirational objective function, like NPV or IRR, to maximize or minimize (instead of historical data to match).

For a detailed description of the algorithmic framework for decision support tools, please refer to Kang et al., 2022.

The sections below cover the basic use of each of the tools. Fundamental functionalities that apply to all three workflows are discussed under Sensitivity Analysis, so even if interested in jumping straight into an Automated History Match we suggest you first read the Sensitivity Analysis section below.

## 10.1 Sensitivity Analysis

Sensitivity Analysis allows the user to rapidly create a population of simulations within specified parameter ranges. The user can choose between different sampling schemes to create their desired distribution.

To launch a Sensitivity Analysis, the user clicks on an existing simulation to serve as the base sim, and chooses “Set Up Sensitivity Analysis”:

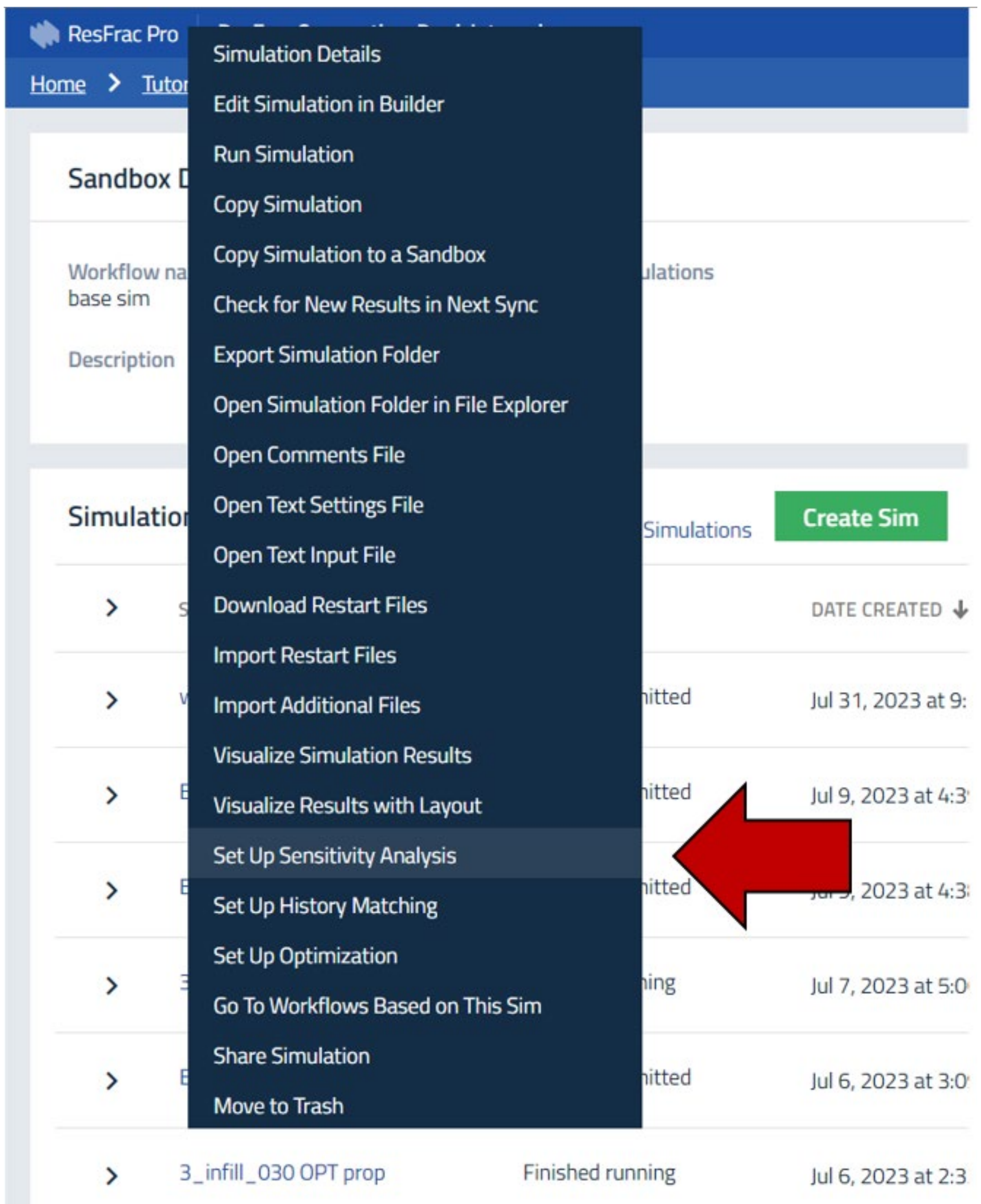

This will create a new Sensitivity Analysis workflow within the project folder. Unlike a Sandbox workflow containing individual simulations, the user cannot create individual simulations within a Sensitivity Analysis workflow. Instead, a set of simulations will be created within the workflow using the Sensitivity Analysis builder:

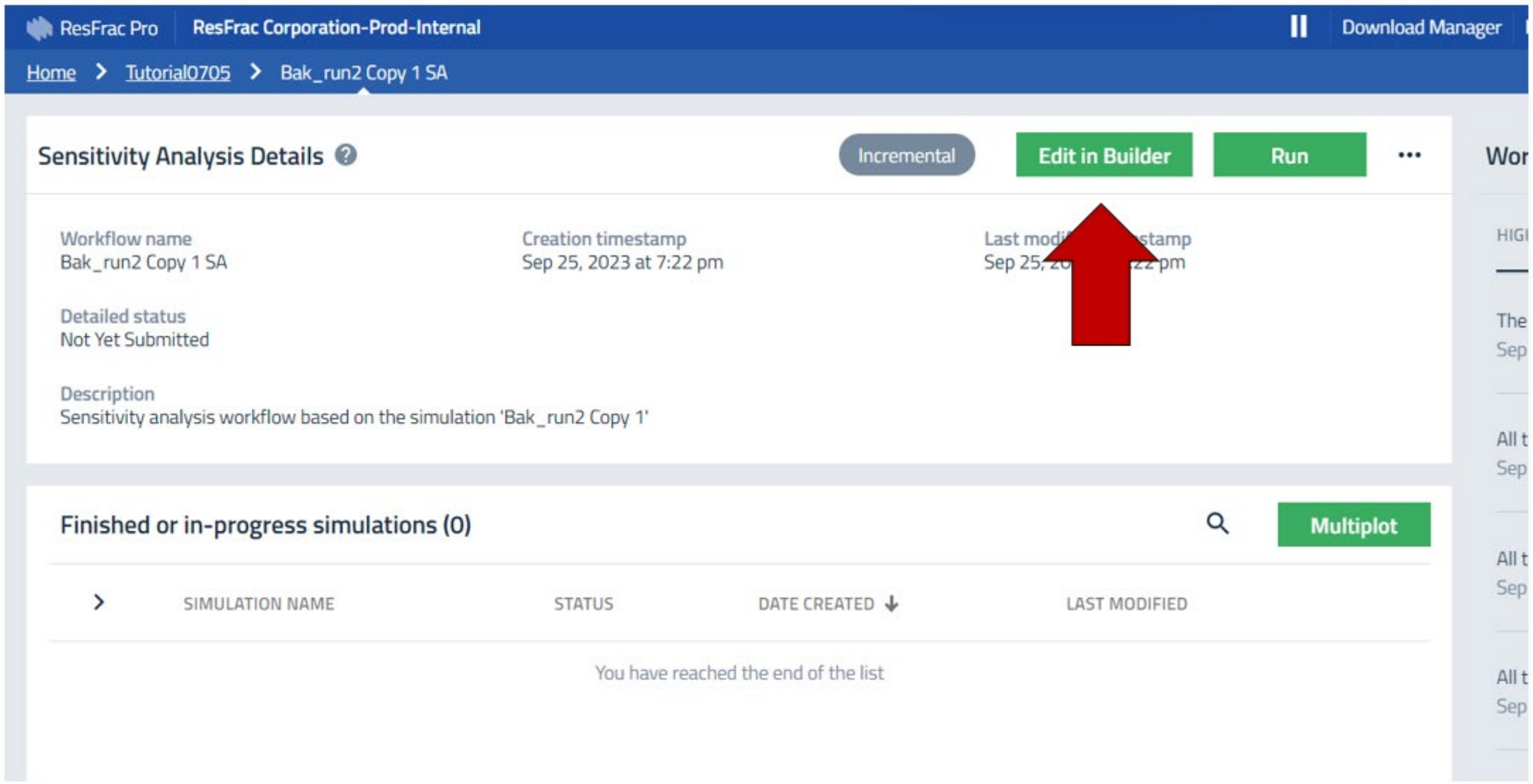

The Sensitivity Analysis Builder opens an interface that looks very similar to the Simulation Builder; however, there is an additional Decision Support panel, and all inputs outside of that panel are read-only in the Sensitivity Analysis Builder. Instead of manually editing simulation inputs, the user can select which inputs throughout the builder they want to change by either right-clicking (for table parameters) or clicking the checkbox next to a parameter (for single-valued parameters):

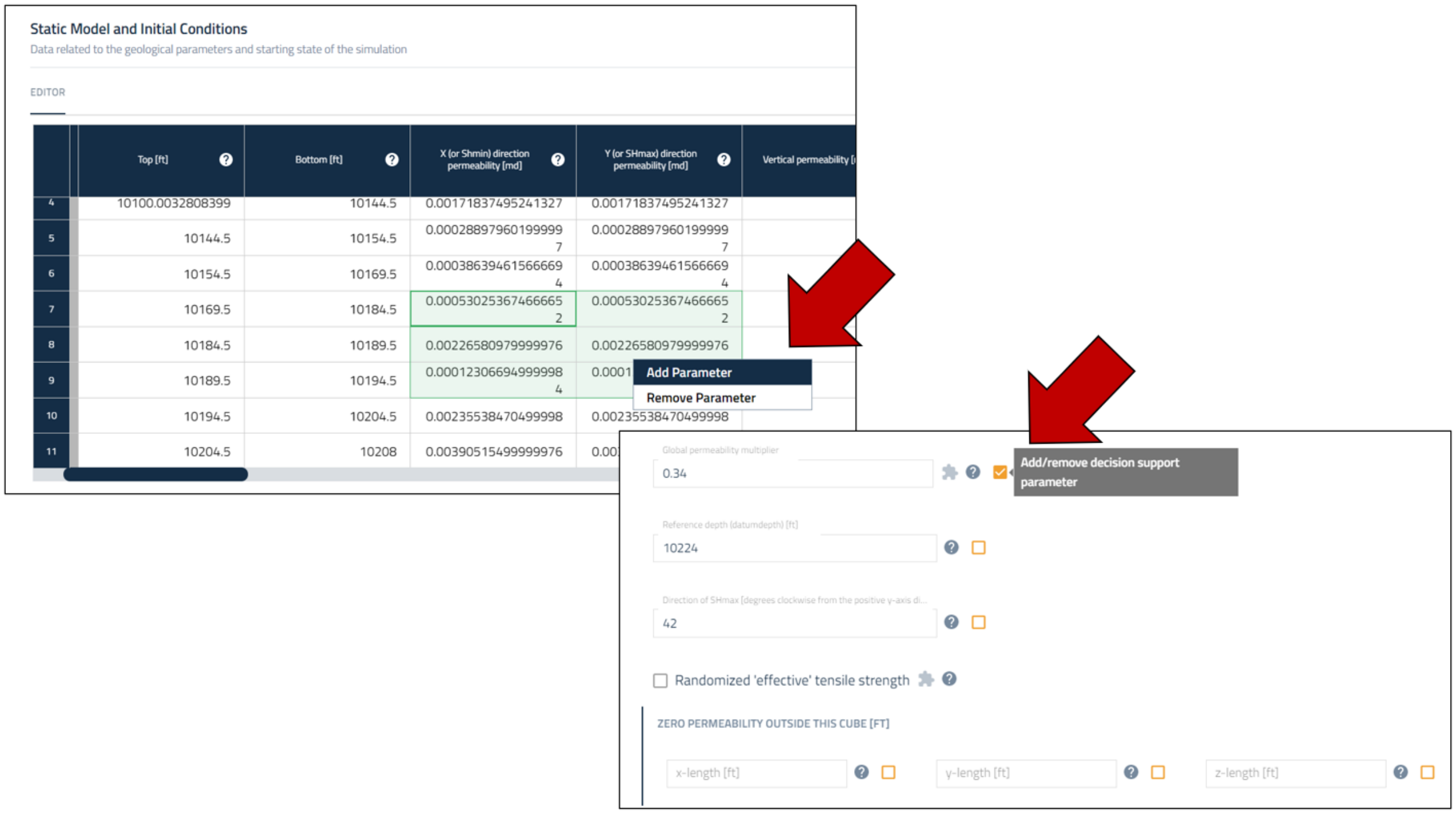

The user may select as many parameters to sensitize on as he or she chooses. Additionally, once parameters are selected they can be coupled such that they vary together in the Decision Support panel. The Decision Support panel has three components:

- Parameter Groups
- Decision Support Parameters
- Sampling Schemes

Users go through a three-step process on the Decision Support panel:

A. Define Parameter Groups and assign individual parameters to a Parameter Group.
Two different parameters assigned to the same parameter group will vary together. For instance, in the example below three parameters have been selected: horizontal fracture toughness, vertical fracture toughness, and relative fracture toughness

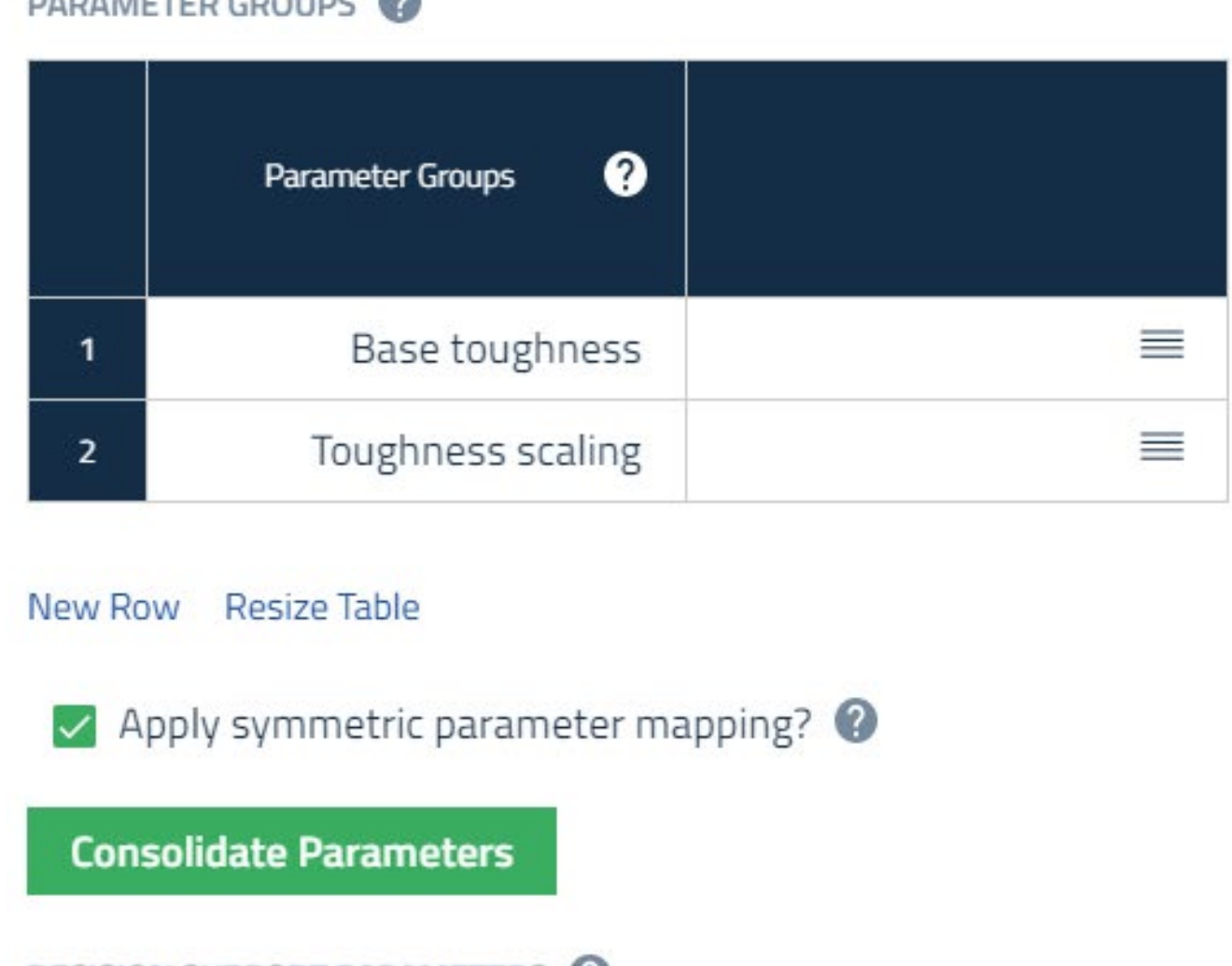
PARAMETER GROUPS

| | Parameter Groups | |
|---|---|---|
| 1 | Base toughness | |
| 2 | Toughness scaling | |

New Row Resize Table

Apply symmetric parameter mapping?

Consolidate Parameters

DECISION SUPPORT PARAMETERS

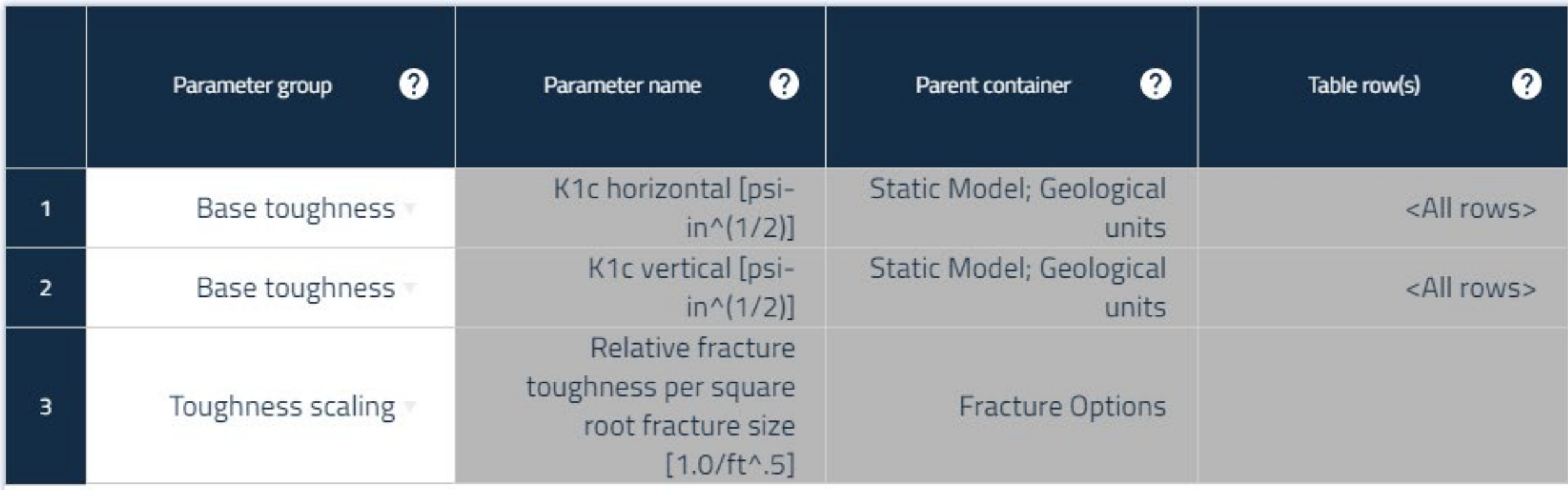

| | Parameter group | Parameter name | Parent container | Table row(s) |
|---|---|---|---|---|
| 1 | Base toughness | K1c horizontal [psi-in^(1/2)] | Static Model; Geological units | <All rows> |
| 2 | Base toughness | K1c vertical [psi-in^(1/2)] | Static Model; Geological units | <All rows> |
| 3 | Toughness scaling | Relative fracture toughness per square root fracture size [1.0/ft^.5] | Fracture Options | |

Two Parameter Groups have been defined: 'Base toughness' and 'Toughness scaling'. Vertical fracture toughness (K1c vertical) and horizontal fracture toughness (K1c horizontal) have both been assigned to the 'Base toughness' parameter group. This means that Vertical fracture toughness and Horizontal fracture toughness will vary together. Relative fracture toughness is assigned to a separate Parameter Group, so Relative fracture toughness can vary independently of the Vertical fracture toughness and Horizontal fracture toughness.

B. Specify ranges for each parameter
Minimum and maximum values for each parameter are defined in the Decision Support

Parameters table using relative references. We use relative references (e.g. “2x the base value”) rather than absolute references (e.g. “100 nd”) because this allows the definition to be fully generalizable. For example, you could select an entire table of heterogeneous permeability values, and simply define your max as 2x the base value. This would preserve the heterogeneity by increasing all rows by the same factor.

In the table below, we select a Relation in column (i). This defines how you specify your minimum and maximum values. The two most used options are ‘linear multiplier’ and ‘linear adder’. A linear multiplier takes the base value and multiplies by the minimum and maximum factors input into columns (ii) and (iii). For example, in the table below, the range for Vertical fracture toughness and Horizontal fracture toughness is from half the base value to twice the base value. Alternatively, linear adders are added to the base value. In the example below, the range for Relative fracture toughness is 0.1 less than the base value to 0.15 more than the base value.

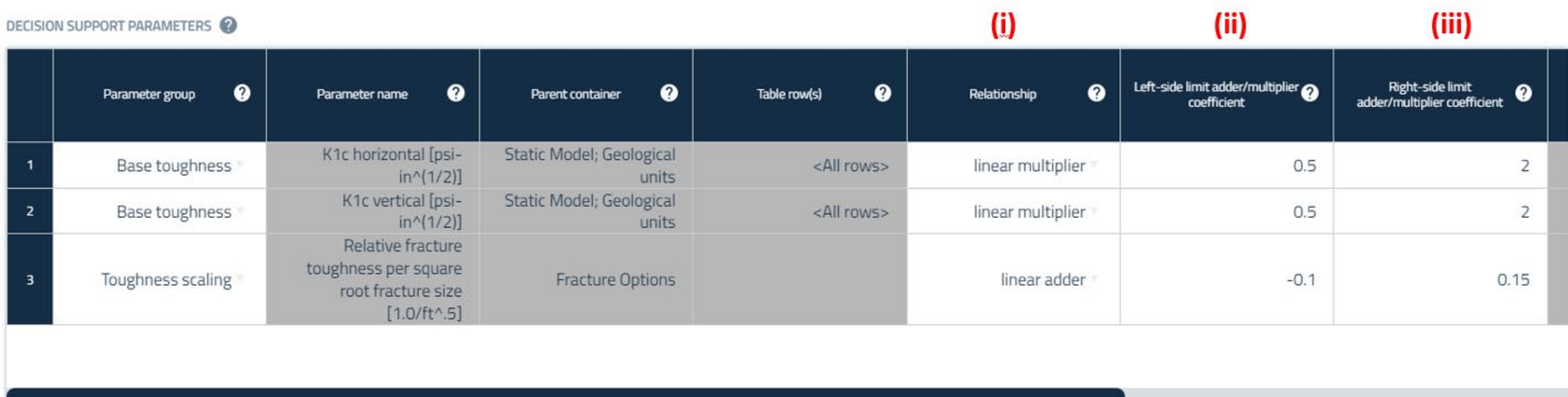

| | Parameter group | Parameter name | Parent container | Table row(s) | Relationship | Left-side limit adder/multiplier coefficient | Right-side limit adder/multiplier coefficient |
|---|---|---|---|---|---|---|---|
| 1 | Base toughness | K1c horizontal [psi-in^(1/2)] | Static Model; Geological units | <All rows> | linear multiplier | 0.5 | 2 |
| 2 | Base toughness | K1c vertical [psi-in^(1/2)] | Static Model; Geological units | <All rows> | linear multiplier | 0.5 | 2 |
| 3 | Toughness scaling | Relative fracture toughness per square root fracture size [1.0/ft^.5] | Fracture Options | | linear adder | -0.1 | 0.15 |

C. Create a Sampling Scheme

The final step in the Decision Support panel is to select and define a Sampling Scheme. Sampling Schemes define the how and how many simulations are created from the parameter ranges specified above.

The user begins by ‘adding’ a sample scheme, shown as (i) below, and then selecting the type of Sampling Scheme to use in (ii). Different Sampling Schemes require different numbers of inputs. For details on all the Sampling Schemes available click the “?” (iii) in image below or watch the decision support video at www.resfrac.com/onlinetraining.

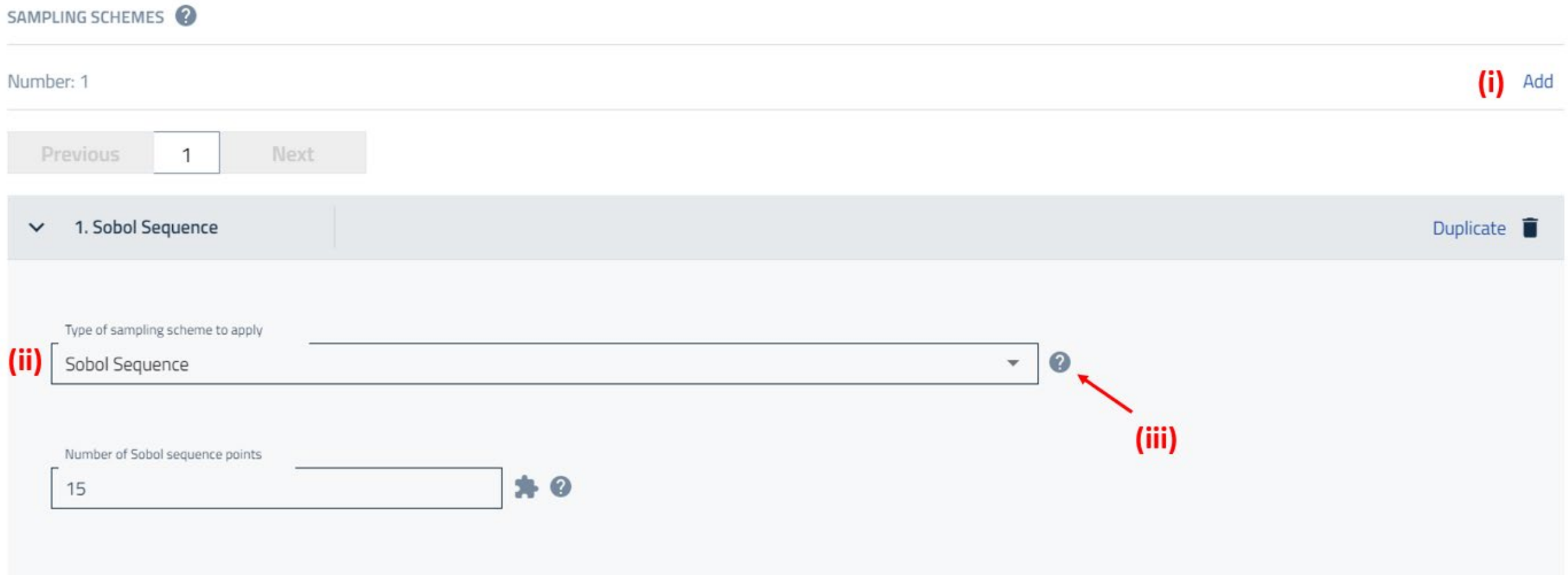

Once a Sampling Scheme is defined, a read-only table showing the cases to be run is displayed below the Sampling Schemes:

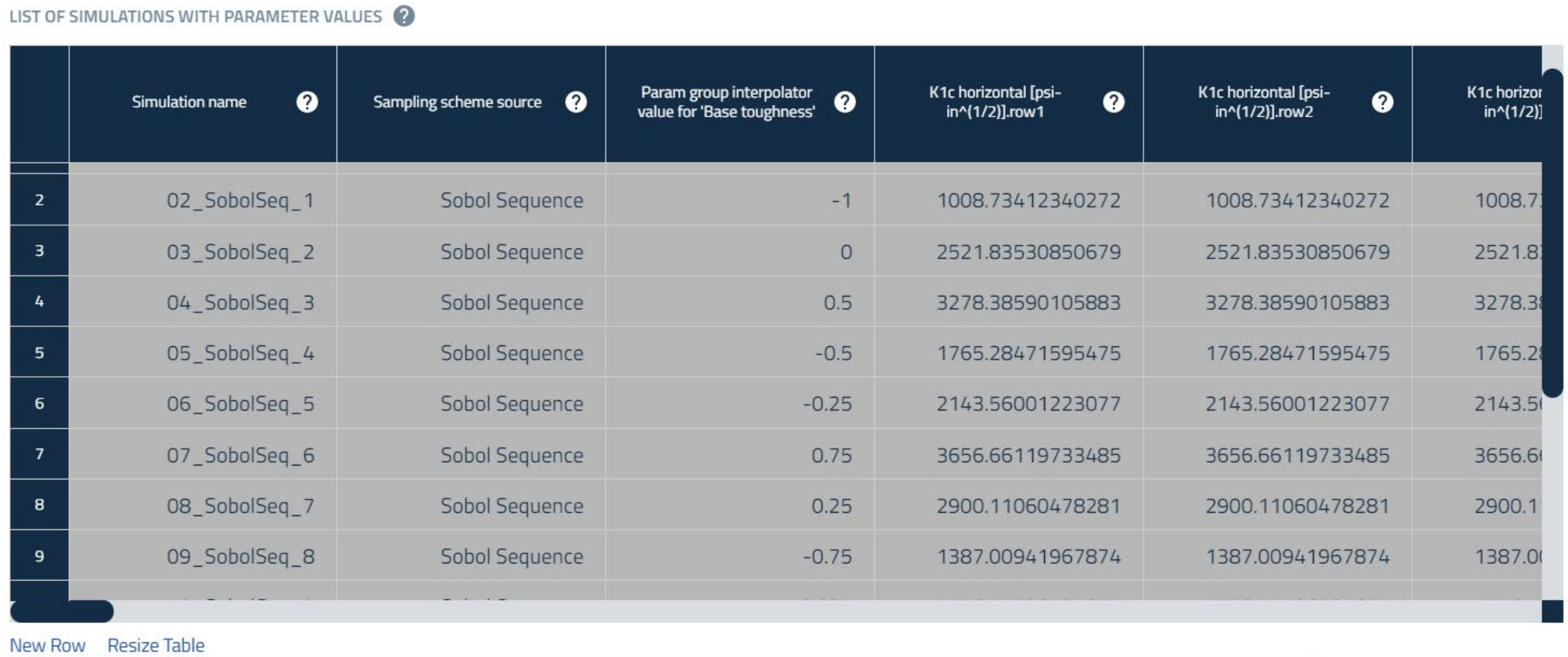

| | Simulation name | Sampling scheme source | Param group interpolator value for 'Base toughness' | K1c horizontal [psi-in^(1/2)].row1 | K1c horizontal [psi-in^(1/2)].row2 | K1c horizor in^(1/2)] |
|---|---|---|---|---|---|---|
| 2 | 02_SobolSeq_1 | Sobol Sequence | -1 | 1008.73412340272 | 1008.73412340272 | 1008.7 |
| 3 | 03_SobolSeq_2 | Sobol Sequence | 0 | 2521.83530850679 | 2521.83530850679 | 2521.8 |
| 4 | 04_SobolSeq_3 | Sobol Sequence | 0.5 | 3278.38590105883 | 3278.38590105883 | 3278.3 |
| 5 | 05_SobolSeq_4 | Sobol Sequence | -0.5 | 1765.28471595475 | 1765.28471595475 | 1765.2 |
| 6 | 06_SobolSeq_5 | Sobol Sequence | -0.25 | 2143.56001223077 | 2143.56001223077 | 2143.5 |
| 7 | 07_SobolSeq_6 | Sobol Sequence | 0.75 | 3656.66119733485 | 3656.66119733485 | 3656.6 |
| 8 | 08_SobolSeq_7 | Sobol Sequence | 0.25 | 2900.11060478281 | 2900.11060478281 | 2900.1 |
| 9 | 09_SobolSeq_8 | Sobol Sequence | -0.75 | 1387.00941967874 | 1387.00941967874 | 1387.0 |

This table displays the interpolator for each parameter as well as the actual values to be used in the simulation (so if your parameter is a table, there will be one column for each value in the selected parameter table). This is a helpful feature to check you've defined your Sampling Scheme as intended!

Once the Decision Support panel is filled out, the Sensitivity Analysis workflow can be run using the “run” button at the top of the workflow:

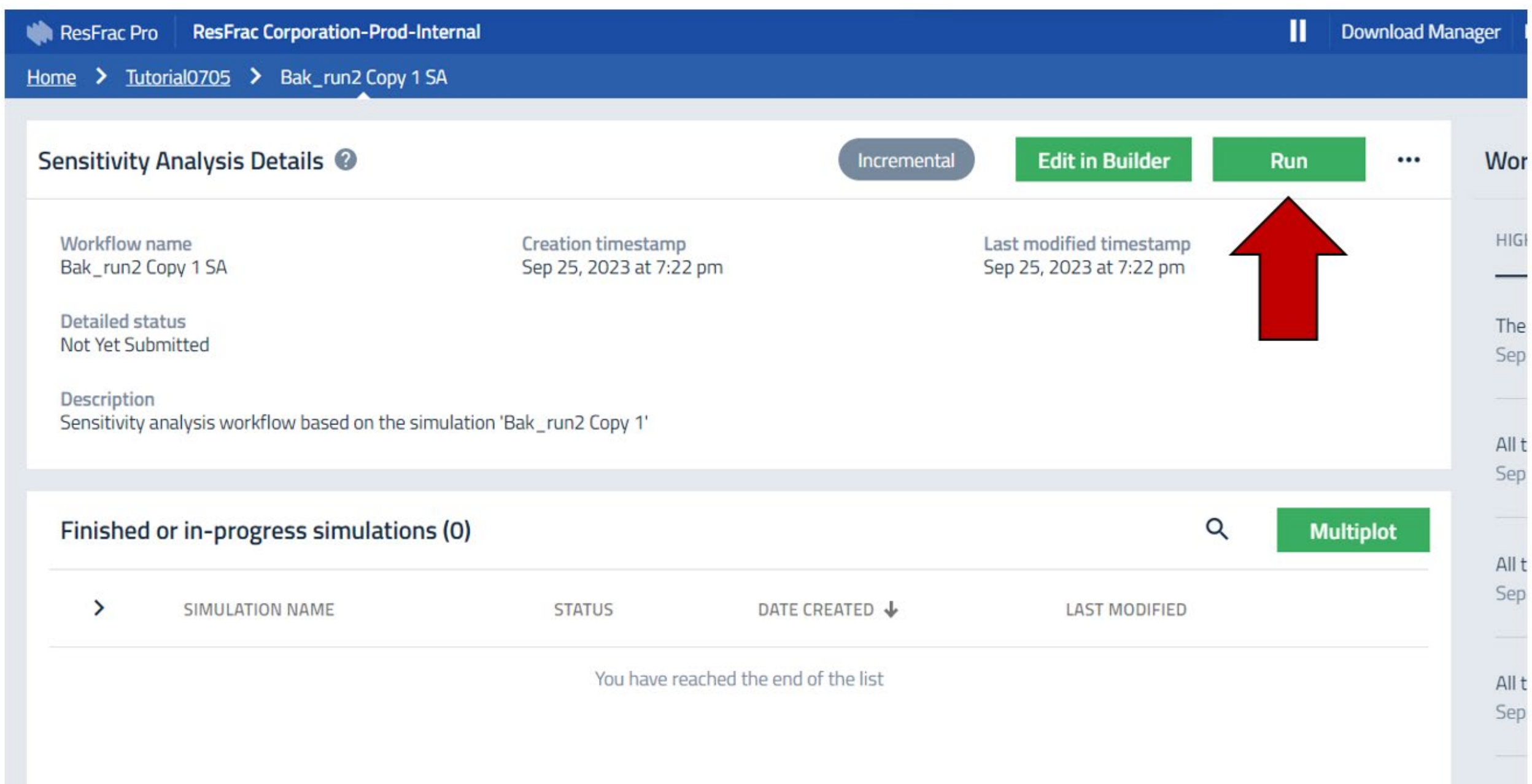

The simulation cases will all run in parallel, and a list of running simulations will populate inside of the workflow:

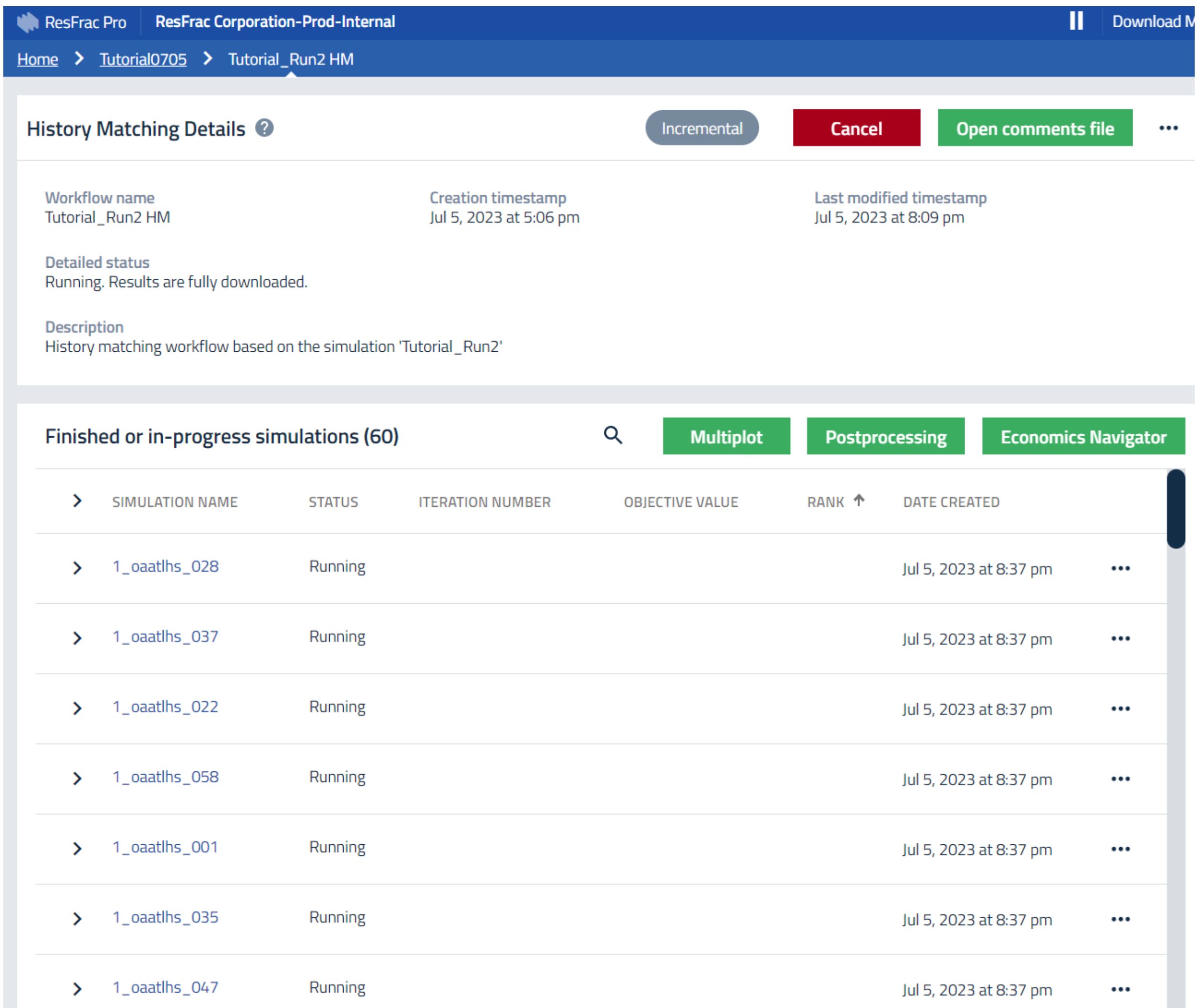

Visualizing results of simulations generated with an automation workflow (like Sensitivity Analysis) can be done one-by-one or using the multiplot functions, similar to reviewing results in a sandbox (see section 5.12 The visualization tool). In addition, there are specific automation workflow post processing tools covered in section 10.3 Decision Support visualization and post-processing tools.

## 10.2 Automated History Matching and Optimization

Automated History Matching and Optimization workflows operate in a similar manner to Sensitivity Analysis (prior section), but with a two key differences:

1) When creating the workflow, instead of specifying a Sampling Scheme, the user specifies Target Functions.
2) When the workflow runs, it will now run multiple generations of simulations (three by default).

### Target Functions

In History Matching and Optimization workflows the user does not specify a Sampling Scheme. Instead, the user defines one or more Target Functions. For History Matching workflows, the Target Functions are usually explicit (matching a specific initial shut-in pressure, a time-series of production over time, a volume-to-first-response between two wells). For Optimization workflows, often the Target Function objective is a minimization or maximization of a parameter (e.g. maximize NPV per Section, maximize IRR, etc.).

To add a target , click "Add" under Target Functions:

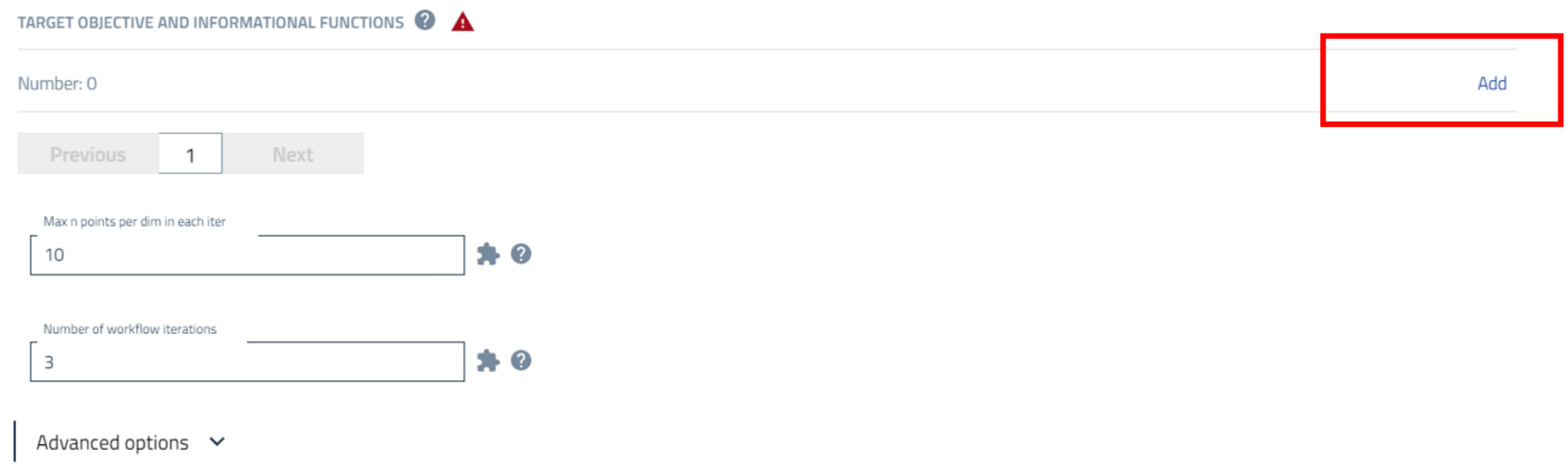

Each Target Function is given a name and an output parameter trend needs to be selected. For instance, the Target Function below is named "CumulativeOil" and the trend "Oil tot prod, Well_MB1" is selected as the data to evaluated (A1, A2 respectively).

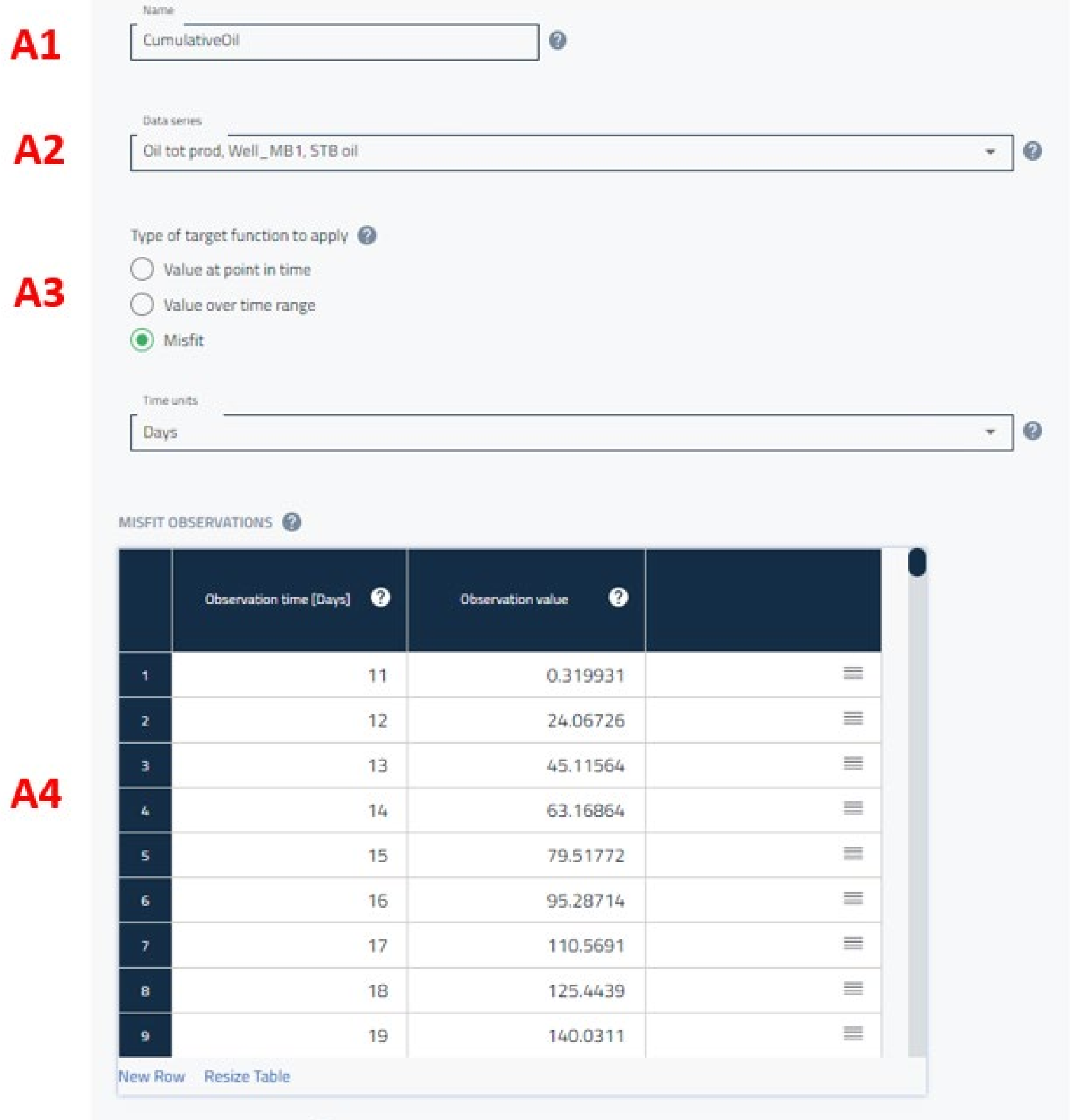

Cumulative oil production could be evaluated at a point in time (e.g. 30 EUR) or over time. Evaluating the objective over a time series is more typical as it provides more definition. In the example image above, selecting a “Misfit” target function type (A3) allows the user to input a table of points in time and objective values (A4). When evaluating the Target Function, ResFrac will compute the error across the entire time series rather than just one point.

After defining the objective data, the user has several options to customize how the target function is evaluated. Most of these options can be left at the default values, however, below we briefly describe each (more detail is given in the built-in user interface help content):

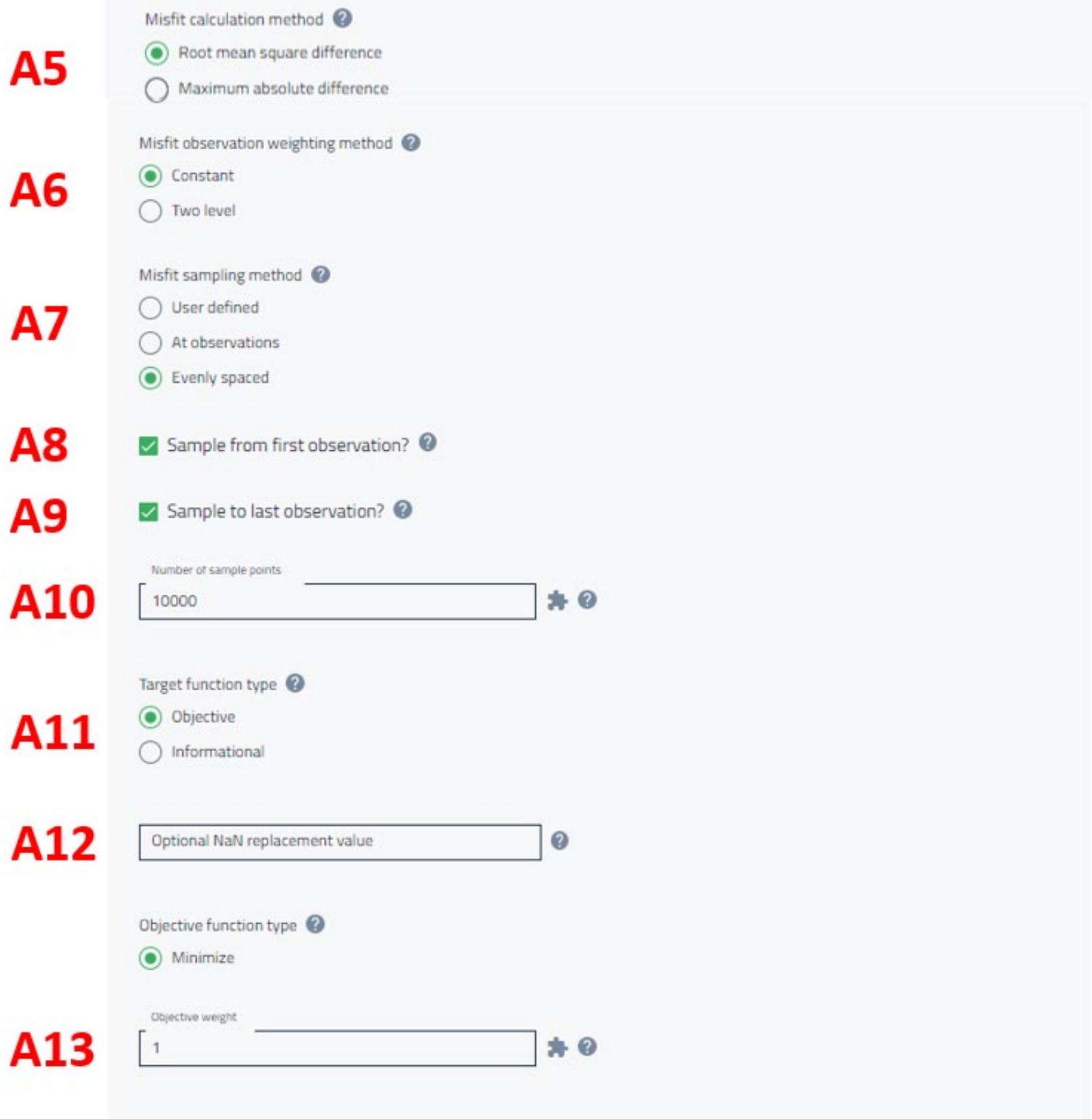

- A5 – Misfit calculation method: user selects type of error function to evaluate the objective data. Root mean square difference is preferrable the vast majority of the time.
- A6 – Misfit observation weighting method: this is used to apply different weights to different intervals of the historical data. For instance, if the user was more concerned with matching the early time production, they could choose a 'two level' weighting method, and assign a weight of 2.0 to the first year of data and a weight of 1.0 to the remaining data. Consequently, the error in the first year will be weighted twice as heavily as the error in the remaining history when evaluating the Target Function.
- A7 – Misfit sampling method: allows user to select how to sample from within their inputted data. Evenly spaced samples typically are the best choice.
- A8 and A9 – Sample from/to controls: if the user wants to eliminate the beginning or end of their inputted historical data they can use this to select a portion of the data in their Misfit Observation table.
- A10 – Number of sample points: This is typically left at the default of 10,000. If there are more sample points than observation points, ResFrac will linearly interpolate between the observation points. There is more computational penalty for having more points, thus this is usually left at 10,0000.
- A11 – Target function type: If the user wants the Target Function evaluation to be included in the ranking and selection of cases, 'Objective' must be selected. An 'Informational' objective has no impact on the ranking or evaluation of workflow simulations.

- A12 – Nan replacement value: if the Target Function could be evaluated to 'nan', then this value will replace the value from the simulation. For example, 'Oil tot prod' is never nan (always 0 or greater), so there is no need for a nan replacement value. However, as covered below, something like volume-to-first-response can be nan, and this value can be used.
- A13 – Objective weight: the user may define as many Target Functions as they desire within a workflow. The Objective weight allows the user to weight each Target Function differently. For example, if the user was more concerned with matching cumulative oil than the ISIP, they might assign a weight of 2.0 to their Target Function evaluating cumulative oil and a weight of 1.0 to their Target Function evaluating ISIP.

Other objectives make more sense to evaluate at a specific point in time. For instance, if evaluating the volume-to-first-response (VFR), it makes the most sense to do so at the end of the target stage. The image below demonstrates how this is set up.

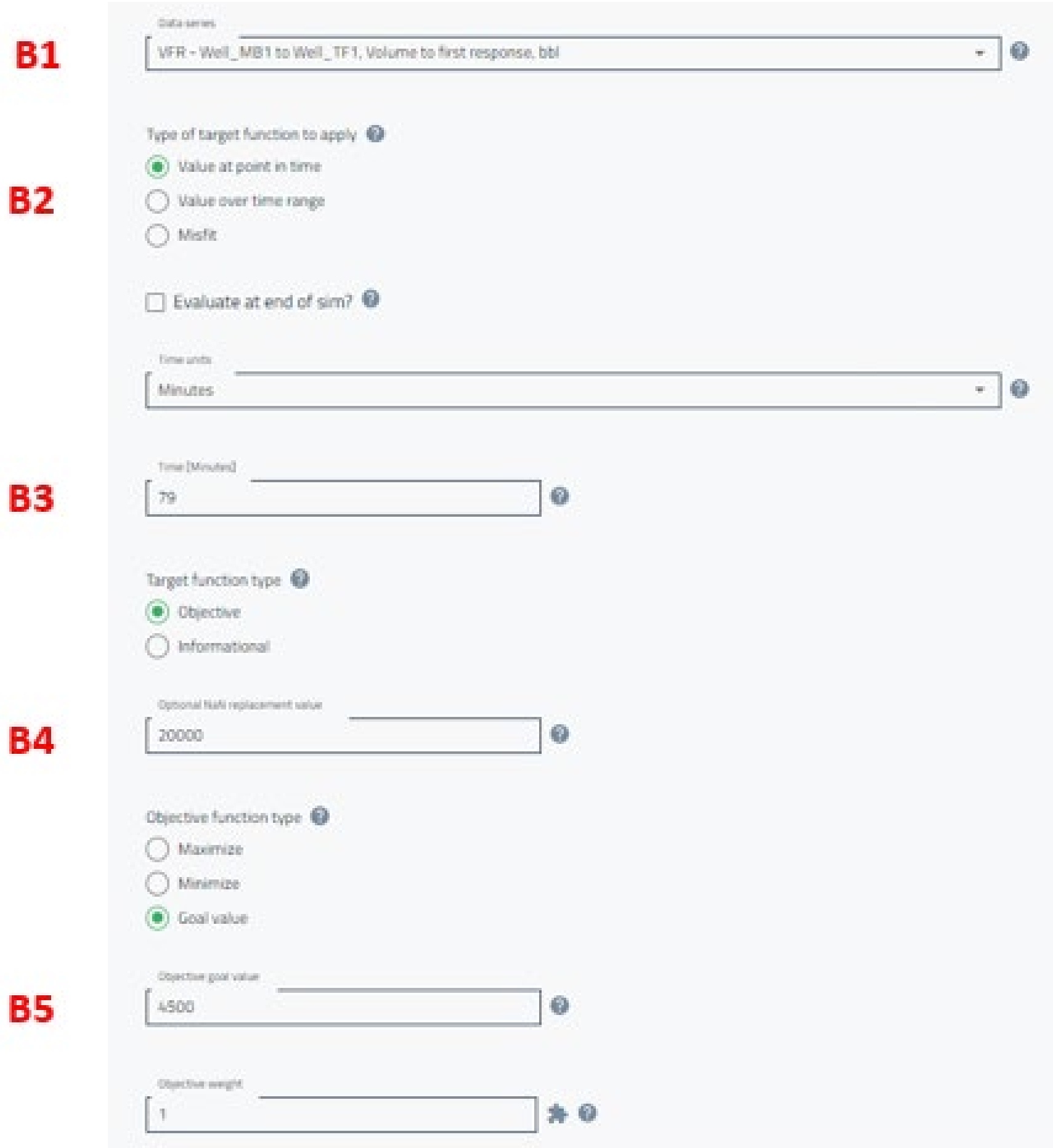

- B1 – The VFR between the target wells is selected as the objective trend.
- B2 – This time, the user evaluates the selected data trend at a specific point in time.
- B3 – The chosen time is just prior to the end of pumping on Well_MB1.
- B4 – If the fractures from Well_MB1 never reach Well_TF1, the data trend 'VFR – Well_MB1 to Well_TF1' returns "NaN" (not-a-number) for all times. The error function cannot calculate a numeric value for the difference between the goal value and nan. Instead, the user specifies a 'Optional NaN replacement value'. If the simulation trend returns NaN, then the replacement value is used in the error function calculations. In the case of VFR, the NaN replacement value should be a value larger than the total volume injected in the stage. In the example above, the total stage volume is 10,0000 barrels, so the NaN replacement value is set to double the entire stage volume.

- B5 – Because the Target Function is being evaluated at a single point in time, the Objective goal value is a single, numerical value. The units of the Objective goal value will be the same as those of the Data values (selected in B1).

In Optimization workflows, as opposed to Automated History Match workflows, the user defines Target Functions in a similar manner with nuanced changes.

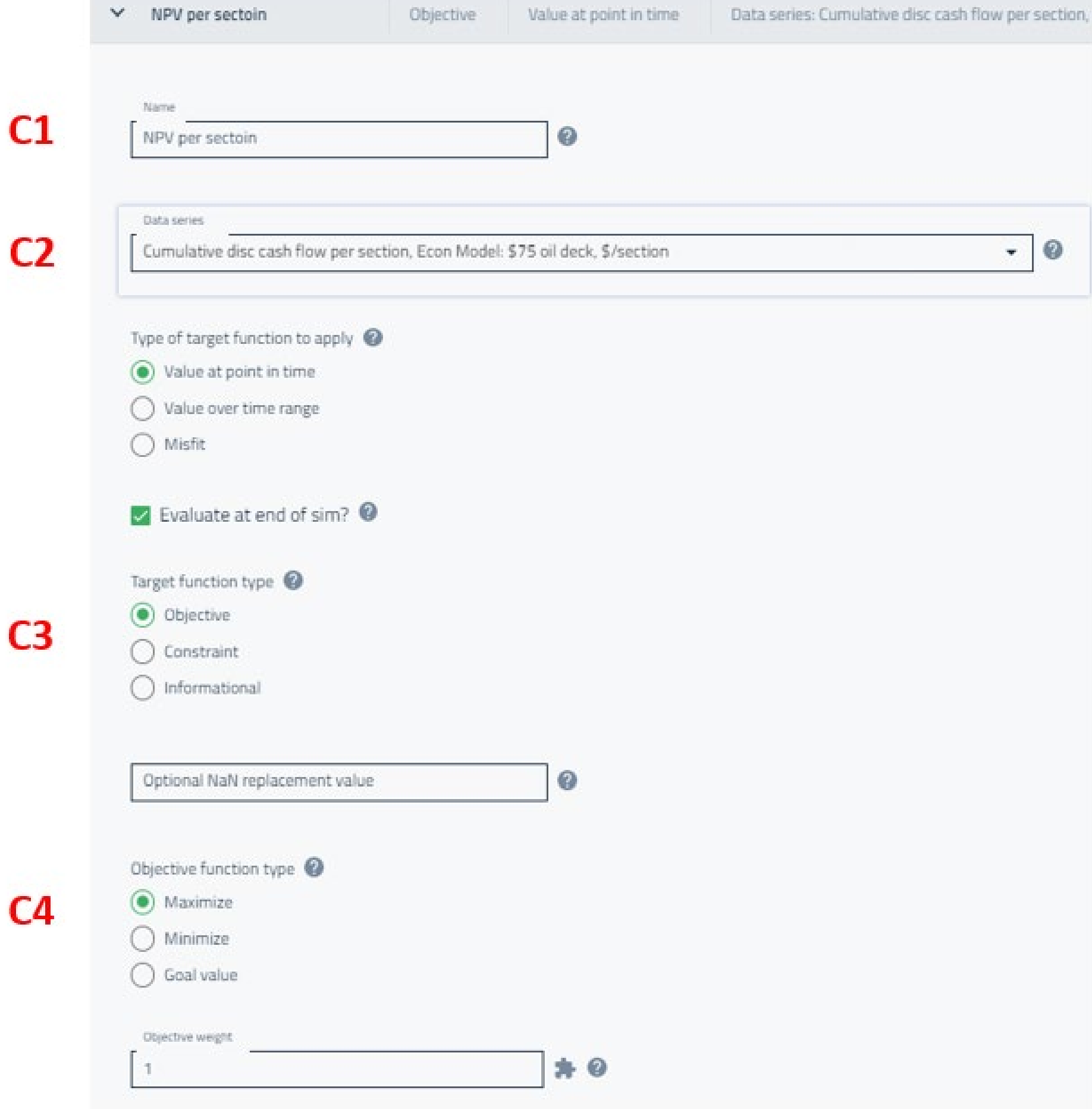

- C1 – Target Functions are given names similar to Automated History Matching workflows.
- C2 – Nearly any data series available in the Sim_track_SIMNAME.csv file (including outputs from the Economic Model) is available. In optimization workflows the most common Data series selections are Cumulative disc cash flow per section (NPV per section), Cumulative disc cash flow (NPV), Instantaneous IRR, Instantaneous DROI, and other Economic Model Outputs.

- C3 – In addition to Objective and Informational Target function types, Optimization workflows add a ‘Constraint’ Target function type. A Constraint target function can be used to constrain the set of simulations generated, for instance, the user could select ‘Cumulative cost’ as the data series and use a Constraint target function type to limit total spending below $28.5M for the pad:

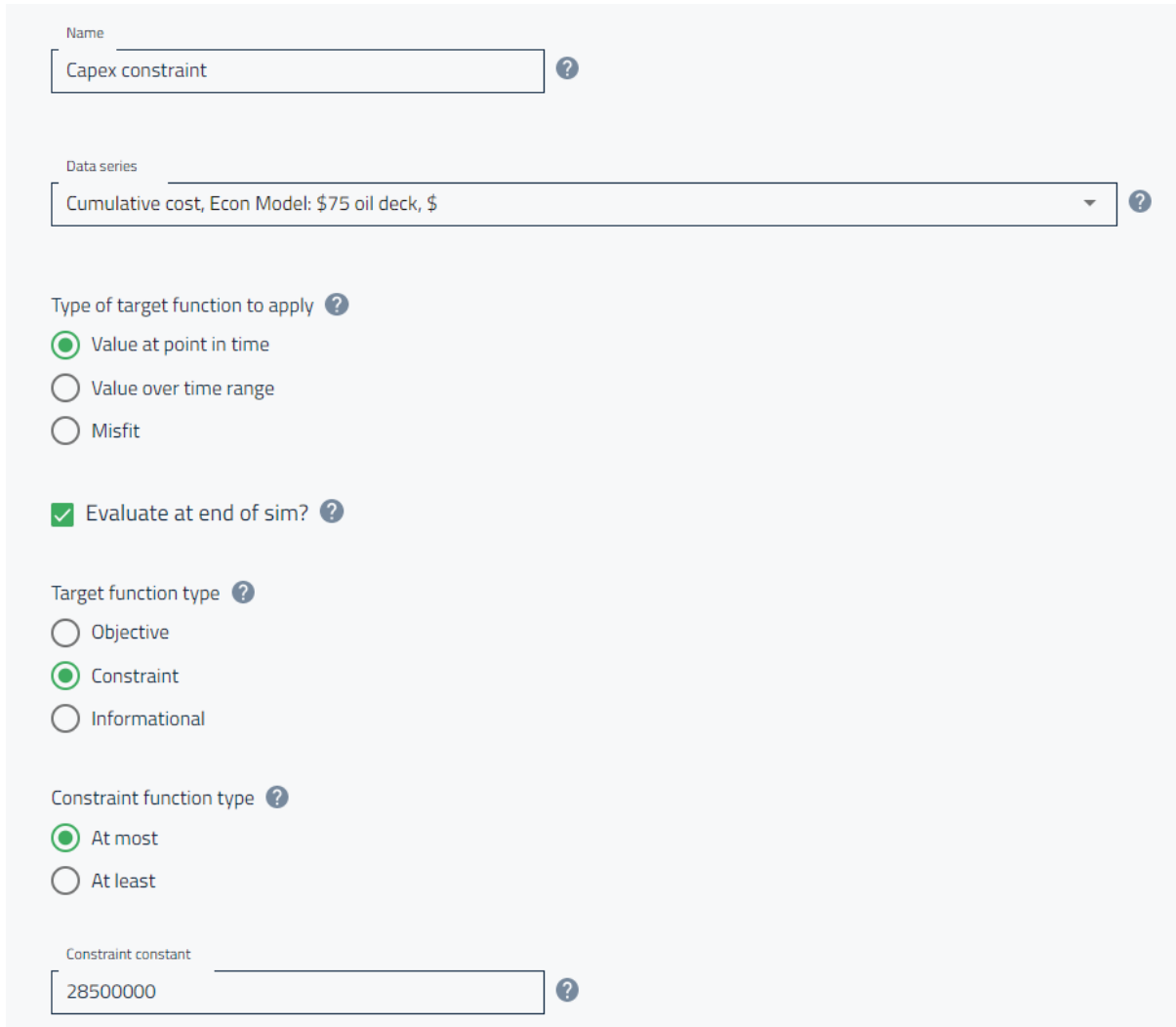

- C4 – If the data series selected is something like NPV or ROI, the Objective function type is usually to Maximize. In select cases, for instance if Cumulative cost was selected as an Objective versus a Constraint, the user could chose to Minimize the Cumulative cost.

## 10.3 Decision Support visualization and post-processing tools

Any simulation run within a workflow can be visualized individually like a simulation run within a normal sandbox. Only the sim_track_SIMNAME.csv file is downloaded for each simulation within an automation workflow by default, and 3D results are kept on the server for 30 days. The 3D results of an individual simulation can be downloaded by selecting ‘Download All Results’ from the simulation menu or the 3D results for all simulations within a workflow can be download from the menu at the top of the workflow.

<u>Download results for individual simulation</u>　　　　<u>Download results for all simulations in workflow</u>

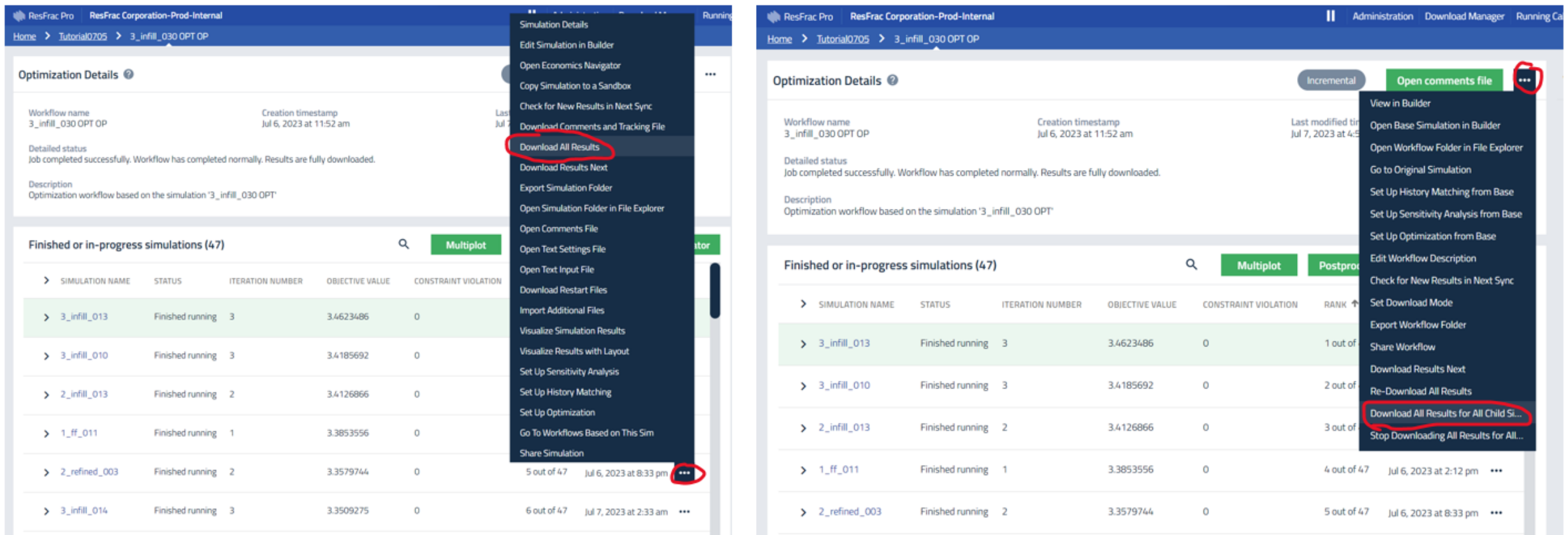

Workflows also provide additional results visualization capabilities accessed through the ‘Postprocessing’ tool:

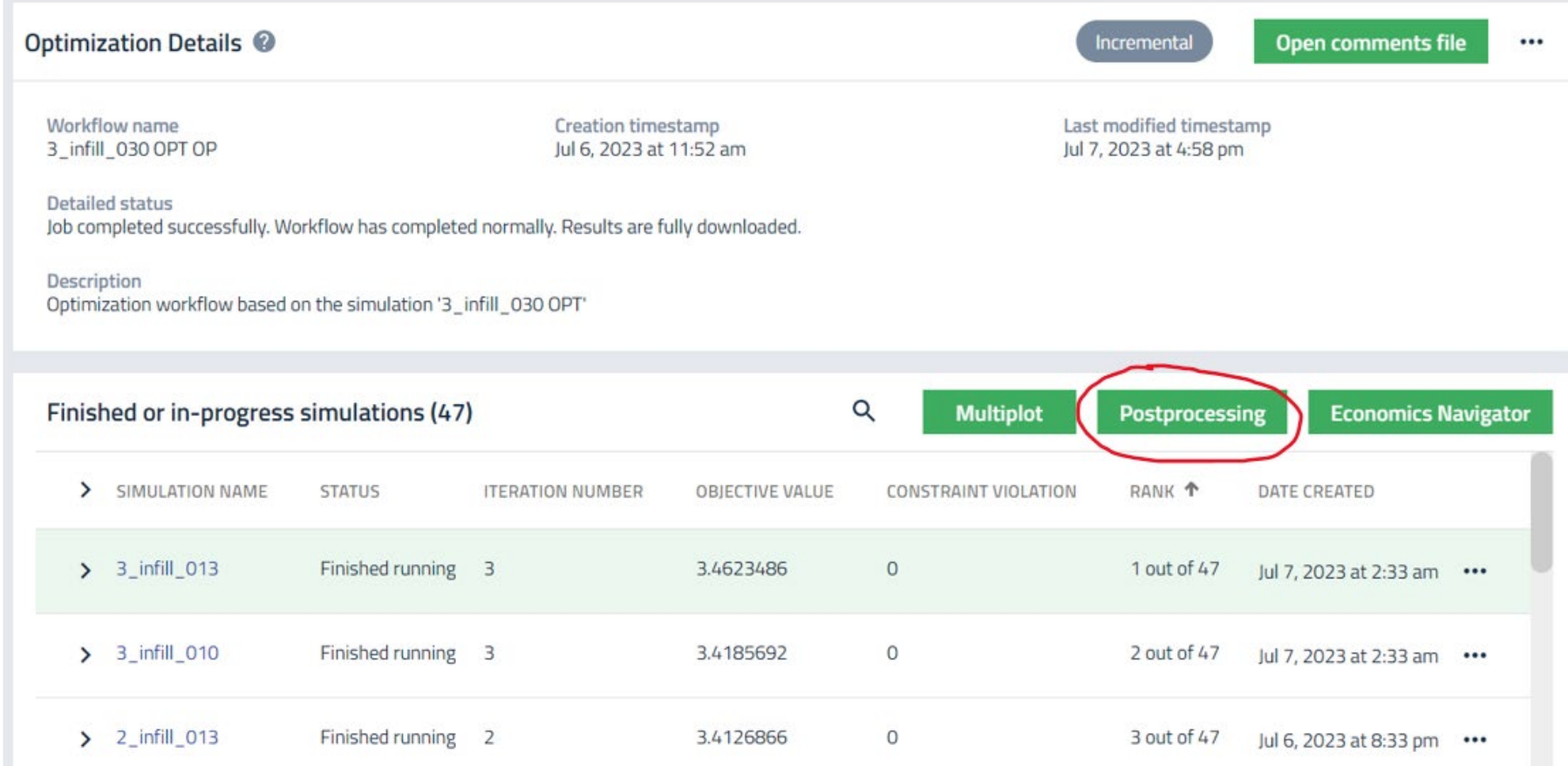

The postprocessing tool is laid out in three sections:

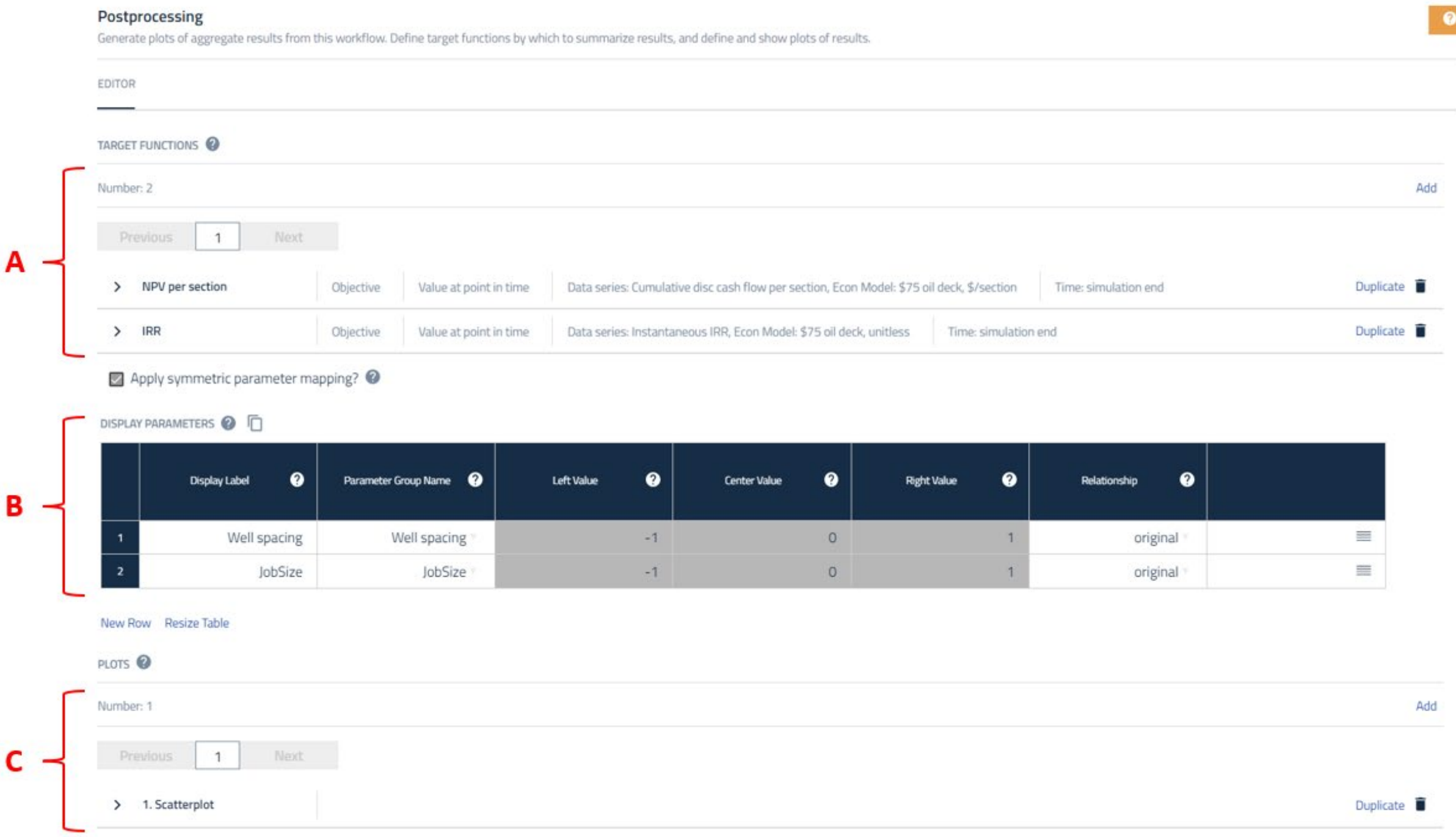

## A. Target Functions

The workflow will automatically load the Target Functions defined in Automated History Matching or Optimization workflows. For sensitivity workflows the Target Functions will be blank to begin, but additional Target Functions can be added (as they can be to History Matching and Optimization workflows as well).

The example image is from an Optimization workflow with NPV per section and IRR as the two objectives. To add 1 year recovery, the user can select “Add” in the upper right of the Target Function area, and define an additional Target Function for 1 year total production across all wells in the model:

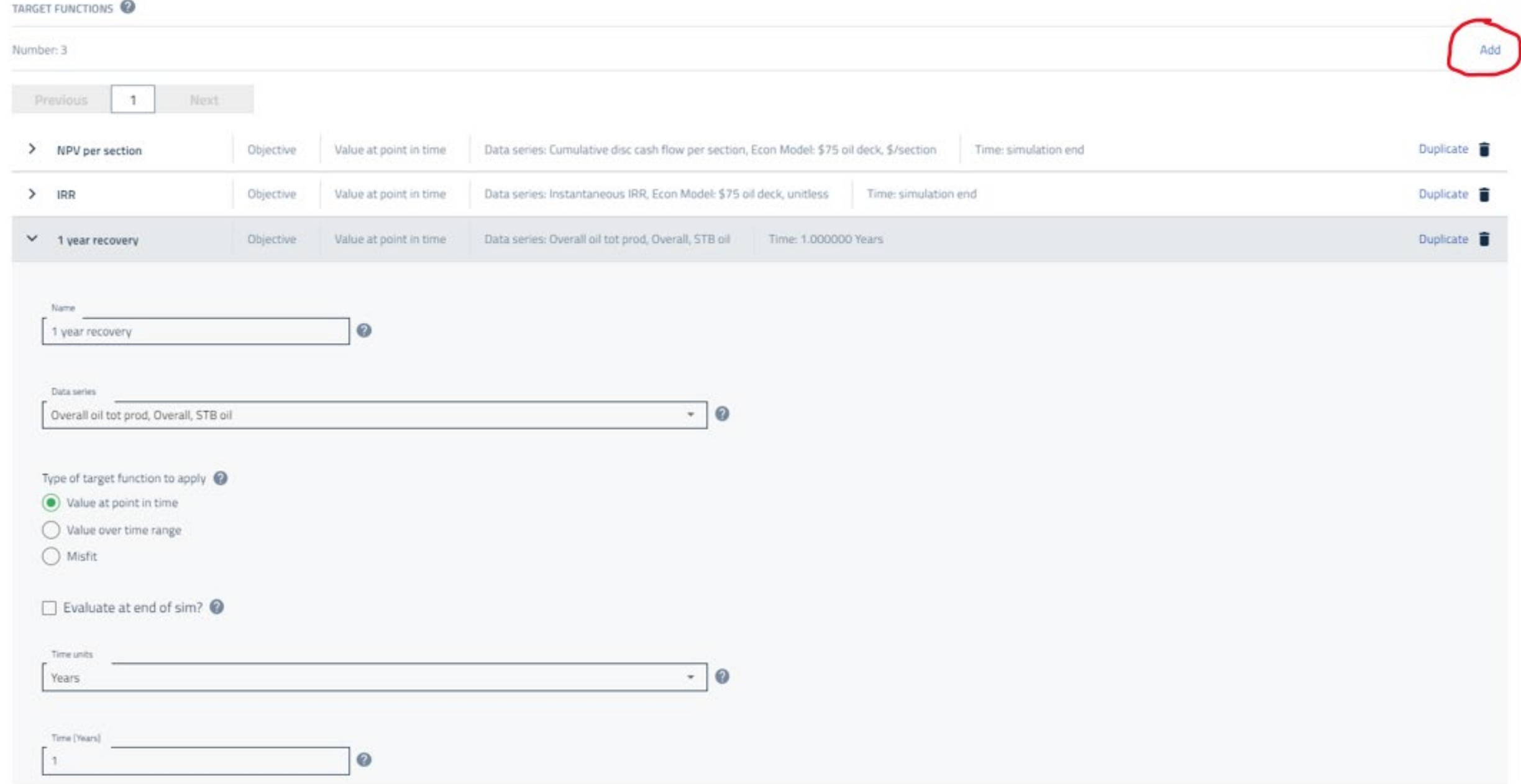

## B. Display Parameters

By default, the parameter groups used within the workflow are visible here. Additional display parameters can be added to transform Parameter Group ranges from parameterized values (-1 to +1) to physical values (like 500 ft to 1500 ft).

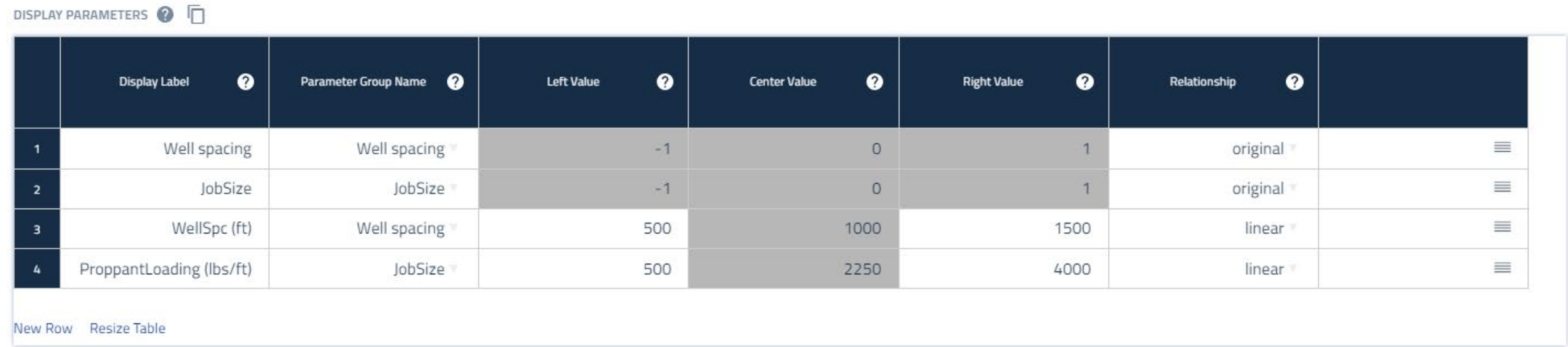

DISPLAY PARAMETERS

| | Display Label | Parameter Group Name | Left Value | Center Value | Right Value | Relationship | |
|---|---|---|---|---|---|---|---|
| 1 | Well spacing | Well spacing | -1 | 0 | 1 | original | |
| 2 | JobSize | JobSize | -1 | 0 | 1 | original | |
| 3 | WellSpc (ft) | Well spacing | 500 | 1000 | 1500 | linear | |
| 4 | ProppantLoading (lbs/ft) | JobSize | 500 | 2250 | 4000 | linear | |

New Row Resize Table

In the example above, the 'Well spacing' and 'JobSize' parameter groups are the parameterized groups used in the workflow calculations. The user then defines two additional Display Parameters ('WellSpc (ft)' and 'ProppantLoading (lbs/ft)') for additional plot detail.

## C. Plots

The postprocessing tool contains several built-in plot types: Scatterplot, Properties vs Time, Histogram, Heat map, and Tornado. The user-interface help content describes the use of each plot type:

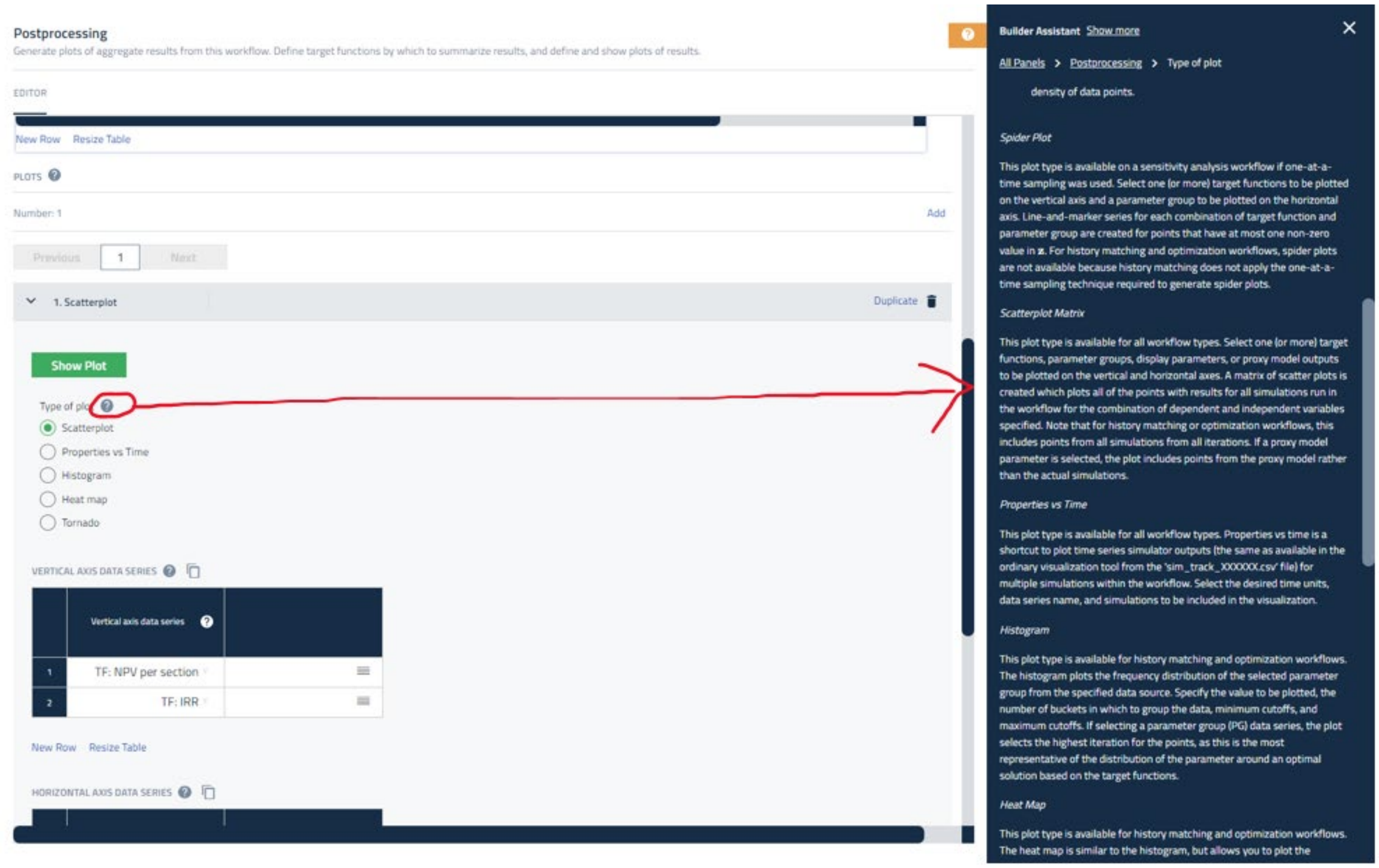

Scatterplots are likely the most frequently used plot. Scatterplots allow the user to place different Target Functions (TF), Proxy Model points (PM), and Display Parameters (DP) on the x and y axes as well as color and size the scatterplot points by the same trends. For example, below shows the proxy model points for NPV (red) and IRR (blue) for the Well Spacing and Proppant Loading optimization from the prior example:

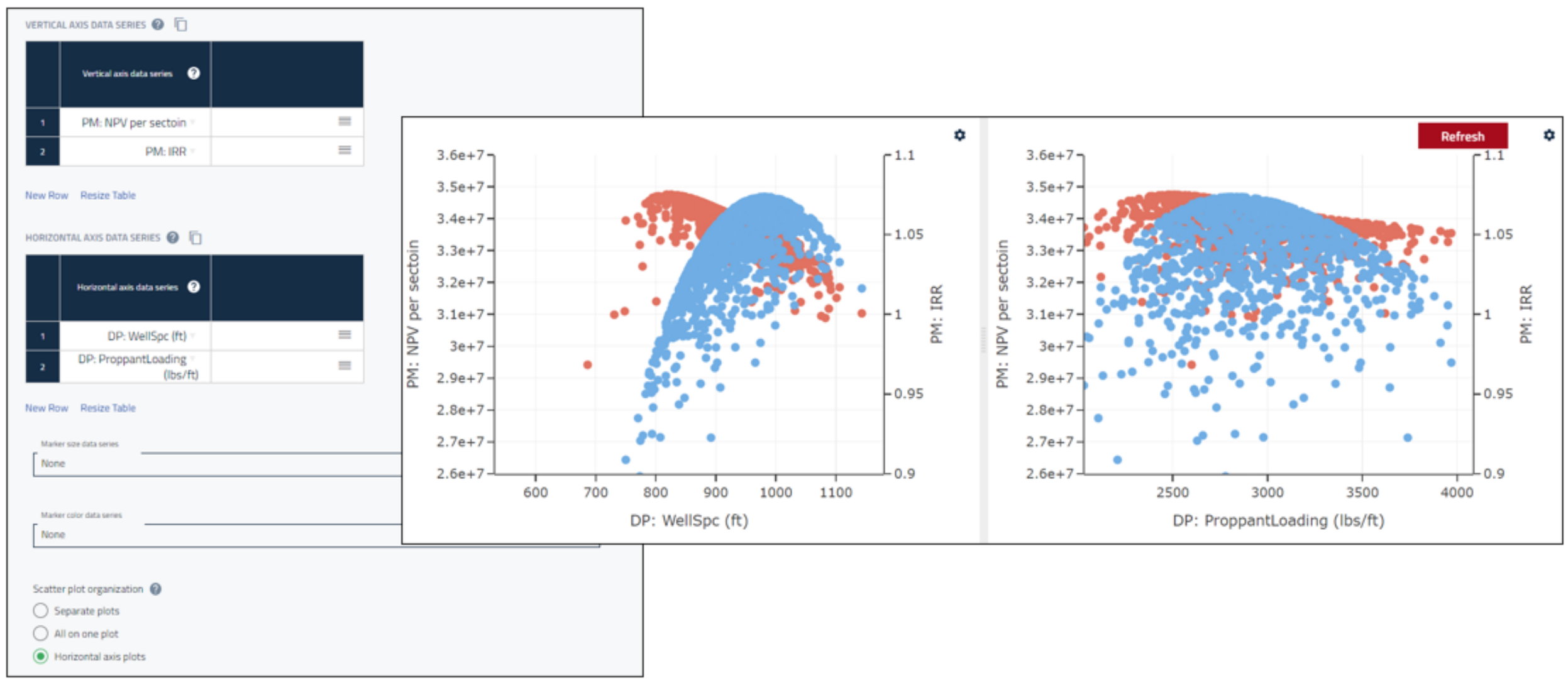

## 10.4 Other useful Decision Support functionalities

### Editing Base Case Simulation

At times you may want to edit or modify the base simulation before running a Decision Support workflow (Sensitivity Analysis, Automated History Match, or Optimization). For instance, you may have history matched your simulation to a five-year production history and you had set your maximum simulation time to five years. When you proceed to optimization, you may want your optimization cases to run to 30 years. To do this, you can edit the base simulation by selecting the ellipses adjacent to the "Run" button:

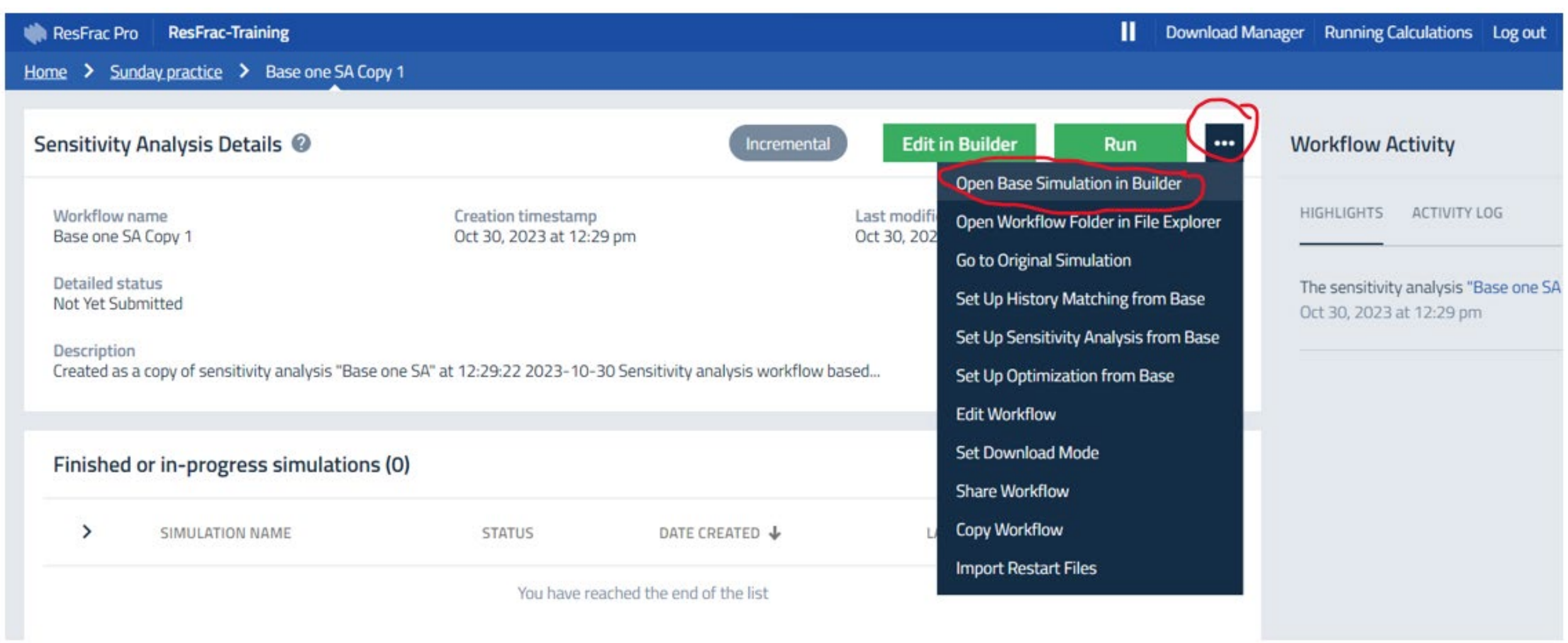

This will open up the regular builder for the simulation, allowing you to edit any of the base parameters (like the maximum simulation time).

### Economics Navigator

The Economics Navigator allows the user to add additional Economic Outputs and settings to simulations *after* the simulation has been run. The Economics Navigator can be used with individual simulations by Accessing the Simulation Menu (clicking the simulation name or the ellipses on the right-hand side) or from within automation workflows by clicking the Economics Navigator button above the list of simulations:

Accessing the Economics Navigator for individual simulations: Accessing the Economics Navigator for workflows:

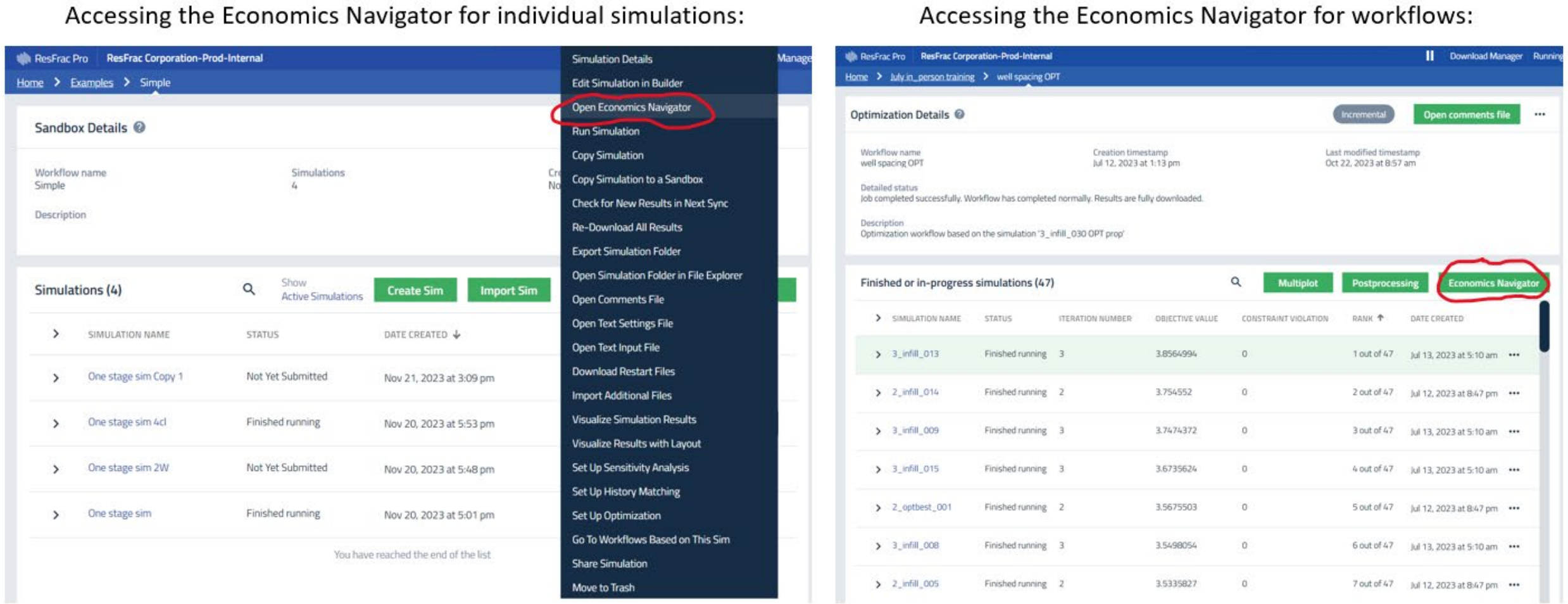

The Economics Navigator opens an interface that looks similar to the Simulation Builder; however, there is only one input panel called “Economics Navigator”.

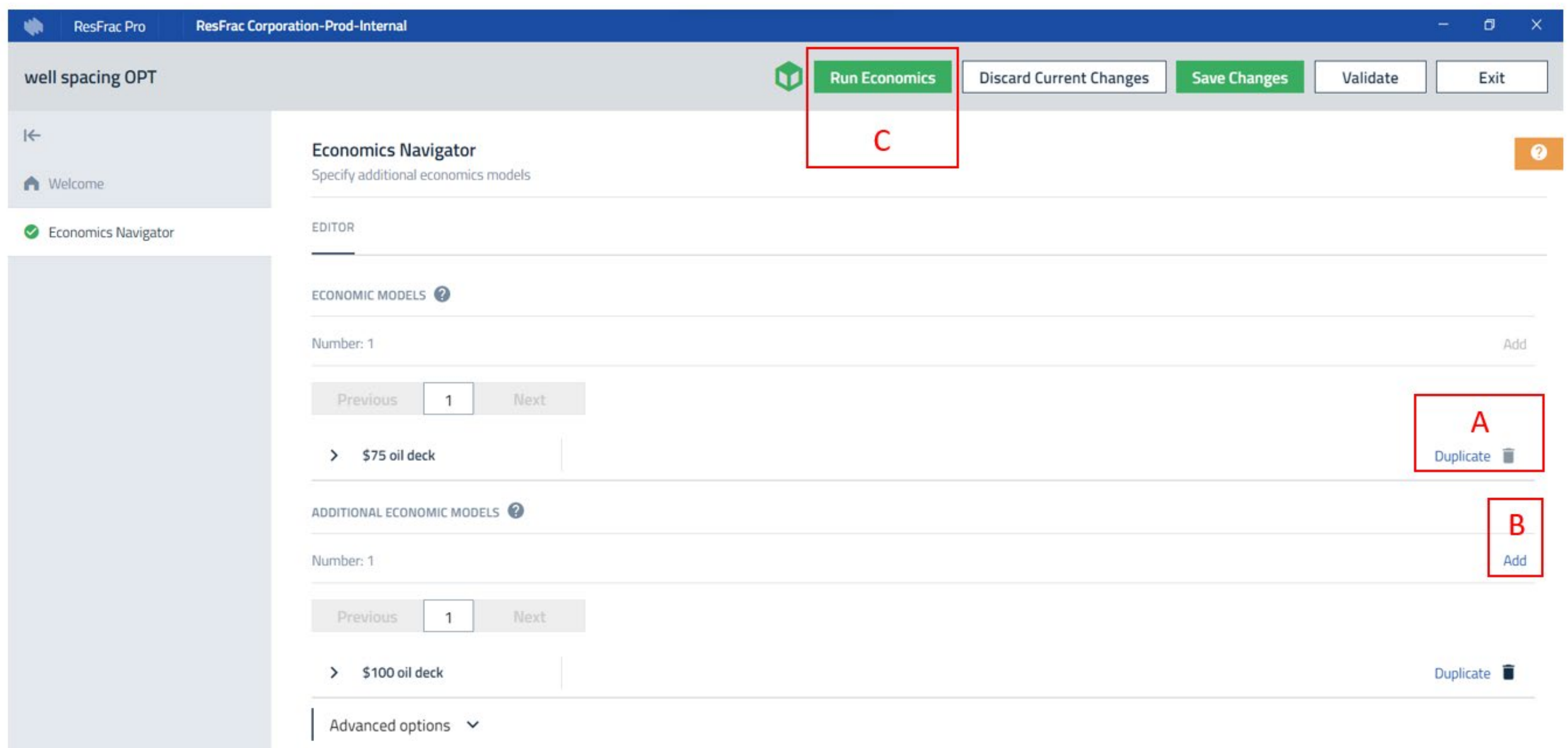

The existing Economic Models, if any, are listed above (labeled “A” in the image above). Additional economic models can be added to the simulation or workflow by clicking the “Add” button (labeled “B” in the image above).

Settings in the new economic model are added as they are in the original simulation. Once all inputs have been added, the user must click “Run Economics” (labeled “C” in image above) and this will append additional columns to the results file (sim_track_SIMNAME.csv) corresponding to the new economics. **Note: the sim_track_SIMNAME.csv cannot be open on the user’s machine when attempting to “Run Economics”.**

### Common parameter coupling within Parameter Groups

Sometimes it makes sense to couple Parameters within a Parameter Group either due to an intrinsic relationship between Parameters (such as porosity and permeability) or to maintain consistency in simulations (for example, maintaining the same degree of limited entry if changing clusters per stage).

To couple Parameters, two or more Parameters are added to the same Parameter Group. The user interface adds one Parameter Group per Parameter, so if multiple Parameters are being added to a Parameter Group, then one or more of the original Parameter Groups need to be removed from the Parameter Group table.

It is common to maintain the same designed limited entry (or equivalently perforation friction) in a Sensitivity or Optimization workflow that is changing the number of clusters per stage. The easiest way to do so is to use the perforation diameter to compensate for fewer or more clusters. In the field, constant limited entry between designs would likely be effectuated by changing the number of perforations; however, in simulations, using the perforation diameter is more convenient because diameter is a *continuous* parameter. The user adds both Clusters per Stage and Perforation Diameter to a parameter group, then uses inverted left and right multipliers for the two parameters. For example, in the image below the base simulation has four clusters, the sensitivity ranges are set such that the perforation friction for the left and right ends of the range (and any position between) remains the same.

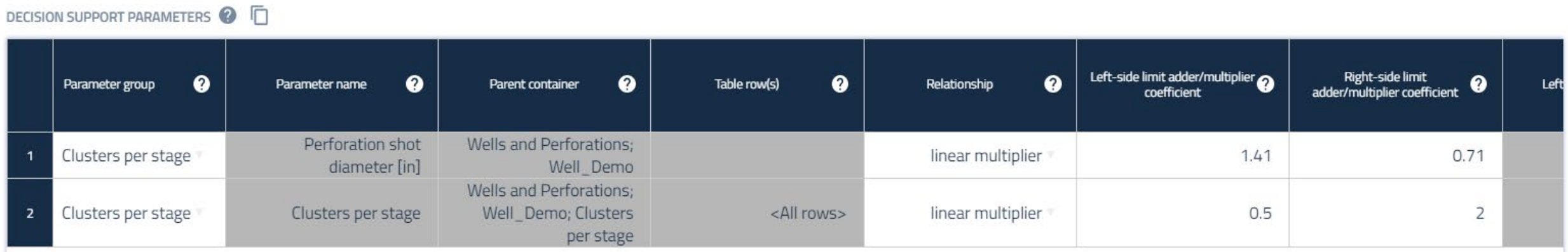
DECISION SUPPORT PARAMETERS

| | Parameter group | Parameter name | Parent container | Table row(s) | Relationship | Left-side limit adder/multiplier coefficient | Right-side limit adder/multiplier coefficient | Left |
|---|---|---|---|---|---|---|---|---|
| 1 | Clusters per stage | Perforation shot diameter [in] | Wells and Perforations; Well_Demo | | linear multiplier | 1.41 | 0.71 | |
| 2 | Clusters per stage | Clusters per stage | Wells and Perforations; Well_Demo; Clusters per stage | <All rows> | linear multiplier | 0.5 | 2 | |

Another common coupling is between Well Position Shift and Average Well Spacing (in the Economic Model). As wells are shifted to adjust well spacing, the Average Well Spacing in the Economic Model needs to be adjusted in tandem such that economics are appropriately calculated.

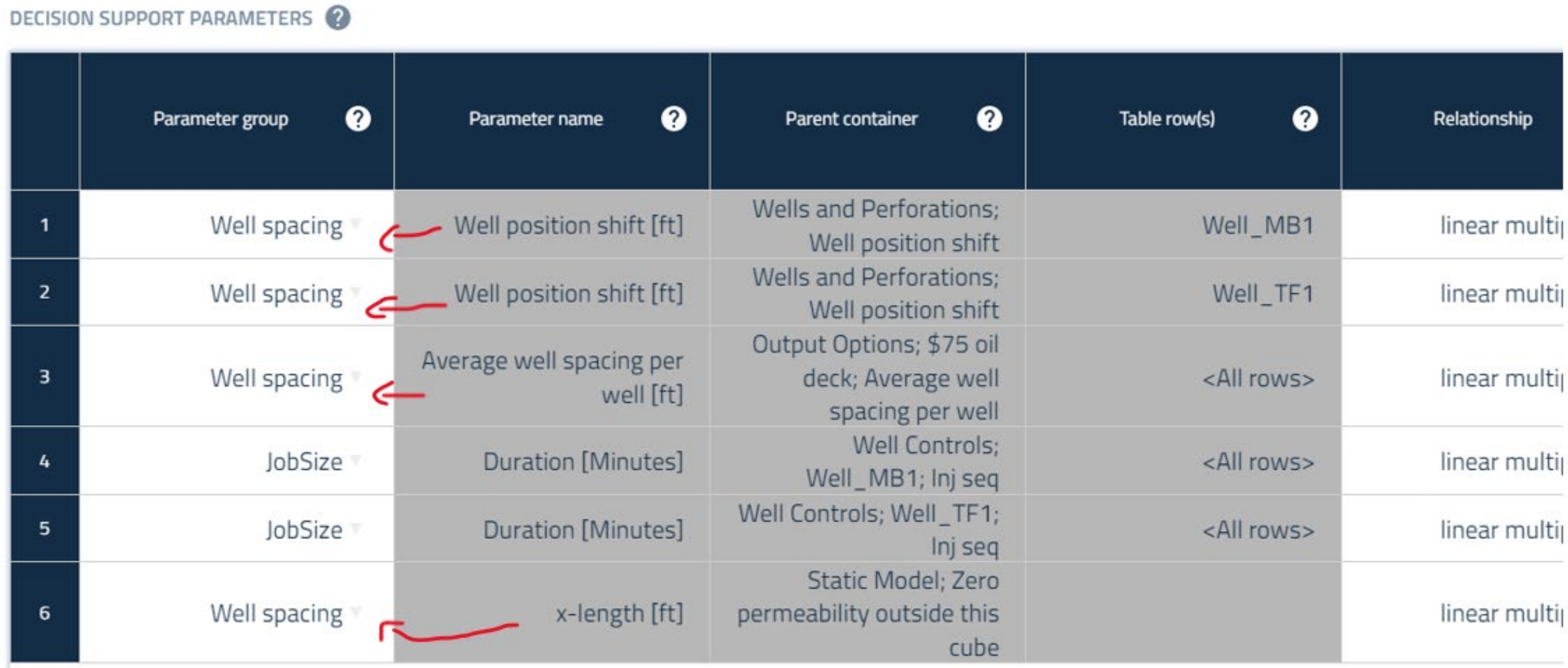
DECISION SUPPORT PARAMETERS

| | Parameter group | Parameter name | Parent container | Table row(s) | Relationship |
|---|---|---|---|---|---|
| 1 | Well spacing | Well position shift [ft] | Wells and Perforations; Well position shift | Well_MB1 | linear multi |
| 2 | Well spacing | Well position shift [ft] | Wells and Perforations; Well position shift | Well_TF1 | linear multi |
| 3 | Well spacing | Average well spacing per well [ft] | Output Options; $75 oil deck; Average well spacing per well | <All rows> | linear multi |
| 4 | JobSize | Duration [Minutes] | Well Controls; Well_MB1; Inj seq | <All rows> | linear multi |
| 5 | JobSize | Duration [Minutes] | Well Controls; Well_TF1; Inj seq | <All rows> | linear multi |
| 6 | Well spacing | x-length [ft] | Static Model; Zero permeability outside this cube | | linear multi |

For additional examples, see www.ResFrac.com/onlinetraining

## 11. ResFrac tutorial – putting the pieces together

In this section we will walk through an example consulting study. The data for this study is available on Dropbox at the following url: https://www.dropbox.com/sh/z8vmsjqeywrxdth/AAA_DXLzR8cyRcTXNunuL-Fja?dl=0. If you have trouble accessing the Dropbox folder, please contact us and we will provide the files in an alternative manner.

## 11.1 Tutorial problem statement

### PROJECT SUMMARY

Your company completed two wells in the Peter Pan Basin three years ago. Your company is now ready to drill the next generation of wells SE of the original pad. The budget has called for four wells, and the contract requires that the stage, fluid, proppant, and cluster design are locked in for these wells. Well spacing can still be changed. Your manager has asked you to advise on recommended well spacing for the upcoming pad.

Your approach to answer this question will be to construct a model of the existing two-well pad and calibrate the model to the three years of production data. That model will then be used as the basis of a sensitivity analysis on well spacing so that you can provide your manager with a recommended spacing as well as a measurement of how sensitive performance is to well spacing.

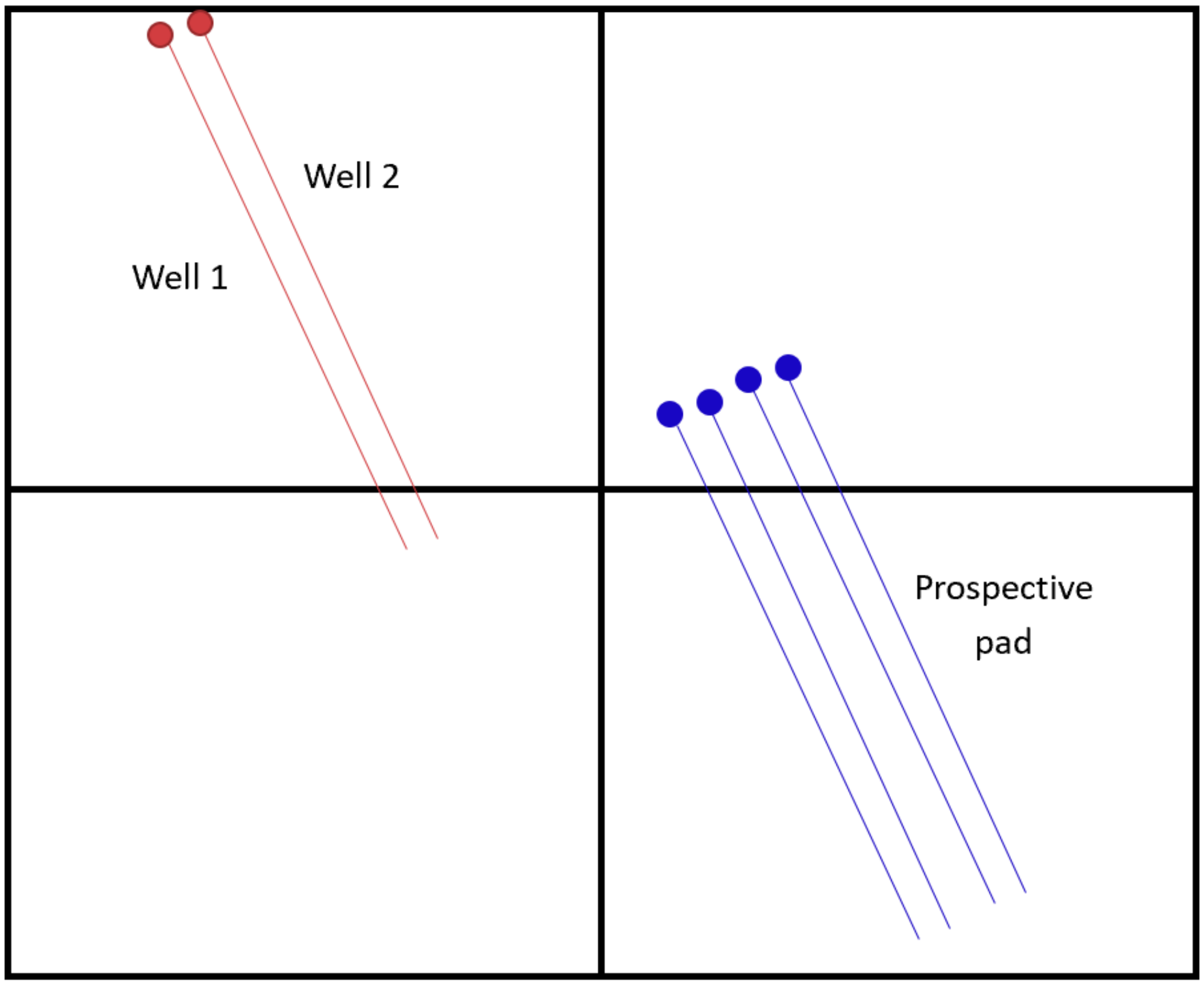

### DATA AVAILABLE

Your team has compiled the necessary information in a convenient Excel file for you. In the Excel sheet, there are tabs for:

- Overview information for Well 1 and Well 2
- Interpreted porosity, water saturation, pore pressure, Poisson's Ratio, stress, Young's Modulus, and permeability logs
- Well surveys
- A black oil model
- Proppant conductivity versus normal stress curves
- Slickwater properties

- Frac stimulation schedule
- Drawdown schedule
- Average ISIP values
- Three years of production data for the two wells

### PROJECT STRUCTURE

Follow the standard ResFrac modeling procedure:

1. Define modeling objectives.
2. Construct base case model and verify setup with key stakeholders.
3. Examine the historical data, list observations, and form and validate hypotheses. Discuss with stakeholders and get feedback before proceeding.
4. Calibrate the inputs to achieve a finalized history match. Share with stakeholders and collectively plan the initial set of sensitivity analysis simulations.
5. Perform initial sensitivity analysis and present to all key stakeholders.
6. Refine and finalize sensitivity analysis based on feedback from stakeholders.

As you go through the modeling process follow these key phases, completing each before moving onto the next. This will ensure a successful and expeditious project.

## 11.2 Define modeling objective and data available

Before starting any modeling project, it behooves you to first think through the project end to end.

First, what is the objective of the study? **Often, our goal is to accelerate the innovation cycle and inform the next design to test in the field.** You will need to communicate with the operator to explain what the design change is, and why it may be better. When it does get implemented, plan to track performance. Is there another pad nearby that is similar and can be compared with this one? Is there diagnostic information that could be gathered to track performance? You may even consider a direct A/B test. Within a six well pad, fracture two of the inner wells with one technique and the other two wells with another, and compare performance.

In the case of the tutorial data provided, our objective is to optimize well spacing in an upcoming pad. We have sufficient data to build and calibrate a model, so we will follow the recommended five phase modeling process in Sections 11.3 to 11.7 below: Construct and verify base case model; Examine the historical data, list observations, and form and validate hypotheses; History match available data; Initial sensitivity analysis; and Refine and finalize sensitivity analysis.

## 11.3 Construct and verify base case model

The first phase in a modeling study is to construct the base case model and verify the setup with all key stakeholders.

The ResFrac simulation builder is laid out in a logical progression such that you can proceed from one tab to the next. If you want more information about a particular setting, click on the ? button in the UI. Also, you can 'control f' for the name of the setting in the ResFrac Technical Writeup.

## Creating a simulation

To create a simulation, select Create Sim from the ResFrac job manager. This will bring up a menu that allows you to name the simulation, use a template to seed initial values (recommended), and choose field or metric units.

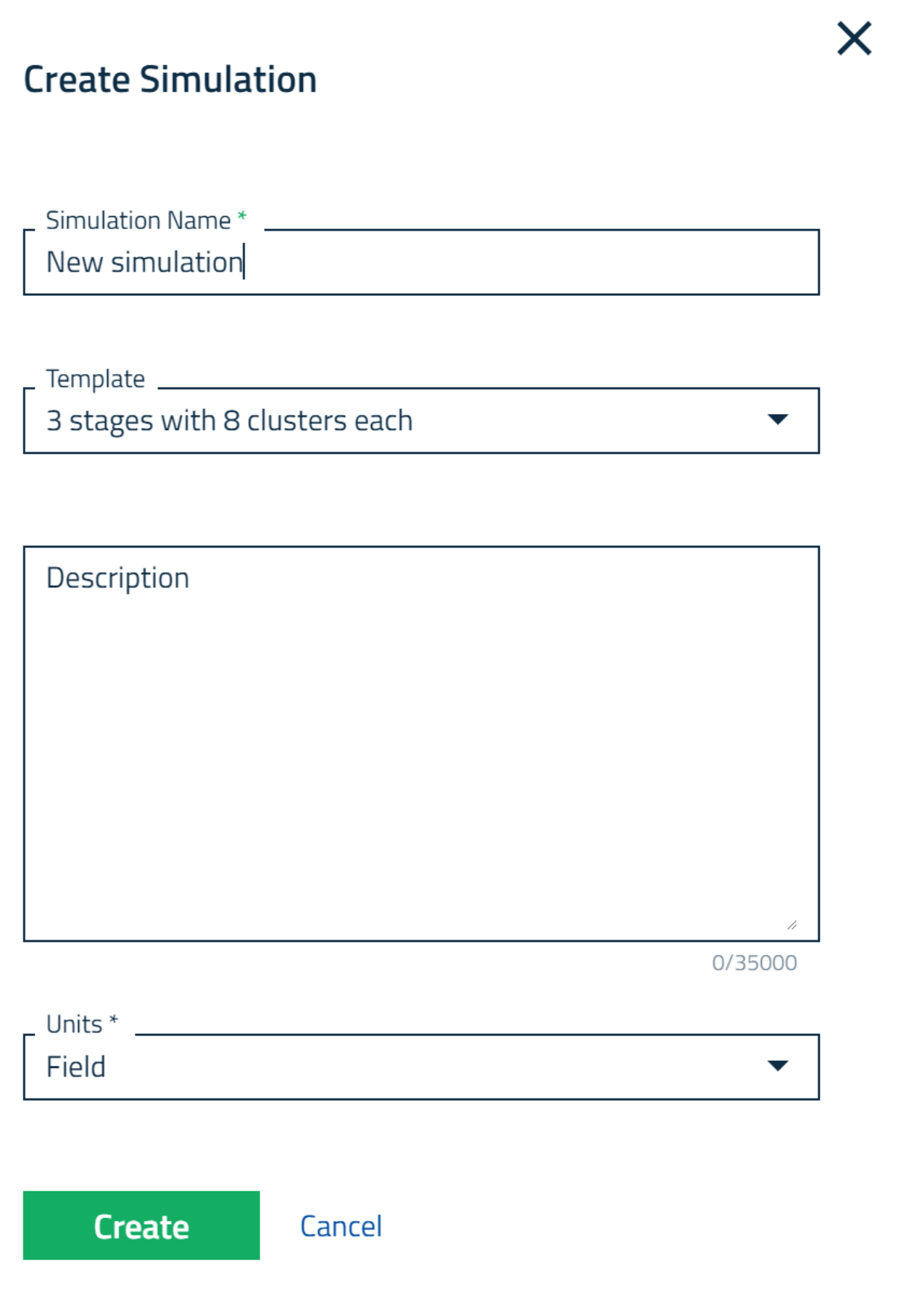

For field-scale simulations (well or well pad versus detailed DFIT simulations), the "3 stages with 8 clusters each" is the best template to use as a starting place. Pressing "Create" will open the simulation builder.

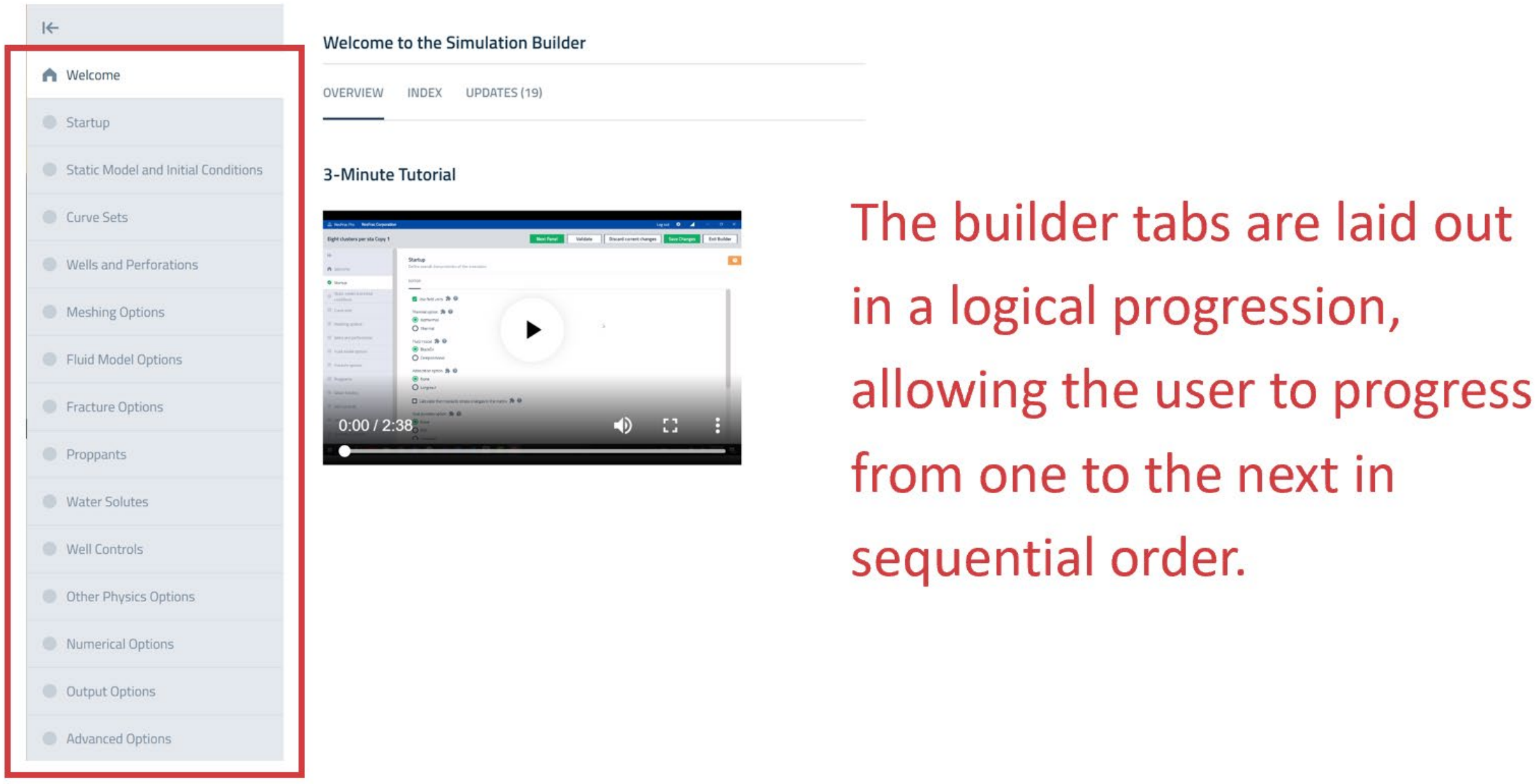

### Startup

The first input panel defines the physics of the simulation to be run. In most cases, the default parameters are sufficient for most simulations. If starting from a copy of a previous simulation, we recommend clicking the “Recommended settings wizard” to update default parameter values to the latest suggestions.

### Static model and initial conditions

On this panel the user specifies the properties of the reservoir being modeled, including geological and geomechanical layering. If you ever have a question about what a parameter is, clicking the “?” next to the parameter will open up a help panel with details about the specific parameter, equations the parameter is used in, and sometime figures or videos to supplement the description.

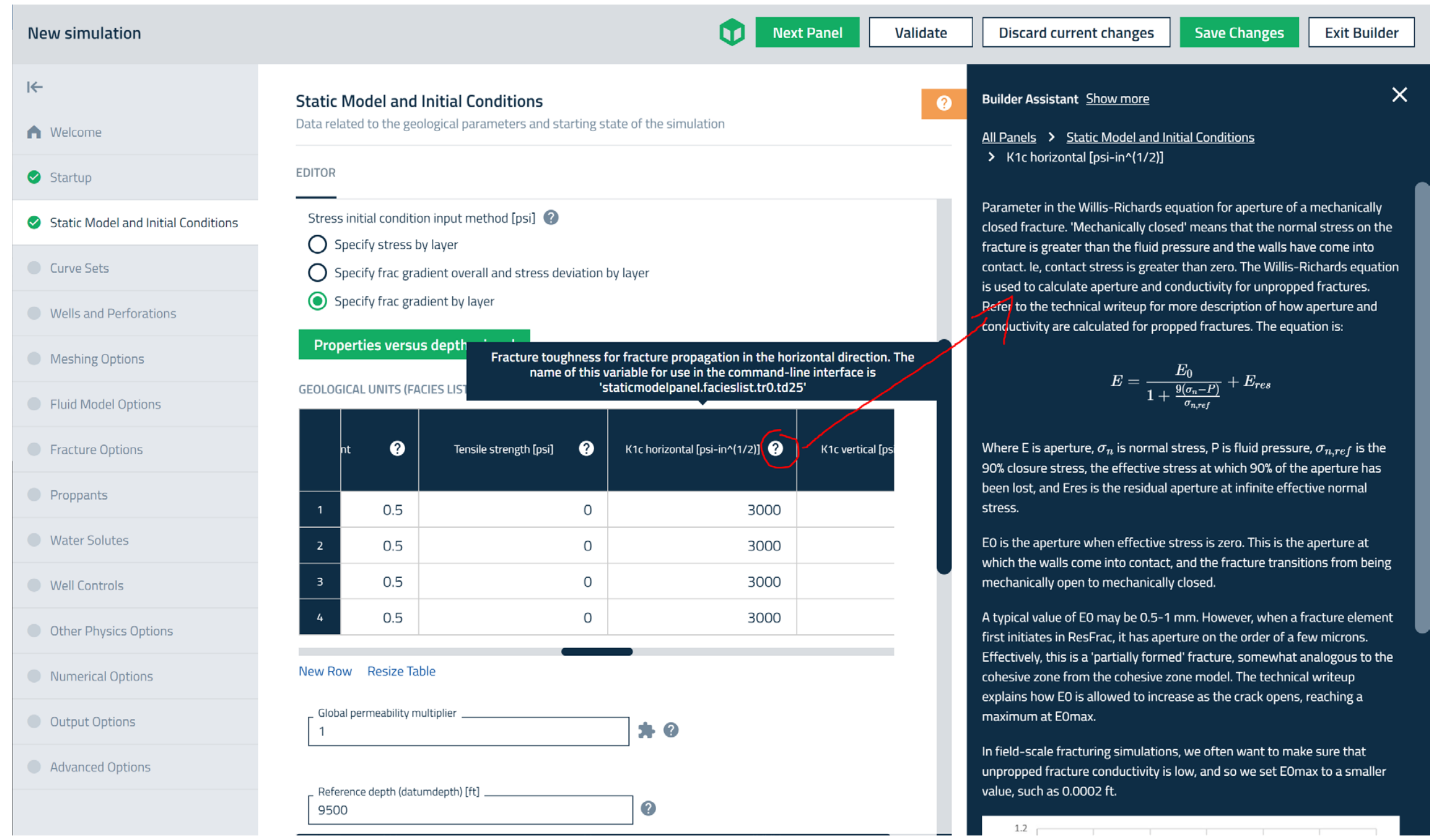

For the tutorial data set provided, geologic and geomechanical properties have been provided on a half foot resolution. To upscale these to an appropriate model-scale, we can use the “Properties versus depth wizard.”

On the logs tab of the Excel sheet provided there are seven properties which we will enter into the properties versus depth wizard. Next we need to select a target layer thickness, which is the maximum layer thickness for the resulting upscaled model. Fifty feet is often a good value to use here. The wizard will subdivide the inputted property profile into layers based on the rate of change of Shmin, permeability, water saturation, and porosity. Finally, you can specify specific depths that you would like as layer boundaries. In the “Well data” tab of the input excel, we see that Pay 1 and Pay 2 have tops of 8243 and 8579 feet, respectively, so we will enter these into the wizard.

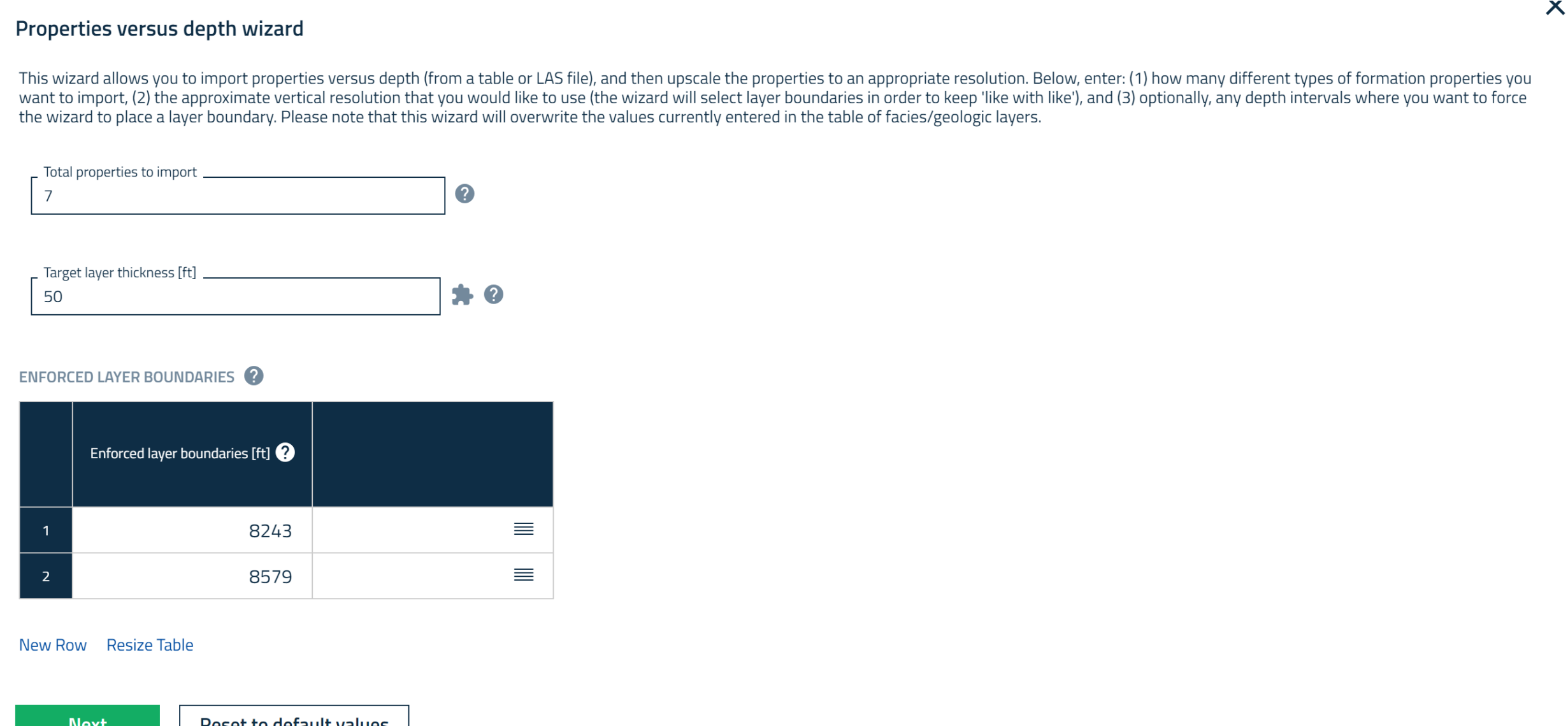

The next wizard prompt will ask you to specify which properties you are inputting.

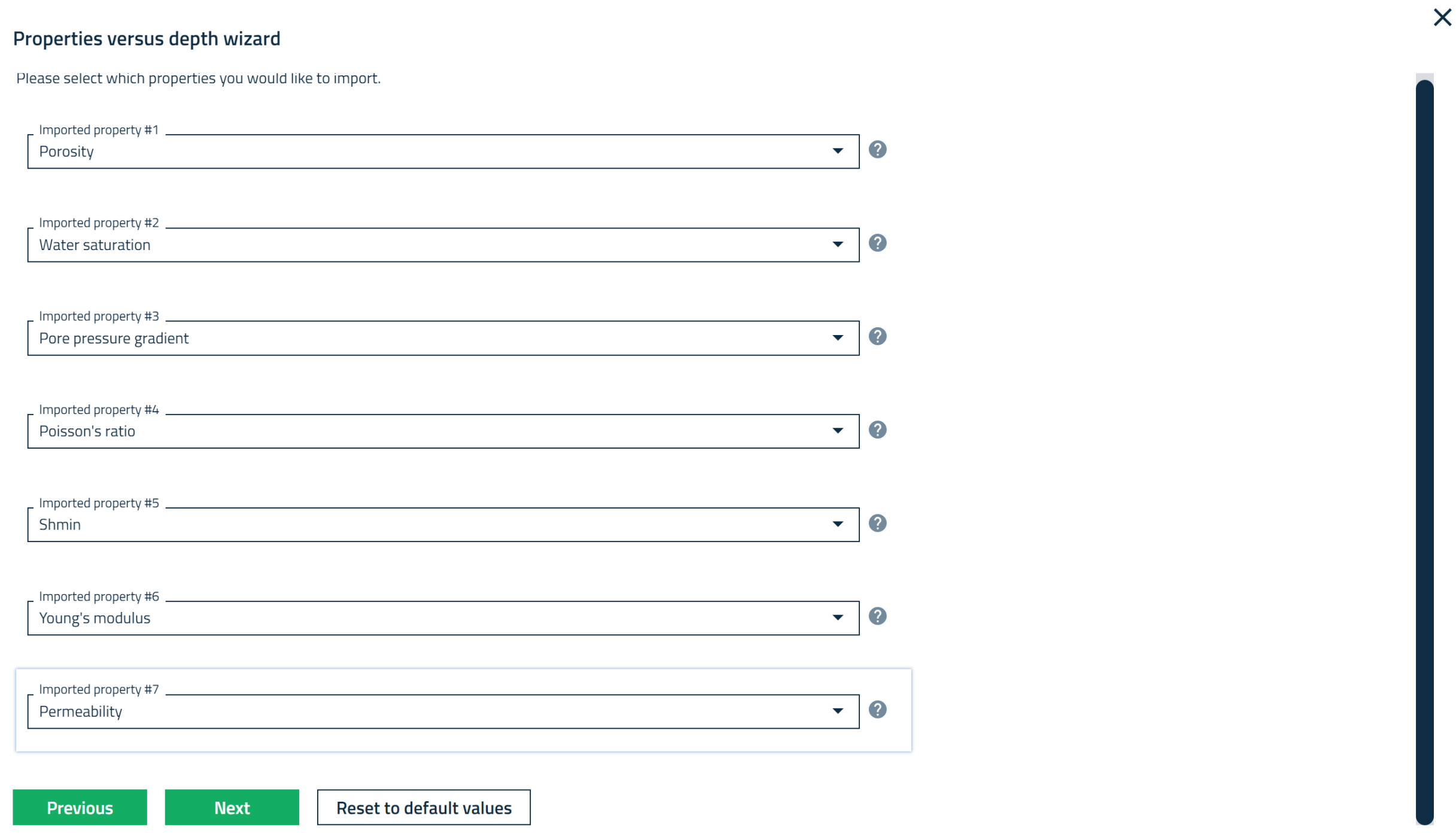

Then, in the final wizard prompt, you will paste in the full half-foot resolution properties.

## Properties versus depth wizard

The table of values for import is below. Please manually modify, as needed, and then press apply to complete the import/upscaling.

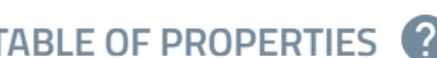

| | Depth [ft] | Porosity (unitless) | Water saturation (unitless) |
|---|---|---|---|
| 1 | 7500 | 0.1016 | 0.9085 |
| 2 | 7500.5 | 0.1004 | 0.8752 |
| 3 | 7501 | 0.0883 | 0.7109 |
| 4 | 7501.5 | 0.1071 | 0.6535 |
| 5 | 7502 | 0.0809 | 0.5896 |
| 6 | 7502.5 | 0.0758 | 0.3297 |
| 7 | 7503 | 0.0896 | 0.2132 |
| 8 | 7503.5 | 0.08 | 0.2378 |
| 9 | 7504 | 0.0879 | 0.2752 |

New Row Resize Table

Previous Apply Reset to default values

Based on these inputs, you will see that the wizard has upscaled the log-resolution property profile to 61 layers. You will also see that there is an error in the table as indicated by the red error triangle above the table. This error is indicating that there are empty cells in your property table.

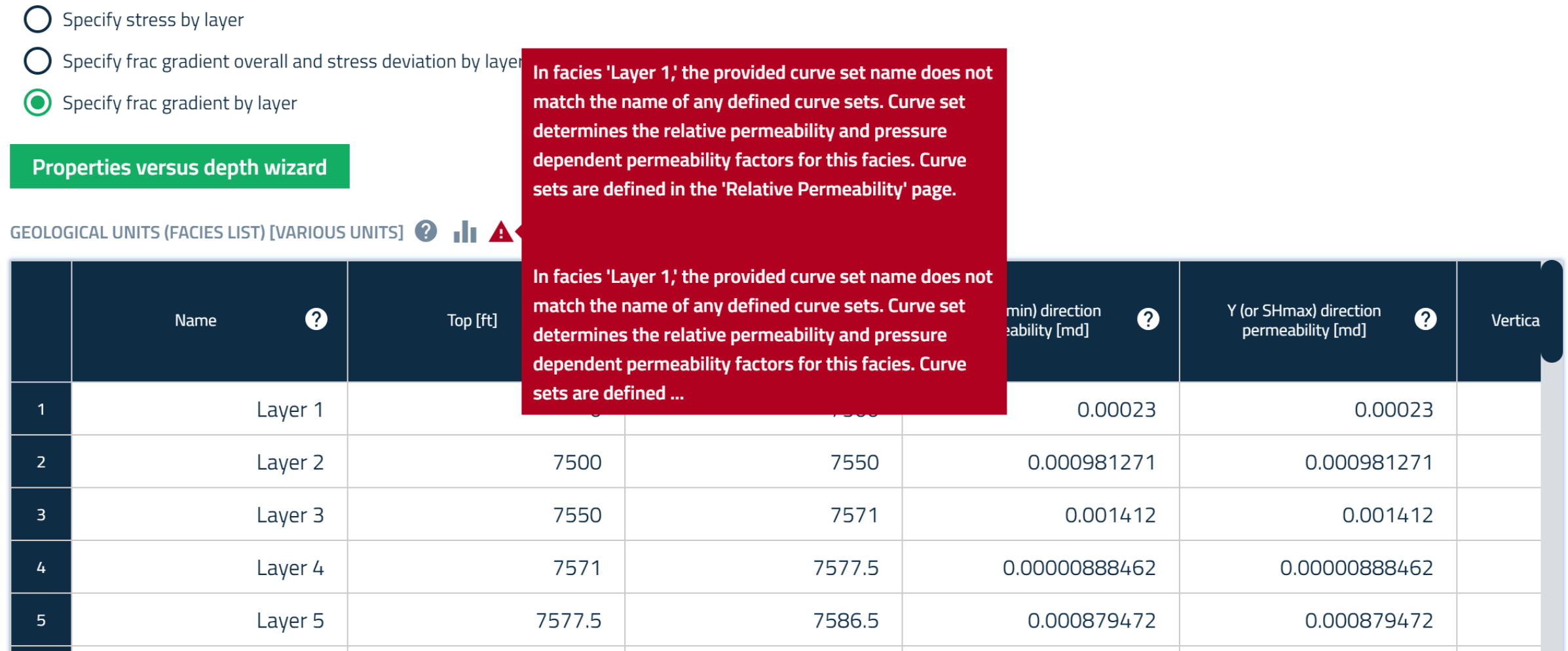

To resolve the error we need to fill in the Curve set name (which tells the simulator which set of relative permeability curves apply to each facies) for each row in the table (we can quickly do this by entering Curve set 1 into the top row then pressing Ctrl + D to copy down).

There are other optional parameters in the static model, like specifying proppant embedment capacity for each layer, that you can explore by using the “?” in the header row.

One very helpful thing to do is to rename particular layers, then mark those as facies tops to ‘show in viz tool’. In the first column of the facies table, you can rename the layers at 8243 and 8579 as Pay 1 and Pay 2, then on the far right of the table select these layer names as visible in the viz tool.

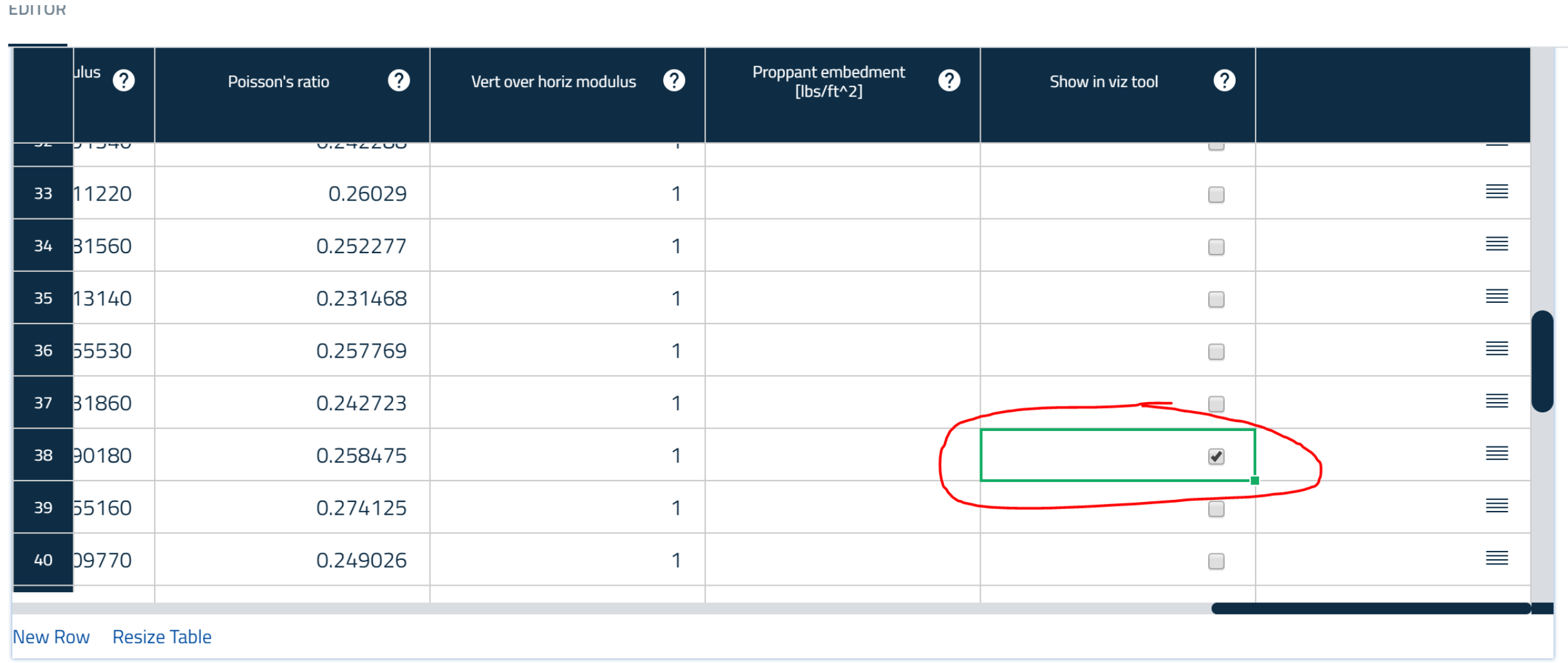

| | ulus | Poisson's ratio | Vert over horiz modulus | Proppant embedment [lbs/ft^2] | Show in viz tool |
|---|---|---|---|---|---|
| 33 | 11220 | 0.26029 | 1 | | ☐ |
| 34 | 31560 | 0.252277 | 1 | | ☐ |
| 35 | 13140 | 0.231468 | 1 | | ☐ |
| 36 | 55530 | 0.257769 | 1 | | ☐ |
| 37 | 31860 | 0.242723 | 1 | | ☐ |
| 38 | 90180 | 0.258475 | 1 | | ☑ |
| 39 | 55160 | 0.274125 | 1 | | ☐ |
| 40 | 09770 | 0.249026 | 1 | | ☐ |

Additionally, once the data is loaded in, a preview of the properties is available by clicking the graph button above the table.

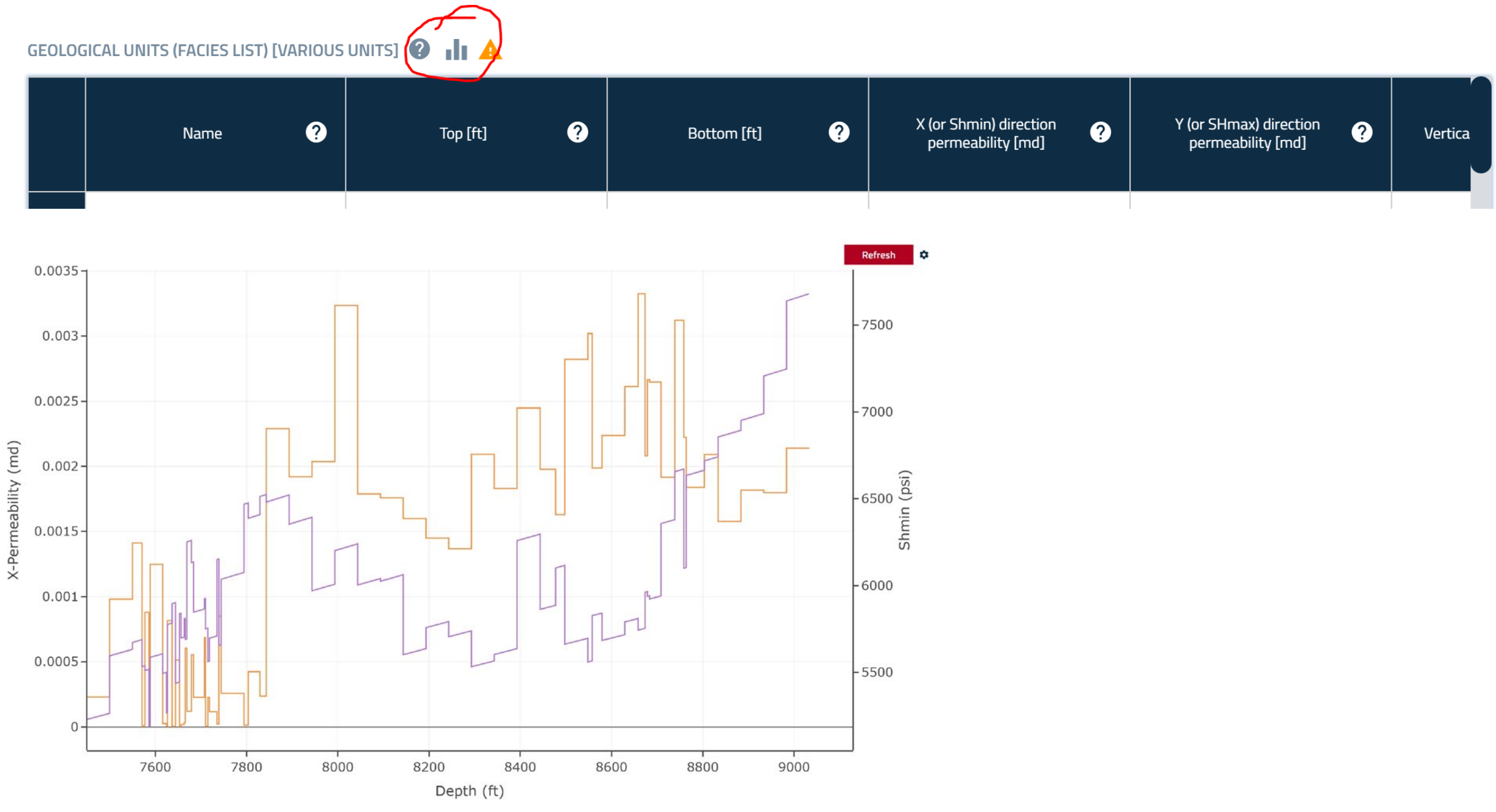

The final input from the data provided is to specify the azimuth of SHmax (which is the azimuth of the hydraulic fractures). This value is entered in a box below the static model property table.

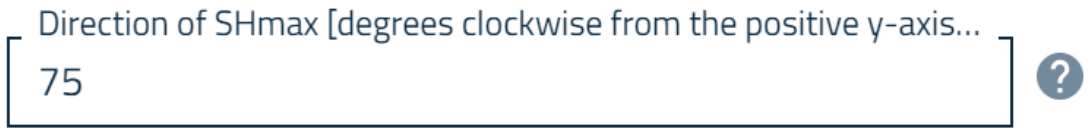

## Curve sets

Each curve set is a set of relative permeability and pressure dependent permeability curves. Every simulation needs at least one curve set, but you can specify multiple if appropriate (for instance, two pay zones that have different relative permeability behavior). In our example, we will use a single curve set across all facies. The Brooks Corey parameters defining the relative permeability are usually used as history matching parameters as initialized at the default values for the first simulation run.

The second component of each curve set is the pressure dependent permeability curve (PDP). We use constitutive relationships in ResFrac to describe phenomena below model-scale. PDP is one of those constitutive relations to approximate the effect of multiple fracture bands and possible accelerated leakoff into natural fracture networks. At the top of the curve sets tab, there is a wizard to help you setup your PDP parameters.

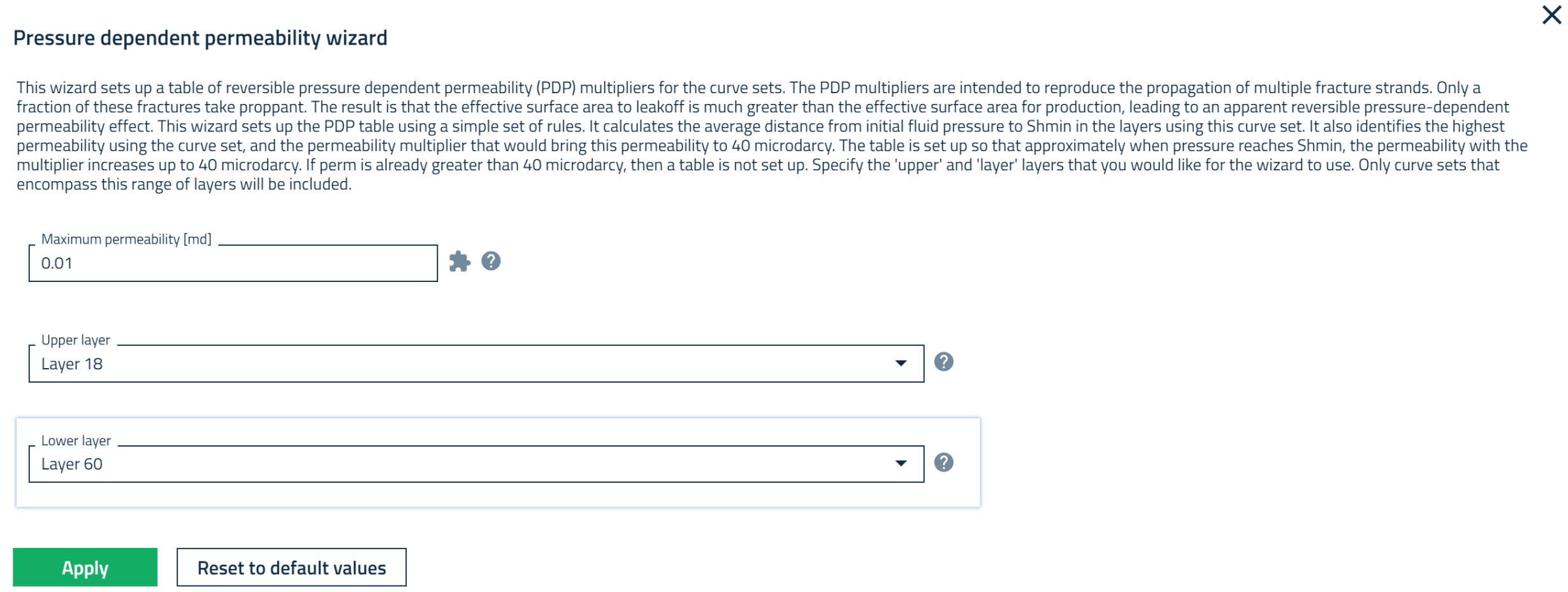

## Wells and perforations

In the wells and perforations tab, the user has multiple options for specifying wells and completion design. For the tutorial data, we import two wellbore surveys and set up our stages and completions using the “in-line method”. Because we started with the template simulation, we first need to delete the existing well by pressing the trash can next to the well name.

To import the two well surveys, we use the Wellbore survey import wizard. In the wizard you need to specify the wellhead location. We are assuming that both wells are drilled from the same surface location so we can use (0,0,0) for the wellhead location of each. On the first well, we can also choose to recenter the model region around a certain MD depth. For the example, we will choose 10,000 feet.

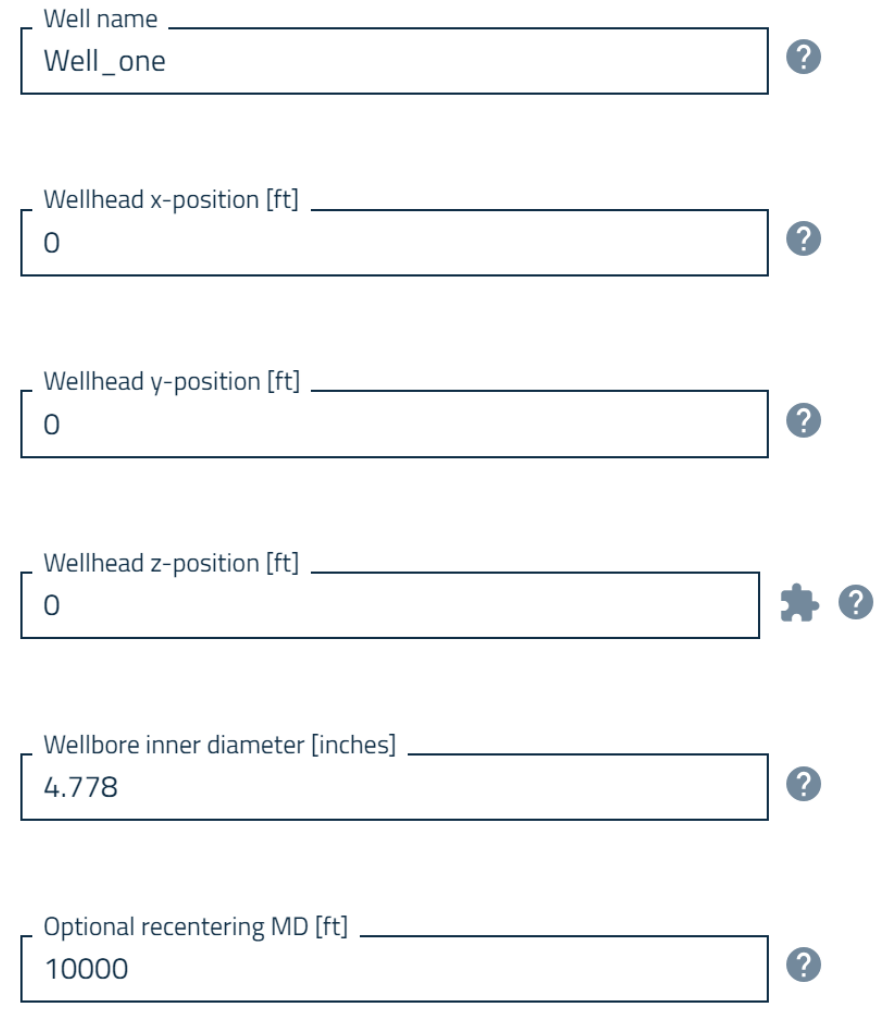

After running the wizard, we will see our well populated in the builder.

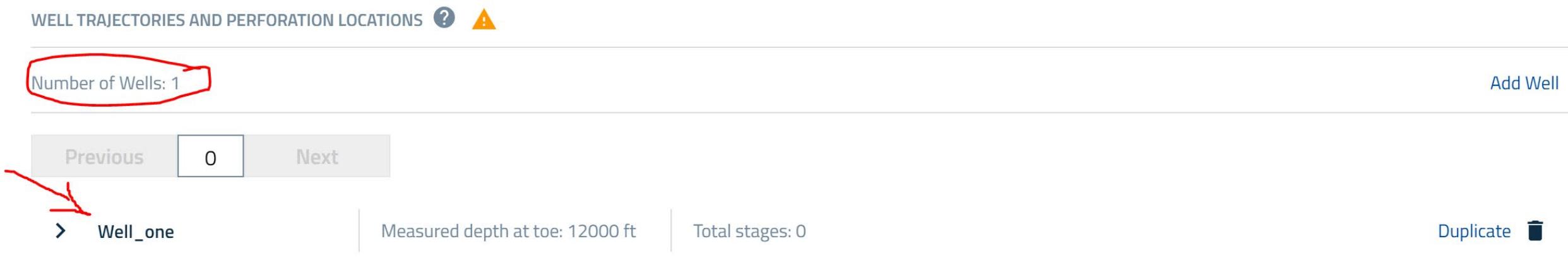

Running the wizard a second time for Well Two, we will follow the same steps, but no longer need to recenter the model around an MD of 10,000 ft.

With both wells imported, we can click the preview button to check the well trajectories.

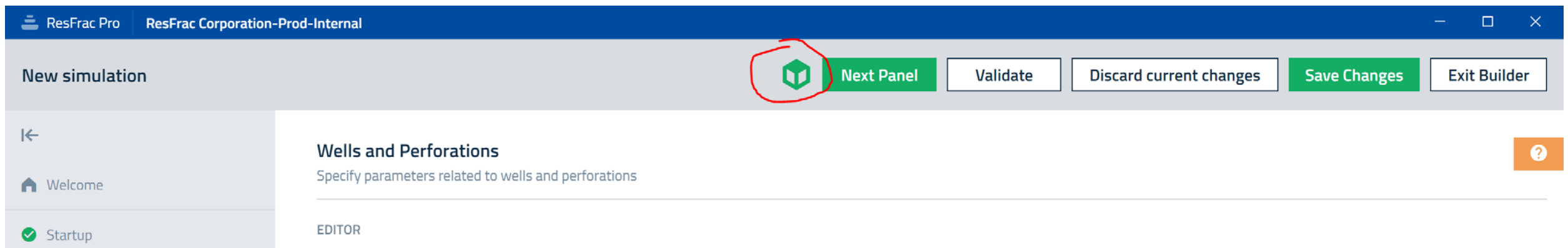

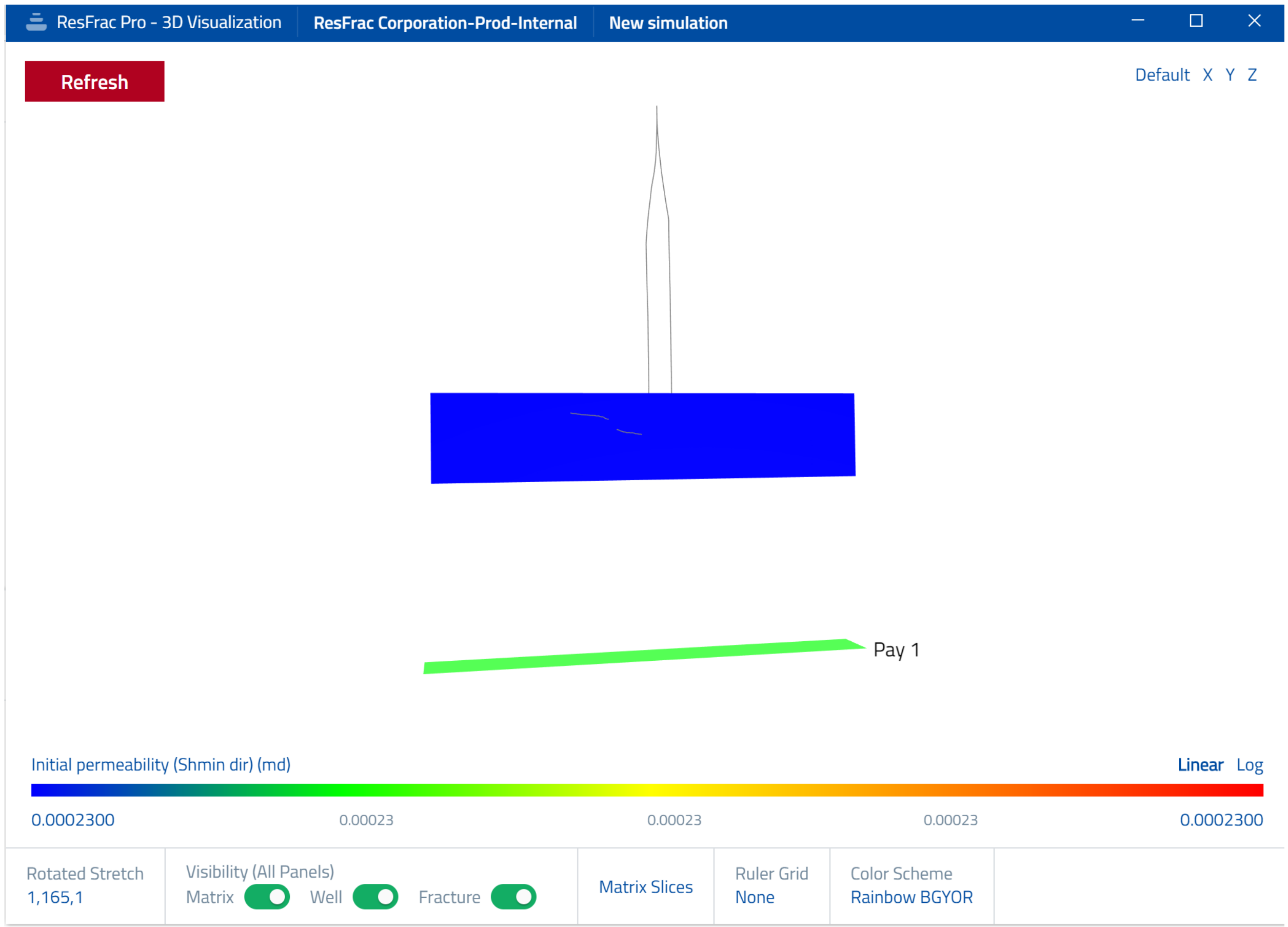

The wellbores are not yet landed in the pay zone/s. This is a common occurrence because formations have dip, and the wellbore surveys and well used for stratigraphy may not be located in exactly the same place. To correct the landing zones, we can use the “Adjust well landing depth wizard” to place Well One in Pay Zone 1 and Well Two in Pay Zone 2.

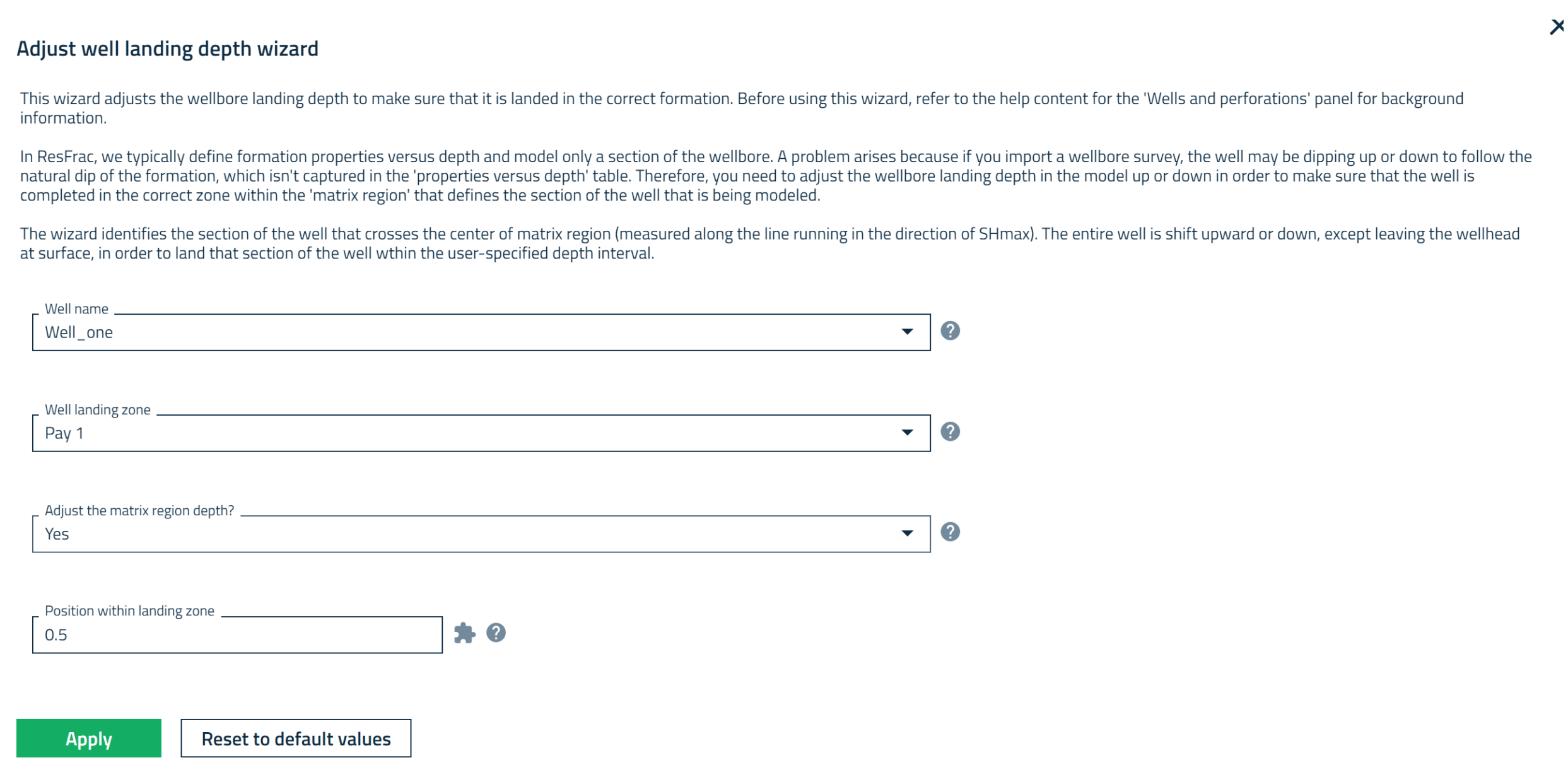

After running the wizard for each well, we can see the well trajectories and landing zones look correct.

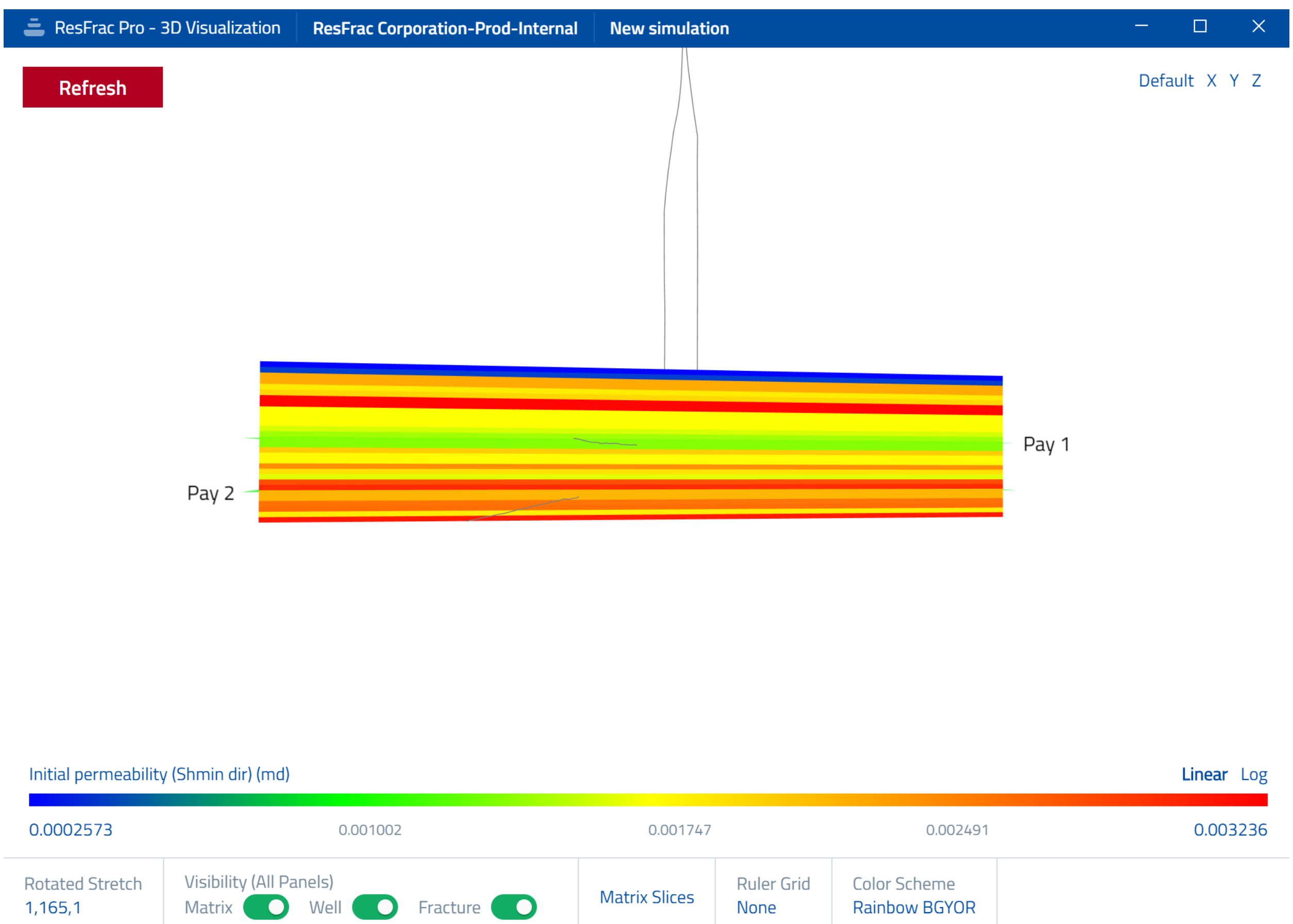

Because we typically only model a portion of the lateral, we use the "Well stages setup wizard" to define the stages we'd like to model along our lateral. The data pack provided in the tutorial indicates that stage length is 250 feet in each of the wells, so to model two stages in each of the wells, we would model 500 feet of lateral.

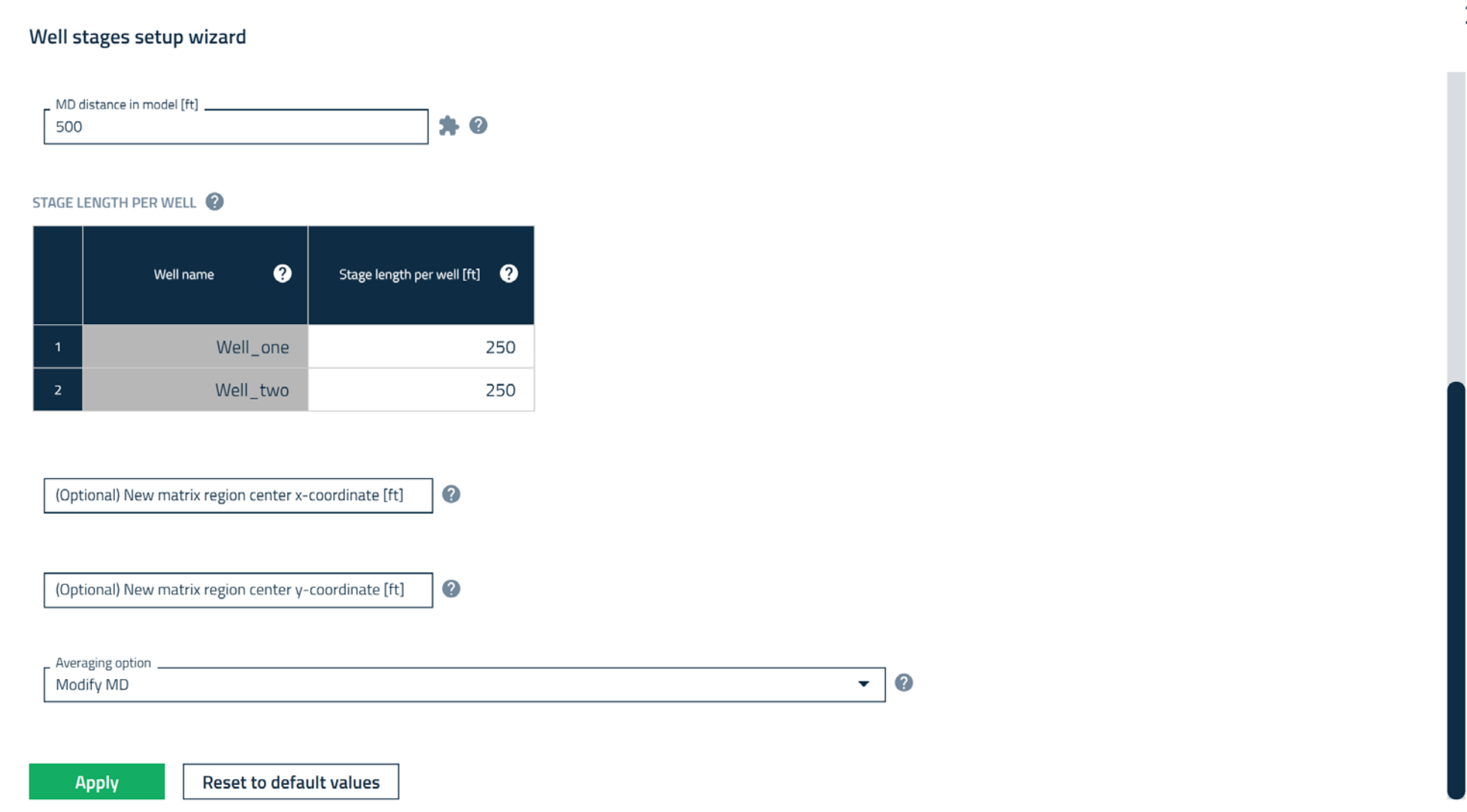

Now, to define the cluster spacing and perforation design, we do so in the parameter list below the well vertices table. Because we have consistent cluster spacing and perforation design along the stage, we are using the “in-line” method. If we had variable perforation designs, such as tapered perforations, we can convert from “in-line” to “MD-specified” using the green button at the top of the Wells and perforation tab.

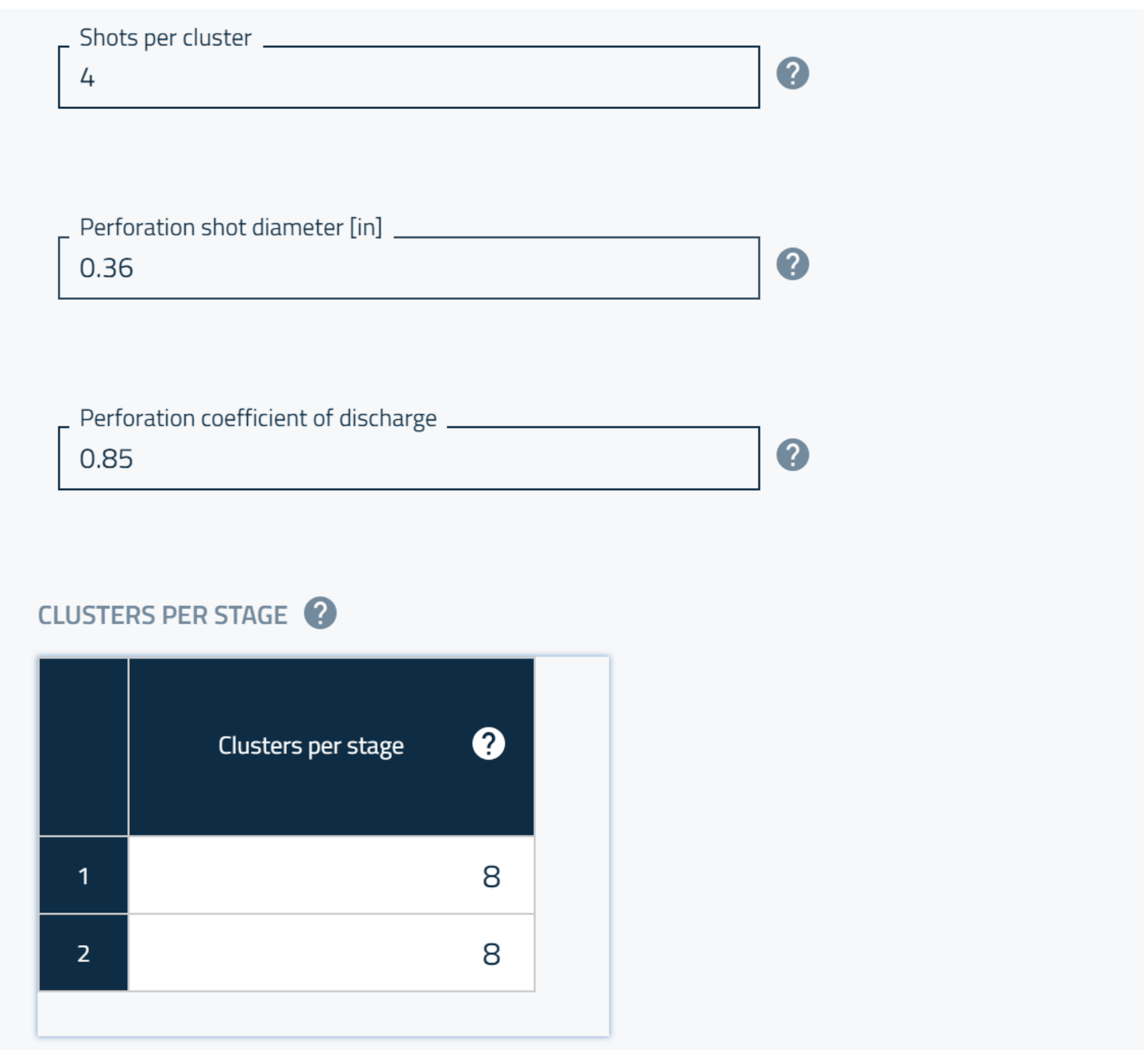

Finally, because we are modeling a section in the middle of the lateral, we want to define external fractures on the toe-side of our model to account for the stress shadow of stages not included in the model. The easiest way to define the coordinates of the external fractures, is to go into the well vertices table and find the vertex corresponding to the beginning of stage 0 (first stage outside of the model).

Set up the external fracture length and height to be equal to approximately the length and height of the fractures created by the model. To select an appropriate value for ‘net pressure’, one strategy is to run a single stage model without any external fracture, and calibrate it to approximately match the actual fracture length from the data. Specify a stress observation plane, and observe the magnitude of the cumulative stress shadow built up along the stage. This can be used as the net pressure for the external fracture. Alternatively, you could set the net pressure equal to the amount of cumulative ISIP buildup seen in the actual data from the first stage to the second, third, and fourth.

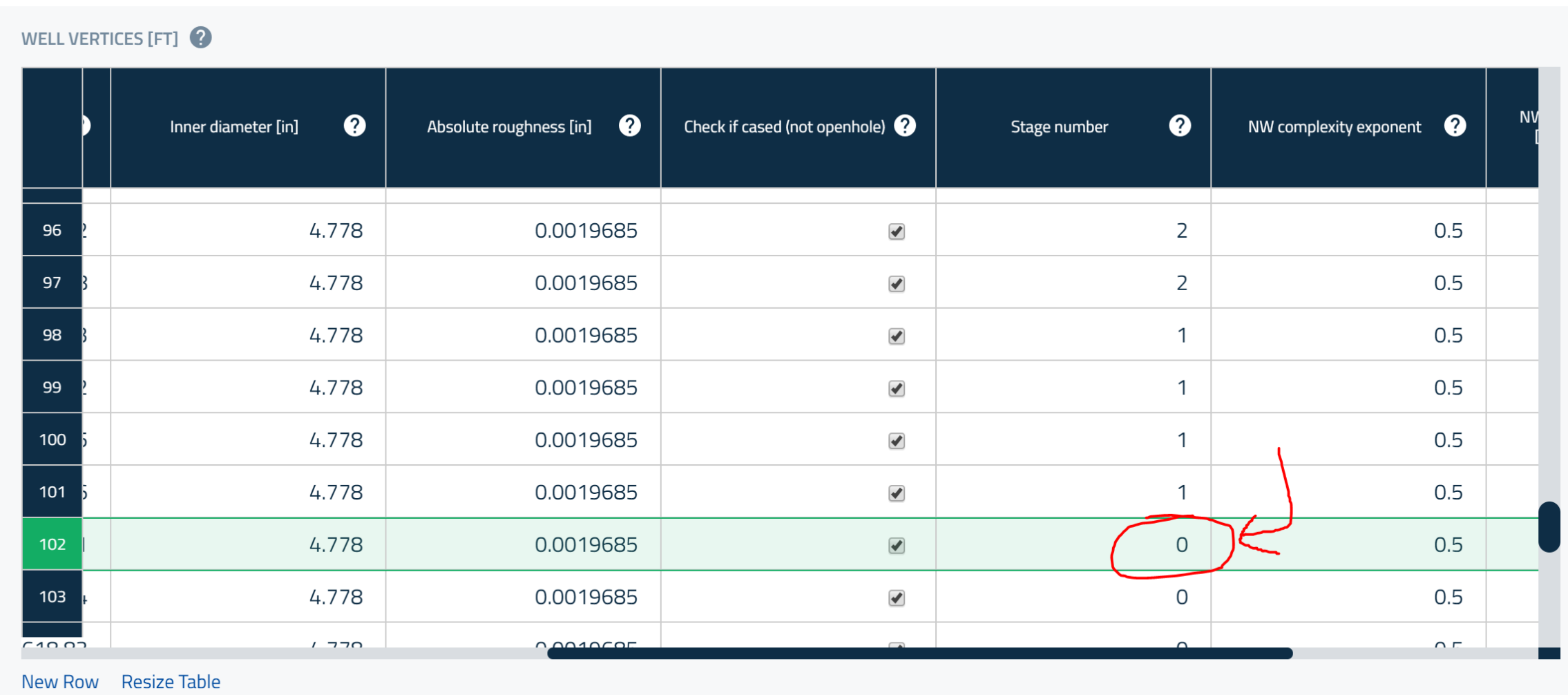
WELL VERTICES [FT]

| | Inner diameter [in] | Absolute roughness [in] | Check if cased (not openhole) | Stage number | NW complexity exponent |
|---|---|---|---|---|---|
| 96 | 4.778 | 0.0019685 | ✔ | 2 | 0.5 |
| 97 | 4.778 | 0.0019685 | ✔ | 2 | 0.5 |
| 98 | 4.778 | 0.0019685 | ✔ | 1 | 0.5 |
| 99 | 4.778 | 0.0019685 | ✔ | 1 | 0.5 |
| 100 | 4.778 | 0.0019685 | ✔ | 1 | 0.5 |
| 101 | 4.778 | 0.0019685 | ✔ | 1 | 0.5 |
| 102 | 4.778 | 0.0019685 | ✔ | 0 | 0.5 |
| 103 | 4.778 | 0.0019685 | ✔ | 0 | 0.5 |

New Row Resize Table

Scrolling to the left of the well vertices table, we can copy the x, y, and z coordinates for the start of stage 0, and paste them into the external fractures table.

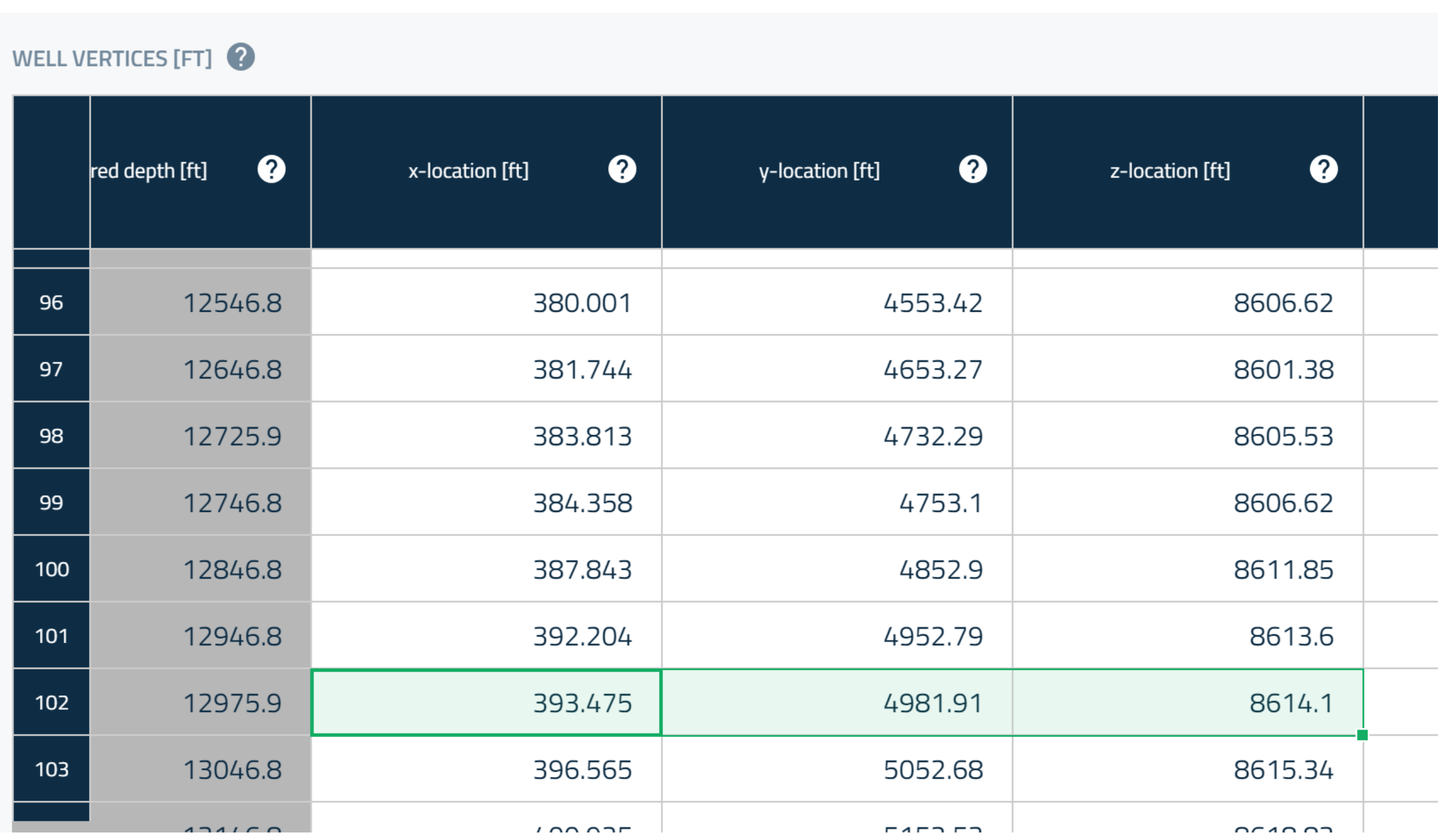
WELL VERTICES [FT]

| | red depth [ft] | x-location [ft] | y-location [ft] | z-location [ft] |
|---|---|---|---|---|
| 96 | 12546.8 | 380.001 | 4553.42 | 8606.62 |
| 97 | 12646.8 | 381.744 | 4653.27 | 8601.38 |
| 98 | 12725.9 | 383.813 | 4732.29 | 8605.53 |
| 99 | 12746.8 | 384.358 | 4753.1 | 8606.62 |
| 100 | 12846.8 | 387.843 | 4852.9 | 8611.85 |
| 101 | 12946.8 | 392.204 | 4952.79 | 8613.6 |
| 102 | 12975.9 | 393.475 | 4981.91 | 8614.1 |
| 103 | 13046.8 | 396.565 | 5052.68 | 8615.34 |

After doing so, sometimes you'll see that the external fracture is still inside the model domain (and it is supposed to be outside).

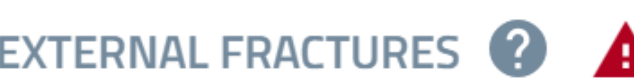

| | Center x-coordinate [ft] | Center y-coordinate [ft] | Center z-coordinate [ft] |
|---|---|---|---|
| 1 | 24.5662 | 4884.22 | 8276.27 |
| 2 | 393.475 | 4981.91 | 8614.1 |

The best way to resolve this is to add a few feet to the x and y coordinates to shift the fracture outside of the model domain.

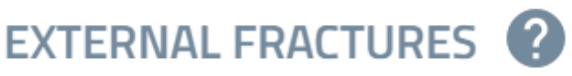

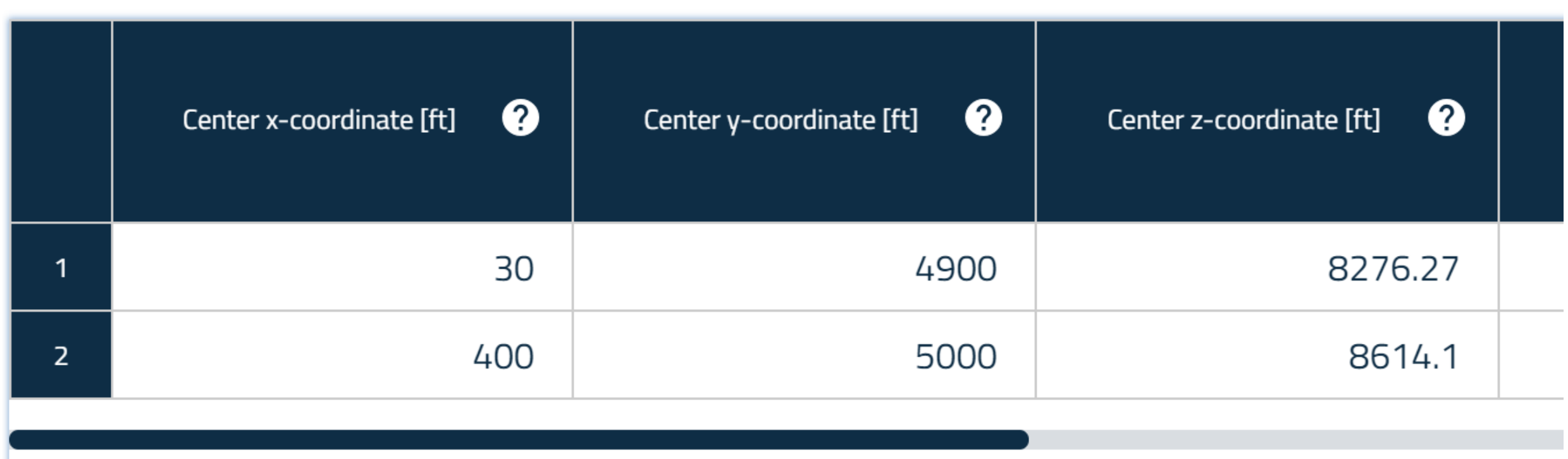

| | Center x-coordinate [ft] | Center y-coordinate [ft] | Center z-coordinate [ft] |
|---|---|---|---|
| 1 | 30 | 4900 | 8276.27 |
| 2 | 400 | 5000 | 8614.1 |

You can check the placement of the external fractures in the 3D preview.

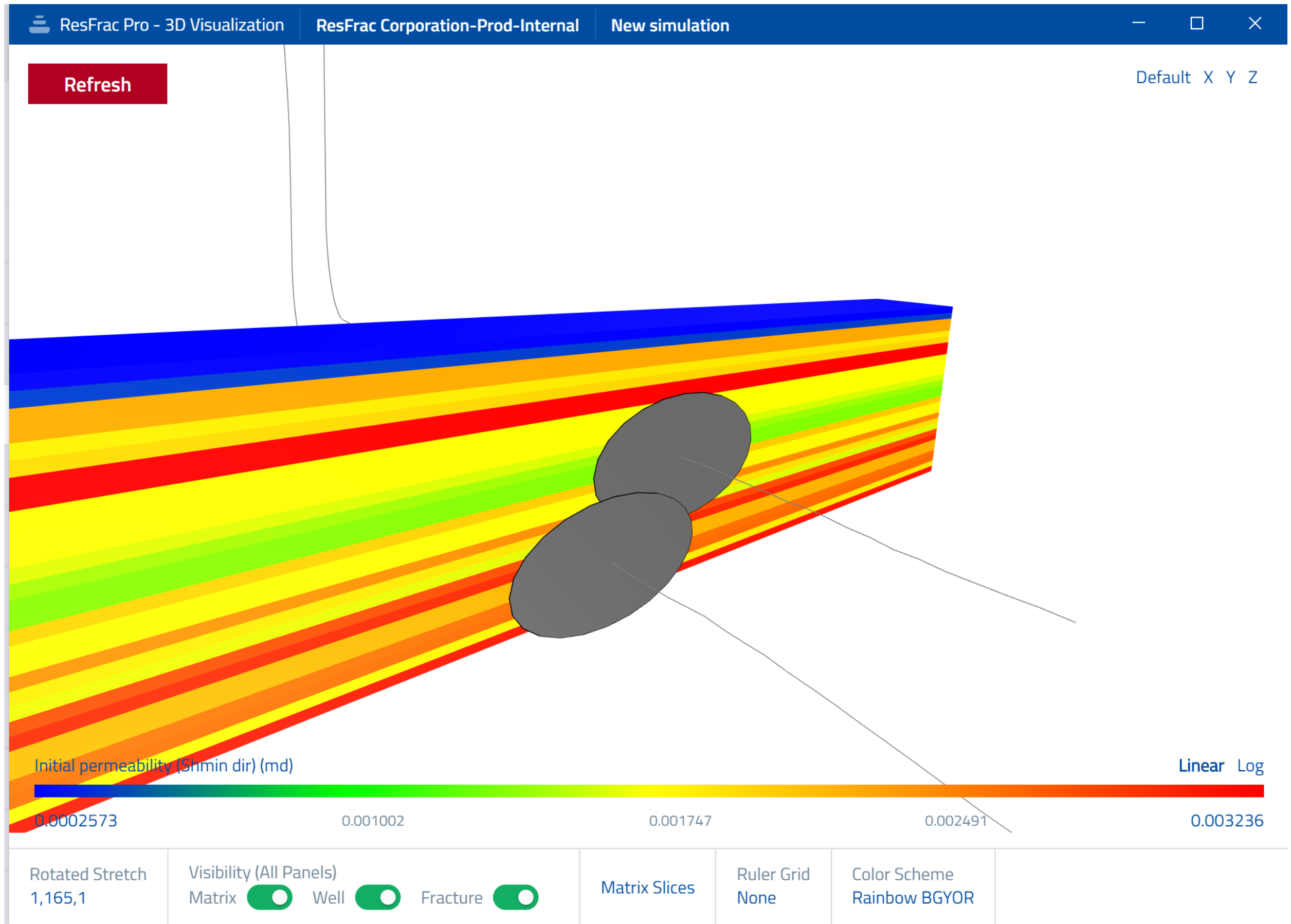

External fracture properties are refined during the history match, however, your history matching diagnostics can provide good starting values. Our matching data indicates that we should expect fracture approximately 1000 ft long and 500 ft high, so we use those same dimensions for our external fractures. Additionally, we sent net pressure to what we think is a good approximation of the stress shadow from a prior stage; in this case, 250 psi.

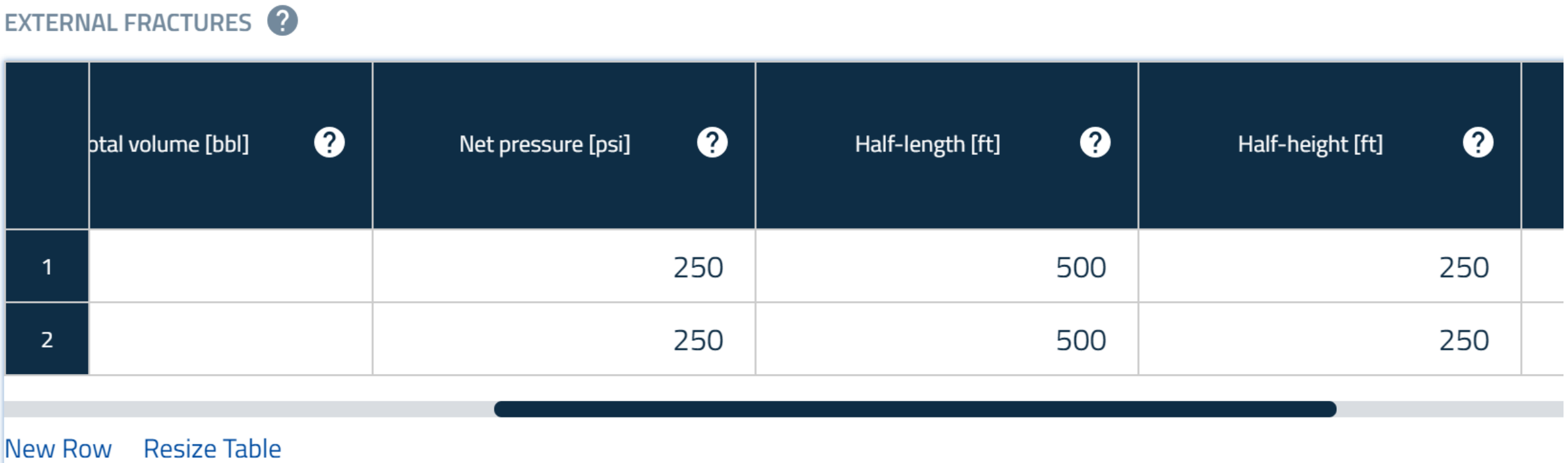

EXTERNAL FRACTURES

| | otal volume [bbl] | Net pressure [psi] | Half-length [ft] | Half-height [ft] |
|---|---|---|---|---|
| 1 | | 250 | 500 | 250 |
| 2 | | 250 | 500 | 250 |

New Row Resize Table

## Meshing

In the meshing tab we define the resolution of the mesh in our simulation. ResFrac has three, non-conforming meshes: the wellbore, the fractures, and the rock matrix. We specify separate dimensions for each. Some general rules of thumb to make setting up the mesh easier are:

- The wellbore element length should be less than your closest cluster spacing. This parameter does not impact runtimes much. Our wells have 31.25 ft cluster spacing, so the default 20 ft wellbore element length is sufficient. However, if we plan to test tighter cluster spacing in the future, we might want to use 10 ft well element length so that we don't need to change it in the future.
- The fracture element size has the largest impact on simulation runtimes of any parameter in the simulation. The ResFrac code can yield a good result with a relatively coarse fracture mesh, and so we can use what may seem like a coarse mesh for the fractures. It is common to use 60 ft or more for fracture element length. There is also an option to apply an aspect ratio (which makes the element height less than the length), though in this case we will chose to use square elements.
- The rock matrix mesh coordinate system is aligned with the Shmin, SHmax, and z dimensions.
    - In the Shmin direction, we should try to have at least one element per cluster. We are modeling two stages of eight clusters each, so we should use at least 16 elements in the Shmin direction. The length in the Shmin direction is set by the Well stages setup wizard.
    - There is not much matrix flow in the direction along fracture strike (SHmax direction), and so we can use coarse elements of 100 ft or more. The length of the mesh in the SHmax direction should be sufficient to capture the entire extent of any fractures. We will initialize the mesh as 5000 ft wide, and can reduce later if fractures do not extend that far.

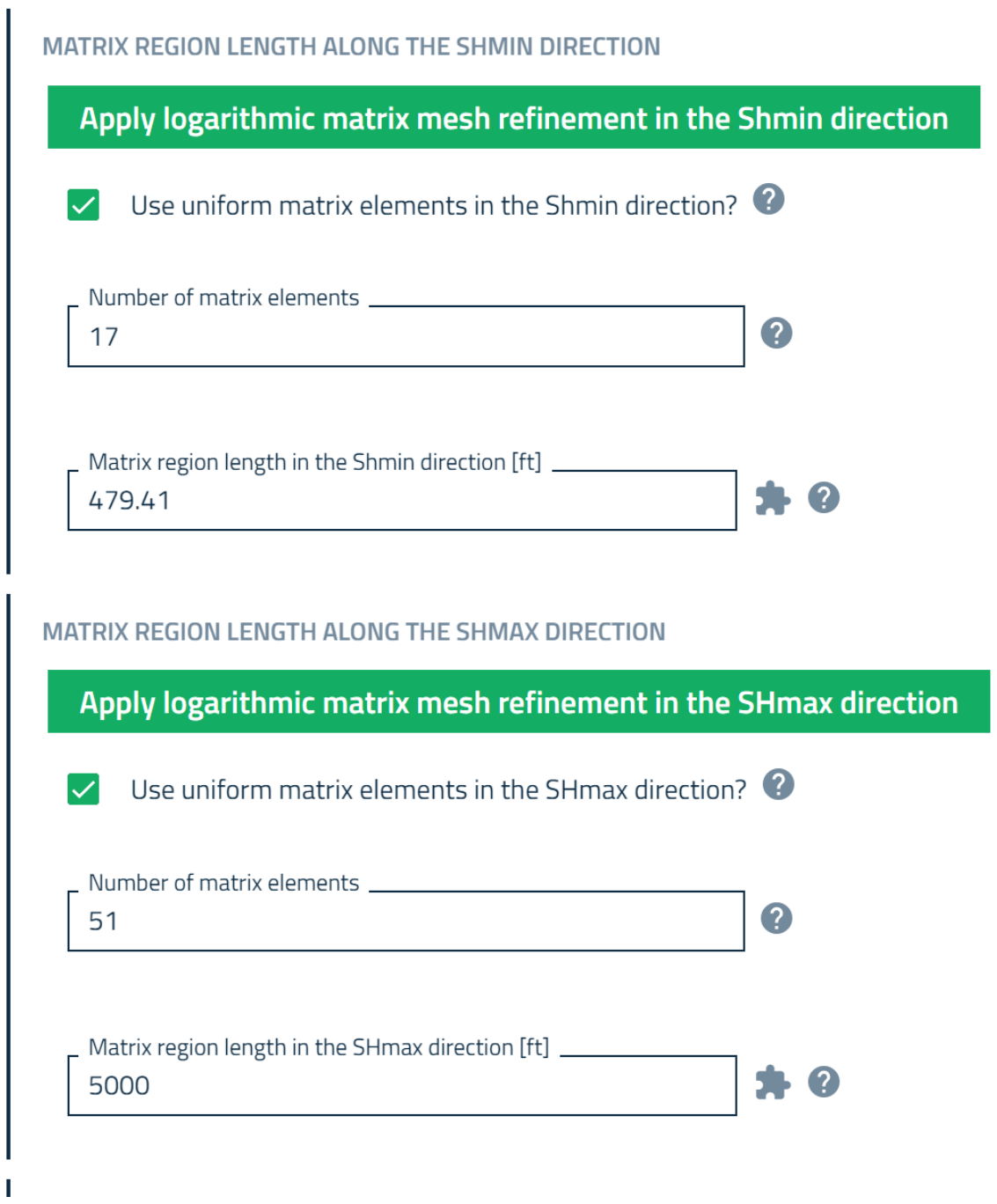

- In the Z direction, the mesh should be aligned with the facies boundaries. There is an alignment wizard that makes this easy.

**Vertical mesh alignment tool**

This wizard aligns the matrix layers in the vertical matrix mesh with the forma
The minimum element thickness when you click 'Apply' is determined by the tl
thin layers. The minimum element thickness generated from running this tool

Matrix Top [ft]
7804

Matrix Bottom [ft]
8999.5

Target Element Length [ft]
50

Apply | Reset to default values

## Fluid model options

The tutorial data in Excel provides a black oil model that we can use directly to define the fluid in our simulation.

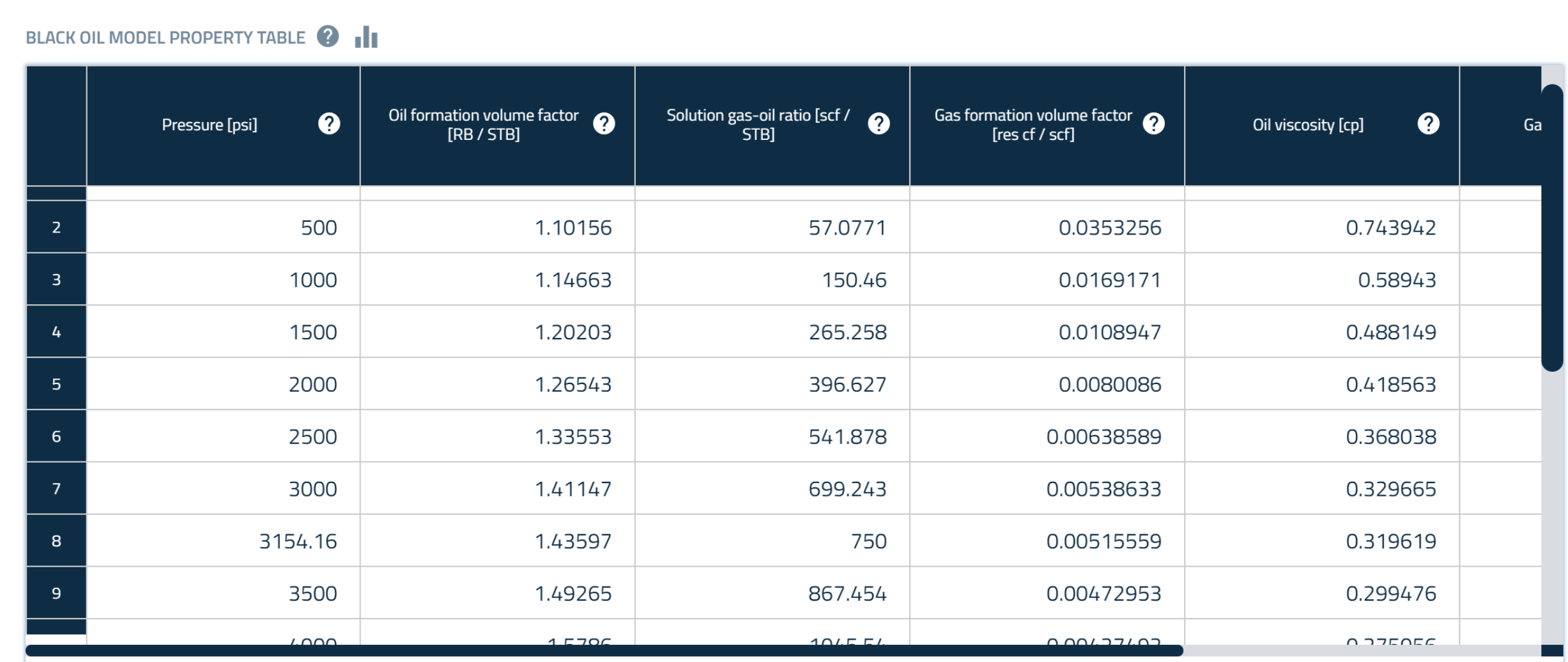

BLACK OIL MODEL PROPERTY TABLE

| | Pressure [psi] | Oil formation volume factor [RB / STB] | Solution gas-oil ratio [scf / STB] | Gas formation volume factor [res cf / scf] | Oil viscosity [cp] | Ga |
|---|---|---|---|---|---|---|
| 2 | 500 | 1.10156 | 57.0771 | 0.0353256 | 0.743942 | |
| 3 | 1000 | 1.14663 | 150.46 | 0.0169171 | 0.58943 | |
| 4 | 1500 | 1.20203 | 265.258 | 0.0108947 | 0.488149 | |
| 5 | 2000 | 1.26543 | 396.627 | 0.0080086 | 0.418563 | |
| 6 | 2500 | 1.33553 | 541.878 | 0.00638589 | 0.368038 | |
| 7 | 3000 | 1.41147 | 699.243 | 0.00538633 | 0.329665 | |
| 8 | 3154.16 | 1.43597 | 750 | 0.00515559 | 0.319619 | |
| 9 | 3500 | 1.49265 | 867.454 | 0.00472953 | 0.299476 | |

Make sure to define the bubble point as 3154.16 psi (as found in the black oil model table). The other parameters we will leave as defaults for this demo.

## Fracture options

The fracture options tab is predominantly used for history matching various observations of fracture geometry and behavior. For our demonstration we will start with a relative fracture toughness scaling parameter of 0.5. In datasets with very large fractures, this parameter may be set to zero. In datasets with short fractures, it may be even higher than 0.5. The second parameter is the random toughness parameter. Anything from 0.01 to 0.1 is reasonable.

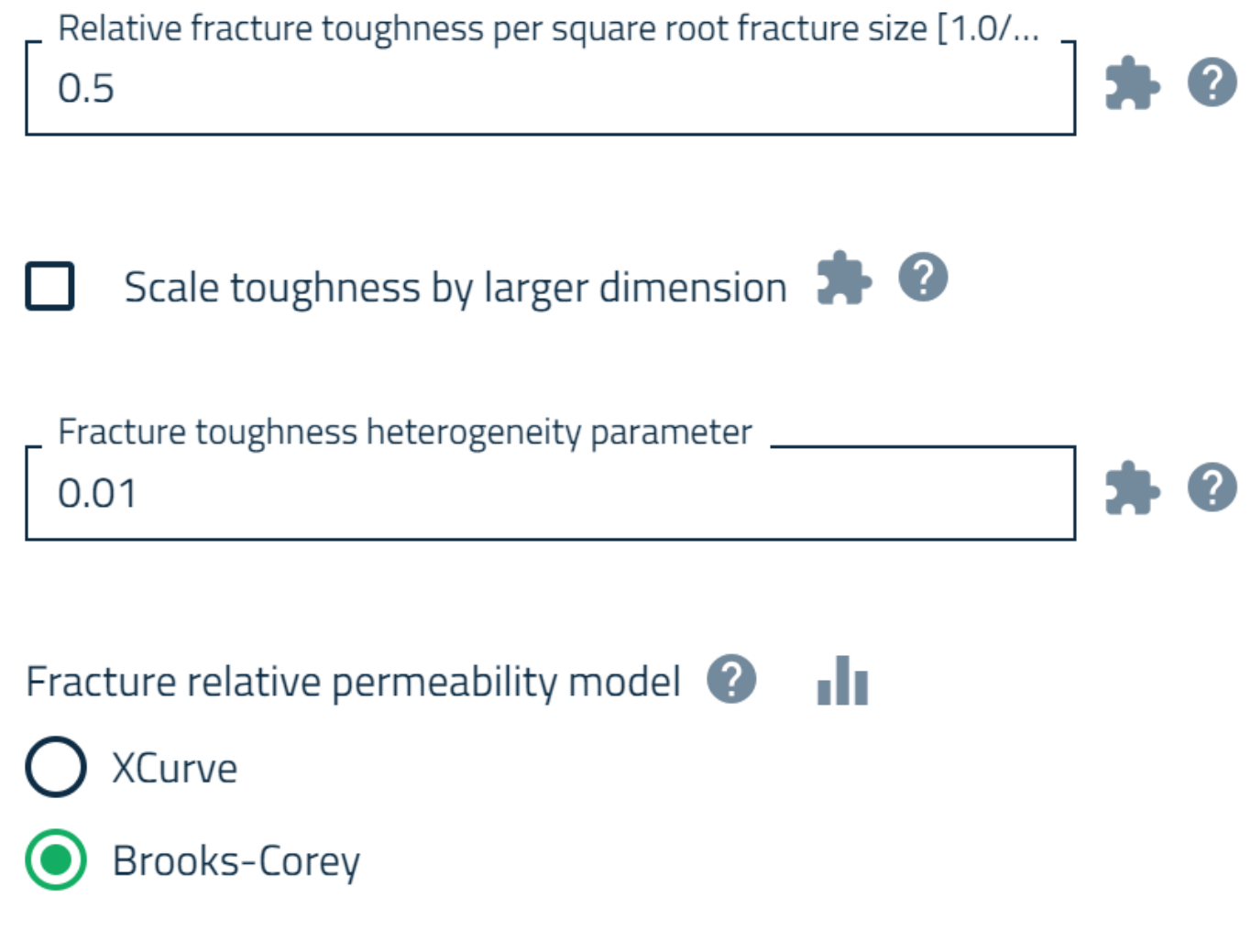

## Proppants

In the proppants tab we define the properties of our proppant/s in the pump schedule. In the pump schedule provided we see that both 100 mesh and 40/70 are used in the stimulation. The Excel input provides a tab with the conductivity versus normal stress for both proppants. We enter these properties into the simulator by first defining our two proppants.

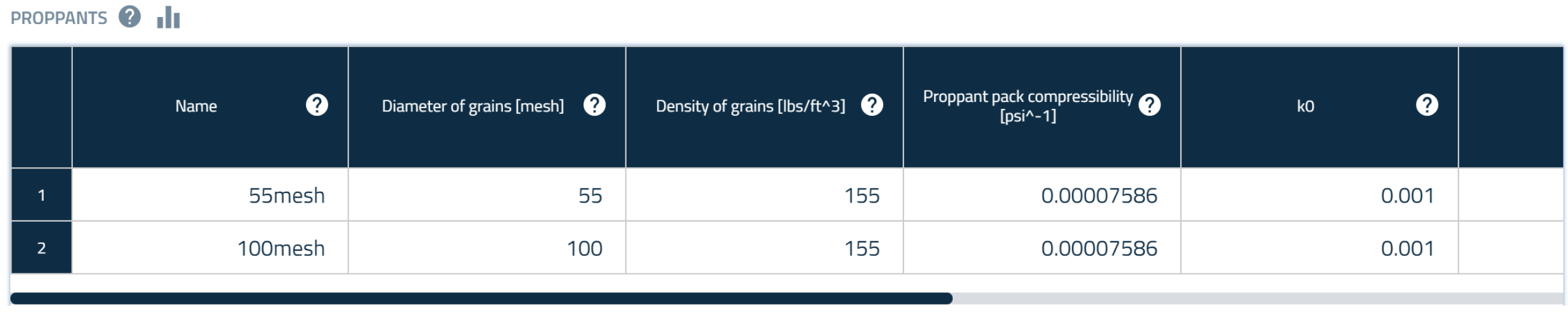
PROPPANTS

| | Name | Diameter of grains [mesh] | Density of grains [lbs/ft^3] | Proppant pack compressibility [psi^-1] | k0 | |
|---|---|---|---|---|---|---|
| 1 | 55mesh | 55 | 155 | 0.00007586 | 0.001 | |
| 2 | 100mesh | 100 | 155 | 0.00007586 | 0.001 | |

Do not worry about the values to the right of grain density – just put in the default values. We will subsequently set those values using the proppant conductivity wizard.

Before running the wizard, define two proppant mixtures, each made up 100% of the respective proppant type.

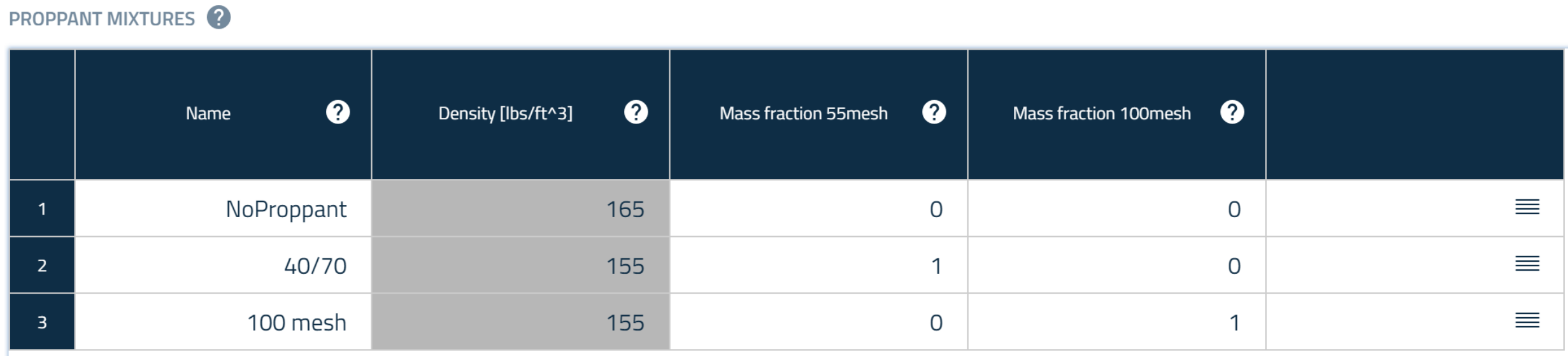
PROPPANT MIXTURES

| | Name | Density [lbs/ft^3] | Mass fraction 55mesh | Mass fraction 100mesh | |
|---|---|---|---|---|---|
| 1 | NoProppant | 165 | 0 | 0 | |
| 2 | 40/70 | 155 | 1 | 0 | |
| 3 | 100 mesh | 155 | 0 | 1 | |

Now, the proppant conductivity wizard can be used to set the proppant properties.

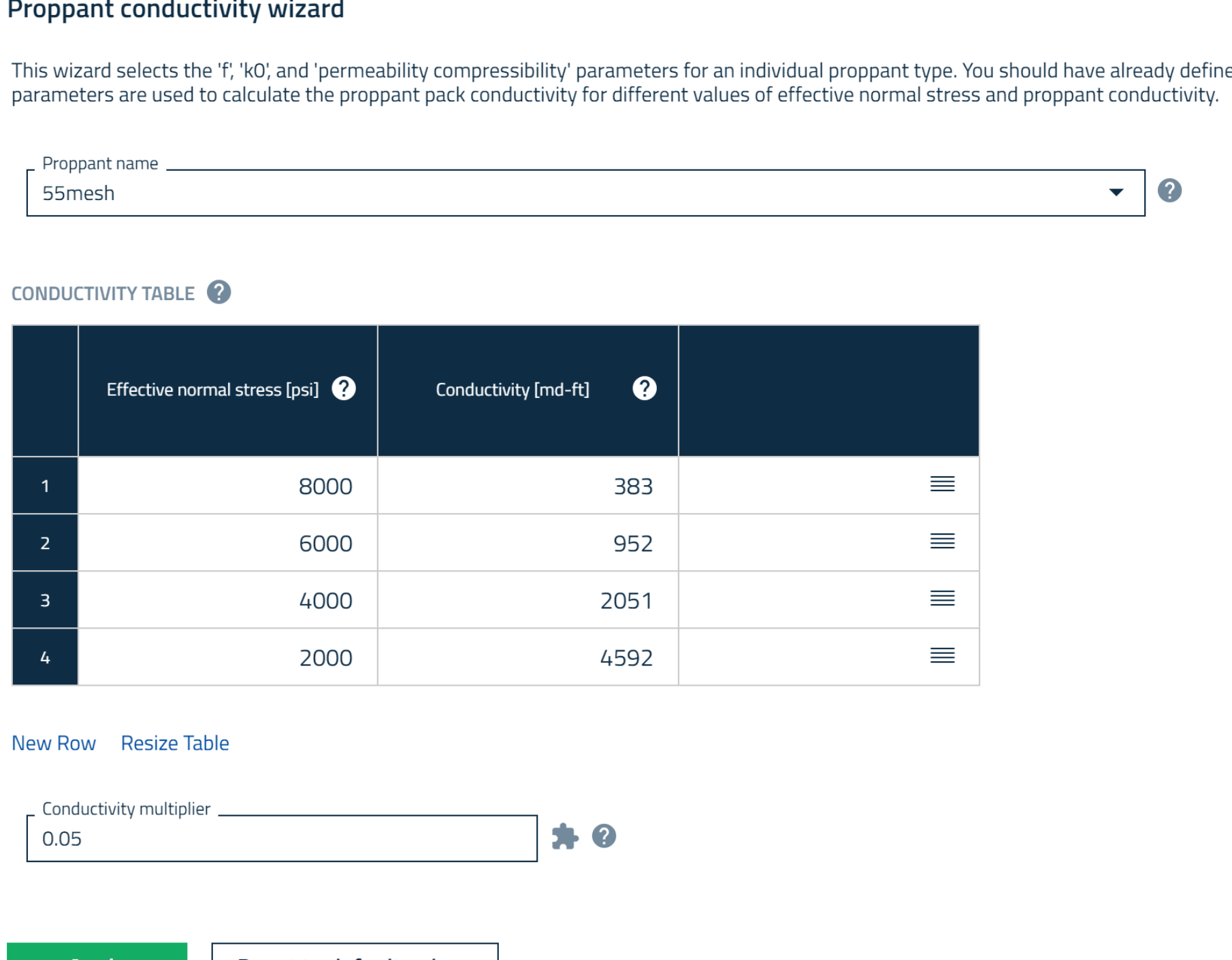

| | Effective normal stress [psi] | Conductivity [md-ft] | |
|---|---|---|---|
| 1 | 8000 | 383 | |
| 2 | 6000 | 952 | |
| 3 | 4000 | 2051 | |
| 4 | 2000 | 4592 | |

Ultimately, we will use the in-situ conductivity of proppant as a history matching parameter, but note that API conductivities are recognized to be optimistic. To reflect this, we generally apply a 95% damage factor (0.05 conductivity multiplier) to the conductivities when using the proppant conductivity wizard.

## Water solutes

Our Excel input data reveals that the slickwater pumped in our well has a viscosity at reservoir conditions of 2 cp at 170 s^-1. To add this fluid to the simulator, we first define a slickwater additive.

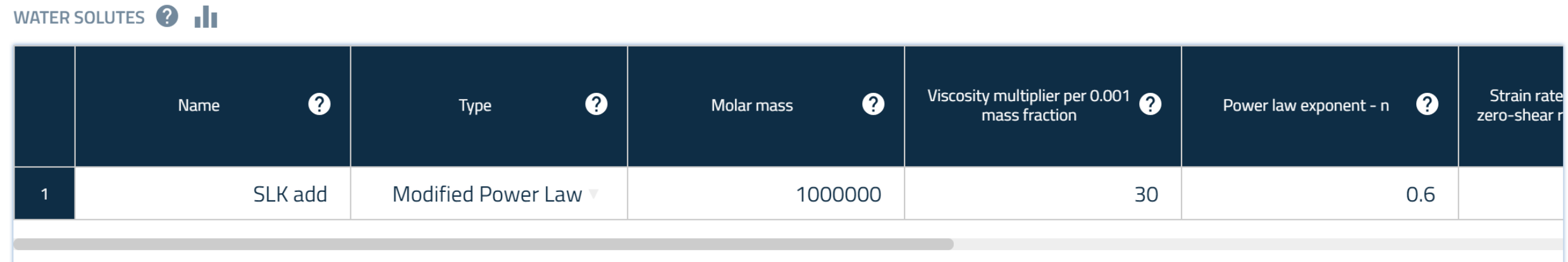

| | Name | Type | Molar mass | Viscosity multiplier per 0.001 mass fraction | Power law exponent - n | Strain rate zero-shear r |
|---|---|---|---|---|---|---|
| 1 | SLK add | Modified Power Law | 1000000 | 30 | 0.6 | |

Similar to setting up the proppants, we do not need to be exact with the viscosity multiplier, as we will set this value with the wizard.

Next, we define out slickwater mixture that contains the SLK additive.

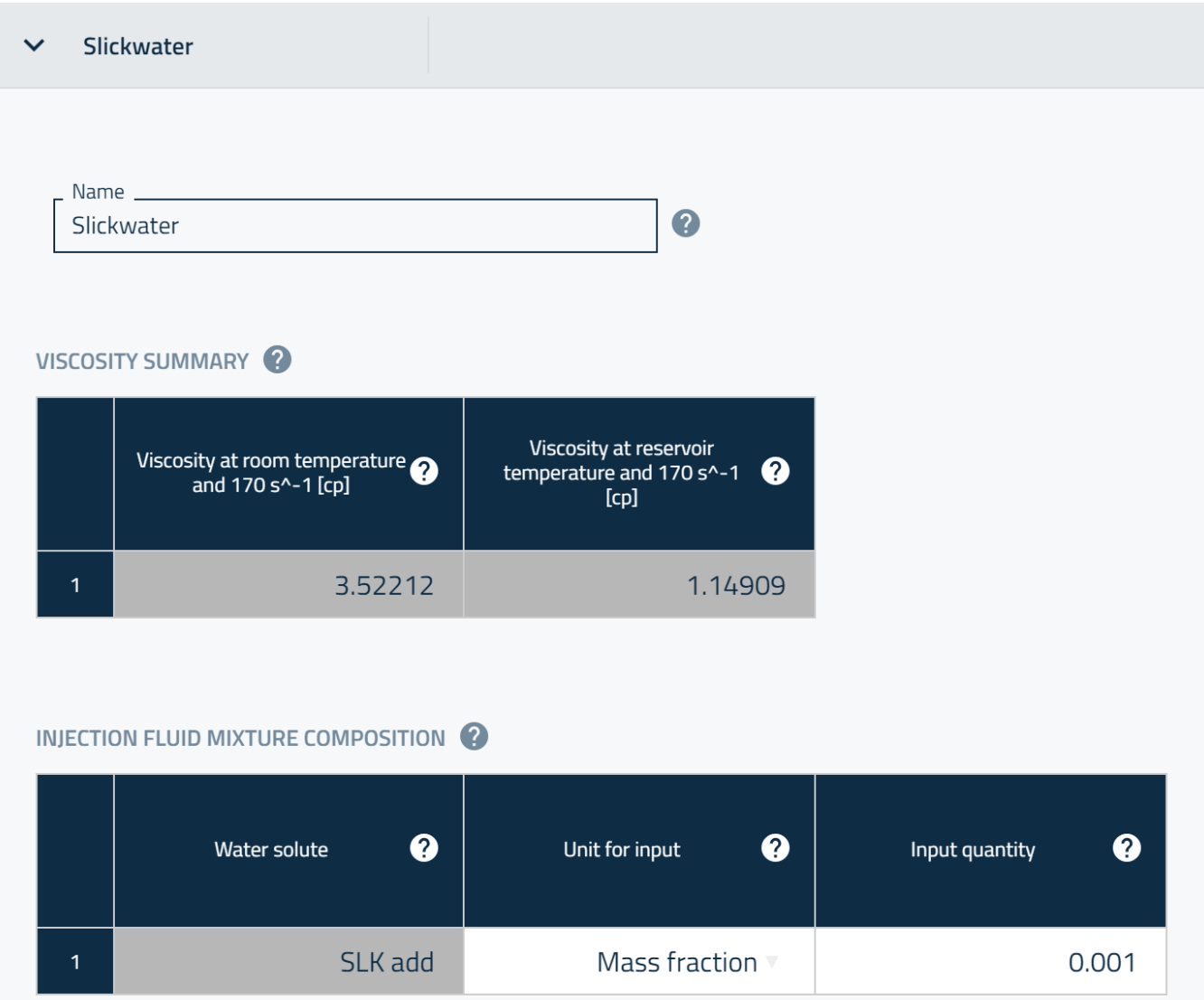

Now, we can run the water solute wizard to adjust the viscosity of the slickwater.

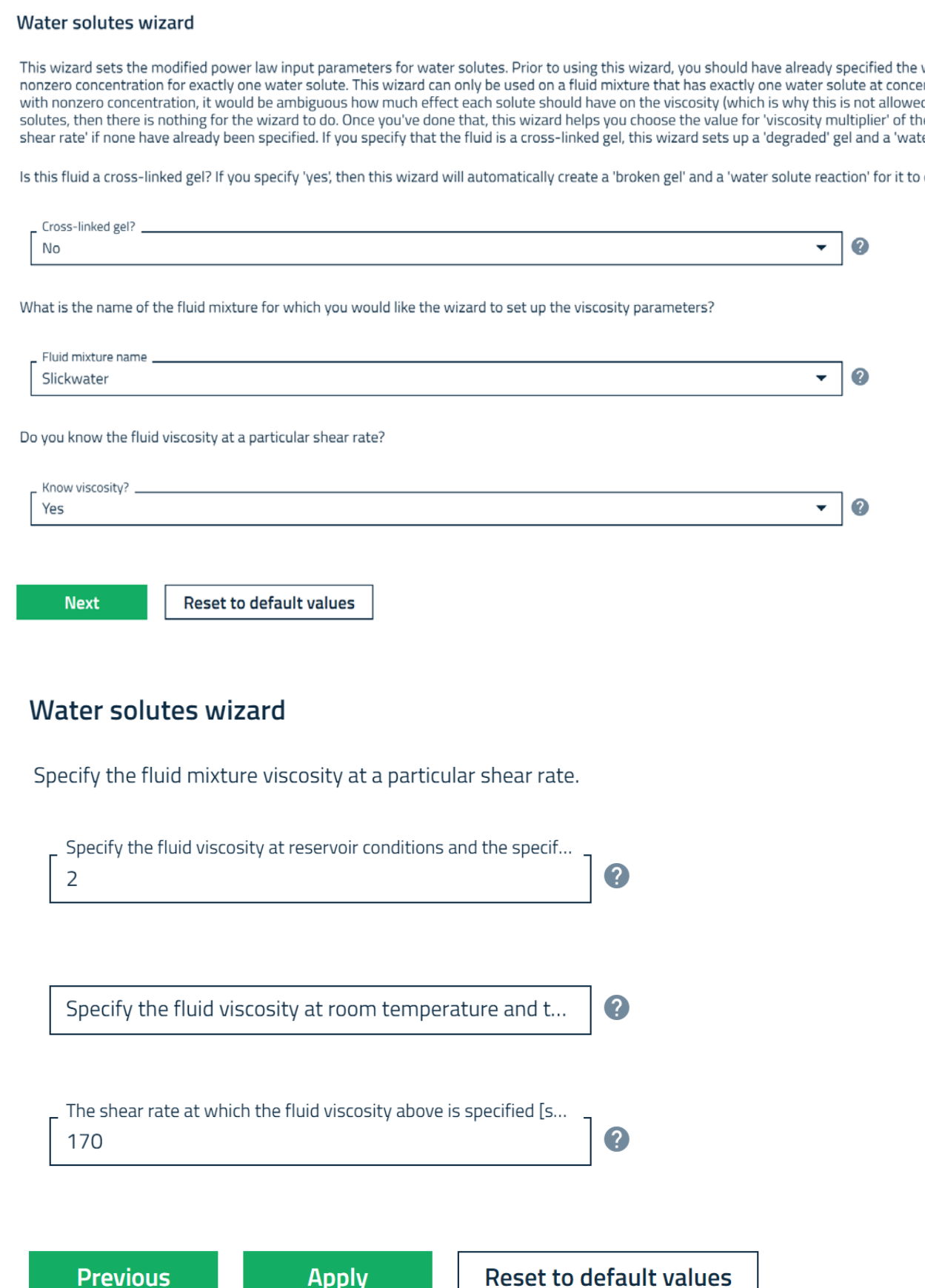

After running the wizard, we see that our slickwater now has the appropriate viscosity at reservoir conditions.

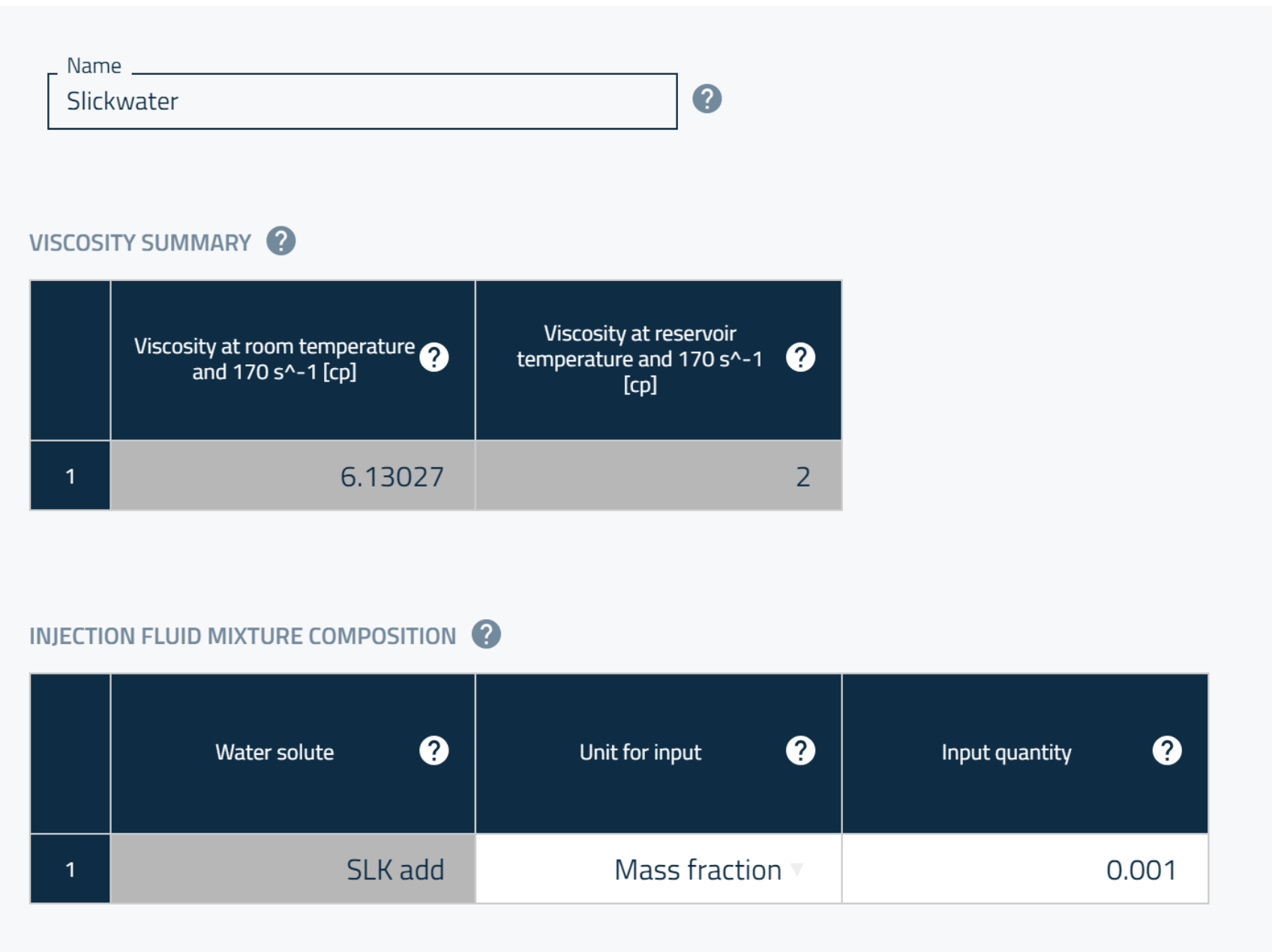

## Well controls

The final required input tab in the ResFrac builder is the well controls tab. On this tab we define the operations of our wells. Well One and Well Two are zippered, so the first step is to define the fracture schedules for the two wells. We start by adding a new control block and selecting an injection sequence. At the top of the injection sequence, in the injection sequence summary table, we select the units that match the frac schedule provided in the Excel input data. Because we want to control injection volume and match pressures, we set the maximum injection pressure arbitrarily high (15,000 psi).

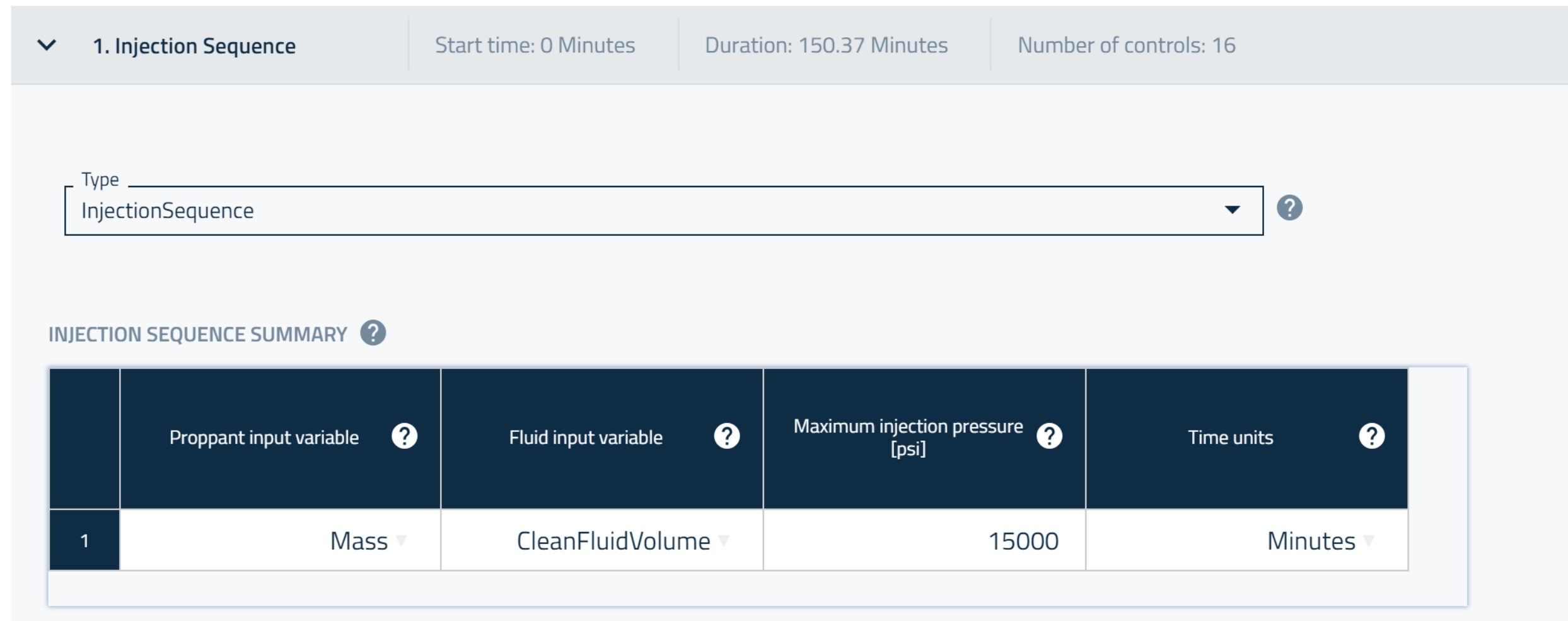

Next, we can paste in the appropriate data into the sequence below.

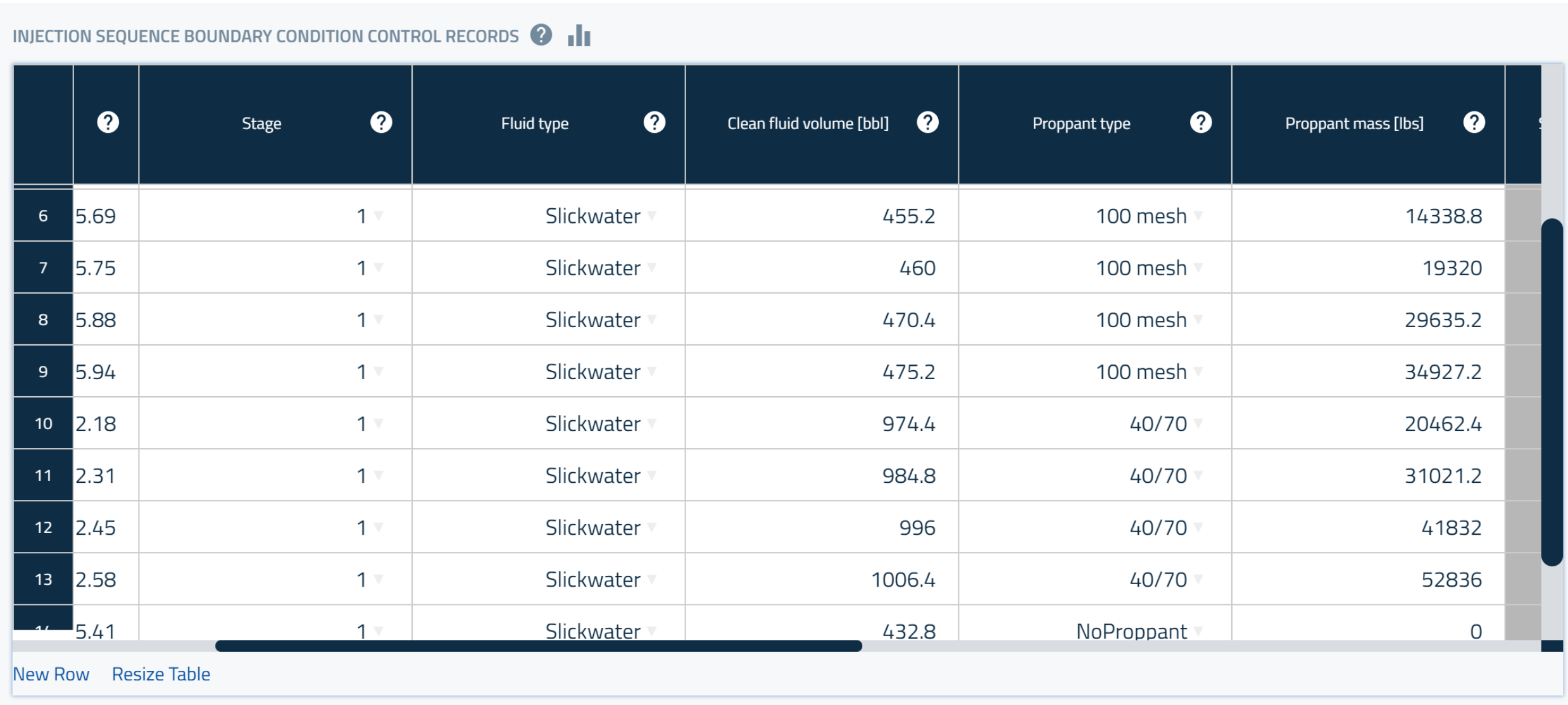
INJECTION SEQUENCE BOUNDARY CONDITION CONTROL RECORDS

| | | Stage | Fluid type | Clean fluid volume [bbl] | Proppant type | Proppant mass [lbs] |
|---|---|---|---|---|---|---|
| 6 | 5.69 | 1 | Slickwater | 455.2 | 100 mesh | 14338.8 |
| 7 | 5.75 | 1 | Slickwater | 460 | 100 mesh | 19320 |
| 8 | 5.88 | 1 | Slickwater | 470.4 | 100 mesh | 29635.2 |
| 9 | 5.94 | 1 | Slickwater | 475.2 | 100 mesh | 34927.2 |
| 10 | 2.18 | 1 | Slickwater | 974.4 | 40/70 | 20462.4 |
| 11 | 2.31 | 1 | Slickwater | 984.8 | 40/70 | 31021.2 |
| 12 | 2.45 | 1 | Slickwater | 996 | 40/70 | 41832 |
| 13 | 2.58 | 1 | Slickwater | 1006.4 | 40/70 | 52836 |
| 14 | 5.41 | 1 | Slickwater | 432.8 | NoProppant | 0 |

New Row Resize Table

Be sure to change the stage number to stage 1. This tells the simulator to only inject into stage 1. Next, we want to add a shut-in between stages. We do this be adding an “isolate wellbore” step at the end of the stage 1 sequence. As Well One and Well Two are being zippered back and forth, we will add a two hour shut-in between stages on each well.

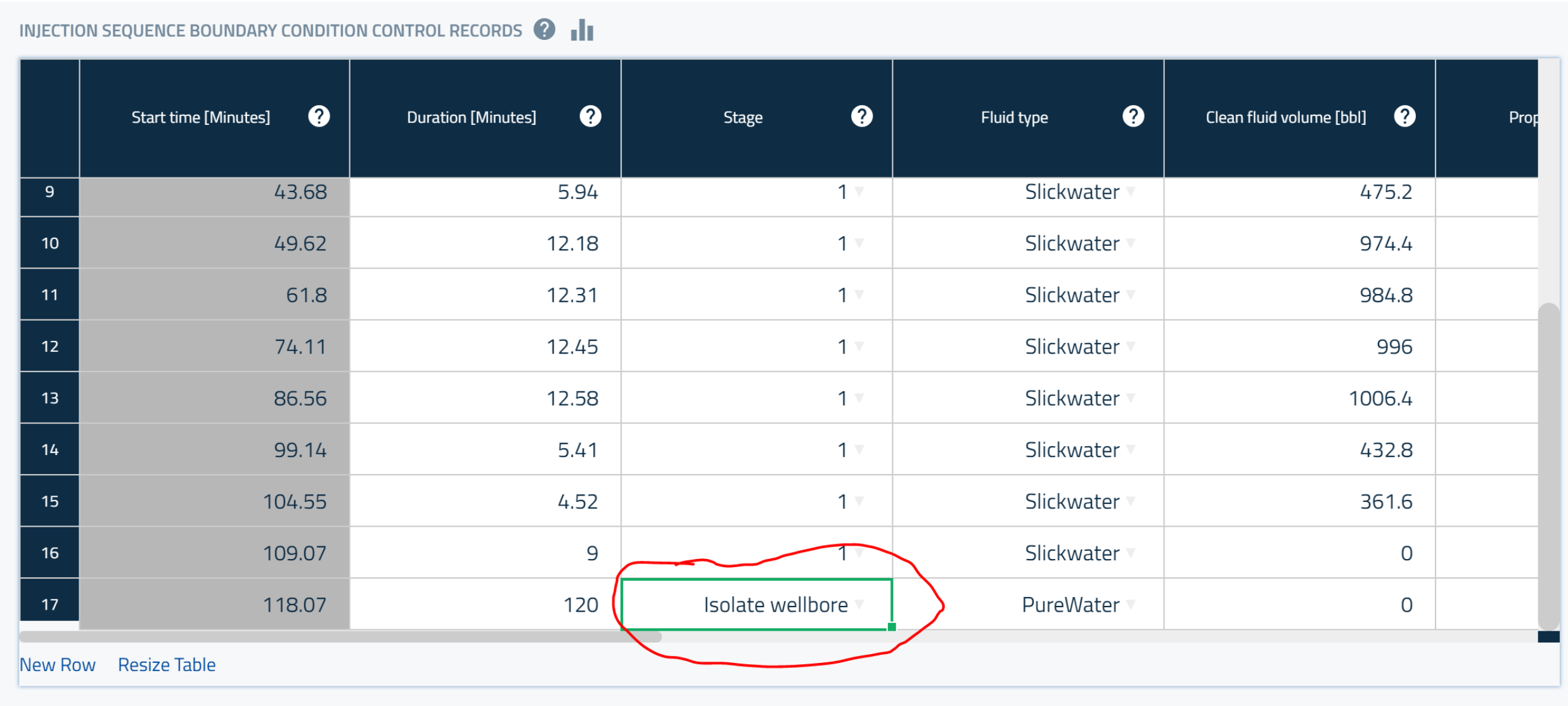
INJECTION SEQUENCE BOUNDARY CONDITION CONTROL RECORDS

| | Start time [Minutes] | Duration [Minutes] | Stage | Fluid type | Clean fluid volume [bbl] | Prop |
|---|---|---|---|---|---|---|
| 9 | 43.68 | 5.94 | 1 | Slickwater | 475.2 | |
| 10 | 49.62 | 12.18 | 1 | Slickwater | 974.4 | |
| 11 | 61.8 | 12.31 | 1 | Slickwater | 984.8 | |
| 12 | 74.11 | 12.45 | 1 | Slickwater | 996 | |
| 13 | 86.56 | 12.58 | 1 | Slickwater | 1006.4 | |
| 14 | 99.14 | 5.41 | 1 | Slickwater | 432.8 | |
| 15 | 104.55 | 4.52 | 1 | Slickwater | 361.6 | |
| 16 | 109.07 | 9 | 1 | Slickwater | 0 | |
| 17 | 118.07 | 120 | Isolate wellbore | PureWater | 0 | |

New Row Resize Table

Next, to stimulate the second stage, we can easily select rows 1 through 17, copy, then paste below by pressing “New Row” and then pasting our stage 1 design into the new row (the table will automatically expand to the size of the pasted region).

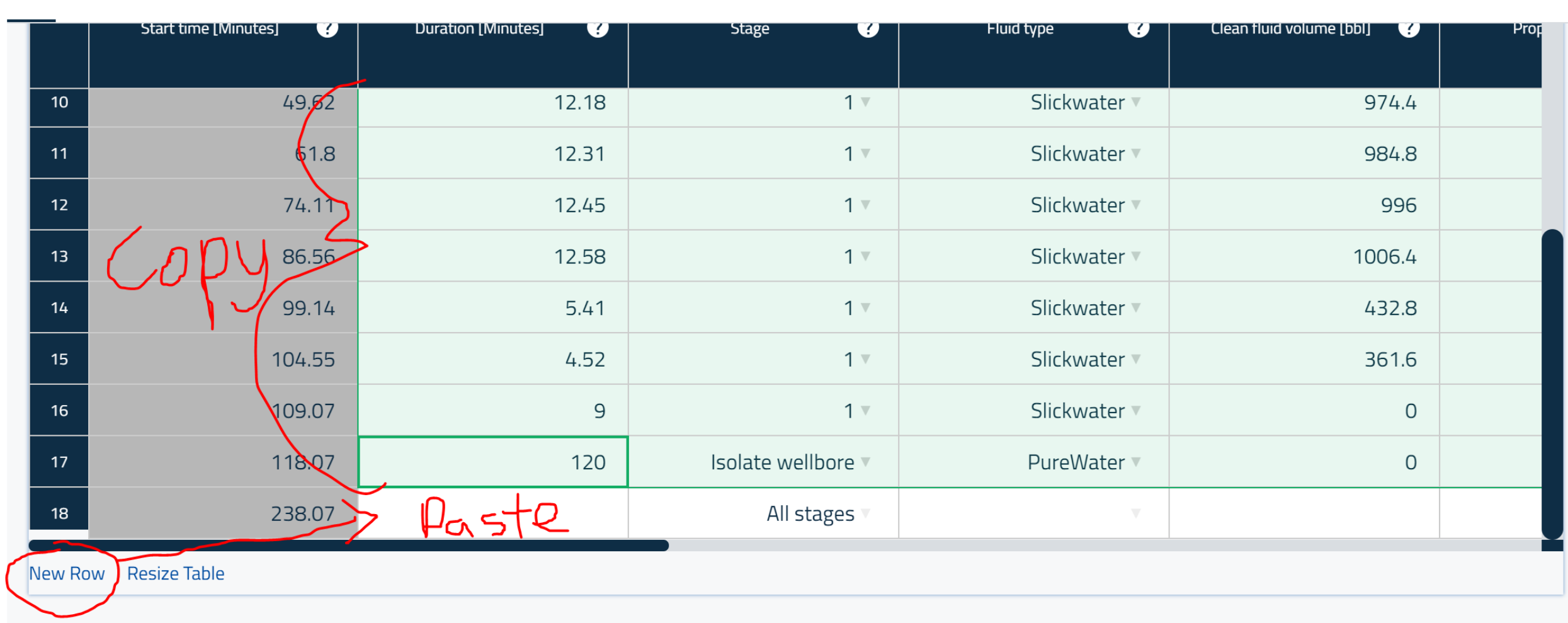

| | Start time [Minutes] | Duration [Minutes] | Stage | Fluid type | Clean fluid volume [bbl] | Prop |
|---|---|---|---|---|---|---|
| 10 | 49.62 | 12.18 | 1 | Slickwater | 974.4 | |
| 11 | 61.8 | 12.31 | 1 | Slickwater | 984.8 | |
| 12 | 74.11 | 12.45 | 1 | Slickwater | 996 | |
| 13 | 86.56 | 12.58 | 1 | Slickwater | 1006.4 | |
| 14 | 99.14 | 5.41 | 1 | Slickwater | 432.8 | |
| 15 | 104.55 | 4.52 | 1 | Slickwater | 361.6 | |
| 16 | 109.07 | 9 | 1 | Slickwater | 0 | |
| 17 | 118.07 | 120 | Isolate wellbore | PureWater | 0 | |
| 18 | 238.07 | | All stages | | | |

The final step is to change the stage number to stage 2 in the newly pasted rows.

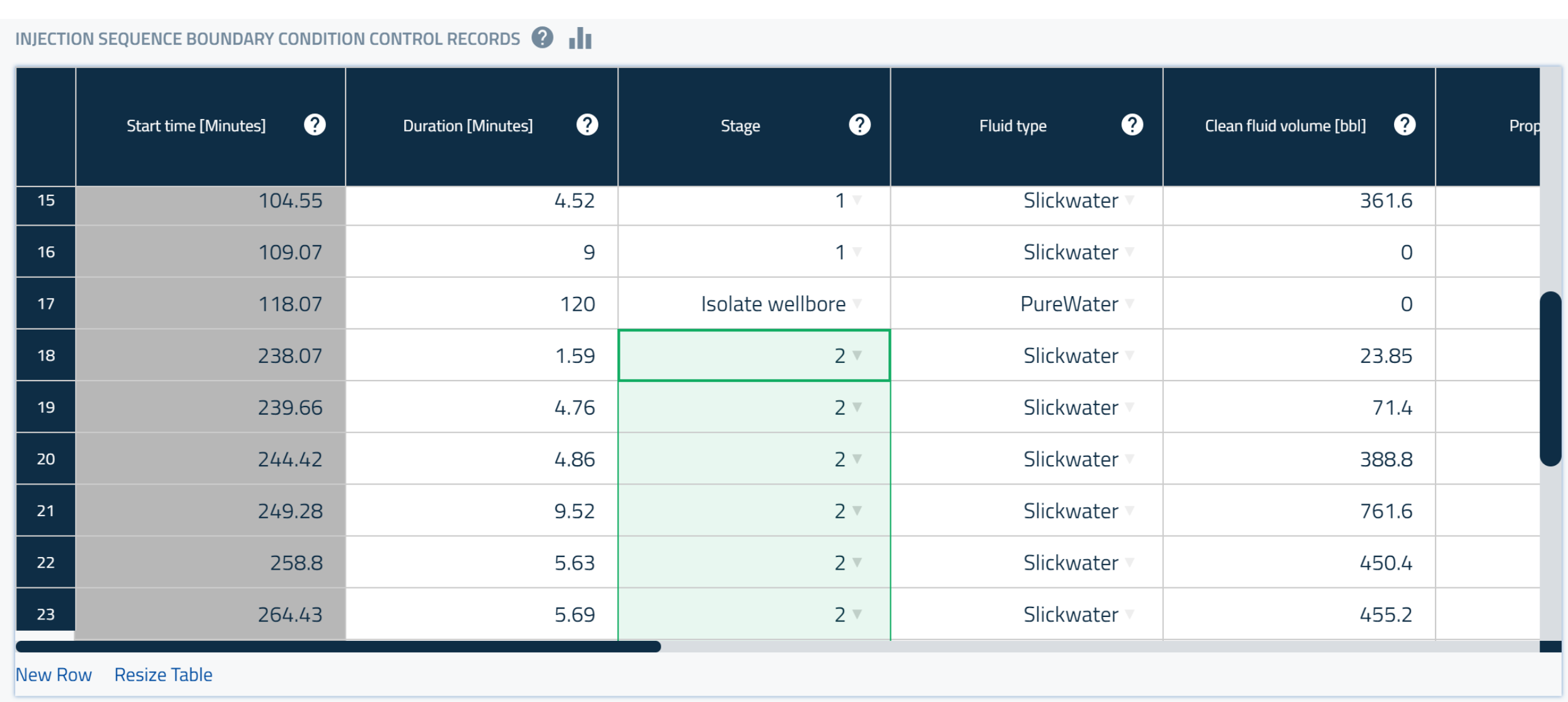

INJECTION SEQUENCE BOUNDARY CONDITION CONTROL RECORDS

| | Start time [Minutes] | Duration [Minutes] | Stage | Fluid type | Clean fluid volume [bbl] | Prop |
|---|---|---|---|---|---|---|
| 15 | 104.55 | 4.52 | 1 | Slickwater | 361.6 | |
| 16 | 109.07 | 9 | 1 | Slickwater | 0 | |
| 17 | 118.07 | 120 | Isolate wellbore | PureWater | 0 | |
| 18 | 238.07 | 1.59 | 2 | Slickwater | 23.85 | |
| 19 | 239.66 | 4.76 | 2 | Slickwater | 71.4 | |
| 20 | 244.42 | 4.86 | 2 | Slickwater | 388.8 | |
| 21 | 249.28 | 9.52 | 2 | Slickwater | 761.6 | |
| 22 | 258.8 | 5.63 | 2 | Slickwater | 450.4 | |
| 23 | 264.43 | 5.69 | 2 | Slickwater | 455.2 | |

Both Well One and Well Two have the same stimulation design, so we can quickly copy Well One's injection sequence to Well Two by using the duplicate function.

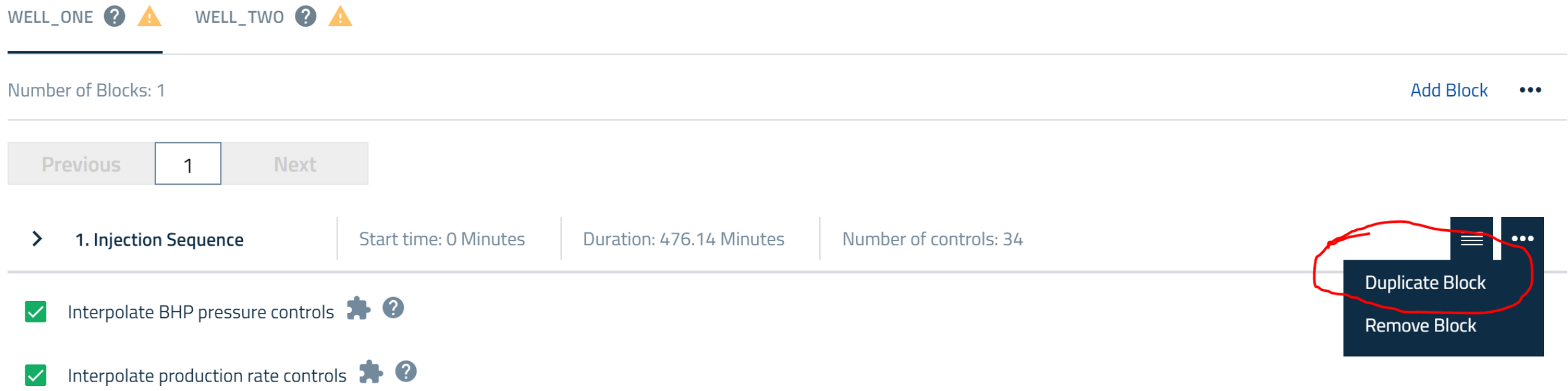

And selecting Well Two.

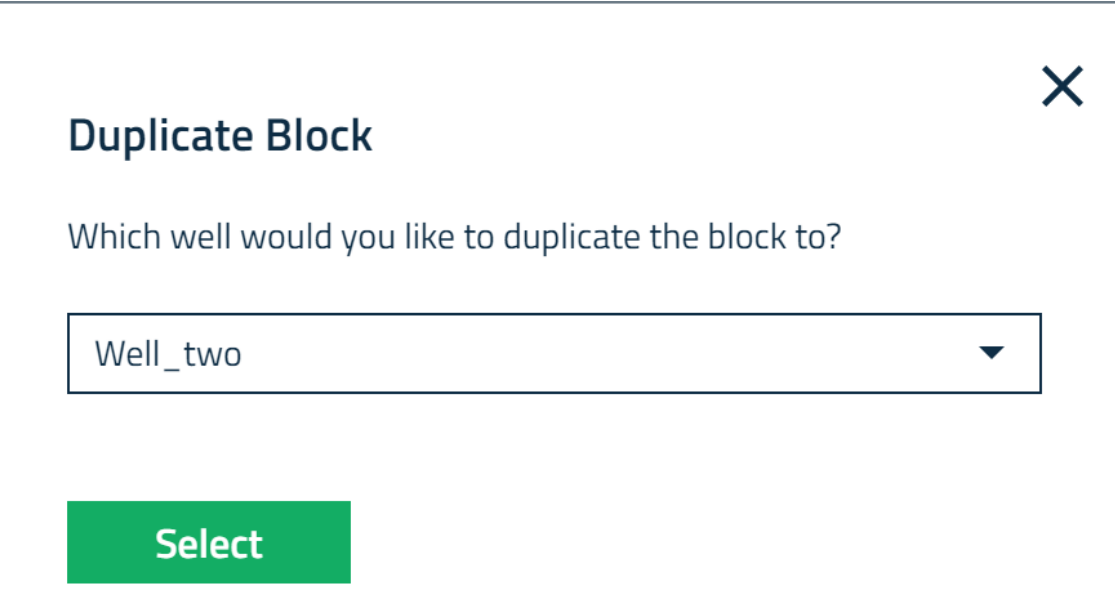

Because we are zippering the two wells, after duplicating the Well One injection sequence, we need to add a two hour shut-in at the beginning of Well Two's control sequence.

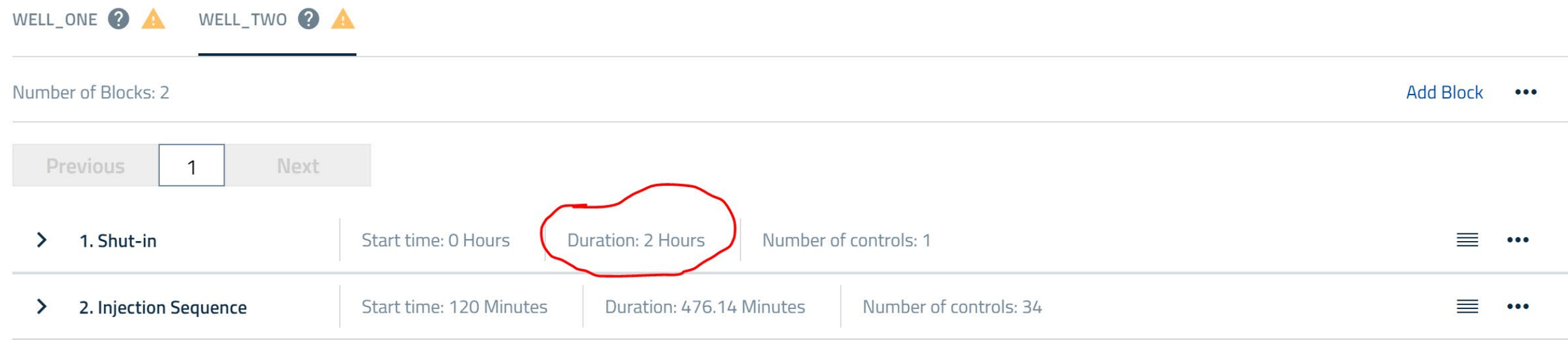

We have two options for that shut-in period, either a "Surface shut-in" or an "Isolate wellbore" control. Isolate wellbore will yield slightly faster compute times, but if you are watching for frac hits on the offsetting well and want to measure the pressure hits from Well One hitting Well Two, you will need to use a surface shut-in. We usually use the isolate wellbore shut-in type.

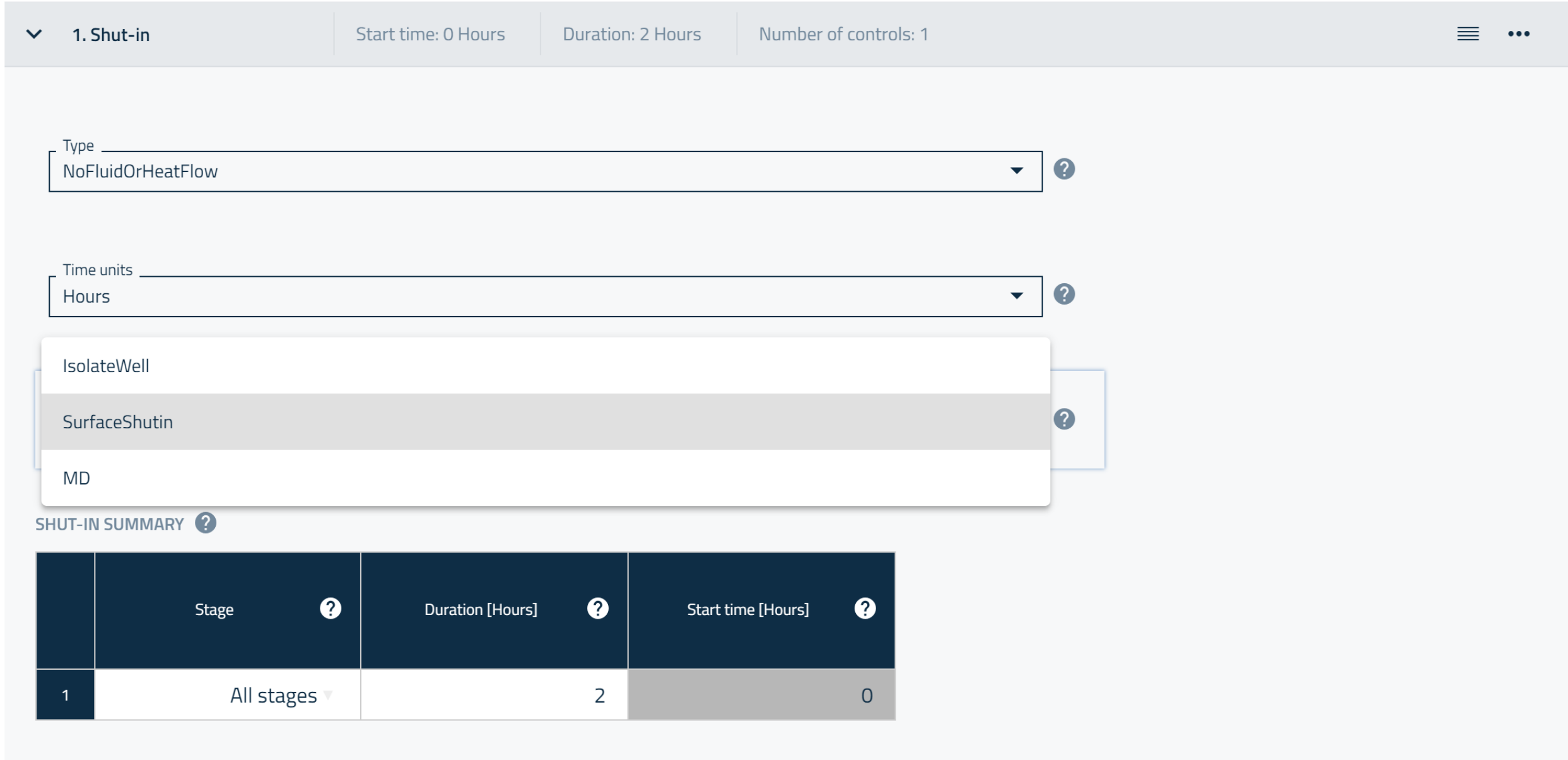

Next, we need to define the production controls along with the shut-in period between stimulation and production. We will do so using the production sequence wizard.

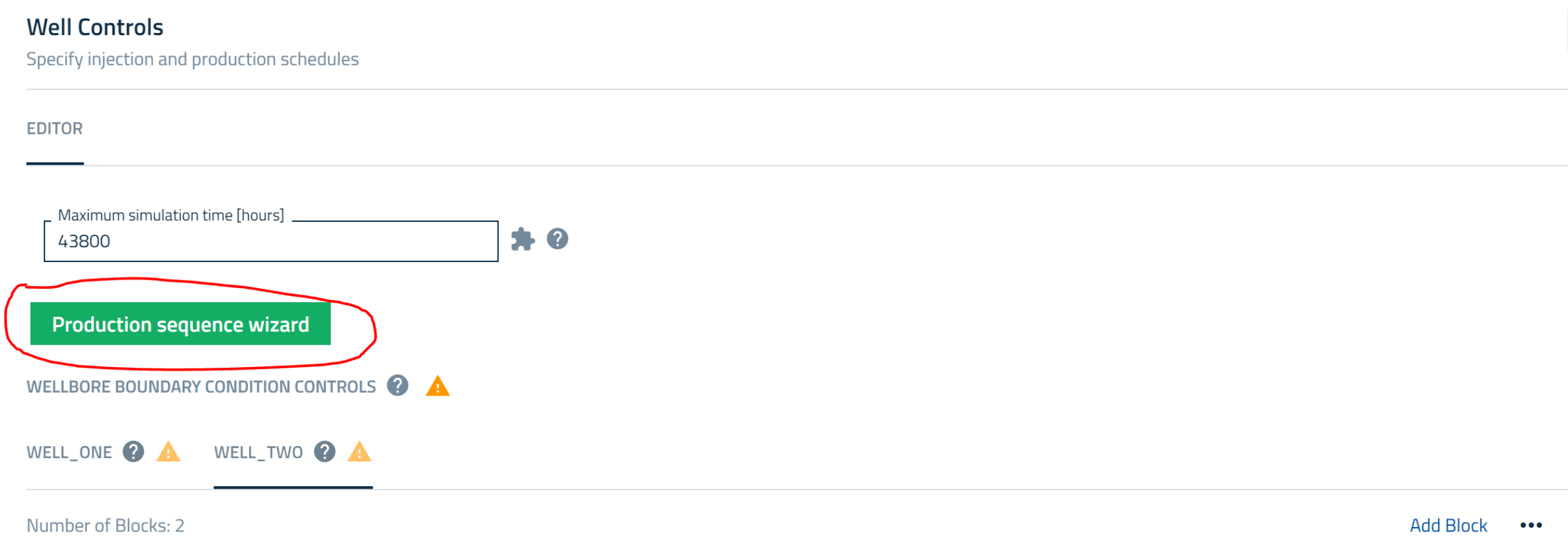

For Well One, we use the wizard to fill in details about the production sequence, then paste in the BHP drawdown schedule from the Excel input data. We set the production start time of production for day 10 in the simulation, and take the bi-daily data provided in the Excel Input and use the wizard to average into 30 day control steps.

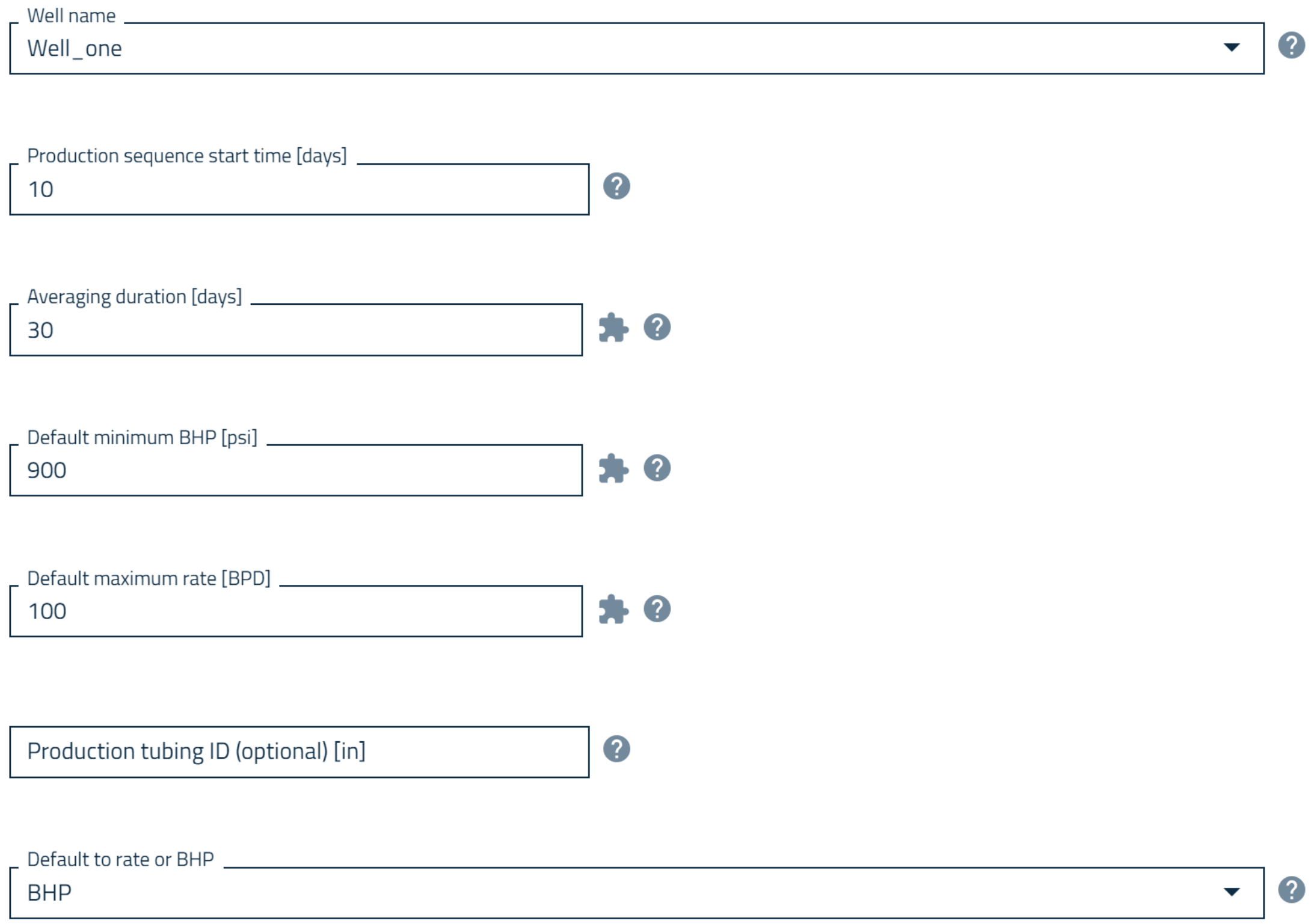

When we do so, we will see that the builder automatically extends the shut-in at the end of the Well One injection sequence to extend until day 10 of the simulation.

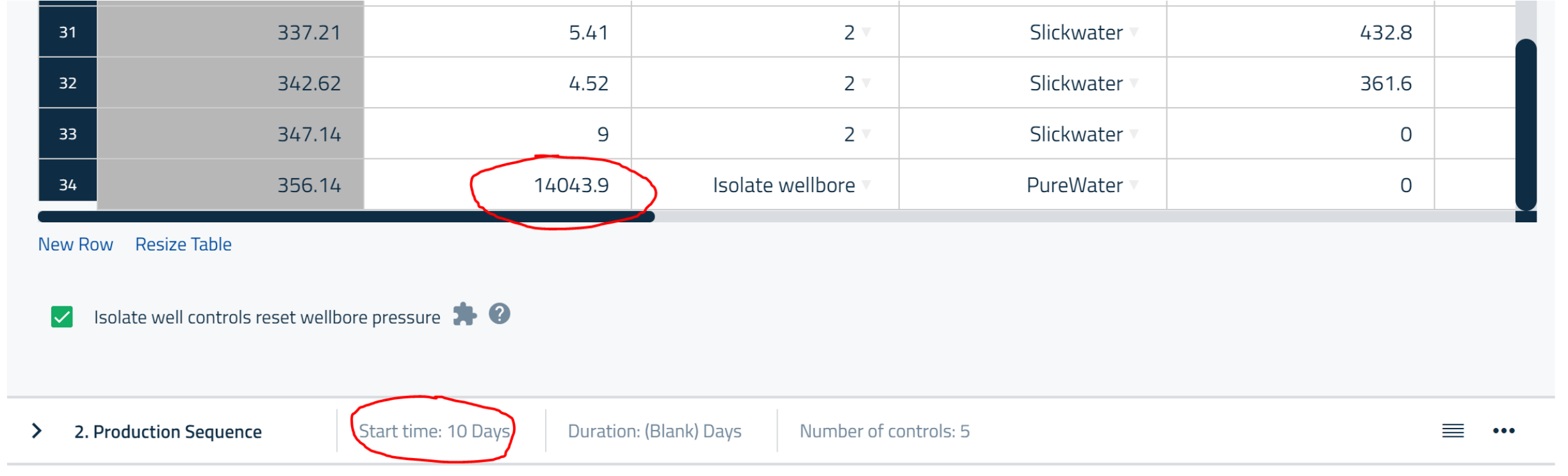

| | | | | | |
|---|---|---|---|---|---|
| 31 | 337.21 | 5.41 | 2 | Slickwater | 432.8 |
| 32 | 342.62 | 4.52 | 2 | Slickwater | 361.6 |
| 33 | 347.14 | 9 | 2 | Slickwater | 0 |
| 34 | 356.14 | 14043.9 | Isolate wellbore | PureWater | 0 |

In the Excel input data, we see that the drawdown schedule for Well Two is the same as for Well One. We can either duplicate the Well One production sequence to Well Two (in which case we would need to add a shut-in period to the end of the Well Two injection sequence such that production commences at day 10, or we can run the Production sequence wizard again, this time for Well Two.

Using the Wizard is convenient, and we can see creates a clean control schedule.

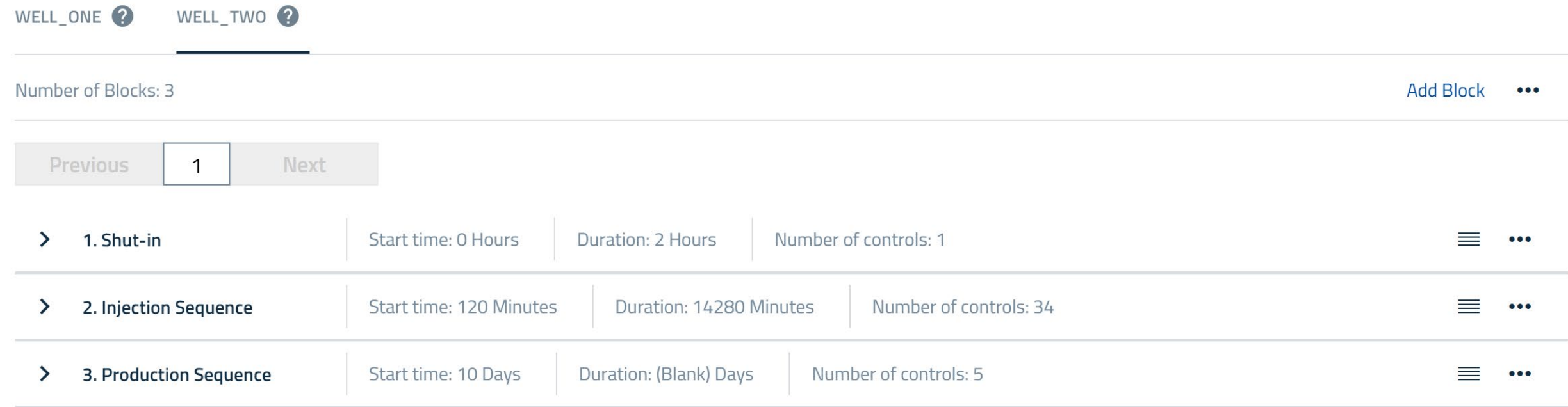

## Output options

Output options allows the user to control outputting parameters and file sizes. For the most part, users are best served with the default outputting options. One standard addition to the outputting options is to add stress observation planes at the elevation of each well in your model to visualize stress shadowing.

From the well vertices table in the Wells and Perforations tab, we can see that Well One and Well Two are landed at ~8265 and ~8606 ft, respectively. From the well vertices table we can also find the x,y coordinates of the center of the modeled portion of the lateral and approximately center our stress observation planes around each.

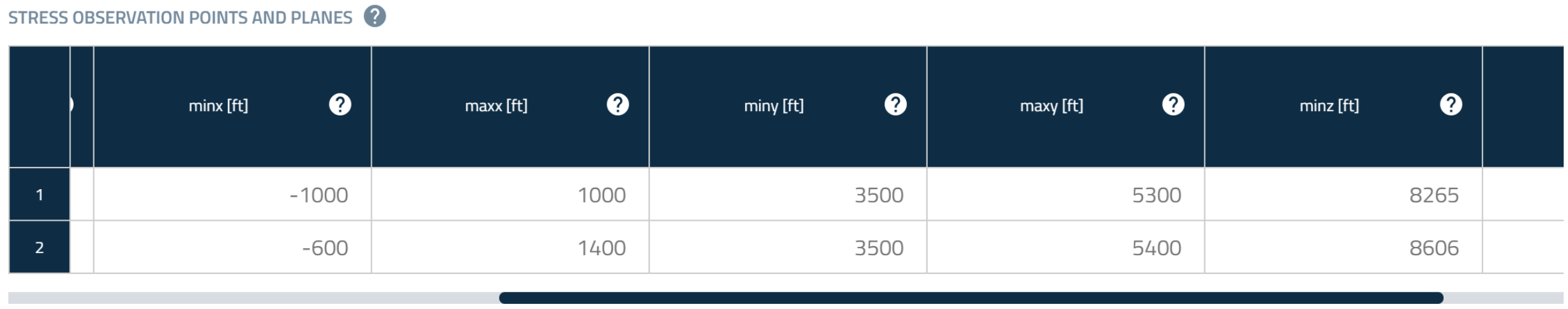

STRESS OBSERVATION POINTS AND PLANES

| | minx [ft] | maxx [ft] | miny [ft] | maxy [ft] | minz [ft] |
|---|---|---|---|---|---|
| 1 | -1000 | 1000 | 3500 | 5300 | 8265 |
| 2 | -600 | 1400 | 3500 | 5400 | 8606 |

When the simulation runs, it will now output the stress shadow through time across each of the planes.

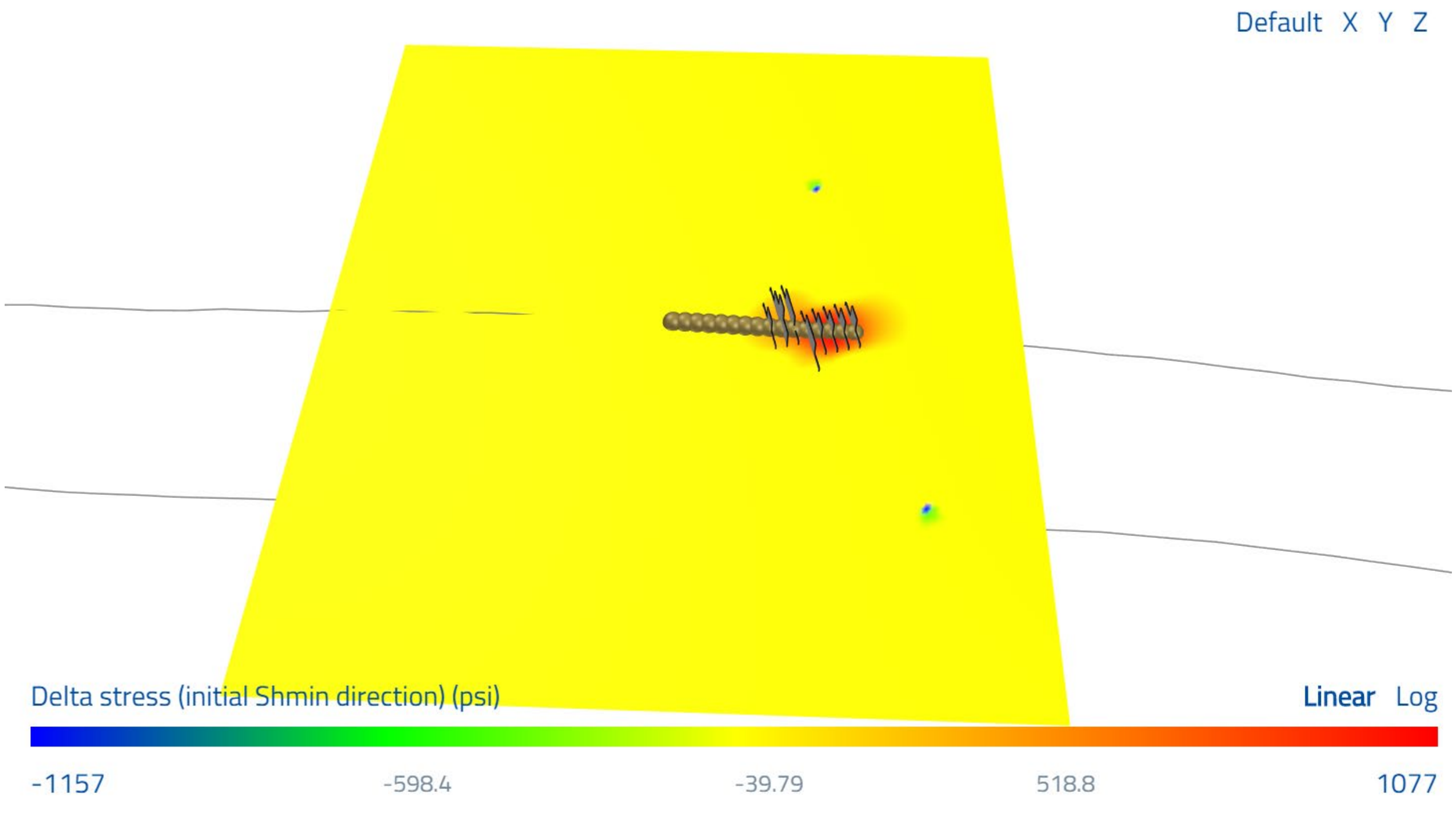

## Other input tabs

The remaining input tabs are optional and advanced options. Other Physics Options allows the use to modify and adjust the physics of the simulation, Numerical Options provides the user with fine tuning control of the numerical methods, and Advanced Options controls parameter visibility and advanced units.

## Present base case model and verify setup

The final step of base case model construction is to present this model to the project stakeholders and confirm model input and setup. The intent of the presentation is to regurgitate all data input into the model back to the key stakeholders and get their confirmation of the data. Also, this presentation serves as a time to address and unanswered questions generated during the model setup process.

Every presentation should start with an abstract slide. When stakeholders review this work in two years, what is it that they should remember from this presentation? *That* is what goes into the abstract.

## Abstract

- Initial model constructed for Peter Pan Well One and Well Two
- Model input data came from ResFrac_Tutorial_Input_Data.xlsx
- Project tracking on schedule

Next, it is helpful to orient everyone by reviewing the project process again.

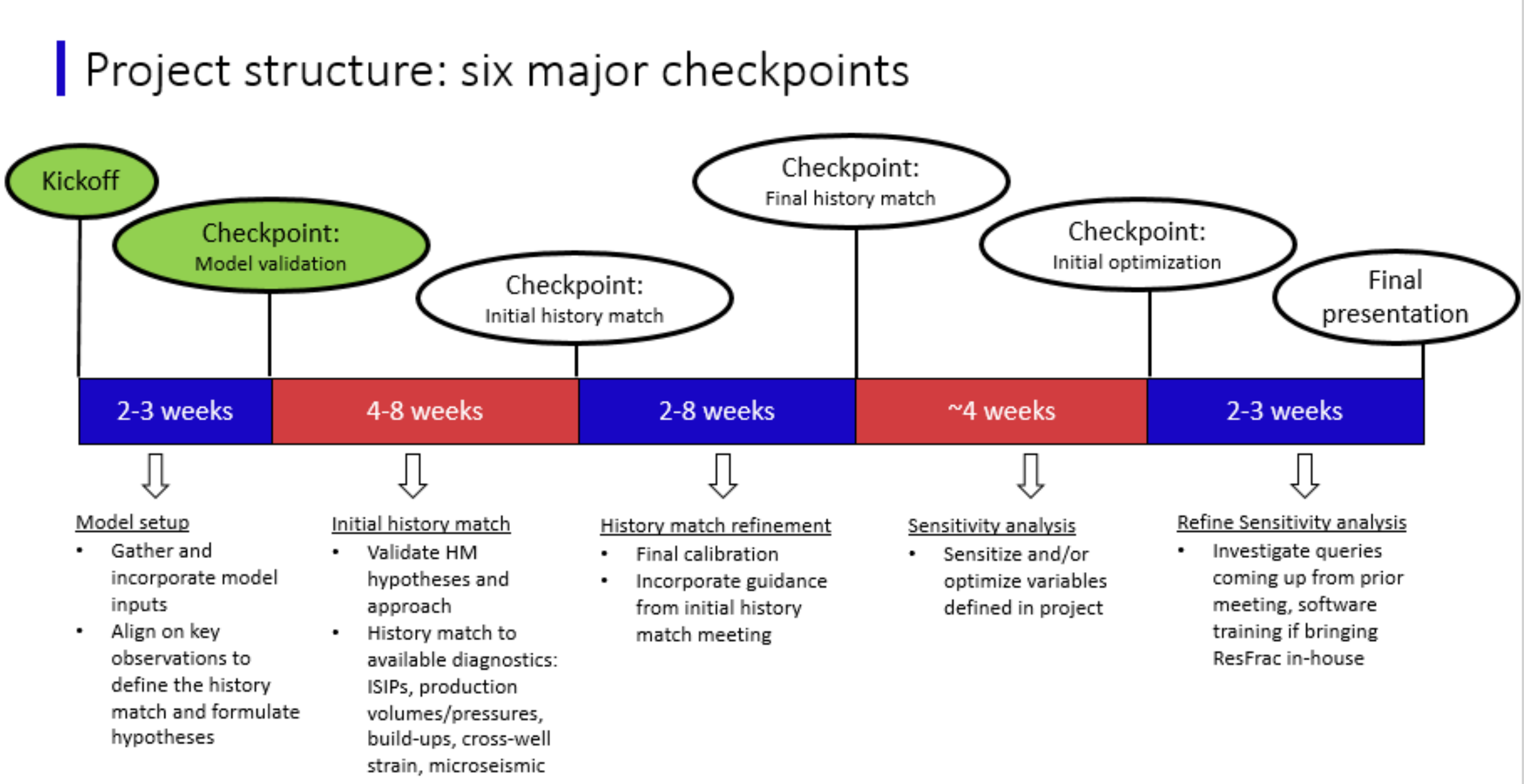

The content of the presentation should review the objective of the project and cover every major input into the simulation: static model, well trajectories and completion style, fluid model, stimulation design, and production schedule.

## Agenda

1. Modeling objectives
2. Static model
3. Well
4. Fluid model
5. Stimulation
6. Schedule

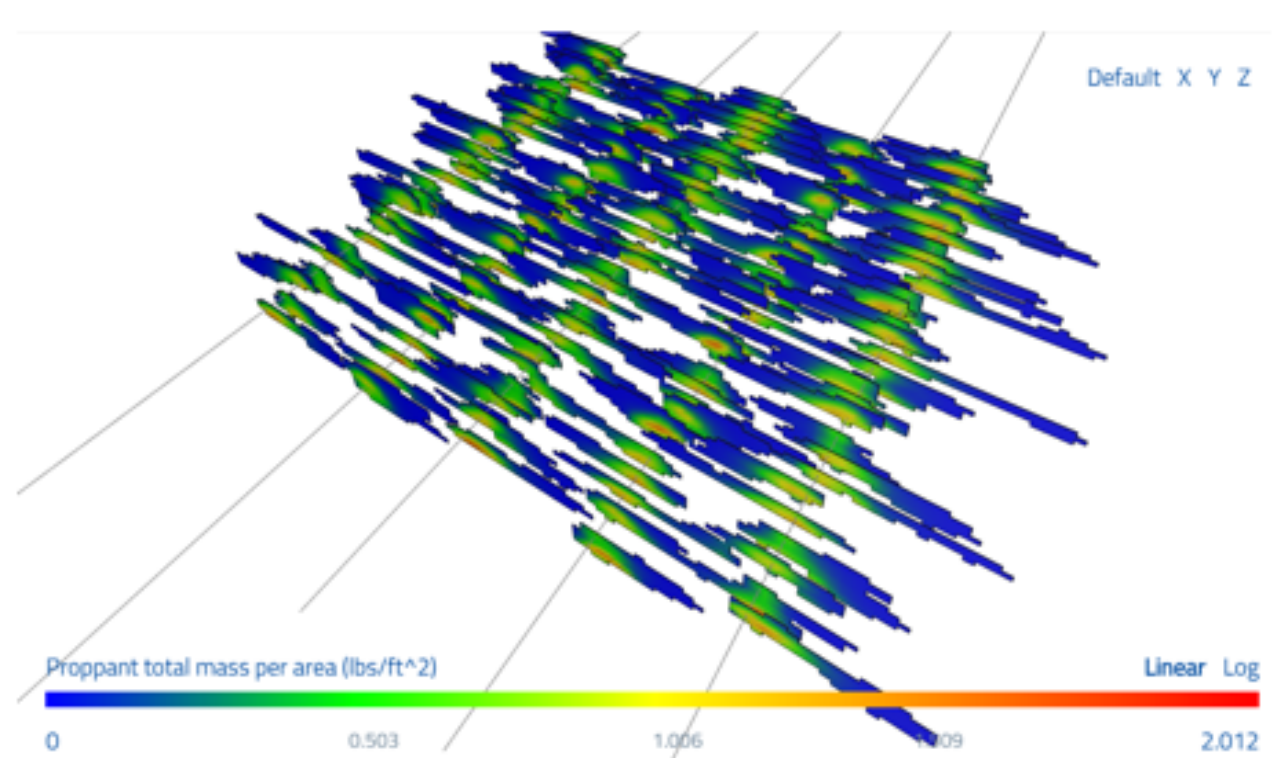

When reviewing objectives, if it often helpful to include a map and/or gunbarrel diagram to contextualize the subject wells for the audience.

## Modeling objectives

Objectives
Static model
Well
Fluid model
Stimulation
Schedule

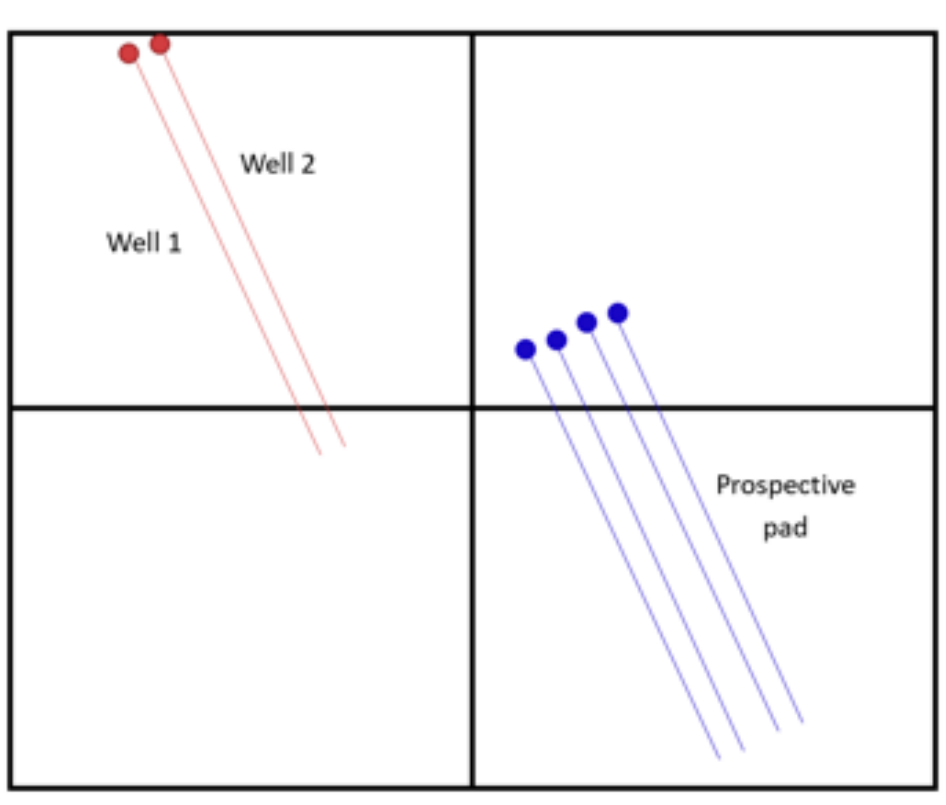

- Well 1 and Well 2 were drilled three years ago
- Will use fully coupled fracture and reservoir ResFrac model to history match the three years of data
- Calibrated model will then be used to optimize well spacing for upcoming Prospective pad

Often the best way to present and review the geologic and geomechanical inputs is graphically (and be sure to mark the zones of interest!).

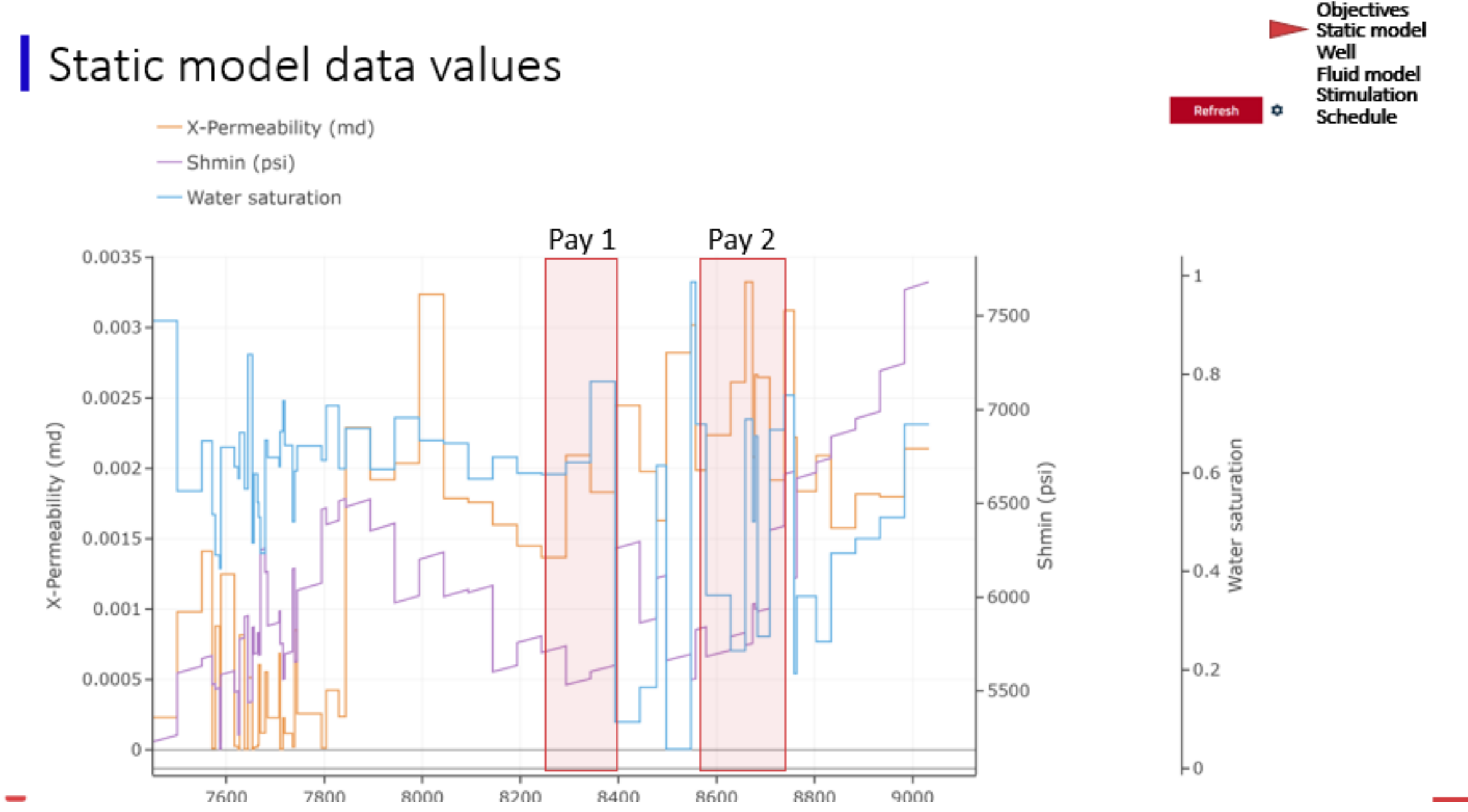

Exact well depths may not be particularly insightful, so presenting the well trajectory as a 3D image with geologic tops marked is most effective.

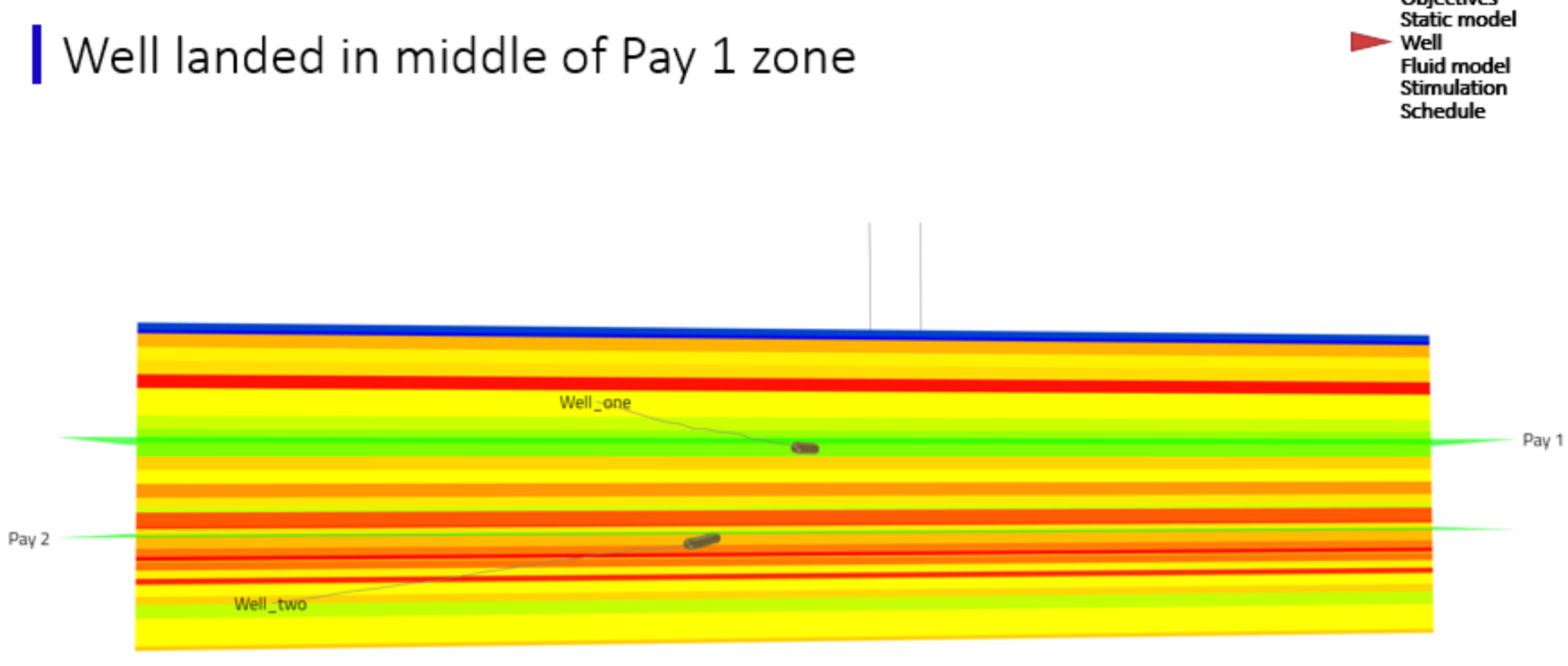

The completion design of each well (cluster and stage spacing, etc.) can easily be presented in tabular form.

# Well completion design

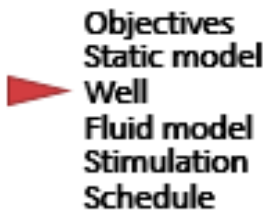

| | Well 1 | Well 2 |
|---|---|---|
| Landing zone | Pay 1 | Pay 2 |
| Cluster spacing (ft) | 31.25 | 31.25 |
| Stage length (ft) | 250 | 250 |
| Perfs/cl | 4 | 4 |
| Perf diameter (in) | 0.36 | 0.36 |
| Limited entry (psi) | 1080 | 1080 |

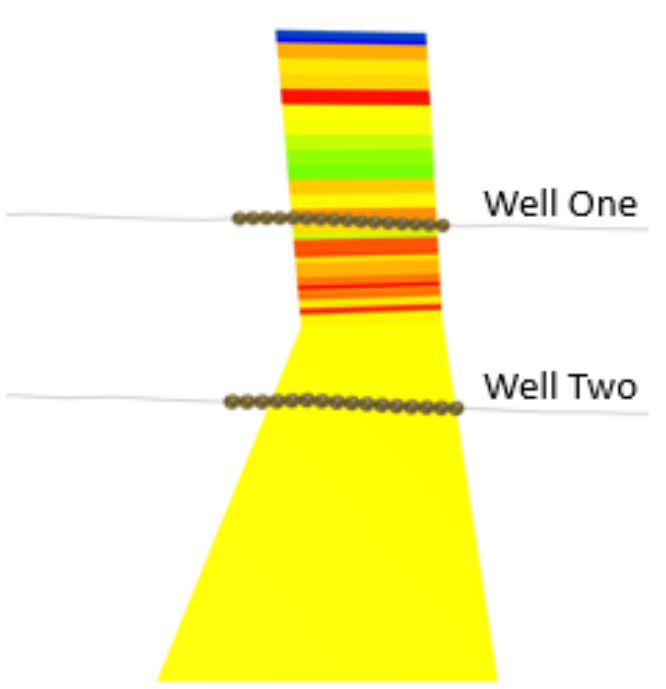

Source: ResFrac_Tutorial_Input_Data.xslx

For the fluid model, stimulation design, and proppant/fluid properties, the preview plots from the simulation builder are a quick and effective away to communicate the information.

# Fluid model

Objectives
Static model
Well
Fluid model
Stimulation
Schedule

- Hydrocarbon is all in oil phase to start
- Bubble point = 3154 psi
- Initial GOR (Rs) = 750 scf/STB
- Source: ResFrac_Tutorial_Input_Data.xslx

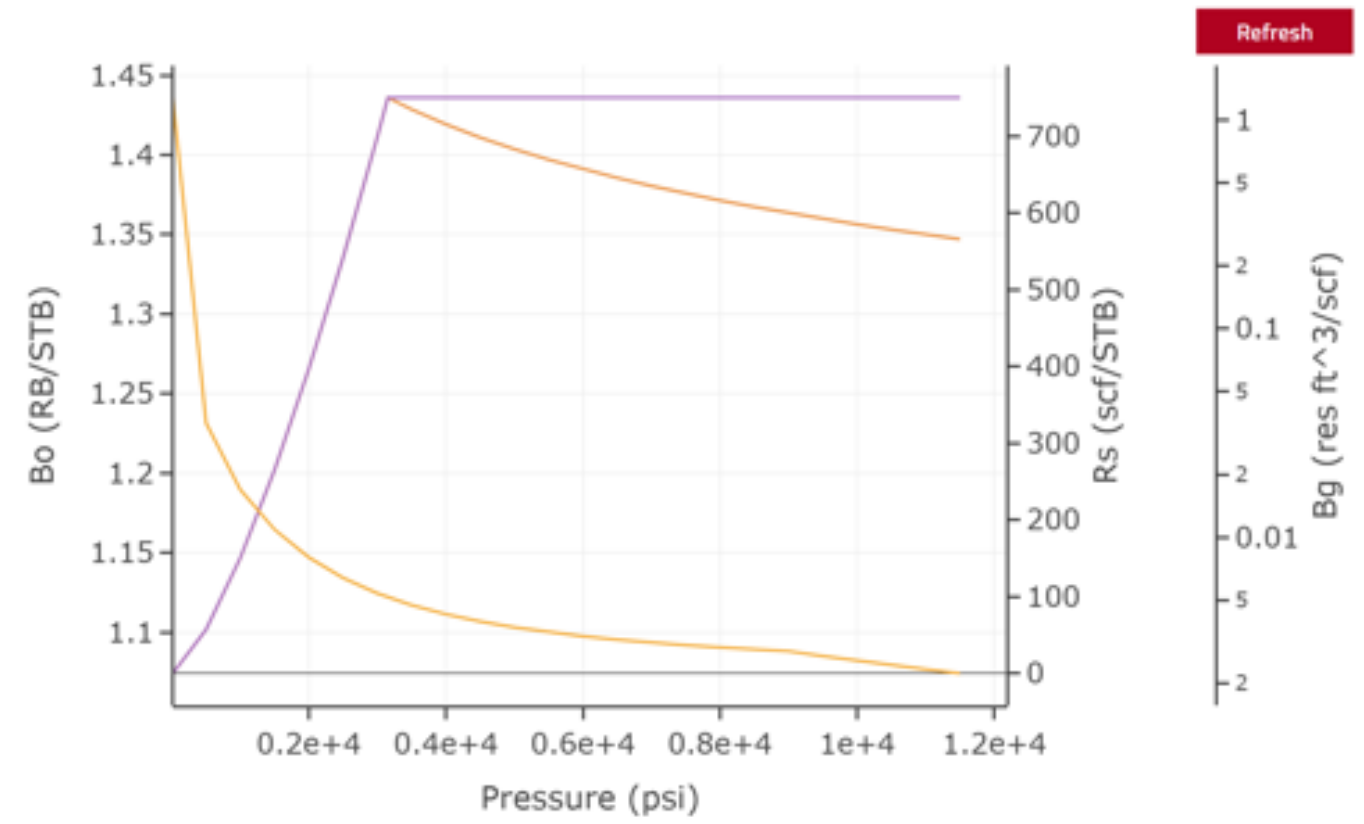

# “Average” stage provided in input data

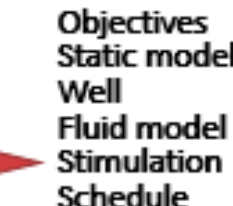

- Used same pump schedule for every stage in all wells
    - 431 lbs/ft 100 mesh proppant
    - 585 lbs/ft 40/70 proppant
    - 1400 gal/ft slickwater
- Source: ResFrac_Tutorial_Input_Data.xslx

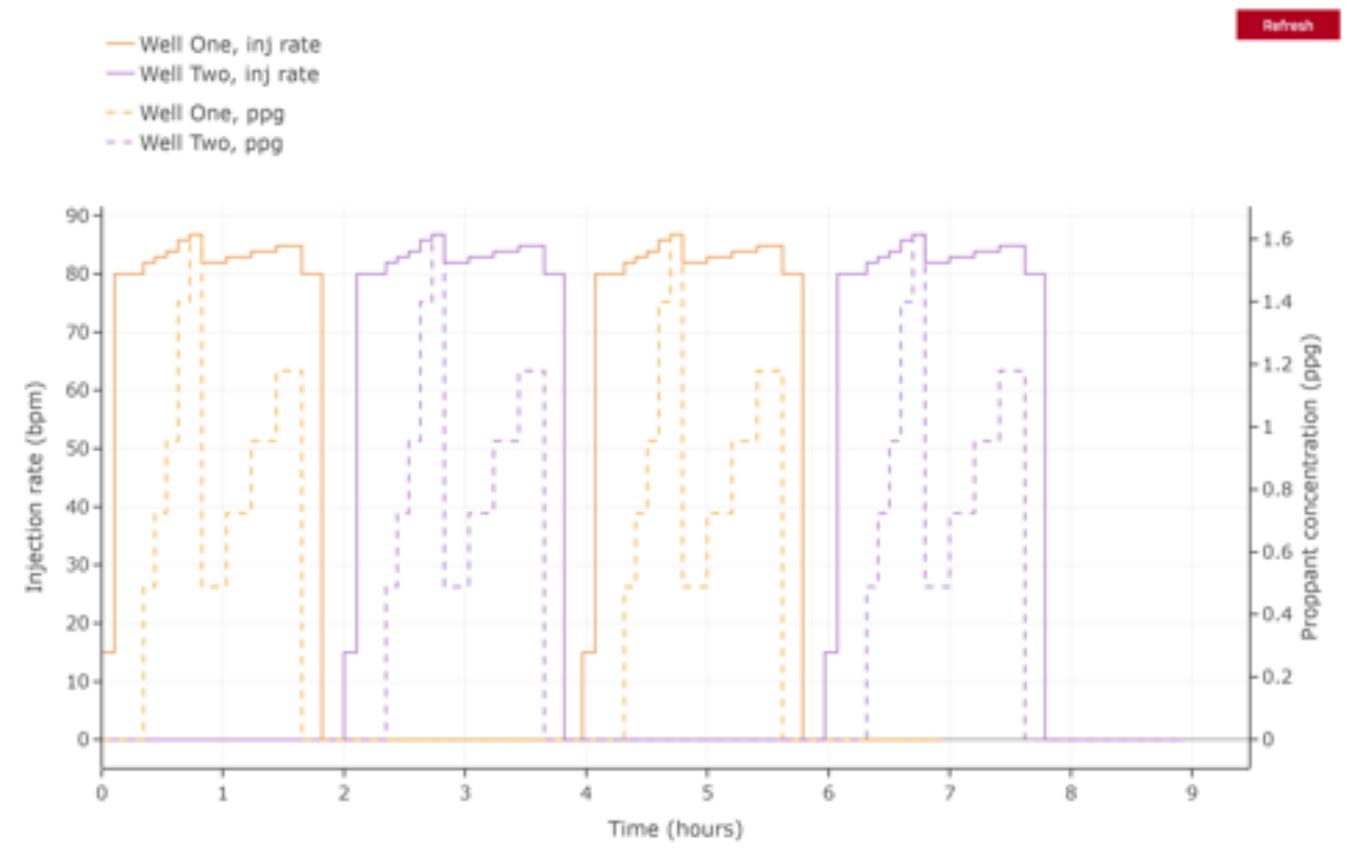

# Fluid and proppant properties

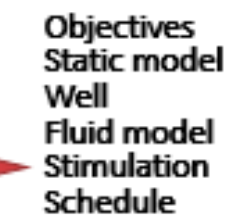

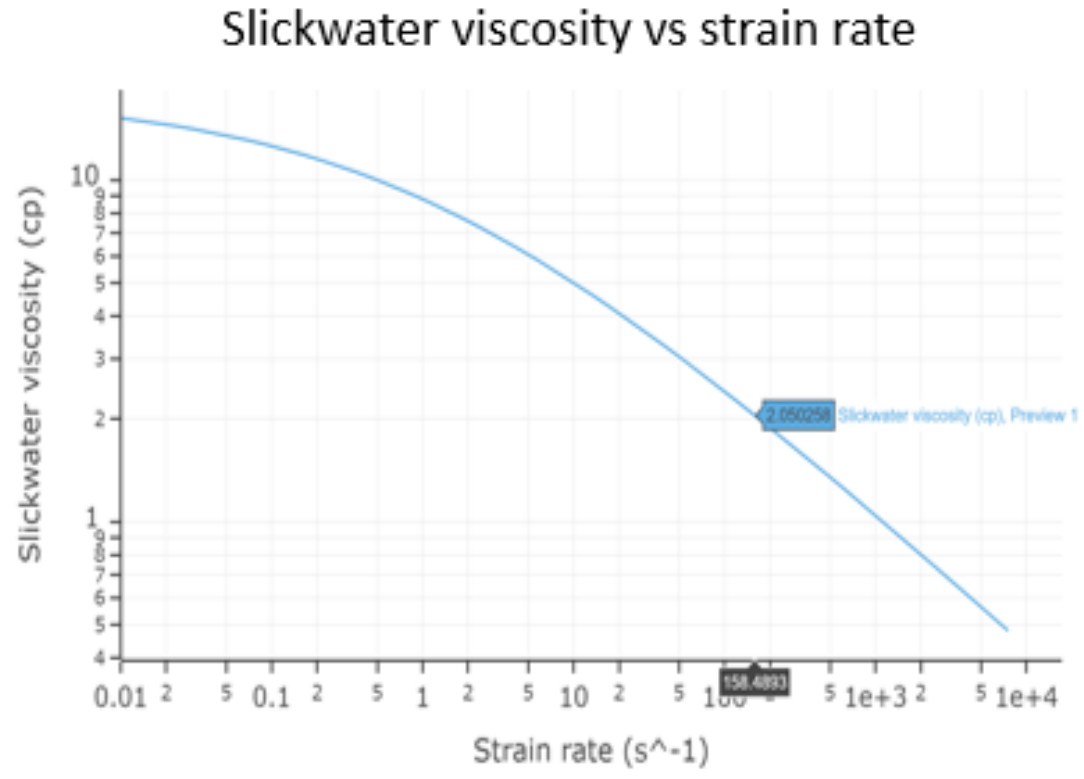

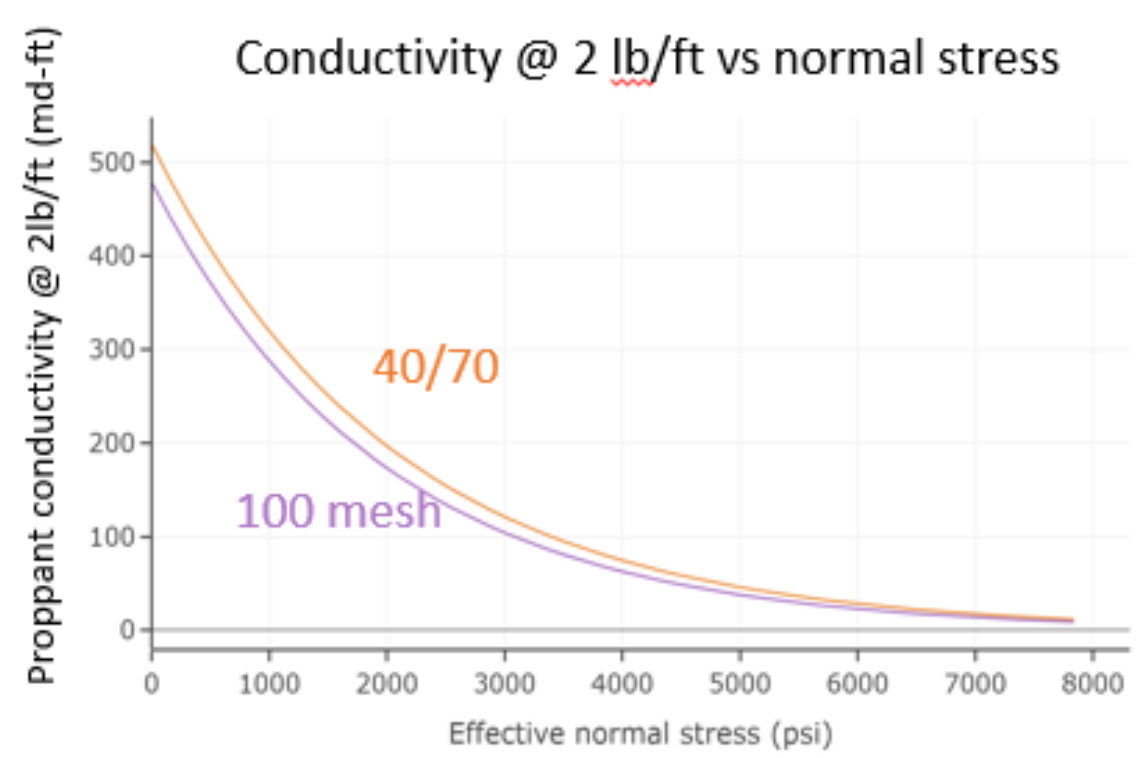

Source: ResFrac Tutorial Input Data.xslx

## Production schedules

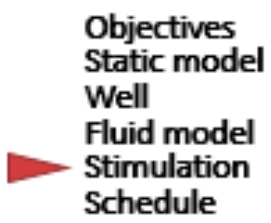

- Both wells are shut-in for 10 days following frac
- Drawdown with same BHP schedule
- Terminal BHP of 950 psi

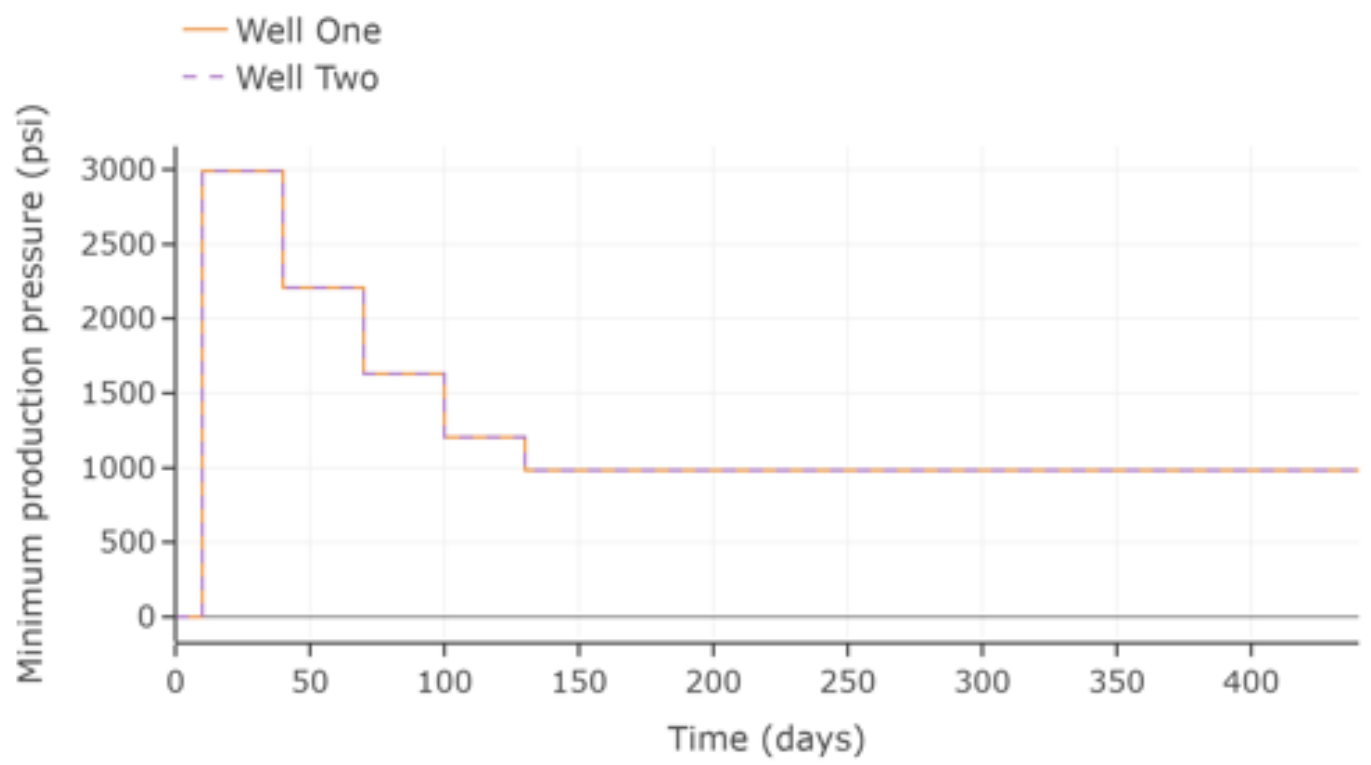

Finally, end the presentation by restating the main takeaways and next steps.

## Conclusions and next steps

- Initial model constructed for Peter Pan Well One and Well Two
- Model input data came from ResFrac_Tutorial_Input_Data.xlsx
- Project tracking on schedule
- Initiating history matching phase following this presentation

## 11.4 Examine the historical data, list observations, and form and validate hypotheses

Before iterating on parameters to history match, it is best to first analyze the data holistically. As ResFrac models are coupled, fracturing and production heuristics should be assessed in tandem. For example, cluster efficiency affects both fracturing and production matching. Be sure to read Sections 8.2.2 and 8.2.3 of this guide on common diagnostics, observations, and the governing parameters.

### 11.4.1 Examining historical data

The input data in the Excel sheet lists some initial observations of the two wells.

| | |
|---|---|
| 17 | **Stimulation observations** |
| 18 | Fractures are expected to be about 1500 ft long (750ft on either side of the wellbore) and 500 ft high |
| 19 | Cluster efficiency was observed to be ~ 75% |
| 20 | Well One ISIPs ~6400 psi |
| 21 | Well Two ISIPs ~6500 psi |

These are standard observations to record and match in a simulation project. Other fracturing observations may be things like frac hits observed on offsetting wells, fracture propagation rate from fiber optic strain gauges, etc. No additional information was provided about our data set, so we next examine the production data. Production data is available in the last tab of the input data.

We can transform the data in the “Daily production” tab into the same format as the external data to load into the ResFrac visualization tool to make observations of the production patterns.

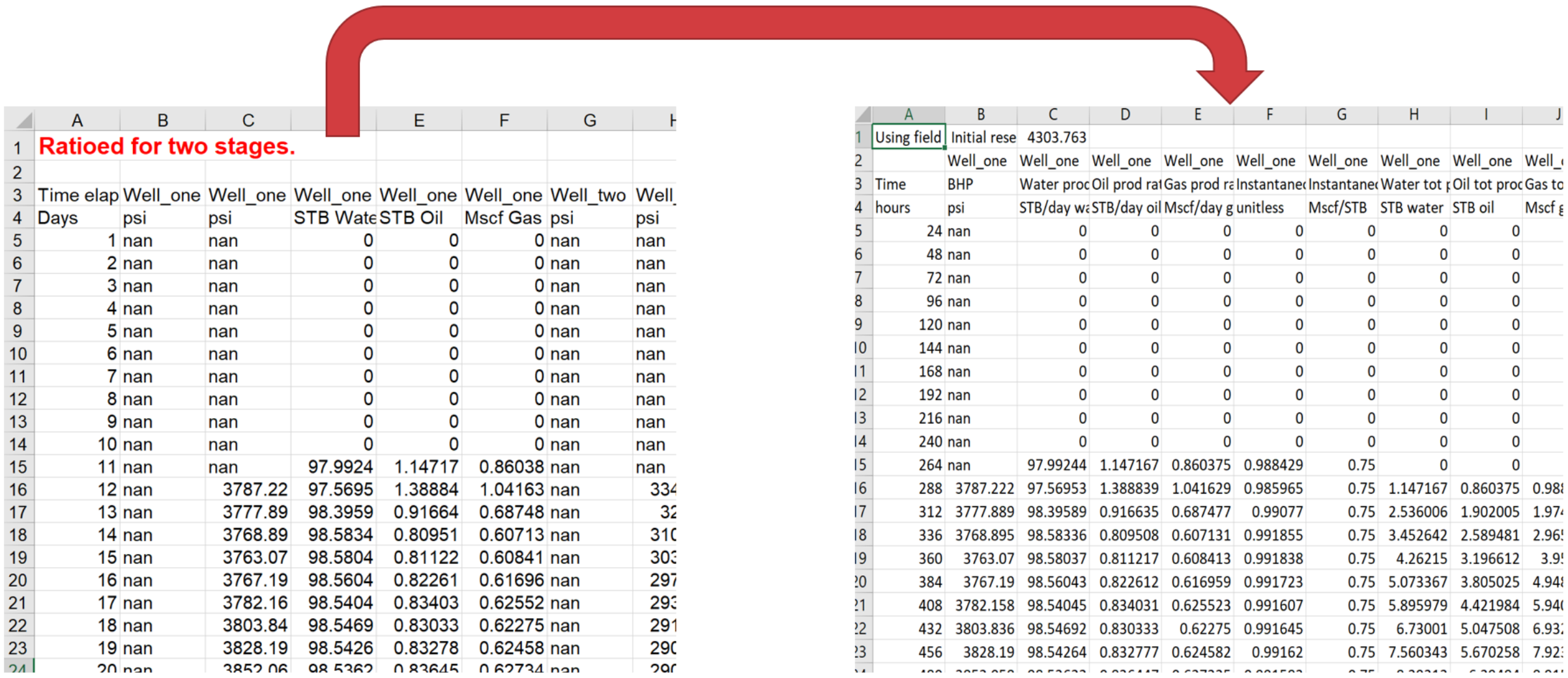

| | A | B | C | D | E | F | G | H |
|---|---|---|---|---|---|---|---|---|
| 1 | Ratioed for two stages. | | | | | | | |
| 2 | | | | | | | | |
| 3 | Time elap | Well_one | Well_one | Well_one | Well_one | Well_one | Well_two | Well_ |
| 4 | Days | psi | psi | STB Wate | STB Oil | Mscf Gas | psi | psi |
| 5 | 1 | nan | nan | 0 | 0 | 0 | nan | nan |
| 6 | 2 | nan | nan | 0 | 0 | 0 | nan | nan |
| 7 | 3 | nan | nan | 0 | 0 | 0 | nan | nan |
| 8 | 4 | nan | nan | 0 | 0 | 0 | nan | nan |
| 9 | 5 | nan | nan | 0 | 0 | 0 | nan | nan |
| 10 | 6 | nan | nan | 0 | 0 | 0 | nan | nan |
| 11 | 7 | nan | nan | 0 | 0 | 0 | nan | nan |
| 12 | 8 | nan | nan | 0 | 0 | 0 | nan | nan |
| 13 | 9 | nan | nan | 0 | 0 | 0 | nan | nan |
| 14 | 10 | nan | nan | 0 | 0 | 0 | nan | nan |
| 15 | 11 | nan | nan | 97.9924 | 1.14717 | 0.86038 | nan | nan |
| 16 | 12 | nan | 3787.22 | 97.5695 | 1.38884 | 1.04163 | nan | 334 |
| 17 | 13 | nan | 3777.89 | 98.3959 | 0.91664 | 0.68748 | nan | 32 |
| 18 | 14 | nan | 3768.89 | 98.5834 | 0.80951 | 0.60713 | nan | 310 |
| 19 | 15 | nan | 3763.07 | 98.5804 | 0.81122 | 0.60841 | nan | 303 |
| 20 | 16 | nan | 3767.19 | 98.5604 | 0.82261 | 0.61696 | nan | 297 |
| 21 | 17 | nan | 3782.16 | 98.5404 | 0.83403 | 0.62552 | nan | 293 |
| 22 | 18 | nan | 3803.84 | 98.5469 | 0.83033 | 0.62275 | nan | 291 |
| 23 | 19 | nan | 3828.19 | 98.5426 | 0.83278 | 0.62458 | nan | 290 |
| 24 | 20 | nan | 3852.06 | 98.5362 | 0.83645 | 0.62734 | nan | 290 |

| | A | B | C | D | E | F | G | H | I | J |
|---|---|---|---|---|---|---|---|---|---|---|
| 1 | Using field | Initial rese | 4303.763 | | | | | | | |
| 2 | | Well_one | Well_one | Well_one | Well_one | Well_one | Well_one | Well_one | Well_one | Well_ |
| 3 | Time | BHP | Water proc | Oil prod rat | Gas prod ra | Instantane | Instantane | Water tot p | Oil tot proc | Gas to |
| 4 | hours | psi | STB/day wa | STB/day oil | Mscf/day g | unitless | Mscf/STB | STB water | STB oil | Mscf g |
| 5 | 24 | nan | 0 | 0 | 0 | 0 | 0 | 0 | 0 | |
| 6 | 48 | nan | 0 | 0 | 0 | 0 | 0 | 0 | 0 | |
| 7 | 72 | nan | 0 | 0 | 0 | 0 | 0 | 0 | 0 | |
| 8 | 96 | nan | 0 | 0 | 0 | 0 | 0 | 0 | 0 | |
| 9 | 120 | nan | 0 | 0 | 0 | 0 | 0 | 0 | 0 | |
| 10 | 144 | nan | 0 | 0 | 0 | 0 | 0 | 0 | 0 | |
| 11 | 168 | nan | 0 | 0 | 0 | 0 | 0 | 0 | 0 | |
| 12 | 192 | nan | 0 | 0 | 0 | 0 | 0 | 0 | 0 | |
| 13 | 216 | nan | 0 | 0 | 0 | 0 | 0 | 0 | 0 | |
| 14 | 240 | nan | 0 | 0 | 0 | 0 | 0 | 0 | 0 | |
| 15 | 264 | nan | 97.99244 | 1.147167 | 0.860375 | 0.988429 | 0.75 | 0 | 0 | |
| 16 | 288 | 3787.222 | 97.56953 | 1.388839 | 1.041629 | 0.985965 | 0.75 | 1.147167 | 0.860375 | 0.98 |
| 17 | 312 | 3777.889 | 98.39589 | 0.916635 | 0.687477 | 0.99077 | 0.75 | 2.536006 | 1.902005 | 1.97 |
| 18 | 336 | 3768.895 | 98.58336 | 0.809508 | 0.607131 | 0.991855 | 0.75 | 3.452642 | 2.589481 | 2.96 |
| 19 | 360 | 3763.07 | 98.58037 | 0.811217 | 0.608413 | 0.991838 | 0.75 | 4.26215 | 3.196612 | 3.9 |
| 20 | 384 | 3767.19 | 98.56043 | 0.822612 | 0.616959 | 0.991723 | 0.75 | 5.073367 | 3.805025 | 4.94 |
| 21 | 408 | 3782.158 | 98.54045 | 0.834031 | 0.625523 | 0.991607 | 0.75 | 5.895979 | 4.421984 | 5.94 |
| 22 | 432 | 3803.836 | 98.54692 | 0.830333 | 0.62275 | 0.991645 | 0.75 | 6.73001 | 5.047508 | 6.93 |
| 23 | 456 | 3828.19 | 98.54264 | 0.832777 | 0.624582 | 0.99162 | 0.75 | 7.560343 | 5.670258 | 7.92 |

Once formatted as an external data set, the production data can be imported into the viz tool and analyzed for observations.

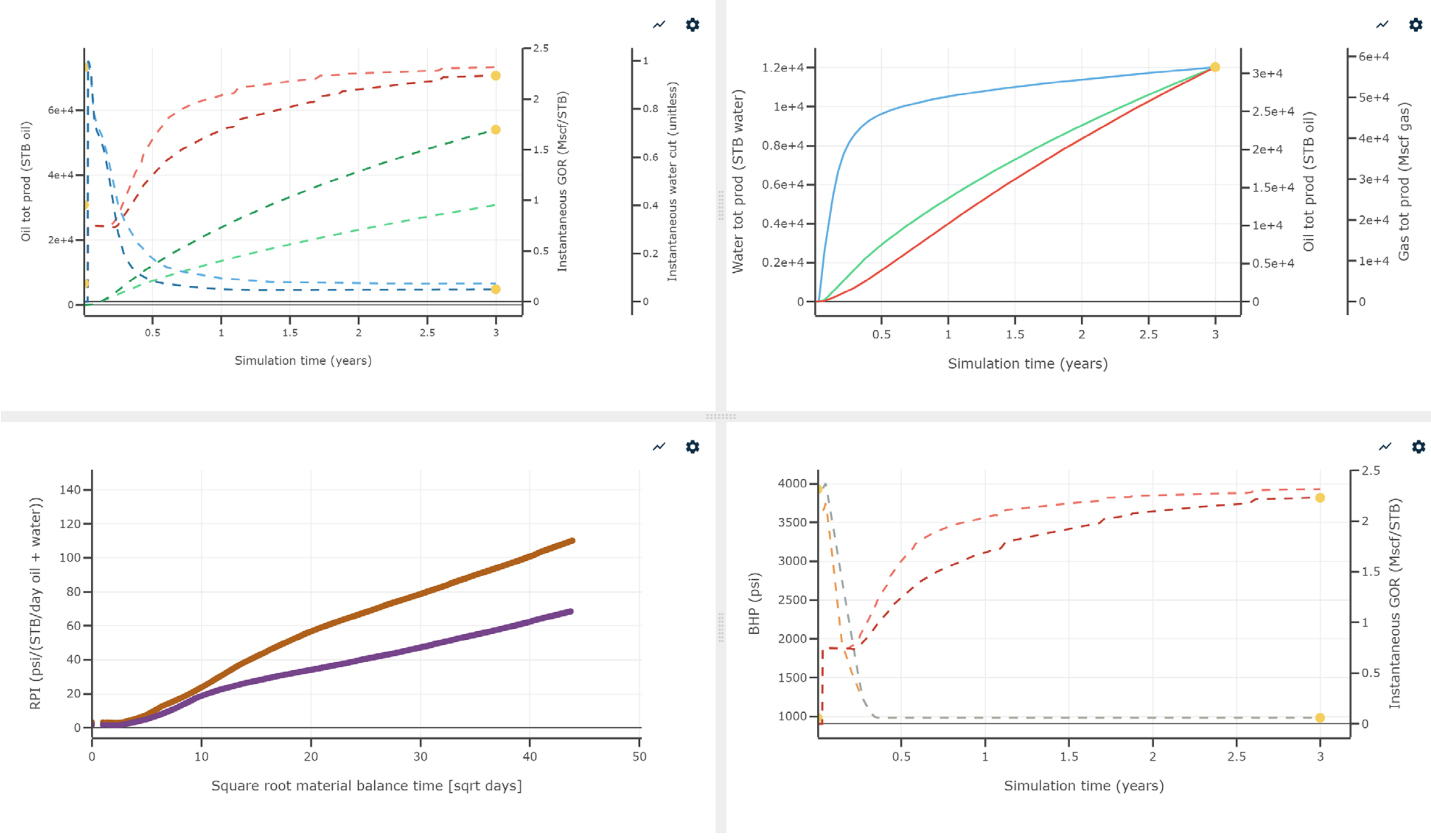

Looking at the production data we make the following observations:

- Well Two is a better producer than Well One
  - Higher cumulative oil
  - Lower water cut
- Water cut starts high, then diminishes with time (for both wells)
- GOR increases soon after BHP drops below Pb, but not immediately, suggesting that the fractures are finite conductivity
- Early-time RTA shows some curvature – maybe from supercharging, but long-term behavior is fairly linear (straight line), indicating that strong well-to-well or cluster-to-cluster interference is not occurring in the first three years, nor is there a strong decline of oil relative perm when dropping below bubble point.

### 11.4.2 Form and validate hypotheses

With our observations made, we organize the observations into a table and form hypotheses for each. When hypothesizing mechanisms to explain the observations, we should also note the model parameters we plan to change in order to match the observation.

| Observation | Hypothesis |
| --- | --- |
| Fractures have expected length of 1500 ft and height of 500 ft | Fracture dimensions largely governed by fracture toughness and Shmin profile. Will tune toughness first to match. |
| ISIP's between 6400-6500 psi for both wells | Fracture toughness will increase ISIP. If dimensions dictate lower toughness, additional ISIP pressure can be a function of near-well complexity. |
| Cluster efficiency ~80% | Cluster efficiency is a function of stress shadowing. Increased toughness will reduce cluster efficiency, and if not enough can use randomized tensile strength to further reduce. |
| Well Two out produces Well One by ~33% | Well Two is landed in higher oil saturation, higher permeability zone, so should produce more due to rel perm. |
| GOR (both wells) starts rising @ ~1700 psi BHP and possible small y-intercept on RTA | Pb is 3100, so delay in GOR breakout suggests some water block in fractures or finite conductivity. Will first iterate on rel perms to get WC, then decrease k0 (prop perm) to get y-intercept. |
| Water cut falls for 1 year, then starts to level out at 5-10%. | This is also a function of rel perm and can be matched using rel perm curves. Well Two exhibits lower WC than Well One, corresponding to higher oil saturations. |

To the extent possible, we want to validate our history matching approach before iterating to finely tune model parameters. In Section 6.3.3 we discuss bracketing the solution: testing the extremes of a parameter range to confirm that the desired solution lies between the extremes. We want to do this for each of our key observations. For instance, our first observation is of average fracture length and height. Our corresponding hypothesis is that we can adjust toughness to match the dimension expectations. To validate the hypothesis, we can quickly run models with very low toughness scaling (maybe 0.01) and very high (maybe 1.5) and confirm that our desired geometry lies somewhere between the two extremes.

### 11.4.3 Present to stakeholders

With our proposed history matching strategy validated, we now want to get the buy-in from the key stakeholders on the history matching strategy. Our objective with the presentation is to align on the defining observations of the data set and strategy to match each observation.

As always, we start with an abstract of the primary presentation messages.

## Abstract

- Frac length, ISIPs matched by varying toughness
- Bracketed parameter ranges to match cluster efficiency, oil production, water cut, and GOR
- Performance differences in Well One and Well Two driven by differences in landing zone properties

Then, we may take a couple slides to review project scope and objectives, before presenting our summary of characteristic observations and hypotheses.

## History matching objectives

| Observation | Hypothesis | Validated | Status |
|---|---|---|---|
| Fractures have expected length of 1500 ft and height of 500 ft | Fracture dimensions largely governed by fracture toughness and Shmin profile. Will tune toughness first to match. | x | Matched |
| ISIP's between 6400-6500 psi for both wells | Fracture toughness will increase ISIP. If dimensions dictate lower toughness, additional ISIP pressure can be a function of near-well complexity. | x | Matched |
| Cluster efficiency ~80% | Cluster efficiency is a function of stress shadowing. Increased toughness will reduce cluster efficiency, and if not enough can use randomized tensile strength to further reduce. | x | Not matched |
| Well Two out produces Well One by ~33% | Well Two is landed in higher oil saturation, higher permeability zone, so should produce more due to rel perm. | x | Not matched |
| GOR (both wells) starts rising @ ~1700 psi BHP and possible small y-intercept on RTA | Pb is 3100, so delay in GOR breakout suggests some water block in fractures or finite conductivity. Will first iterate on rel perms to get WC, then decrease k0 (prop perm) to get y-intercept. | x | Not matched |
| Water cut falls for 1 year, then starts to level out at 5-10%. | This is also a function of rel perm and can be matched using rel perm curves. Well Two exhibits lower WC than Well One, corresponding to higher oil saturations. | x | Not matched |

Next, we review each observation and hypothesis validation. In the case of fracture geometries and ISIPs, we were able to match those observations during our solution-bracketing, so we indicate as such with green checkmarks.

# Obs 1: Fracture geometry

- Fracture geometry expectation falls within bounds of low and high toughness runs
- Will be able to match geometry using medium toughness scaling factor

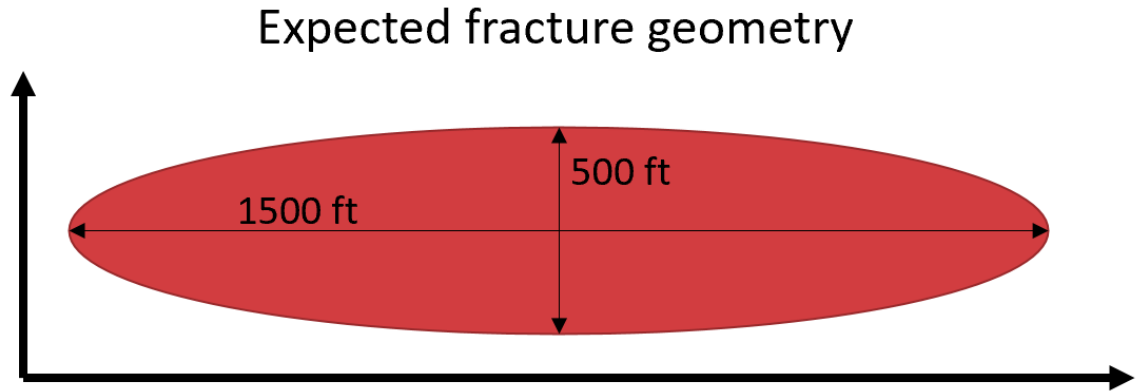

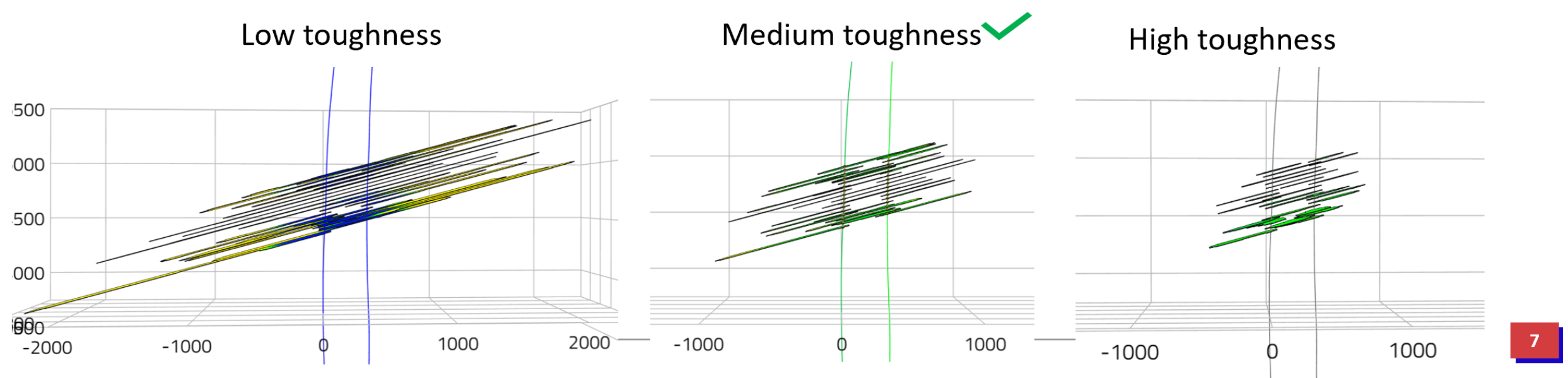

# Obs 2: ISIPs

- ISIPs for both wells are ~6400-6500 psi
- Well Two ISIPs trending slightly higher that Well One
- Medium toughness scaling value that matches geometries also matches ISIPs

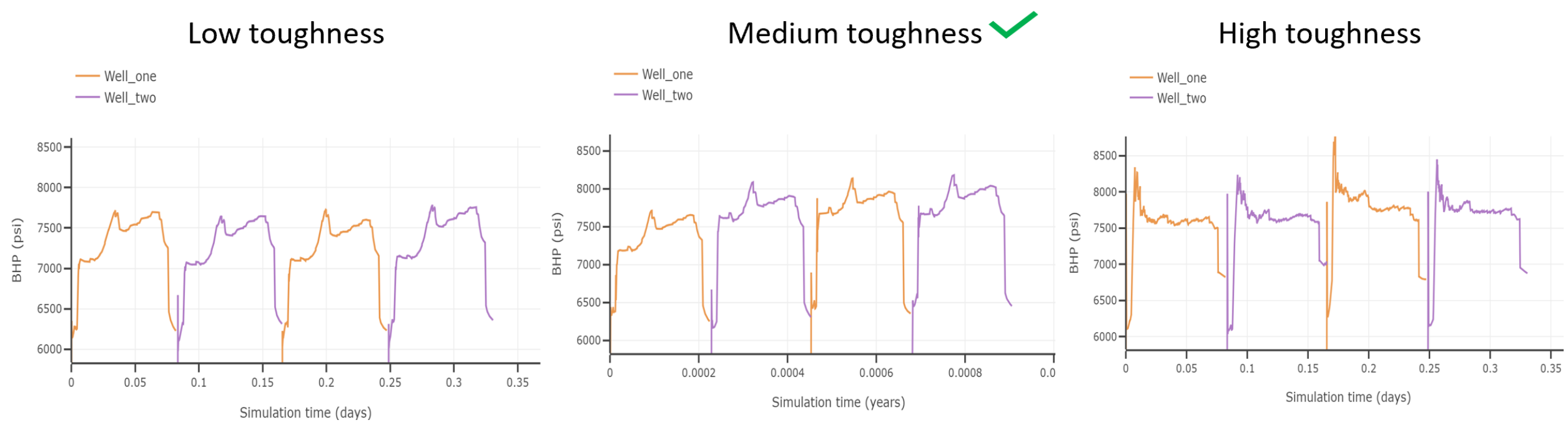

# Obs 3: Cluster efficiency

- Based on field diagnostics, cluster efficiency is believed to be ~80%
- Fracture geometries and ISIPs match with lower toughness values (prior two slides), will increase tensile strength slightly two reduce cluster efficiency. Some further refinement to toughness may be necessary after.
- Solution lies between low and high strength images below.

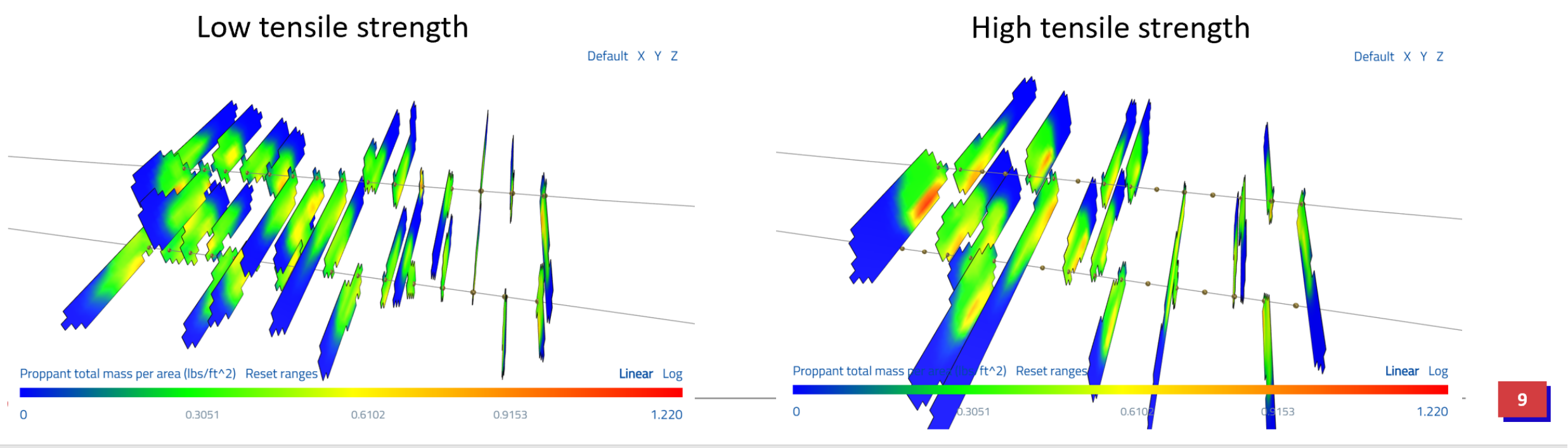

Sometimes, it may be helpful to review the observation in more detail prior to presenting the hypothesis, as we've done with the production data.

# Historical production

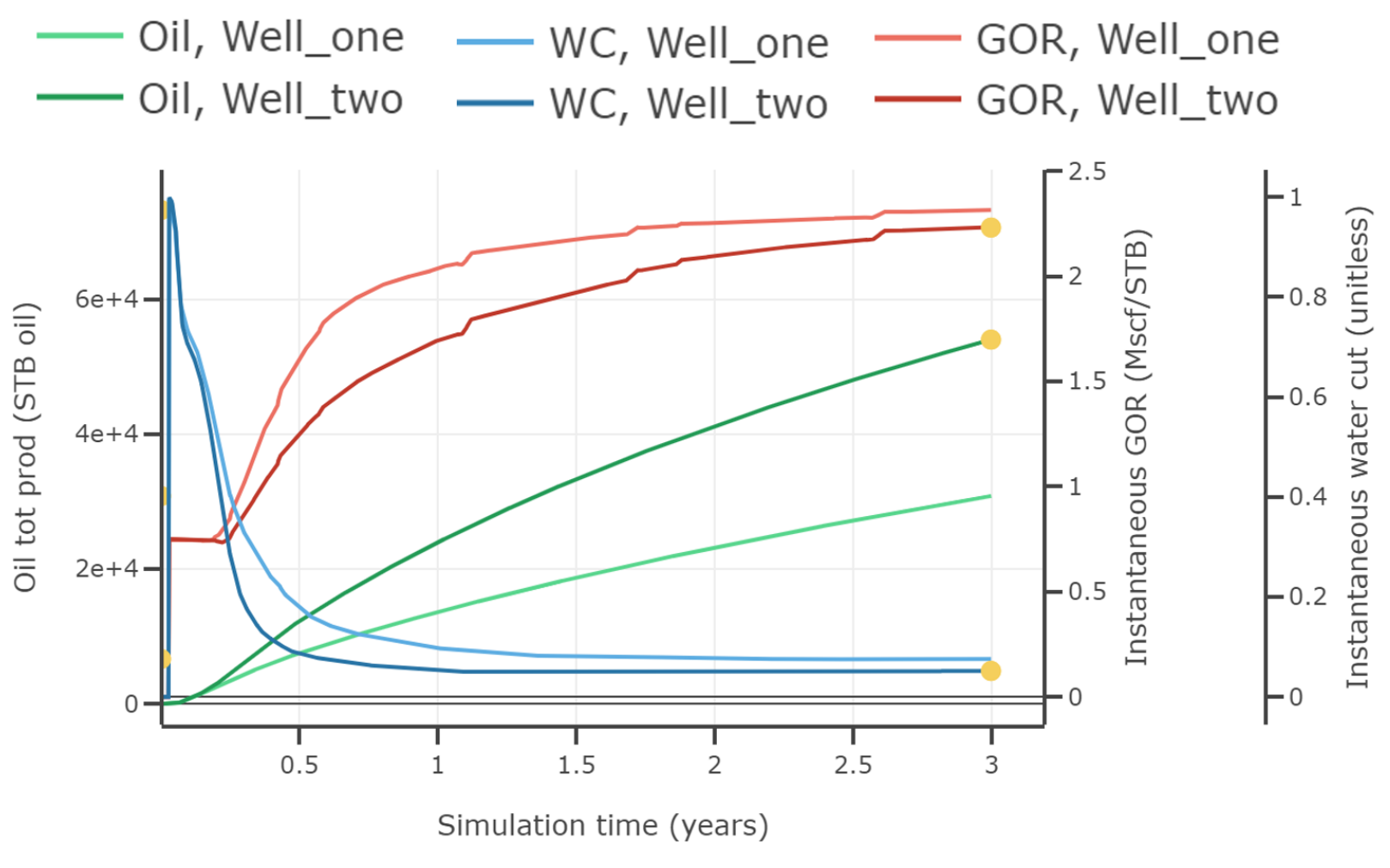

- Well Two has lower WC and GOR than Well One, but produces ~33% more oil
- GOR for both wells is flat for first ~3 months, then starts to rise (Well One > Well Two)
- WC levels off after ~1 year at 5-10%

## RTA of actual data

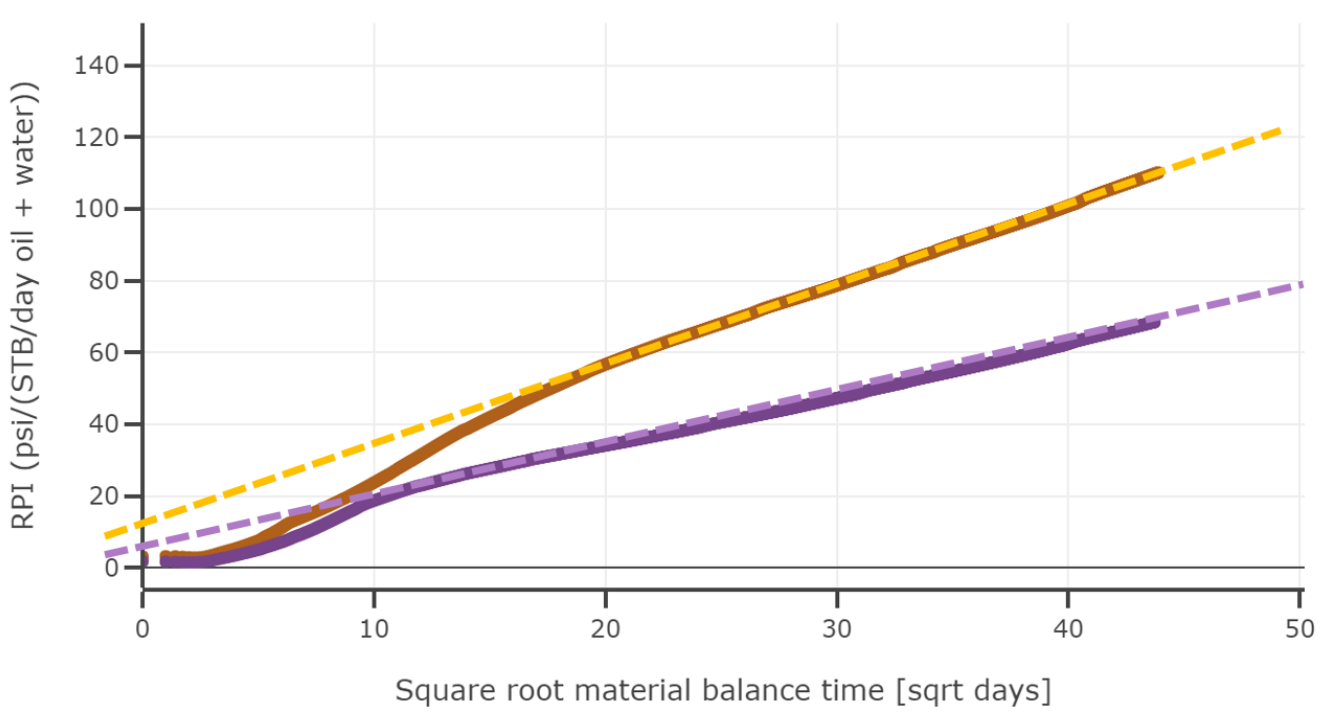

- Fairly linear RTA trends
- Possible small y-intercept for both wells
- Well One RTA trend slope ~1.4x steeper than Well Two, suggesting 2x lower perm.
- Calculated perm ratio matches perm differential in the static model

An effective way of validating permeability is to compare the slopes of RTA trends.

## Obs 4: Production volumes

- Will be able to history match production volumes by reducing bulk permeability
- Permeability reduction likely in 70-90% range (95% reduction resulted in too much reduction in production)

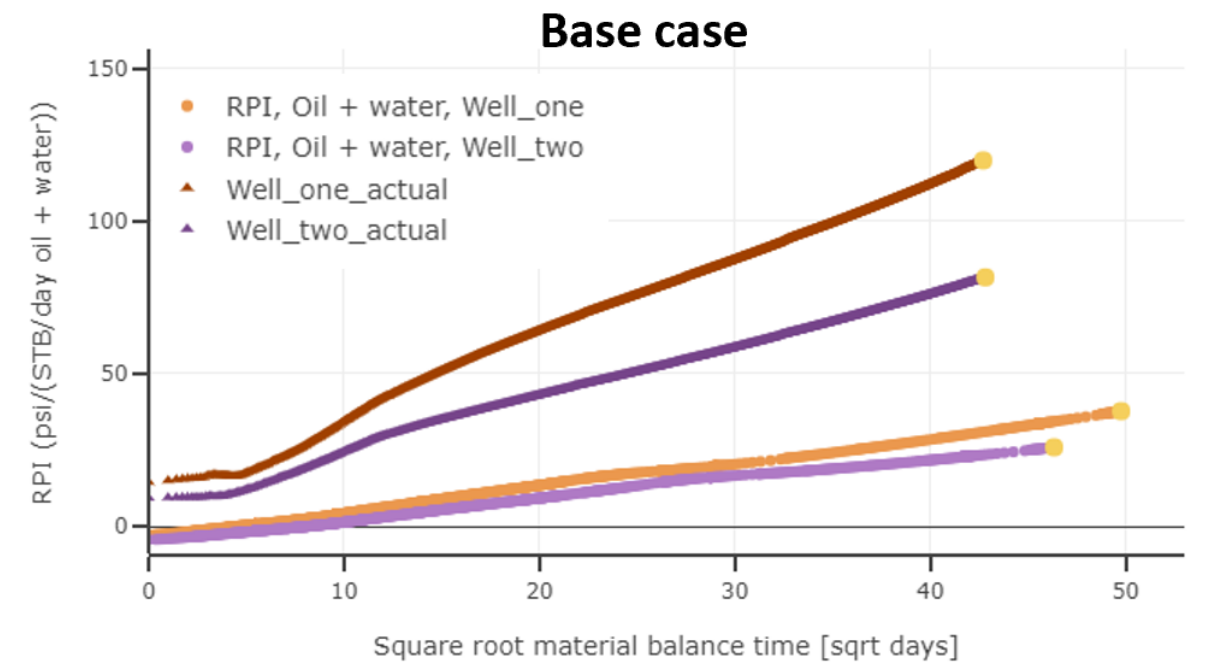

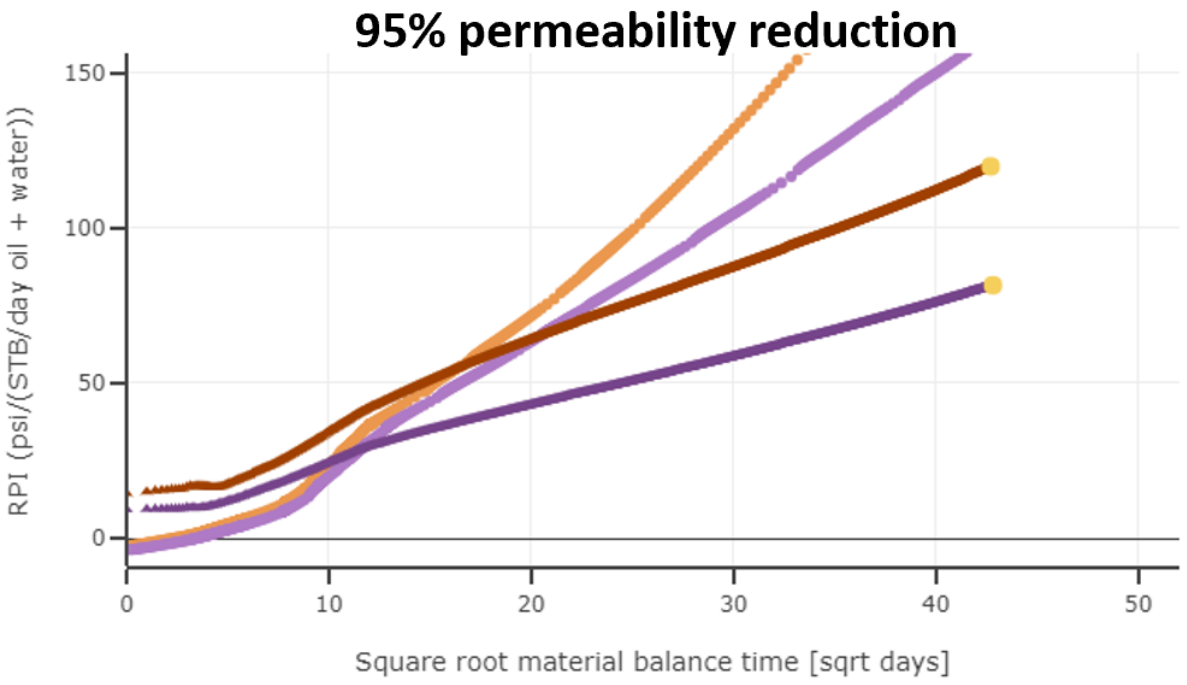

# Obs 5: Fracture conductivity and GOR

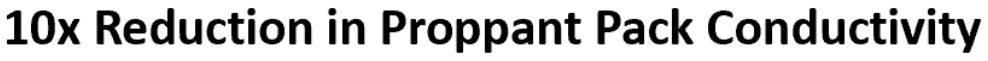

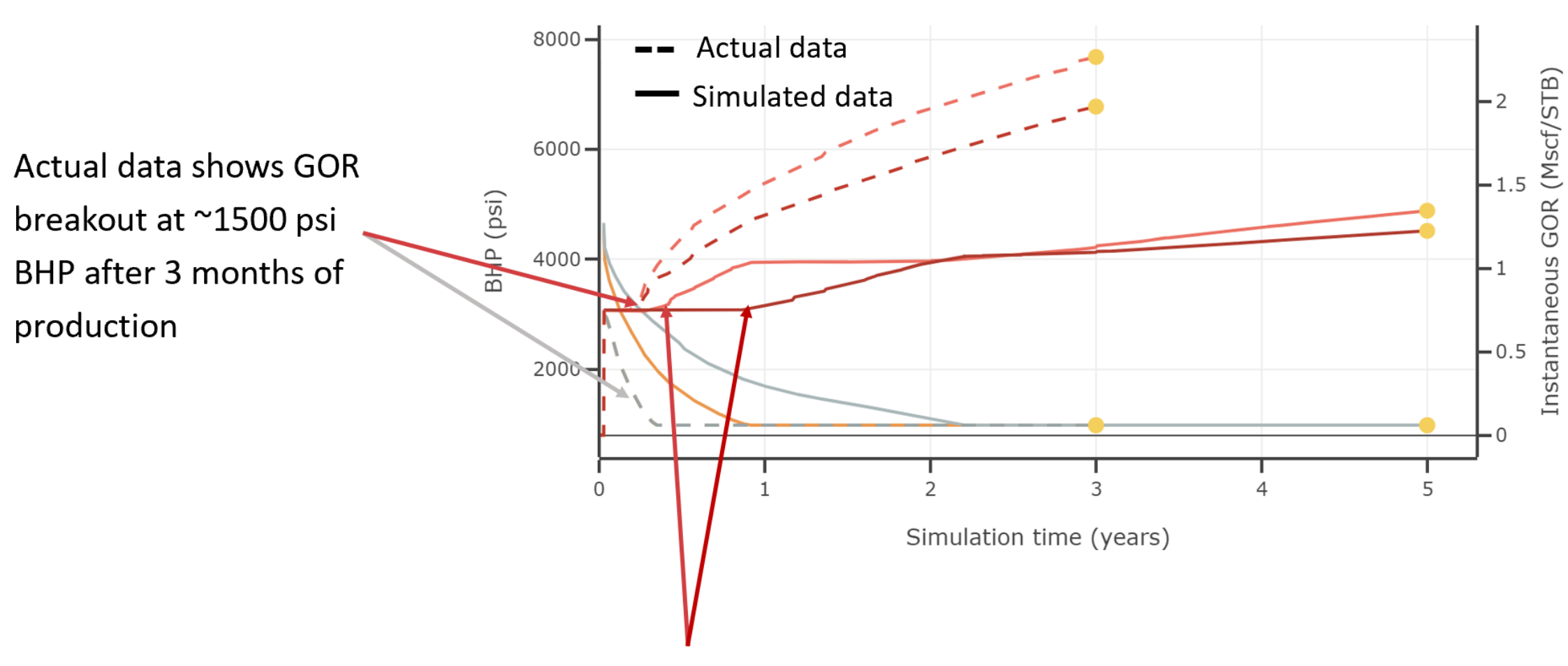

# Obs 5: Fracture conductivity and GOR

**10x Reduction in Proppant Pack Conductivity**

- When fracture conductivity is too low, fracture pressure is not pulled down to BHP, and the matrix pressure does not go below Pb as quickly, delaying GOR rise
- Varying conductivity will allow us to match GOR trend and y-intercept on RTA plot.

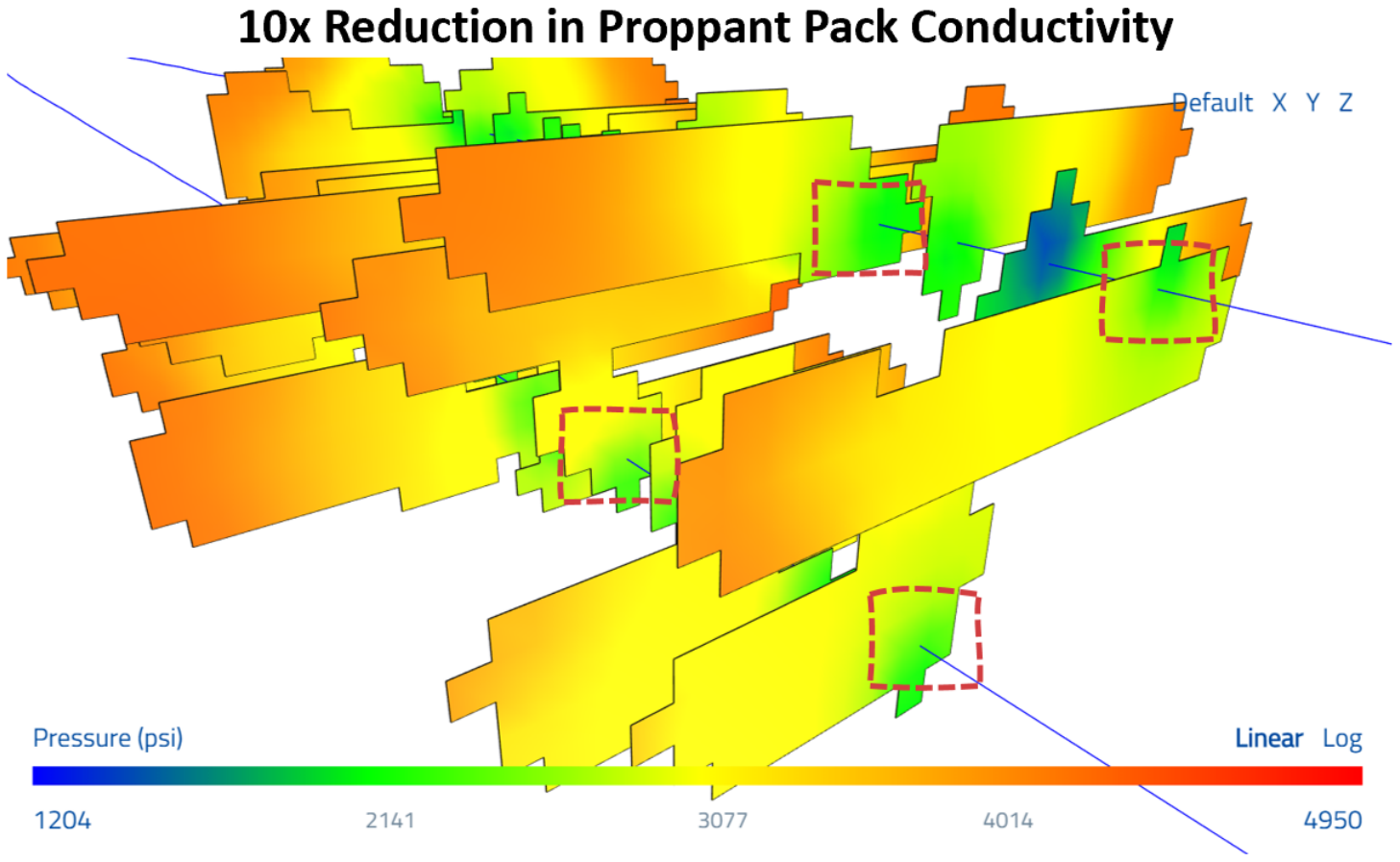

## Obs 6: Water cut trends

- Water cut predominantly controlled by relative permeability curve and producing layers
- Low and high water cut scenarios bracket the solution
- Relative ordering of Well One and Well Two is correct

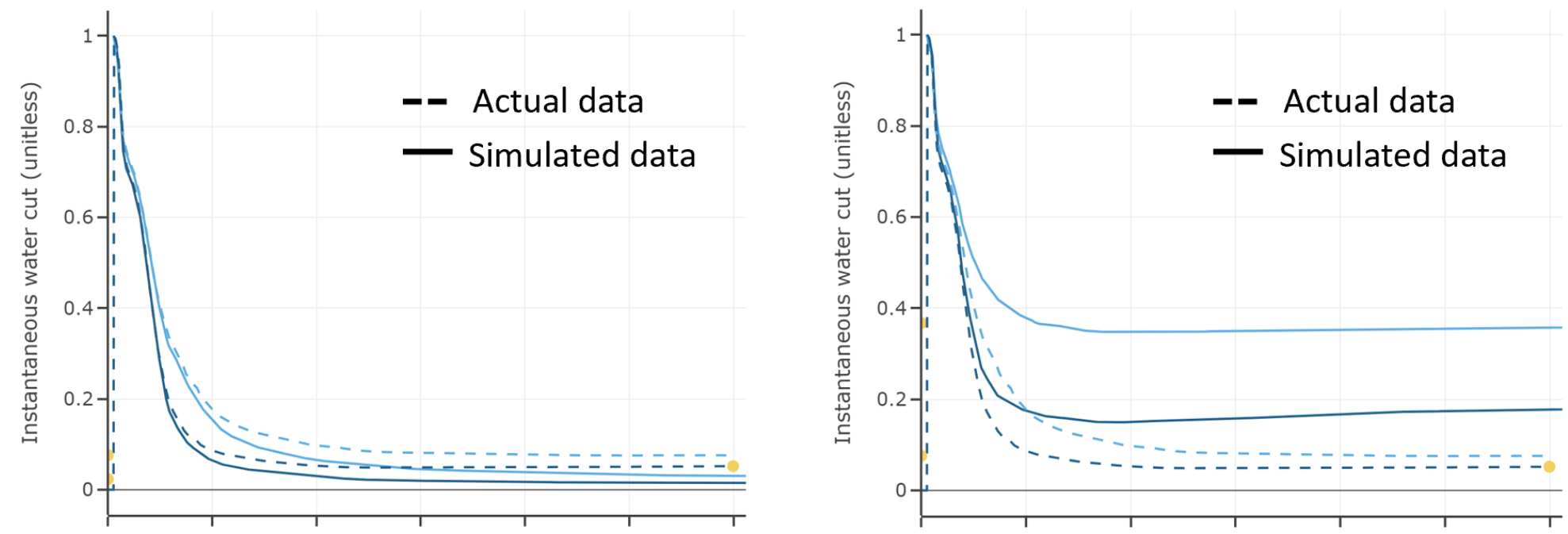

And finally, we always end with summarizing the main points again and aligning on next steps.

## Conclusions

- Frac length, ISIPs matched by varying toughness
- Bracketed parameter ranges to match cluster efficiency, oil production, water cut, and GOR
- Performance differences in Well One and Well Two driven by differences in landing zone properties
- Next step is to iterate on identified parameters and refine history match

### 11.5 History match available data

With the modeling approach validated, we can now fine tune the parameters selected for the history match. Refer to Section 8.4 for a detailed walkthrough of history matching tactics.

It is rare to ever match all the data "perfectly", and to do so could result in overfitting the data. As you present your final history match to the key stakeholders, it is important to set expectations accordingly.

When presenting the final history match, we begin with an abstract and then reminding the audience of the key observations we were trying to history match.

## Abstract

- Key observations of frac length, ISIPs, cluster efficiency, oil production, water cut, and GOR all matched
- Performance differences in Well One and Well Two driven by differences in landing zone properties: permeability and water saturation
- Model is ready for sensitivity analysis

## History matching objectives

| Observation | Hypothesis | Validated | Status |
| --- | --- | --- | --- |
| Fractures have expected length of 1500 ft and height of 500 ft | Fracture dimensions largely governed by fracture toughness and Shmin profile. Will tune toughness first to match. | x | Matched |
| ISIP's between 6400-6500 psi for both wells | Fracture toughness will increase ISIP. If dimensions dictate lower toughness, additional ISIP pressure can be a function of near-well complexity. | x | Matched |
| Cluster efficiency ~80% | Cluster efficiency is a function of stress shadowing. Increased toughness will reduce cluster efficiency, and if not enough can use randomized tensile strength to further reduce. | x | Matched |
| Well Two out produces Well One by ~33% | Well Two is landed in higher oil saturation, higher permeability zone, so should produce more due to rel perm. | x | Matched |
| GOR (both wells) starts rising @ ~1700 psi BHP and possible small y-intercept on RTA | Pb is 3100, so delay in GOR breakout suggests some water block in fractures or finite conductivity. Will first iterate on rel perms to get WC, then decrease k0 (prop perm) to get y-intercept. | x | Matched |
| Water cut falls for 1 year, then starts to level out at 5-10%. | This is also a function of rel perm and can be matched using rel perm curves. Well Two exhibits lower WC than Well One, corresponding to higher oil saturations. | x | Matched |

Then, we go through each observation and the resulting match in the model.

# Geometry, ISIPs, and perf efficiency matched

- Fracturing observations matched by adjust toughness and tensile strength
- Final values within reasonable and expected range

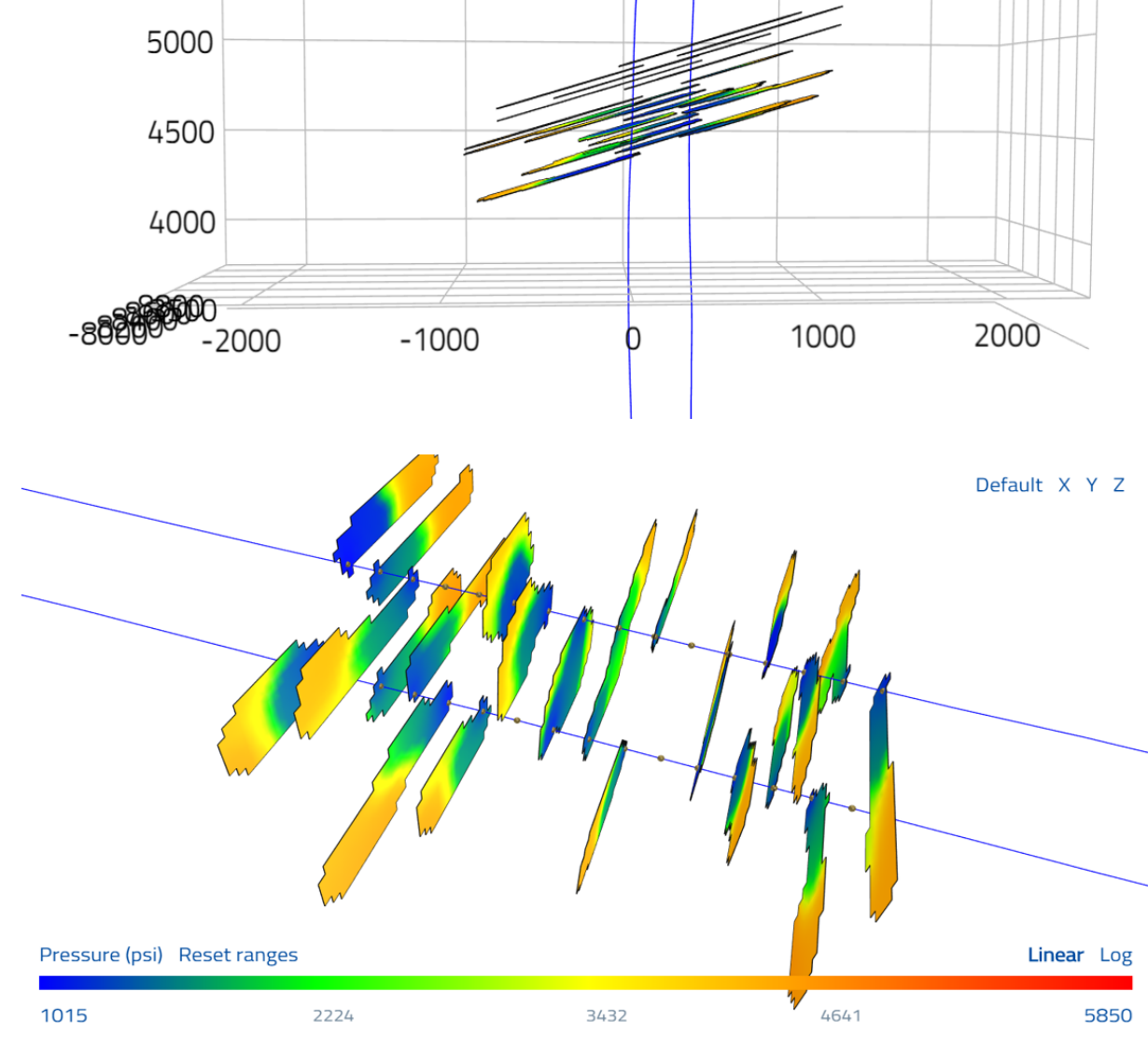

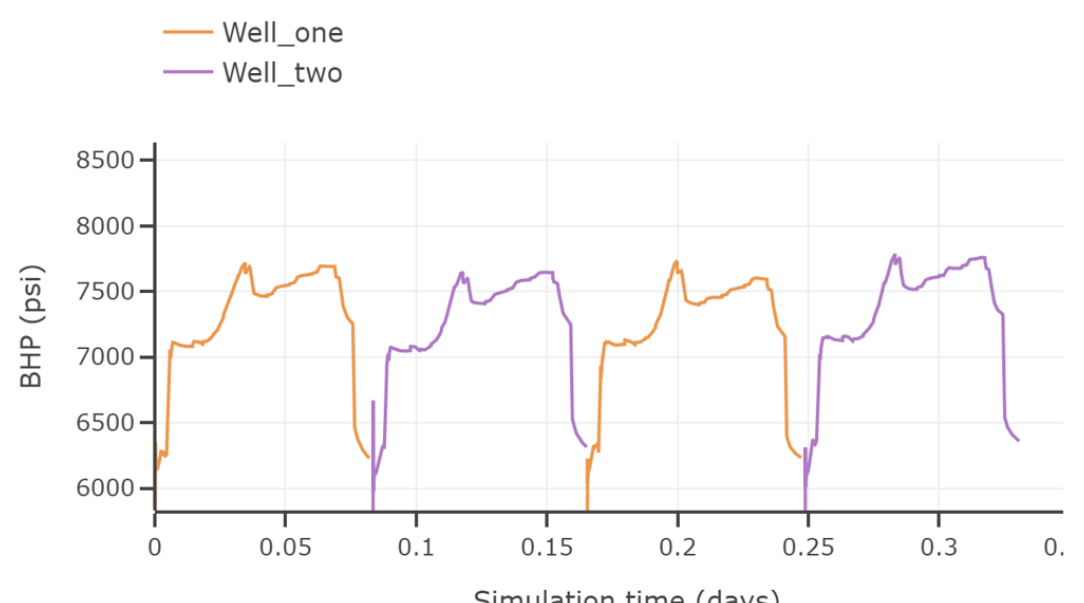

# Final history match

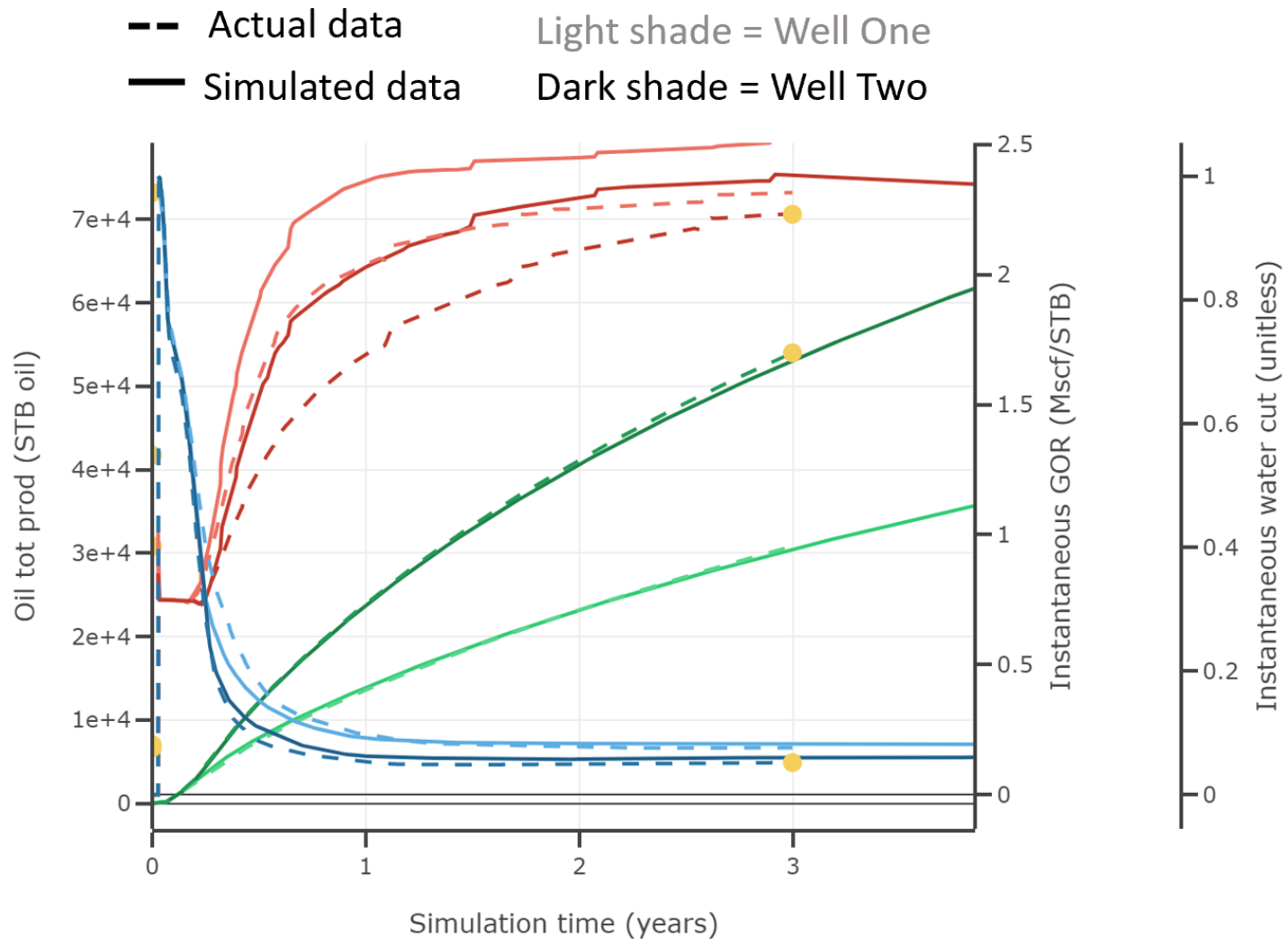

- Good history match
- Very tight match to oil
- Further iteration could bring gas production down some.

## Final history match

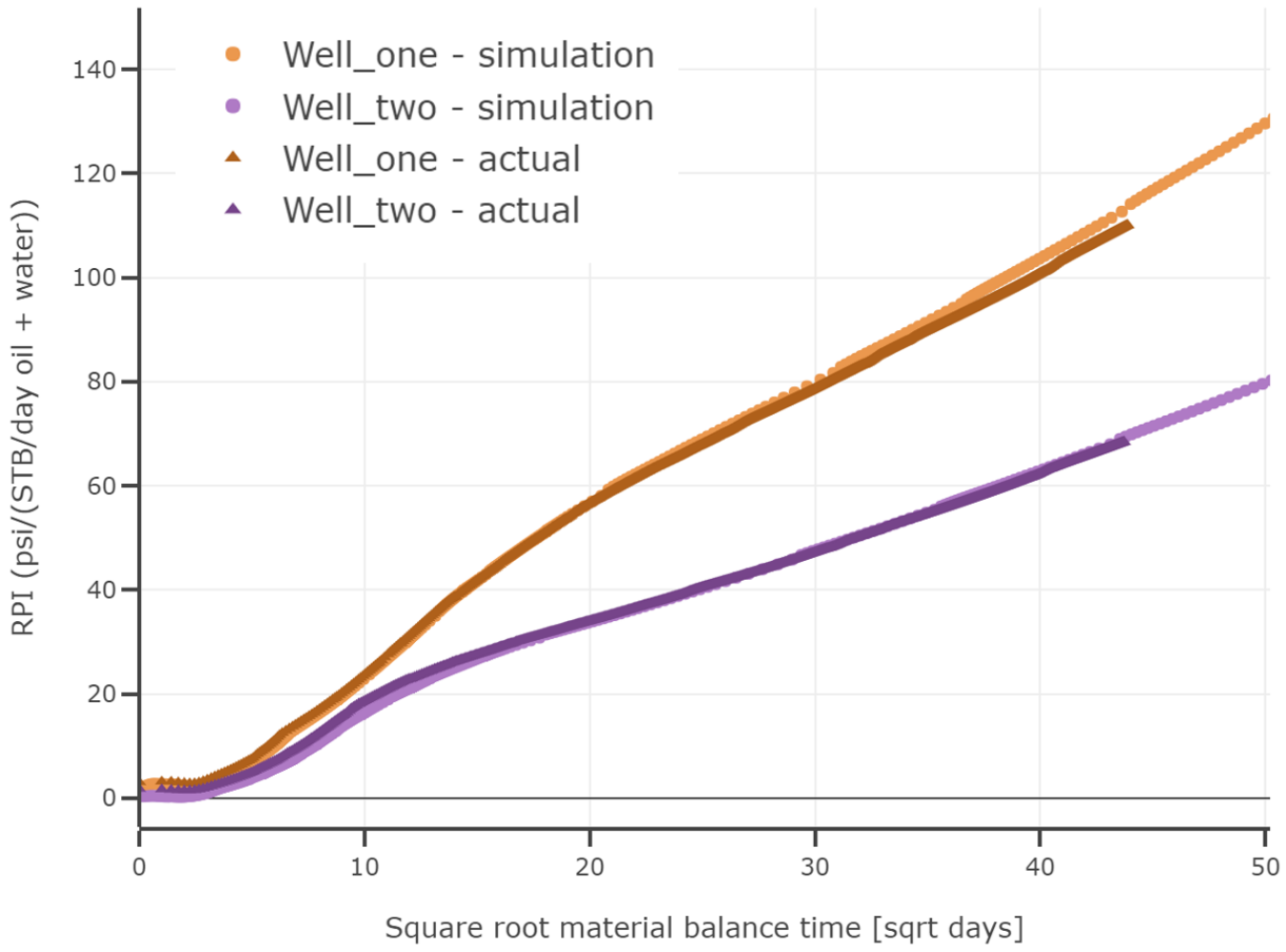

- Very good match to RTA indicating that productivity of the system is well represented

During or after presenting the matches to the key observations, it can be helpful to include slides that illustrate important physical processes in the model. For instance, in the slide below, we show how finite conductivity fractures produce a pressure differential between the BHP and the fracture pressure.

## Fracture conductivity and GOR escalation

- Fractures are finite conductivity, resulting in a pressure differential between BHP and pressure in the fracture
- Pressure differential means there is a lag between BHP going below Pb and GOR increasing

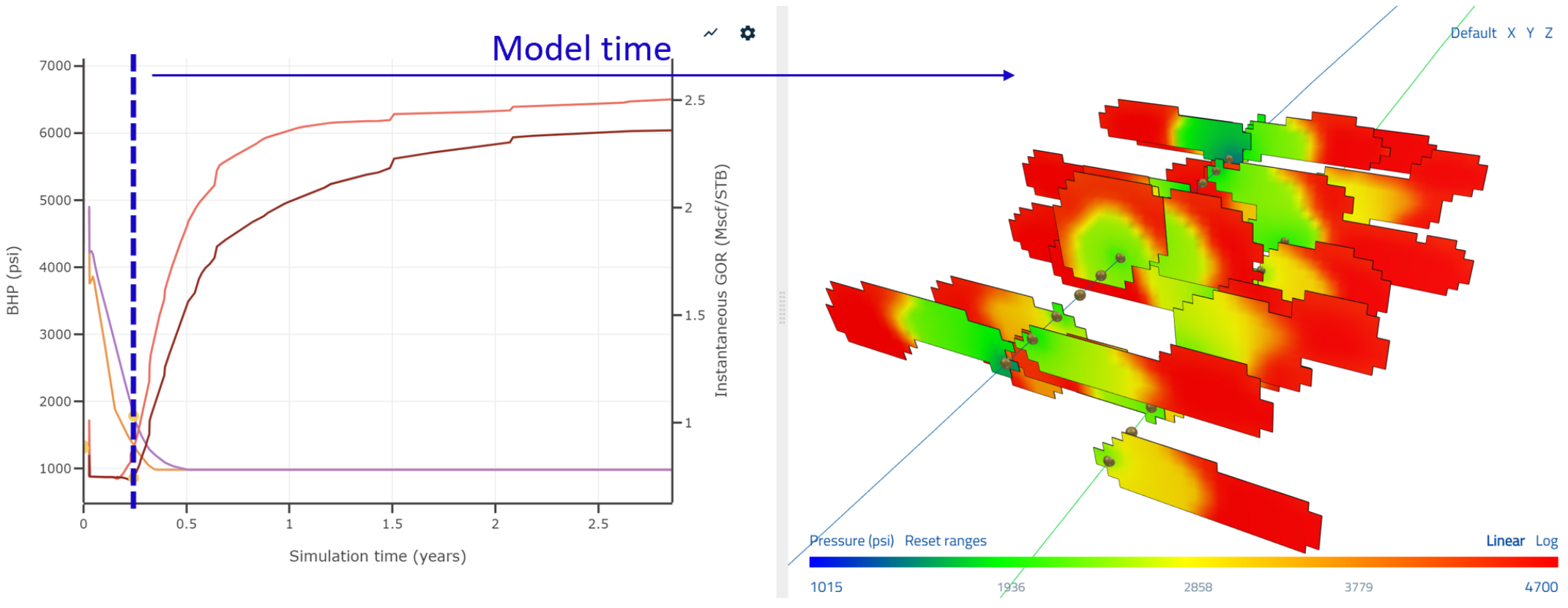

And again, we conclude by reinforcing the takeaways and aligning on next steps.

## Conclusion

- Key observations of frac length, ISIPs, cluster efficiency, oil production, water cut, and GOR all matched
- Performance differences in Well One and Well Two driven by differences in landing zone properties: permeability and water saturation
- Model is ready for sensitivity analysis

## 11.6 Initial sensitivity analysis

As defined in the problem statement, your manager has asked you to assess well spacing for the upcoming pad. To quickly setup different well spacing scenarios, we will:

1. Copy Base Case simulation
2. Duplicate Well One and Well Two
3. Use the Well Shift Wizard to adjust the spacing between wells
4. Adjust external fractures
5. Check matrix region
6. Adjust well controls to reflect timing

### 11.6.1. Copy Base Case simulation

We start by making a copy of the base case simulation (select Copy Simulation from the simulation menu in the job manager).

### 11.6.2. Duplicate Well One and Well Two

In the Wells and Perforations tab of the new simulation, use the Duplicate function to create copies of Well One and Well Two.

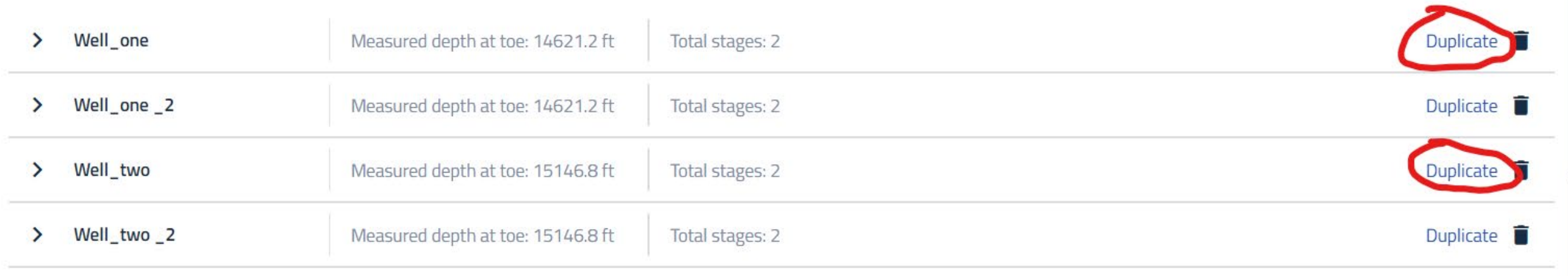

### 11.6.3. Use Well Shift Wizard to adjust spacing

We can now use the Well Shift Wizard to adjust the lateral spacing of the wells as we desire. For instance, here is the case with 500 ft spacing in each bench:

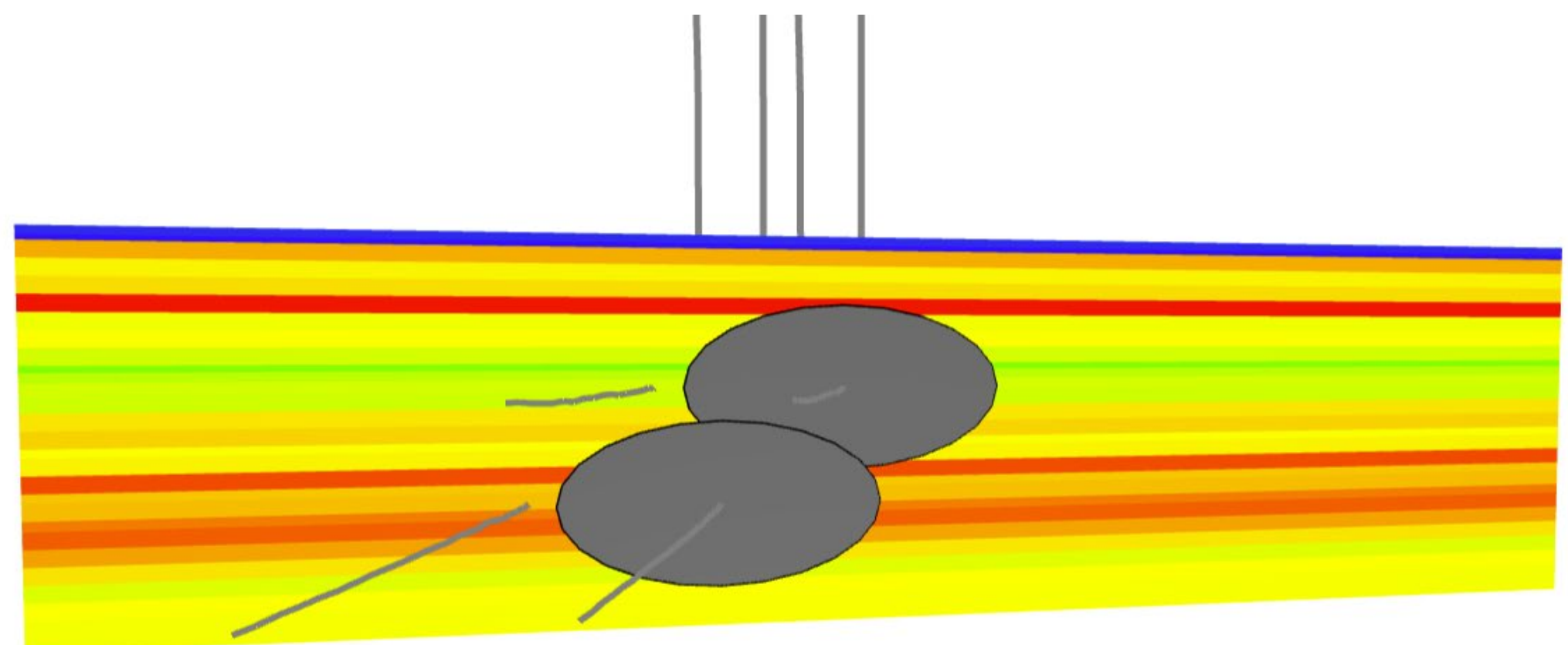

One thing that we notice is that the two new wells do not yet have external fractures, so we need to add those. To do so, add two more rows to the external fracture table and copy down the parameters used for the existing external fractures:

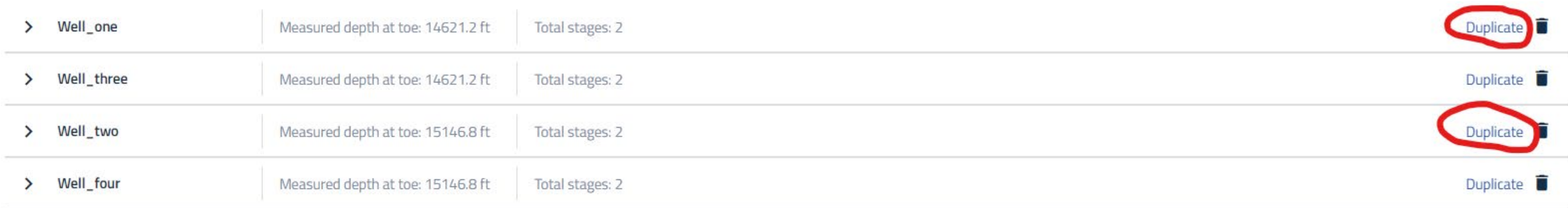

### 11.6.9. Adjust well controls to reflect timing

The final step to set up the simulation is to adjust the well controls. We will zipper Well One and Well Two together and Well Three and Well Four together. We can easily do this by adding a two hour shut-in to Well Two and Well Four, and subtracting 120 minutes from the shut-in at the end of the injection sequence for each.

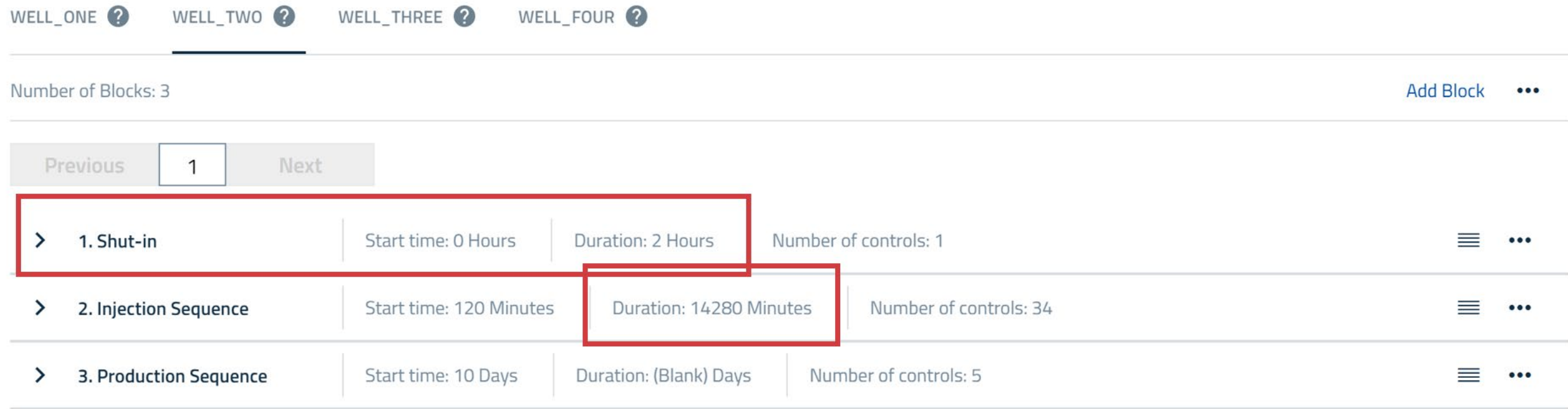

## 11.6.10. Add zero perm cube if desired

Zero permeability cubes are a way of adding no-flow boundaries in ResFrac models. In the case of well spacing sensitivities, we typically add a zero permeability cube such that there is a no-flow boundary at half-a-well-spacing on either side of the edge wells, as pictured below.

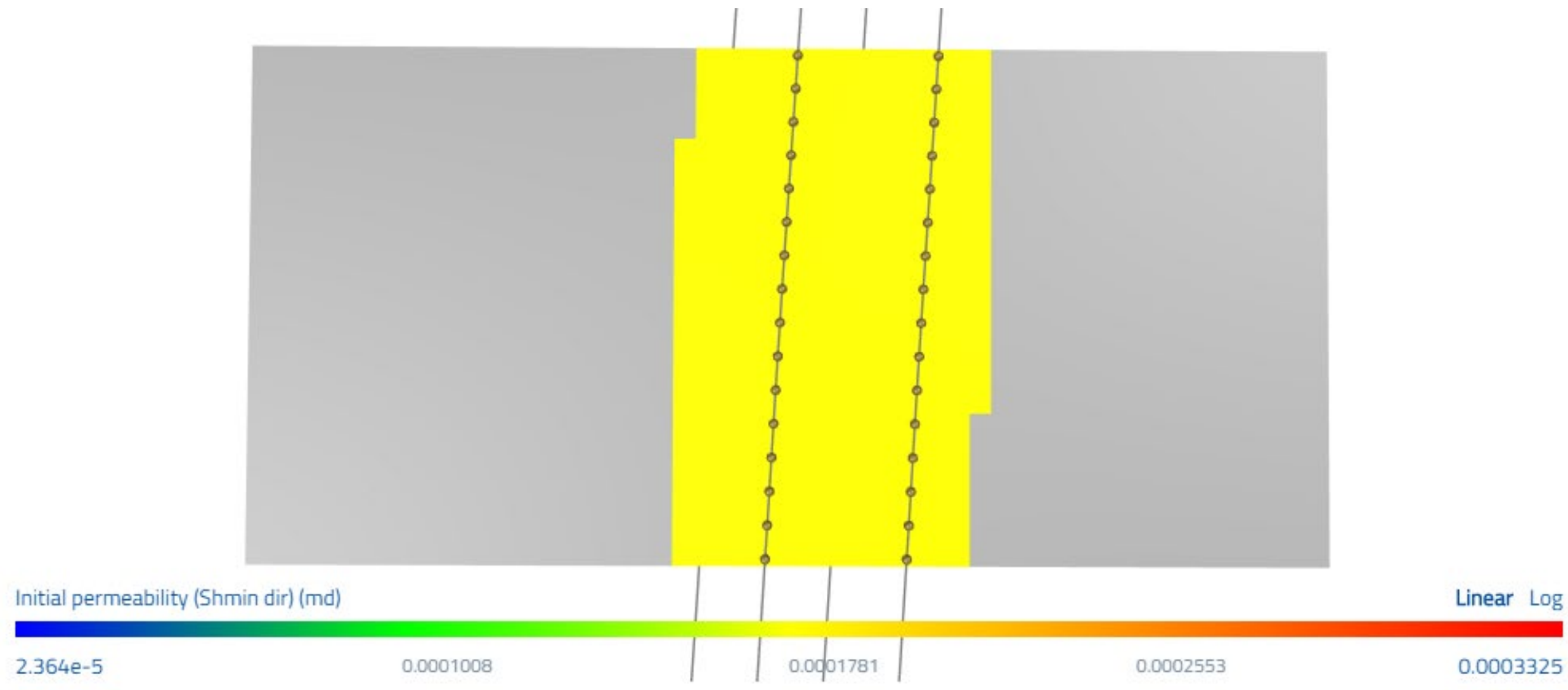

These zero perm boundaries prohibit the edge wells from draining reservoir regions that would otherwise be drained by an adjacent well. In other words, using the zero permeability cube setup above approximates the production characteristics as if that same well spacing pattern continued across the entire acreage (i.e. 10s or 100s of wells, versus four as modeled). If you do not use a 'zero perm cube,' the outer wells are likely to outperform the inner wells.

If the stage lengths on adjacent wells are mismatched, you can use the 'zero perm *inside* cube' setting to adjust. For example, the figure below shows two wells: one with three stages of length 200 ft each, and another with two stages of 250 ft each. The model width needs to be 600 ft for the first well and 500 ft for the second. So, two 'zero perm inside cubes' have been specified, shown as the two grey strips along the edge of the model. They effectively 'cut out' a section of the matrix region so that it is not included, so that the mismatched stage lengths can be simulated in a single model.

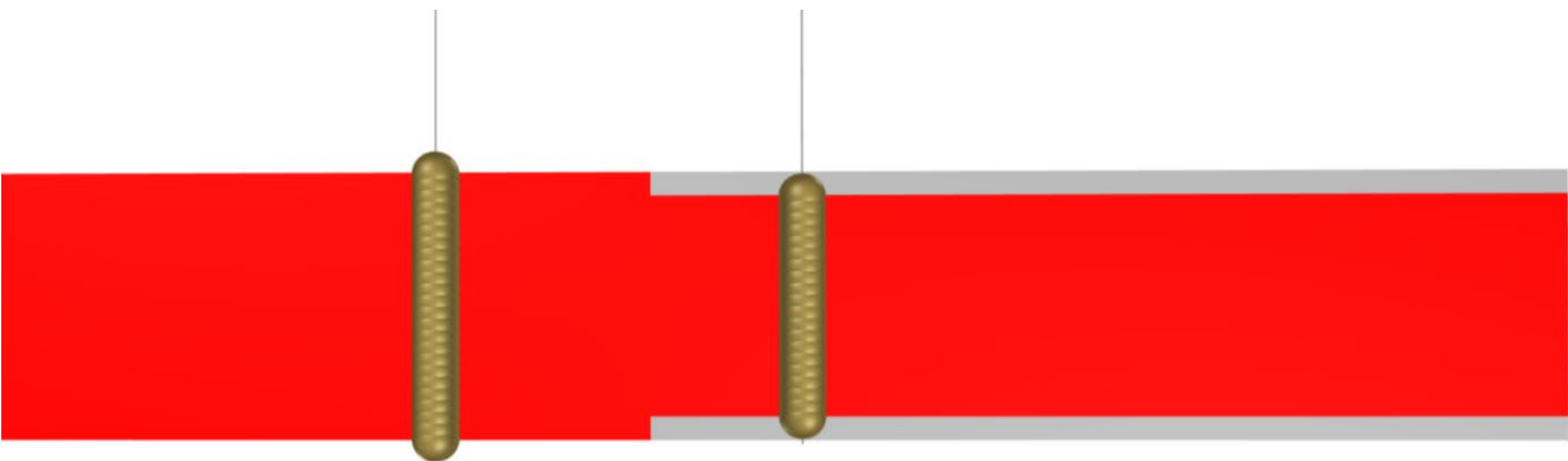

### 11.6.11. Present initial sensitivity analysis

The first round of sensitivity analysis often informs a later round. Nevertheless, the first round of conclusions are important and warrant a presentation to the project stakeholders.

We recommend starting all presentations with an abstract. Then each slide/section of the presentation supports the abstract points.

## Abstract

- Optimal well spacing is a function of stimulation design, geology, and fluid properties.
- For the two-bench development in Peter Pan Basin, optimal well spacing is 500 ft (1000 ft within a zone).
- 500 ft optimal assumes that wells are bounded. If wells are unbounded, optimal shifts closer because outside wells can drain further out.

After the abstract, it is helpful to start with a few slides repeating project purpose, and reviewing the history match, before moving on to slides that support the key points listed in the abstract slide.

In our example presentation, the first slide summarizes the primary finding: optimal well spacing is 500 ft. Note that because of the wine rack, this is 1000 ft for wells within the same zone, and 500 ft for adjacent wells in different zones.

## Optimal well spacing is 500 ft

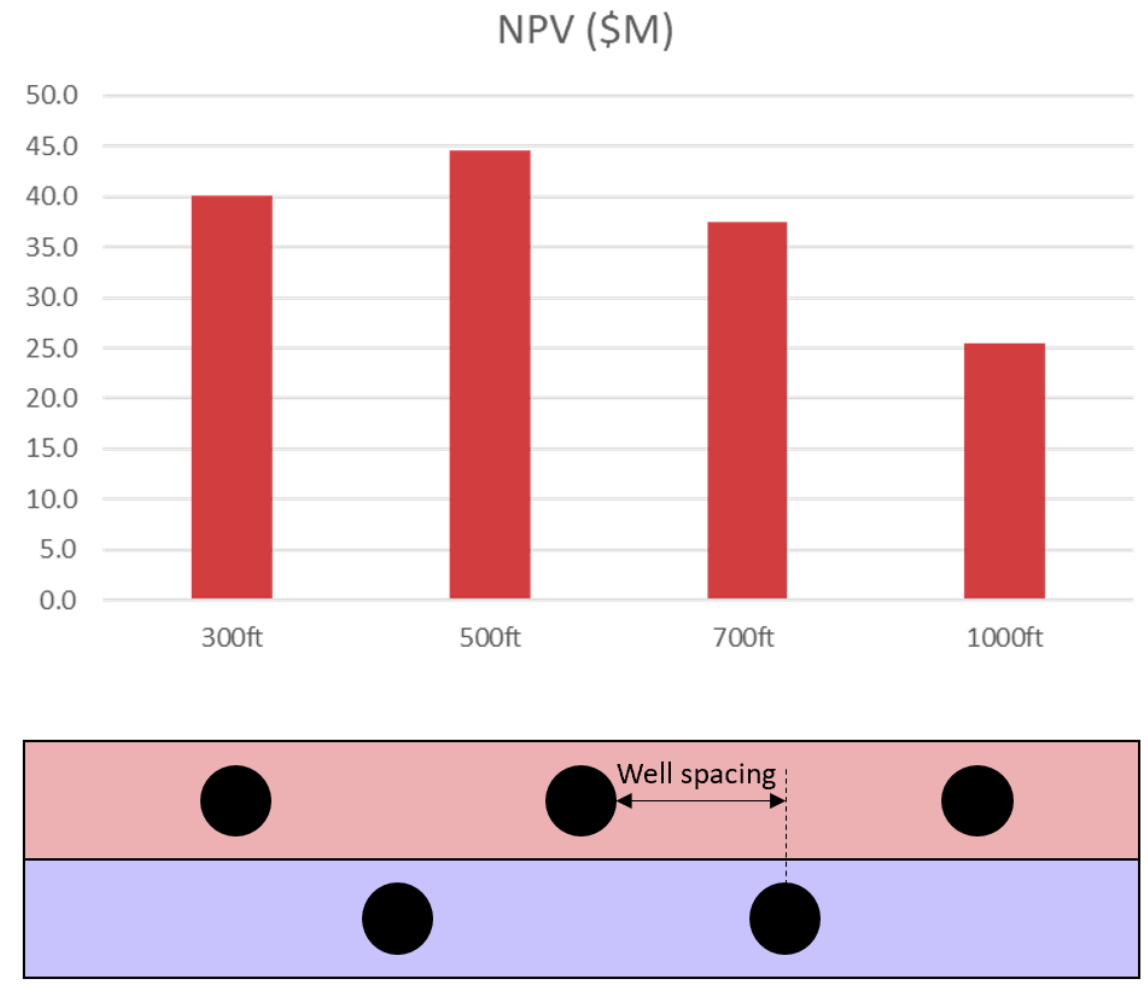

- Assumptions
  - \$37/bbl realized oil price (net of royalties, discounts, etc.)
  - No market for gas
  - \$1/bbl water disposal cost
  - \$13M well and completion cost
  - 15% discount rate

Also included on this slide are a simple gun barrel to show what 500 ft means and the assumptions that went into the economic analysis.

The next slides then explain this primary finding.

## Tighter well spacing yields more interference between wells

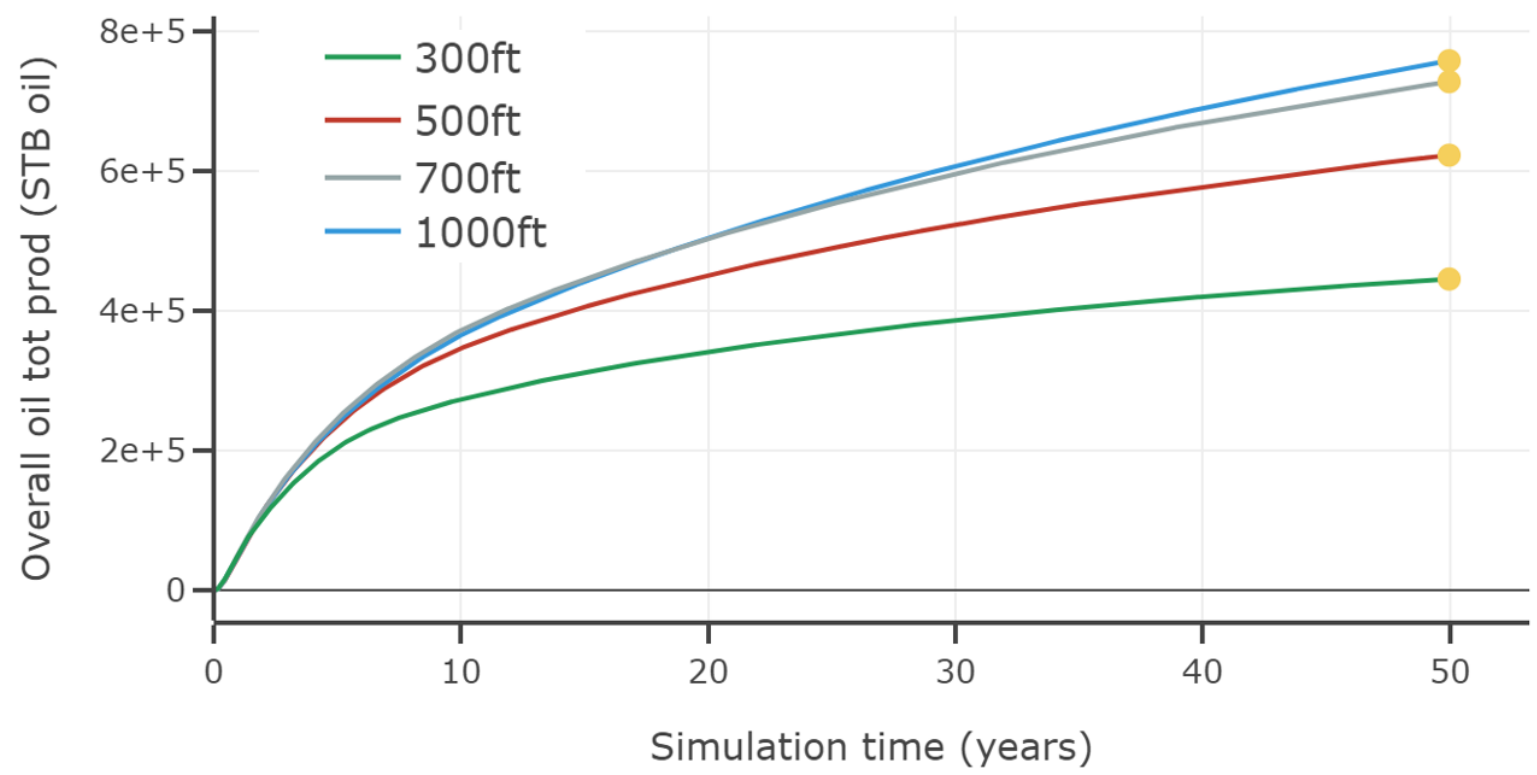

- At 300ft, wells interfere severely.
- There are diminishing returns to increased well spacing, with little incremental benefit of going from 700 to 1000 ft.

# 30-year pressure depletion

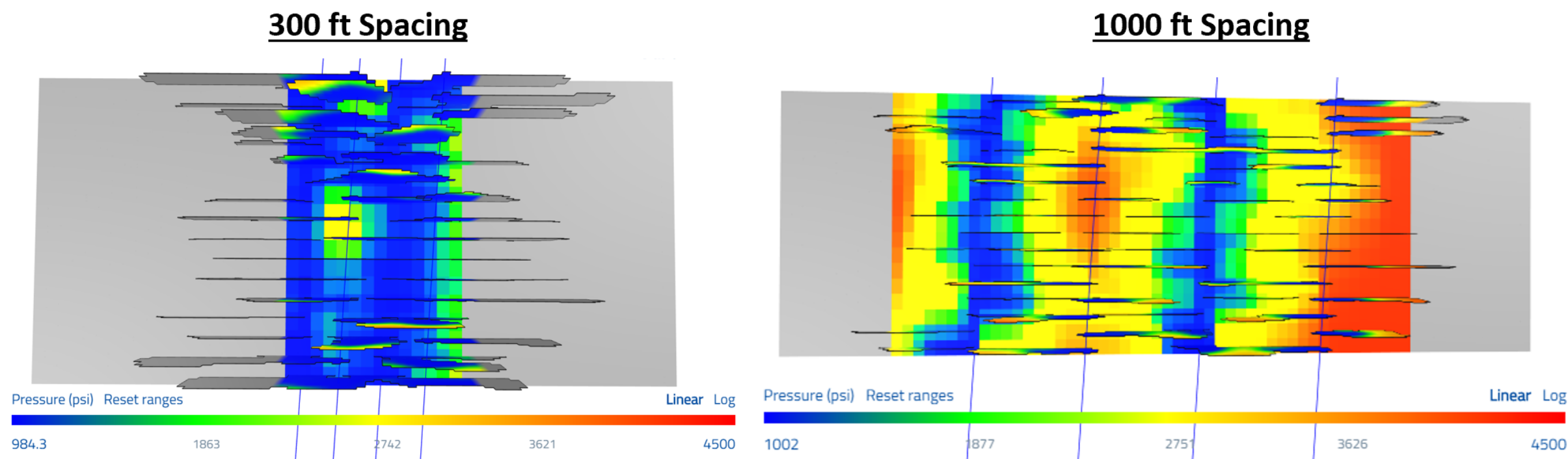

- 300 ft spacing scenario shows nearly uniform, severe pressure depletion causing severe inference between wells
- 1000 ft spacing scenario shows separation of well drainage regions, particularly between Pay 1 and Pay 2 zones

# RTA to diagnose when interference between wells occurs

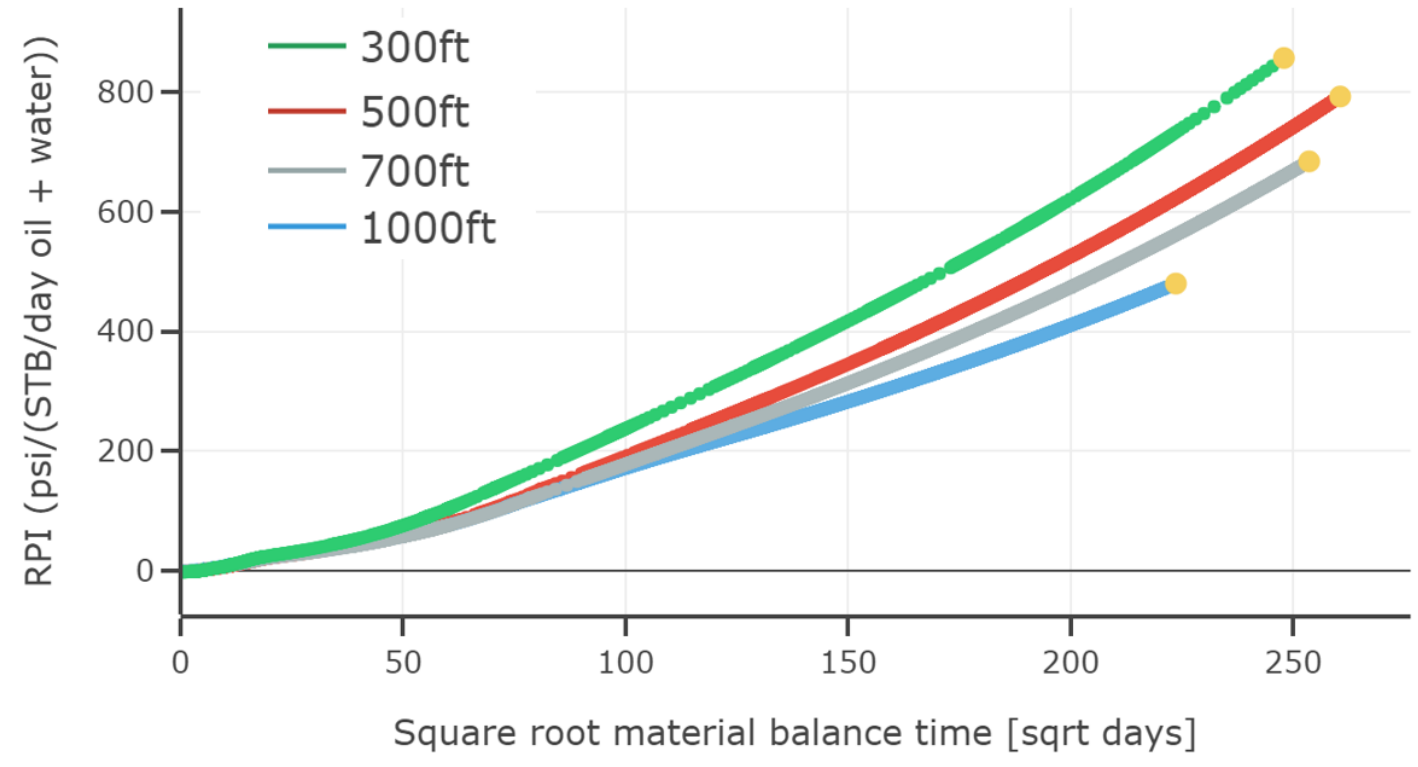

- Deviation from linear flow can be indicative of inference
- Progressively steeper slopes with tighter well spacings indicates interference is occurring early and is more severe

Next, in our abstract slide we had a secondary point: that at the edge of the development, optimal well spacing may be different. To support this point, we include a couple slides showing the effect of removing the zero permeability cube.

## Unbounded results in better performance at tighter spacing

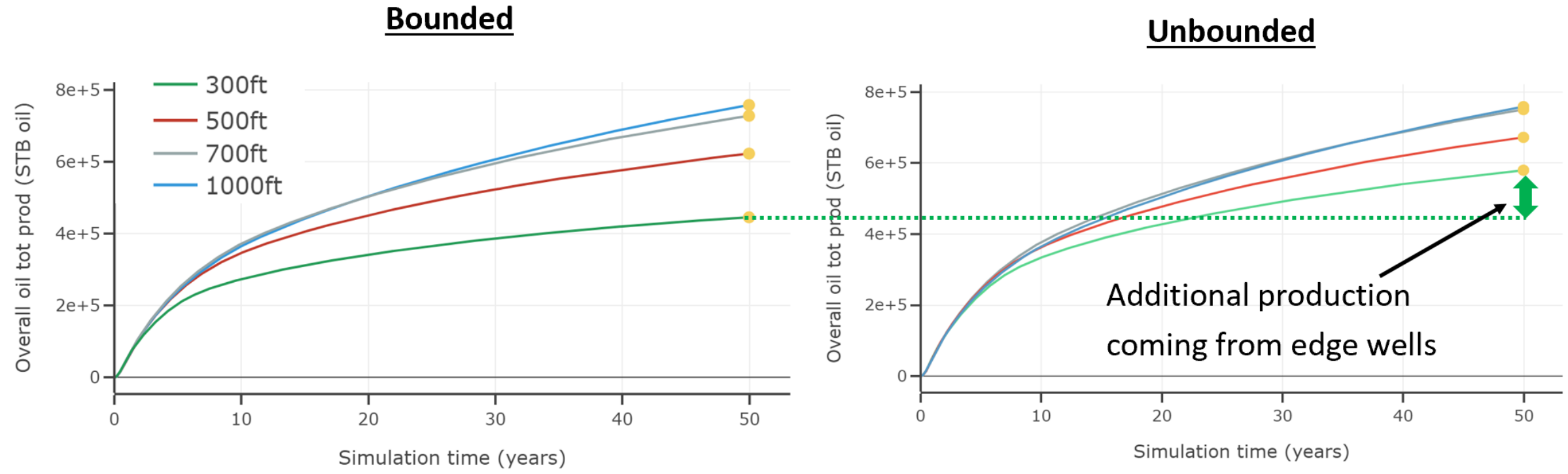

- In unbounded configuration, outside wells can drain from further away from the wellbore
- On edge of development, likely optimal to go to spacing slightly less than 500 ft due to this effect

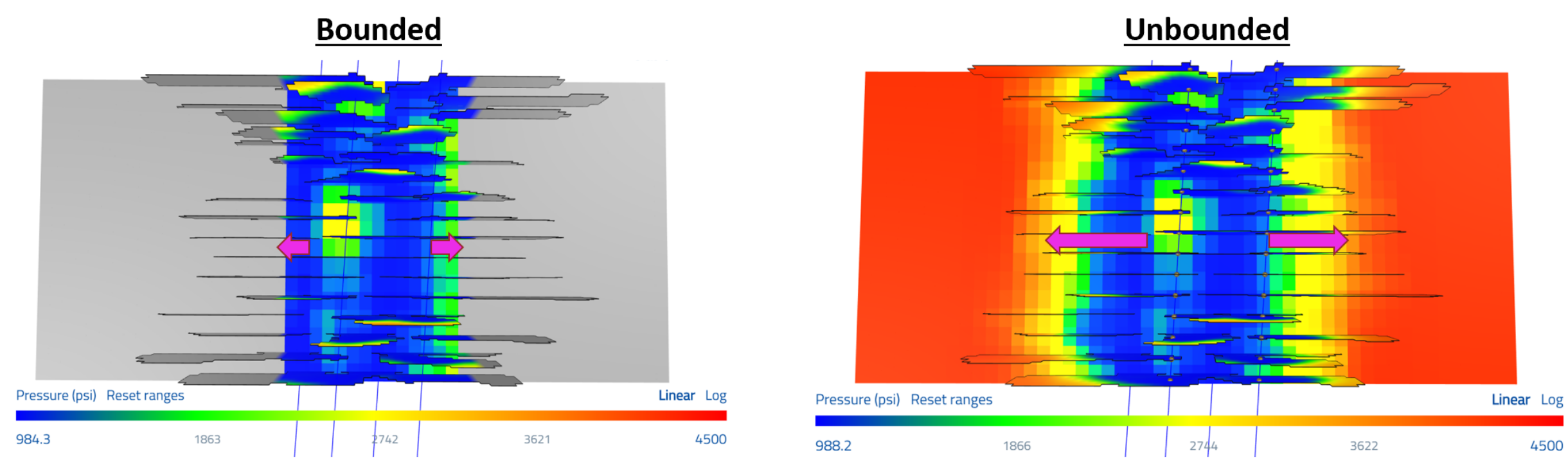

Finally, we conclude by restating our main points from the abstract.

## Conclusions

- Optimal well spacing is a function of stimulation design, geology, and fluid properties.
- For the two-bench development in Peter Pan Basin, optimal well spacing is 500 ft (1000 ft within a zone).
- 500ft optimal assumes that wells are bounded. If wells are unbounded, optimal shifts closer because outside wells can drain further out.

### 11.7 Refine and finalize sensitivity analysis

Very often, new ideas arise during the presentation of the sensitivity analysis results. If these new ideas are tractable given the resources allocated to the project, they should be investigated in a second round of sensitivities analyses.

In the example of the well spacing sensitivity presented above, additional ideas could be:

- Pay 2 wells produce better than Pay 1. Is it more economic to only drill Pay 2 wells? What would the optimal spacing be?
- Should Pay 1 and Pay 2 wells be spaced differently?
- There is some uncertainty on fluid properties in offsetting acreage. Does optimal spacing change if oil API is higher?
- How does optimal well spacing change if cluster spacing / proppant loading is changed as well?

These and many more questions could all warrant investigation. Be sure to consult the recommendations and workflows in Section 9 as you pursue these and other sensitivities, as well as the philosophies in Section 3. Good luck!

## 12. ResFrac template simulations and workflows

Each installation of ResFrac includes template simulations and simulation workflows. These templates are designed to either a) provide generic basin-appropriate input parameters or b) demonstrate specific simulation settings.

Simulation templates are accessible from within the 'Create Simulation' dialogue:

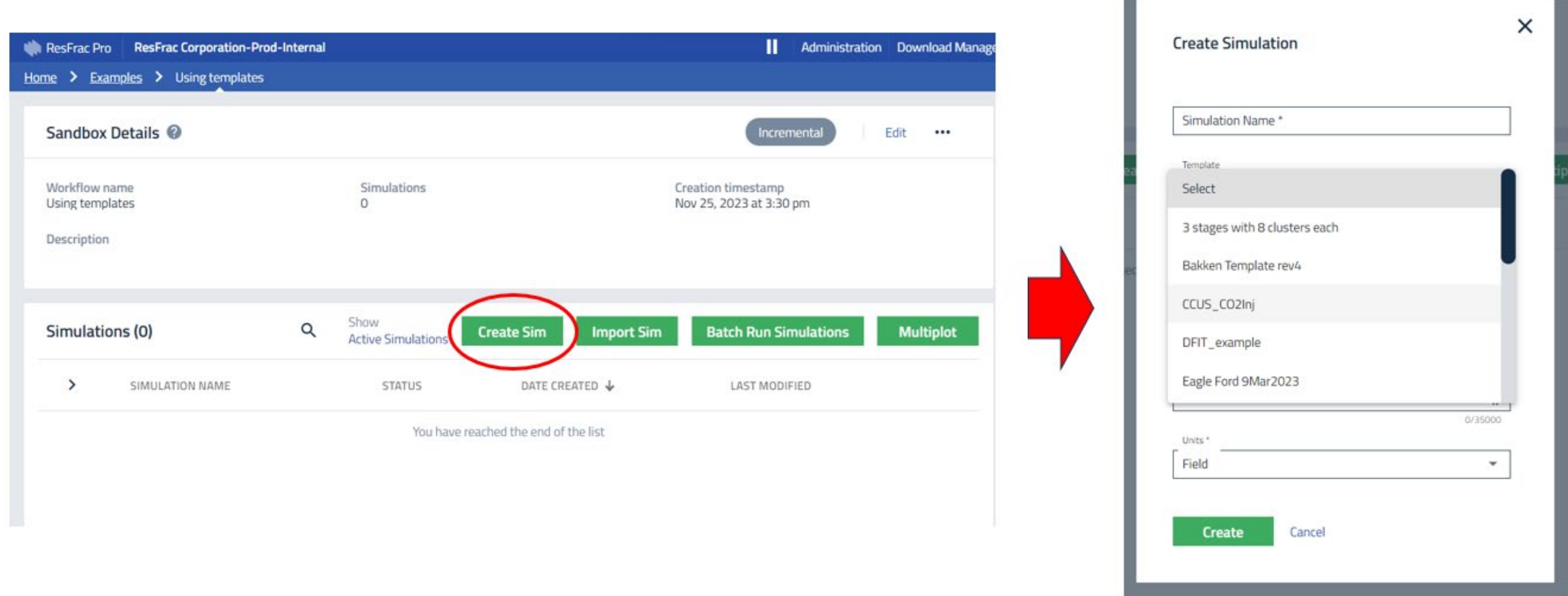

Workflow templates are accessible from within the 'Create Workflow From Template' dialogue:

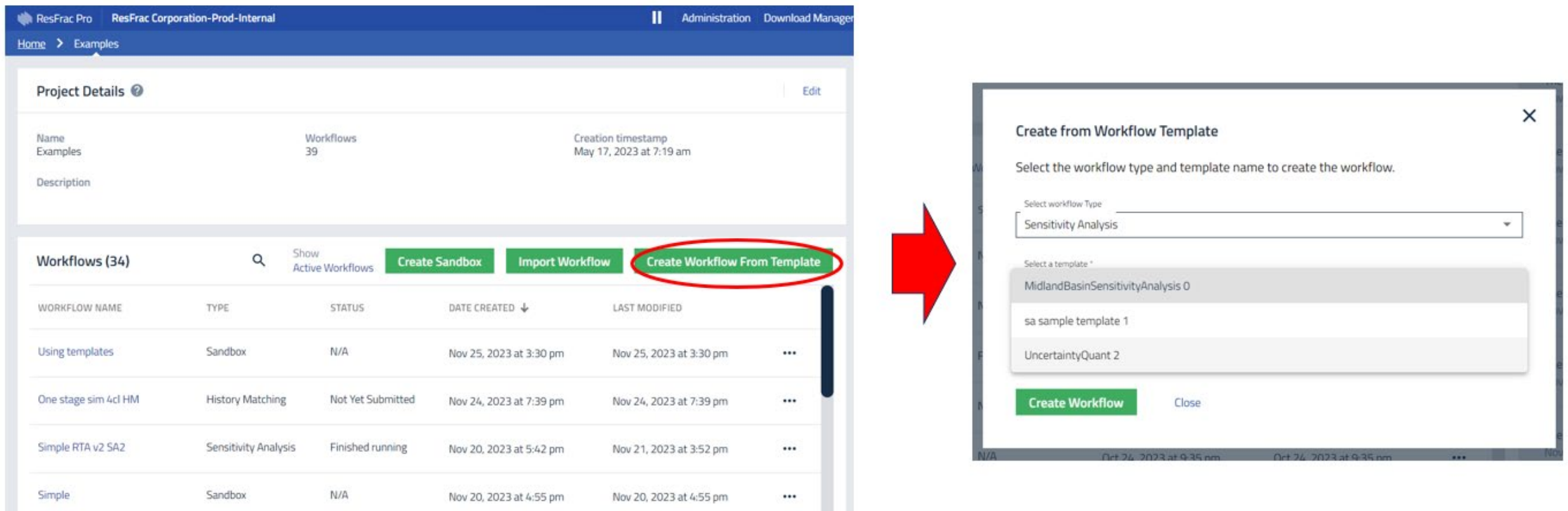

There are separate template workflows for Sensitivity Analysis, History Matching, and Optimization workflows.

Each template, whether a simulation or workflow template, includes documentation pdf located inside the folder where ResFracPro is installed:

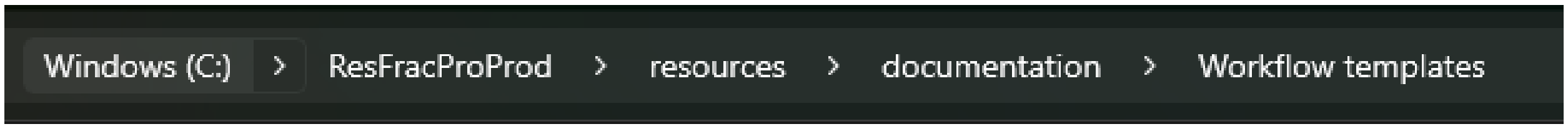

These pdfs include citations for the source data (for basin templates) and walk through the settings changes required to model specialized processes:

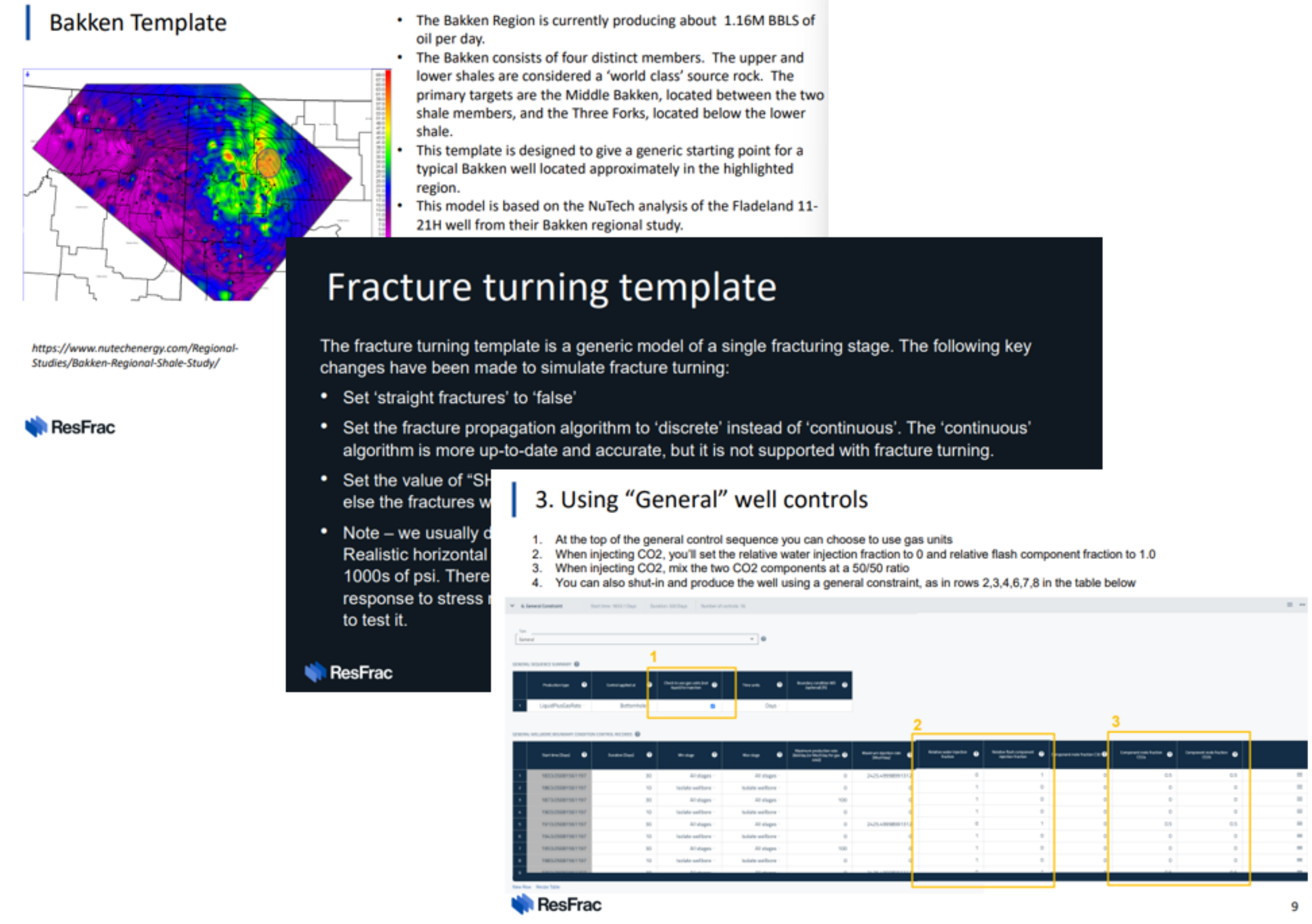

Templates are continually being added to ResFrac so be sure to check the release notes with each update to see if any new templates are included.

# 13. ResFrac toolbox

ResFrac has many powerful tools and features. We highlight a few in this section along with Excel-based tools to help manipulate input and output files.

- Using restart files
- Optimizing simulation runtime
- Using Raw Results
- Fracture heatmap spreadsheet

## 13.1 Using restart files

### Why use restart files?

Simulation compute times are often on the order of 10s of hours. And while we always suggest optimizing your model for speed (Section 11.2), you can also often save time by using restart files during history matching.

Restart files save the exact state of a simulation. For example, if you download a restart file from timestep 12350, that file contains the exact state of the simulation at timestep 12350. If you restart the simulation with the exact set of parameters you originally ran, the result will be exactly the same. However, we often want to *change* parameters in a restart.

It is common to perform a restart from after fracturing the well/s in the model, but prior to production. Because fracturing usually takes most of the compute time, using restart files can save you a lot of time while you iterate on production parameters. ResFrac is a fully coupled code, and so many parameters impact both *fracturing* and *production*; however, many parameters affect only the production response (e.g. relative permeability) and have little impact on fracturing. These are great parameters to modify in a restart simulation.

## How to download restart files and run a simulation from a restart

Restart files can be downloaded from the menu visible when clicking on a simulation in the job manager:

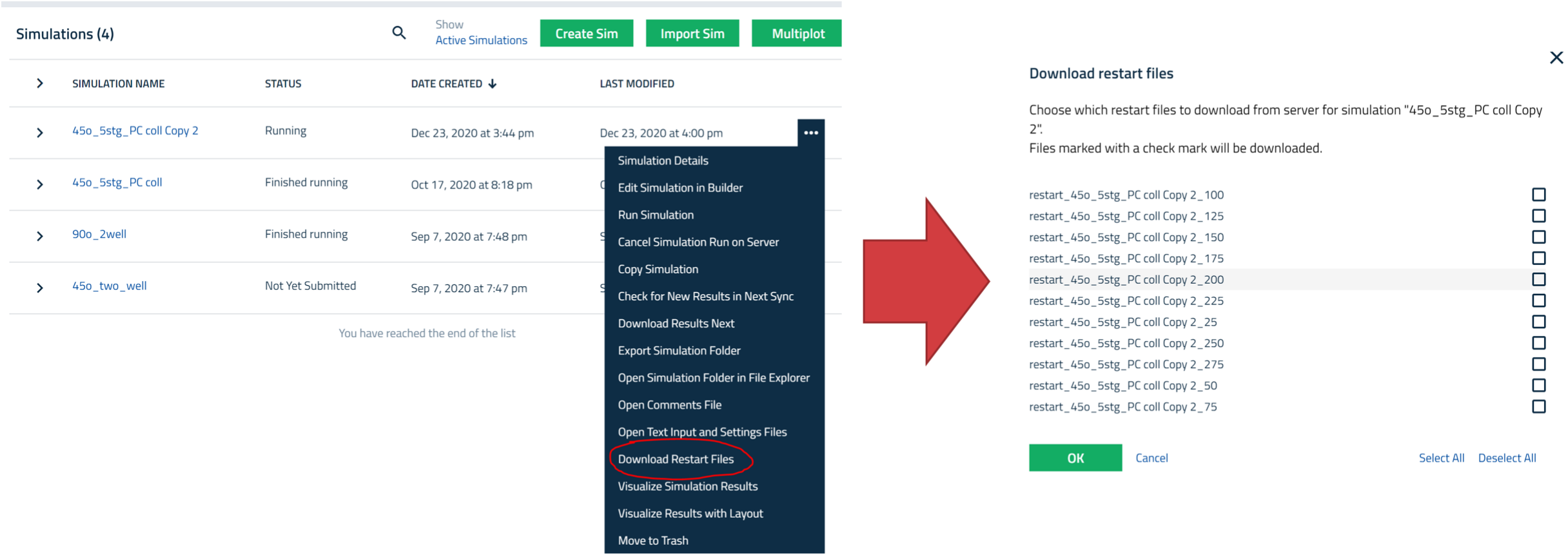

Clicking 'Download Restart Files' will bring up a selection menu where you can select one or more restart files to download. How do you know which restart file is the one you want? The numbers at the end of the restart file name is the timestep number. If you open up the comments file, you can see the correspondence between timestep number and the time in the simulation:

```
comments_45o_5stg_PC coll Copy 2.txt - Notepad
File  Edit  Format  View  Help
Timestep duration will be limited by a 'maxnextdt' set somewhere earlier in the timestep.
Timestep limiting factor would have been pressure in element 25876. Value changed from 5366.8741654268815 psi to 5468.759023717
Volumes in place: 9.77482e+06 STB water, 1.96667e+06 STB oil, 7.63841e+07 Mscf gas.
Pore volume: 63986872.182691336 bbl. Average porosity: 0.0287383. Fracture volume: 141.54077218924817 bbl.
Total simulation wall-clock time: 259.559 s. Cumulative iterations: 713.

Beginning timestep #156...
Max flow cfl: 33.001775278924377
Fracture max flow cfl: 0.31294873309371629
Mechanically open fracture elements: 60, closed fracture elements: 0.
Implicit composition elements: 3119, explicit: 22761.
Implicit water solute elements: 3820, explicit: 22060.
Implicit proppant elements: 3368, explicit: 22512.
Implicit water solute frac elements: 60, explicit: 0.
Implicit proppant frac elements: 5, explicit: 55.
Implicit pressure elements: 3882, explicit: 21998.
Starting reform system
Solver step size multiplied by factor: 0.59707. Iteration number: 1. Equation in element: 25878, element type: fracture, type:
Solver step size multiplied by factor: 0.542295. Iteration number: 2. Equation in element: 25878, element type: fracture, type:
Solver step size multiplied by factor: 0.602985. Iteration number: 3. Equation in element: 25878, element type: fracture, type:
Solver step size multiplied by factor: 0.774717. Iteration number: 4. Equation in element: 25878, element type: fracture, type:
Convergence in 7 iterations.
Before poro KSP
Before poro solve
Timestep duration: 0.00025399999961391999 hours. Cumulative time elapsed: 23.090895017884396 hours.
Final spot in OutputTo
Well_P1 is injecting at specified rate from the wellhead. Wellhead pressure: 4237.1355878488639 psi. Bottomhole pressure: 7168.
Well_P2 is set to an 'isolated wellbore' boundary condition.
Well_C1 is set to an 'isolated wellbore' boundary condition.
Well_C2 is set to an 'isolated wellbore' boundary condition.
Well_C3 is set to an 'isolated wellbore' boundary condition.
Timestep #156 successful. Number of discarded timesteps due to excessive variable change: 0; due to numerical problems: 2.
Timestep limiting factor is pressure in element 25868. Value changed from 5751.2113318238644 psi to 5546.4961315530463 psi.
Volumes in place: 9.77482e+06 STB water, 1.96667e+06 STB oil, 7.63841e+07 Mscf gas.
Pore volume: 63986873.122821443 bbl. Average porosity: 0.0287383. Fracture volume: 142.46286696519317 bbl.
Total simulation wall-clock time: 261.45 s. Cumulative iterations: 720.
```

Note that the comments file reports what every well in the simulation is doing at every timestep. For example, in the image above, Well_P1 is injecting while the other four wells are shut-in.

Once you find the timestep in the comments file that you would like to use as a restart point, go back to the restart selection menu and choose the last restart *before* the timestep that you noted in the comments file. Restart files are available every 25 to 300 timesteps, depending on how long your simulation is (restart frequency decreases after about 20,000 timesteps have occurred).

Once downloaded, the easiest way to restart a simulation from that restart file is to:

- Make a copy of the simulation
- Change the parameters of interest
- Select the restart file when submitting the copy:

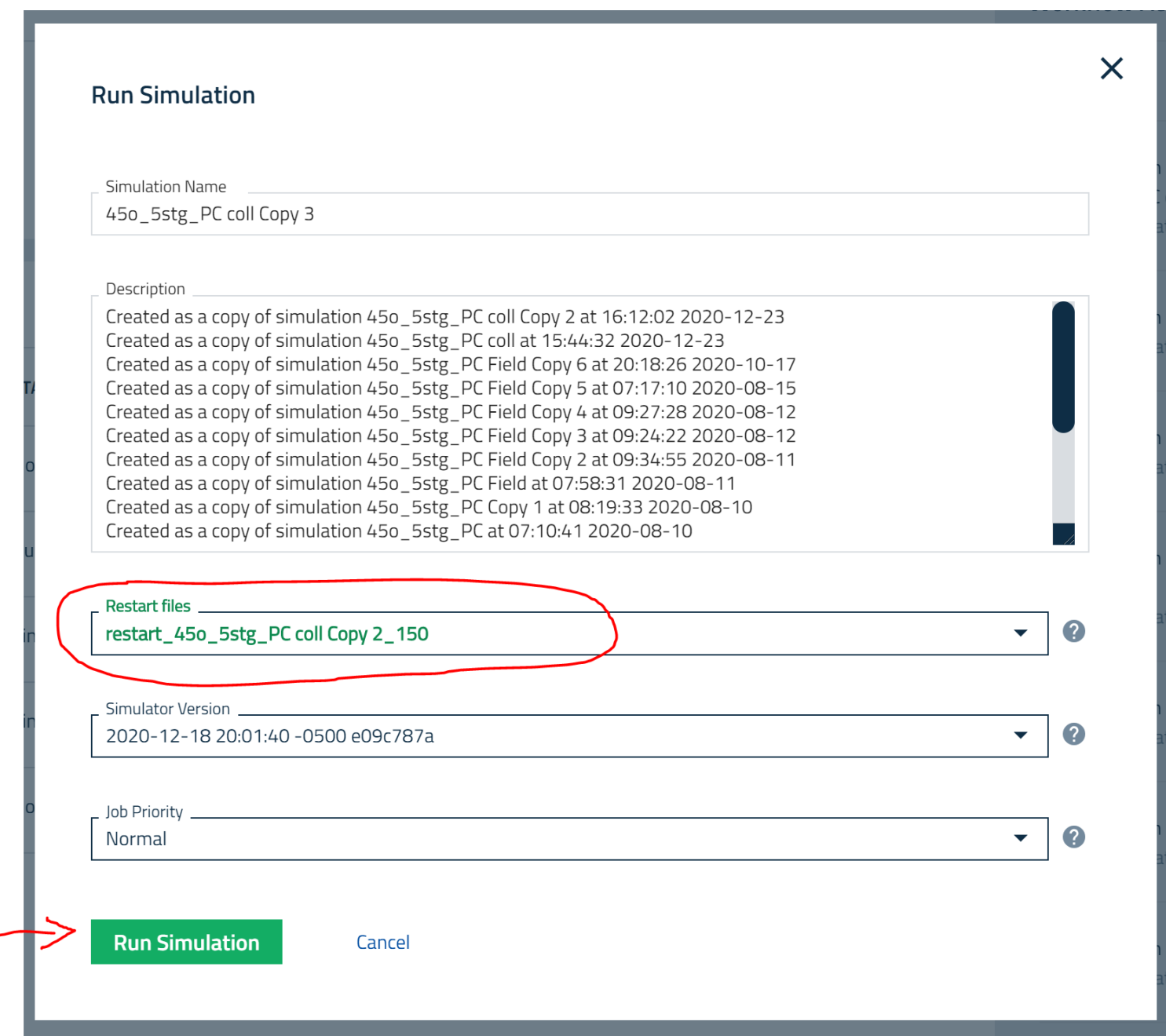

Note that when you run a simulation from a restart, the UI needs to upload the restart file back to the server (along with the simulation input/settings files), so it may take several minutes to fully upload from the time you click 'Run Simulation.'

What if you made a copy of the simulation before downloading the restart file? Click on the name of the simulation in the job manager, and select Import Restart Files from the menu. Then select "Import from existing simulation." Select the simulation that you want to import the restart file from, and then select the restart file.

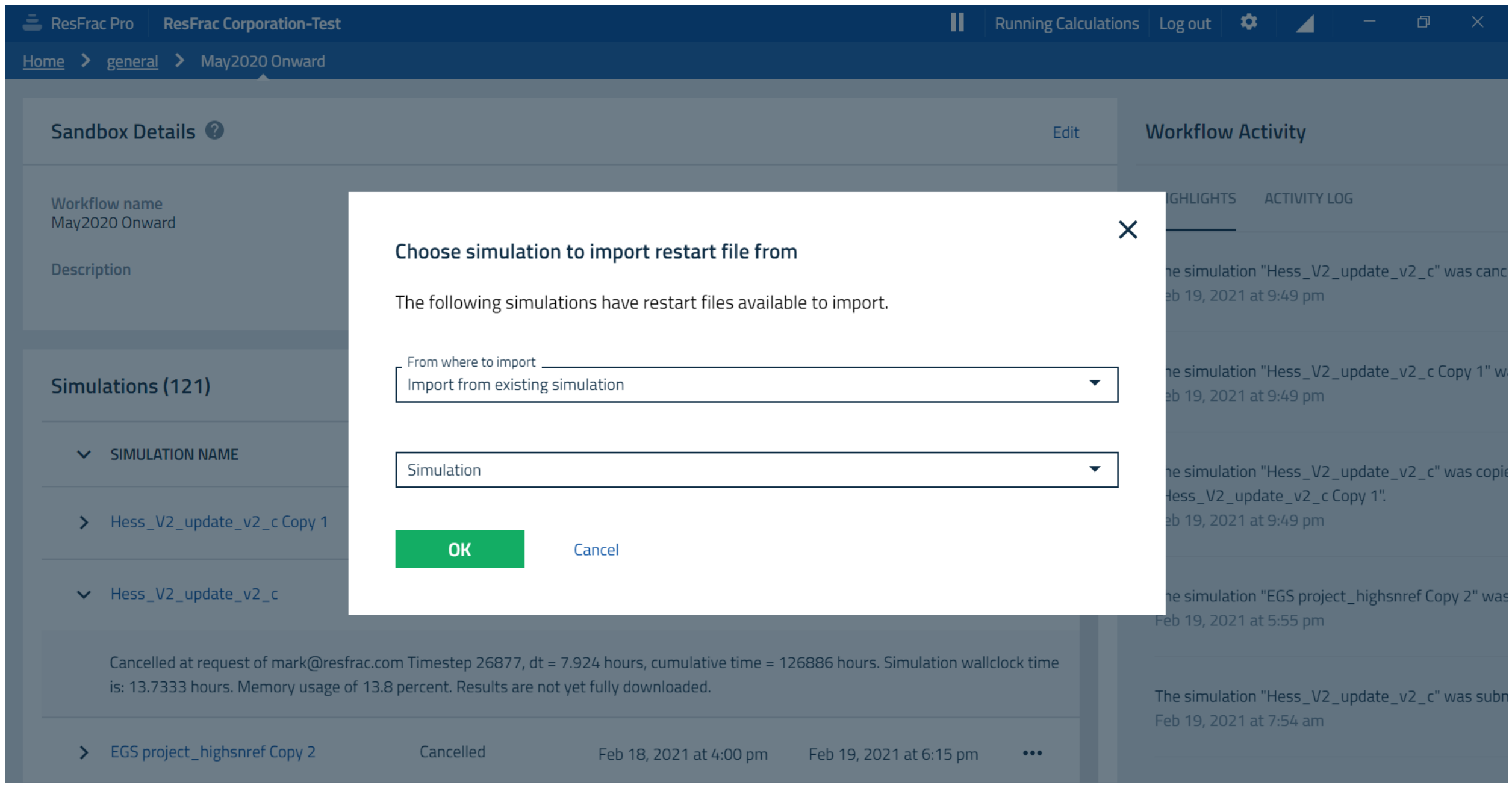

## What to change and what not to change in a restart

Typically, you only want to make changes in a restart file that will not impact the simulation up to the point at which the restart is taken. For example, it is unlikely that you would want to download a restart file in between frac'ing two wells, and then change the toughness. Toughness usually has a dominant impact on fracture geometry, so frac'ing two wells or two stages with different toughness values would likely impact ultimate predictivity of that model.

Below are some common parameters changed in restarts.

- Relative permeability
  Relative permeability curves are one of the primary parameters used to match production data, and they have negligible impact on fracture propagation. It is very common to download a restart prior to production and change rel perm curves.
- k0 (in the Proppants table)
  k0 is a linear multiplier to the conductivity of the proppant pack. Changing this parameter will not have an impact on fracture propagation or proppant transport, but will impact production and especially things like GOR.
- Global permeability multiplier
  Global permeability multiplier is another dominant parameter for production and commonly used to history match production data. Modifying global permeability *will* impact leakoff and can subtly change fracture propagation, but is generally not the primary parameter governing fracture geometry. It is ok to change in a restart (especially if changing by less than an order of magnitude), though be aware that you should rerun the simulation from the start to confirm geometries don't change more than desired. If you want to change permeability during production, but keep 'effective permeability to leakoff' during injection constant, you can modify the permeability multiplier in the PDP downward at the same time that you increase the global permeability multiplier.

- Permeability by zone
  Similar to global permeability multiplier, you may change permeability of various layers of your static model in order to match production by zone.
- Pressure Dependent Permeability
  Pressure Dependent Permeability (PDP) is used in several ways in ResFrac simulations. For delta pressure above zero (fracture pressure > matrix pressure), it is common to use multipliers greater than one to accelerate leakoff and account for the presence of multiple fracture strands. If using PDP in this sense (for example to match initial water flowback), the simulation should be run from the start. PDP is also used to model stress dependent permeability reduction (fracture pressure < matrix pressure), where multipliers less than one are used. Often this phenomenon is observed in over-pressured gas shales. For these simulations, it *would* be appropriate to modify the negative delta pressure portion of your PDP table in a restart file.
- Well controls
  If you want to change a well control, restarts are a great tool. Just make sure to download a restart file prior to the time at which you are making a change.
- Drilling new wells
  If you have multiple generations of wells in your model, you can download a restart prior to the completion of a second or latter generation well, and then change the location/completion design of that well in the restart simulation. Be sure to use the 'Wellbore Drilling Times' option in the advanced section of Wells and Perforations to do so:

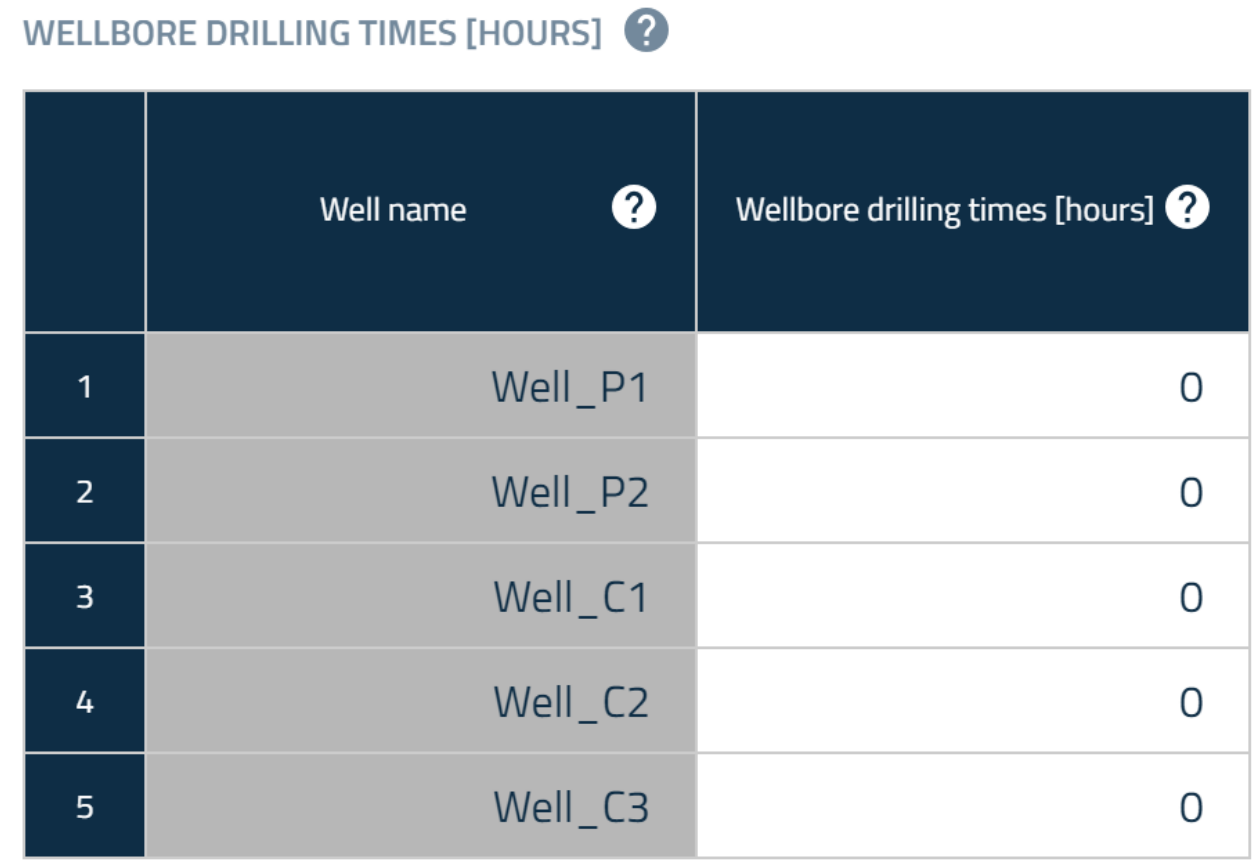

WELLBORE DRILLING TIMES [HOURS]

| | Well name | Wellbore drilling times [hours] |
|---|---|---|
| 1 | Well_P1 | 0 |
| 2 | Well_P2 | 0 |
| 3 | Well_C1 | 0 |
| 4 | Well_C2 | 0 |
| 5 | Well_C3 | 0 |

Parameters you *should not* change in a restart (and yes, sometimes there are exceptions):

- Meshing
  Pretty much everything on the Meshing tab is a no-go. Mesh size, mesh center, meshing scheme, etc. cannot be changed in a restart.

- Perforation locations or well locations (unless using the wellbore drilling times option above)
- Anything that defines the “initial state” of the model
    - Saturations
    - Random toughness
    - Orientation of SHmax
    - Fluid model options
        - You occasionally may want to change fluid properties. If you do so, you will need to start the simulation from the beginning.
- Everything on the ‘Startup’ panel (the fluid model, thermal/isothermal, etc.)
- Physics and correlation selections on the ‘Other Physics Options’ and ‘Numerical Options’ panels

## 13.2 Optimizing simulation runtime

### Model scale

Simulation model compute times are often on the order of 10s of hours. Often, we find the sweet spot for balancing model resolution with runtime leaves us with models that run in 12 to 24 hours. To understand the scale and resolution of the model required, we need to first consider the questions we are trying to answer with the model. Trying to optimize fracture sequencing across a large, multi-well pad versus understanding near-fracture leakoff characteristics of a DFIT require different resolution and a fundamentally different setup.

The first consideration for computational speed of a model is the number of fracture elements that will need to be created. Typically, the creation of fracture elements is the most computationally intensive process of a simulation and takes roughly 80% of the compute time, with production accounting for the other 20%. Considerations impacting the number of fracture elements (and therefore compute speed of fracturing):

- Number of wells included
- Number of stages included
- Fracture element size

If you are simulating a large well pad with many wells, then you’ll likely want to pare down the number of stages per well that you model. Remember, you can use external fractures to replicate stress shadowing from stages *not* included in your model, and ‘zero permeability cubes’ can then be used to account for additional offsetting wells.

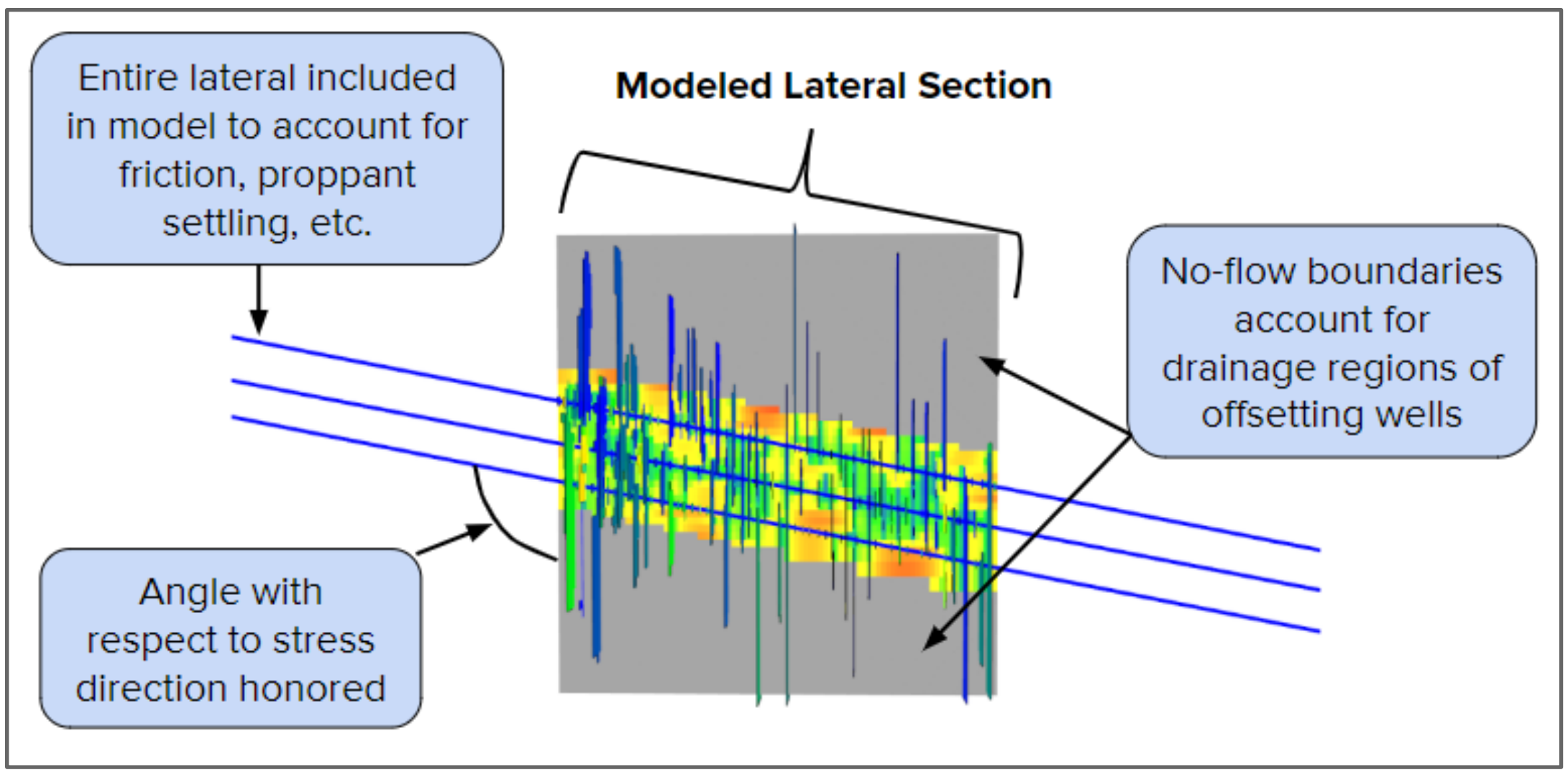

Three-well model above uses "zero permeability outside this cube" to account for offsetting wells above and below (in the image) from the three wells being modeled. The modeled well is used for evaluation criteria.

See Section 9.1.1 for examples on how to use zero permeability cubes to create units of symmetry in your simulation.

To reduce the number of stages you need to model, try using external fractures. Found on the 'Wells and Perforations' tab, External Fractures allow the user to specify a fracture *outside* of the matrix domain with a geometry and time-dependent net pressure (ie fracture net pressure can be zero until a specified time, peak to some value, then decay to a residual net pressure - just as stress shadowing observed in an offsetting stage).

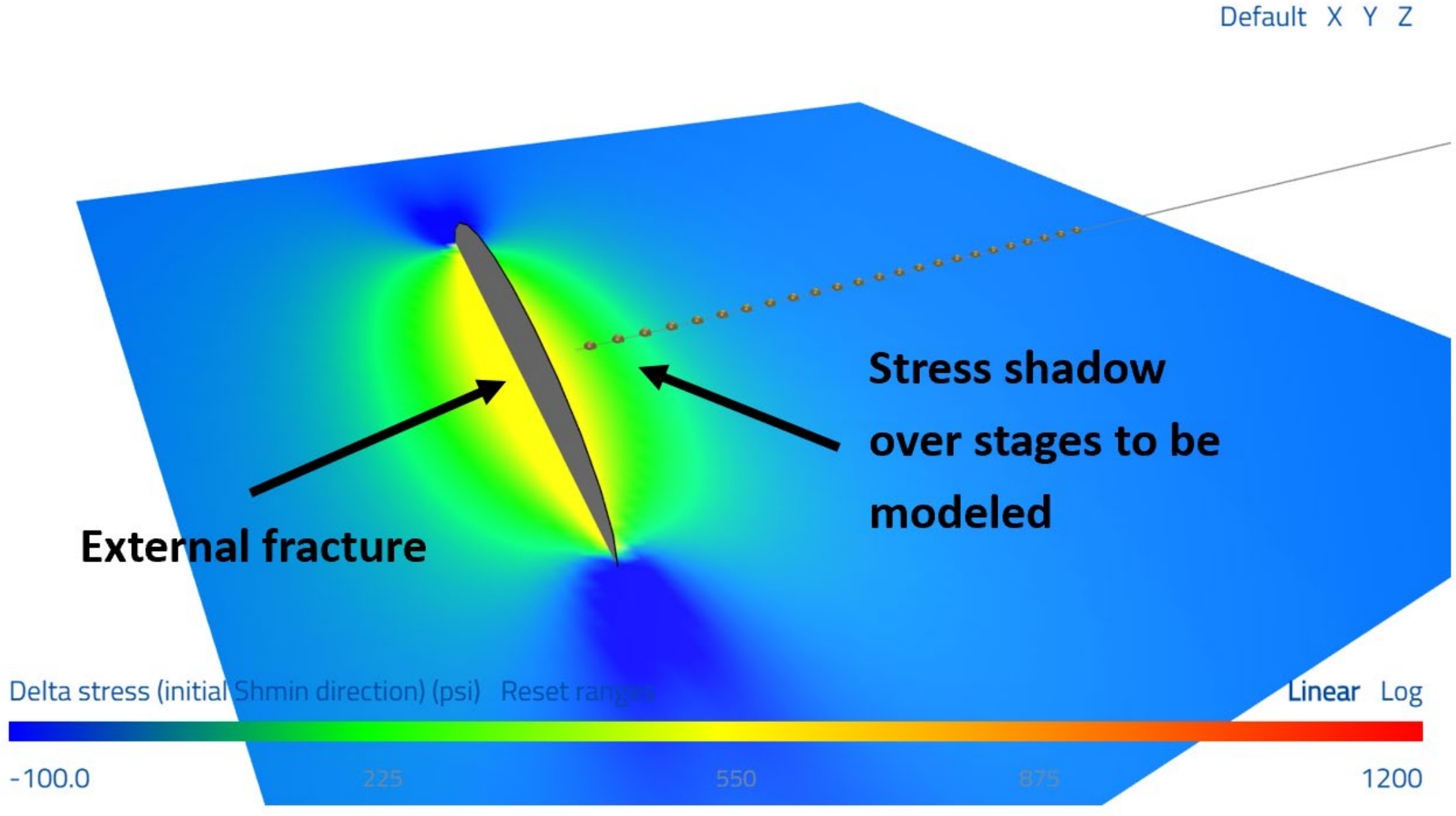

Finally, mesh resolution will impact simulation runtimes. There are three non-conforming meshes in a ResFrac simulation: the wellbore, fractures, and matrix. The wellbore element length has negligible impact on runtimes, so just ensure that the element length is less than the cluster spacing.

Fracture element size will have a BIG impact on simulation runtime. Typical element lengths vary from 50 to 100 feet. This may seem large, but the code is doing local refinement and capturing the (possibly) finer layering of the static model mesh. By default, fracture elements are square; however, by increasing the aspect ratio to greater than 1.0, the user can make rectangular elements.

Finally, the matrix mesh size has a moderate impact on simulation runtimes. The builder will produce warnings for when the matrix mesh contains too many elements, but certain models may require finer resolution or larger scale. Typical matrix block widths (perpendicular to fracture face) range from 5 to 25 feet, and typical lengths (parallel to fracture face) range from 30 to 150 feet. Matrix block height should be refined to correspond to the variable layer heights defined in your static model (and meshing wizard will do this for you). Like the fracture elements, these mesh sizes may seem large, but this is accounted for with the numerical methods employed.

Picking the initial layout can be challenging. Don’t hesitate to reach out and run ideas past the ResFrac team!

## Diagnosing numerical problems with the comments file

Occasionally, a numerical problem causes ResFrac to run too slowly. Simulations that should run in a few hours may run for many days - or so slowly that they never finish.

If you encounter a genuine numerical problem, then you should report the simulation to ResFrac support. We will diagnose the problem and fix, usually within 24-48 hours.

However, timestep failures are a natural part of the operation of the code. It is rare to run a simulation that has zero timestep failures. How can you tell whether timestep failures indicate a genuine numerical problem or not?

The key is to read the comments file to track through the log of what the simulator is doing.

When things are changing quickly in the simulation, the simulator is forced to take small timesteps. It adaptively tries to guess what the timestep duration should be. But sometimes, it picks a timestep that is too large, and this causes a convergence failure. If dt is too large, convergence failure happens, and if too small, then it wastes effort taking more timesteps than necessary.

During normal operation, the simulator tends to increase timestep duration gradually until a failure. After a failure, it cuts dt, and for the next 30 timesteps or so, it only allows dt to increase very gradually. This means that during normal operation, it's pretty common for it to have a timestep failure roughly every 30 timesteps. That's ok - the periodic failures indicate that the code is being close to as aggressive as possible on increasing dt.

However, sometimes, there really is a code bug causing failures to occur when they shouldn't be. This can cause simulations to take virtually forever, as the code is forced to take super short timesteps. This can be identified from the comments file.

In each timestep, the simulator prints a line like this one:

Timestep limiting factor would have been composition in element 145816. Value changed from 0.48173245035044887 to 0.43236025694869484.

The code has looked at every element and looked at how much different things changed - pressure, composition, etc. This line is telling us the element where change was greatest, and by how much. In this case, we can see that composition changed by 0.05. That's a fairly substantial change, which indicates that the timestepping is working correctly.

Conversely, what if it said something like this?

Timestep limiting factor would have been composition in element 145816. Value changed from 0.48173245035044887 to 0.48273245035044887.

In that case, we'd see that the 'max change' in the timestep was very small. If we see this happening persistently, and it is surrounded by a lot of timestep failures, this is an indication that something is wrong. The solver ought to be able to converge through timesteps with greater change than that.

A counter example - I was recently looking at a simulation where production BHP was changed every 24 hours. The timesteps were persistently small, and a substantial number of failures. However, looking at the successful timesteps, we see that the max change in each timestep is large. Thus, even though there are a lot of timesteps and failures, the code is working correctly. The issue is that by changing BHP every 24 hours for a long time you're imposing a lot of sudden change on the system. This causes ripple effects where the distribution of pressure, composition, etc., is changing throughout the system. This causes rapid change. No way to avoid that taking a long time in the code. Other than, of course, to change BHP less frequently, like every 2 or 4 weeks, or even less often than that.

Some simulations will inherently have more timestep failures than others, and the cause may not always be obvious. So it's a judgement call, sometimes it's ambiguous whether something is 'normal operation' or a numerical problem.

Timestep failures can happen because of failure of convergence of the numerical solver, or because of a simple bug. For example, maybe there's a line of code like this:

```
double a = b/c;
```

If the variable 'c' is ever 0, then a would return 'not a number'. That would propagate through the code and eventually trigger an error message, and a timestep failure.

Either way, with convergence failure, or other types of bugs, the end result is an excessive number of timestep failures, and that will be apparent because the 'max change' in variables is persistently low.

There is one last type of numerical problem that I should mention. This is 'instability'. In this problem, the simulator is converging, but the answer is cycling. For example, in one timestep, you might see this:

Timestep limiting factor would have been composition in element 145816. Value changed from 0.48173245035044887 to 0.43236025694869484.

And then the next timestep, you see this:

Timestep limiting factor would have been composition in element 145816. Value changed from 0.43236025694869484 to 0.48173245035044887.

Note that the same element's composition is cycling back and forth. This can cause a simulation to take a long time because the cycling means that things are constantly 'changing fast', and so the timestep is always forced to be low. In other words, you might see that in every timestep, the 'timestep limiting factor' indicates that something changed quickly somewhere. But it's cycling - not really progressing through the simulation.

On the other hand, maybe it's taking a ton of timesteps because the simulation simply is complicated and a lot of things are changing a lot. Another way to tell would be to visualize the results. If you can see that things are changing in the 3D simulation output, then the simulation is progressing. If it's just cycling due to instability, then it might be hard to see much change in a movie of the results, even as the simulator is churning, taking tons of timesteps. Also, be aware that outputting itself is based on the amount that things change, so if the code is cycling, it might take tons of timesteps without outputting many or any visualization snapshots.

Numerical instability occurs rarely within ResFrac, but does crop occasionally due to a bug.

To recap, there are three general types of problems that can cause slow simulator performance: convergence failure, simple bugs, and instability. For all three, there are systematic ways for me to identify and resolve them. However, it is hard to anticipate all of them in advance, so that's why we occasionally still run into one! Also, you may have noticed, when we roll out new features, there tends to be a period when bugs in those features are more likely.

Again, if you have a simulation that is running too slowly, and it appears to be related to numerical issues, please share with us, so that we can resolve the problem.

If you have a simulation that is running too slowly, but does not appear to be having any numerical problems, check the number of fracture elements (which is printed in the comments file). Runtime usually scales with the number of fracture elements. If you have much more than around 5000, this may be the cause of slower runtime. Also, you check your boundary condition controls. If you have a huge number of production rate/BHP changes, this can slow down the simulator. This problem can be resolved by checking the boxes for spline interpolation of BC controls and the box for 'Do not align timesteps with production sequence control points'.

## 13.4 Using external grid files with ResFrac

### Exporting grid files from Petrel to ResFrac

Petrel is a common geomodelling software used to create grids for reservoir and hydraulic fracture simulation. You are able to directly import your grid from Petrel into ResFrac. The individual grid cell properties that are supported for use within ResFrac are listed below. If other values are included in the Petrel files, those values will not be imported into ResFrac.

1. PRESSURE - Initial pore pressure
2. SWAT - Water saturation
3. SOIL - Oil saturation
4. PERMX - X-Permeability
5. PERMY - Y-Permeability
6. PERMZ - Z-Permeability
7. BIOTC - Biot coefficient
8. TOTALSTRESS - Stress
9. PORO - Initial porosity
10. YOUNGMOD - Young's modulus
11. POISSONR - Poisson's ratio
12. FRACTURETOUGHNESS - Horizontal fracture toughness
13. VERTICAL_FRACTURE_TOUGHNESS - Vertical fracture toughness
14. NTG - Net to gross
15. TEMPERATURE - Temperature
16. ACTNUM - Active cells

In addition to the grid cell properties, the grid itself and the wells defined in the Petrel case can be imported into ResFrac using the ZCORN/COORD/SPECGRID keywords or the DX/DY/DZ/TOPS/SPECGRID keywords in the Petrel exported file. SPECGRID is needed to define the size of the grid in terms of grid block numbers in the x, y and z directions and ZCORN/COORD/ DX/DY/DZ/TOPS define the pillar locations for the grid.

The wells from the Petrel file are imported using the COMPDAT keyword and rely on the grid block number to determine the location of the well. An alternative to using the COMPDAT keyword in the Petrel file is to use the "*well import wizard"* in the *"Wells and Perforations"* tab within ResFrac. The well coordinates should match the coordinates given during the import of the Petrel grid.

Once all of the properties and grid has been created in Petrel, the "*.GRDECL*" format of the grid should be exported from Petrel. In the processes tab, select "*Define simulation case"* in order to select the properties and format to export.

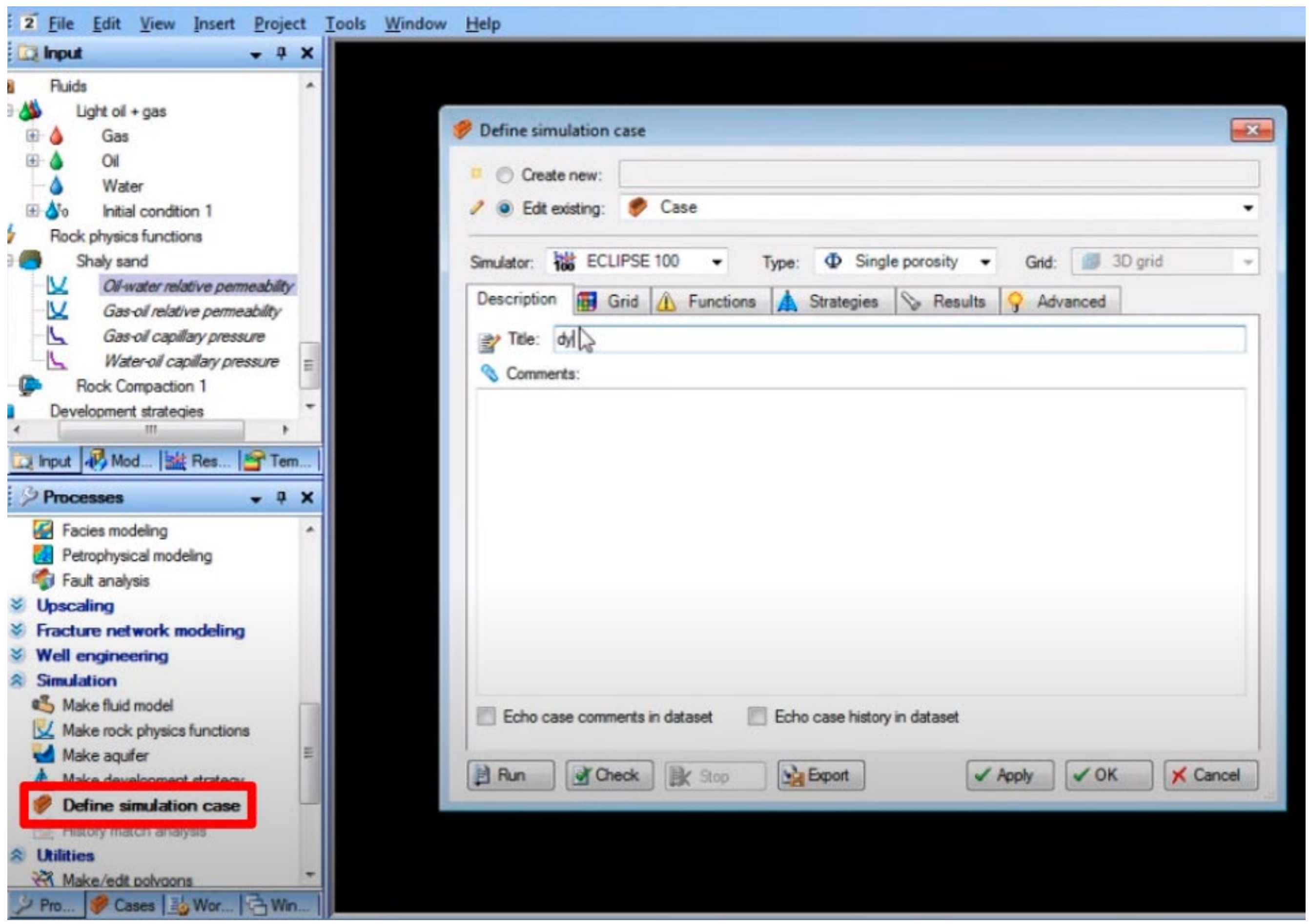

Once the "*Define simulation case*" pop up has appeared, select either create new or edit existing and name the grid that you are about to export on the "*Description*" tab. You can then select the "*Advanced*" tab that will allow you to select the type of grid to export.

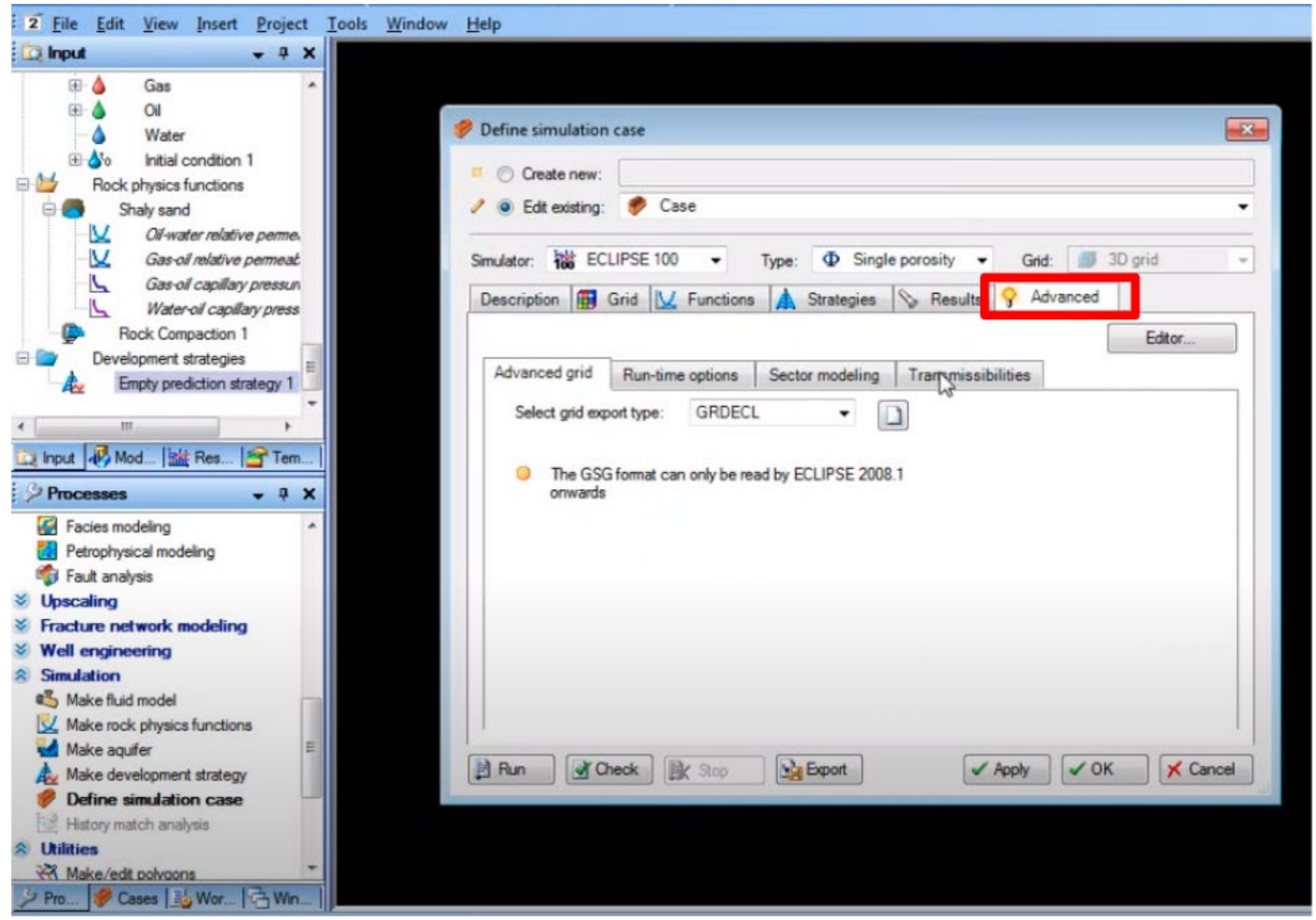

Please select "*GRDECL*" in the "*Select grid export type*" drop down menu in order to create the correct format for ResFrac. Then select the "*Grid*" tab in the pop up screen.

In the "*Grid*" tab, you are then able to select the grid properties to export to the "*.GRDECL*" file. Some things to note, please ensure that the correct keyword is placed with the correct property, the properties that you are exporting are from the same geological realization and that the I, J and K permeabilities are all defined if one of the permeabilities have been defined.

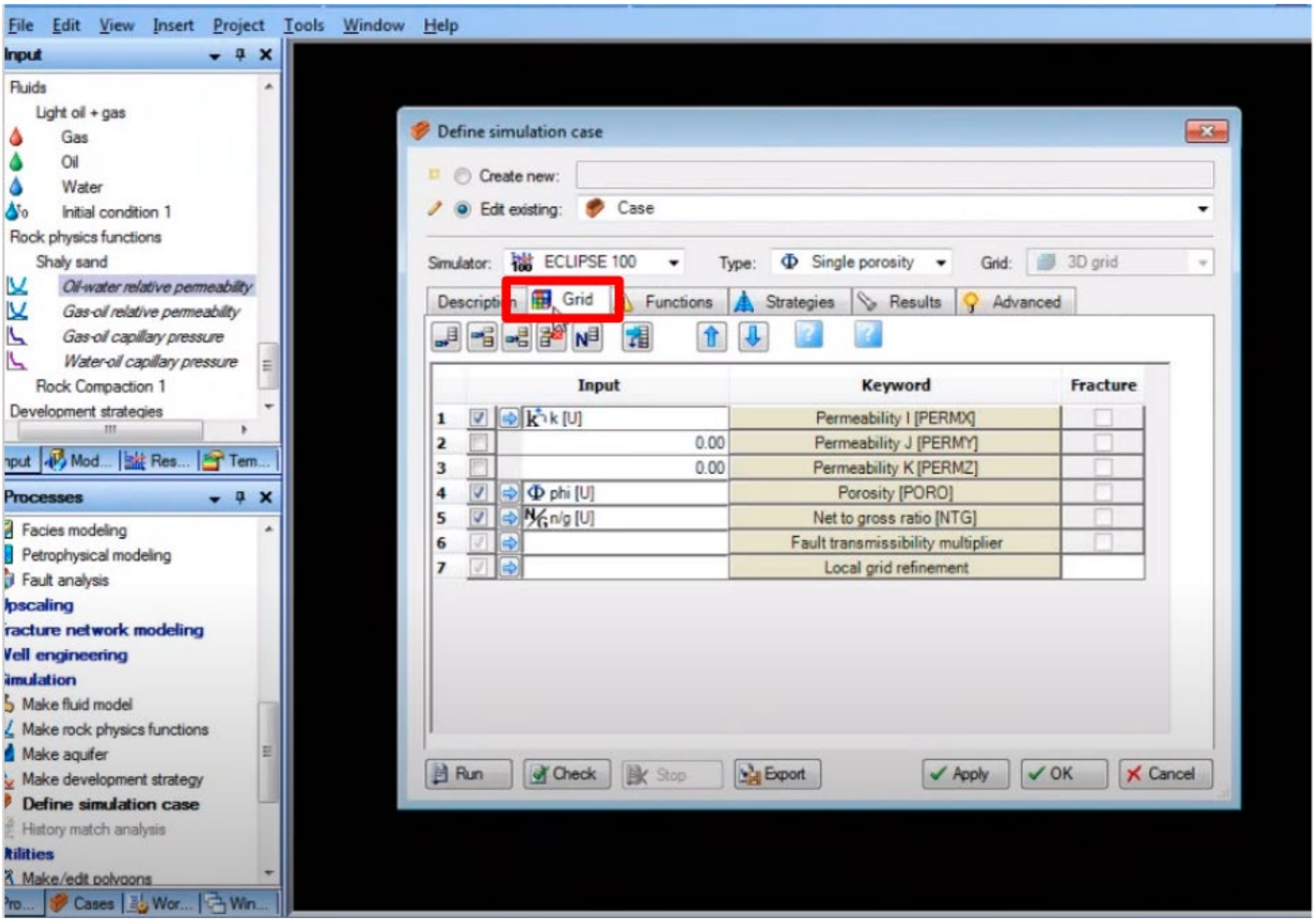

To insert a new property into the exported grid further from the list given in the default screen, there are insertion and deletion tabs above the properties table. Use these to ensure that the grid created has the correct keywords and input values for importation into ResFrac.

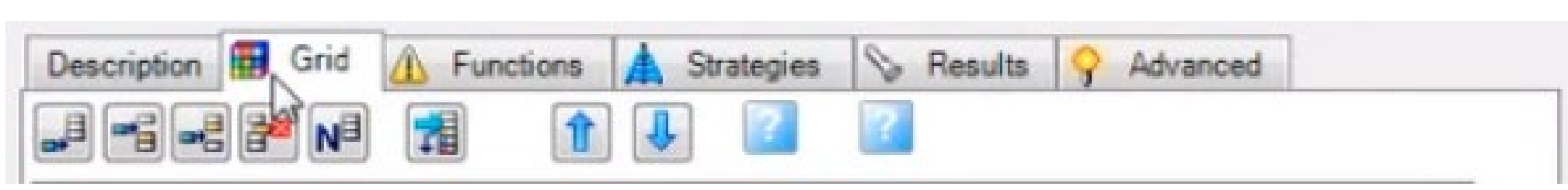

After the grid properties have been selected, pressing the "*Export*" button at the bottom of the pop up window will prompt you to save the file. This "*.GRDECL*" file will be used to import both the grid and the model properties into ResFrac.

## Exporting grid files From Leapfrog to ResFrac

Leapfrog is a common geomodeling software used to create grids for reservoir and hydraulic fracture simulation. With the introduction of corner point modeling, you are able to directly import your grid from Leapfrog into ResFrac. The individual grid cell properties that are supported for use within ResFrac are listed below along with their corresponding keywords. If other values are included in the Leapfrog files, those values will not be imported into ResFrac.

1. PRESSURE - Initial pore pressure

2. SWAT - Water saturation

3. SOIL - Oil saturation

4. PERMX - X-Permeability

5. PERMY - Y-Permeability

6. PERMZ - Z-Permeability

7. BIOTC - Biot coefficient

8. TOTALSTRESS - Stress

9. PORO - Initial porosity

10. YOUNGMOD - Young's modulus

11. POISSONR - Poisson's ratio

12. FRACTURETOUGHNESS - Horizontal fracture toughness

13. VERTICAL_FRACTURE_TOUGHNESS - Vertical fracture toughness

14. NTG - Net to gross

15. TEMPERATURE - Temperature

16. ACTNUM - Active cells

In addition to the grid cell properties, the grid itself and the wells defined in the Leapfrog case can be imported into ResFrac using the ZCORN/COORD/SPECGRID keywords or the DX/DY/DZ/TOPS/SPECGRID keywords in the Leapfrog exported file. SPECGRID is needed to define the size of the grid in terms of grid block numbers in the x, y and z directions and ZCORN/COORD or DX/DY/DZ/TOPS define the pillar locations for the grid.

The wells from the Leapfrog file are imported using the COMPDAT keyword and rely on the grid block number to determine the location of the well. An alternative to using the COMPDAT keyword in the Leapfrog file is to use the *"well import wizard"* in the *"Wells and Perforations"* tab within ResFrac. The well coordinates should match the coordinates given during the import of the Leapfrog grid.

To export the model from Leapfrog into the Eclipse format required by ResFrac, right click on the flow model in Leapfrog and select *"Export to Reservoir Simulator"*.

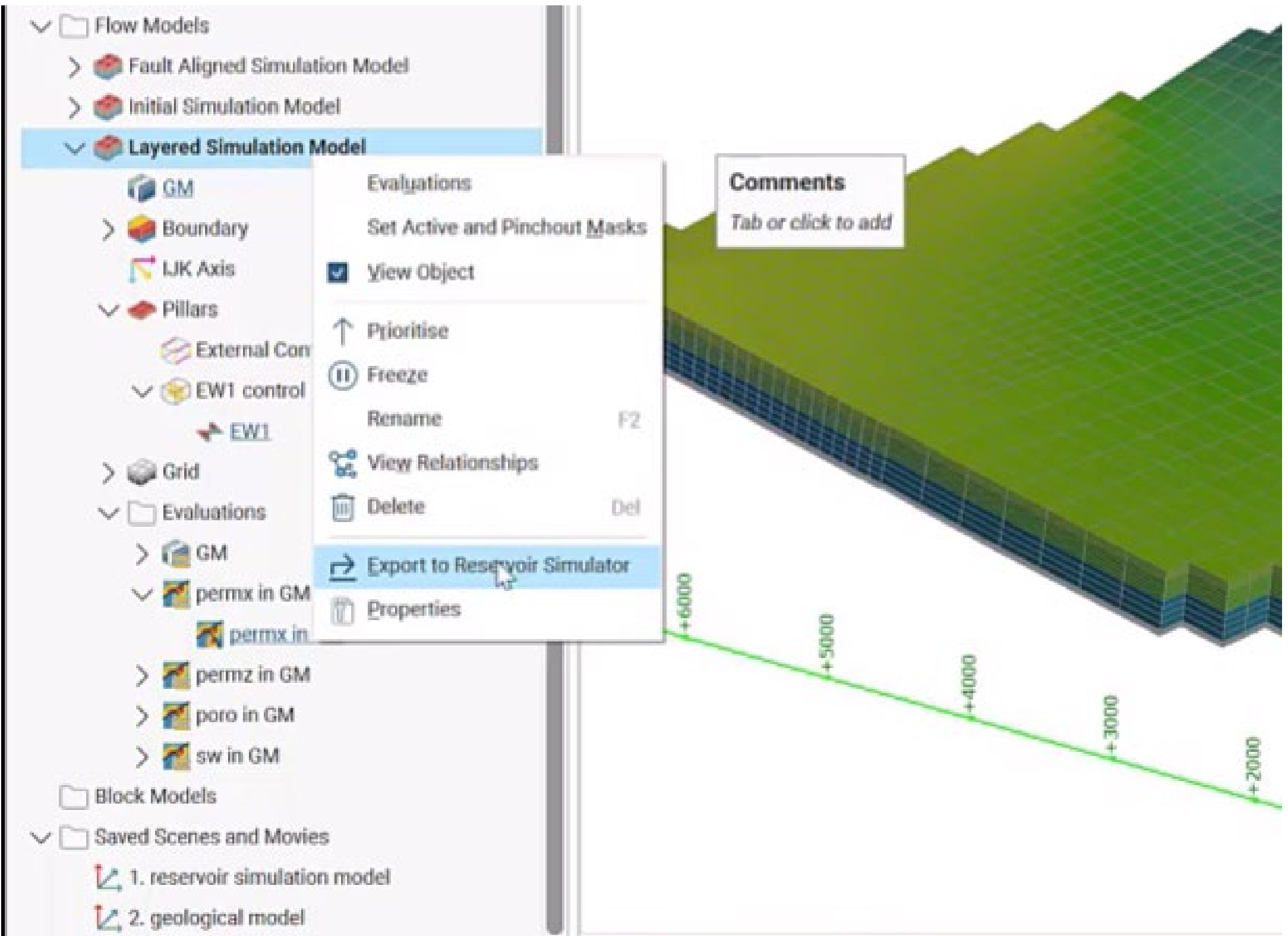

In the screen that pops up, select the simulator as *ECLIPSE* and set the other options as desired.

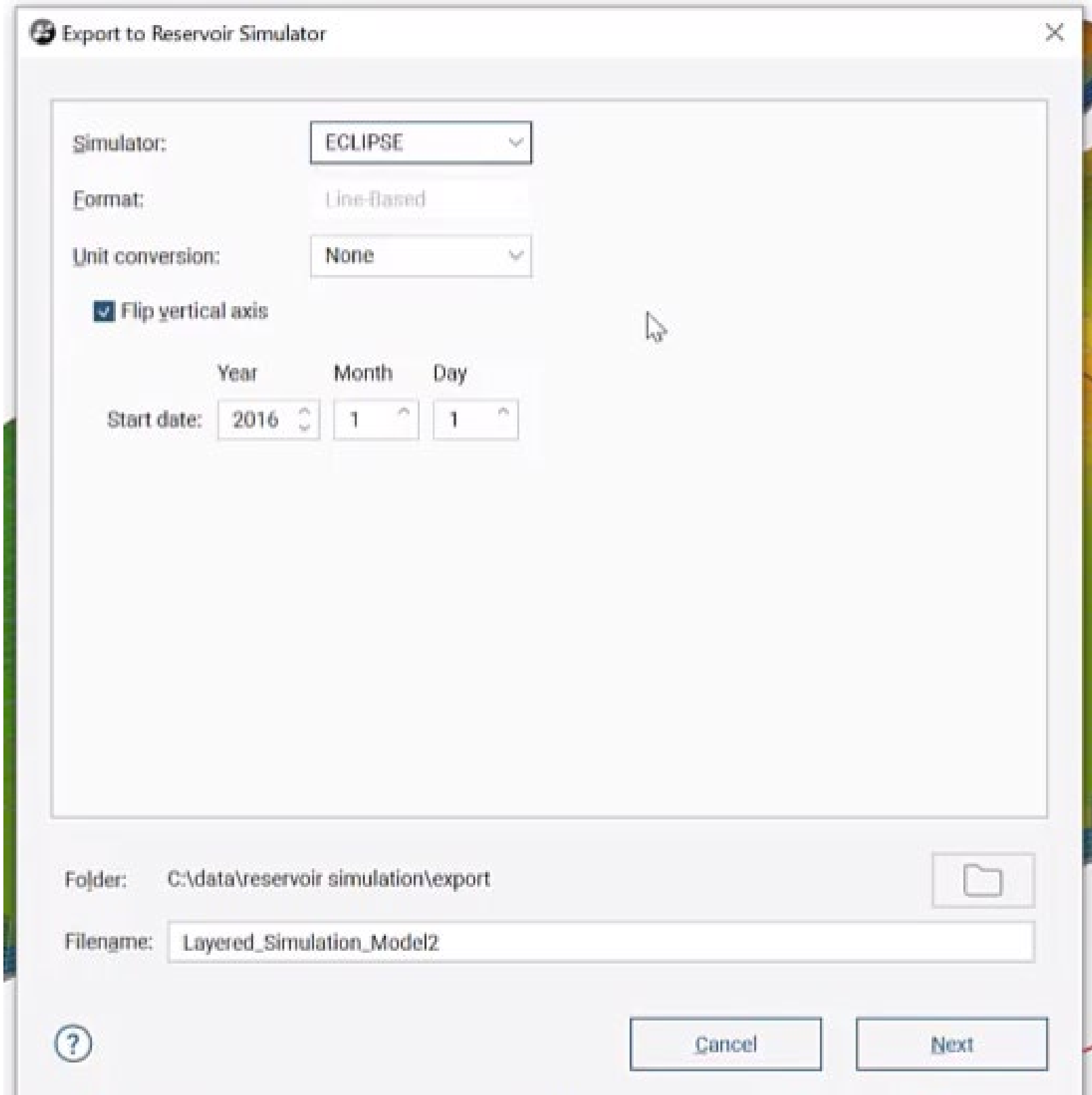

On the following screen, select the appropriate properties that you would like to export with the model using the naming conventions as above and click on the *Export* button.

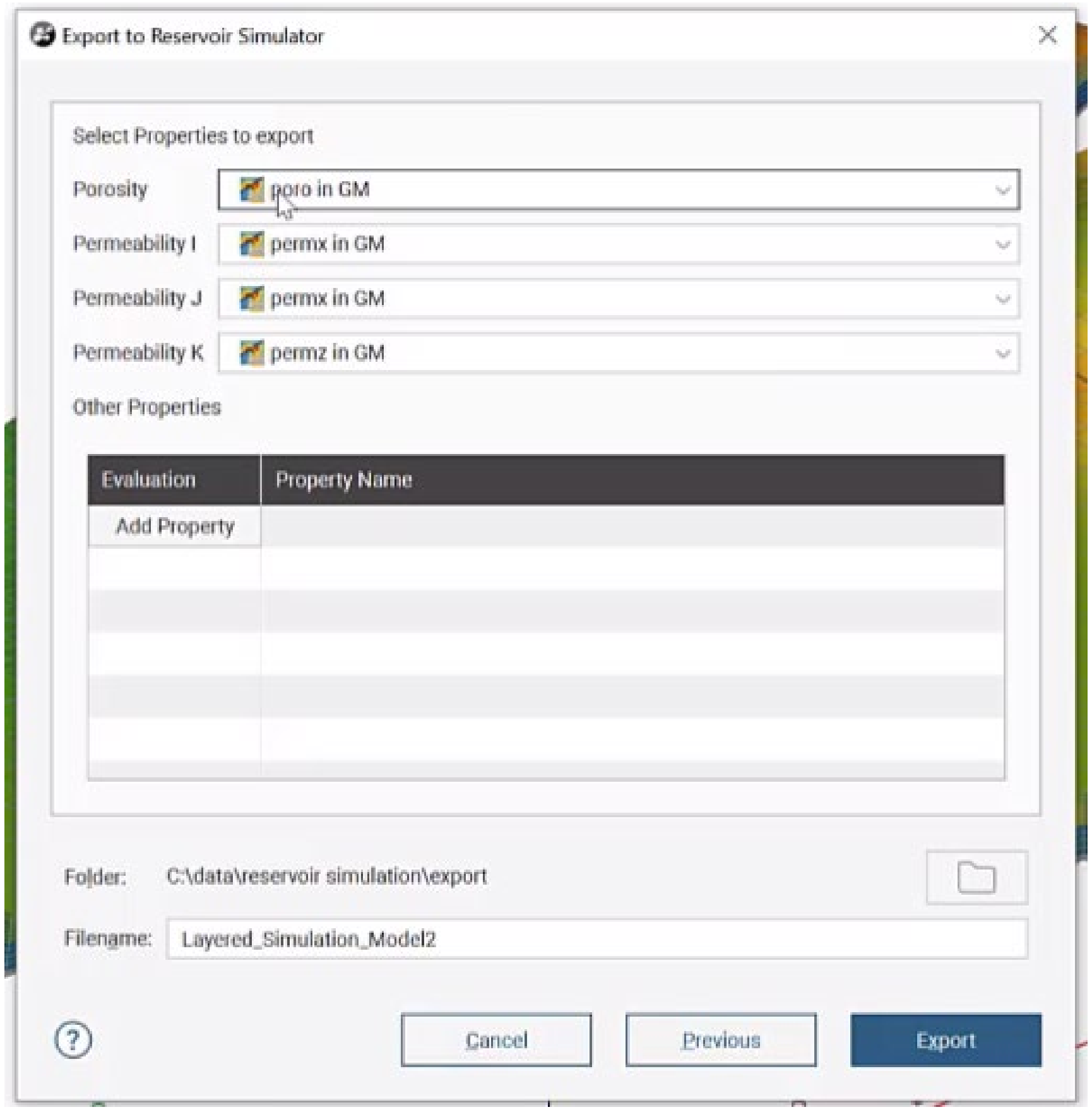

The model should have been exported to the file specified by the user and ready for import to ResFrac. For further information on Leapfrog grid import and export, please refer to this video on Seequent's website (https://www.seequent.com/see-petrel-data-import-and-data-export-to-eclipse-and-cmg-with-leapfrog-energy-2021-2/).

Once all of the properties and grid have been created in Leapfrog, the *".GRDECL"* format of the grid should be exported from Leapfrog. This *".GRDECL"* file will be used to import both the grid and the model properties into ResFrac.

## Importing external files into ResFrac

To import the exported "*.GRDECL*" file from Petrel or Leapfrog into ResFrac, navigate to the "*Static Model and Initial Conditions*" panel. At the top of the panel, there is a check box labelled "*Check to use either cornerpoint or generalized rectilinear grid; uncheck to use a standard rectilinear grid extension*", check this box to reveal the following screen.

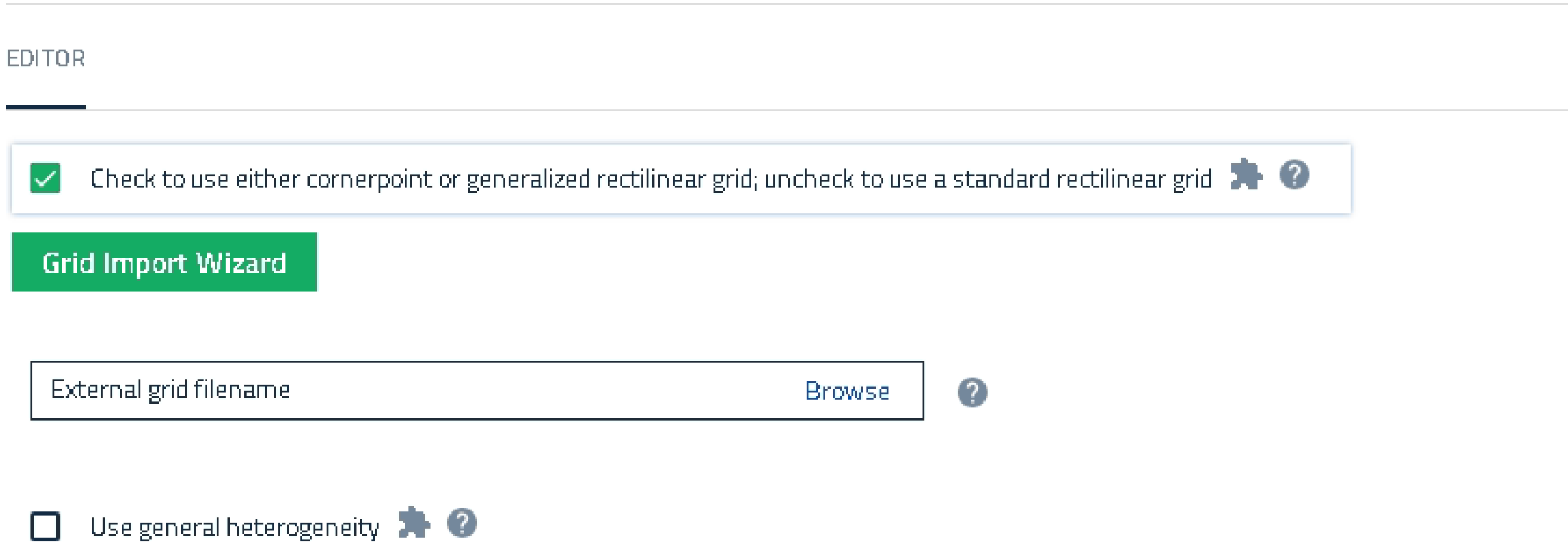

For Petrel files, select the “*Grid Import Wizard*” box, this will allow you to select a folder to import the grid from.

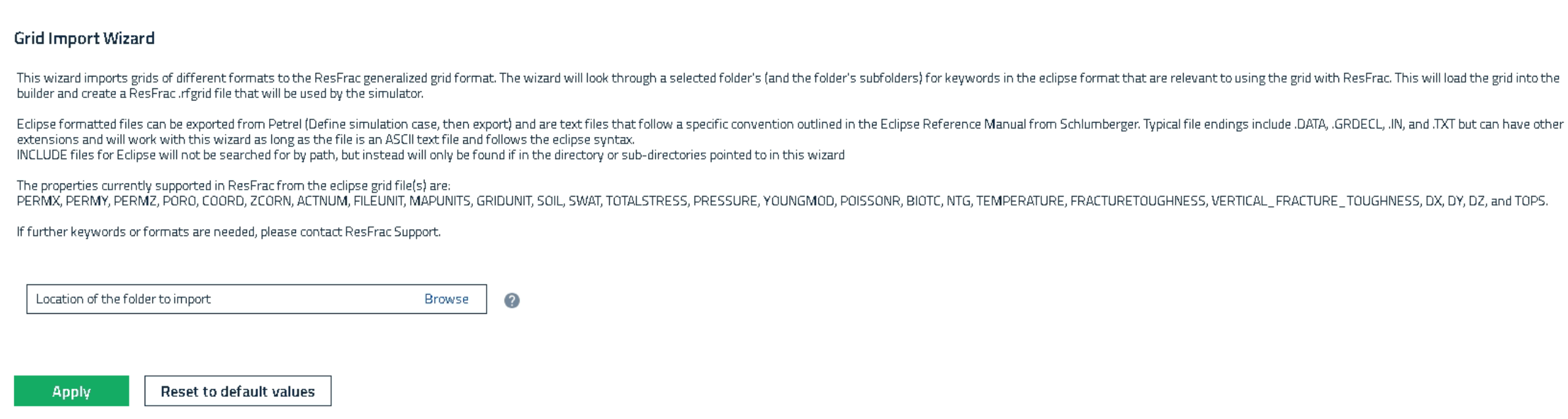

Click the “*Browse*” button to allow navigation to and selection of the grid folder.

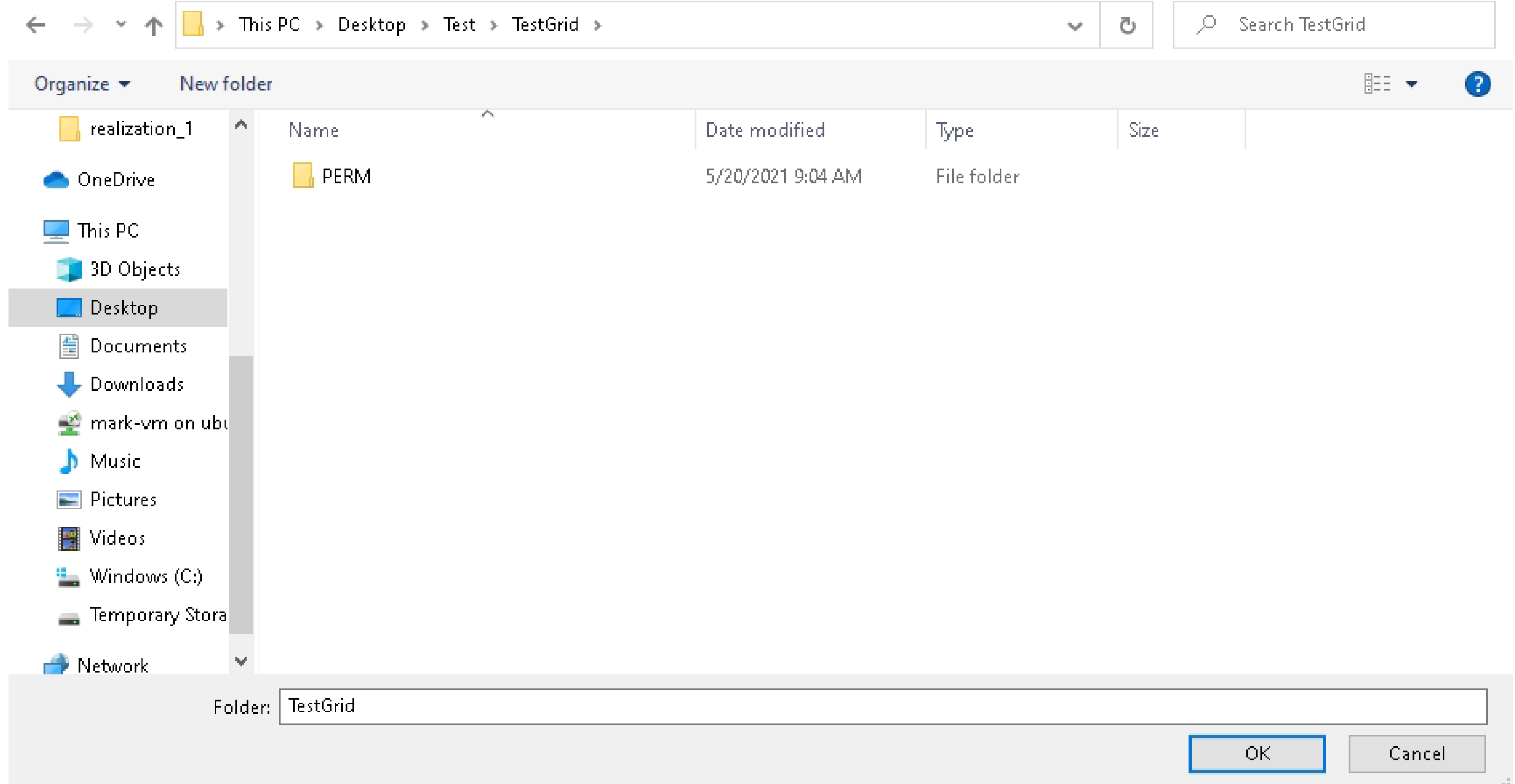

For this example folder, it contains "*.inc*" files, as well as the "*.GRDECL*" file and a folder for the permeability files labelled "*PERM*". ResFrac will search in all of the files in all of the sub folders in order to import the grid and the properties. An explorer view of the example folder is shown here.

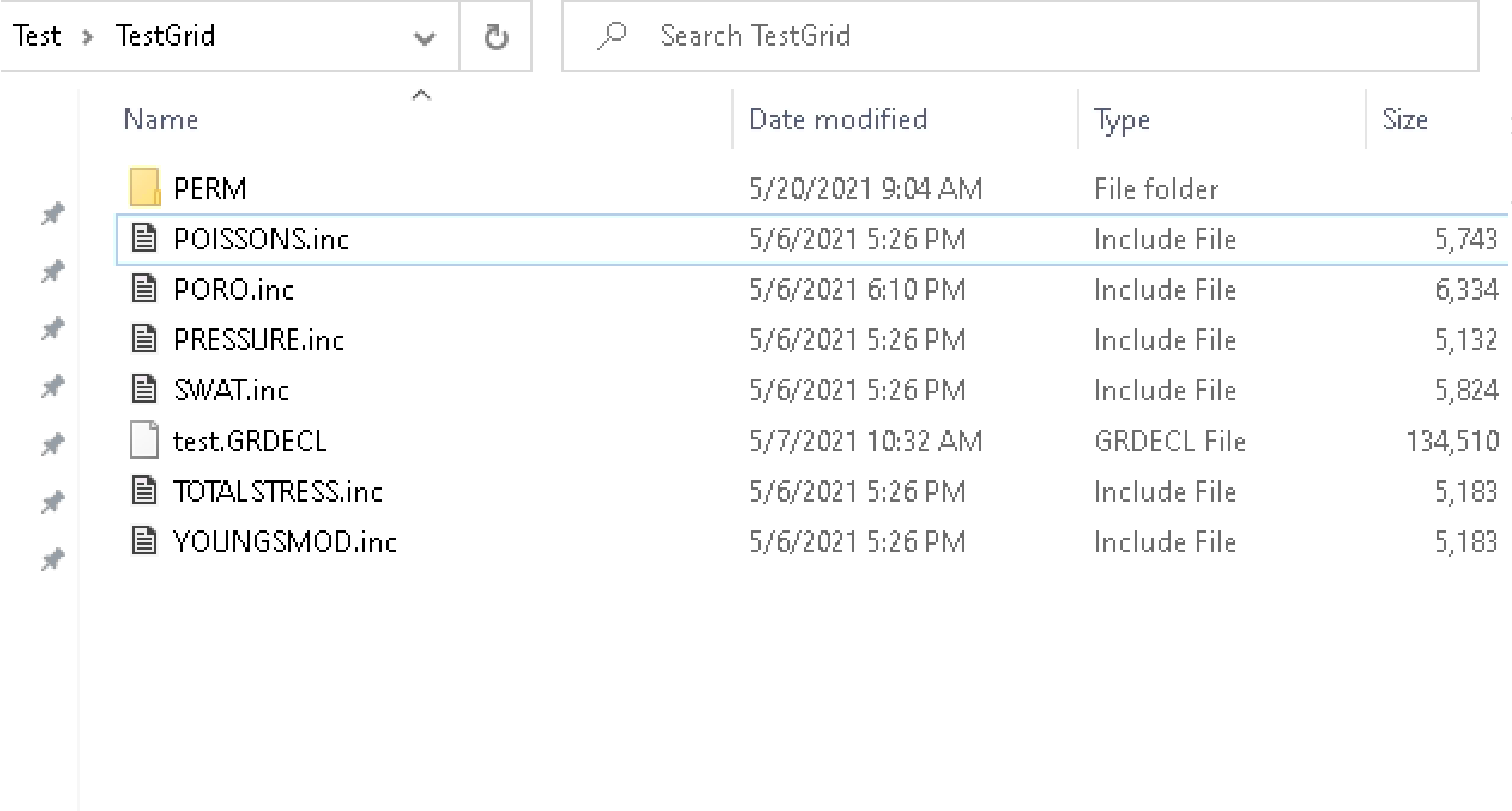

Once the folder has been selected, click OK in the folder selection dialogue and then the "*Apply*" button in the "*Grid Import Wizard*" pop up window.

## Grid Import Wizard

This wizard imports grids of different formats to the ResFrac generalized grid format. The wizard will lo
builder and create a ResFrac .rfgrid file that will be used by the simulator.

Eclipse formatted files can be exported from Petrel (Define simulation case, then export) and are text fil
extensions and will work with this wizard as long as the file is an ASCII text file and follows the eclipse s
INCLUDE files for Eclipse will not be searched for by path, but instead will only be found if in the director

The properties currently supported in ResFrac from the eclipse grid file(s) are:
PERMX, PERMY, PERMZ, PORO, COORD, ZCORN, ACTNUM, FILEUNIT, MAPUNITS, GRIDUNIT, SOIL, SWAT

If further keywords or formats are needed, please contact ResFrac Support.

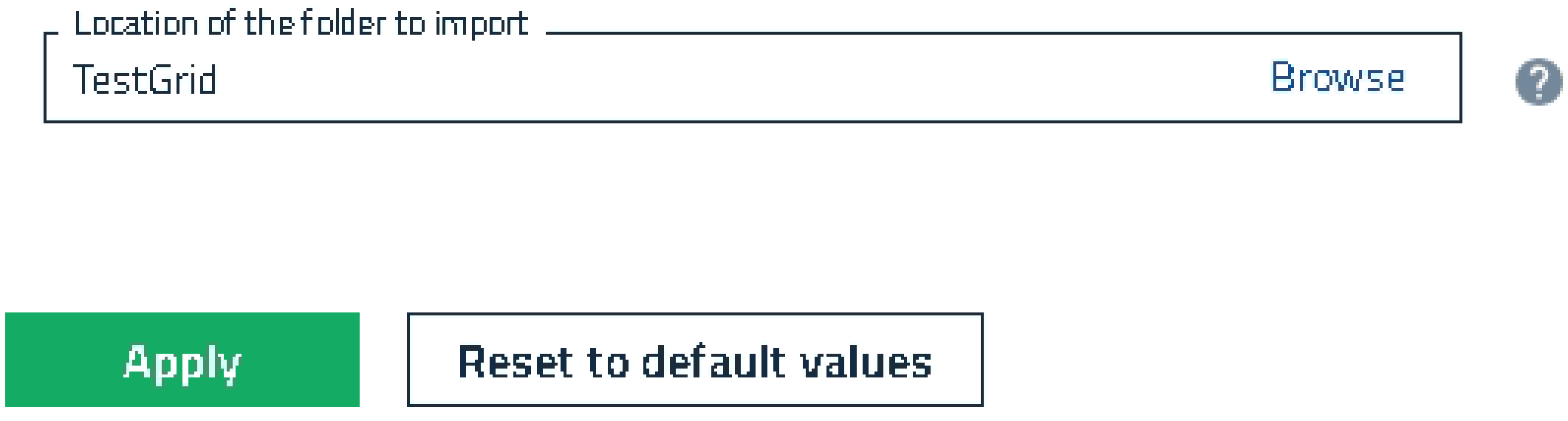

This will then create a ResFrac corner point grid format file "*.rfgrid*" and a ResFrac general heterogeneity file with the imported heterogeneity properties listed in the table in ResFrac.

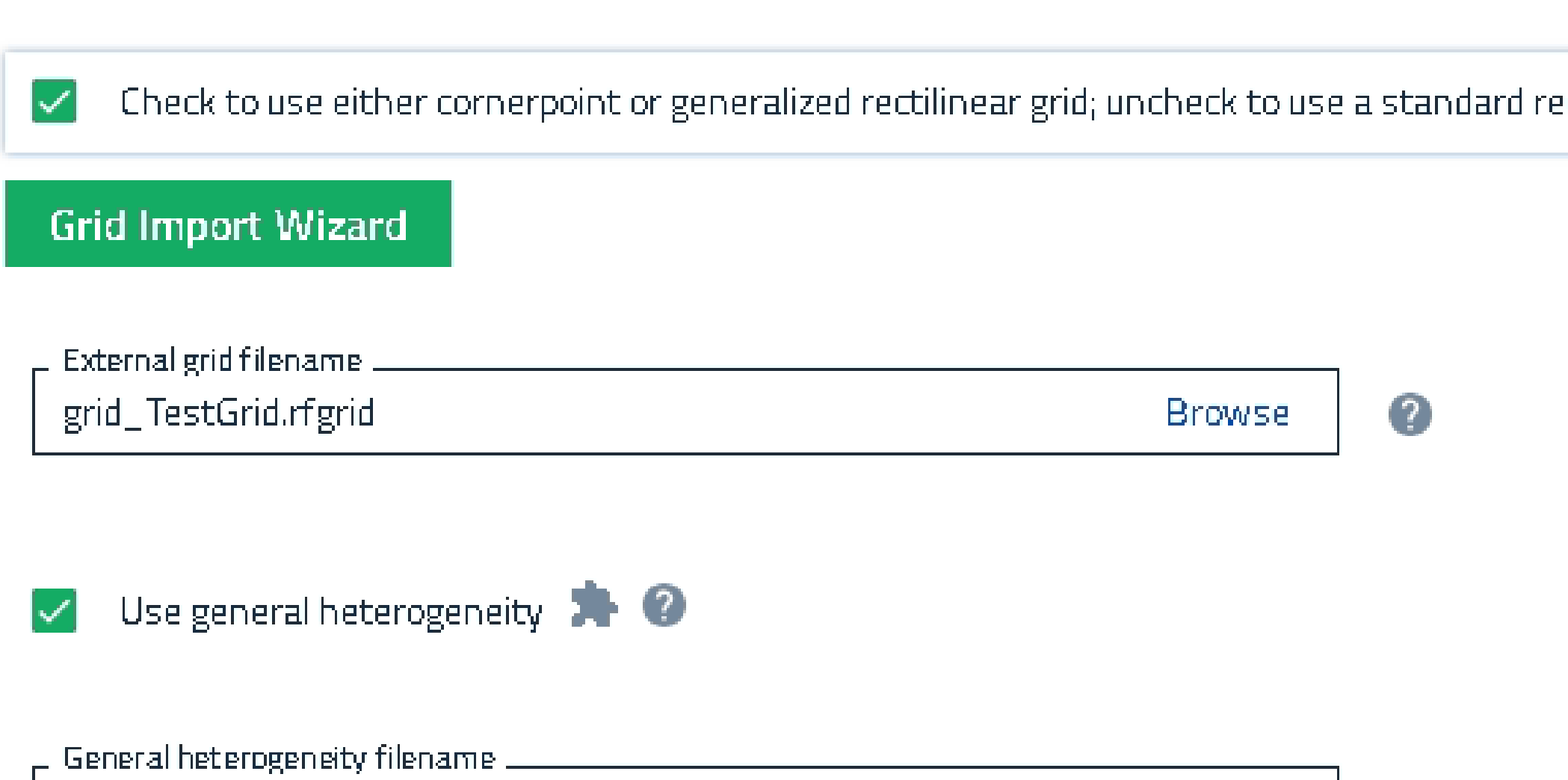

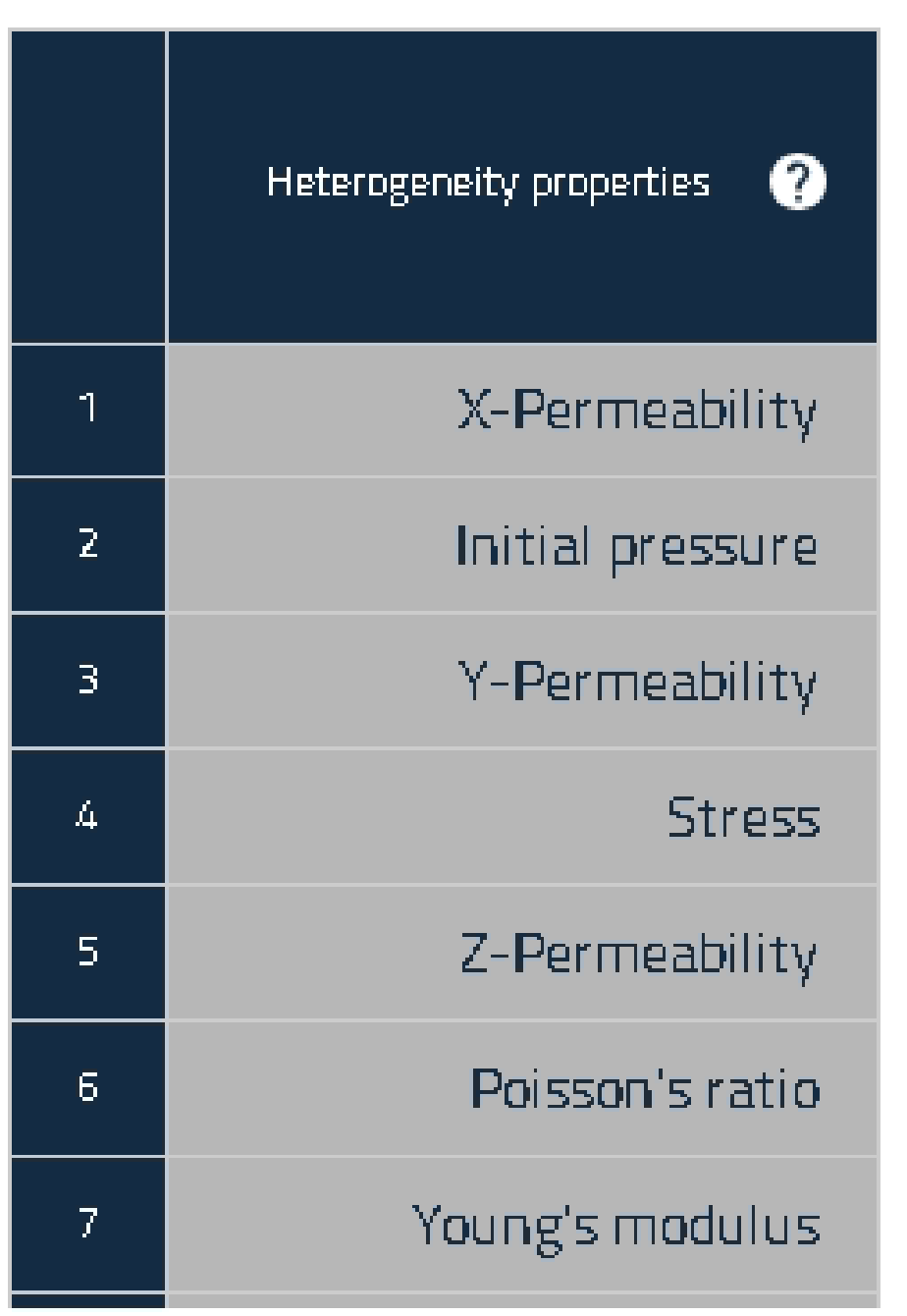

When the grid and general heterogeneity files are created in the wizard, they will be saved within the "Additional_Files" folder in the simulation folder. If you are not importing these files from the Petrel format, you can use the browse button in the *"External grid filename"* and *"General heterogeneity filename"* boxes to select the file.

Also, when importing simulation folders into the user interface, you can select the location of the "Additional_Files" folder when choosing the appropriate input and settings files in the import window.

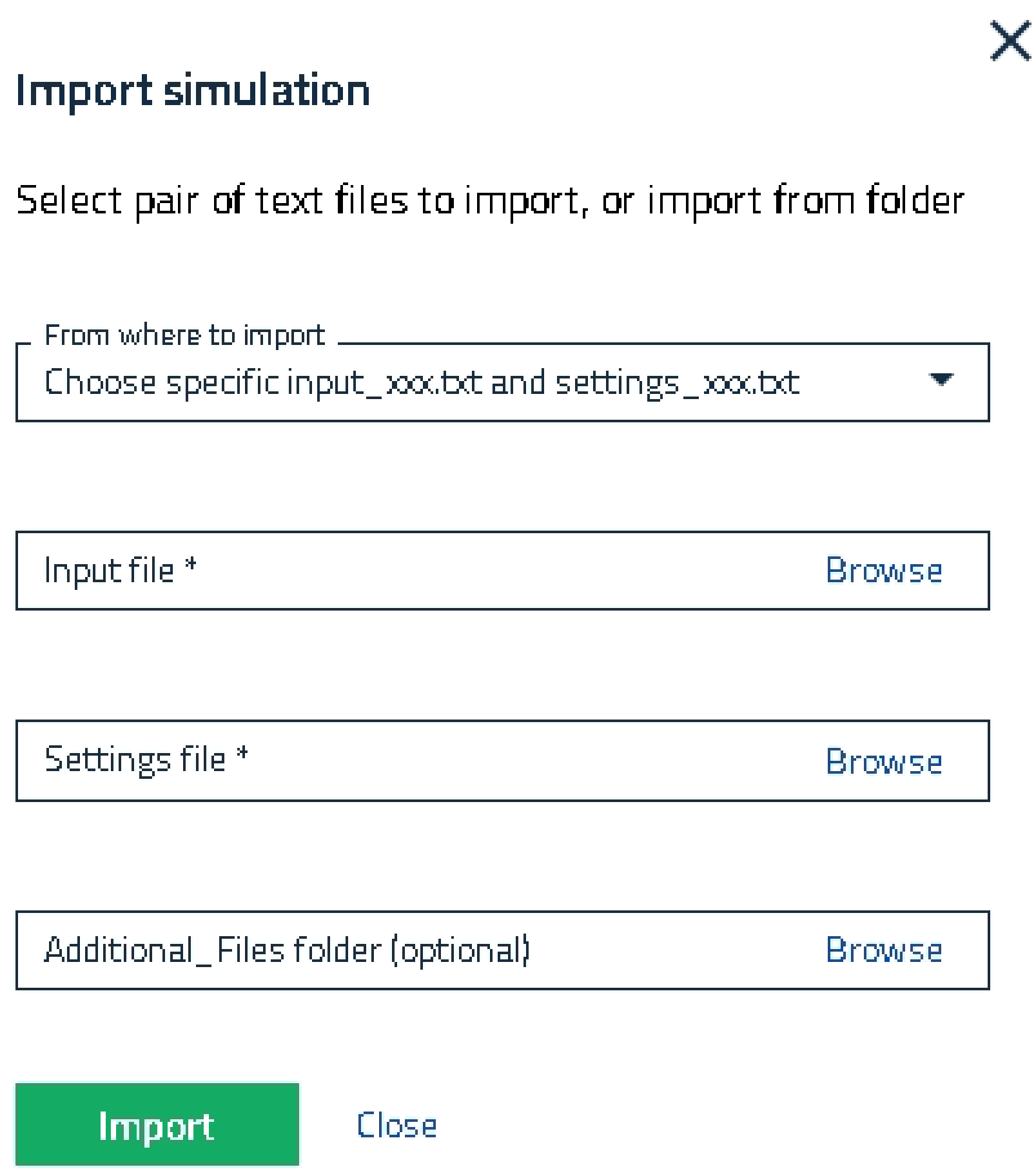

After the import of the geomodel, it may be necessary to cut out a sector of the model or create more refined grid blocks around the wells in order to conduct a numerically accurate and stable fracture simulation. The overall matrix size and the size of the grid blocks can be changed within two wizards. One being the *"Corner point meshing wizard"* on the *"Meshing Options"* panel that allows the user to define the new sizes, along with the center of the new grid, to create a sector model or change the grid block size. Similarly, the *"Well stages setup wizard"* on the *"Wells and Perforations"* panel allows the user similar functionality while also creating the stages for the wells that have been previously defined.

## 13.5 Importing external data

### Importing a black oil table of tables from a petrel file

Black oil models are sometimes given in a 'table of tables' format. At each saturation pressure, a separate table is provided to output unsaturated properties.

To import the exported table of tables "*.GRDECL*" file from Petrel into ResFrac, navigate to the "*Fluid model options*" panel. There is a drop menu labelled "*Unsaturated oil black options", select the 'Table of tables' option*.

Input all the required values on this panel. One of them is the bubble point pressure. Input the value for "Initial bubble/dew point [psi]", making sure this exact value is included as one of the saturated pressures in the table of tables file.

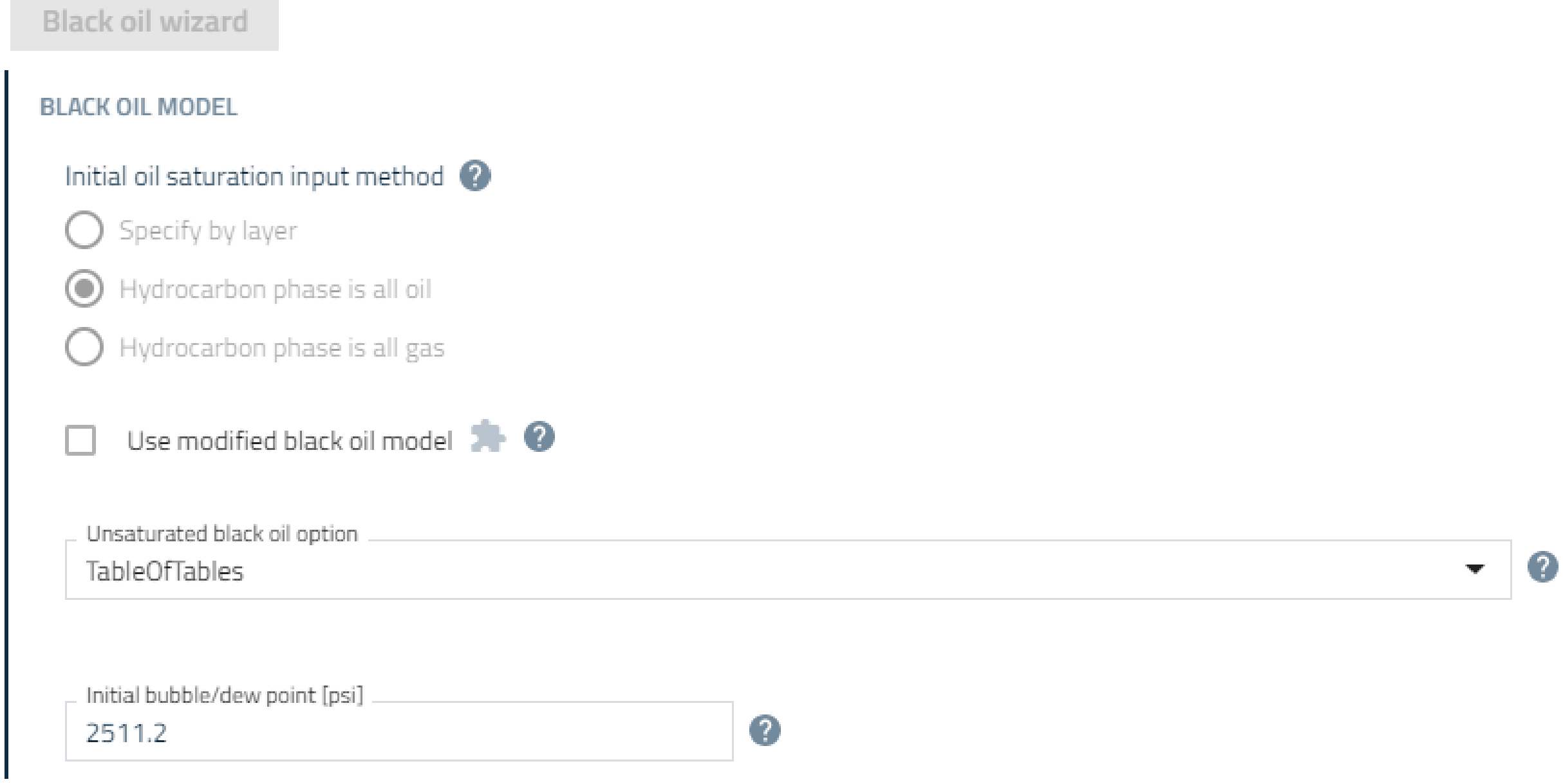

**Understanding the format of the *GRDECL file**

The data file *GRDECL received from petrel has a specific format, its variables and units need to be understood prior to input in ResFrac. Here is an example and some tips about how to import a table of tables file from Petrel.

- Identify the section labeled PVTO in the Petrel file, this section has multiple tables for various gas solution ratios and saturation pressures. The first row should have four columns, the first value on this first row, should correspond to the first gas solution ratio (Rs with units [mscf/STB]) for the first table of tables. The second value, in the first row, corresponds to the first saturation pressure (psi) for the first table of tables, and it's also the first value of pressure for column 1 in this table.

The subsequent values on the first row, correspond to column 2 and column 3, first row of the table of tables. The order and units of the columns in the file should be: Column1=Pressure (psi), column 2= volumetric oil factor (Bo with units RB/STB), and column 3= oil viscosity (Vo with units Cp). Note that RB: reservoir barrels.

- The following rows in this first table should have only 3 columns: Column1=Pressure, column 2=(Bo), and column 3=(Vo).
- The next table of tables could be identified when the rows increase the number of columns from 3 to 4. This would be the next table of tables for that specific gas solution ratio (mscf/STB) and saturation pressure (psi)

```
PVTW                                   -- Generated : Petrel
        3148.8       1.0445   3.3421E-06      0.24765   7.4069E-06 /

PVTO                                   -- Generated : Petrel
                  0                 15          1.04627         1.00416
              371.6            1.04326          1.03535
              728.2            1.04044           1.0657
             1084.8             1.0378          1.09528
             1441.4            1.03532          1.12411
               1798            1.03298          1.15224
             2154.6            1.03077          1.17969
             2511.2            1.02868          1.20649
             2867.9            1.02669          1.23268
             3148.8             1.0252          1.25289
             3224.5            1.02481          1.25827
             3581.1            1.02301          1.28329
             3937.7             1.0213          1.30776
             4294.3            1.01967           1.3317
             4650.9             1.0181          1.35513
             5007.5            1.01661          1.37807
             5364.1            1.01518          1.40053
             5720.7             1.0138          1.42254
             6077.3            1.01248           1.4441
             6433.9            1.01121          1.46524
             6790.5            1.00999          1.48596
             7147.1            1.00882          1.50628
             7503.7            1.00768           1.5262
             7860.4            1.00659          1.54575
               8217            1.00554          1.56494
             8573.6            1.00452          1.58377
             8930.2            1.00353          1.60225
             9286.8            1.00258           1.6204
             9643.4            1.00165          1.63823
              10000            1.00076          1.65573 /
           0.097086              371.6          1.13155        0.433547
              728.2            1.12522         0.452824
             1084.8            1.11943          0.47169
             1441.4             1.1141         0.490171
               1798            1.10916         0.508289
             2154.6            1.10458         0.526064
             2511.2            1.10031         0.543512
             2867.9            1.09631         0.560647
```

- Next is to identify the section labeled PVDG in the Petrel file, this section contains the information for the gas of the fluid model. This PVDG section should have only one table that will be the same for all the tables and different gas solution ratios and saturation pressures. The order ot the columns should be: column 1= pressure (psi), column 2= gas volumetric factor (Bg with units RB/mscf), and column 3 = gas viscosity (Vg with units Cp). Note that RB: reservoir barrels.

```
PVDG                                          -- Generated : Petrel
                 15          237.87189              0.0116
              371.6            8.90998              0.0132
              728.2             4.3402              0.0141
             1084.8            2.80902              0.0151
             1441.4            2.05436              0.0165
               1798            1.61196              0.0184
             2154.6            1.32593               0.021
             2511.2            1.12933              0.0244
             2867.9            0.98886              0.0288
             3148.8             0.9056              0.0331
             3224.5               0.89              0.0338
             3581.1            0.82756              0.0369
             3937.7            0.77894              0.0398
             4294.3            0.73999              0.0424
             4650.9            0.70805              0.0449
             5007.5            0.68133              0.0472
             5364.1            0.65861              0.0494
             5720.7            0.63903              0.0514
             6077.3            0.62194              0.0533
             6433.9            0.60688              0.0552
             6790.5            0.59349              0.0569
             7147.1            0.58149              0.0586
             7503.7            0.57066              0.0602
             7860.4            0.56084              0.0618
               8217            0.55187              0.0634
             8573.6            0.54365              0.0649
             8930.2            0.53607              0.0663
             9286.8            0.52907              0.0678
             9643.4            0.52257              0.0692
              10000            0.51651              0.0706
 /
```

**Input the table of tables data in ResFrac**

It is suggested to build a new table of tables with the same format and units as is required in ResFrac, then copy and paste it in ResFrac in the "Black oil model property table".

Using the data from the *.GRDECL file from petrel, unit conversions will be required to match the required units in ResFrac. This new table of tables should have seven columns as follows:

- Column 1= saturation pressure (psi), column 2= pressure (psi), column 3= oil volumetric factor Bo (RB/STB), column 4= gas solution ration Rs (scf/STB), column 5= gas volumetric factor Bg (reservoir cf/scf), column 6= oil viscosity (Cp), and column 7= gas viscosity (Cp).
- Build the table of tables keeping saturation pressure and gas solution ratio constant for each table. The specific saturation pressure and gas solution ratio for a given table table will be increasing on the subsequent table. Note that its important to have a table for the saturation pressure equal to bubble or dew point pressure.

- Note the following two variables needed unit conversion from Petrel units to ResFrac units: gas solution ratio (scf/STB) and Bg (reservoir scf/scf)
- Unit conversions:
  - Rs (scf/STB) = Rs (mscf/STB) * 1000
  - Bg (res scf/scf) = Bg (RB/scf) * 0.0282793 / (1000*0.00503676)

Extract of the built table of tables with format and units required in ResFrac

| saturation pressure | pressure | Bo | Rs | Bg | Vo | Vg |
|---|---|---|---|---|---|---|
| psi | psi | RB/STB | scf/STB | Reservoir cf/scf | Cp | Cp |
| 15 | 14.69 | 1.04627 | 0 | 1.335551136 | 1.00416 | 0.0116 |
| 15 | 371.6 | 1.04326 | 0 | 0.05002581 | 1.03535 | 0.0132 |
| 15 | 728.2 | 1.04044 | 0 | 0.024368407 | 1.0657 | 0.0141 |
| 15 | 1084.8 | 1.0378 | 0 | 0.015771472 | 1.09528 | 0.0151 |
| 15 | 1441.4 | 1.03532 | 0 | 0.011534372 | 1.12411 | 0.0165 |
| 15 | 1798 | 1.03298 | 0 | 0.009050481 | 1.15224 | 0.0184 |
| 15 | 2154.6 | 1.03077 | 0 | 0.007444542 | 1.17969 | 0.021 |
| 15 | 2511.2 | 1.02868 | 0 | 0.006340715 | 1.20649 | 0.0244 |
| 15 | 2867.9 | 1.02669 | 0 | 0.005552035 | 1.23268 | 0.0288 |
| 15 | 3148.8 | 1.0252 | 0 | 0.005084565 | 1.25289 | 0.0331 |
| 15 | 3224.5 | 1.02481 | 0 | 0.004996978 | 1.25827 | 0.0338 |
| 15 | 3581.1 | 1.02301 | 0 | 0.004646403 | 1.28329 | 0.0369 |
| 15 | 3937.7 | 1.0213 | 0 | 0.004373422 | 1.30776 | 0.0398 |
| 15 | 4294.3 | 1.01967 | 0 | 0.004154734 | 1.3317 | 0.0424 |
| 15 | 4650.9 | 1.0181 | 0 | 0.003975404 | 1.35513 | 0.0449 |
| 15 | 5007.5 | 1.01661 | 0 | 0.003825383 | 1.37807 | 0.0472 |
| 15 | 5364.1 | 1.01518 | 0 | 0.00369782 | 1.40053 | 0.0494 |
| 15 | 5720.7 | 1.0138 | 0 | 0.003587886 | 1.42254 | 0.0514 |
| 15 | 6077.3 | 1.01248 | 0 | 0.003491933 | 1.4441 | 0.0533 |
| 15 | 6433.9 | 1.01121 | 0 | 0.003407377 | 1.46524 | 0.0552 |
| 15 | 6790.5 | 1.00999 | 0 | 0.003332198 | 1.48596 | 0.0569 |
| 15 | 7147.1 | 1.00882 | 0 | 0.003264823 | 1.50628 | 0.0586 |
| 15 | 7503.7 | 1.00768 | 0 | 0.003204017 | 1.5262 | 0.0602 |
| 15 | 7860.4 | 1.00659 | 0 | 0.003148882 | 1.54575 | 0.0618 |
| 15 | 8217 | 1.00554 | 0 | 0.003098519 | 1.56494 | 0.0634 |
| 371.6 | 371.6 | 1.13155 | 97.086 | 0.05002581 | 0.433547 | 0.0132 |
| 371.6 | 728.2 | 1.12522 | 97.086 | 0.024368407 | 0.452824 | 0.0141 |
| 371.6 | 1084.8 | 1.11943 | 97.086 | 0.015771472 | 0.47169 | 0.0151 |
| 371.6 | 1441.4 | 1.1141 | 97.086 | 0.011534372 | 0.490171 | 0.0165 |
| 371.6 | 1798 | 1.10916 | 97.086 | 0.009050481 | 0.508289 | 0.0184 |
| 371.6 | 2154.6 | 1.10458 | 97.086 | 0.007444542 | 0.526064 | 0.021 |
| 371.6 | 2511.2 | 1.10031 | 97.086 | 0.006340715 | 0.543512 | 0.0244 |
| 371.6 | 2867.9 | 1.09631 | 97.086 | 0.005552035 | 0.560647 | 0.0288 |
| 371.6 | 3148.8 | 1.09333 | 97.086 | 0.005084565 | 0.573934 | 0.0331 |
| 371.6 | 3224.5 | 1.09256 | 97.086 | 0.004996978 | 0.577481 | 0.0338 |
| 371.6 | 3581.1 | 1.08903 | 97.086 | 0.004646403 | 0.594028 | 0.0369 |
| 371.6 | 3937.7 | 1.08569 | 97.086 | 0.004373422 | 0.610296 | 0.0398 |
| 371.6 | 4294.3 | 1.08255 | 97.086 | 0.004154734 | 0.626295 | 0.0424 |
| 371.6 | 4650.9 | 1.07956 | 97.086 | 0.003975404 | 0.642034 | 0.0449 |
| 371.6 | 5007.5 | 1.07673 | 97.086 | 0.003825383 | 0.657521 | 0.0472 |
| 371.6 | 5364.1 | 1.07404 | 97.086 | 0.00369782 | 0.672763 | 0.0494 |
| 371.6 | 5720.7 | 1.07147 | 97.086 | 0.003587886 | 0.687766 | 0.0514 |
| 371.6 | 6077.3 | 1.06902 | 97.086 | 0.003491933 | 0.702538 | 0.0533 |
| 371.6 | 6433.9 | 1.06669 | 97.086 | 0.003407377 | 0.717085 | 0.0552 |
| 371.6 | 6790.5 | 1.06445 | 97.086 | 0.003332198 | 0.731412 | 0.0569 |
| 371.6 | 7147.1 | 1.06231 | 97.086 | 0.003264823 | 0.745526 | 0.0586 |
| 371.6 | 7503.7 | 1.06025 | 97.086 | 0.003204017 | 0.75943 | 0.0602 |
| 371.6 | 7860.4 | 1.05828 | 97.086 | 0.003148882 | 0.773131 | 0.0618 |
| 371.6 | 8217 | 1.05639 | 97.086 | 0.003098519 | 0.786632 | 0.0634 |
| 728.2 | 728.2 | 1.22439 | 241.205 | 0.024368407 | 0.344428 | 0.0141 |
| 728.2 | 1084.8 | 1.21608 | 241.205 | 0.015771472 | 0.360916 | 0.0151 |
| 728.2 | 1441.4 | 1.20853 | 241.205 | 0.011534372 | 0.37709 | 0.0165 |
| 728.2 | 1798 | 1.20162 | 241.205 | 0.009050481 | 0.392972 | 0.0184 |
| 728.2 | 2154.6 | 1.19526 | 241.205 | 0.007444542 | 0.408582 | 0.021 |
| 728.2 | 2511.2 | 1.18939 | 241.205 | 0.006340715 | 0.423935 | 0.0244 |

BLACK OIL MODEL PROPERTY TABLE

| | Saturation pressure [psi] | Pressure [psi] | Oil formation volume factor [RB / STB] | Solution gas-oil ratio [scf / STB] | Gas formation volume factor [res cf / scf] | Oil viscosity [cp] | Gas viscosity [cp] |
|---|---|---|---|---|---|---|---|
| 1 | 14.69 | 14.69 | 1.04627 | 0 | 1.33555 | 1.00416 | 0.0116 |
| 2 | 14.69 | 371.6 | 1.04326 | 0 | 0.050026 | 1.03535 | 0.0132 |
| 3 | 14.69 | 728.2 | 1.04044 | 0 | 0.024368 | 1.0657 | 0.0141 |
| 4 | 14.69 | 1084.8 | 1.0378 | 0 | 0.015771 | 1.09528 | 0.0151 |
| 5 | 14.69 | 1441.4 | 1.03532 | 0 | 0.011534 | 1.12411 | 0.0165 |
| 6 | 14.69 | 1798 | 1.03298 | 0 | 0.00905 | 1.15224 | 0.0184 |
| 7 | 14.69 | 2154.6 | 1.03077 | 0 | 0.007445 | 1.17969 | 0.021 |
| 8 | 14.69 | 2511.2 | 1.02868 | 0 | 0.006341 | 1.20649 | 0.0244 |
| 9 | 14.69 | 2867.9 | 1.02669 | 0 | 0.005552 | 1.23268 | 0.0288 |

## 13.6 Output definitions

For any questions on what various visualization parameters are showing (either line plot or 3D) please reference the <ResFrac simulation outputs.xlsx> in the resources > documentation folder of your ResFrac installation.

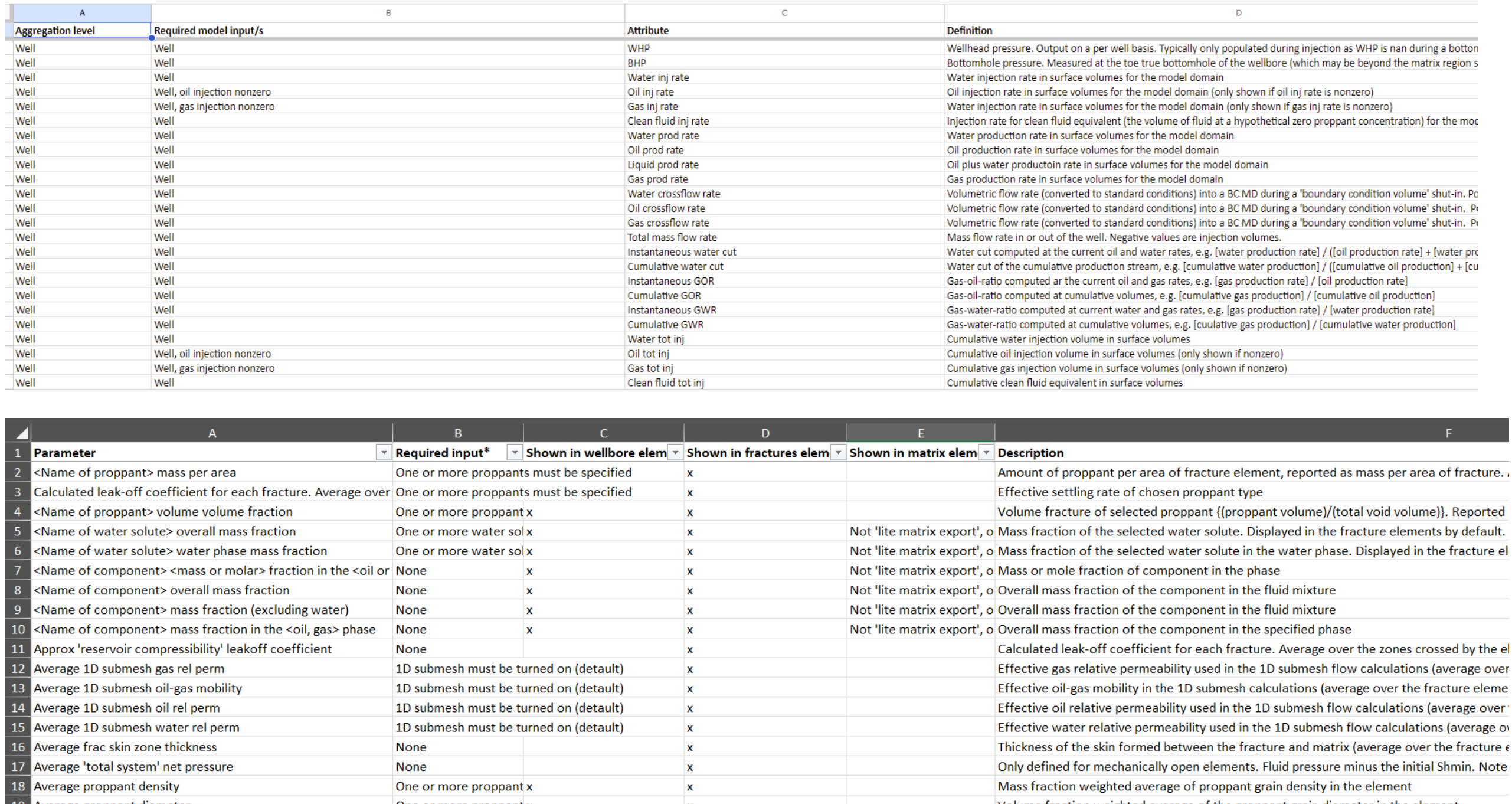

| Aggregation level | Required model input/s | Attribute | Definition |
|---|---|---|---|
| Well | Well | WHP | Wellhead pressure. Output on a per well basis. Typically only populated during injection as WHP is nan during a botton |
| Well | Well | BHP | Bottomhole pressure. Measured at the toe true bottomhole of the wellbore (which may be beyond the matrix region s |
| Well | Well | Water inj rate | Water injection rate in surface volumes for the model domain |
| Well | Well, oil injection nonzero | Oil inj rate | Oil injection rate in surface volumes for the model domain (only shown if oil inj rate is nonzero) |
| Well | Well, gas injection nonzero | Gas inj rate | Water injection rate in surface volumes for the model domain (only shown if gas inj rate is nonzero) |
| Well | Well | Clean fluid inj rate | Injection rate for clean fluid equivalent (the volume of fluid at a hypothetical zero proppant concentration) for the moc |
| Well | Well | Water prod rate | Water production rate in surface volumes for the model domain |
| Well | Well | Oil prod rate | Oil production rate in surface volumes for the model domain |
| Well | Well | Liquid prod rate | Oil plus water productoin rate in surface volumes for the model domain |
| Well | Well | Gas prod rate | Gas production rate in surface volumes for the model domain |
| Well | Well | Water crossflow rate | Volumetric flow rate (converted to standard conditions) into a BC MD during a 'boundary condition volume' shut-in. Pc |
| Well | Well | Oil crossflow rate | Volumetric flow rate (converted to standard conditions) into a BC MD during a 'boundary condition volume' shut-in. P |
| Well | Well | Gas crossflow rate | Volumetric flow rate (converted to standard conditions) into a BC MD during a 'boundary condition volume' shut-in. P |
| Well | Well | Total mass flow rate | Mass flow rate in or out of the well. Negative values are injection volumes. |
| Well | Well | Instantaneous water cut | Water cut computed at the current oil and water rates, e.g. [water production rate] / ([oil production rate] + [water prc |
| Well | Well | Cumulative water cut | Water cut of the cumulative production stream, e.g. [cumulative water production] / ([cumulative oil production] + [cu |
| Well | Well | Instantaneous GOR | Gas-oil-ratio computed ar the current oil and gas rates, e.g. [gas production rate] / [oil production rate] |
| Well | Well | Cumulative GOR | Gas-oil-ratio computed at cumulative volumes, e.g. [cumulative gas production] / [cumulative oil production] |
| Well | Well | Instantaneous GWR | Gas-water-ratio computed at current water and gas rates, e.g. [gas production rate] / [water production rate] |
| Well | Well | Cumulative GWR | Gas-water-ratio computed at cumulative volumes, e.g. [cuulative gas production] / [cumulative water production] |
| Well | Well | Water tot inj | Cumulative water injection volume in surface volumes |
| Well | Well, oil injection nonzero | Oil tot inj | Cumulative oil injection volume in surface volumes (only shown if nonzero) |
| Well | Well, gas injection nonzero | Gas tot inj | Cumulative gas injection volume in surface volumes (only shown if nonzero) |
| Well | Well | Clean fluid tot inj | Cumulative clean fluid equivalent in surface volumes |

| | Parameter | Required input* | Shown in wellbore elem | Shown in fractures elem | Shown in matrix elem | Description |
|---|---|---|---|---|---|---|
| 2 | <Name of proppant> mass per area | One or more proppants must be specified | | x | | Amount of proppant per area of fracture element, reported as mass per area of fracture. . |
| 3 | Calculated leak-off coefficient for each fracture. Average over | One or more proppants must be specified | | x | | Effective settling rate of chosen proppant type |
| 4 | <Name of proppant> volume volume fraction | One or more proppant | x | x | | Volume fracture of selected proppant {(proppant volume)/(total void volume)}. Reported |
| 5 | <Name of water solute> overall mass fraction | One or more water sol | x | x | Not 'lite matrix export', o | Mass fraction of the selected water solute. Displayed in the fracture elements by default. |
| 6 | <Name of water solute> water phase mass fraction | One or more water sol | x | x | Not 'lite matrix export', o | Mass fraction of the selected water solute in the water phase. Displayed in the fracture el |
| 7 | <Name of component> <mass or molar> fraction in the <oil or | None | x | x | Not 'lite matrix export', o | Mass or mole fraction of component in the phase |
| 8 | <Name of component> overall mass fraction | None | x | x | Not 'lite matrix export', o | Overall mass fraction of the component in the fluid mixture |
| 9 | <Name of component> mass fraction (excluding water) | None | x | x | Not 'lite matrix export', o | Overall mass fraction of the component in the fluid mixture |
| 10 | <Name of component> mass fraction in the <oil, gas> phase | None | x | x | Not 'lite matrix export', o | Overall mass fraction of the component in the specified phase |
| 11 | Approx 'reservoir compressibility' leakoff coefficient | None | | x | | Calculated leak-off coefficient for each fracture. Average over the zones crossed by the el |
| 12 | Average 1D submesh gas rel perm | 1D submesh must be turned on (detault) | | x | | Effective gas relative permeability used in the 1D submesh flow calculations (average over |
| 13 | Average 1D submesh oil-gas mobility | 1D submesh must be turned on (detault) | | x | | Effective oil-gas mobility in the 1D submesh calculations (average over the fracture eleme |
| 14 | Average 1D submesh oil rel perm | 1D submesh must be turned on (detault) | | x | | Effective oil relative permeability used in the 1D submesh flow calculations (average over |
| 15 | Average 1D submesh water rel perm | 1D submesh must be turned on (detault) | | x | | Effective water relative permeability used in the 1D submesh flow calculations (average o |
| 16 | Average frac skin zone thickness | None | | x | | Thickness of the skin formed between the fracture and matrix (average over the fracture e |
| 17 | Average 'total system' net pressure | None | | x | | Only defined for mechanically open elements. Fluid pressure minus the initial Shmin. Note |
| 18 | Average proppant density | One or more proppant | x | x | | Mass fraction weighted average of proppant grain density in the element |
| 19 | Average proppant diameter | One or more proppant | x | x | | Volume fraction weighted average of the proppant grain diameter in the element. |
| 20 | Average velocity | None | | x | | Shown in fracture elements. Average fluid velocity. |

This document details each of the sim_track (line plot) and 3D output variables. The document also lists the required attributes in order for the parameter to populate (for instance, if the output must be turned on in “Output options” in the simulation builder).

# 14. ResFracPro installation and configuration

## 14.1 Installing ResFracPro

ResFracPro should be installed on your local drive (usually your C drive). This will ensure the best performance.

If downloading the installer from an email link, you may receive a message warning you that the installer is an unrecognized app from the internet. If you press ‘More info’, you will be given an option to run the application.

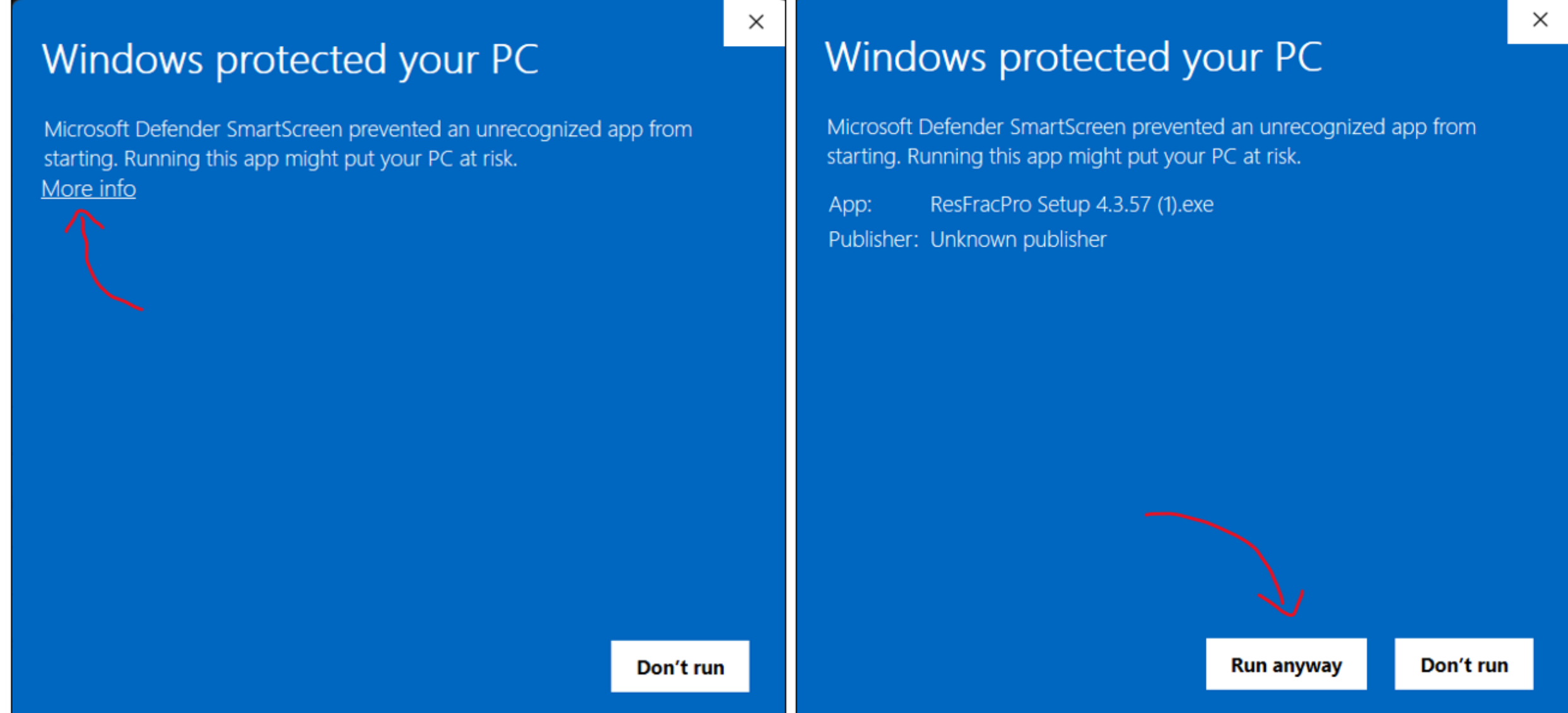

We recommend installing directly on your C drive:

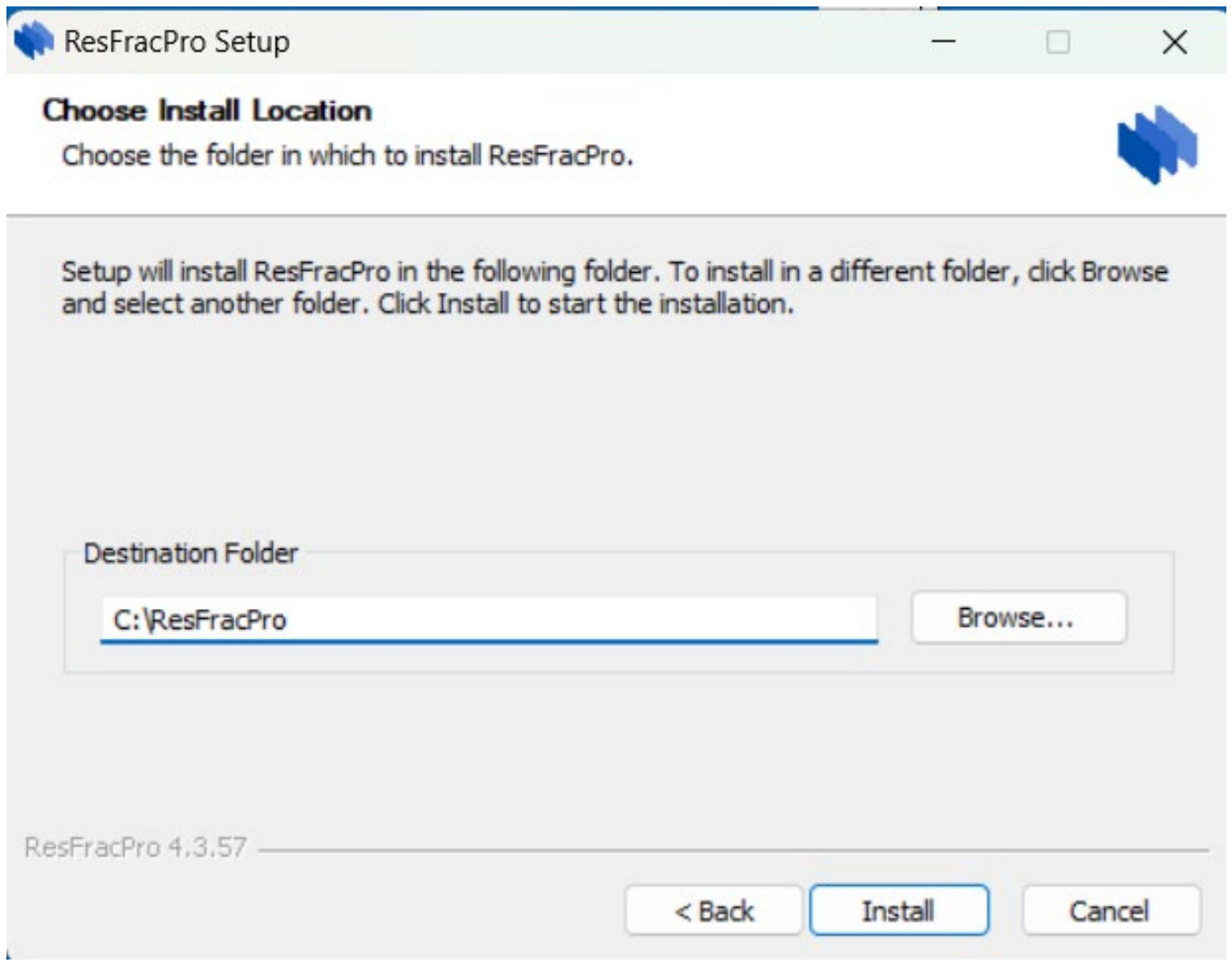

Once the installation is complete, you can open ResFracPro from within the ResFrac folder in the installation location:

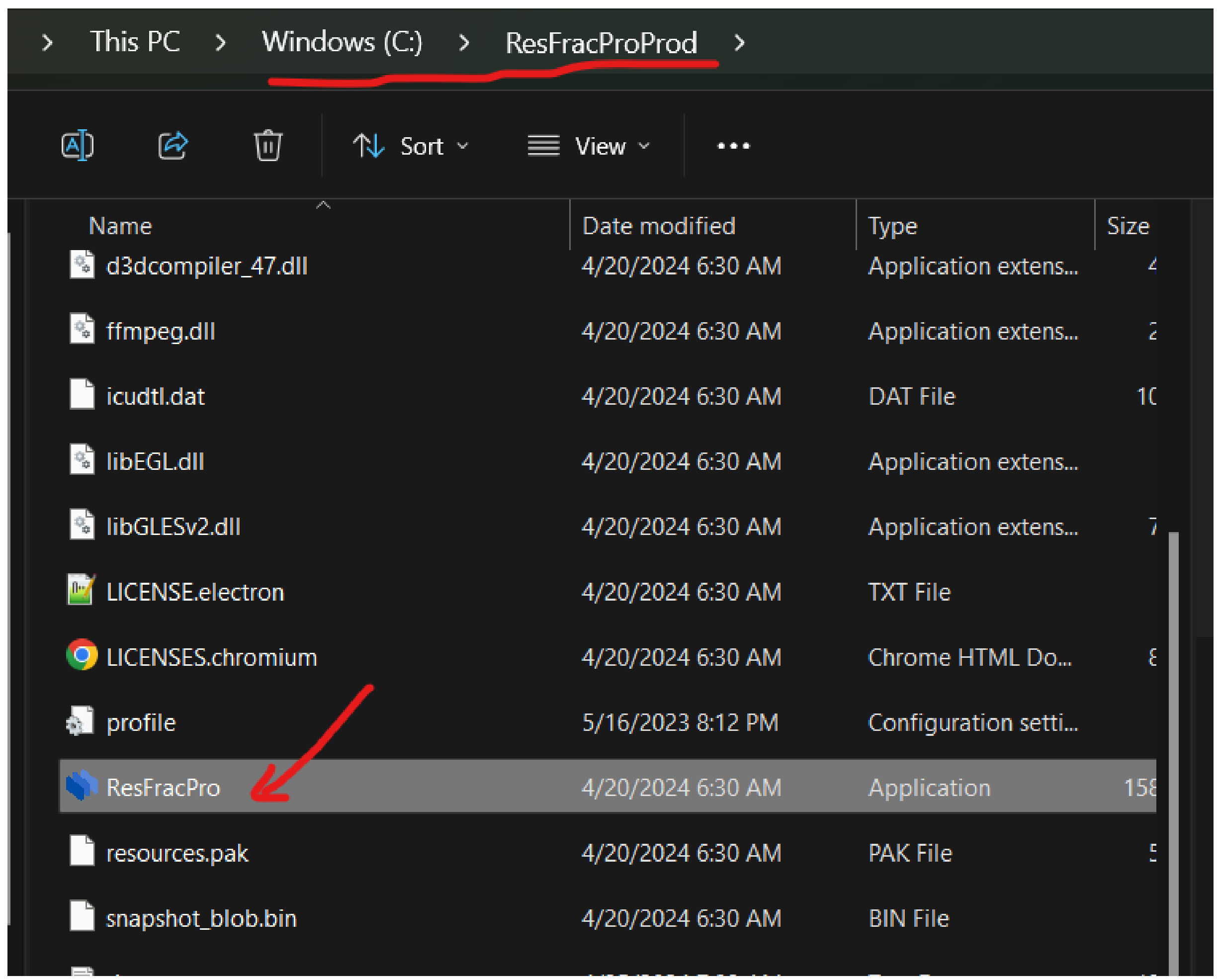

We recommend creating a shortcut to the application so that you can conveniently open in the future.

*Note: if you are installing an update, be sure to install over the same folder of the original installation. Update installers are designed to preserve the data of the existing installation.

## 14.2 Login options

All ResFrac users must login and authenticate their credentials prior to submitting or downloading simulations. For users who are already logged in, an automatic logout is triggered after three days, requiring the user to log in again.

Users may choose to authenticate using either Google or Microsoft. Both of these methods use the OAuth 2.0 protocol to give the user access to ResFrac.

### Microsoft authentication

1. ResFracPro connects to client organization Azure Active Directory using Application credentials (Application Name: resfraclogin)
2. "resfraclogin" application needs to be approved by Microsoft Entra ID (Azure AD) administrators for use by their users (one time approval and once approved for the first user, it works for all users in the same domain).

ResFracPro uses OAuth2 protocol to login using Microsoft Entra ID (formerly known as Azure AD). The ResFracPro application includes a client configuration that needs to be approved by Entra ID (Azure AD) administrators of the licensee organization. This approval is a one-time process and once approved for first user, the approval is in place for all other users of the licensee organization and they will usually be able to login without additional approvals from IT.

Here is an article that explains the admin consent workflow in Entra ID (Azure AD): https://learn.microsoft.com/en-us/azure/active-directory/manage-apps/configure-admin-consent-workflow

If you face any issues, we recommend connecting ResFrac Support with your Entra ID (Azure AD) team so that we can explain the requirement and provide them with necessary details of resfraclogin client application that needs to be approved.

### Google authentication

Google authentication is supported for personal and corporate (Google Workspace) accounts. Google's implementation of OAuth 2.0 is such that most clients can directly use their Google accounts to log in to ResFrac; the admin consent flow required with Microsoft is rarely encountered. If you do run into any problems, please reach out to ResFrac support.

## 14.3 Running ResFracPro with detailed logging for debugging

The ResFracPro user interface has a feature that causes the program to do detailed logging of everything it does. This way, if you have a problem, you can provide the log file to ResFrac so we can track down the issue. Turning on detailed logging will create a file called "logs.log" in the "logs" folder of the installed instance of ResFracPro. After you turn on detailed logging, leave the UI open for 15 minutes to accumulate some logs before sharing with ResFrac. The default path to this file is "C:\ResFracPro\logs\logs.log".

To turn on detailed logging, go to the gear icon in the job manager and open the settings. Switch over to the User Settings tab, and select the box for turn on detailed logging:

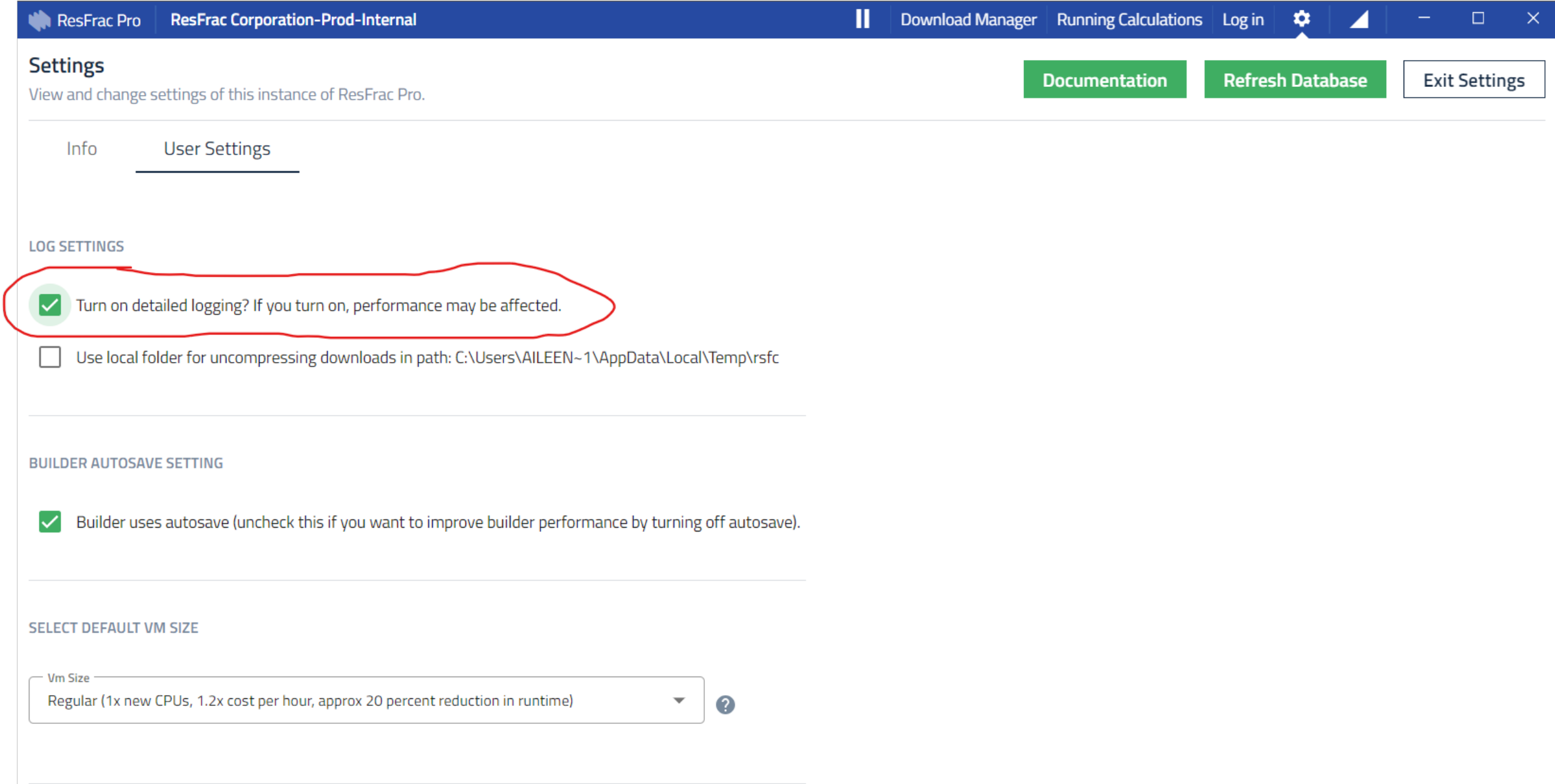

In most cases, this is all you need to do to turn on detailed logging. You can turn detailed logging back off by unchecking this box.

The log file that gets created will be placed at C:\ResFracPro\logs\logs.log. After turning on detailed logging, you typically want to do whatever action that you are having trouble with. If your problem is simulation status or file sync related, wait at least 15 minutes to accumulate some results before sharing the file with ResFrac support. If your problem is with logging in, you will want to do a couple full log out / log in cycles before sharing with ResFrac.

We do not turn on detailed logging by default, because detailed logging sometimes attracts the interest of the Windows antivirus software, causing ResFracPro to run very slowly. If you would like to send us a log, and you experience a slowdown in the UI after turning on detailed logging, you can use the steps below to tell the Windows antivirus software to back off on ResFracPro. These steps are not necessary but can be convenient if you want to continue using ResFracPro and are being hindered by the slowdown. Note that some companies prevent their users from modifying their security settings, so the below steps may or may not be available to you depending on how your company's IT configuration is set up.

Go to the settings for Windows Security:

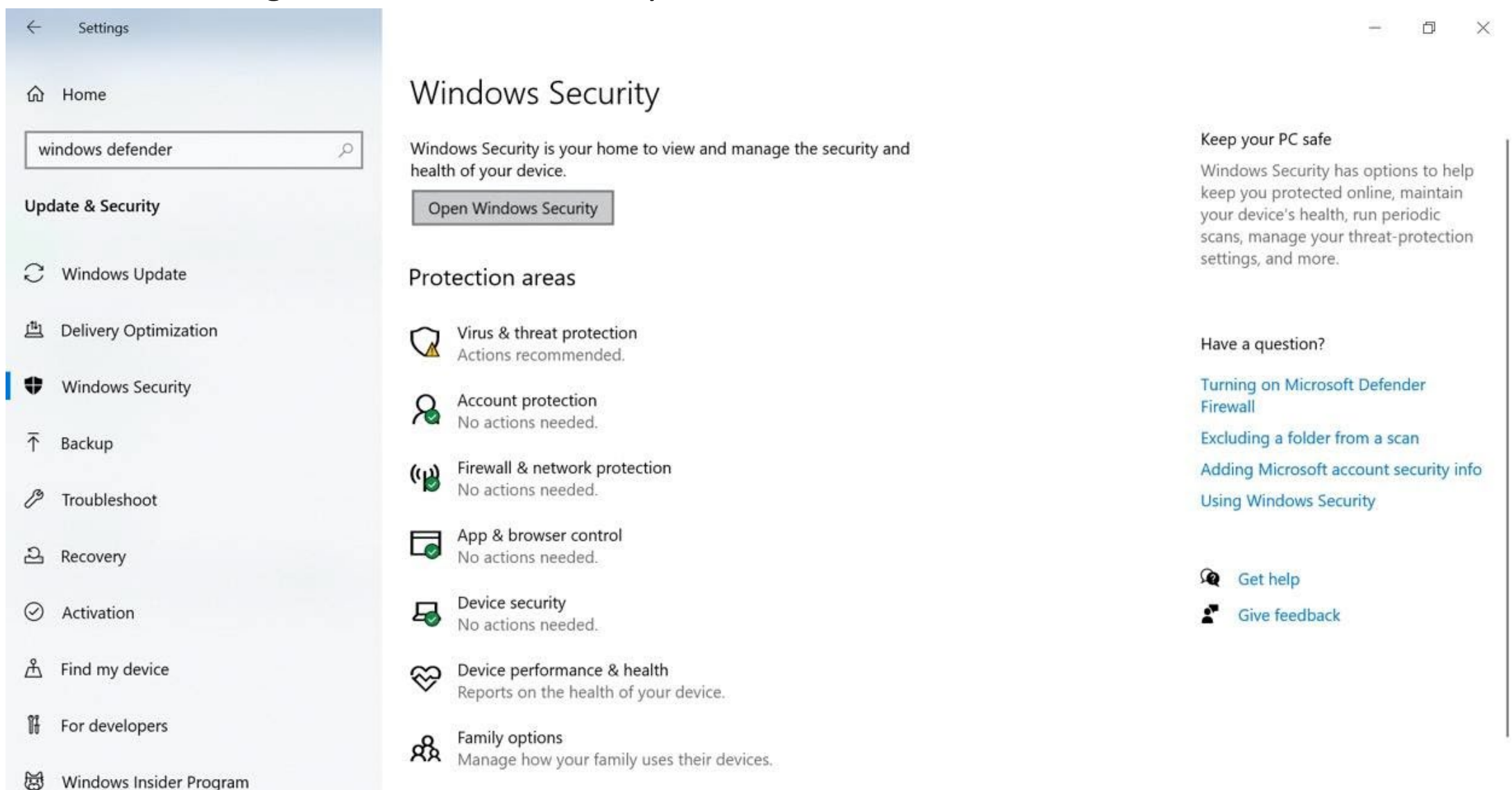

Go to Virus and threat protection, and click on manage settings:

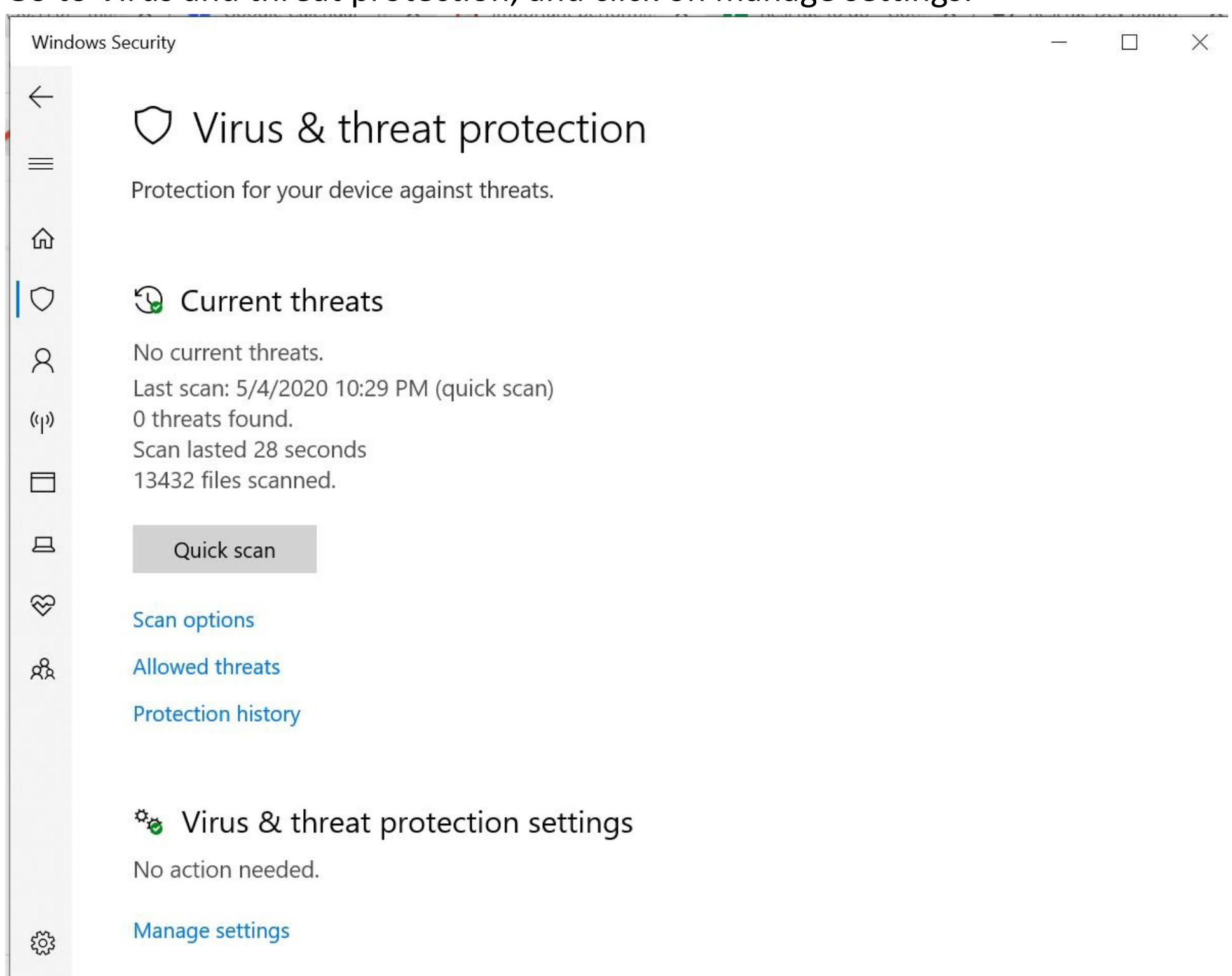

Scroll down and select exclusions:

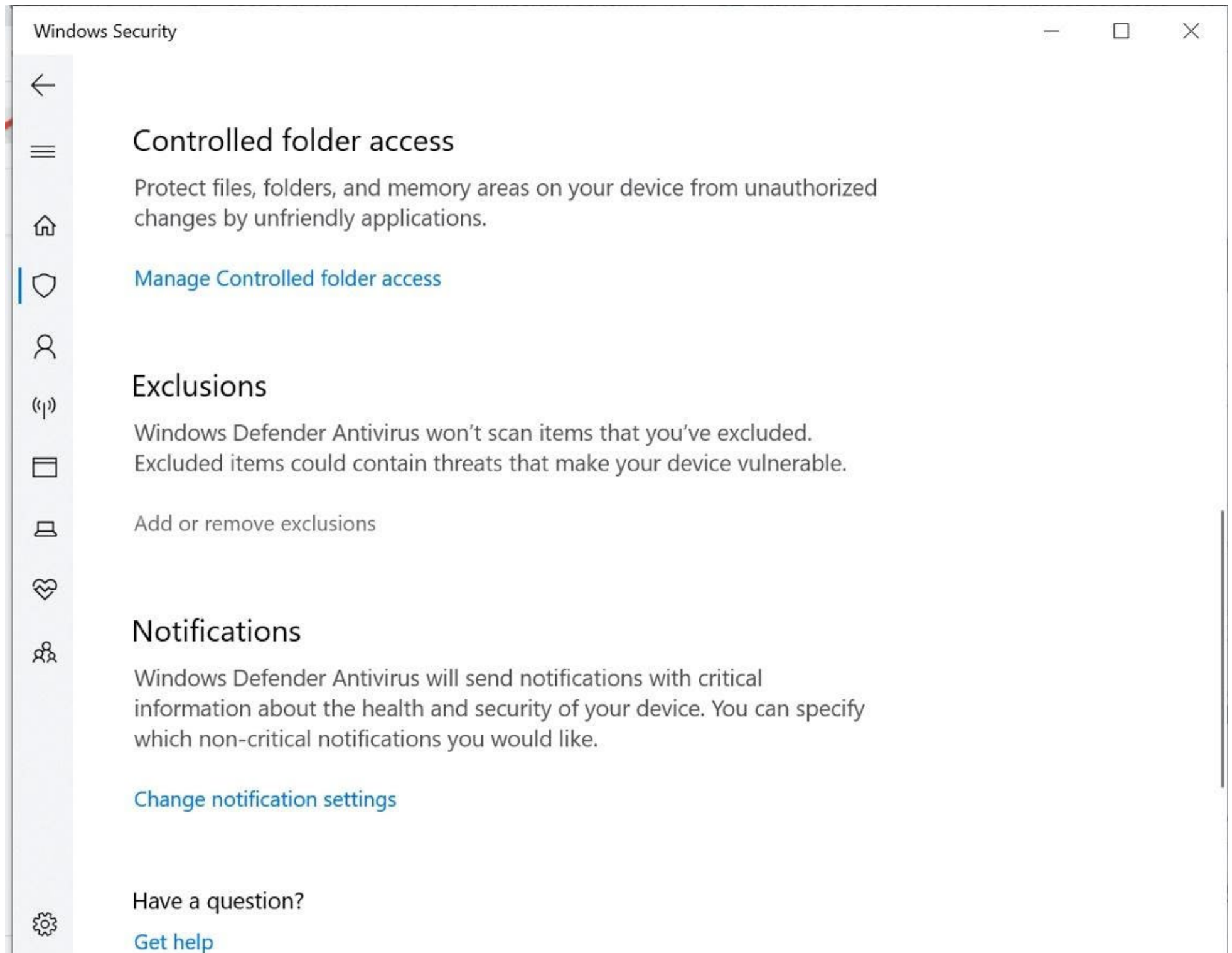

Add an exclusion for a process, and type in ResFracPro.exe

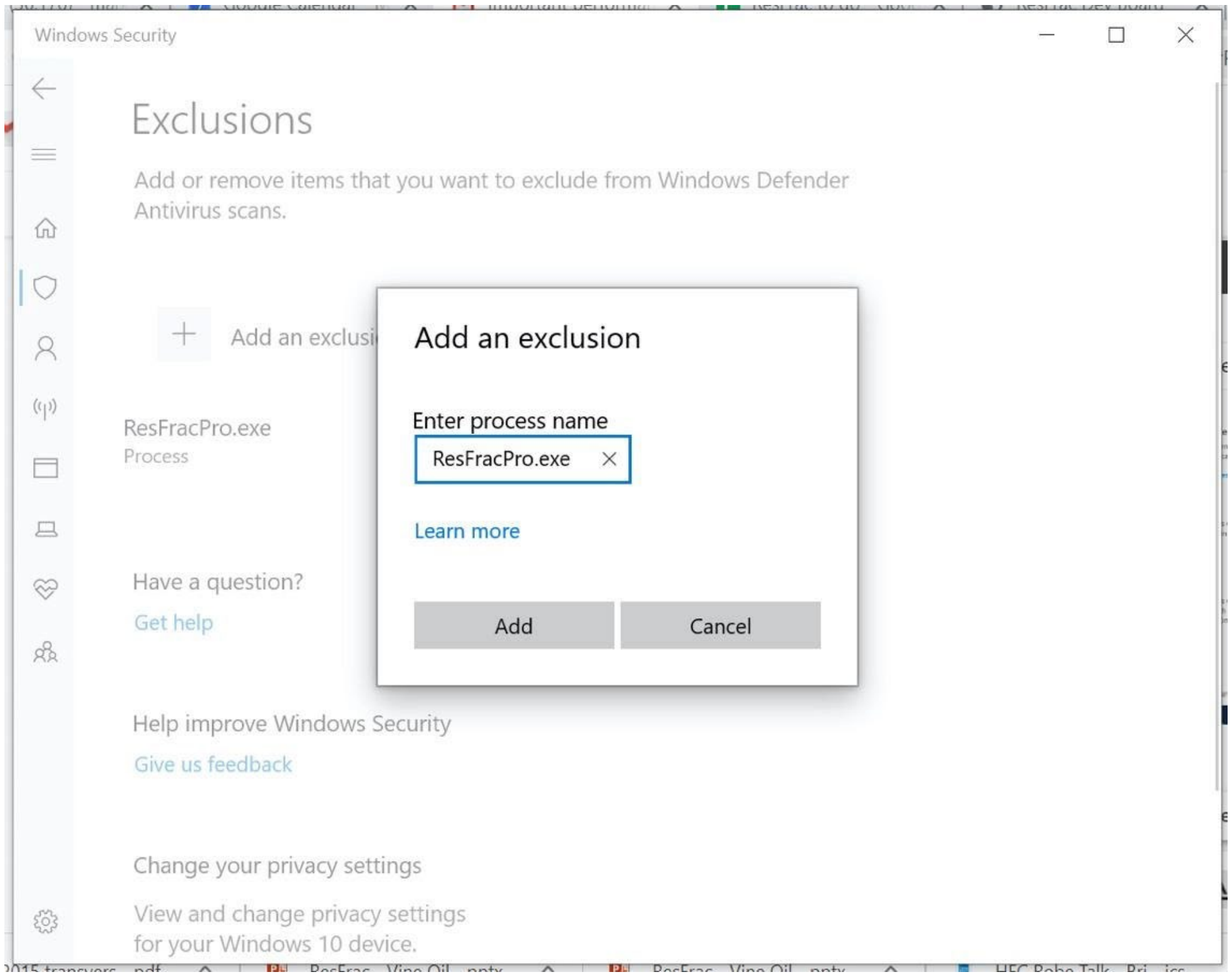

It should look like this:

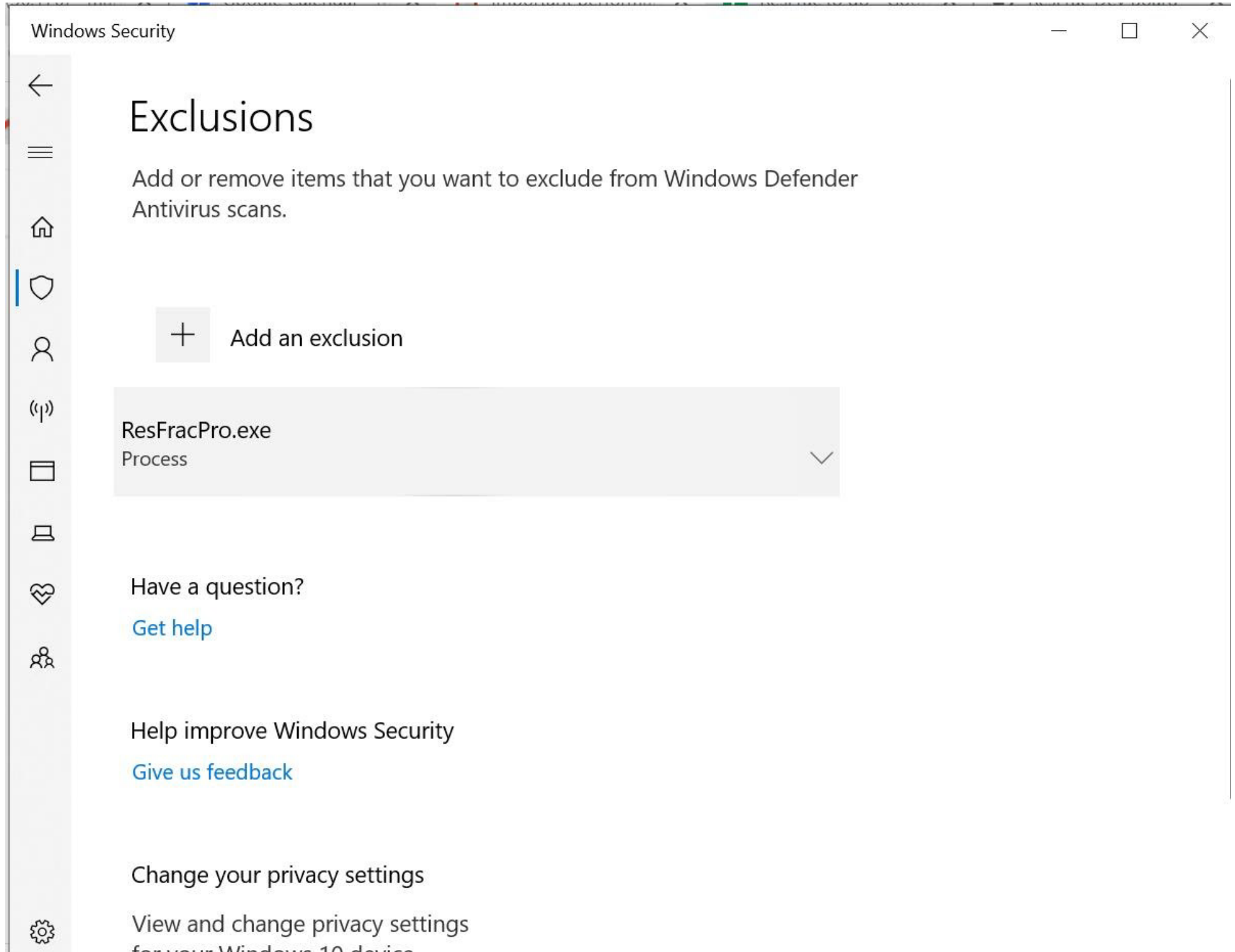

If the exclusion works properly, then ResFracPro will no longer feel sluggish. If the Windows antivirus is slowing down your machine, you will see a process in the Windows task manager called “Antimalware Service” with high CPU usage. If that happens, try repeating the steps above and make sure you have the exclusion on the “process” ResFracPro.exe.

## 14.4 Saving simulations to a network drive

ResFracPro is typically installed on the C drive of the client’s machine, and simulations are saved locally. Due to space constraints, it is possible to set the data folder location to be a network drive. This should only be done if the network drive is accessed over a high-speed connection. If the network connection is slow, ResFracPro will feel sluggish - particularly during operations within the visualization tool.

If you want to move the data folder to a network location, here is the recommended procedure:

1. Exit ResFracPro
2. Copy or move the data folder from the default location (C:\ResFracPro\data) to the new location (let's call it N:\data) to continue working with the same simulations in the new drive.
3. In profile.ini (in main ResFracPro folder), update the path for the data folder to the new location (N:\data).
4. Reopen ResFracPro. Check to see that everything shows up as expected.
5. (optional) Check the checkbox on the Settings screen > User Settings tab (accessed via the gear icon at the top right of the main window) for "Use local folder for uncompressing downloads."
6. (optional) Remove the original data folder.

In the procedure above, step 2 is only necessary if you want to continue working with the same simulations you have from before. If you do not copy the data folder to the new location, a new empty data folder will be created there and ResFracPro will display no simulations. The optional step 5 can help with performance in certain cases when the data folder is placed on a network location, depending on the details of how the client computer is set up.

If you're running out of space, a good alternative to moving the entire data file to a network location would be to periodically export old simulations to a network drive for archival purposes and delete them from the local data folder. Later, if you need to access old simulations, you can import them from the network location as required.

# Glossary

## Acronyms

BHP – Bottom hole pressure

DFIT – Diagnostic fracture injection test

DP – Display parameters

EGS – Enhanced geothermal systems

EOR – Enhanced oil recovery

GOR – Gas to oil ratio

IRR – Internal rate of return

ISIP – Initial shut-in pressure

NPV – Net present value

NW – Near-wellbore

PDP – Pressure dependent permeability

PM – Proxy model

PPG – Pounds per gallon concentration

QC – Quality control

ROI – Return on investment

RPI – Reciprocal productivity index

RTA – Rate transient analysis

TF – Target functions

UI – User interface

VFR – Volume-to-first-response

WC – Water cut

WHP – Wellhead pressure

## Parameter Abbreviations

E – Young's Modulus

fw - Fractional flow of water

k – permeability

k0 - ??

krg – relative permeability of gas

kro – relative permeability of oil

krw - relative permeability of water

K1c – Fracture toughness

Mu - Viscosity

Pb – Bubble point

Phi – Porosity

Rs - Solution gas-oil-ratio

Sg – Gas saturation

So – Oil saturation

Srg – Residual gas saturation

Sro – Residual oil saturation

Srw – Residual water saturation

Sw – Water saturation

Shmin – Minimum horizontal stress

SHmax – Maximum horizontal stress

Sv - Vertical stress